# Dynamic Programming

Volume I: Finite States

Thomas J. Sargent and John Stachurski

January 18, 2024

This book is dedicated to the memory of Robert E. Lucas, Jr.

# Contents

# Preface

This book is about dynamic programming and its applications in economics, finance, and adjacent fields. It brings together recent innovations in the theory of dynamic programming and provides applications and code that can help readers approach the research frontier. The book is aimed at graduate students and researchers, although most chapters are accessible to undergraduate students with solid quantitative backgrounds.

The book contains classical results on dynamic programming that are found in texts such as Bellman (1957), Denardo (1981), Puterman (2005), and Stokey and Lucas (1989), as well as extensions created by researchers and practitioners during the last few decades as they wrestled with how to formulate and solve dynamic models that can explain patterns observed in data. These extensions include recursive preferences, robust control, continuous time models, and time varying-discount rates. Such settings often fail to satisfy contraction-mapping restrictions on which traditional methods are based. To accommodate these applications, the key theoretical chapters of this book (Chapters 8–9) adopt and extend the abstract framework of Bertsekas (2022b). This approach provides great generality while also offering transparent proofs.

Chapters 1–3 provide motivation and background material on solving fixed point problems and computing lifetime valuations. Chapters 4 and 5 cover optimal stopping and Markov decision processes, respectively. Chapter 6 extends the Markov decision framework to settings where discount rates vary over time. Chapter 7 treats recursive preferences. The main theoretical results on dynamic programming from Chapters 4–6 are special cases of more general results in Chapters 8–9. A brief discussion of continuous time models can be found in Chapter 10.

Mathematically inclined readers with some background in dynamic programming might prefer to start with the general results in Chapters 8–9. Indeed, it is possible to read the text in reverse, jumping to Chapters 8–9 and then moving back to cover special cases according to interests. However, our teaching experience tells us that most students find the general results challenging on first pass, but considerably easier after they have practiced dynamic programming through the earlier chapters. This is why we have started the presentation with special cases and ended it with general results.

Instructors wishing to use this book as a text for undergraduate students can start with Chapter 1, skim through Chapter 2, cover Chapters 3–5 in depth, optionally include Chapter 6 and skip Chapters 7–10 entirely.

This book focuses on dynamic programs with finite state spaces, leaving more general settings to Volume II. Restricting attention to finite states involves some costs, since there are specific settings where continuous state models are simpler (one example being Gaussian linear-quadratic models). Moreover, many continuous state models allow us to unleash calculus, one of humanity's most useful inventions.

Nevertheless, finite state models are extremely useful. Computational representations are always implemented using finitely many floating point numbers, and many workhorse models in economics and finance are already discrete. In addition, focusing on problems with finite state spaces allows us to avoid using function-analytic and measure-theoretic machinery and imposing associated auxillary conditions required to ensure measurability and existence of extrema. Without these distractions, the core theory of dynamic programming is especially simple.

For these reasons, we believe that even for sophisticated readers, a good approach dynamic programming begins with a thorough analysis of the finite state case. This is the task that we have tackled in Volume I.

Computer code is a first-class citizen in this book. Code is written in Julia and can be found at

<div align="center">

https://github.com/QuantEcon/book-dp1

</div>

We chose Julia because it is open source and because Julia allows us to write computer code that is as close as possible to the relevant mathematical equations. Julia code in the text is written to maximize clarity rather than speed.

We have also written matching Python code, which can be found in the same repository. When combined with appropriate scientific libraries, Python is very practical and efficient for dynamic programming, but implementations tend to be library specific and are sometimes not as clean as those in Julia. That is why we chose Julia for programs embedded in the text.

We have tried to mix rigorous theory with exciting applications. Despite the various layers of abstractions used to unify the theory, the results are practical, being motivated by important optimization problems from economics and finance.

This book is one of several being written in partnership with the QuantEcon organization, with funding generously provided by Schmidt Futures (see acknowledgments below). There is some overlap with the first book in the series, Sargent and Stachurski (2023b), particularly on the topic of Markov chains. Although repetition is sometimes undesirable, we decided that some overlap would be useful, since it saves readers from having to jump between two documents.

We are greatly indebted to Jim Savage and Schmidt Futures for generous financial support, as well as to Shu Hu, Smit Lunagariya, Maanasee Sharma and Chien Yeh for outstanding research assistance. We are grateful to Makoto Nirei for hosting John Stachurski at the University of Tokyo in June and July 2022, where significant progress was made.

We also thank Alexis Akira Toda, Quentin Batista, Fernando Cirelli, Chase Coleman, Yihong Du, Ippei Fujiwara, Saya Ikegawa, Fazeleh Kazemian, Yuchao Li, Dawie van Lill, Qingyin Ma, Simon Mishricky, Pietro Monticone, Shinichi Nishiyama, Flint O'Neil, Zejin Shi, Akshay Shanker, Arnav Sood, Alexis Akira Toda, Natasha Watkins, Jingni Yang and Ziyue (Humphrey) Yang for many important fixes, comments and suggestions. Yuchao Li read the entire manuscript, from cover to cover, and his input and deep knowledge of dynamic programming helped us immensely. Jesse Perla provided insightful comments on our code.

# Common Symbols

| | |
|---|---|
| $\mathbb{1}\{P\}$ | indicator, equal to 1 if statement $P$ is true and 0 otherwise |
| $\alpha := 1$ | $\alpha$ is defined as equal to $1$ |
| $f \equiv 1$ | function $f$ is everywhere equal to $1$ |
| $\wp(A)$ | the power set of $A$; that is, the collection of all subsets of set $A$ |
| $[n]$ | $\{1, \ldots, n\}$ |
| $\mathbb{N}$, $\mathbb{Z}$, $\mathbb{R}$ and $\mathbb{C}$ | the natural, integer, real and complex numbers respectively |
| $\mathbb{Z}_+$, $\mathbb{R}_+$, etc. | the nonnegative elements of $\mathbb{Z}$, $\mathbb{R}$, etc. |
| $|x|$ for $x \in \mathbb{R}$ | the absolute value of $x$ |
| $|\lambda|$ for $\lambda \in \mathbb{C}$ | the modulus of $\lambda$ (i.e., $\sqrt{a^2 + b^2}$ when $\lambda = a + ib$) |
| $|B|$ for set $B$ | the cardinality of $B$ |
| $\mathbb{R}^n$ | all $n$-tuples of real numbers |
| $x \leqslant y$ $(x, y \in \mathbb{R}^n)$ | $x_i \leqslant y_i$ for $i = 1, \ldots n$ (pointwise partial order) |
| $x \ll y$ $(x, y \in \mathbb{R}^n)$ | $x_i < y_i$ for $i = 1, \ldots n$ |
| $\mathcal{D}(F)$ | the set of distributions on $F$ |
| $\mathbb{R}^M$ | the set of all functions from set M to $\mathbb{R}$ |
| $i\mathbb{R}^M$ | the set of increasing functions in $\mathbb{R}^M$ |
| $\mathcal{L}(\mathsf{X})$ | the set of linear operators on $\mathbb{R}^{\mathsf{X}}$ |
| $\mathcal{M}(\mathsf{X})$ | the set of Markov operators in $\mathcal{L}(\mathsf{X})$ (see §2.3.3.4) |
| $\langle a, b \rangle$ | the inner product of vectors $a$ and $b$ |
| $\vee_{\alpha \in A} u_\alpha$ | supremum of $\{u_\alpha\}_{\alpha \in A}$ |
| $\wedge_{\alpha \in A} u_\alpha$ | infimum of $\{u_\alpha\}_{\alpha \in A}$ |
| IID | independent and identically distributed |
| $X \stackrel{d}{=} Y$ | $X$ and $Y$ have the same distribution |
| $X \sim F$ | $X$ has distribution $F$ |
| $F \preceq_{\mathrm{F}} G$ | $G$ first order stochastically dominates $F$ |

# Common Abbreviations

| | |
|---|---|
| SDF | Stochastic discount factor (see §6.3.1.2 and §6.3.1.3) |
| MDP | Markov decision process (see §5.1.1) |
| VFI | Value function iteration (see §5.1.4.1) |
| HPI | Howard policy iteration (see §5.1.4.2) |
| OPI | Optimistic policy iteration (see §5.1.4.4) |
| RDP | Recursive decision process (see §8.1.1) |
| ADP | Abstract dynamic program (see §9.1.2) |
| SDP | Stable dynamic program (see §9.1.4.1) |

# Chapter 1

# Introduction

The temporal structure of a typical dynamic program is

---

1   an initial state $X_0$ is given
2   $t \leftarrow 0$
3   **while** $t < T$ **do**
4      the controller of the system observes the current **state** $X_t$
5      the controller chooses an **action** $A_t$
6      the controller receives a **reward** $R_t$ that depends on the current state and
       action
7      the state updates to $X_{t+1}$
8      $t \leftarrow t + 1$
9   **end**

---

The state $X_t$ is a vector listing current values of variables deemed relevant to choosing the current action. The action $A_t$ is a vector describing choices of a set of decision variables. If $T < \infty$ then the problem has a **finite horizon**. Otherwise it is an **infinite horizon** problem. Figure 1.1 illustrates the first two rounds of a dynamic program. As shown in the figure, a rule for updating the state depends on the current state and action.

Dynamic programming provides a way to maximize expected *lifetime* reward of a decision maker who receives a prospective reward sequence $(R_t)_{t \geqslant 0}$ and who confronts a system that maps today's state and control into next period's state. A **lifetime reward** is an aggregation of the individual period rewards $(R_t)_{t \geqslant 0}$ into a single value. An example of lifetime reward is an expected discounted sum $\mathbb{E} \sum_{t \geqslant 0} \beta^t R_t$ for some $\beta \in (0, 1)$.

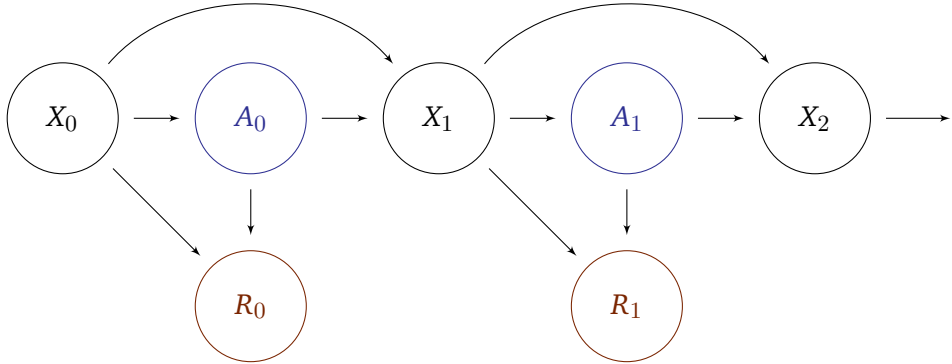

Figure 1.1: A dynamic program

**Example 1.0.1.** A manager wants to set prices and inventories to maximize a firm's **expected present value** (EPV), which, given interest rate $r$, is defined as

$$\mathbb{E}\left[\pi_0 + \frac{1}{1+r}\pi_1 + \left(\frac{1}{1+r}\right)^2 \pi_2 + \cdots\right]. \tag{1.1}$$

Here $X_t$ will be a vector that quantifies the size of the inventories, prices set by competitors and other factors relevant to profit maximization. The action $A_t$ sets current prices and orders of new stock. The current reward $R_t$ is current profit $\pi_t$, and the profit stream $(\pi_t)_{t \geqslant 0}$ is aggregated into a lifetime reward via (1.1).

Dynamic programming has a *vast* array of applications, from robotics and artificial intelligence to the sequencing of DNA. Dynamic programming is used every day to control aircraft, route shipping, test products, recommend information on media platforms and solve research problems. Some companies produce specialized computer chips that are designed for specific dynamic programs.

Within economics and finance, dynamic programming is applied to topics including unemployment, monetary policy, fiscal policy, asset pricing, firm investment, wealth dynamics, inventory control, commodity pricing, sovereign default, the division of labor, natural resource extraction, human capital accumulation, retirement decisions, portfolio choice, and dynamic pricing. We discuss some of these applications in chapters below.

The core theory of dynamic programming is relatively simple and concise. But implementation can be computationally demanding. That situation provides one of the major challenges facing the field of dynamic programming.

**Example 1.0.2.** To illustrate how computationally demanding problems can be, consider again Example 1.0.1. Suppose that, for each book, a book retailer chooses to hold between 0 and 10 copies. If there are 100 books to choose from, then the number of possible combinations for her inventories is $11^{100}$, about 20 orders of magnitude larger than the number of atoms in the known universe. In reality there are probably many more books to choose from, as well as other factors in the business environment that affect choices of a retailer.

In this book we discuss fundamental theory, traditional economic applications and recent applications with computationally demanding environments. We also cover recent trends towards more sophisticated specifications of lifetime rewards, often called recursive preferences. Throughout the book, theory and computation are combined, since, for interesting problems, brute-force computation is futile, while theory alone provides limited insight. The interplay between interesting applications, fundamental theory, computational methods and evolving hardware capability makes dynamic programming exciting.

## 1.1 Bellman Equations

In this section we introduce the recursive structure of dynamic programming in a simple setting. After solving a finite-horizon model, we consider an infinite-horizon version and explain how it produces a system of nonlinear equations. Then we turn to methods for solving such systems.

### 1.1.1 Finite-Horizon Job Search

We begin with a celebrated model of job search created by McCall (1970). McCall analyzed the decision problem of an unemployed worker in terms of current and prospective wage offers, impatience, and the availability of unemployment compensation. Here we study a simple version of the model in which essential ideas of dynamic programming are particularly clear.

Readers who are familiar with Bellman equations can skim this section quickly and proceed directly to §1.2.

#### 1.1.1.1 A Two-Period Problem

Imagine someone who begins her working life at time $t = 1$ without employment. While unemployed, she receives a new job offer paying wage $W_t$ at each date $t$. She can accept the offer and work *permanently* at that wage level or reject the offer, receive unemployment compensation $c$, and draw a new offer next period. We assume that the wage offer sequence is IID and nonnegative, with distribution $\varphi$. In particular,

- W $\subset \mathbb{R}_+$ is a finite set of possible wage outcomes and

- $\varphi \colon$ W $\to [0, 1]$ is a probability distribution on W, assigning a probability $\varphi(w)$ to each possible wage outcome $w$.

The worker is impatient. Impatience is parameterized by a time discount factor $\beta \in (0, 1)$, so that the present value of a next-period payoff of $y$ dollars is $\beta y$. Since $\beta < 1$, the worker will be tempted to accept reasonable offers, rather than to wait for better ones. A key question is how long to wait.

Suppose as a first step that working life is just two periods. To solve our problem we work backwards, starting at the final date $t = 2$, after $W_2$ has been observed.[1] If she is already employed, the worker has no decision to make: she continues working at her current wage. If she is unemployed, then she should take the largest of $c$ and $W_2$.

Now we step back to $t = 1$. At this time, having received offer $W_1$, the unemployed worker's options are (a) accept $W_1$ and receive it in both periods or (b) reject it, receive unemployment compensation $c$, and then, in the second period, choose the maximum of $W_2$ and $c$.

Let's assume that the worker seeks to maximize expected present value. The EPV of option (a) is $W_1 + \beta W_1$, which is also called the **stopping value**. The EPV of option (b), also called the **continuation value**, is $h_1 := c + \beta \, \mathbb{E} \max\{c, W_2\}$. More explicitly,

$$h_1 = c + \beta \sum_{w' \in \mathsf{W}} v_2(w')\varphi(w') \quad \text{where} \quad v_2(w) := \max\{c, w\}. \tag{1.2}$$

The optimal choice at $t = 1$ is now clear: accept the offer if $W_1 + \beta W_1 \geqslant h_1$ and reject otherwise. A decision tree is shown in Figure 1.2.

### 1.1.1.2 Comments on Information

In determining the optimal choice above, we assumed that the worker (a) cares about expected values and (b) knows how to compute them.

In Chapters 7–8 we discuss how to extend or weaken these assumptions. Some of these extensions allow decision makers to focus on measurements that differ from from expected values. Other extensions assume that the decision maker does not know underlying probability distributions. For now we put these issues aside and return to the set up discussed in the previous section.

---

[1] The procedure of solving the last period first and then working back in time is called **backward induction**. Starting with the last period makes sense because there is no future to consider.

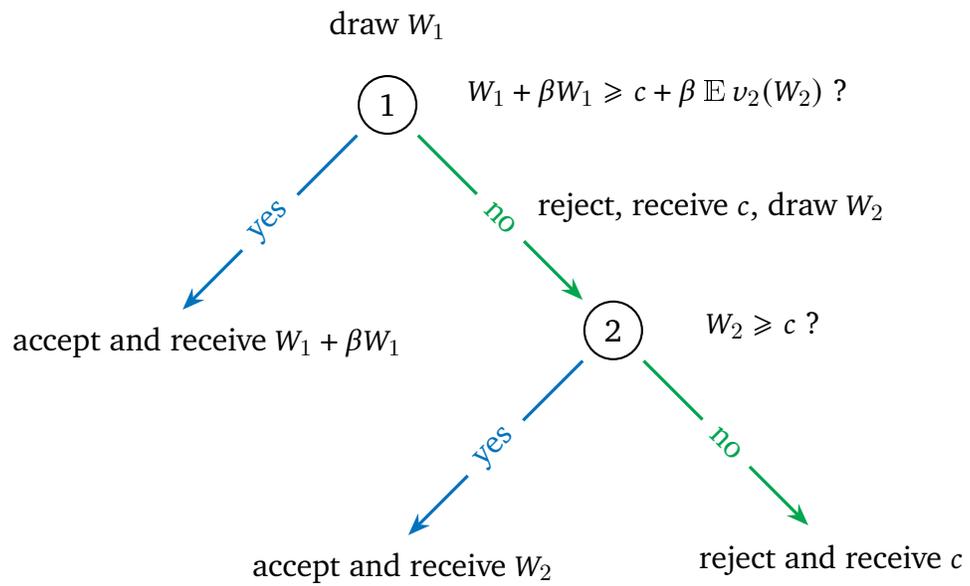

Figure 1.2: Decision tree for a two period problem

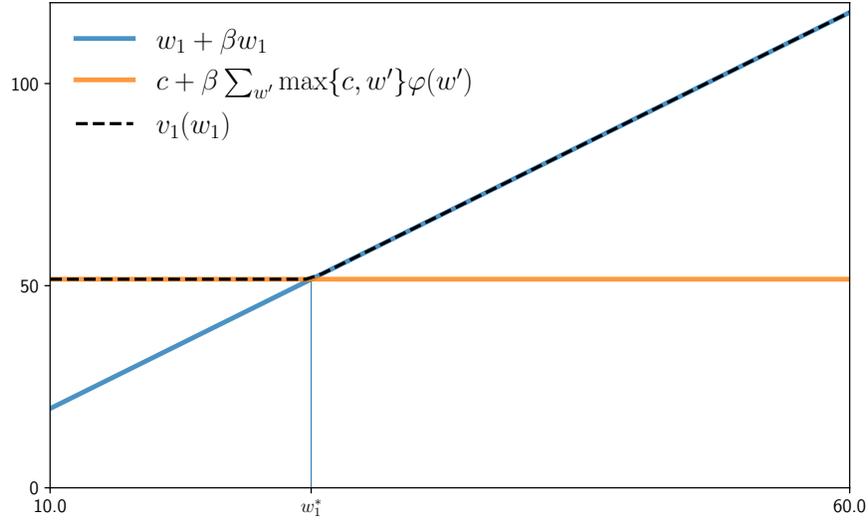

Figure 1.3: The value function $v_1$ and the reservation wage

### 1.1.1.3 Value Functions

A key idea in dynamic programming is to use "value functions" to track maximal lifetime rewards from a given state at a given time. The **time 2 value function** $v_2$ defined in (1.2) returns the maximum value obtained in the final stage for each possible realization of the time 2 wage offer. The **time 1 value function** $v_1$ evaluated at $w \in \mathsf{W}$ is

$$v_1(w) := \max\left\{ w + \beta w, \; c + \beta \sum_{w' \in \mathsf{W}} v_2(w')\varphi(w') \right\}. \tag{1.3}$$

It represents the present value of expected lifetime income after receiving the first offer $w$, conditional on choosing optimally in both periods.

The value function is shown in Figure 1.3. Figure 1.3 also shows the **reservation wage**

$$w_1^* := \frac{h_1}{1+\beta}. \tag{1.4}$$

It is the $w$ that solves the indifference condition

$$w + \beta w = c + \beta \sum_{w' \in W} v_2(w')\varphi(w'),$$

and equates the value of stopping to the value of continuing. For an offer $W_1$ above $w_1^*$, the stopping value exceeds the continuation value. For an offer below the reservation wage, the reverse is true. Hence, the optimal choice for the worker at $t = 1$ is completely described by the reservation wage.

Parameters and functions underlying the figure are shown in Listing 1.

Figure (1.4) is instructive. We can see that higher unemployment compensation $c$ shifts up the continuation value $h_1$ and increases the reservation wage. As a result, the worker will, on average, spend more time unemployed when unemployment compensation is higher.

EXERCISE 1.1.1. If unemployment compensation increases unemployment duration, should we conclude that increasing such compensation is detrimental to society? Provide some thoughts on this question in the context of the McCall model.

### 1.1.1.4  Three Periods

Now let's suppose that the worker works in period $t = 0$ as well as $t = 1, 2$. Figure 1.4 shows the decision tree for the three periods. Notice that the subtree containing nodes 1 and 2 is just the decision tree for the two-period problem in Figure 1.2. We will use this to find optimal actions.

At $t = 0$, the value of accepting the current offer $W_0$ is $W_0 + \beta W_0 + \beta^2 W_0$, while maximal value of rejecting and waiting, is $c$ plus, after discounting by $\beta$, the maximum value that can be obtained by behaving optimally from $t = 1$. We have already calculated this value: it is just $v_1(W_1)$, as given in (1.3)!

Maximal time zero value $v_0(w)$ is the maximum of the value of these two options, given $W_0 = w$, so we can write

$$v_0(w) = \max\left\{ w + \beta w + \beta^2 w, \, c + \beta \sum_{w' \in W} v_1(w')\varphi(w') \right\}. \tag{1.5}$$

By plugging $v_1$ from (1.3) into this expression, we can determine $v_0$, as well as the optimal action, the one that achieves the largest value in the max term in (1.5).

```julia
using Distributions

"Creates an instance of the job search model, stored as a NamedTuple."
function create_job_search_model(;
        n=50,           # wage grid size
        w_min=10.0,     # lowest wage
        w_max=60.0,     # highest wage
        a=200,          # wage distribution parameter
        b=100,          # wage distribution parameter
        β=0.96,         # discount factor
        c=10.0          # unemployment compensation
    )
    w_vals = collect(LinRange(w_min, w_max, n+1))
    φ = pdf(BetaBinomial(n, a, b))
    return (; n, w_vals, φ, β, c)
end

" Computes lifetime value at t=1 given current wage w_1 = w. "
function v_1(w, model)
    (; n, w_vals, φ, β, c) = model
    h_1 = c + β * max.(c, w_vals)'φ
    return max(w + β * w, h_1)
end

" Computes reservation wage at t=1. "
function res_wage(model)
    (; n, w_vals, φ, β, c) = model
    h_1 = c + β * max.(c, w_vals)'φ
    return h_1 / (1 + β)
end
```

Listing 1: Computing $v_1$ and $w_1^*$ (`two_period_job_search.jl`)

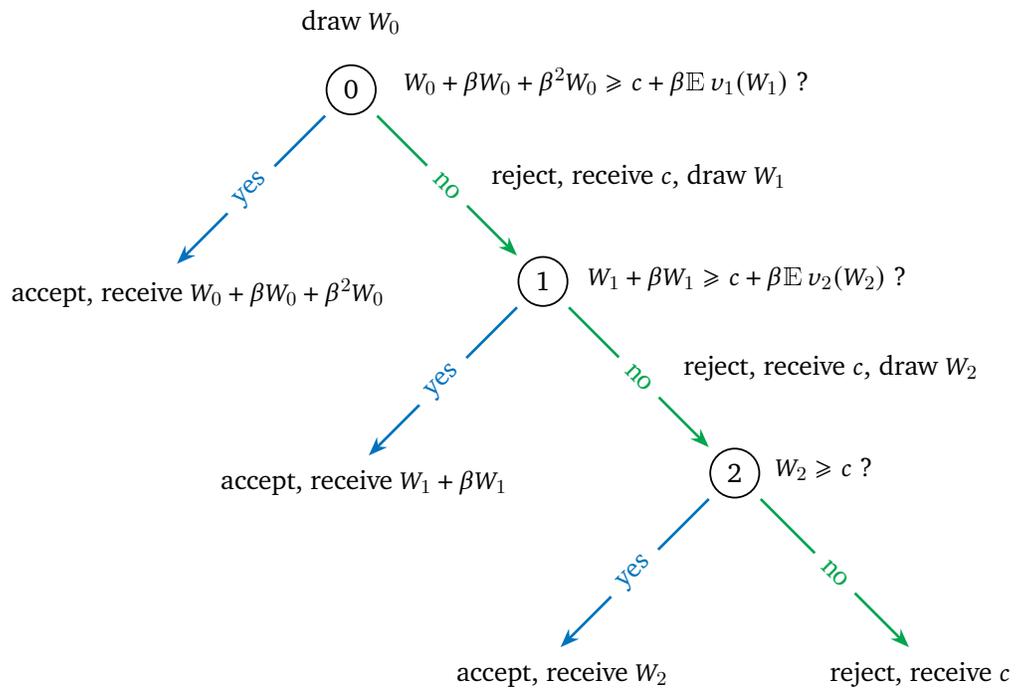

Figure 1.4: Decision tree for job seeker

Figure 1.4 illustrates how the backward induction process works. The last period value function $v_2$ is trivial to obtain. With $v_2$ in hand we can compute $v_1$. With $v_1$ in hand we can compute $v_0$. Once all the value functions are available, we can calculate whether to accept or reject at each point in time.

EXERCISE 1.1.2. The optimal action at time $t = 0$ is determined by a time zero reservation wage $w_0^*$. The worker should accept the time zero wage offer if and only if $W_0$ exceeds $w_0^*$. Calculate $w_0^*$ for this problem, by analogy with $w_1^*$ in (1.4).

Notice how we subdivided the three period problem down into a pair of two period problems, given by the two equations (1.3) and (1.5). Breaking many-period problems down into a sequence of two period problems is the essence of dynamic programming. The recursive relationships between $v_0$ and $v_1$ in (1.5), as well as between $v_1$ and $v_2$ in (1.3), are examples of what are called **Bellman equations**. We will see many other examples.

EXERCISE 1.1.3. Extend the above arguments to $T$ time periods, where $T$ can be any finite number. Using Julia or another programming language, write a function that takes $T$ as an argument and returns $(w_0^*, \ldots, w_T^*)$, the sequence of reservation wages for each period.

## 1.1.2 Infinite Horizon

Next we consider an infinite horizon problem that in some ways is more challenging but in other ways simpler. On one hand, the lack of a terminal period means that backward induction requires a subtler justification. On the other hand, the infinite horizon means that the worker always faces an infinite future, so that we only have to study a single value function and need not keep track of the number of remaining periods in the problem. This will become clearer as the section unfolds.[2]

With the above discussion in mind, let us consider a worker who aims to maximize

$$\mathbb{E} \sum_{t=0}^{\infty} \beta^t R_t, \tag{1.6}$$

where $R_t \in \{c, W_t\}$ is earnings at time $t$. As before, jobs are permanent, so accepting a job at a given wage means earning that wage in every subsequent period.

---

[2]Incidentally, imposing an infinite horizon is not the same as assuming humans live forever. Rather, it corresponds to the idea that humans have no specific "termination" date. More generally, we can understand an infinite horizon as an approximation to a finite horizon in which observations are recorded at relatively high frequency and no clear termination date exists.

Let's clarify our assumptions:

**Assumption 1.1.1.** The wage process satisfies $(W_t)_{t \geqslant 0} \overset{\text{IID}}{\sim} \varphi$ where $\varphi \in \mathcal{D}(W)$ and $W \subset \mathbb{R}_+$ is finite. The parameters $c$ and $\beta$ are positive and $\beta < 1$.

Here and below, for any finite or countable set $F$, the symbol $\mathcal{D}(F)$ indicates the set of distributions on $F$.

As with the finite state case, infinite-horizon dynamic programming involves a two step procedure that first assigns values to states and then deduces optimal actions given those values. We begin with an informal discussion and then formalize the main ideas.

To trade off current and future rewards optimally, we need to compare current payoffs we get from our two choices with the states that those choices lead to and the maximum value that can be extracted from those states. But how do we calculate the maximum value that can be extracted from each state when lifetime is infinite?

Consider first the present expected lifetime value of being employed with wage $w \in W$. This case is easy because, under the current assumptions, workers who accept a job are employed forever. Lifetime payoff is

$$w + \beta w + \beta^2 w + \cdots = \frac{w}{1 - \beta}. \tag{1.7}$$

How about maximum present expected lifetime value attainable when entering the current period unemployed with wage offer $w$ in hand? Denote this (as yet unknown) value by $v^*(w)$. We call $v^*$ the **value function**. While $v^*$ is not trivial to pin down, the task is not impossible. Our first step in the right direction is to observe that it satisfies the **Bellman equation**

$$v^*(w) = \max \left\{ \frac{w}{1 - \beta}, \, c + \beta \sum_{w' \in W} v^*(w')\varphi(w') \right\} \tag{1.8}$$

at every $w \in W$. (Here $w'$ is the offer next period.)

Our reasoning is as follows: The first term inside the max operation is the **stopping value**, or lifetime payoff from accepting current offer $w$. The second term inside the max operation is the **continuation value**, or current expected value of rejecting and behaving optimally thereafter. Maximal value is obtained by selecting the largest of these two alternatives.

Note the similarity between (1.8) and our finite horizon Bellman equations (1.3) and (1.5). The only real difference is that the value function is no longer time-dependent. This is because the worker always looks forward toward an infinite horizon, regardless of the current date.

Equation (1.8) is to be solved for a function $v^* \in \mathbb{R}^W$, the set of all functions from W to $\mathbb{R}$. Once we have solved for $v^*$ (assuming this is possible), optimal choices can be made by observing current $w$ and then choosing the largest of the two alternatives on the right-hand side of (1.8), just as we did in the finite horizon case. This idea – that optimal choices can be made by computing the value function and maximizing the right-hand side of the Bellman equation – is called **Bellman's principle of optimality**, and will be a cornerstone of what follows. Later we prove it in a general setting.

To solve for $v^*$, use fixed point theory, our topic in the next section. Later, in §1.3, we return to the job search problem and apply fixed point theory to solve for $v^*$.

## 1.2 Stability and Contractions

In this section we cover enough fixed point theory to solve an infinite horizon job search problem. (In Chapter 2 we consider more general results.) Readers who are familiar with the Neumann series lemma and Banach's fixed point theorem can skim this section and proceed to §1.3.

### 1.2.1 Vector Space

To begin, we recall some fundamental properties of real numbers, finite-dimensional vector space, basic topology and equivalence of norms.

#### 1.2.1.1 Real and Complex Vectors

For the most part, we are interested in vectors whose elements are real numbers (as distinguished from complex numbers). Before investigating such vectors, let's provide some useful language about the real line $\mathbb{R}$. (You might want to review some elementary concepts from real analysis in Appendix §A, such as suprema, infima, minima, maxima, and convergence.)

Given $a, b \in \mathbb{R}$, let $a \vee b := \max\{a, b\}$ and $a \wedge b := \min\{a, b\}$. The **absolute value** of $a \in \mathbb{R}$ is defined as $|a| := a \vee (-a)$.

A **real-valued vector** $u = (u_1, \ldots, u_n)$ is a finite real sequence with $u_i \in \mathbb{R}$ as the $i$-th element. The set of all real vectors of length $n$ is denoted by $\mathbb{R}^n$. The **inner product** of $n$-vectors $(u_1, \ldots, u_n)$ and $(v_1, \ldots, v_n)$ is $\langle u, v \rangle := \sum_{i=1}^n u_i v_i$.

The set $\mathbb{C}$ of complex numbers is defined in the appendix to Sargent and Stachurski (2023b) and many other places; as is the set $\mathbb{C}^n$ of all complex-valued $n$-vectors. We assume readers know what complex numbers are and how to compute the modulus of a complex number.

EXERCISE 1.2.1. Let $\alpha$, $s$ and $t$ be real numbers. Show that $\alpha \vee (s + t) \leqslant s + \alpha \vee t$ whenever $s \geqslant 0$.

### 1.2.1.2 Norms

The **Euclidean norm** on a real vector space is defined as

$$\|u\| := \sqrt{\langle u, u \rangle} \qquad (u \in \mathbb{R}^n).$$

Because they provide more flexibility when checking conditions that underlie various results, some alternative norms on $\mathbb{R}^n$ are important for applications of fixed point theory.

As a first step, recall that a function $\|\cdot\| \colon \mathbb{R}^n \to \mathbb{R}$ is called a **norm** on $\mathbb{R}^n$ if, for any $\alpha \in \mathbb{R}$ and $u, v \in \mathbb{R}^n$,

(a) $\|u\| \geqslant 0$                                (nonnegativity)

(b) $\|u\| = 0 \iff u = 0$                     (positive definiteness)

(c) $\|\alpha u\| = |\alpha| \|u\|$ and                    (absolute homogeneity)

(d) $\|u + v\| \leqslant \|u\| + \|v\|$                (triangle inequality)

The Euclidean norm on $\mathbb{R}^n$ satisfies the **Cauchy–Schwarz inequality**

$$|\langle u, v \rangle| \leqslant \|u\| \cdot \|v\| \quad \text{for all } u, v \in \mathbb{R}^n.$$

This inequality can be used to prove that the triangle inequality holds for the Euclidean norm (see, e.g., Kreyszig (1978)).

**Example 1.2.1.** The $\ell_1$ **norm** of a vector $u = (u_1, \ldots, u_n) \in \mathbb{R}^n$ is defined by

$$\|u\|_1 := \sum_{i=1}^{n} |u_i|. \tag{1.9}$$

In machine learning applications, $\|\cdot\|_1$ is sometimes called the "Manhattan norm," and $d_1(u, v) := \|u - v\|_1$ is called the "Manhattan distance" or "taxicab distance" between vectors $u$ and $v$. We will refer to it as the $\ell_1$ **distance** or $\ell_1$ **deviation**.

EXERCISE 1.2.2. Verify that the $\ell_1$ norm on $\mathbb{R}^n$ satisfies (a)–(d) above.

EXERCISE 1.2.3. Fix $p \in \mathbb{R}^n$ with $p_i > 0$ for all $i \in [n]$ and $\sum_i p_i = 1$. Show that $\|u\|_{1,p} := \sum_{i=1}^n |u_i| p_i$ is a norm on $\mathbb{R}^n$.

The $\ell_1$ norm and the Euclidean norm are special cases of the so-called **$\ell_p$ norm**, which is defined for $p \geqslant 1$ by

$$\|u\|_p := \left( \sum_{i=1}^n |u_i|^p \right)^{1/p}. \tag{1.10}$$

It can be shown that $u \mapsto \|u\|_p$ is a norm for all $p \geqslant 1$, as suggested by the name (see, e.g., Kreyszig (1978)). For this norm, the subadditivity asserted in (d) is called **Minkowski's inequality**.

Since the Euclidean case is obtained by setting $p = 2$, the Euclidean norm is also called the $\ell_2$ norm, and we write $\|\cdot\|_2$ rather than $\|\cdot\|$ when extra clarity is required.

EXERCISE 1.2.4. Prove that the **supremum norm** (or **$\ell_\infty$ norm**), defined by $\|u\|_\infty := \max_{i=1}^n |u_i|$, is also a norm on $\mathbb{R}^n$.

(The symbol $\|u\|_\infty$ is used because, for all $u \in \mathbb{R}^n$, we have $\|u\|_p \to \|u\|_\infty$ as $p \to \infty$.)

For the next exercise, we recall that the **indicator function** of logical statement $P$, denoted here by $\mathbb{1}\{P\}$, takes value 1 (resp., 0) if $P$ is true (resp., false). For example, if $x, y \in \mathbb{R}$, then

$$\mathbb{1}\{x \leqslant y\} = \begin{cases} 1 & \text{if } x \leqslant y \\ 0 & \text{otherwise.} \end{cases}$$

If $A \subset S$, where $S$ is any set, then $\mathbb{1}_A(x) := \mathbb{1}\{x \in A\}$ for all $x \in S$.

EXERCISE 1.2.5. The so-called $\ell_0$ "norm" $\|u\|_0 := \sum_{i=1}^n \mathbb{1}\{u_i \neq 0\}$ used in some data science applications is *not* a norm on $\mathbb{R}^n$. Prove this.

### 1.2.1.3   Equivalence of Vector Norms

An important property of a finite-dimensional normed vector space is that all norms are "equivalent." Let's review this result and discuss why it matters.

To begin, recall that when $u$ and $(u_m) := (u_m)_{m \in \mathbb{N}}$ are all elements of $\mathbb{R}^n$, we say that $(u_m)$ **converges** to $u$ and write $u_m \to u$ if

$$\|u_m - u\| \to 0 \text{ as } m \to \infty \text{ for some norm } \| \cdot \| \text{ on } \mathbb{R}^n.$$

It might seem that this definition is imprecise. Don't we need to clarify that the convergence is with respect to a particular norm?

No we don't. This is because any two norms $\| \cdot \|_a$ and $\| \cdot \|_b$ on $\mathbb{R}^n$ are **equivalent** in the sense that there exist finite positive constants $M, N$ such that

$$M\|u\|_a \leqslant \|u\|_b \leqslant N\|u\|_a \quad \text{for all } u \in \mathbb{R}^n. \tag{1.11}$$

(See, e.g., Kreyszig (1978).)

EXERCISE 1.2.6. Let us write $\| \cdot \|_a \sim \| \cdot \|_b$ if there exist finite $M, N$ such that (1.11) holds. Prove that $\sim$ is an equivalence relation (see §A.1) on the set of all norms on $\mathbb{R}^n$.

EXERCISE 1.2.7. Let $\| \cdot \|_a$ and $\| \cdot \|_b$ be any two norms on $\mathbb{R}^n$. Given a point $u$ in $\mathbb{R}^n$ and a sequence $(u_m)$ in $\mathbb{R}^n$, use (1.11) to confirm that $\|u_m - u\|_a \to 0$ implies $\|u_m - u\|_b \to 0$ as $m \to \infty$.

The next exercise tells us that pointwise convergence and norm convergence are the same thing in finite dimensions.

EXERCISE 1.2.8. Let $\| \cdot \|$ be any norm on $\mathbb{R}^n$. Fixing a point $u$ in $\mathbb{R}^n$ and a sequence $(u_m)$ in $\mathbb{R}^n$, let $u^i$ and $u_m^i$ be the $i$-th component of $u$ and $u_m$ respectively. Show that $u_m^i \to u^i$ for all $i \in \{1, \ldots, n\}$ if and only if $\|u_m - u\| \to 0$.

Recall that a set $C \subset \mathbb{R}^n$ is called **bounded** if there exists an $M \in \mathbb{N}$ with $\|x\| \leqslant M$ for all $x \in C$; and **closed** in $\mathbb{R}^n$ if, for all $u \in \mathbb{R}^n$ and sequences $(u_m) \subset C$ such that $u_m \to u$ as $m \to \infty$, we also have $u \in C$. A set $G \subset \mathbb{R}^n$ is called **open** in $\mathbb{R}^n$ if $G^c$ is closed in $\mathbb{R}^n$. A set $N$ is called a **neighborhood** of $u \in \mathbb{R}^n$ if there exists an open set $G \subset \mathbb{R}^n$ with $u \in G \subset N$. A map $T$ from $U \subset \mathbb{R}^n$ to $\mathbb{R}^k$ is called **continuous at** $u \in U$ if $Tu_m \to Tu$ for any $(u_m) \subset U$ with $u_m \to u$; and **continuous** if $T$ is continuous at every $u \in U$. These notions apply for any norm, since convergence does not depend on a choice of norm.

### 1.2.1.4   Matrices and Neumann Series

Next we discuss geometric series in matrix space, along with the Neumann series lemma, one of many useful results in applied and numerical analysis.

Before starting we recall that if $A = (a_{ij})$ is an $n \times n$ matrix with $i, j$-th element $a_{ij}$, then the definition of matrix multiplication tells us that for $u \in \mathbb{R}^n$, the $i$-th element of $Au$ is $\sum_{j=1}^n a_{ij} u_j$, while the $j$-th element of $u^\top A$ is $\sum_{i=1}^n a_{ij} u_i$. Think of $u \mapsto Au$ and $u \mapsto u^\top A$ is two different mappings, each of which takes an $n$-vector and produces a new $n$-vector.

**Remark 1.2.1.** In this book, we adopt a convention that a vector in $\mathbb{R}^n$ is just an $n$-tuple of real values. This coincides with the viewpoint of languages like Julia and Python: vectors are just "flat" arrays. But when we use vectors in matrix algebra, they should be understood as column vectors unless we state otherwise.

Just as we considered norms of vectors in §1.2.1.2, we will find it helpful to have a notion of norms of matrices. A real-valued map defined on $\mathbb{R}^{n \times n}$, the set of real $n \times n$ matrices, is called a **matrix norm** if it has the following properties: for any $\alpha \in \mathbb{R}$ and any $n \times n$ matrices $A, B$,

(a) $\|A\| \geqslant 0$,

(b) $\|A\| = 0 \iff A = 0$,

(c) $\|\alpha A\| = |\alpha| \|A\|$,

(d) $\|A + B\| \leqslant \|A\| + \|B\|$, and

These are called nonnegativity, positive definiteness, absolute homogeneity and the triangle inequality, analogous to the norms on $\mathbb{R}^n$ discussed in §1.2.1.2.

An example of a matrix norm is the so-called **operator norm**

$$\|B\|_o := \max_{\|u\|=1} \|Bu\|. \tag{1.12}$$

Here $B$ is $n \times n$, $u$ is in $\mathbb{R}^n$ and the norm on the right-hand side is the Euclidean norm over the $n$-vector $Bu$. Another example of a matrix norm is the supremum norm defined as

$$\|B\|_\infty := \max_{1 \leqslant i, j \leqslant n} |b_{ij}|, \quad \text{where } b_{ij} \text{ is the } i, j\text{-th element of } B. \tag{1.13}$$

Some matrix norms have the **submultiplicative** property, which means that, for all $A, B \in \mathbb{R}^{n \times n}$, we have $\|AB\| \leqslant \|A\| \|B\|$.

EXERCISE 1.2.9. Show that the operator $\|\cdot\|_o$ is submultiplicative on $\mathbb{R}^{n \times n}$. Provide a counterexample to the claim that $\|\cdot\|_\infty$ is submultiplicative.

In what follows we often use the operator norm as our choice of matrix norm (partly because of its attractive submultiplicative property). Hence, by convention, an expression such as $\|A\|$ refers to the operator norm $\|A\|_o$ of $A$.

Analogous to the vector case, we say that a sequence $(A_k)$ of $n \times n$ matrices converges to an $n \times n$ matrix $A$ and write $A_k \to A$ if $\|A_k - A\| \to 0$ as $k \to \infty$. Just as with vectors, this form of norm convergence holds if and only if each element of $A_k$ converges to the corresponding element of $A$. The proof is similar to the solution to Exercise 1.2.8.

If $A$ is an $n \times n$ matrix, then $\lambda \in \mathbb{C}$ is called an **eigenvalue** of $A$ if there exists a nonzero $e \in \mathbb{C}^n$ such that $Ae = \lambda e$. (Here $\mathbb{C}$ is the set of complex numbers and $\mathbb{C}^n$ is the set of complex $n$-vectors.) A vector $e$ satisfying this equality is called an **eigenvector** of $A$ and $(\lambda, e)$ is called an **eigenpair**.

In Julia, we can compute the eigenvalues of a square matrix $A$ via `eigvals(A)`. The code

```julia
using LinearAlgebra
A = [0 -1;
     1  0]
println(eigvals(A))
```

produces

```
2-element Vector{ComplexF64}:
 0.0 - 1.0im
 0.0 + 1.0im
```

Here `im` stands for $i$, the imaginary unit, so the eigenvalues of $A$ are $-i$ and $i$.

Turning to geometric series, let us begin in one dimension. Consider the one-dimensional linear equation $u = au + b$, where $a, b$ are given and $u$ is unknown. Its solution $u^*$ satisfies

$$|a| < 1 \quad \implies \quad u^* = \frac{b}{1-a} = \sum_{k \geqslant 0} a^k b. \tag{1.14}$$

This scalar result extends naturally to vectors. To show this we suppose that $u$ and $b$ are column vectors in $\mathbb{R}^n$, and that $A$ is an $n \times n$ matrix. We consider the vector equation $u = Au + b$. For the next result, we recall that the **spectral radius** of $A$ is defined as

$$\rho(A) := \max\{|\lambda| : \lambda \text{ is an eigenvalue of } A\} \tag{1.15}$$

Here $|\lambda|$ indicates the modulus of complex number $\lambda$.

With $I$ as the $n \times n$ identity matrix, we can state the following result.

**Theorem 1.2.1** (Neumann Series Lemma). *If $\rho(A) < 1$, then $I - A$ is nonsingular and*

$$(I - A)^{-1} = \sum_{k \geqslant 0} A^k.$$

It follows directly that the vector system $u = Au + b$ has a unique solution $u^* = (I - A)^{-1}b = \sum_{k \geqslant 0} A^k b$ whenever $\rho(A) < 1$. This is the multivariate extension of (1.14).

The code in Listing 2 shows how to compute the spectral radius of an arbitrary matrix $A$ in Julia. The print statement produces `0.5828`, so, for this matrix, $\rho(A) < 1$.

```julia
using LinearAlgebra
ρ(A) = maximum(abs(λ) for λ in eigvals(A))   # Spectral radius
A = [0.4 0.1;                                 # Test with arbitrary A
     0.7 0.2]
print(ρ(A))
```

Listing 2: Computing a spectral radius (`compute_spec_rad.jl`)

EXERCISE 1.2.10. Prove that $\rho(\alpha B) = |\alpha| \, \rho(B)$ for all $\alpha \in \mathbb{R}$.

The rest of this section works through the proof of the Neumann series lemma, with several parts left as exercises. An informal proof of the lemma runs as follows. If $S := \sum_{k \geqslant 0} A^k$, then

$$I + AS = I + A \sum_{k \geqslant 0} A^k = I + A + A^2 + \cdots = S.$$

Rearranging $I + AS = S$ gives $S = (I - A)^{-1}$, which matches the claim in the Neumann series lemma.

This informal argument lacks rigor. To make it rigorous, we must prove (a) that the sum $\sum_{k \geqslant 0} A^k$ converges and (b) that the matrix $I - A$ is invertible.

**Lemma 1.2.2.** *If B is any square matrix and $\|\cdot\|$ is any matrix norm, then*

$$\rho(B)^k \leqslant \|B^k\| \text{ for all } k \in \mathbb{N} \qquad and \qquad \|B^k\|^{1/k} \to \rho(B) \text{ as } k \to \infty.$$

A proof of Lemma 1.2.2 can be found in Chapter 12 of Bollobás (1999). The second result is sometimes called **Gelfand's formula**.

EXERCISE 1.2.11. Using Lemma 1.2.2, show that

(i) $\|B^k\| \to 0$ as $k \to \infty$ if and only if $\rho(B) < 1$.

(ii) $\rho(B) > 1$ implies $\|B^k\| \to \infty$ as $k \to \infty$.

EXERCISE 1.2.12. Prove: If $A$ and $B$ are square matrices that commute (i.e., $AB = BA$), then $\rho(AB) \leqslant \rho(A)\rho(B)$. [Hint: Show $(AB)^k = A^k B^k$ and use Gelfand's formula.]

EXERCISE 1.2.13. Prove: $\rho(A) < 1$ implies that the series $\sum_{k \geqslant 0} A^k$ converges, in the sense that every element of the matrix $S_K := \sum_{k=0}^{K} A^k$ converges as $K \to \infty$.

From this last result, one can show that $(I - A)^{-1}$ exists by computing it:

EXERCISE 1.2.14. Prove this claim by showing that, when $\sum_{k \geqslant 0} A^k$ exists, the inverse of $I - A$ exists and indeed $(I - A)^{-1} = \sum_{k \geqslant 0} A^k$.[3]

Listing 3 helps illustrate the result in Exercise 1.2.14, although we truncate the infinite sum $\sum_{k \geqslant 0} A^k$ at 50.

The output `5.621e-12` is close enough to zero for many practical purposes.

## 1.2.2 Nonlinear Systems

While the Neumann series lemma is a powerful tool for solving linear systems, it doesn't help us with *non*linear problems. In this section, we present Banach's fixed point theorem, one of a variety of techniques for handling nonlinear systems. (Chapter 2 introduces other methods.)

---

[3]Hint: To prove that $A$ is invertible and $B = A^{-1}$, it suffices to show that $AB = I$.

```
3   # Primitives
4   A = [0.4 0.1;
5        0.7 0.2]
6
7   # Method one: direct inverse
8   B_inverse = inv(I - A)
9
10  # Method two: power series
11  function power_series(A)
12      B_sum = zeros((2, 2))
13      A_power = I
14      for k in 1:50
15          B_sum += A_power
16          A_power = A_power * A
17      end
18      return B_sum
19  end
20
21  # Print maximal error
22  print(maximum(abs.(B_inverse - power_series(A))))
```

Listing 3: Matrix inversion vs power series (`power_series.jl`)

#### 1.2.2.1 Fixed Points

A standard approach to solving an equation is to formulate it as a fixed point problem. This section provides the basic definitions and some simple results from fixed point theory.

Let $U$ be any nonempty set. We call $T$ a **self-map** on $U$ if $T$ is a function from $U$ into itself. For a self-map $T$ on $U$, a point $u^* \in U$ is called a **fixed point** of $T$ in $U$ if $Tu^* = u^*$. (In fixed point theory, it is common to write $Tu$ for the image of $u$ under $T$, rather than $T(u)$.)

**Example 1.2.2.** Let $U = \mathbb{R}^n$ and let $T$ be defined by $Tu = Au + b$, where $A$ and $b$ are as in §1.2.1.4. Since $u$ is a fixed point of $T$ if and only if $u = Au + b$, solving the equation $u = Au + b$ is the same as searching for the fixed point of $T$. By the Neumann series lemma, $T$ has unique fixed point $u^* := (I - A)^{-1}b$ in $U$ whenever $\rho(A) < 1$.

**Example 1.2.3.** Every $u$ in set $U$ is fixed under the identity map $I \colon u \mapsto u$.

**Example 1.2.4.** If $U = \mathbb{N}$ and $Tu = u + 1$, then $T$ has no fixed point.

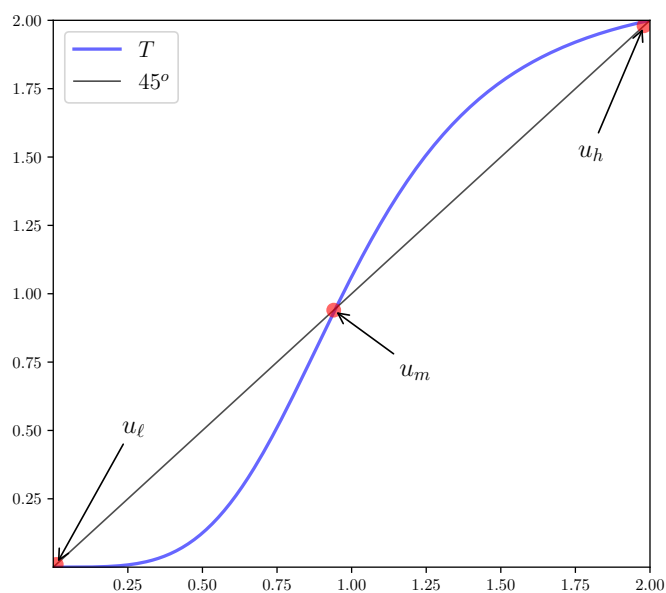

Figure 1.5: Graph and fixed points of $T\colon u \mapsto 2.125/(1+u^{-4})$

Figure 1.5 shows another example, for a self-map $T$ on $U := [0, 2]$. Fixed points are numbers $u \in [0, 2]$ where $T$ meets the 45 degree line. In this case there are three.

EXERCISE 1.2.15. Let $U$ be any set and let $T$ be a self-map on $U$. Suppose there exists an $\bar{u} \in U$ and an $m \in \mathbb{N}$ such that $T^k u = \bar{u}$ for all $u \in U$ and $k \geqslant m$. Prove that, under this condition, $\bar{u}$ is the unique fixed point of $T$ in $U$.

EXERCISE 1.2.16. Let $T$ be a self-map on $U \subset \mathbb{R}^d$. Prove the following: If $T^m u \to u^*$ as $m \to \infty$ for some pair $u, u^* \in U$ and, in addition, $T$ is continuous at $u^*$, then $u^*$ is a fixed point of $T$.

When considering fixed points, given a self-map $T$ on $U$, we typically seek conditions on $T$ and $U$ under which the following properties hold:

- $T$ has at least one fixed point on $U$ (existence)

- $T$ has at most one fixed point on $U$ (uniqueness)

- $T$ has a fixed point on $U$ and the fixed point can be computed numerically.

### 1.2.2.2 Global Stability

A self-map $T$ on $U$ is called **globally stable** on $U$ if $T$ has a unique fixed point $u^*$ in $U$ and $T^k u \to u^*$ as $k \to \infty$ for all $u \in U$. Here $T^k$ indicates $k$ compositions of $T$ with itself. Global stability is a desirable property in the setting of dynamic programming. A number of our results rely on it.

EXERCISE 1.2.17. As in Example 1.2.2, let $U = \mathbb{R}^n$ and let $T$ be defined by $Tu = Au + b$. Using induction, prove that

$$T^k u = A^k u + A^{k-1} b + A^{k-2} b + \cdots + Ab + b \tag{1.16}$$

for all $u \in U$ and $k \in \mathbb{N}$. Next, show that $T$ is globally stable on $U$ whenever $\rho(A) < 1$.

Let $T$ be a self-map on $U \subset \mathbb{R}^n$. We call $T$ **invariant** on $C \subset U$ and call $C$ an **invariant set** if $T$ is also a self-map on $C$; that is, if $u \in C$ implies $Tu \in C$.

EXERCISE 1.2.18. Let $T$ be a globally stable self-map on $U \subset \mathbb{R}^n$, with fixed point $u^*$. Prove the following: If $C$ is closed and $T$ is invariant on $C$, then $u^* \in C$.

### 1.2.2.3 Banach's Fixed Point Theorem

Next we present the Banach fixed point theorem, a workhorse for analyzing nonlinear operators.

Let $U$ be a nonempty subset of $\mathbb{R}^n$ and let $\| \cdot \|$ be a norm on $\mathbb{R}^n$. A self-map $T$ on $U$ is called a **contraction** on $U$ with respect to $\| \cdot \|$ if there exists a $\lambda < 1$ such that

$$\|Tu - Tv\| \leqslant \lambda \|u - v\| \quad \text{for all} \quad u, v \in U. \tag{1.17}$$

The constant $\lambda$ is called the **modulus of contraction**.

EXERCISE 1.2.19. Let $T$ be a contraction on $U$ with respect to a norm $\| \cdot \|$. Show that, $T$ is continuous on $U$ and has at most one fixed point in $U$.

EXERCISE 1.2.20. Let $U = \mathbb{R}^n$ and let $Tx = Ax + b$, where $A$ is $n \times n$ and $b$ is $n \times 1$. Prove that $T$ is a contraction of modulus $\|A\|$ on $U$ (see (1.12) for the definition) whenever $\|A\| < 1$.

The following theorem features a contraction.

**Theorem 1.2.3** (Banach's contraction mapping theorem)**.** *If $U$ is closed in $\mathbb{R}^n$ and $T$ is a contraction of modulus $\lambda$ on $U$ with respect to some norm $\|\cdot\|$ on $\mathbb{R}^n$, then $T$ has a unique fixed point $u^*$ in $U$ and*

$$\|T^k u - u^*\| \leqslant \lambda^k \|u - u^*\| \quad \text{for all } k \in \mathbb{N} \text{ and } u \in U. \tag{1.18}$$

*In particular, $T$ is globally stable on $U$.*

We prove Theorem 1.2.3 in stages that build on the following exercises.

EXERCISE 1.2.21. Let $U$ and $T$ have the properties stated in Theorem 1.2.3. Fix $u_0 \in U$ and let $u_m := T^m u_0$. Show that

$$\|u_m - u_k\| \leqslant \sum_{i=m}^{k-1} \lambda^i \|u_0 - u_1\|$$

holds for all $m, k \in \mathbb{N}$ with $m < k$.

EXERCISE 1.2.22. Using the results in Exercise 1.2.21, prove that $(u_m)$ is a Cauchy sequence in $\mathbb{R}^n$. (A sequence $(v_m) \subset \mathbb{R}^n$ is called a **Cauchy sequence** if, for any $\varepsilon > 0$, there exists an $N \in \mathbb{N}$ such that $m, n \geqslant N$ implies $\|v_m - v_n\| < \varepsilon$.)

A fundamental property of $\mathbb{R}^n$ is that if $(v_m)$ is a Cauchy sequence in $\mathbb{R}^n$, then there exists a $\bar{v} \in \mathbb{R}^n$ such that $(v_m)$ converges to $\bar{v}$. (This property is called **completeness** of the vector space $\mathbb{R}^n$. See, for example, Çınlar and Vanderbei (2013).) Hence it follows from Exercise 1.2.22 that $(u_m)$ has a limit $u^* \in \mathbb{R}^n$.

EXERCISE 1.2.23. Prove that $u^* \in U$.

*Proof of Theorem 1.2.3.* The preceding exercises established existence of a point $u^* \in U$ such that $T^m u \to u^*$. The fact that $u^*$ is a fixed point of $T$ now follows from Exercise 1.2.16 and Exercise 1.2.19. Uniqueness is implied by Exercise 1.2.19. The bound (1.18) follows from iteration on the contraction inequality (1.17) while setting $v = u^*$. □

EXERCISE 1.2.24. Let $T$ be a contraction of modulus $\beta$ on $\mathbb{R}^n$ and fixed point $\bar{u}$. Consider the **damped** or **relaxed** iteration scheme $u_{n+1} = (1 - \alpha)u_n + \alpha T u_n$. Show that, for any choice of $u_0$, these iterates converge to $\bar{u}$ whenever $0 < \alpha \leqslant 1$.

### 1.2.3 Successive Approximation

Consider a self-map $T$ on $U \subset \mathbb{R}^n$. We seek algorithms that compute fixed points of $T$ whenever they exist.

#### 1.2.3.1 Iteration

If $T$ is globally stable on $U$, then a natural algorithm for approximating the unique fixed point $u^*$ of $T$ in $U$ is to pick any $u \in U$ and iterate with $T$ for some finite number of steps:

---

```
1  fix u₀ ∈ U and τ > 0
2  k ← 0
3  ε ← τ + 1
4  while ε > τ do
5      u_{k+1} ← Tu_k
6      ε ← ‖u_{k+1} − u_k‖
7      k ← k + 1
8  end
9  return u_k
```

---

By the definition of global stability, $(u_k)_{k \geqslant 0}$ converges to $u^*$. The algorithm just described is called either **successive approximation** or **fixed point iteration**. Listing 4 provides a function that implements this procedure. Distances between points are measured with the $\ell_\infty$ norm.

Listing 5 applies successive approximation to the map $Tu = Au + b$ using the function defined in s_approx.jl. Figure 1.6 shows the sequence of iterates generated by four runs of the successive approximation algorithm, each with a different starting condition $u_0$. The map and parameters are the same as in Listing 5. It is clear from the figure that a good choice of initial condition (i.e., one that is close to the fixed point) accelerates convergence.

Of course for $Tu = Au + b$ with $\rho(A) < 1$, there is a more direct method to compute the fixed point: the Neumann series lemma tells us that $u^* = (I - A)^{-1}b$ so we can apply a numerical linear equation solver. However, even for this case, sometimes successive approximation is used instead. One reason is that $(I - A)^{-1}$ can be very large, making application of a linear solver problematic. Another is that we might be satisfied with a quick approximation of the fixed point, computed with a few iterations of $T$. Both of these situations can arise in dynamic programming.

```julia
"""
Computes an approximate fixed point of a given operator T
via successive approximation.

"""
function successive_approx(T,                    # operator (callable)
                           u_0;                  # initial condition
                           tolerance=1e-6,       # error tolerance
                           max_iter=10_000,      # max iteration bound
                           print_step=25)        # print at multiples

    u = u_0
    error = Inf
    k = 1

    while (error > tolerance) & (k <= max_iter)

        u_new = T(u)
        error = maximum(abs.(u_new - u))

        if k % print_step == 0
            println("Completed iteration $k with error $error.")
        end

        u = u_new
        k += 1
    end

    if error <= tolerance
        println("Terminated successfully in $k iterations.")
    else
        println("Warning: hit iteration bound.")
    end

    return u
end
```

Listing 4: Successive approximation (`s_approx.jl`)

```julia
include("s_approx.jl")
using LinearAlgebra

# Compute the fixed point of Tu = Au + b via linear algebra
A, b = [0.4 0.1; 0.7 0.2], [1.0; 2.0]
u_star = (I - A) \ b  # compute (I - A)^{-1} * b

# Compute the fixed point via successive approximation
T(u) = A * u + b
u_0 = [1.0; 1.0]
u_star_approx = successive_approx(T, u_0)

# Test for approximate equality (prints "true")
print(isapprox(u_star, u_star_approx, rtol=1e-5))
```

Listing 5: Using successive approximations to compute $u^*$ (`linear_iter.jl`)

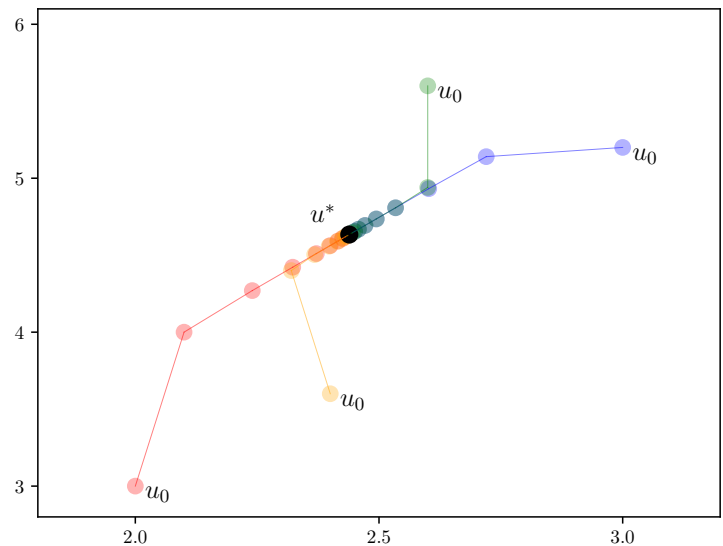

Figure 1.6: Successive approximations from different initial conditions

### 1.2.3.2   A One-Dimensional Example

To illustrate successive approximations in a nonlinear setting, we use the Solow–Swan growth model, which is a good place to begin presenting a theory of economic growth. A fixed point for the Solow-Swan model can be computed with pencil and paper. The model also provides a good laboratory for studying how successive approximations might converge to a fixed point.

One version of the Solow–Swan growth dynamics is

$$k_{t+1} = sf(k_t) + (1 - \delta)k_t, \qquad t = 0, 1, \ldots, \tag{1.19}$$

where $k_t$ is capital stock per worker, $f \colon (0, \infty) \to (0, \infty)$ is a production function, $s > 0$ is a saving rate and $\delta \in (0, 1)$ is a rate of depreciation. If we set $g(k) := sf(k) + (1 - \delta)k$, then iterating with $g$ from a starting point $k_0$ (i.e., setting $k_{t+1} = g(k_t)$ for all $t \geqslant 0$) generates the sequence in (1.19). We can also understand this process as using successive approximation to compute the fixed point of $g$.

EXERCISE 1.2.25. Let $f(k) = Ak^\alpha$ with $A > 0$ and $0 < \alpha < 1$. Show that, while the Solow-Swan map $g(k) = sAk^\alpha + (1 - \delta)k$ sends $U := (0, \infty)$ into itself, $g$ is *not* a contraction on $U$. [Hint: use the definition of the derivative of $g$ as a limit and consider the derivative $g'(k)$ for $k$ close to zero.]

Although the model specified in Exercise 1.2.25 does not generate a contraction, it is globally stable. The next exercise asks you to prove this.

EXERCISE 1.2.26. Show that, in the setting of Exercise 1.2.25, the unique fixed point of $g$ in $U$ is

$$k^* := \left(\frac{sA}{\delta}\right)^{1/(1-\alpha)}$$

Prove that, for $k \in U$,

  (i)  $k \leqslant k^*$ implies $k \leqslant g(k) \leqslant k^*$ and

  (ii)  $k^* \leqslant k$ implies $k^* \leqslant g(k) \leqslant k$.

Conclude that $g$ is globally stable on $U$. (Why?)

Figure 1.7 illustrates the dynamics in a 45 degree diagram when $f(k) = Ak^\alpha$. In the top subfigure, $A = 2.0$, $\alpha = 0.3$, $s = 0.3$ and $\delta = 0.4$. The function $g$ is plotted alongside the 45 degree line. When $g(k_t)$ lies strictly above the 45 degree line, then $k_{t+1} = g(k_t) > k_t$ and so capital per worker rises. If $g(k_t) < k_t$ then it falls. A trajectory $(k_t)_{t \geqslant 0}$ that is produced by starting from a particular choice of $k_0$ is traced out in the figure.

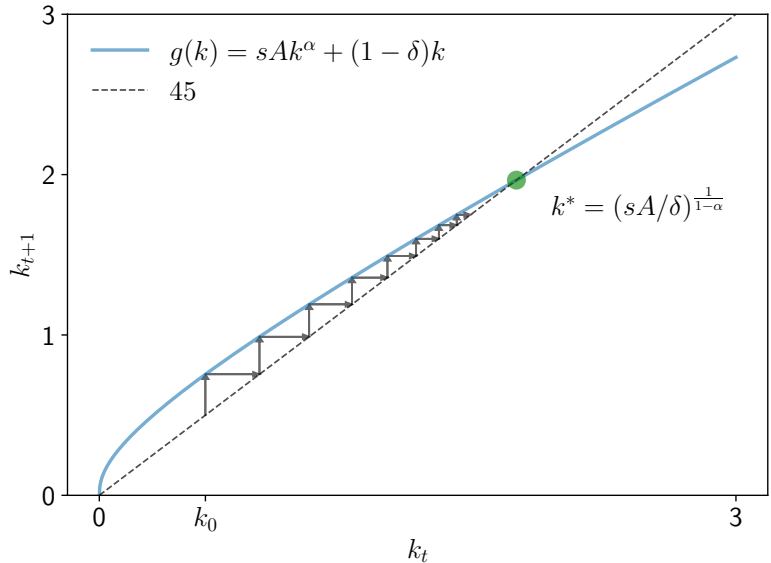

Figure 1.7: Successive approximation for the Solow–Swan model

The figure illustrates that $k^*$ is the unique fixed point of $g$ in $U$ and all sequences converge to it. The second statement can be rephrased as: successive approximation successfully computes the fixed point of $g$ by stepping through the time path of capital.

### 1.2.4  Finite-Dimensional Function Space

In §1.1.2 we introduced a Bellman equation for the infinite horizon job search problem. The unknown object in the Bellman equation is a function $v^*$ defined on the set W of possible wage offers. Below we discuss how to solve for this unknown function.

Since the set of wage offers is finite we can write W as $\{w_1, \dots, w_n\}$ for some $n \in \mathbb{N}$. If we adopt this convention and also write $v^*(w_i)$ as $v_i^*$, then we can view $v^*$ as a vector $(v_1^*, \dots, v_n^*)$ in $\mathbb{R}^n$. The vector interpretation is useful when coding, since vectors (numerical arrays) are an efficient data type.

Nevertheless, for mathematical exposition, we usually find it more convenient to express function-like objects (e.g., value functions) as functions rather than vectors. Thus, we typically write $v^*(w)$ instead of $v_i^*$.

**Remark 1.2.2.** There is a deeper reason that we usually work with functions rather than vectors: when we shift to general state and action spaces in Volume II, objects such as value functions can no longer be represented by finite-dimensional vectors. Instead we must use the language of functional analysis. By adopting this language now, the leap to general spaces will be smoother, since terminology and notation will mostly be unchanged.

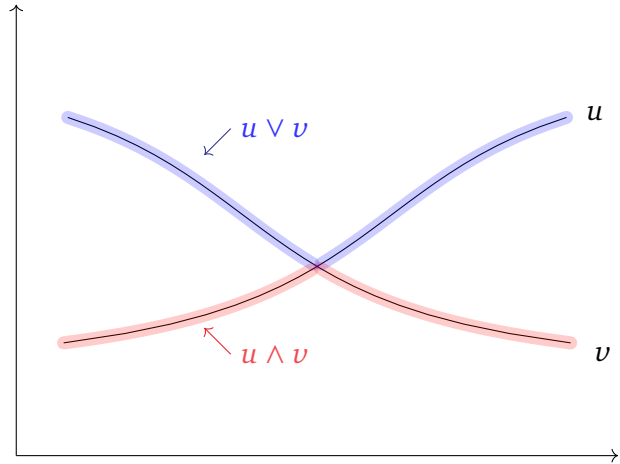

Figure 1.8: Functions $u \vee v$ and $u \wedge v$

The next section clarifies our notation with respect to functions and vectors.

### 1.2.4.1   Pointwise Operations on Functions

If X is any set and $u$ maps X to $\mathbb{R}$, then we call $u$ a **real-valued function** on X and write $u\colon \mathsf{X} \to \mathbb{R}$. Throughout, the symbol $\mathbb{R}^{\mathsf{X}}$ denotes the set of all real-valued functions on X. This is a special case of the symbol $B^A$ that represents the set of all functions from $A$ to $B$, where $A$ and $B$ are sets.

If $u, v \in \mathbb{R}^{\mathsf{X}}$ and $\alpha, \beta \in \mathbb{R}$, then the expressions $\alpha u + \beta v$ and $uv$ also represent elements of $\mathbb{R}^{\mathsf{X}}$, defined at $x \in \mathsf{X}$ by

$$(\alpha u + \beta v)(x) = \alpha u(x) + \beta v(x) \quad \text{and} \quad (uv)(x) = u(x)v(x). \tag{1.20}$$

Similarly, $|u|$, $u \vee v$, and $u \wedge v$ are real-valued functions on X defined by

$$|u|(x) = |u(x)|, \quad (u \vee v)(x) = u(x) \vee v(x) \quad \text{and} \quad (u \wedge v)(x) = u(x) \wedge v(x). \tag{1.21}$$

Figure 1.8 illustrates functions $u \vee v$ and $u \wedge v$ when X is a subset of $\mathbb{R}$.

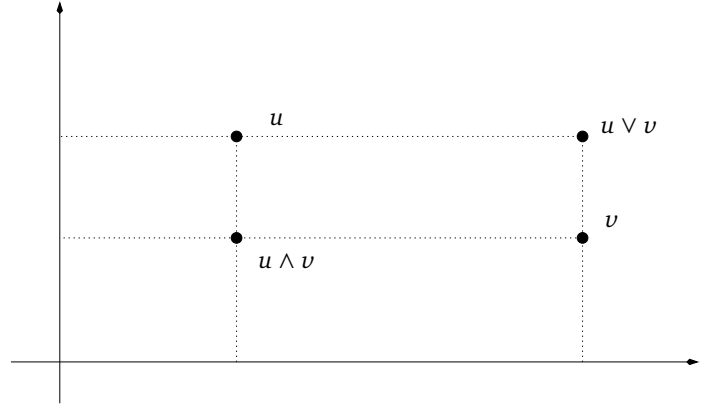

Figure 1.9: The vectors $u \vee v$ and $u \wedge v$ in $\mathbb{R}^2$

Similarly, if $u = (u_i)_{i=1}^n$ and $v = (v_i)_{i=1}^n$ are vectors in $\mathbb{R}^n$, then

$$|u| := (|u_i|)_{i=1}^n, \quad u \wedge v := (u_i \wedge v_i)_{i=1}^n \quad \text{and} \quad u \vee v := (u_i \vee v_i)_{i=1}^n. \tag{1.22}$$

Figure 1.9 illustrates in $\mathbb{R}^2$.

### 1.2.4.2 Functions vs Vectors

Let $\mathsf{X}$ be finite, so that $\mathsf{X} = \{x_1, \ldots, x_n\}$ for some $n \in \mathbb{N}$. The set $\mathbb{R}^{\mathsf{X}}$ is, in essence, the vector space $\mathbb{R}^n$ expressed in different notation. The next lemma clarifies.

**Lemma 1.2.4.** *If* $\mathsf{X} = \{x_1, \ldots, x_n\}$, *then*

$$\mathbb{R}^{\mathsf{X}} \ni u \quad \longleftrightarrow \quad (u(x_1), \ldots, u(x_n)) \in \mathbb{R}^n \tag{1.23}$$

*is a one-to-one correspondence between the function space* $\mathbb{R}^{\mathsf{X}}$ *and the vector space* $\mathbb{R}^n$.

The claim in Lemma 1.2.4 is obvious: a real-valued function $u$ on $\mathsf{X}$ is uniquely identified by the set of values that it takes on $\mathsf{X}$, which is an $n$-tuple of real numbers.

Throughout the text, whenever the supporting set $\mathsf{X}$ is finite, we freely use the identification in (1.23). For example, if $\| \cdot \|$ is any norm on $\mathbb{R}^n$, then $\| \cdot \|$ extends to $\mathbb{R}^{\mathsf{X}}$ via the identification in (1.23). That is, for $u \in \mathbb{R}^{\mathsf{X}}$, the value $\|u\|$ is given by the norm of the vector $(u(x_1), \ldots, u(x_n)) \in \mathbb{R}^n$.

We say that a subset of $\mathbb{R}^{\mathsf{X}}$ is **closed** (resp., **open**, **compact**, etc.) if the corresponding subset of $\mathbb{R}^n$ is closed (resp., open, compact, etc.)

With these conventions, the Neumann series lemma and Banach's contraction mapping theorem extend directly from $\mathbb{R}^n$ to $\mathbb{R}^X$. For example, if $|X| = n$, $C$ is closed in $\mathbb{R}^X$ and $T$ is a contraction on $C \subset \mathbb{R}^X$, in the sense that $T \colon C \to C$ and

$$\text{there exists a } \lambda \in [0, 1) \text{ s.t. } \quad \|Tf - Tg\| \leqslant \lambda \|f - g\| \quad \text{for all} \quad f, g \in C,$$

then $T$ has a unique fixed point $f^*$ in $C$ and

$$\|T^n f - f^*\| \leqslant \lambda^n \|f - f^*\| \quad \text{for all } n \in \mathbb{N} \text{ and } f \in \mathbb{R}^X.$$

Incidentally, in the preceding paragraph $T$ is a function that sends functions into functions (e.g., sends $f$ into $Tf$). To help distinguish $T$ from the functions that it acts on, $T$ in this setting is often called an **operator** rather than a function. This is a convention rather than a formal distinction: from a mathematical perspective, an operator is just a function.

A foundational class of operators acting on $\mathbb{R}^X$ is the set of linear operators. There is a strong sense in which linear operators are just matrices. We investigate these ideas in §2.3.3. At the same time, when studying dynamic programming we also use many operators that are not linear. One example is the "Bellman operator," which we start to investigate in §1.3.1.2.

### 1.2.4.3 Distributions

Given a set $X$ with $n$ elements, the set of probability **distributions** on $X$ is written as $\mathcal{D}(X)$ and contains all $\varphi \in \mathbb{R}_+^X$ with $\sum_{x \in X} \varphi(x) = 1$. Since we can identify any $f \in \mathbb{R}^X$ with a corresponding vector in $\mathbb{R}^n$, the set $\mathcal{D}(X)$ can also be thought of as a subset of $\mathbb{R}^n$. This collection of vectors (i.e., the nonnegative vectors that sum to unity) is also called the **unit simplex**. Given $X_0 \subset X$ and $\varphi \in \mathcal{D}(X)$, we say that $\varphi$ is **supported** on $X_0$ if $\varphi(x) > 0$ implies $x \in X_0$.

Fix $h \in \mathbb{R}^X$ and $\varphi \in \mathcal{D}(X)$. Let $X$ be a random variable with distribution $\varphi$, so that $\mathbb{P}\{X = x\} = \varphi(x)$ for all $x \in X$. The **expectation** of $h(X)$ is

$$\mathbb{E}h(X) := \sum_{x \in X} h(x)\varphi(x) = \langle h, \varphi \rangle .$$

EXERCISE 1.2.27. Fix $h \in \mathbb{R}^X$. Show that $\varphi^* \in \operatorname{argmax}_{\varphi \in \mathcal{D}(X)} \langle h, \varphi \rangle$ if and only if $\varphi^*$ is supported on $\operatorname{argmax}_{x \in X} h(x)$.

If $X \subset \mathbb{R}$, then the **cumulative distribution function** (CDF) corresponding to $\varphi$ is the map $\Phi$ from $X$ to $\mathbb{R}$ given by

$$\Phi(x) := \mathbb{P}\{X \leqslant x\} = \sum_{x' \in X} \mathbb{1}\{x' \leqslant x\}\varphi(x').$$

If $\tau \in [0, 1]$, then the $\tau$-th **quantile** of $X$ is

$$Q_\tau X := \min\{x \in X : \Phi(x) \geqslant \tau\}. \tag{1.24}$$

If $\tau = 1/2$, then $Q_\tau X$ is called the **median** of $X$.

**Example 1.2.5.** Suppose $X = \{x_1, x_2, x_3\}$. If $\varphi = (0.5, 0.0, 0.5)$ and $X \sim \varphi$, then $\Phi = (0.5, 0.5, 1)$ and $Q_{1/2}(X) = x_1$. The min in (1.24) allows us to select a unique median (even though $x_2$ is also a reasonable choice).

Evidently, if the median of $X$ is $x$, then the median of $X + \alpha$ will be $x + \alpha$. This same logic carries over to arbitrary quantiles, as the next exercise asks you to show.

EXERCISE 1.2.28. Prove that the quantile function is additive over constants. That is, for any $\tau \in [0, 1]$, random variable $X$ on $X$ and $\alpha \in \mathbb{R}$, we have $Q_\tau(X + \alpha) = Q_\tau(X) + \alpha$.

# 1.3   Infinite-Horizon Job Search

Armed with fixed point methods, we return to the job search problem discussed in §1.1.2.

## 1.3.1   Values and Policies

In this section we solve for the value function of an infinite horizon job search problem and associated optimal choices.

### 1.3.1.1   Optimal Choices

Let's recall the strategy for solving the infinite-horizon job search problem we proposed in §1.1.2. The first step is to compute the optimal value function $v^*$ that solves

the Bellman equation

$$v^*(w) = \max \left\{ \frac{w}{1-\beta},\, c + \beta \sum_{w' \in \mathsf{W}} v^*(w')\varphi(w') \right\} \qquad (w \in \mathsf{W}). \qquad (1.25)$$

Suppose for a moment that we can compute $v^*$, and let

$$h^* := c + \beta \sum_{w'} v^*(w')\varphi(w') \qquad (1.26)$$

be the infinite-horizon **continuation value** that equals the maximal lifetime value that the worker can receive, contingent on deciding to continue being unemployed today.

With $h^*$ in hand, the optimal decision at any given time, facing current wage draw $w \in \mathsf{W}$, is as follows:

 (i)  If $w/(1-\beta) \geqslant h^*$, then accept the job offer.

 (ii)  If not, then reject and wait for the next offer.

This decision maximizes lifetime value given the current offer.

   (We will prove below that this decision process is optimal as claimed.  For now, however, we focus on computing $v^*$ and $h^*$.)

### 1.3.1.2   The Bellman Operator

The method proposed above requires that we solve for $v^*$.  To do so, we introduce a **Bellman operator** $T$ defined at $v \in \mathbb{R}^{\mathsf{W}}$ that is constructed to assure that any fixed point of $T$ solves the Bellman equation and vice versa:

$$(Tv)(w) = \max \left\{ \frac{w}{1-\beta},\, c + \beta \sum_{w' \in \mathsf{W}} v(w')\varphi(w') \right\} \qquad (w \in \mathsf{W}). \qquad (1.27)$$

Let $V := \mathbb{R}_+^{\mathsf{W}}$ and let $\|\cdot\|_\infty$ be the supremum norm on $V$.  We measure distance between two elements $f, g$ of $V$ by $\|f-g\| = \max_{w \in \mathsf{W}} |f(w) - g(w)|$.  Under this distance, we have the following result.

**Proposition 1.3.1.** *$T$ is a contraction of modulus $\beta$ on $V$.*

   A proof of Proposition 1.3.1 is given below.  An implication of the proposition is that $T^k v \to v^*$ as $k \to \infty$ for any $v \in V$, so we can compute $v^*$ to any required degree of accuracy by successive approximation.

Our proof of Proposition 1.3.1 uses the elementary bound

$$|\alpha \vee x - \alpha \vee y| \leqslant |x - y| \qquad (\alpha, x, y \in \mathbb{R}) \tag{1.28}$$

EXERCISE 1.3.1. Verify that (1.28) always holds. [Exercise 1.2.1 might be helpful.]

*Proof of Proposition 1.3.1.* Take any $f, g$ in $V$ and fix any $w \in \mathsf{W}$. Apply the bound in (1.28) to get

$$|(Tf)(w) - (Tg)(w)| \leqslant \left| c + \beta \sum_{w'} f(w')\varphi(w') - \left( c + \beta \sum_{w'} g(w')\varphi(w') \right) \right|$$

$$= \beta \left| \sum_{w'} [f(w') - g(w')]\varphi(w') \right|.$$

Apply the triangle inequality to obtain

$$|(Tf)(w) - (Tg)(w)| \leqslant \beta \sum_{w'} |f(w') - g(w')|\varphi(w') \leqslant \beta \|f - g\|_\infty.$$

Taking the supremum over all $w$ on the left hand side of this expression leads to

$$\|Tf - Tg\|_\infty \leqslant \beta \|f - g\|_\infty.$$

Since $f, g$ were arbitrary elements of $V$, the contraction claim is verified. □

### 1.3.1.3 Optimal Policies

A dynamic program seeks optimal policies. We briefly introduce the notion of a policy and relate it to the job search application.

In general, for a dynamic program, choices by the controller aim to maximize lifetime rewards and consist of a state-contingent sequence $(A_t)_{t \geqslant 0}$ specifying how the agent acts at each point in time. Workers do not know what the future will bring, so it is natural to assume that $A_t$ can depend on present and past events but not future ones. Hence $A_t$ is a function of the current state $X_t$ and past state-action pairs $(A_{t-i}, X_{t-i})$ for $i \geqslant 1$. That is,

$$A_t = \sigma_t(X_t, A_{t-1}, X_{t-1}, A_{t-2}, X_{t-2}, \ldots, A_0, X_0)$$

for some function $\sigma_t$; $\sigma_t$ is called a time $t$ **policy function**.

A key insight of dynamic programming is that some problems can be set up so that *the optimal current action can be expressed as a function of the current state $X_t$.*

**Example 1.3.1.** In Example 1.0.1, the retailer chooses stock orders and prices in each period. Every quantity relevant to this decision belongs in the current state. It might include not just the level of current inventories and various measures of business conditions, but also information about rates at which inventories have changed over each of the past six months.

If the current state $X_t$ is enough to determine a current optimal action, then policies are just maps from states to actions. So we can write $A_t = \sigma(X_t)$ for some function $\sigma$. A policy function that depends only on the current state is often called a **Markov policy**. Since all policies we consider will be Markov policies, we refer to them more concisely as "policies."

**Remark 1.3.1.** In the last paragraph, we dropped the time subscript on $\sigma$ with no loss of generality because we can always include the date $t$ in the current state; i.e., if $Y_t$ is the state without time, then we can set $X_t = (t, Y_t)$). Whether this is necessary depends on the problem at hand. For the job search model with finite horizon, the date matters because opportunities for future earnings decrease with the passage of time. For the infinite horizon version of the problem, in which an agent always looks forward toward an infinite horizon, the only current information that matters to the agent at time $t$ is the wage offer $W_t$. As a result, the calendar date $t$ does not affect the agent's decision at time $t$, so there is no need to include time in the state. (In §8.1.3.5, we will formalize this argument.)

In the job search model, the state is the current wage offer and possible actions are to accept or to reject the current offer. With $0$ interpreted as reject and $1$ understood as accept, the action space is $\{0, 1\}$, so a policy is a map $\sigma$ from W to $\{0, 1\}$. Let $\Sigma$ be the set of all such maps.

A policy is an "instruction manual": for an agent following $\sigma \in \Sigma$, if current wage offer is $w$, the agent always responds with $\sigma(w) \in \{0, 1\}$. The policy dictates whether the agent accepts or rejects at any given wage.

For each $v \in V$, a $v$-**greedy policy** is a $\sigma \in \Sigma$ satisfying

$$\sigma(w) = \mathbb{1}\left\{ \frac{w}{1-\beta} \geqslant c + \beta \sum_{w' \in W} v(w')\varphi(w') \right\} \quad \text{for all } w \in W. \qquad (1.29)$$

Equation (1.29) says that an agent accepts if $w/(1-\beta)$ exceeds the continuation value computed using $v$ and rejects otherwise. Our discussion of optimal choices in §1.3.1.1 can now be summarized as the recommendation

$$\text{Adopt a } v^*\text{-greedy policy.}$$

This statement is sometimes called Bellman's principle of optimality.

Inserting $v^*$ into (1.29) and rearranging, we can express a $v^*$-greedy policy via

$$\sigma^*(w) = \mathbb{1}\{w \geqslant w^*\} \quad \text{where } w^* := (1 - \beta)h^*. \tag{1.30}$$

The quantity $w^*$ in (1.30) is called the **reservation wage**, and parallels the reservation wage that we introduced for the finite-horizon problem. Equation (1.30) states that value maximization requires accepting an offer if and only if it exceeds the reservation wage. Thus, $w^*$ provides a scalar description of an optimal policy.

## 1.3.2 Computation

Let's turn to computation. In §1.3.2.1, we apply a standard dynamic programming method, called value function iteration. In §1.3.2.2, we apply a more specialized method that uses the structure of the job search problem to accelerate computation.

### 1.3.2.1 Value Function Iteration

Recall that, by Proposition 1.3.1, we can compute an approximate optimal policy by applying successive approximation via the Bellman operator. In the language of dynamic programming, this is called **value function iteration**. Algorithm 1.1 provides a full description.

While $T^k v$ rarely attains $v^*$ for $k < \infty$, we can obtain a close approximation by monitoring distances between successive iterates, waiting until they become small enough. Later we will study how these distances depend on $k$, the number of iterations, as well as on parameters defining rewards and opportunities.

Listing 6 implements value function iteration for the infinite-horizon job search model, using the function for successive approximation from Listing 4.

Figure 1.10 shows a sequence of iterates $(T^k v)_k$ when $v \equiv 0$ and parameters are as given in Listing 1 (page 8). Iterates $0, 1$ and $2$ are shown, in addition to iterate $1000$, which we take as a good approximation to the limiting function. If you experiment with different initial conditions, you will see that they all converge to the same limit.

---

**Algorithm 1.1:** Value function iteration for job search

---

**1** input $v_0 \in V$, an initial guess of $v^*$
**2** input $\tau$, a tolerance level for error
**3** $\varepsilon \leftarrow \tau + 1$
**4** $k \leftarrow 0$
**5** **while** $\varepsilon > \tau$ **do**
**6**     **for** $w \in \mathsf{W}$ **do**
**7**         $v_{k+1}(w) \leftarrow (Tv_k)(w)$
**8**     **end**
**9**     $\varepsilon \leftarrow \|v_k - v_{k+1}\|_\infty$
**10**     $k \leftarrow k + 1$
**11** **end**
**12** Compute a $v_k$-greedy policy $\sigma$
**13** **return** $\sigma$

---

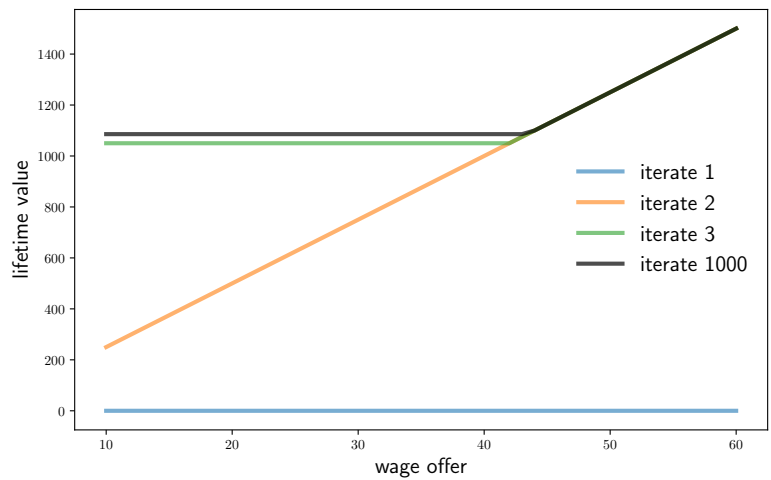

Figure 1.10:  A sequence of iterates of the Bellman operator

```julia
include("two_period_job_search.jl")
include("s_approx.jl")

" The Bellman operator. "
function T(v, model)
    (; n, w_vals, φ, β, c) = model
    return [max(w / (1 - β), c + β * v'φ) for w in w_vals]
end

" Get a v-greedy policy. "
function get_greedy(v, model)
    (; n, w_vals, φ, β, c) = model
    σ = w_vals ./ (1 - β) .>= c .+ β * v'φ   # Boolean policy vector
    return σ
end

" Solve the infinite-horizon IID job search model by VFI. "
function vfi(model=default_model)
    (; n, w_vals, φ, β, c) = model
    v_init = zero(model.w_vals)
    v_star = successive_approx(v -> T(v, model), v_init)
    σ_star = get_greedy(v_star, model)
    return v_star, σ_star
end
```

Listing 6: Value function iteration (`iid_job_search.jl`)

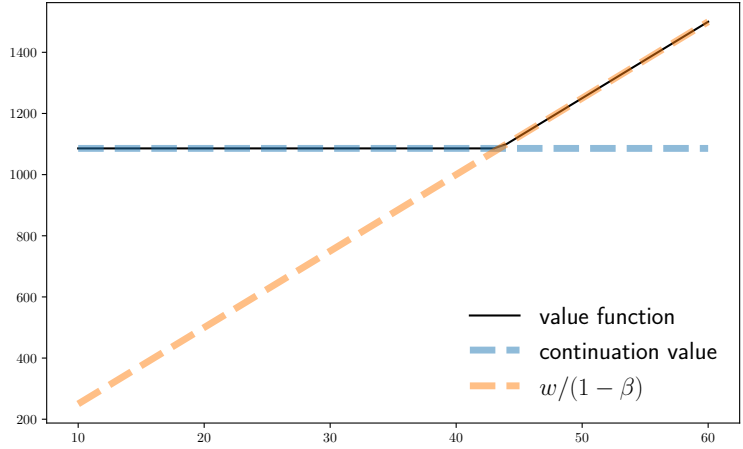

Figure 1.11: The approximate value function for job search

Figure 1.11 shows an approximation of $v^*$ computed using the code in Listing 6, along with the stopping reward $w/(1-\beta)$ and the corresponding continuation value (1.26). As anticipated, the value function is the pointwise supremum of the stopping reward and the continuation value. The worker chooses to accept an offer only when that offer exceeds some value close to 43.5.

### 1.3.2.2 Computing the Continuation Value Directly

The technique we employed to solve the job search model in §1.3.1 follows a standard approach to dynamic programming. But for this particular problem, there is an easier way to compute the optimal policy that sidesteps calculating the value function. This section explains how.

Recall that the value function satisfies Bellman equation

$$v^*(w) = \max\left\{\frac{w}{1-\beta},\ c + \beta \sum_{w'} v^*(w')\varphi(w')\right\} \qquad (w \in \mathsf{W}), \qquad (1.31)$$

and that the continuation value is given by (1.26). We can use $h^*$ to eliminate $v^*$ from (1.31). First we insert $h^*$ on the right hand side of (1.31) and then we replace $w$ with $w'$, which gives $v^*(w') = \max\{w'/(1-\beta),\ h^*\}$. Then we take mathematical

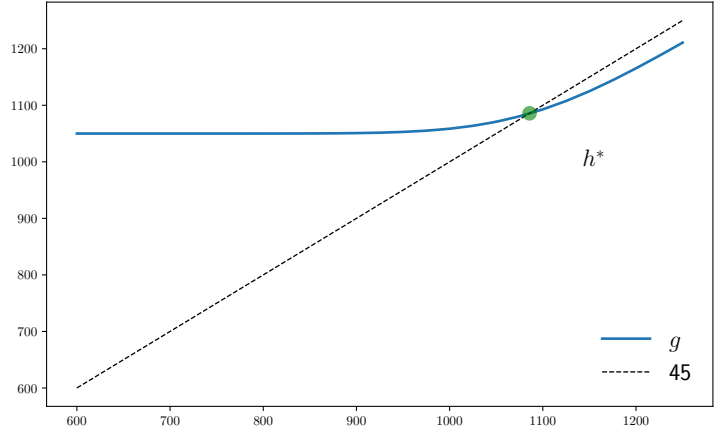

Figure 1.12: Computing the continuation value as the fixed point of $g$

expectations of both sides, multiply by $\beta$ and add $c$ to obtain

$$h^* = c + \beta \sum_{w'} \max \left\{ \frac{w'}{1-\beta}, h^* \right\} \varphi(w'). \tag{1.32}$$

To obtain the unknown value $h^*$, we introduce the mapping $g \colon \mathbb{R}_+ \to \mathbb{R}_+$ defined by

$$g(h) = c + \beta \sum_{w'} \max \left\{ \frac{w'}{1-\beta}, h \right\} \varphi(w'). \tag{1.33}$$

By construction, $h^*$ solves (1.32) if and only if $h^*$ is a fixed point of $g$.

EXERCISE 1.3.2. Show that $g$ is a contraction map on $\mathbb{R}_+$. Conclude that $h^*$ is the unique fixed point of $g$ in $\mathbb{R}_+$.

Figure 1.12 shows the function $g$ using the discrete wage offer distribution and parameters as adopted previously. The unique fixed point is $h^*$.

Exercise 1.3.2 implies that we can compute $h^*$ by choosing arbitrary $h \in \mathbb{R}_+$ and iterating with $g$. Doing so produces a value of approximately 1086. (The associated reservation wage is $w^* = (1-\beta)h^* \approx 43.4$.) Computation of $h^*$ using this method is much faster than value function iteration because the fixed point problem is in $\mathbb{R}_+$ rather than $\mathbb{R}_+^n$.

With $h^*$ in hand we have solved the dynamic programming problem, since a policy

$\sigma^*$ is $v^*$-greedy if and only if it satisfies

$$\sigma^*(w) = \mathbb{1}\left\{\frac{w}{1-\beta} \geqslant h^*\right\} \qquad (w \in \mathbb{R}_+).\tag{1.34}$$

EXERCISE 1.3.3. As a computational exercise, compare the value function $v^*$ computed via

$$v^*(w) = \max\left\{\frac{w}{1-\beta},\, h^*\right\}$$

with our previous result, shown in Figure 1.11. You should find them essentially identical.

## 1.4 Chapter Notes

Dynamic programming is often attributed to Richard Bellman (1920–1984). Both the term "dynamic programming" and the technique were popularized by Bellman (1957). According to his autobiography, Bellman chose the name dynamic programming to avoid giving the impression that he was conducting mathematical research within RAND Corporation. His ultimate boss, Secretary of Defense Charles Wilson, apparently disliked such research (Bellman (1984)).

For treatments of dynamic programming from the perspective of economics and finance, see, for example, Sargent (1987), Stokey and Lucas (1989), Van and Dana (2003), Bäuerle and Rieder (2011), or Stachurski (2022).

The job search model was introduced by McCall (1970). The McCall model and its extensions transformed economists' way of thinking about labor markets (see, e.g., Lucas (1978b)). Influential extensions to the job search model include Burdett (1978), Jovanovic (1979), Pissarides (1979), Jovanovic (1984), Mortensen (1986), Ljungqvist (2002) and Chetty (2008). Rogerson et al. (2005) provides a useful survey.

For elementary real analysis, the book by Bartle and Sherbert (2011) is excellent. Ok (2007) is a superb treatment of real analysis and how it is used throughout economic theory. Discussions of Banach's theorem and the Neumann series lemma can be found in Cheney (2013) and Atkinson and Han (2005). Martins-da Rocha and Vailakis (2010) provides an extension to Banach's theorem that requires only local contractivity.

# Chapter 2

# Operators and Fixed Points

This chapter discusses techniques that underlie the optimization and fixed point methods used throughout the book. Many of these techniques relate to order. Order-theoretic concepts will prove valuable not only for fixed point methods but also for understanding the main concepts in dynamic programming. Chapter 8 will show core components of dynamic programming can be expressed in terms of simple order-theoretic constructs.

## 2.1   Stability

In this section we discuss algorithms for computing fixed points and analyze their convergence.

### 2.1.1   Conjugate Maps

First we treat a technique for simplifying analysis of stability and fixed points that we'll apply in applications.

To illustrate the idea, suppose that we want to study dynamics induced by a self-map $T$ on $U \subset \mathbb{R}^n$. We might want to know if a unique fixed point of $T$ exists and if iterates of $T$ converge to a fixed point. One approach is to apply fixed point theory to $T$.

However sometimes there is an easier approach: transform $T$ into a "simpler" map $\hat{T}$ and study its the fixed point properties. For this to work, we need to be sure that useful properties we discover about $\hat{T}$ will transmit themselves back to properties of $T$, the map that actually interests us.

This section explains a notion of conjugacy that formalizes these ideas. The study of conjugate relationships originated in the field of dynamical systems theory. Later we will apply this approach to operators that arise in contexts of dynamic programming and recursive preferences.

### 2.1.1.1 Conjugacy

A **dynamical system** is a pair $(U, T)$, where $U$ is any set and $T$ is a self-map on $U$. Two dynamical systems $(U, T)$ and $(\hat{U}, \hat{T})$ are said to be **conjugate** under $\Phi$ if $\Phi$ is a bijection from $U$ into $\hat{U}$ such that $T = \Phi^{-1} \circ \hat{T} \circ \Phi$ on $U$.

Conjugacy of $(U, T)$ and $(\hat{U}, \hat{T})$ under $\Phi$ can be understood as follows: shifting a point $u \in U$ to $Tu$ via $T$ is equivalent to moving $u$ into $\hat{U}$ via $\hat{u} = \Phi u$, applying $\hat{T}$, and then moving the result back using $\Phi^{-1}$:

$$
\begin{array}{ccc}
u & \xrightarrow{\ T\ } & Tu \\
\Big\downarrow{\scriptstyle \Phi} & {\scriptstyle \Phi^{-1}} & \Big\uparrow \\
\hat{u} & \xrightarrow{\ \hat{T}\ } & \hat{T}\hat{u}
\end{array}
$$

**Example 2.1.1.** Let $A$ be $n \times n$ **diagonalizable**, meaning that there exists a diagonal matrix $D$ and a matrix $P$ such that $A = P^{-1}DP$. We regard $A$ as a self-map on $\mathbb{R}^n$, $D$ as a self-map on $\mathbb{C}^n$, and $P$ as a map from $\mathbb{R}^n$ to $\mathbb{C}^n$. The identity $A = P^{-1}DP$ implies that the dynamical systems $(A, \mathbb{R}^n)$ and $(D, \mathbb{C}^n)$ are conjugate.

The next two exercises illustrate benefits of establishing a conjugate relationship between two dynamical systems.

EXERCISE 2.1.1. Show that if $(U, T)$ and $(\hat{U}, \hat{T})$ are conjugate under $\Phi$, then $u \in U$ is a fixed point of $T$ on $U$ if and only if $\Phi u \in \hat{U}$ is a fixed point of $\hat{T}$ on $\hat{U}$.

EXERCISE 2.1.2. Extending Exercise 2.1.1, let $(U, T)$ and $(\hat{U}, \hat{T})$ be dynamical systems and let fix$(T)$ and fix$(\hat{T})$ be the set of fixed points of $T$ and $\hat{T}$, respectively. Show that $\Phi$ is a bijection from fix$(T)$ to fix$(\hat{T})$.

The next result summarizes the most important consequences of our findings.

**Proposition 2.1.1.** *If $(U, T)$ and $(\hat{U}, \hat{T})$ are conjugate dynamical systems, then*

(i) *$u$ is a fixed point of $T$ if and only if $\Phi u$ is a fixed point of $\hat{T}$,*

(ii) *$\hat{u}$ is a fixed point of $\hat{T}$ if and only if $\Phi^{-1}\hat{u}$ is a fixed point of $T$, and*

(iii) *the set of fixed points of $T$ and the set of fixed points of $\hat{T}$ have the same cardinality.*

In particular, if $T$ has a unique fixed point on $U$ if and only if $\hat{T}$ has a unique fixed point on $\hat{U}$.

### 2.1.1.2 Topological Conjugacy

Let $U$ and $\hat{U}$ be two subsets of $\mathbb{R}^n$. A function $\Phi$ from $U$ to $\hat{U}$ is called a **homeomorphism** if it is continuous, bijective, and its inverse $\Phi^{-1}$ is also continuous.

**Example 2.1.2.** The map $\Phi u = \ln u$ from $(0, \infty)$ to $\mathbb{R}$ is a homeomorphism, with continuous inverse $\Phi^{-1}y = \exp(y)$.

**Example 2.1.3.** Let $\Phi$ be an $n \times n$ matrix. We can regard $\Phi$ as a map sending column vector $u$ into column vector $\Phi u$. This map is a homeomorphism from $\mathbb{R}^n$ to itself if and only if $\Phi$ is nonsingular.

Assume again that $U$ and $\hat{U}$ are subsets of $\mathbb{R}^n$. In this setting, we say that dynamical systems $(U, T)$ and $(\hat{U}, \hat{T})$ are **topologically conjugate** under $\Phi$ if $(U, T)$ and $(\hat{U}, \hat{T})$ are conjugate under $\Phi$ and, in addition, $\Phi$ is a homeomorphism.

EXERCISE 2.1.3. Let $U := (0, \infty)$ and $\hat{U} := \mathbb{R}$. Let $Tu = Au^{\alpha}$, where $A > 0$ and $\alpha \in \mathbb{R}$, and let $\hat{T}\hat{u} = \ln A + \alpha \hat{u}$. Show that $T$ and $\hat{T}$ are topologically conjugate under $\Phi := \ln$.

EXERCISE 2.1.4. Consider again the setting of Exercise 2.1.1, but now suppose that $(U, T)$ and $(\hat{U}, \hat{T})$ are topologically conjugate under $\Phi$, Fixing $u, u^* \in U$, show that $\lim_{k \to \infty} T^k u = u^*$ if and only if $\lim_{k \to \infty} \hat{T}^k \Phi u = \Phi u^*$.

The next exercise asks you to show that topologically conjugacy is an equivalence relation, as defined in §A.1.

EXERCISE 2.1.5. Let $\mathbf{U}$ be the set of all dynamical systems $(U, T)$ with $U \subset \mathbb{R}^n$. Show that topologically conjugacy is an equivalence relation on $\mathbf{U}$.

From the preceding exercises we can state the following useful result:

**Proposition 2.1.2.** *If $(U, T)$ and $(\hat{U}, \hat{T})$ are topologically conjugate, then*

   (i) *$T$ is globally stable on $U$ if and only if $\hat{T}$ is globally stable on $\hat{U}$, and*

   (ii) *the unique fixed points $u^* \in U$ and $\hat{u}^* \in \hat{U}$ satisfy $\hat{u}^* = \Phi u^*$.*

## 2.1.2 Local Stability

In §1.2.2.2 we investigated global stability. Here we introduce local stability and provide a sufficient condition for situations in which the map is smooth.

Let $U$ be a subset of $\mathbb{R}^n$ and let $T$ be a self-map on $U$. A fixed point $u^*$ of $T$ in $U$ is called **locally stable** for the dynamical system $(U, T)$ if there exists an open set $O \subset U$ such that $u^* \in O$ and $T^k u \to u^*$ as $k \to \infty$ for every $u \in O$. In other words, the domain of attraction for $u^*$ contains an open neighborhood of $u^*$.

**Example 2.1.4.** Consider the self-map $g$ on $\mathbb{R}$ defined by $g(x) = x^2$. The fixed point 1 is not stable (for example, $g^t(x) \to \infty$ for any $x > 1$). However, 0 is locally stable, because $-1 < x < 1$ implies that $g^t(x) \to 0$ as $t \to \infty$.

EXERCISE 2.1.6. Returning to the setting of Exercise 2.1.4, let $(U, T)$ and $(\hat{U}, \hat{T})$ be topologically conjugate and let $u^*$ be a fixed point of $T$ in $U$. Show that $u^*$ is locally stable for $(U, T)$ if and only if $\Phi u^*$ is locally stable for $(\hat{U}, \hat{T})$.

For an interior fixed point $x^*$ of a smooth self-map $g$ on an interval of $\mathbb{R}$, local stability holds whenever $|g'(x^*)| < 1$. The proof strategy proceeds as follows: When $|g'(x^*)| < 1$, the first-order linear approximation

$$\hat{g}(x) := g(x^*) + g'(x^*)(x - x^*) = x^* + g'(x^*)(x - x^*)$$

is a contraction of modulus $|g'(x^*)|$ with unique fixed point $x^*$. Hence all trajectories of $\hat{g}$ converge to $x^*$. Moreover, since $g$ and $\hat{g}$ are similar in a neighborhood of $x^*$, the same is true for trajectories of $g$ starting close to $x^*$.

The next theorem formalizes this line of argument and extends it to multiple dimensions. In stating the theorem, we take $T$ to be a self-map on $U$ with fixed point $u^*$ in $U$ and assume that $T$ is continuously differentiable on $U$. Recall that the **Jacobian** of $T$ at $u \in U$ is

$$J_T(u) := \begin{pmatrix} \frac{\partial T_1}{\partial u_1}(u) & \cdots & \frac{\partial T_1}{\partial u_n}(u) \\ & \cdots & \\ \frac{\partial T_n}{\partial u_1}(u) & \cdots & \frac{\partial T_n}{\partial u_n}(u) \end{pmatrix} \quad \text{where} \quad Tu = \begin{pmatrix} T_1 u \\ \vdots \\ T_n u \end{pmatrix},$$

and let $\hat{T}$ be the first-order approximation to $T$ at $u^*$:

$$\hat{T}u = u^* + J_T(u^*)(u - u^*) \qquad (u \in U).$$

**Theorem 2.1.3** (Hartman–Grobman). *If $J_T(u^*)$ is nonsingular and contains no eigenvalues on the unit circle in $\mathbb{C}$, then there exists an open neighborhood $O$ of $u^*$ such that $(O, T)$ and $(O, \hat{T})$ are topologically conjugate.*

Combining this theorem with the result of Exercise 2.1.6, we see that, under the conditions of the theorem, $u^*$ is globally stable for $(O, T)$, and hence locally stable for $(U, T)$, whenever $(O, \hat{T})$ is globally stable. By the Neumann series lemma, the first-order approximation will be globally stable whenever $J_T(u^*)$ has spectral radius less than one. Thus, we have

**Corollary 2.1.4.** *Under the conditions of Theorem 2.1.3, the fixed point $u^*$ is locally stable whenever $\rho(J_T(u^*)) < 1$.*

### 2.1.3   Convergence Rates

To discuss relative rates of convergence we fix a norm $\|\cdot\|$ on $\mathbb{R}^n$ and take a sequence $(u_k)_{k \geqslant 0} \subset \mathbb{R}^n$ converging to $u^* \in \mathbb{R}^n$. Set $e_k := \|u_k - u^*\|$ for all $k$. We say that $(u_k)$ converges to $u^*$ **at rate at least** $q$ if $q \geqslant 1$ and, for some $\beta \in (0, \infty)$ and $N \in \mathbb{N}$, we have

$$e_{k+1} \leqslant \beta e_k^q \quad \text{for all } k \geqslant N.$$

We say that convergence occurs **at rate** $q$ if, in addition,

$$\limsup_{k \to \infty} \frac{e_{k+1}}{e_k^q} = \beta.$$

When

- $q = 2$ then we say that convergence is (at least) **quadratic**

- $q = 1$ and $\beta < 1$, then we say that convergence is (at least) **linear**.

**Example 2.1.5.** Let $T$ be a contraction of modulus $\lambda$ on a closed set $U \subset \mathbb{R}^n$. If $u^*$ is the unique fixed point of $T$ in $U$ and $u_k := T^k u_0$, then $(u_k)$ converges at least linearly to $u^*$, since

$$e_{k+1} = \|u_{k+1} - u^*\| = \|Tu_k - Tu^*\| \leqslant \lambda e_k.$$

Orders of convergence are studied in the neighborhood of zero, implying that higher orders are faster. For example, suppose $\varepsilon_k := \|u_k - u^*\|$ is the size of the error and that $u_k$ converges to $u^*$ quadratically. If, say, $\varepsilon_k = 10^{-5}$, then $\varepsilon_{k+1} \approx \beta 10^{-10}$. Provided that $\beta$ is not large, the number of accurate digits roughly doubles at each step.

Successive approximations typically converge at a linear rate. To see this in one dimension, try the following exercise.

EXERCISE 2.1.7. Let $T \colon U \to U$ where $T$ is twice continuously differentiable and $U$ is an open interval in $\mathbb{R}$. Suppose that $T$ has a fixed point $u^* \in U$ and that $u_k := T^k u_0$ converges to $u^*$ as $k \to \infty$. Prove that the rate of convergence is linear whenever $0 < |T'u^*| < 1$. In completing the proof, you might find it helpful to use the fact that, by a second order Taylor expansion, there is a $v_k \in (u_k, u^*)$ such that

$$Tu_k = u^* + T'u^*(u_k - u^*) + \frac{T''v_k}{2}(u_k - u^*)^2. \tag{2.1}$$

(The restriction that $0 < |T'u^*| < 1$ in Exercise 2.1.7 is mild. For example, given convergence of successive approximation to the fixed point, we expect $|T'u^*| < 1$, since this inequality implies that $u^*$ is locally stable.)

### 2.1.4   Gradient-Based Methods

While successive approximation always converges when global stability holds, faster fixed point algorithms can often be obtained by leveraging extra information, such as gradients. Newton's method is an important gradient-based technique. (As we discuss in §5.1.4.2, Newton's method is a key component of algorithms for solving dynamic programs.)

While Newton's method is often used to solve for roots of a given function, here we use it to find fixed points.

#### 2.1.4.1   Newton Fixed Point Iteration

Suppose first that $T$ is a differentiable self-map on an open set $U \subset \mathbb{R}^n$ and that we want to find a fixed point of $T$. Our plan is to start with a guess $u_0$ of the fixed point and then update it to $u_1$. To do this we use the first-order approximation $\hat{T}$ of $T$ around $u_0$ and solve for the fixed point of $\hat{T}$ – which we can do exactly since $\hat{T}$ is linear. We take this new point as $u_1$ and then continue.

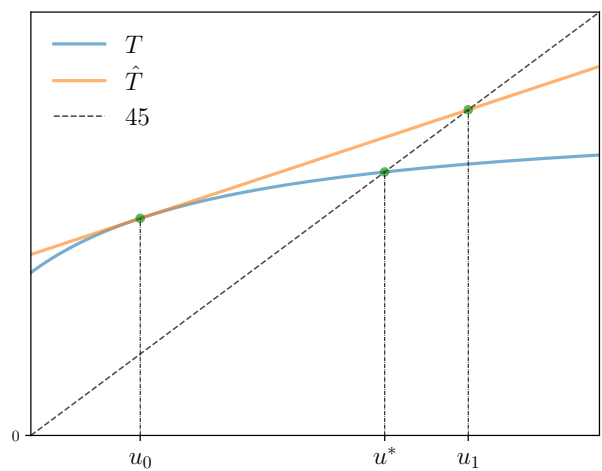

Figure 2.1: First step of Newton's method applied to $T$

If $T$ is one-dimensional then $\hat{T}u := Tu_0 + T'u_0(u - u_0)$. For $n > 1$ we replace $T'u_0$ with the Jacobian of $T$ at $u_0$, which we write as $J_T(u_0)$. We then solve $\hat{T}u_1 = u_1$ for $u_1$, which gives

$$u_1 = (I - J_T(u_0))^{-1}(Tu_0 - J_T(u_0)u_0) \qquad (I \text{ is the } n \times n \text{ identity}).$$

Figure 2.1 shows $u_0$ and $u_1$ when $n = 1$ and $Tu = 1 + u/(u + 1)$ and $u_0 = 0.5$. The value $u_1$ is the fixed point of the first-order approximation $\hat{T}$. It is closer to the fixed point of $T$ than $u_0$, as desired.

**Newton's (fixed point) method** continues in the same way, from $u_1$ to $u_2$ and so on, leading to the sequence of points

$$u_{k+1} = Qu_k \quad \text{where} \quad Qu := (I - J_T(u))^{-1}(Tu - J_T(u)u) \qquad k = 0, 1, \ldots \qquad (2.2)$$

We need not write a new solver, since the successive approximation function in Listing 4 can be applied to $Q$ defined in (2.2).

### 2.1.4.2 Rates of Convergence

Figure 2.2 shows both the Newton approximation sequence and the successive approximation sequence applied to computing the fixed point of the Solow–Swan model from §1.2.3.2. We use two different initial conditions (top and bottom subfigures). Both sequences converge, but the Newton sequences converge faster.

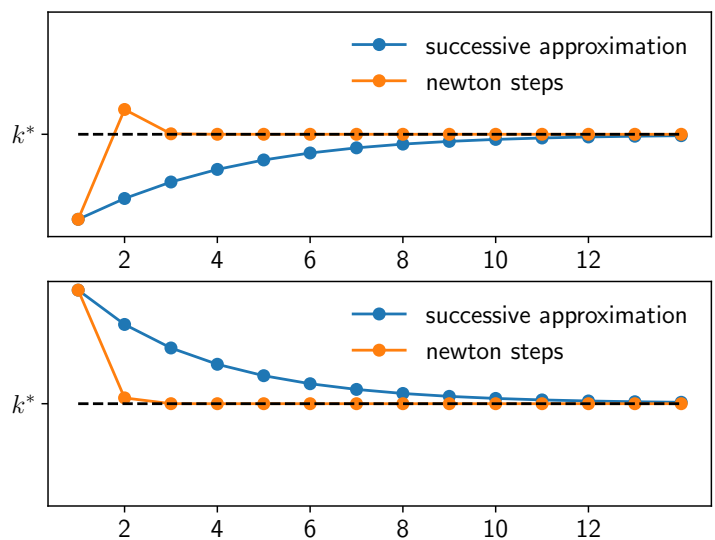

Figure 2.2: Newton's method applied to the Solow–Swan update rule

A fast rate of convergence for Newton scheme can be confirmed theoretically: under mild conditions, there exists a neighborhood of the fixed point within which the Newton iterates converge quadratically. See, for example, Theorem 5.4.1 of Atkinson and Han (2005). Some dynamic programming algorithms take advantage of this fast rate of convergence (see §5.1.4.3).

### 2.1.4.3 Speed vs Robustness

Sometimes we can accelerate computations by exploiting a problem's special structure (e.g., differentiability, convexity, monotonicity). But we often face a trade-off between speed and robustness to details of problem specification. More robust methods impose less structure.

Relative to other algorithms, successive approximation tends to be robust but slow. We saw one illustration of the relatively slow rate of convergence in Figure 2.2. But we can also see its relatively strong robustness properties via the same example, by inspecting Figure 2.3, which compares the update rule of successive approximation (the function $g$) with the update rule for Newton's method (the function $Q$ in (2.2)). Also plotted is the dashed 45 degree line.

The parameterization is the same as for the top subfigure in Figure 1.7. As previously discussed, the shape of $g$ implies global convergence of successive approxi-mation. However, $Q$ is well-behaved near the fixed point (i.e., very flat and hence strongly contractive) but poorly behaved away from the fixed point. This illustrates that Newton's method is fast but generally less robust.

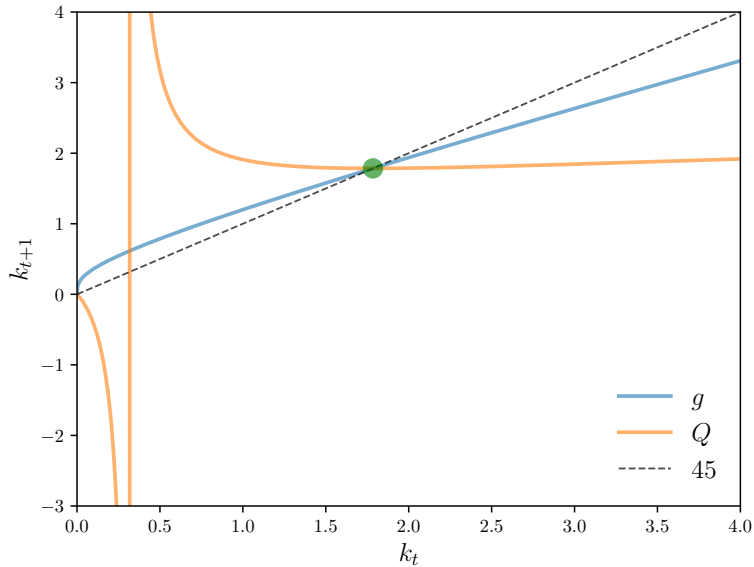

Figure 2.3: Robustness of successive approximation vs Newton's method

### 2.1.4.4 Parallelization

We have discussed rates of convergence for fixed point methods. Mathematicians and computer scientists also analyze algorithms via **worst case complexity**, which measures the number of fundamental operations (e.g., addition and multiplication of floating point numbers) when an algorithm acts on data that is least favorable for good performance. These measures are attractive because they are independent of the software and hardware platforms on which algorithms are implemented.

Software and hardware matter not just for absolute performance of algorithms but also for *relative* performance. For example, although a single update step in successive approximation can often be partially parallelized, the algorithm is inherently serial, in the sense that the $(k + 1)$-th iterate cannot be computed until iterate $k$ is available. Moreover, because the rate of convergence is typically slow (i.e., linear), there can be many small serial steps. This limits parallelization.

Newton's method is also serial to some degree, since we are just iterating with a different map (the operator $Q$ in (2.2)). However, because it involves inverting ma-trices of possibly high dimension, each step is computationally intensive. At the same time, since the rate of convergence is faster, we have to take fewer steps. In this sense, the algorithm is less serial – it involves a smaller number of more expensive steps. Because it is less serial, Newton's method offers far more potential for parallelization. Thus, the speed gain associated with Newton's method can become very large when using effective parallelization.

## 2.2  Order

This section reviews key concepts from order theory.

### 2.2.1  Partial Orders

We define partial orders and examine some of their basic properties.

#### 2.2.1.1  Partially Ordered Sets

A **partial order** on a nonempty set $P$ is a relation $\preceq$ on $P \times P$ that, for any $p, q, r$ in $P$, satisfies

| | |
|---|---|
| $p \preceq p$ | (reflexivity) |
| $p \preceq q$ and $q \preceq p$ implies $p = q$ and | (antisymmetry) |
| $p \preceq q$ and $q \preceq r$ implies $p \preceq r$ | (transitivity) |

The pair $(P, \preceq)$ is called a **partially ordered set**. For convenience, we sometimes write $P$ for $(P, \preceq)$ and $q \succeq p$ for $p \preceq q$. The statement $p \preceq q \preceq r$ means $p \preceq q$ and $q \preceq r$.

**Example 2.2.1.** The usual order $\leqslant$ on $\mathbb{R}$ is a partial order on $\mathbb{R}$. For example, $a \leqslant b$ and $b \leqslant a$ implies $a = b$.

EXERCISE 2.2.1. Let $P$ be any set and consider the relation induced by equality, so that $p \preceq q$ if and only if $p = q$. Show that this relation is a partial order on $P$.

EXERCISE 2.2.2. Let $M$ be any set. Show that set inclusion $\subset$ induces a partial order on $\wp(M)$, the set of all subsets of $M$.

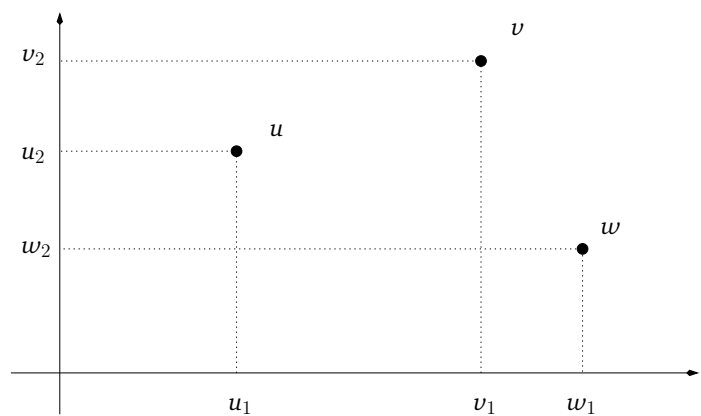

Figure 2.4: Pointwise we have $u \leqslant v$ and $u \ll v$ but not $w \leqslant v$

**Example 2.2.2** (Pointwise partial order)**.** Fix an arbitrary nonempty set X. The **pointwise order** $\leqslant$ on the set $\mathbb{R}^X$ of all functions from X to $\mathbb{R}$ is defined as follows:

$$\text{given } u, v \text{ in } \mathbb{R}^X, \text{ set } u \leqslant v \text{ if } u(x) \leqslant v(x) \text{ for all } x \in X.$$

EXERCISE 2.2.3. Show that the pointwise order $\leqslant$ is a partial order on $\mathbb{R}^X$.

In what follows, for $u, v \in \mathbb{R}^X$, we write $u \ll v$ if $u(x) < v(x)$ for all $x \in X$.

EXERCISE 2.2.4. Show that the relation $\ll$ is *not* a partial order on $\mathbb{R}^X$.

The preceding pointwise concepts extend immediately to vectors, since vectors are just real-valued functions under the identification asserted in Lemma 1.2.4 (page 30). In particular, for vectors $u = (u_1, \ldots, u_n)$ and $v = (v_1, \ldots, v_n)$ in $\mathbb{R}^n$, we write

- $u \leqslant v$ if $u_i \leqslant v_i$ for all $i \in [n]$ and
- $u \ll v$ if $u_i < v_i$ for all $i \in [n]$.

Statements $u \geqslant v$ and $u \gg v$ are defined analogously. Figure 2.4 illustrates. Naturally, $\leqslant$ is called the **pointwise order** on $\mathbb{R}^n$.

EXERCISE 2.2.5. Limits in $\mathbb{R}$ preserve weak inequalities. Use this property to prove that the same is true in $\mathbb{R}^n$. In particular, show that, for vectors $a, b \in \mathbb{R}^n$ and sequence $(u_k)$ in $\mathbb{R}^n$ with $a \leqslant u_k \leqslant b$ for all $k \in \mathbb{N}$ and $u_k \to u \in \mathbb{R}^n$, we have $a \leqslant u \leqslant b$.

**Example 2.2.3** (Pointwise order over matrices)**.** Analogous to vectors, for $n \times k$ matrices $A = (a_{ij})$ and $B = (b_{ij})$, we write

- $A \leqslant B$ if $a_{ij} \leqslant b_{ij}$ for all $i, j$.

- $A \ll B$ if $a_{ij} < b_{ij}$ for all $i, j$.

We call $\leqslant$ the **pointwise order** over matrices.

EXERCISE 2.2.6. Explain why the pointwise order introduced in Example 2.2.3 is also a special case of the pointwise order over functions.

EXERCISE 2.2.7. Prove the next two facts:

(i) If $B$ is $m \times k$ and $B \geqslant 0$, then $|Bu| \leqslant B|u|$ for all $k \times 1$ column vectors $u$.

(ii) If $A$ is $n \times n$ with $A \geqslant 0$ and $(u_k)$ is a sequence in $\mathbb{R}^n$ satisfying $u_{k+1} \leqslant Au_k$ for all $k \geqslant 0$, then $u_k \leqslant A^k u_0$.

EXERCISE 2.2.8. Let $A$ be $n \times k$ and let $u$ and $v$ be $k$-vectors. Prove that $A \gg 0$, $u \leqslant v$ and $u \neq v$ implies $Au \ll Av$.

A partial order $\preceq$ on $P$ is called **total** if, for all $p, q \in P$, either $p \preceq q$ or $q \preceq p$.

**Example 2.2.4.** The usual order $\leqslant$ on $\mathbb{R}$ is a total order, as is the same order on $\mathbb{N}$.

**Example 2.2.5.** Figure 2.4 shows that the pointwise order $\leqslant$ is not a total order on $\mathbb{R}^n$. For example, neither $v \leqslant w$ nor $w \leqslant v$, since $w_1 > v_1$ but $w_2 < v_2$.

EXERCISE 2.2.9. Is the partial order defined in Exercise 2.2.2 a total order? Either prove that it is or provide a counterexample.

### 2.2.1.2 Least and Greatest Elements

Given a partially ordered set $(P, \preceq)$ and $A \subset P$, we say that $g \in P$ is a **greatest element** of $A$ if $g \in A$ and, in addition, $a \in A \implies a \preceq g$. We call $\ell \in P$ a **least element** of $A$ if $\ell \in A$ and, in addition, $a \in A \implies \ell \preceq a$.

If $A$ is totally ordered, then a greatest element $g$ of $A$ is also called a **maximum** of $A$, while a least element $\ell$ of $A$ is also called a **minimum**. See Appendix A for more about maxima and minima.

**Remark 2.2.1.** Elementary optimization problems have real-valued objectives, which means that we seek maxima and minima. In contrast, the objective in dynamic programming is to maximize a lifetime value function (or minimize a lifetime cost function), a *function* over a state space. Thus, the objective takes values in a partially ordered set and we seek greatest (or least) elements.

EXERCISE 2.2.10. Let $P$ be any partially ordered set and fix $A \subset P$. Prove that $A$ has at most one greatest element and at most one least element.

EXERCISE 2.2.11. Let $M$ be a nonempty set and let $\wp(M)$ be the set of all subsets of $M$, partially ordered by $\subset$. Let $\{A_i\} = \{A_i\}_{i \in I}$ be a subset of $\wp(M)$, where $I$ is an arbitrary nonempty index set. Show that $S := \bigcup_i A_i$ is the greatest element of $\{A_i\}$ if and only if $S \in \{A_i\}$.

EXERCISE 2.2.12. Adopt the setting of Exercise 2.2.11 and suppose that $\{A_i\}$ is the set of bounded subsets of $\mathbb{R}^n$. Prove that $\{A_i\}$ has no greatest element.

### 2.2.1.3 Sup and Inf

Concepts of suprema and infima on the real line (Appendix A) extend naturally to partially ordered sets. Given a partially ordered set $(P, \preceq)$ and a nonempty subset $A$ of $P$, we call $u \in P$ an **upper bound** of $A$ if $a \preceq u$ for all $a$ in $A$. Letting $U_P(A)$ be the set of all upper bounds of $A$ in $P$, we call $\bar{u} \in P$ a **supremum** of $A$ if

$$\bar{u} \in U_P(A) \quad \text{and} \quad \bar{u} \preceq u \quad \text{for all} \quad u \in U_P(A).$$

Thus, $\bar{u}$ is the least element (see §2.2.1.2) of the set of upper bounds $U_P(A)$, whenever it exists.

EXERCISE 2.2.13. Prove that $A$ has at most one supremum in $P$.

If $P \subset \mathbb{R}$ and $\preceq$ is $\leqslant$, then the notion of supremum on a partially ordered set reduces to the elementary definition of the supremum for subsets of the real line discussed in Appendix A.

Letting $A$ be a subset of partially ordered space $P$,

- the supremum of $A$ is typically denoted $\bigvee A$.

- If $A = \{a_i\}_{i \in I}$ for some index set $I$, we also write $\bigvee A$ as $\bigvee_i a_i$.

- If $A = \{a, b\}$, then $\bigvee A$ is also written as $a \vee b$.

Suprema and greatest elements are clearly related. The next exercise clarifies this.

EXERCISE 2.2.14. Prove the following statements in the setting described above:

(i) If $\bar{a} = \bigvee A$ and $\bar{a} \in A$, then $\bar{a}$ is a greatest element of $A$.

(ii) If $A$ has a greatest element $\bar{a}$, then $\bar{a} = \bigvee A$.

**Remark 2.2.2.** In view of Exercise 2.2.14, when $A$ has a greatest element, we can refer to it by $\bigvee A$. This notation is used frequently throughout the book.

We call $\ell \in P$ a **lower bound** of $A$ if $a \geq \ell$ for all $a$ in $A$. An element $\bar{\ell}$ of $P$ is called a **infimum** of $A$ if $\bar{\ell}$ is a lower bound of $A$ and $\bar{\ell} \geq \ell$ for every lower bound $\ell$ of $A$. We use analogous notation to denote the infimum. For example, if $A = \{a, b\}$, then $\bigwedge A$ is also written as $a \wedge b$.

EXERCISE 2.2.15. Let $(P, \leq)$ be a partially ordered set and let $A$ be a subset of $P$. Prove that if $\ell$ is a least element of $A$, then $\ell = \bigwedge A$.

EXERCISE 2.2.16. Let $M$ be a nonempty set and let $\wp(M)$ be the set of all subsets of $M$, partially ordered by $\subset$. Let $\{A_i\}_{i \in I}$ be a subset of $\wp(M)$. Prove that $\bigvee_i A_i = \cup_i A_i$ and $\bigwedge_i A_i = \cap_i A_i$.

EXERCISE 2.2.17. Even when $P$ is totally ordered, existence of suprema and infima for an abstract partially ordered set $(P, \leq)$ can fail. Provide an example of a totally ordered set $P$ and a subset $A$ of $P$ that has no supremum in $P$.

## 2.2.2 The Case of Pointwise Order

For us, the pointwise partial order $\leq$ introduced in Example 2.2.2 is especially useful. In this section we review some properties of this order. Throughout, $X$ is an arbitrary finite set.

### 2.2.2.1 Suprema and Infima under a Pointwise Order

Given $u, v \in \mathbb{R}^X$, the symbol $u \wedge v$ is possibly ambiguous because we used the symbol both for a pointwise minimum in §1.2.4.1 and an infimum of $\{u, v\}$ in §2.2.1.3. Fortunately, for elements of the partially ordered set $(\mathbb{R}^X, \leqslant)$, these two definitions coincide. Indeed, if $f(x) := \min\{u(x), v(x)\}$ for all $x \in X$, then

(i) $f$ is a lower bound for $\{u, v\}$ in $(\mathbb{R}^X, \leqslant)$, and

(ii) $g \leqslant u$ and $g \leqslant v$ implies $g \leqslant f$.

Hence $f$ is the infimum of $\{u, v\}$ in $(\mathbb{R}^X, \leqslant)$.

EXERCISE 2.2.18. Prove that the supremum $u \vee v$ of $\{u, v\}$ in $(\mathbb{R}^X, \leqslant)$ is the pointwise maximum $f(x) := \max\{u(x), v(x)\}$.

A subset $V$ of $\mathbb{R}^X$ is called a **sublattice** of $\mathbb{R}^X$ if

$$u, v \in V \text{ implies } u \vee v \in V \text{ and } u \wedge v \in V.$$

**Example 2.2.6.** The sets

$$V_1 := \{f \in \mathbb{R}^X : f \geqslant 0\}, \quad V_2 := \{f \in \mathbb{R}^X : f \gg 0\} \text{ and } V_3 := \{f \in \mathbb{R}^X : |f| \leqslant 1\}$$

are all sublattices of $\mathbb{R}^X$.

Above we discussed the fact that, for a pair of functions $\{u, v\}$, the supremum in $(\mathbb{R}^X, \leqslant)$ is the pointwise maximum, while the infimum in $(\mathbb{R}^X, \leqslant)$ is the pointwise minimum. The same principle holds for finite collections of functions. Thus, if $\{v_i\} := \{v_i\}_{i \in I}$ is a finite subset of $\mathbb{R}^X$, then, for all $x \in X$,

$$\left( \bigvee_i v_i \right)(x) := \max_{i \in I} v_i(x) \quad \text{and} \quad \left( \bigwedge_i v_i \right)(x) := \min_{i \in I} v_i(x).$$

EXERCISE 2.2.19. Verify these claims.

EXERCISE 2.2.20. Show that if $V$ is a sublattice and $\{v_i\}$ is a finite collection of functions in $V$, then $\bigvee_i v_i$ and $\bigwedge_i v_i$ are also in $V$.

The next example discusses greatest elements in the setting of pointwise order.

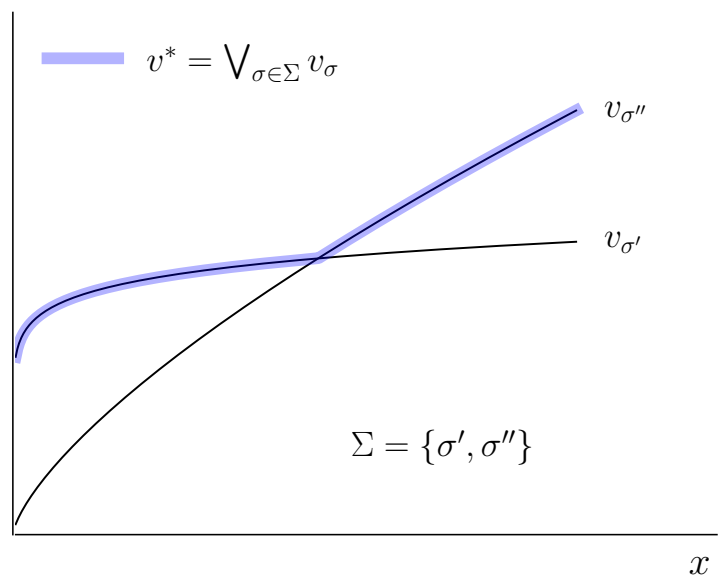

Figure 2.5: $v^*$ is the upper envelope of $\{v_\sigma\}_{\sigma \in \Sigma}$

**Example 2.2.7.** Let $\mathsf{X}$ be nonempty and fix $V \subset \mathbb{R}^{\mathsf{X}}$. Let $V$ be partially ordered by the pointwise order $\leqslant$. Let $\{v_\sigma\} := \{v_\sigma\}_{\sigma \in \Sigma}$ be a finite subset of $V$ and let $v^* := \vee_\sigma v_\sigma \in \mathbb{R}^{\mathsf{X}}$ be the pointwise maximum. If $v^* \in \{v_\sigma\}$, then $v^*$ is the greatest element of $\{v_\sigma\}$. If not, then $\{v_\sigma\}$ has no greatest element.

Figure 2.5 helps illustrate Example 2.2.7. In this case, $v^*$ is not in $\{v_\sigma\}$ and $\{v_\sigma\}$ has no greatest element (since neither $v_{\sigma'} \leqslant v_{\sigma''}$ nor $v_{\sigma''} \leqslant v_{\sigma'}$).

**EXERCISE 2.2.21.** Prove the two claims at the end of Example 2.2.7.

Given a partially ordered set $(P, \preceq)$ and $a, b \in P$, the **order interval** $[a, b]$ is defined as all $p \in P$ such that $a \preceq p \preceq b$. (If $a \preceq b$ fails, the order interval is empty.)

**EXERCISE 2.2.22.** Let $V$ be a sublattice of $\mathbb{R}^{\mathsf{X}}$. Show that the intersection of any two order intervals in $V$ is an order interval in $V$.

### 2.2.2.2 Inequalities and Identities

In this section we note some useful inequalities and identities related to the pointwise partial order on $\mathbb{R}^{\mathsf{X}}$. As before, $\mathsf{X}$ is any finite set.

**Lemma 2.2.1.** *For $f, g, h \in \mathbb{R}^X$, the following statements are true:*

  (i) $|f + g| \leqslant |f| + |g|$.

  (ii) $(f \wedge g) + h = (f + h) \wedge (g + h)$ *and* $(f \vee g) + h = (f + h) \vee (g + h)$.

  (iii) $(f \vee g) \wedge h = (f \wedge h) \vee (g \wedge h)$ *and* $(f \wedge g) \vee h = (f \vee h) \wedge (g \vee h)$.

  (iv) $|f \wedge h - g \wedge h| \leqslant |f - g|$.

  (v) $|f \vee h - g \vee h| \leqslant |f - g|$.

These results follow immediately from proofs of corresponding claims when $f, g, h \in \mathbb{R}$. For example, by the usual triangle inequality for scalars, we have $|f(x) + g(x)| \leqslant |f(x)| + |g(x)|$ for all $x \in X$. This is equivalent to the statement $|f + g| \leqslant |f| + |g|$ in (i). Similarly, inequality (v) follows directly from a corresponding scalar inequality that was already proved in Exercise 1.3.1, page 34.

A complete proof of lemma 2.2.1 can be found with Theorem 30.1 of Aliprantis and Burkinshaw (1998).

It is also true that, if $f, g, h \in \mathbb{R}^X_+$, then

$$(f + g) \wedge h \leqslant (f \wedge h) + (g \wedge h). \tag{2.3}$$

EXERCISE 2.2.23. Prove: If $a, b, c \in \mathbb{R}_+$, then $|a \wedge c - b \wedge c| \leqslant |a - b| \wedge c$.

We note the following useful inequality.

**Lemma 2.2.2.** *Let $D$ be a finite set. If $f$ and $g$ are elements of $\mathbb{R}^D$, then*

$$\left| \max_{z \in D} f(z) - \max_{z \in D} g(z) \right| \leqslant \max_{z \in D} |f(z) - g(z)|. \tag{2.4}$$

*Proof.* Fixing $f, g \in \mathbb{R}^D$, we have

$$f = f - g + g \leqslant |f - g| + g$$

$$\therefore \quad \max f \leqslant \max(|f - g| + g) \leqslant \max |f - g| + \max g$$

$$\therefore \quad \max f - \max g \leqslant \max |f - g|$$

Reversing the roles of $f$ and $g$ proves the claim.     ☐

The inequality in Lemma 2.2.2 helps with dynamic programming problems that involve maximization. The next exercise below concerns minimization.

EXERCISE 2.2.24. Prove that, in the setting of Lemma 2.2.2, we have

$$|\min_{z \in D} f(z) - \min_{z \in D} g(z)| \leqslant \max_{z \in D} |f(z) - g(z)|. \tag{2.5}$$

We end this section with a discussion of upper envelopes. To frame the discussion, we take $\{T_\sigma\} := \{T_\sigma\}_{\sigma \in \Sigma}$ to be a finite family of self-maps on a sublattice $V$ of $\mathbb{R}^X$. Consider some properties of the operator $T$ on $V$ defined by

$$Tv = \bigvee_{\sigma \in \Sigma} T_\sigma v \qquad (v \in V).$$

It follows from the sublattice property that $T$ is a self-map on $V$. In some sources, $T$ is called the **upper envelope** of the functions $\{T_\sigma\}$. The following lemma will be useful for dynamic programming.

**Lemma 2.2.3.** *If, for each $\sigma \in \Sigma$, the operator $T_\sigma$ is a contraction of modulus $\lambda_\sigma$ under the supremum norm, then $T$ is a contraction of modulus $\max_\sigma \lambda_\sigma$ under the same norm.*

*Proof.* Let the stated conditions hold and fix $u, v \in V$. Applying Lemma 2.2.2, we get

$$\|Tu - Tv\|_\infty = \max_x |\max_\sigma (T_\sigma u)(x) - \max_\sigma (T_\sigma v)(x)|$$

$$\leqslant \max_x \max_\sigma |(T_\sigma u)(x) - (T_\sigma v)(x)|$$

$$= \max_\sigma \max_x |(T_\sigma u)(x) - (T_\sigma v)(x)|.$$

$$\therefore \quad \|Tu - Tv\|_\infty \leqslant \max_\sigma \|T_\sigma u - T_\sigma v\|_\infty \leqslant \max_\sigma \lambda_\sigma \|u - v\|_\infty.$$

Hence $T$ is a contraction of modulus $\max_\sigma \lambda_\sigma$ on $V$, as claimed. □

## 2.2.3 Order-Preserving Maps

Order-preserving maps appear throughout the theory of dynamic programming. Here we define them and state a condition for contractivity that requires the order preserving property.

### 2.2.3.1 Definition

Given two partially ordered sets $(P, \preceq)$ and $(Q, \trianglelefteq)$, a map $T$ from $P$ to $Q$ is called **order-preserving** if, given $p, p' \in P$, we have

$$p \preceq p' \quad \implies \quad Tp \trianglelefteq Tp'. \tag{2.6}$$

$T$ is called **order-reversing** if, instead,

$$p \preceq p' \quad \implies \quad Tp' \trianglelefteq Tp. \tag{2.7}$$

**Example 2.2.8.** Let $\leqslant$ be the pointwise order on $\mathbb{R}^n$. If $A$ is $n \times n$ with $A \geqslant 0$, then $T \colon \mathbb{R}^n \to \mathbb{R}^n$ defined by $Tu = Au + b$ is order preserving on $\mathbb{R}^n$, since $u \leqslant v$ implies $v - u \geqslant 0$, and hence $A(v - u) \geqslant 0$. But then $Au \leqslant Av$ and hence $Tu \leqslant Tv$.

**Example 2.2.9.** Given $a \leqslant b$ in $\mathbb{R}$, let $C[a, b]$ be all continuous functions from $[a, b]$ to $\mathbb{R}$ and let $\leqslant$ be the pointwise order on $C[a, b]$. Let

$$I(f) := \int_a^b f(x)dx \qquad (f \in C[a, b]).$$

Since $f \leqslant g$ implies $\int_a^b f(x)dx \leqslant \int_a^b g(x)dx$, the map $I$ is order-preserving on $C[a, b]$.

EXERCISE 2.2.25. Let $P, Q$ be partially ordered sets and let $F$ be an order-preserving map from $P$ to $Q$. Suppose that $\{u_i\} \subset P$ has a greatest and a least element. Prove that, in this setting, both $\bigvee_i Fu_i$ and $\bigwedge_i Fu_i$ exist in $Q$, and, moreover,

$$F \bigvee_i u_i = \bigvee_i Fu_i \quad \text{and} \quad F \bigwedge_i u_i = \bigwedge_i Fu_i.$$

EXERCISE 2.2.26. Let $(P, \preceq)$ be a partially ordered set and let $A$ be an order-preserving self-map on $P$. Prove that $A^k$ is order-preserving on $P$ for any $k \in \mathbb{N}$.

EXERCISE 2.2.27. Let $A$ be $n \times k$ with $A \geqslant 0$. Show that the map $u \mapsto Au$ is order-preserving on $\mathbb{R}^k$ under the pointwise order.

EXERCISE 2.2.28. Let $A$ and $B$ be $n \times n$ with $0 \leqslant A \leqslant B$. Prove that $A^k \leqslant B^k$ for all $k \in \mathbb{N}$ and, in addition, that $\rho(A) \leqslant \rho(B)$.

### 2.2.3.2   Increasing and Decreasing Functions

Regarding the definitions in (2.6)–(2.7), when $(Q, \trianglelefteq) = (\mathbb{R}, \leqslant)$, it is common to say "increasing" instead of order-preserving, and "decreasing" instead of order-reversing. We adopt this terminology. In particular, given partially ordered set $(P, \trianglelefteq)$, we call $h \in \mathbb{R}^P$

- **increasing** if $p \trianglelefteq p'$ implies $h(p) \leqslant h(p')$ and

- **decreasing** if $p \trianglelefteq p'$ implies $h(p) \geqslant h(p')$.

We use the symbol $i\mathbb{R}^P$ for the set of increasing functions in $\mathbb{R}^P$.

**Example 2.2.10.** If $P = \{1, \ldots, n\}$ and $\trianglelefteq$ is the usual order $\leqslant$ on $\mathbb{R}$, then $x \mapsto 2x$ and $x \mapsto \mathbb{1}\{2 \leqslant x\}$ are in $i\mathbb{R}^P$ but $x \mapsto -x$ and $x \mapsto \mathbb{1}\{x \leqslant 2\}$ are not.

**Remark 2.2.3.** Instead of adopting the fancy terms "order-preserving" and "order-reversing", why not just use "increasing" and "decreasing"? A short answer is that for a general partial order the concepts of order-preserving and order-reversing can be very different from usual notions of increasing and decreasing functions.

EXERCISE 2.2.29. Prove: If $P$ is any partially ordered set and $f, g \in i\mathbb{R}^P$, then

(i) $\alpha f + \beta g \in i\mathbb{R}^P$ whenever $\alpha, \beta \geqslant 0$.

(ii) $f \vee g \in i\mathbb{R}^P$ and $f \wedge g \in i\mathbb{R}^P$.

EXERCISE 2.2.30. Given finite $P$, show that $i\mathbb{R}^P$ is closed in $\mathbb{R}^P$.

EXERCISE 2.2.31. Let $X$ be a random variable taking values in finite $\mathsf{X}$. Define $\ell : \mathbb{R}^\mathsf{X} \to \mathbb{R}$ by $\ell h = \mathbb{E} h(X)$. Show that $\ell$ is increasing when $\mathbb{R}^\mathsf{X}$ has the pointwise order.

The next exercise shows that, in a totally ordered setting, an increasing function can be represented as the sum of increasing binary functions.

EXERCISE 2.2.32. Let $\mathsf{X} = \{x_1, \ldots, x_n\}$ where $x_k \leq x_{k+1}$ for all $k$. Show that, for any $u \in i\mathbb{R}^\mathsf{X}$, there exist $s_1, \ldots, s_n$ in $\mathbb{R}_+$ such that $u(x) = \sum_{k=1}^n s_k \mathbb{1}\{x \geq x_k\}$ for all $x \in \mathsf{X}$.

As usual, if $h \colon P \to Q$ and $P, Q \subset \mathbb{R}$, then we will call $h$

- **strictly increasing** if $x < y$ implies $h(x) < h(y)$, and

- **strictly decreasing** if $x < y$ implies $h(x) > h(y)$.

### 2.2.3.3  Order Isomorphisms

Let $(P, \leq)$ and $(Q, \trianglelefteq)$ be two partially ordered sets. A map $F$ from $P$ onto $Q$ is called an

- **order isomorphism** if $p \leq p' \iff Fp \trianglelefteq Fp'$, and an
- **order anti-isomorphism** if $p \leq p' \iff Fp' \trianglelefteq Fp$.

(Note that $F$ is required to be onto, so each $q \in Q$ has a preimage under $F$.)

**Example 2.2.11.** Let $P = Q = \mathbb{R}_+^n$ with the usual pointwise order. Consider $F$ mapping $p \in P$ to $p^2 \in Q$, where the operation $p \mapsto p^2$ acts pointwise. $F$ is clearly onto and, for $p \geq 0$, we have $p \leq p'$ if and only if $p^2 \leq (p')^2$. Hence $F$ is an order isomorphism.

Every order isomorphism $F$ from $P$ to $Q$ is a bijection. To see this, observe that, by reflexivity, $Fp = Fp'$ implies $Fp \trianglelefteq Fp'$ and $Fp' \trianglelefteq Fp$. Since $F$ is order isomorphic, this yields $p \leq p'$ and $p' \leq p$. Using antisymmetry, we get $p = p'$. Hence $F$ is one-to-one as well as onto, and therefore bijective.

EXERCISE 2.2.33. Show that every order anti-isomorphism is a bijection.

EXERCISE 2.2.34. Let $F$ be a bijection from $(P, \leq)$ to $(Q, \trianglelefteq)$. Show that

(i) $F$ is an order isomorphism if and only if $F$ and $F^{-1}$ are order-preserving, and

(ii) $F$ is an order anti-isomorphism if and only if $F$ and $F^{-1}$ are order-reversing.

EXERCISE 2.2.35. Let $P$ and $Q$ be order-isomorphic; i.e., suppose that there exists an order isomorphism from $P$ to $Q$. Prove: If $P$ is totally ordered, then so is $Q$.

### 2.2.3.4  Blackwell's Condition

Our discussion of Banach's theorem in §1.2.2.3 showed the usefulness of contractivity. For an order-preserving operator on a subset of $\mathbb{R}^{\mathsf{X}}$, the following condition often simplifies establishing this property. In the statement of the lemma, $U$ is a subset of $\mathbb{R}^{\mathsf{X}}$, partially ordered by $\leq$, and $\mathsf{X}$ is finite. Also, $U$ has the property that $u \in U$ and $c \in \mathbb{R}_+$ implies $u + c \in U$.

**Lemma 2.2.4.** *If $T$ is an order preserving self-map on $U$ and there exists a constant $\beta \in (0,1)$ such that*

$$T(u + c) \leqslant Tu + \beta c \quad \text{for all } u \in U \text{ and } c \in \mathbb{R}_+, \tag{2.8}$$

*then $T$ is a contraction of modulus $\beta$ on $U$ with respect to the supremum norm.*

*Proof.* Let $U, T$ have the stated properties and fix $u, v \in U$. We have

$$Tu = T(v + u - v) \leqslant T(v + \|u - v\|_\infty) \leqslant Tv + \beta \|u - v\|_\infty.$$

Rearranging gives $Tu - Tv \leqslant \beta \|u - v\|_\infty$. Reversing roles of $u$ and $v$ proves the claim. □

### 2.2.4 Stochastic Dominance

So far we have discussed partial orders over vectors, functions and sets. It is also useful to have a partial order over distributions that tells us when one distribution is in some sense "larger" than another. In this section we introduce a partial order over some distributions commonly used economics and finance.

Let's start with an example. Recall that a random variable $X$ is binomial $B(n, 0.5)$ if it counts the number of heads in $n$ flips of a fair coin. Figure 2.6 shows two distributions, $\varphi \stackrel{d}{=} X \sim B(10, 0.5)$ and $\psi \stackrel{d}{=} Y \sim B(18, 0.5)$. Since $Y$ counts over more flips, we expect it to take larger values in some sense, and we also expect its distribution $\psi$ to reflect this. How can we make these thoughts precise?

A standard order over distributions that captures this idea is defined as follows: Given finite set X partially ordered by $\preceq$ and $\varphi, \psi \in \mathcal{D}(\mathsf{X})$, we say that $\psi$ **stochastically dominates** $\varphi$ and write $\varphi \preceq_{\mathrm{F}} \psi$ if

$$\sum_x u(x)\varphi(x) \leqslant \sum_x u(x)\psi(x) \quad \text{for every } u \text{ in } i\mathbb{R}^{\mathsf{X}} \tag{2.9}$$

The relation $\preceq_{\mathrm{F}}$ is also called **first order stochastic dominance** to differentiate it from other forms of stochastic order.

**Example 2.2.12.** If $\varphi$ and $\psi$ are the binomial distributions defined above and $\mathsf{X} = \{0, \dots, 18\}$, then $\varphi \preceq_{\mathrm{F}} \psi$ holds. Indeed, if $W_1, \dots, W_{18}$ are IID binary random variables with $\mathbb{P}\{W_i = 1\} = 0.5$ for all $i$, then $X := \sum_{i=1}^{10} W_i$ has distribution $\varphi$ and $Y := \sum_{i=1}^{18} W_i$ has distribution $\psi$. In addition, $X \leqslant Y$ with probability one (i.e., for any outcome of the draws $W_1, \dots, W_{18}$). It follows that, for any given $u \in i\mathbb{R}^{\mathsf{X}}$, we have $u(X) \leqslant u(Y)$ with probability one. Hence $\mathbb{E}u(X) \leqslant \mathbb{E}u(Y)$ holds, which is the same statement as (2.9).

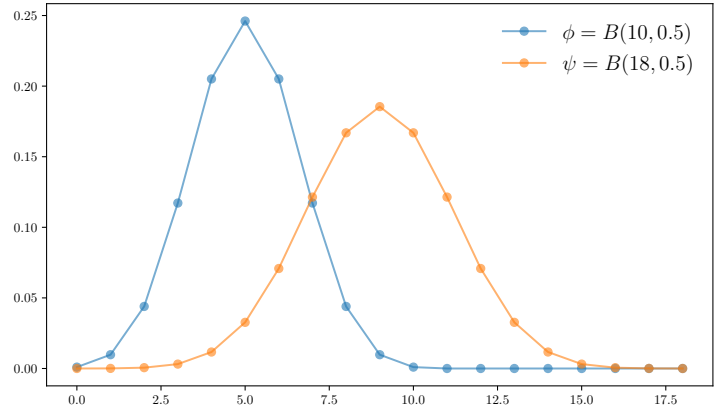

Figure 2.6: Two binomial distributions

A good way to interpret first order stochastic dominance is to suppose that an agent has preferences over outcomes in $\mathsf{X}$ described by a utility function $u \in \mathbb{R}^{\mathsf{X}}$. Suppose in addition that the agent prefers more to less, in the sense that $u \in i\mathbb{R}^{\mathsf{X}}$, and that the agent ranks lotteries over $\mathsf{X}$ according to expected utility, so that the agent evaluates $\varphi \in \mathcal{D}(\mathsf{X})$ according to $\sum_x u(x)\varphi(x)$. Then the agent (weakly) prefers $\psi$ to $\varphi$ whenever $\varphi \preceq_{\mathrm{F}} \psi$.

We can say more. Consider the class $\mathcal{A}$ of *all* agents who (a) have preferences over outcomes in $\mathsf{X}$, (b) prefer more to less, and (c) rank lotteries over $\mathsf{X}$ according to expected utility. Then $\varphi \preceq_{\mathrm{F}} \psi$ if and only if every agent in $\mathcal{A}$ prefers $\psi$ to $\varphi$.

**Remark 2.2.4.** The last paragraph helps explain the pervasiveness of stochastic dominance in economics. It is standard to assume that economic agents have increasing utility functions and use expected utility to rank lotteries. In such environments, an upward shift in a lottery, as measured by stochastic dominance, makes all agents better off.

EXERCISE 2.2.36. A simple setting in which we can study stochastic dominance is where $\mathsf{X} = \{1, 2\}$ and $\mathsf{X}$ is partially ordered by $\leqslant$. In this case, $\varphi \preceq_{\mathrm{F}} \psi$ if and only $\varphi$ puts more mass on 1 than $\psi$, and, equivalently, less mass on 2. That is,

$$\varphi \preceq_{\mathrm{F}} \psi \iff \psi(1) \leqslant \varphi(1) \iff \varphi(2) \leqslant \psi(2).$$

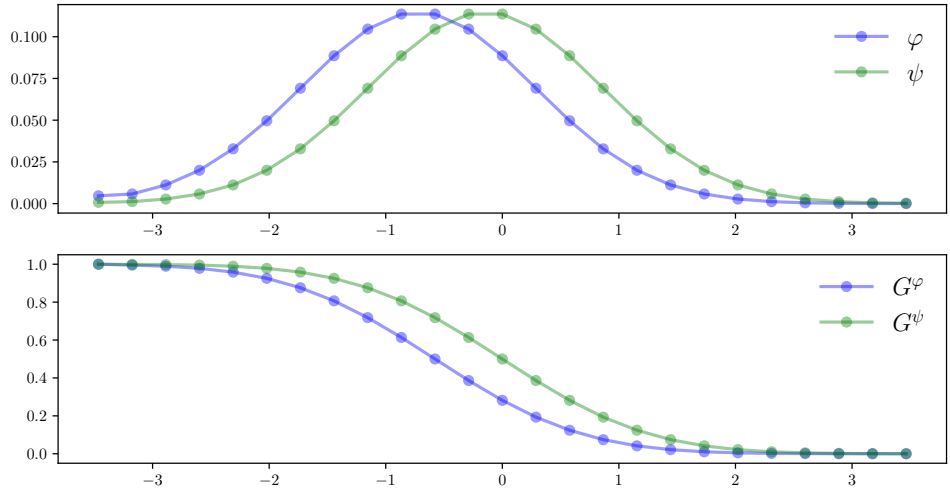

Figure 2.7: Visualization of $\varphi \preceq_F \psi$

Verify the equivalence of these statements.

To state another useful perspective on stochastic dominance, we introduce the notation

$$G^\varphi(y) := \sum_{x \geq y} \varphi(x) \qquad (\varphi \in \mathcal{D}(\mathsf{X}), \ y \in \mathsf{X}).$$

For a given distribution $\varphi$, the function $G^\varphi$ is sometimes called the **counter CDF** (counter cumulative distribution function) of $\varphi$.

**Lemma 2.2.5.** *For each $\varphi, \psi \in \mathcal{D}(\mathsf{X})$, the following statements hold:*

(i) $\varphi \preceq_F \psi \implies G^\varphi \leqslant G^\psi$.

(ii) *If $\mathsf{X}$ is totally ordered by $\leq$, then $G^\varphi \leqslant G^\psi \implies \varphi \preceq_F \psi$.*

The proof is given on page 363. Figure 2.7 helps to illustrate. Here $\mathsf{X} \subset \mathbb{R}$ and $\varphi$ and $\psi$ are distributions on $\mathsf{X}$. We can see that $\varphi \preceq_F \psi$ because the counter CDFs are ordered in the sense that $G^\varphi \leqslant G^\psi$ pointwise on $\mathsf{X}$.

**Lemma 2.2.6.** *Stochastic dominance is a partial order on $\mathcal{D}(\mathsf{X})$.*

Exercise 2.2.37. Prove the transitivity component of Lemma 2.2.6, i.e., prove that $\preceq_F$ is transitive on $\mathcal{D}(\mathsf{X})$.

EXERCISE 2.2.38. Fix $\tau \in (0, 1]$ and let $Q_\tau$ be the quantile function defined on page 32. Choose $\varphi, \psi \in \mathcal{D}(\mathsf{X})$ and let $X, Y$ be $\mathsf{X}$-valued random variables with distributions $\varphi$ and $\psi$ respectively. Prove that $\varphi \leq_F \psi$ implies $Q_\tau(X) \leqslant Q_\tau(Y)$.

### 2.2.5  Parametric Monotonicity

We are often interested in whether a change in a parameter shifts an outcome up or down. For example, a parameter might appear in a central bank decision rule for pegging an interest rate, and we want to know whether increasing that parameter will increase steady state inflation. By providing sufficient conditions for monotone shifts in fixed points, results in this section can help answer such questions.

Let $(P, \preceq)$ be a partially ordered set. Given two self-maps $S$ and $T$ on a set $P$, we write $S \preceq T$ if $Su \preceq Tu$ for every $u \in P$ and say that $T$ **dominates** $S$ on $P$.

**Example 2.2.13.** Let $P = (\mathbb{R}^n_+, \leqslant)$, let $Su = Au + b$ and $Tu = Bu + b$, where $b \in P$ and $A$ and $B$ are $n \times n$ with $0 \leqslant A \leqslant B$. For any $u \in P$, we have $Au \leqslant Bu$. Hence $Su \leqslant Tu$ and $T$ dominates $S$ on $P$.

EXERCISE 2.2.39. Let $(P, \preceq)$ be a partially ordered set and let $S$ and $T$ be order-preserving self-maps such that $S \preceq T$. Show that $S^k \preceq T^k$ holds for all $k \in \mathbb{N}$.

EXERCISE 2.2.40. Let $(P, \preceq)$ be a partially ordered set, let $\mathbb{S}$ be the set of all self-maps on $P$ and, as above, write $S \preceq T$ if $T$ dominates $S$ on $P$. Show that $\preceq$ is a partial order on $\mathbb{S}$.

One might assume that, in a setting where $T$ dominates $S$, the fixed points of $T$ will be larger. This can hold, as in Figure 2.8, but it can also fail, as in Figure 2.9. A difference between these two situations is that in Figure 2.8 the map $T$ is globally stable. This leads us to our next result.

**Proposition 2.2.7.** *Let $S$ and $T$ be self-maps on $M \subset \mathbb{R}^n$ and let $\leqslant$ be the pointwise order. If $T$ dominates $S$ on $M$ and, in addition, $T$ is order-preserving and globally stable on $M$, then its unique fixed point dominates any fixed point of $S$.*

*Proof of Proposition 2.2.7.* Assume the conditions of the proposition and let $u_T$ be the unique fixed point of $T$. Let $u_S$ be any fixed point of $S$. Since $S \leqslant T$, we have $u_S = Su_S \leqslant Tu_S$. Applying $T$ to both sides of this inequality and using the order-preserving property of $T$ and transitivity of $\leqslant$ gives $u_S \leqslant T^2 u_S$. Continuing in this fashion yields $u_S \leqslant T^k u_S$ for all $k \in \mathbb{N}$. Taking the limit in $k$ and using the fact that $\leqslant$ is closed under limits gives $u_S \leqslant u_T$. □

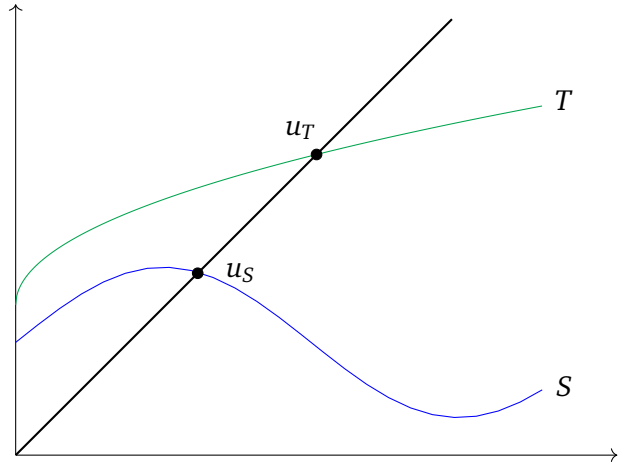

Figure 2.8: Ordered fixed points when global stability holds

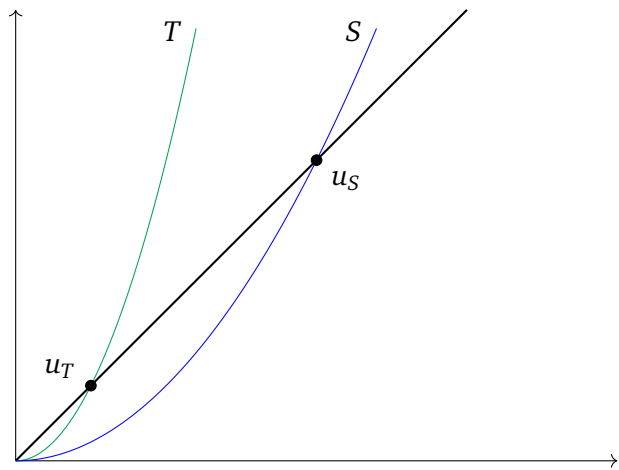

Figure 2.9: Reverse-ordered fixed points when global stability fails

As an application of Proposition 2.2.7, consider again the Solow–Swan growth model $k_{t+1} = g(k_t) \coloneqq sf(k_t) + (1-\delta)k_t$. We saw in §1.2.3.2 that if $f(k) = Ak^\alpha$ where $A > 0$ and $\alpha \in (0, 1)$, then $g$ is globally stable on $M \coloneqq (0, \infty)$. Clearly $k \mapsto g(k)$ is order-preserving on $M$. If we now increase, say, the savings rate $s$, then $g$ will be shifted up everywhere, implying, via Proposition 2.2.7, that the fixed point also rises. Exercise 2.2.41 asks you to step through the details.

EXERCISE 2.2.41. Let $g(k) = sAk^\alpha + (1-\delta)k$ where all parameters are strictly positive, $\alpha \in (0, 1)$ and $\delta \leqslant 1$. Let $k^*(s, A, \alpha, \delta)$ be the unique fixed point of $g$ in $M$. Without using the expression we derived for $k^*$ previously (Exercise 1.2.26), show that

(i) $k^*(s, A, \alpha, \delta)$ is increasing in $s$ and $A$.

(ii) $k^*(s, A, \alpha, \delta)$ is decreasing in $\delta$.

Figure 2.10 helps illustrate the results of Exercise 2.2.41. The top left sub-figure shows a baseline parameterization, with $A = 2.0$, $s = \alpha = 0.3$ and $\delta = 0.4$. The other sub-figures show how the steady state changes as parameters deviate from that baseline.

EXERCISE 2.2.42. In (1.33) on page 40, we defined a map $g$ such that the optimal continuation value $h^*$ is a fixed point. Using this construction, prove that $h^*$ is increasing in $\beta$.

Figure 2.11 gives an illustration of the result in Exercise 2.2.42. Here an increase in $\beta$ leads to a larger continuation value. This seems reasonable, since larger $\beta$ indicates more concern about outcomes in future periods.

While the preceding examples of parametric monotonicity are all one-dimensional, we will soon see that Proposition 2.2.7 can also be applied in high-dimensional settings.

## 2.3 Matrices and Operators

Many aspects of dynamic programming are most clearly framed using operator theory. In this section, we discuss linear operators and their connections to matrices.

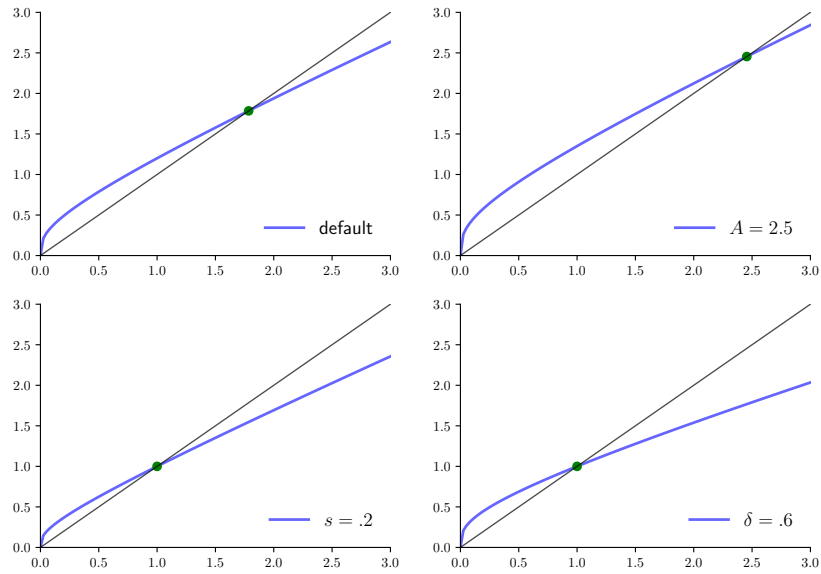

Figure 2.10: Parametric monotonicity for the Solow-Swan model

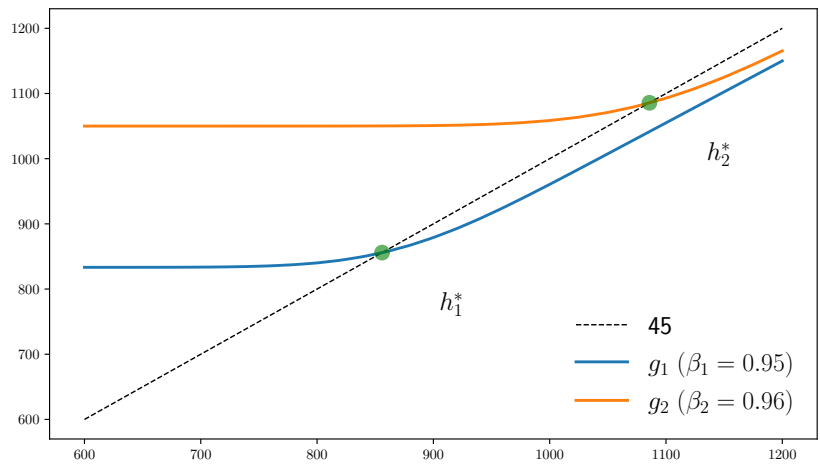

Figure 2.11: Parametric monotonicity in $\beta$ for the continuation value

We emphasize nonnegative matrices and so-called positive linear operators that arise naturally in dynamic programming.

### 2.3.1 Nonnegative Matrices

We begin by reviewing basic properties of nonnegative matrices. The Perron–Frobenius theorem is a key result.

#### 2.3.1.1 Nonnegative Matrices and their Powers

We call a matrix $A$ **nonnegative** and write $A \geqslant 0$ if all elements of $A$ are nonnegative. We call $A$ **everywhere positive** and write $A \gg 0$ if all elements of $A$ are strictly positive. A square matrix $A$ is called **irreducible** if $A \geqslant 0$ and $\sum_{k=1}^{\infty} A^k \gg 0$. An interpretation in terms of connected networks is given in Chapter 1 of Sargent and Stachurski (2023b).

Let $A$ be $n \times n$. It is not always true that the spectral radius $\rho(A)$ is an eigenvalue of $A$.[1] However, when $A \geqslant 0$, the spectral radius is always an eigenvalue. The following theorem states this result and several extensions.

**Theorem 2.3.1** (Perron–Frobenius). *If $A \geqslant 0$, then $\rho(A)$ is an eigenvalue of $A$ with nonnegative, real-valued right and left eigenvectors. In particular, we can find a nonnegative, nonzero column vector $e$ and a nonnegative, nonzero row vector $\varepsilon$ such that*

$$Ae = \rho(A)e \quad \text{and} \quad \varepsilon A = \rho(A)\varepsilon. \tag{2.10}$$

*If $A$ is irreducible, then the right and left eigenvectors are everywhere positive and unique. Moreover, if $A$ is everywhere positive, then with $e$ and $\varepsilon$ normalized so that $\langle \varepsilon, e \rangle = 1$, we have*

$$\rho(A)^{-t} A^t \to e\,\varepsilon \qquad (t \to \infty). \tag{2.11}$$

The convergence in (2.11) provides a sharp characterization of large powers of $A$ that will prove useful in what follows. The assumption that $A$ is everywhere positive can be weakened without affecting this convergence. A complete statement and full proof of the Perron–Frobenius theorem can be found in Meyer (2000).

**Remark 2.3.1.** Note that, in general, if $\nu$ is an everywhere positive real-valued eigenvector for $A$, then so is $\alpha\nu$ for all $\alpha > 0$. Hence the uniqueness asserted in the Perron–Frobenius theorem is up to positive multiples. It tells us that if $e$ is the right eigenvector corresponding to $\rho(A)$ and $\hat{e}$ is another positive vector satisfying $A\hat{e} = \rho(A)\hat{e}$, then $\hat{e} = \alpha e$ for some $\alpha > 0$. A similar statement holds for the left eigenvalue $\varepsilon$.

---

[1]For example, eigenvalues of $A = \mathrm{diag}(-1, 0)$ are $\{-1, 0\}$. Hence $\rho(A) = |-1| = 1$, which is not an eigenvalue of $A$.

We can use the Perron–Frobenius theorem to provide bounds on the spectral radius of a nonnegative matrix. Fix $n \times n$ matrix $A = (a_{ij})$ and set

- $\mathrm{rowsum}_i(A) := \sum_j a_{ij}$ = the $i$-th row sum of $A$ and

- $\mathrm{colsum}_j(A) := \sum_i a_{ij}$ = the $j$-th column sum of $A$.

**Lemma 2.3.2.** *If $A \geqslant 0$, then*

(i) $\min_i \mathrm{rowsum}_i(A) \leqslant \rho(A) \leqslant \max_i \mathrm{rowsum}_i(A)$ *and*

(ii) $\min_j \mathrm{colsum}_j(A) \leqslant \rho(A) \leqslant \max_j \mathrm{colsum}_j(A)$.

EXERCISE 2.3.1. Prove Lemma 2.3.2. (Hint: Since $e$ and $\varepsilon$ are nonnegative and nonzero, and since eigenvectors are defined only up to nonzero multiples, you can assume that both of these vectors sum to 1.)

### 2.3.1.2 A Local Spectral Radius Result

Let $A$ be an $n \times n$ matrix. We know from Gelfand's formula (page 19) that if $\|\cdot\|$ is any matrix norm, then $\|A^k\|^{1/k} \to \rho(A)$ as $k \to \infty$. While useful, this lemma can be difficult to apply because it involves matrix norms. Fortunately, when $A$ is nonnegative, we have the following variation, which only involves vector norms.

**Lemma 2.3.3.** *Let $\|\cdot\|$ be any norm on $\mathbb{R}^n$. If $A$ is nonnegative and $h \in \mathbb{R}^n$ obeys $h \gg 0$, then*

$$\|A^k h\|^{1/k} \to \rho(A) \quad \text{as } k \to \infty. \tag{2.12}$$

The expression on the left of (2.12) is sometimes called the **local spectral radius** of $A$ at $h$. Lemma 2.3.3 gives one set of conditions under which a local spectral radius equals the spectral radius. This result will be useful when we examine state-dependent discounting in Chapter 6.

For a proof of Lemma 2.3.3 see Theorem 9.1 of Krasnosel'skii et al. (1972).

### 2.3.1.3 Markov Matrices

An $n \times n$ matrix $P$ is called a **stochastic matrix** or **Markov matrix** if

$$P \geqslant 0 \quad \text{and} \quad P\mathbb{1} = \mathbb{1}$$

where $\mathbb{1}$ is a column vector of ones, so that $P$ is nonnegative and has unit row sums. The Perron–Frobenius theorem will be useful for the following exercise.

EXERCISE 2.3.2. Let $P, Q$ be $n \times n$ Markov matrices. Prove the following facts.

(i) $PQ$ is also a Markov matrix.

(ii) $\rho(P) = 1$.

(iii) There exists a row vector $\psi \in \mathbb{R}_+^n$ such that $\psi \mathbb{1} = 1$ and $\psi P = \psi$.

(iv) If $P$ is irreducible, then the vector $\psi$ in (iii) is everywhere positive and unique, in the sense that no other vector $\psi \in \mathbb{R}_+^n$ satisfies $\psi \mathbb{1} = 1$ and $\psi P = \psi$.

The vector $\psi$ in part (iii) of Exercise 2.3.2 is called a **stationary distribution** for $P$. Such distributions play an important role in the theory of Markov chains. We discuss their interpretation and significance in §3.1.2.

EXERCISE 2.3.3. Given Markov matrix $P$ and constant $\varepsilon > 0$, prove the following result: There exists no $h \in \mathbb{R}^X$ with $Ph \geqslant h + \varepsilon$.

## 2.3.2 A Lake Model

We illustrate the power of the Perron–Frobenius theorem by showing how it helps us analyze a model of employment and unemployment flows in a large population.

The model is sometimes called a "lake model" because there are two pools of workers: those who are currently employed and those who are currently unemployed but still seeking work. The flows between states are as follows:

- Workers exit the labor market at rate $d$.

- New workers enter the labor market at rate $b$.

- Employed workers separate from their jobs and become unemployed at rate $\alpha$.

- Unemployed workers find jobs at rate $\lambda$.

We assume that all parameters lie in $(0, 1)$. New workers are initially unemployed.

Transition rates between two pools appear in Figure 2.12. For example, the rate of flow from employment to unemployment is $\alpha(1 - d)$, which equals the fraction of employed workers who remained in the labor market and separated from their jobs.

Let $e_t$ and $u_t$ be the number of employed and unemployed workers at time $t$ respectively. The total population (of workers) is $n_t := e_t + u_t$. In view of the rates stated above, the number of unemployed workers evolves according to

$$u_{t+1} = (1 - d)\alpha e_t + (1 - d)(1 - \lambda)u_t + bn_t.$$

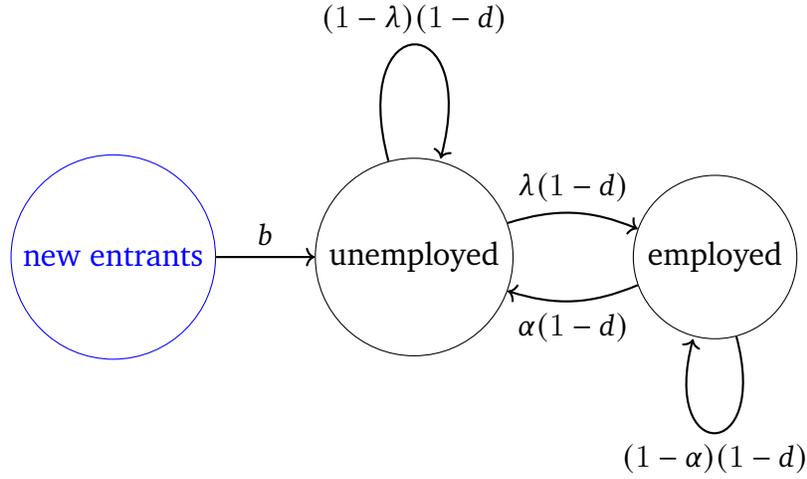

Figure 2.12:  Lake model transition dynamics

The three terms on the right correspond to the newly unemployed (due to separation), the unemployed who failed to find jobs last period, and new entrants into the labor force. The number of employed workers evolves according to

$$e_{t+1} = (1-d)(1-\alpha)e_t + (1-d)\lambda u_t.$$

Evolution of the time series for $u_t$, $e_t$ and $n_t$ is illustrated in Figure 2.13. We set parameters to $\alpha = 0.01$, $\lambda = 0.1$, $d = 0.02$, and $b = 0.025$. The initial population of unemployed and employed workers are $u_0 = 0.6$ and $e_0 = 1.2$, respectively. The series grow over the long run due to net population growth.

Can we say more about the dynamics of this system? For example, what long run unemployment rate should we expect? Also, do long run outcomes depend heavily on the initial conditions $u_0$ and $e_0$? Can we make some general statements that hold regardless of the initial state?

To begin to address these questions, we first organize the linear system for $(e_t)$ and $(u_t)$ by setting

$$x_t := \begin{pmatrix} u_t \\ e_t \end{pmatrix} \quad \text{and} \quad A := \begin{pmatrix} (1-d)(1-\lambda) + b & (1-d)\alpha + b \\ (1-d)\lambda & (1-d)(1-\alpha) \end{pmatrix}. \qquad (2.13)$$

With these definitions, we can write the dynamics as $x_{t+1} = Ax_t$. As a result, $x_t = A^t x_0$, where $x_0 = (u_0 \; e_0)^\top$.

The overall growth rate of the total labor force is $g = b - d$, in the sense that $n_{t+1} = (1+g)n_t$ for all $t$.

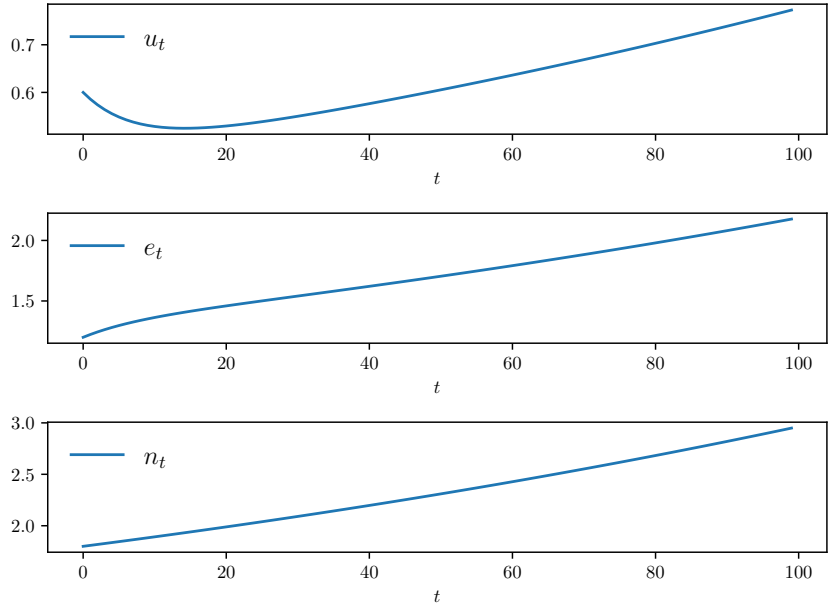

Figure 2.13:  Time series for $e_t$, $u_t$ and $n_t$, (`lake_2.jl`)

EXERCISE 2.3.4. Confirm this claim by using the equation $x_{t+1} = Ax_t$.

EXERCISE 2.3.5. Prove that $\rho(A) = 1+g$.  [Hint: Use one of the results in §2.3.1.1.]

EXERCISE 2.3.6. By the Perron-Frobenius theorem, $1+g$ is an eigenvalue (in fact the dominant eigenvalue) of $A$. Show that $\mathbb{1}^\top := (1\ 1)$ is a left eigenvector corresponding to this eigenvalue.

EXERCISE 2.3.7. Prove that the unique right eigenvector $\bar{x}$ satisfying $A\bar{x} = \rho(A)\bar{x}$ and $\mathbb{1}^\top \bar{x} = 1$ is given by

$$\bar{x} := \begin{pmatrix} \bar{u} \\ \bar{e} \end{pmatrix} \quad \text{with} \quad \bar{u} := \frac{1+g-(1-d)(1-\alpha)}{1+g-(1-d)(1-\alpha)+(1-d)\lambda} \tag{2.14}$$

and $\bar{e} := 1 - \bar{u}$.

In the language of Perron–Frobenius theory, the right eigenvector $\bar{x}$ is called the **dominant eigenvector**, since it corresponds to the dominant (i.e., largest) eigenvalue $\rho(A)$. This eigenvector plays an important role in determining long run outcomes. In the remainder of this section we illustrate this fact.

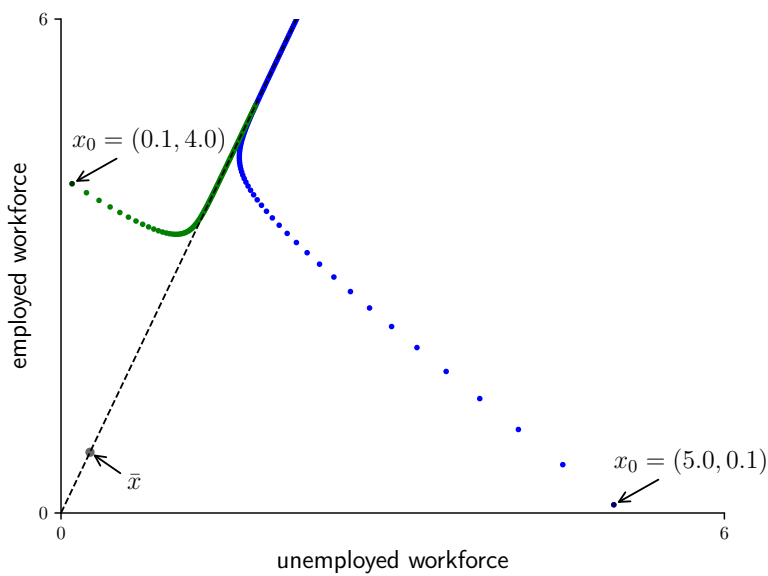

Figure 2.14: Time paths $x_t = A^t x_0$ for two choices of $x_0$ (`lake_1.jl`)

To begin, recall that $\alpha \bar{x}$ is also a right eigenvector corresponding to the eigenvalue $\rho(A)$ when $\alpha > 0$. The set $D := \{x \in \mathbb{R}^2 : x = \alpha \bar{x}$ for some $\alpha > 0\}$ is shown as a dashed black line in Figure 2.14. The figure also shows two time paths, each of the form $(x_t)_{t \geqslant 0} = (A^t x_0)_{t \geqslant 0}$, generated from two different initial conditions. In both cases, we see that both paths converge to $D$ over time. The figure suggests that paths share strong similarities in the long run that are determined by the dominant eigenvector $\bar{x}$.

To see why this is so, we return (2.11) from to the Perron–Frobenius theorem, which tells us that, since $A \gg 0$, we have

$$A^t \approx \rho(A)^t \cdot \bar{x} \mathbb{1}^\top = (1+g)^t \begin{pmatrix} \bar{u} & \bar{u} \\ \bar{e} & \bar{e} \end{pmatrix} \quad \text{for large } t.$$

As a result, for any initial condition $x_0 = (u_0 \; e_0)^\top$, we have

$$A^t x_0 \approx (1+g)^t \begin{pmatrix} \bar{u} & \bar{u} \\ \bar{e} & \bar{e} \end{pmatrix} \begin{pmatrix} u_0 \\ e_0 \end{pmatrix} = (1+g)^t (u_0 + e_0) \begin{pmatrix} \bar{u} \\ \bar{e} \end{pmatrix} = n_t \bar{x},$$

where $n_t = (1+g)^t n_0$ and $n_0 = u_0 + e_0$. This says that, regardless of the initial condition, the state $x_t$ scales along $\bar{x}$ at the rate of population growth. This is precisely what we saw in Figure 2.14.

We can provide additional interpretations to the components $\bar{u}$ and $\bar{e}$ of $\bar{x}$. Since $n_t$ is the size of the workforce at time $t$, the rate of unemployment is $u_t/n_t$. As just shown, for large $t$ this is close to $(n_t\bar{u})/n_t = \bar{u}$. Hence $\bar{u}$ is the long term rate of unemployment along the stable growth path. Similarly, the other component $\bar{e}$ of the dominant eigenvector is the long run employment rate.

In summary, the dominant eigenvector provides with both the long-run rate of unemployment and the stable growth path, to which all trajectories with positive initial conditions converge over time.

**Remark 2.3.2.** A more thorough analysis would require us to think carefully about how the underlying rates $\alpha$, $\lambda$, $b$ and $d$ are determined. For the hiring rate $\lambda$, we could use the job search model to fix the rate at which workers are matched to jobs. In particular, with $w^*$ as the reservation wage, we could set

$$\lambda = \mathbb{P}\{w_t \geqslant w^*\} = \sum_{w \geqslant w^*} \varphi(w)$$

Doing so would allow us to study determinants of $\lambda$ that could include unemployment compensation and workers' impatience.

### 2.3.3   Linear Operators

There are two ways to think about a matrix. In one definition, an $n \times k$ matrix $A$ is an $n \times k$ array of (real) numbers. In the second, $A$ is a linear operator from $\mathbb{R}^k$ to $\mathbb{R}^n$ that takes a vector $u \in \mathbb{R}^k$ and sends it to $Au$ in $\mathbb{R}^n$. Let's clarify these ideas in a setting where $n = k$. While the matrix representation is important, the linear operator representation is more fundamental and more general.

#### 2.3.3.1   Matrices vs Linear Operators

A **linear operator** on $\mathbb{R}^n$ is a map $L$ from $\mathbb{R}^n$ to $\mathbb{R}^n$ such that

$$L(\alpha u + \beta v) = \alpha Lu + \beta Lv \quad \text{for all } u, v \in \mathbb{R}^n \text{ and } \alpha, \beta \in \mathbb{R}. \tag{2.15}$$

(We write $Lu$ instead of $L(u)$, etc.) For example, if $A$ is an $n \times n$ matrix, then the map from $u$ to $Au$ defines a linear operator, since the rules of matrix algebra yield $A(\alpha u + \beta v) = \alpha Au + \beta Av$.

We just showed that each matrix can be regarded as a linear operator. In fact the converse is also true:

**Theorem 2.3.4.** *If $L$ is a linear operator on $\mathbb{R}^n$, then there exists an $n \times n$ matrix $A = (a_{ij})$ such that $Lu = Au$ for all $u \in \mathbb{R}^n$.*

A proof of Theorem 2.3.4 can be found in Kreyszig (1978) and many other sources.

Why introduce linear operators if they are essentially the same as matrices? One reason is that, while a one-to-one correspondence between linear operators and matrices holds in $\mathbb{R}^n$, the concept of linear operators is far more general. Linear operators can be defined over many different kinds of sets whose elements have vector-like properties. This is related to the point that we made about function spaces in Remark 1.2.2 on page 28.

Another reason is computational: the matrix representation of a linear operator can be tedious to construct and difficult to instantiate in memory in large problems. We illustrate this point in §2.3.3.3 below.

### 2.3.3.2 Linear Operators on Function Space

The definition of linear operators on $\mathbb{R}^n$ extends naturally to linear operators on $\mathbb{R}^{\mathsf{X}}$ when $\mathsf{X} = \{x_1, \ldots, x_n\}$: A **linear operator** on $\mathbb{R}^{\mathsf{X}}$ is a map $L$ from $\mathbb{R}^{\mathsf{X}}$ to itself such that, for all $u, v \in \mathbb{R}^{\mathsf{X}}$ and $\alpha, \beta \in \mathbb{R}$, we have $L(\alpha u + \beta v) = \alpha L u + \beta L v$. In what follows,

$$\mathcal{L}(\mathbb{R}^{\mathsf{X}}) := \text{ the set of all linear operators on } \mathbb{R}^{\mathsf{X}}.$$

Let $L$ be a function from $\mathsf{X} \times \mathsf{X}$ to $\mathbb{R}$. This function induces an operator $L$ from $\mathbb{R}^{\mathsf{X}}$ to itself via

$$(Lu)(x) = \sum_{x' \in \mathsf{X}} L(x, x') u(x') \qquad (x \in \mathsf{X},\ u \in \mathbb{R}^{\mathsf{X}}). \tag{2.16}$$

We use the same symbol $L$ on both sides of the equals sign because both represent essentially the same object (in the sense that a matrix $A$ can be viewed as a collection of numbers $(A_{ij})$ or as a linear map $u \mapsto Au$).

The function $L$ on the right-hand side of (2.16) is sometimes called the "kernel" of the operator $L$. However, we will call it a matrix in what follows, since $L(x, x') = L(x_i, x_j)$ is just an $n \times n$ array of real numbers. When more precision is required, we will call it the **matrix representation** of $L$.

In essence, the operation in (2.16) is just matrix multiplication: $(Lu)(x)$ is row $x$ of the matrix product $Lu$.

EXERCISE 2.3.8. Confirm that $L$ on the left-hand side of (2.16) is in fact a linear operator (i.e., an element of $\mathcal{L}(\mathbb{R}^{\mathsf{X}})$).

The eigenvalues and eigenvectors of the linear operator $L$ are defined as the eigenvalues and eigenvectors of its matrix representation. The spectral radius $\rho(L)$ of $L$ is defined analogously.

We used the same symbol for the operator $L$ on the left-hand side of (2.16) and its matrix representation on the right because these two objects are in one-to-one correspondence. In particular, every $L \in \mathcal{L}(\mathbb{R}^{\mathsf{X}})$ can be expressed in the form of (2.16) for a suitable choice of matrix $(L(x, x'))$. Readers who are comfortable with these claim can skip ahead to §2.3.3.3. The next lemma provides more details.

**Lemma 2.3.5.** *When* $\mathsf{X} = \{x_1, \ldots, x_n\}$, *the following sets are in one-to-one correspondence:*

(a) *The set of all $n \times n$ real matrices.*

(b) *The set of all linear operators on $\mathbb{R}^n$.*

(c) *The set $\mathcal{L}(\mathbb{R}^{\mathsf{X}})$ of linear operators on $\mathbb{R}^{\mathsf{X}}$.*

(d) *The set of all functions from $\mathsf{X} \times \mathsf{X}$ to $\mathbb{R}$.*

Lemma 2.3.5 needs no formal proof. Theorem 2.3.4 already tells us that (a) and (b) are in one-to-one correspondence. Also, (b) and (c) are in one-to-one correspondence because each $L \in \mathcal{L}(\mathbb{R}^{\mathsf{X}})$ can be identified with a linear operator $u \mapsto Lu$ on $\mathbb{R}^n$ by pairing $u, Lu \in \mathbb{R}^{\mathsf{X}}$ with its vector representation in $\mathbb{R}^n$ (see §1.2.4.2). Finally, (d) and (a) are in one-to-one correspondence under the identification $L(x_i, x_j) \leftrightarrow L_{ij}$.

### 2.3.3.3 Computational Issues

At the end of §2.3.3.1 we claimed that working with linear operators brings some computational advantages vis-à-vis working with matrices. This section fills in some details (Readers who prefer not to think about computational issues at this point can skip ahead to §2.3.3.4.)

To illustrate the main idea, consider a setting where the state space $\mathsf{X}$ takes the form $\mathsf{X} = \mathsf{Y} \times \mathsf{Z}$ with $|\mathsf{Y}| = j$ and $|\mathsf{Z}| = k$. A typical element of $\mathsf{X}$ is $x = (y, z)$. As we shall see, this kind of setting arises naturally in dynamic programming.

Let $Q$ be a map from $\mathsf{Z} \times \mathsf{Z}$ to $\mathbb{R}$ (i.e., a $k \times k$ matrix) and consider the operator sending $u \in \mathbb{R}^{\mathsf{X}}$ to $Lu \in \mathbb{R}^{\mathsf{X}}$ according to the rule

$$(Lu)(x) = (Lu)(y, z) = \sum_{z' \in \mathsf{Z}} u(y, z') Q(z, z') \tag{2.17}$$

EXERCISE 2.3.9. Prove that $L \in \mathcal{L}(\mathbb{R}^X)$.

Since $L$ is a linear operator on $\mathbb{R}^X$, Lemma 2.3.5 tells us that $L$ can be represented as an $n \times n$ matrix $(L(x_i, x_j)) = (L_{ij})$, where $n = |X| = j \times k$. To construct this matrix, we first need to "flatten" $Y \times Z$ into a set $X = \{x_1, \ldots, x_n\}$ with a single index. There are two natural ways to do this. Considering $Y \times Z$ as a two-dimensional array with typical element $(y_i, z_j)$, we can (a) stack all $k$ columns vertically into one long column, or (b) concatenate all $j$ rows into one long row. The first arrangement is called **column-major ordering** and is the default for languages such as Julia and Fortran. The second is called **row-major ordering** and is the default for languages such as Python and C. Either way we obtain a set of elements indexed by $1, \ldots, n$.

After adopting one of these conventions, Lemma 2.3.5 assures us we can construct a uniquely defined $n \times n$ matrix that represents $L$. Once we decide how to construct this matrix, we can instantiate it in computer memory and compute the operation $u \mapsto Lu$ by matrix multiplication.

There are, however, several disadvantages to implementing $L$ using this matrix-based approach. One is that constructing the matrix representation is tedious. Another is that confusion can arise when swapping between column- and row-major orderings in order to shift between languages or to communicate with colleagues. A third is that differences are introduced between computer code and the natural representation (2.17), which can be a source of bugs. A fourth issue is that an $n \times n$ matrix has to be instantiated in memory, even though the linear operation in (2.17) is only an inner product in $\mathbb{R}^k$. The last issue can be alleviated in most languages by employing sparse matrices, but doing so adds boilerplate and can be a source of inefficiency.

Because of these issues, most modern scientific computing environments support linear operators directly, as well as actions on linear operators such as inverting linear maps. These considerations encourage us to take an operator-based approach.

### 2.3.3.4 Positive Operators and Markov Operators

Having agreed on the benefits of an operator-theoretic exposition, let us now describe some kinds of linear operators. We continue to assume that $X$ is a finite set with $n$ elements.

The set $\mathbb{R}^X_+$ of all $u \in \mathbb{R}^X$ with $u \geqslant 0$ is called the **positive cone** of $\mathbb{R}^X$. An operator $L \in \mathcal{L}(\mathbb{R}^X)$ is called **positive** if $L$ is invariant on the positive cone; that is, if

$$u \geqslant 0 \implies Lu \geqslant 0. \qquad (2.18)$$

**Example 2.3.1.** The operator $L \in \mathcal{L}(\mathbb{R}^X)$ defined in (2.17) is positive whenever $Q \geqslant 0$. This is because

$$u \geqslant 0 \implies \sum_{z'} u(y, z') Q(z, z') \geqslant 0 \text{ for all } x = (y, z) \text{ in } X.$$

**Lemma 2.3.6.** *An operator $L \in \mathcal{L}(\mathbb{R}^X)$ is positive if and only if its matrix representation is a nonnegative matrix.*

EXERCISE 2.3.10. Prove Lemma 2.3.6.

**Remark 2.3.3.** The Lemma 2.3.6 characterization suggests that we should really call a linear operator satisfying (2.18) "nonnegative" rather than positive. Nevertheless, the "positive" terminology is standard (see, e.g., Zaanen (2012)).

EXERCISE 2.3.11. Given $L \in \mathcal{L}(\mathbb{R}^X)$, prove the following statement: $L$ is positive if and only if $L$ is order-preserving on $\mathbb{R}^X$ under the pointwise order.

An operator $P \in \mathcal{L}(\mathbb{R}^X)$ is called a **Markov operator** on $\mathbb{R}^X$ if $P$ is positive and $P\mathbb{1} = \mathbb{1}$. We let

$$\mathcal{M}(\mathbb{R}^X) := \text{ the set of all Markov operators on } \mathbb{R}^X$$

Viewed as matrices, elements of $\mathcal{M}(\mathbb{R}^X)$ are nonnegative matrices whose rows sum to one. The next exercise asks you to confirm this.

EXERCISE 2.3.12. Fix $P \in \mathcal{L}(\mathbb{R}^X)$ and let $P(x, x')$ be the matrix representation. Prove that $P \in \mathcal{M}(\mathbb{R}^X)$ if and only if $P(x, x') \geqslant 0$ for all $x, x' \in X$ and $\sum_{x' \in X} P(x, x') = 1$ for all $x \in X$.

EXERCISE 2.3.13. Prove: If $P \in \mathcal{M}(\mathbb{R}^X)$ and $v \in \mathbb{R}^X$ with $v \gg 0$, then $Pv \gg 0$.

In the next exercise, you can think of $\varphi$ as a row vector and $\varphi P$ as premuliplying the matrix $P$ by this row vector. Chapter 3 uses the map $\varphi \mapsto \varphi P$ to update marginal distributions generated by Markov chains.

EXERCISE 2.3.14. Fix $P \in \mathcal{L}(\mathbb{R}^X)$. Prove that $P \in \mathcal{M}(\mathbb{R}^X)$ if and only if the function $\varphi P$ defined by

$$(\varphi P)(x') = \sum_{x \in X} P(x, x') \varphi(x) \qquad (x' \in X) \tag{2.19}$$

is in $\mathcal{D}(\mathsf{X})$ whenever $\varphi \in \mathcal{D}(\mathsf{X})$.

Markov operators are important for us because they generate Markov dynamics, a foundation of dynamic programming. Thus, (2.19) is a rule for updating distributions by one period under the Markov dynamics specified by $P$. We'll use it often in the next chapter.

## 2.4 Chapter Notes

Davey and Priestley (2002) provide a good introduction to partial orders and order-theoretic concepts. Our favorite books on fixed points and analysis include Ok (2007), Zhang (2012), Cheney (2013) and Atkinson and Han (2005). Good background material on order-theoretic fixed point methods can be found in Guo et al. (2004) and Zhang (2012).

# Chapter 3

# Markov Dynamics

To prepare to analyze dynamic programs, we now study stochastic processes generated by Markov chains. These processes are widely used to construct economic and financial models.

At the end of this chapter we return to the job search problem from Chapter 1 and allow wage draws to be correlated over time (rather than IID). We use a Markov chain to generated serially correlated wage draws.

Throughout this chapter, the symbol $\mathsf{X}$ represents a finite set.

## 3.1 Foundations

This section describes elementary properties of Markov models.

### 3.1.1 Markov Chains

Let's start with a definition and some simple examples.

#### 3.1.1.1 Defining Markov Chains

Fix $\mathsf{X} = \{x_1, \ldots, x_n\}$ and $P \in \mathcal{M}(\mathbb{R}^{\mathsf{X}})$. We interpret $P(x, x')$ as the probability that a random process moves from $x$ to $x'$ over one unit of time. For this interpretation to make sense we need $P(x, x')$ to be nonnegative and $\sum_{x' \in \mathsf{X}} P(x, x')$ to equal $1$ for every $x \in \mathsf{X}$, since we want the chain to stay somewhere in the state space after each update. These are exactly the properties guaranteed by the assumption $P \in \mathcal{M}(\mathbb{R}^X)$ (see Exercise 2.3.12).

To formalize ideas, let $(X_t) \coloneqq (X_t)_{t \geqslant 0}$ be a sequence of random variables taking values in X and call $(X_t)$ a **Markov chain** on **state space** X if there exists a $P \in \mathcal{M}(\mathbb{R}^X)$ such that

$$\mathbb{P}\{X_{t+1} = x' \mid X_0, X_1, \ldots, X_t\} = P(X_t, x') \quad \text{for all} \quad t \geqslant 0, \ x' \in X. \tag{3.1}$$

To simplify terminology, we also call $(X_t)$ $P$-**Markov** when (3.1) holds. We call either $X_0$ or its distribution $\psi_0$ the **initial condition** of $(X_t)$, depending on context. $P$ is also called the **transition matrix** of the Markov chain.

The definition of a Markov chain says two things:

(i) When updating to $X_{t+1}$ from $X_t$, earlier states are not required.

(ii) $P$ encodes all of the information required to perform the update, given the current state $X_t$.

One way to think about Markov chains is algorithmically: Fix $P \in \mathcal{M}(\mathbb{R}^X)$ and let $\psi_0$ be an element of $\mathcal{D}(X)$. Now generate $(X_t)$ via Algorithm 3.1. The resulting sequence is $P$-Markov with initial condition $\psi_0$.

---

**Algorithm 3.1:** Generation of $P$-Markov $(X_t)$ with initial condition $\psi_0$

---
**1** $t \leftarrow 0$
**2** $X_t \leftarrow$ a draw from $\psi_0$
**3** **while** $t < \infty$ **do**
**4** $\quad$ $X_{t+1} \leftarrow$ a draw from the distribution $P(X_t, \cdot)$
**5** $\quad$ $t \leftarrow t + 1$
**6** **end**

---

### 3.1.1.2 Application: S-s Dynamics

As an example, consider a firm whose inventory of some product follows *S-s* dynamics, meaning that the firm waits until its inventory falls below some level $s > 0$ and then immediately replenishes by ordering $S$ units. This pattern of decisions can be rationalized if ordering requires paying a fixed cost. Thus, in §5.2.1, we will show that *S-s* behavior is optimal in a setting where fixed costs exist and the firm's aim is to maximize its present value.

To represent $S$-$s$ dynamics, we suppose that a firm's inventory $(X_t)_{t \geqslant 0}$ of a given product obeys

$$X_{t+1} = \max\{X_t - D_{t+1}, 0\} + S\mathbb{1}\{X_t \leqslant s\},$$

where

- $(D_t)_{t \geqslant 1}$ is an exogenous IID demand process with $D_t \overset{d}{=} \varphi \in \mathcal{D}(\mathbb{Z}_+)$ for all $t$ and

- $S$ is the quantity ordered when $X_t \leqslant s$.

For the distribution $\varphi$ of demand we take the geometric distribution, so that $\varphi(d) = \mathbb{P}\{D_t = d\} = p(1-p)^d$ for $d \in \mathbb{Z}_+$.

EXERCISE 3.1.1. Confirm the following claim: An appropriate state space for this model is $\mathsf{X} := \{0, \dots, S + s\}$, since

$$X_t \in \mathsf{X} \implies \mathbb{P}\{X_{t+1} \in \mathsf{X}\} = 1.$$

If we define $h(x, d) := \max\{x - d, 0\} + S\mathbb{1}\{x \leqslant s\}$, so that $X_{t+1} = h(X_t, D_{t+1})$ for all $t$, then the transition matrix can be expressed as

$$P(x, x') = \mathbb{P}\{h(x, D_{t+1}) = x'\} = \sum_{d \geqslant 0} \mathbb{1}\{h(x, d) = x'\}\varphi(d) \qquad ((x, x') \in \mathsf{X} \times \mathsf{X}).$$

Listing 7 provides code that simulates inventory paths and computes other objects of interest. Since the state space $\mathsf{X} = \{x_1, \dots, x_n\}$ corresponds to $\{0, \dots, S + s\}$ and Julia indexing starts at 1, we set $x_i = i - 1$. This convention is used when computing `P[i, j]`, which corresponds to $P(x_i, x_j)$. The code in the listing is used to produce the simulation of inventories in Figure 3.1.

The function `compute_mc` returns an instance of a `MarkovChain` object that can store both the state $\mathsf{X}$ and the transition probabilities. The `QuantEcon.jl` library defines this data type and provides functions that simulate a Markov chains, compute a stationary distribution, and perform related tasks.

### 3.1.1.3   Higher Order Transition Matrices

Given a finite state space $\mathsf{X}$, $k \geqslant 0$ and $P \in \mathcal{M}(\mathbb{R}^{\mathsf{X}})$, let $P^k$ be the $k$-th power of $P$. (If $k = 0$, then $P^k$ is the identity matrix.) Since $\mathcal{M}(\mathbb{R}^{\mathsf{X}})$ is closed under multiplication (Exercise 2.3.2), $P^k$ is in $\mathcal{M}(\mathbb{R}^{\mathsf{X}})$ for all $k \geqslant 0$. In this context, $P^k$ is sometimes called the **$k$-step transition matrix** corresponding to $P$. In what follows, $P^k(x, x')$ denotes the $(x, x')$-th element of the matrix representation of $P^k$.

```julia
using Distributions, QuantEcon, IterTools

function create_inventory_model(; S=100,   # Order size
                                  s=10,    # Order threshold
                                  p=0.4)   # Demand parameter
    ϕ = Geometric(p)
    h(x, d) = max(x - d, 0) + S * (x <= s)
    return (; S, s, ϕ, h)
end

"Simulate the inventory process."
function sim_inventories(model; ts_length=200)
    (; S, s, ϕ, h) = model
    X = Vector{Int32}(undef, ts_length)
    X[1] = S  # Initial condition
    for t in 1:(ts_length-1)
        X[t+1] = h(X[t], rand(ϕ))
    end
    return X
end

"Compute the transition probabilities and state."
function compute_mc(model; d_max=100)
    (; S, s, ϕ, h) = model
    n = S + s + 1  # Size of state space
    state_vals = collect(0:(S + s))
    P = Matrix{Float64}(undef, n, n)
    for (i, j) in product(1:n, 1:n)
        P[i, j] = sum((h(i-1, d) == j-1) * pdf(ϕ, d) for d in 0:d_max)
    end
    return MarkovChain(P, state_vals)
end

"Compute the stationary distribution of the model."
function compute_stationary_dist(model)
    mc = compute_mc(model)
    return mc.state_values, stationary_distributions(mc)[1]
end
```

Listing 7: An implementation of *S-s* inventory dynamics (`inventory_sim.jl`)

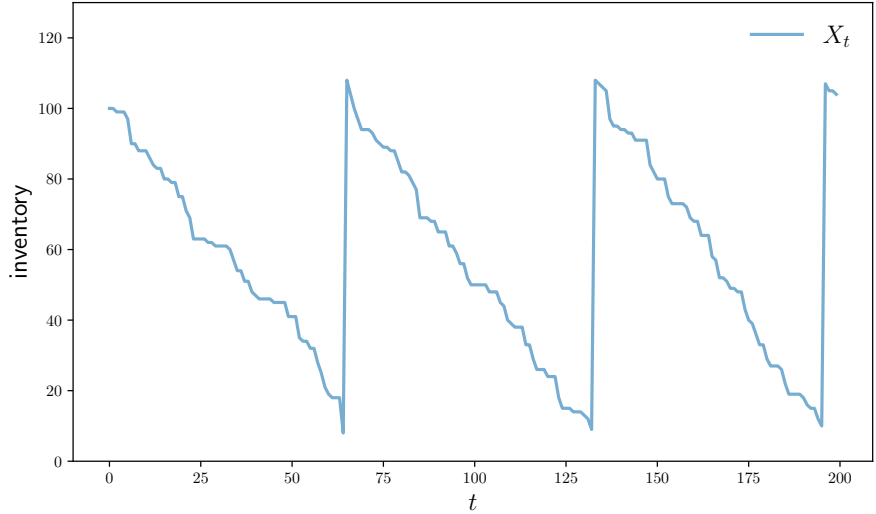

Figure 3.1: Inventory simulation (`inventory_sim.jl`)

The $k$-step transition matrix has the following interpretation: If $(X_t)$ is $P$-Markov, then for any $t, k \in \mathbb{Z}_+$ and $x, x' \in \mathsf{X}$,

$$P^k(x, x') = \mathbb{P}\{X_{t+k} = x' \mid X_t = x\}. \tag{3.2}$$

Thus, $P^k$ provides the $k$-step transition probabilities for the $P$-Markov chain $(X_t)$.

Exercise 3.1.2. Prove the claim in the last sentence via induction.

We can now give the following useful characterization of irreducibility:

**Lemma 3.1.1.** *Given $P \in \mathcal{M}(\mathbb{R}^{\mathsf{X}})$, the following statements are equivalent:*

(i) *$P$ is irreducible.*

(ii) *If $(X_t)$ is $P$-Markov and $x, x' \in \mathsf{X}$, then there exists a $k \geqslant 0$ such that*

$$\mathbb{P}\{X_k = x' \mid X_0 = x\} > 0.$$

Thus, irreducibility of $P$ means that the $P$-Markov chain eventually visits any state from any other state with positive probability.

*Proof of Lemma 3.1.1.* Fix $P \in \mathcal{M}(\mathbb{R}^{\mathsf{X}})$. $P$ is irreducible if and only if $\sum_{k \geqslant 0} P^k \gg 0$. This is equivalent to the statement that for each $(x, x') \in \mathsf{X} \times \mathsf{X}$, there exists a $k \geqslant 0$ such that $P^k(x, x') > 0$, which is in turn equivalent to part (ii) of Lemma 3.1.1. □

```
using QuantEcon
P = [0.1 0.9;
     0.0 1.0]
mc = MarkovChain(P)
print(is_irreducible(mc))
```

Listing 8: Testing irreducibility (`is_irreducible.jl`)

EXERCISE 3.1.3. Using Lemma 3.1.1, prove that the stochastic matrix associated with the *S*-*s* inventory dynamics in §3.1.1.2 is irreducible.

Several libraries have code for testing irreducibility, including `QuantEcon.jl`. See Listing 8 for an example of a call to this functionality. In this case, irreducibility fails because state 2 is an **absorbing state**. Once entered, the probability of ever leaving that state is zero. (A subset Y of X with this property is called an **absorbing set**.)

## 3.1.2 Stationarity and Ergodicity

Next we review aspects of Markov dynamics, including stationarity and ergodicity.

Fix $P \in \mathcal{M}(\mathbb{R}^{\mathsf{X}})$ and let $(X_t)$ be a *P*-chain. Let $\psi_t$ be the distribution of $X_t$. Marginal distributions $\psi_t$ evolve according to

$$\psi_{t+1}(x') = \sum_x P(x, x')\psi_t(x) \quad \text{for all } x' \in \mathsf{X} \text{ and } t \geqslant 0. \tag{3.3}$$

To verify (3.3), rewrite it as $\mathbb{P}\{X_{t+1} = x'\} = \sum_x \mathbb{P}\{X_{t+1} = x' \mid X_t = x\}\mathbb{P}\{X_t = x\}$, which is true by the law of total probability. With each $\psi_t$ regarded as a row vector, (3.3) can also be written as

$$\psi_{t+1} = \psi_t P. \tag{3.4}$$

Equation (3.4) tells us that dynamics of marginal distributions for Markov chains are generated by deterministic linear difference equations in distribution space. This is remarkable because the dynamics that drive $(X_t)$ are stochastic and can be arbitrarily nonlinear.

Iterating on (3.4), we get $\psi_t = \psi_0 P^t$ for all $t$. In summary,

$$(X_t)_{t \geqslant 0} \text{ is } P\text{-Markov with } X_0 \stackrel{d}{=} \psi_0 \implies X_t \stackrel{d}{=} \psi_0 P^t \text{ for all } t \geqslant 0. \tag{3.5}$$

For (3.5) and $\psi_{t+1} = \psi_t P$ to hold, each $\psi_t$ must be a row vector. In what follows, we always treat the distributions $(\psi_t)_{t \geqslant 0}$ of $(X_t)_{t \geqslant 0}$ as row vectors.

EXERCISE 3.1.4. Let $(X_t)$ be $P$-Markov on X with $X_0 \stackrel{d}{=} \psi_0$. Show that

$$\mathbb{E} h(X_t) = \psi_0 P^t h = \langle \psi_0 P^t, h \rangle \quad \text{for all } t \in \mathbb{N} \text{ and } h \in \mathbb{R}^{\mathsf{X}}. \tag{3.6}$$

Consistent with our definition of stationary distributions in §2.3.1.3, a marginal distribution $\psi^* \in \mathcal{D}(\mathsf{X})$ is called **stationary** for $P$ if

$$\sum_x P(x, x') \psi^*(x) = \psi^*(x') \quad \text{for all } x \in \mathsf{X}.$$

In vector form this is $\psi^* P = \psi^*$. By this definition and (3.3), if $\psi^*$ is stationary and $X_t$ has distribution $\psi^*$, then so does $X_{t+k}$ for all $k \geqslant 1$.

We saw in Exercise 2.3.2 that every irreducible $P \in \mathcal{M}(\mathbb{R}^{\mathsf{X}})$ has exactly one stationary distribution in $\mathcal{D}(\mathsf{X})$. The following **ergodic property** holds under the same assumptions.

**Theorem 3.1.2.** *If $P$ is irreducible with stationary distribution $\psi^*$, then, for any $P$-Markov chain $(X_t)$ and any $x \in \mathsf{X}$, we have*

$$\mathbb{P} \left\{ \lim_{k \to \infty} \frac{1}{k} \sum_{t=0}^{k-1} \mathbb{1}\{X_t = x\} = \psi^*(x) \right\} = 1. \tag{3.7}$$

A proof of (3.7) can be found in Brémaud (2020).

Property (3.7) tells us that, with probability one (i.e., for almost every $P$-Markov chain that we generate), the fraction of time that the chain spends in any given state is, in the limit, equal to the probability assigned to that state by the stationary distribution. Markov chains with this property are sometimes said to be **ergodic**.

Since the $S$-$s$ inventory model from §3.1.1.2 is irreducible, the ergodicity result from Theorem 3.1.2 applies. In particular, the process has only one stationary distribution $\psi^*$ in $\mathcal{D}(\mathsf{X})$, where $\mathsf{X} = \{0, \ldots, S + s\}$, and (3.7) is valid. Figure 3.2 illustrates this by plotting both the stationary distribution $\psi^*$ (which is computed using the code in Listing 7), and the value $m(y) := \frac{1}{k} \sum_{t=0}^{k-1} \mathbb{1}\{X_t = y\}$ at each $y \in \mathsf{X}$ for $k$ set to $1,000,000$. As predicted by the theorem, the fraction of time spent by the chain in each state is close to the probability assigned by $\psi^*$.

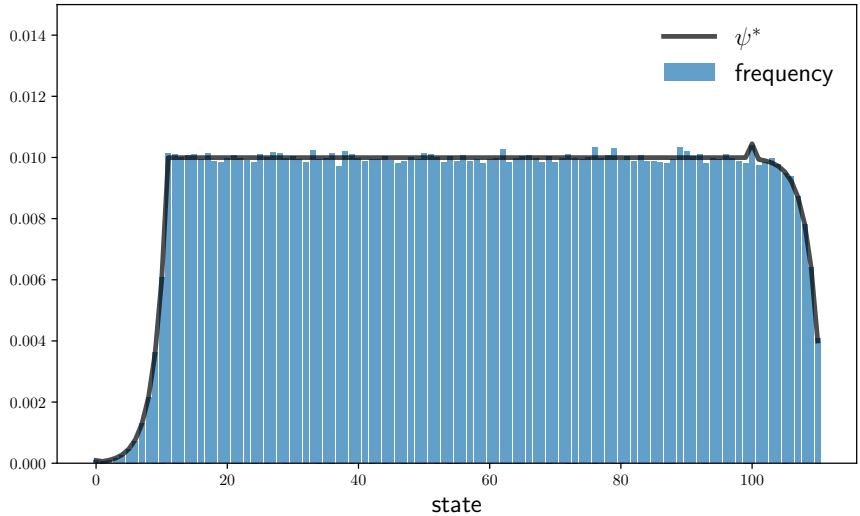

Figure 3.2: Ergodicity (`inventory_sim.jl`)

### 3.1.2.1 Application: Day Laborer

Suppose that a day laborer is either unemployed ($X_t = 1$) or employed ($X_t = 2$) in each period. In state 1 he is hired with probability $\alpha \in (0, 1)$. In state 2 he is fired with probability $\beta \in (0, 1)$. The corresponding state space and transition matrix are

$$\mathsf{X} = \{1, 2\} \quad \text{and} \quad P = \begin{pmatrix} 1 - \alpha & \alpha \\ \beta & 1 - \beta \end{pmatrix}. \tag{3.8}$$

Listing 9 provides a function to update from $X_t$ to $X_{t+1}$, using the fact that `rand()` generates a draw from the uniform distribution on $[0, 1)$.

EXERCISE 3.1.5. Explain why Listing 9 updates the current state according to the probabilities in $P$.

EXERCISE 3.1.6. Because $P$ is everywhere positive, it must be irreducible, so $P$ has the unique stationary distribution in $\psi^* \in \mathcal{D}(\mathsf{X})$. Show that $\psi^*$ is given by

$$\psi^* = \frac{1}{\alpha + \beta} \begin{pmatrix} \beta & \alpha \end{pmatrix}. \tag{3.9}$$

It is also true that $\psi P^t \to \psi^*$ as $t \to \infty$ for any $\psi \in \mathcal{D}(\mathsf{X})$. Thus, the operator $P$

```julia
function create_laborer_model(; α=0.3, β=0.2)
    return (; α, β)
end

function laborer_update(x, model)   # update X from t to t+1
    (; α, β) = model
    if x == 1
        x′ = rand() < α ? 2 : 1
    else
        x′ = rand() < β ? 1 : 2
    end
    return x′
end
```

Listing 9: Updating the state of the day laborer (`laborer_sim.jl`)

when understood as the mapping $\psi \mapsto \psi P$, is globally stable on $\mathcal{D}(\mathsf{X})$

EXERCISE 3.1.7. Prove this using the Perron–Frobenius theorem. More generally, show that this global stability result holds for any $P \in \mathcal{M}(\mathbb{R}^{\mathsf{X}})$ with $P \gg 0$.

EXERCISE 3.1.8. Fix $\alpha = 0.3$ and $\beta = 0.2$. Compute the sequence $(\psi P^t)$ for different choices of $\psi$ and confirm that your results are consistent with the claim that $\psi P^t \to \psi^*$ as $t \to \infty$ for any $\psi \in \mathcal{D}(\mathsf{X})$.

EXERCISE 3.1.9. Since $P$ is irreducible, ergodicity property (3.7) holds. Simulate a long realization Markov of a $P$-Markov chain from an arbitrary initial condition and confirm that your results are consistent with (3.7).

### 3.1.3 Approximation

To simplify numerical calculations, we sometimes approximate a continuous state Markov process with a Markov chain. For example, consider a **linear Gaussian AR(1)** model, where $(X_t)_{t \geqslant 0}$ evolves in $\mathbb{R}$ according to

$$X_{t+1} = \rho X_t + b + \nu \varepsilon_{t+1}, \quad |\rho| < 1, \quad (\varepsilon_t) \overset{\text{IID}}{\sim} N(0, 1). \tag{3.10}$$

The model (3.10) has a unique **stationary distribution** $\psi^*$ given by

$$\psi^* = N(\mu_x, \sigma_x^2) \quad \text{with} \quad \mu_x := \frac{b}{1-\rho} \quad \text{and} \quad \sigma_x^2 := \frac{\nu^2}{1-\rho^2}.$$

This means that

$$X_t \overset{d}{=} \psi^* \text{ and } X_{t+1} = \rho X_t + b + \nu \varepsilon_{t+1} \text{ implies } X_{t+1} \overset{d}{=} \psi^*.$$

EXERCISE 3.1.10. Suppose that $X_t \overset{d}{=} \psi^*$, $\varepsilon_{t+1} \overset{d}{=} N(0, 1)$ and $X_t$ and $\varepsilon_{t+1}$ are independent. Prove that $\rho X_t + b + \nu \varepsilon_{t+1}$ has distribution $\psi^*$. Is this still true if we drop the independence assumption made above?

Process (3.10) is also ergodic in a similar sense to (3.7) on page 88: on average, realizations of the process spend most of their time in regions of the state where the stationary distribution puts high probability mass. (You can check this via simulations if you wish.) Hence, in the discretization that follows, we shall put the discrete state space in this area.

EXERCISE 3.1.11. Set $b = 0$ in (3.10) and let $F$ be the CDF of $N(0, \nu^2)$. Show that

$$\mathbb{P}\{t - \delta < X_{t+1} \leqslant t + \delta \mid X_t = x\} = F(t - \rho x + \delta) - F(t - \rho x - \delta) \qquad (3.11)$$

for all $\delta, t \in \mathbb{R}$.

To discretize (3.10) we use **Tauchen's method**, starting with the case $b = 0$.[1] As a first step, we choose $n$ as the number of states for the discrete approximation and $m$ as an integer that sets the width of the state space. Then we create a state space $\mathsf{X} := \{x_1, \ldots, x_n\} \subset \mathbb{R}$ as an equispaced grid that brackets the stationary mean on both sides by $m$ standard deviations:

- set $x_1 = -m \, \sigma_x$,

- set $x_n = m \, \sigma_x$ and

- set $x_{i+1} = x_i + s$ where $s = (x_n - x_1)/(n-1)$ and $i$ in $[n-1]$.

The next step is to create an $n \times n$ matrix $P$ that approximates the dynamics in (3.10). For $i, j \in [n]$,

---

[1]Tauchen's method (Tauchen (1986)) is simple but sub-optimal in some cases. For a more general discretization method and a survey of the literature, see Farmer and Toda (2017).

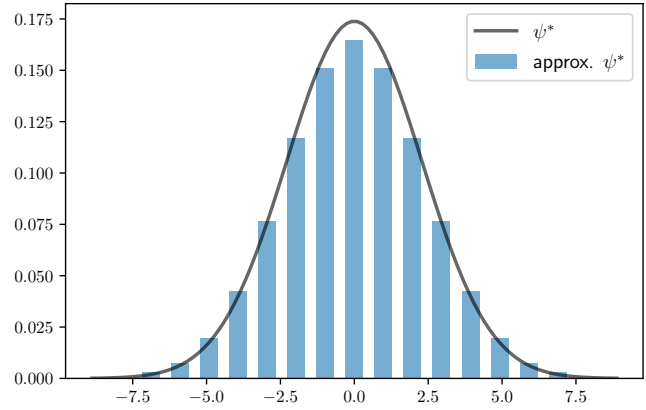

Figure 3.3: Comparison of $\psi^* = N(\mu_x, \sigma_x^2)$ and its discrete approximant

(i) if $j = 1$, then set $P(x_i, x_j) = F(x_1 - \rho x_i + s/2)$.

(ii) If $j = n$, then set $P(x_i, x_j) = 1 - F(x_n - \rho x_i - s/2)$.

(iii) Otherwise, set $P(x_i, x_j) = F(x_j - \rho x_i + s/2) - F(x_j - \rho x_i - s/2)$.

The first two are boundary rules and the third applies Exercise 3.1.11.

EXERCISE 3.1.12. Prove that $\sum_{j=1}^{n} P(x_i, x_j) = 1$ for all $i \in [n]$.

Finally, if $b \neq 0$, then we shift the state space to center it on the mean $\mu_x$ of the stationary distribution $N(\mu_x, \sigma_x^2)$. This is done by replacing $x_i$ with $x_i + \mu_x$ for each $i$.

Julia routines that compute X and $P$ can be found in the library QuantEcon.jl.

Figure 3.3 compares the continuous stationary distribution $\psi^*$ and the unique stationary distribution of the discrete approximation when X and $P$ are constructed as above when $\rho = 0.9$, $b = 0.0$, $\nu = 1.0$ and the discretization parameters are $n = 15$ and $m = 3$.

## 3.2 Conditional Expectations

In this section we discuss how to compute conditional expectations for Markov chains. The theory will be essential for the study of finite MDPs, since, in these models, lifetime rewards are mathematical expectations of flow reward functions of Markov states.

### 3.2.1 Mathematical Expectations

We begin with mathematical expectations of functions of Markov states.

#### 3.2.1.1 Conditional Expectations

Fix $P \in \mathcal{M}(\mathbb{R}^{\mathsf{X}})$. For each $h \in \mathbb{R}^{\mathsf{X}}$, we define

$$(Ph)(x) = \sum_{x' \in \mathsf{X}} h(x') P(x, x') \qquad (x \in \mathsf{X}). \tag{3.12}$$

Noting that $P(x, \cdot)$ is the distribution of $X_{t+1}$ given $X_t = x$, we can write

$$(Ph)(x) = \mathbb{E}[h(X_{t+1}) \,|\, X_t = x], \tag{3.13}$$

where $(X_t)$ is any $P$-Markov chain on X. In terms of matrix algebra, viewing $h$ has an $n \times 1$ column vector, the expression $(Ph)(x)$ is one element of the vector $Ph$ obtained by premultiplying $h$ by $P$.

The interpretation in (3.13) extends to powers of $P$. In particular, we have

$$(P^k h)(x) = \sum_{x'} h(x') P^k(x, x') = \mathbb{E}[h(X_{t+k}) \,|\, X_t = x]. \tag{3.14}$$

EXERCISE 3.2.1. Show that

(i) Every constant function $h \in \mathbb{R}^{\mathsf{X}}$ is a fixed point of $P$ (i.e., $Ph = h$).

(ii) $\max_x |Ph(x)| \leqslant \max_x |h(x)|$ for all $h \in \mathbb{R}^{\mathsf{X}}$.

#### 3.2.1.2 The Law of Iterated Expectations

The **law of iterated expectations** is a workhorse in economics and finance. A common version of the law states that if $X$ and $Y$ are two random variables, then $\mathbb{E}[\mathbb{E}[Y \,|\, X]] = \mathbb{E}[Y]$. Let's show how this law applies for Markov chains.

Let $(X_t)$ be $P$-Markov with $X_0 \stackrel{d}{=} \psi_0$. Fix $t, k \in \mathbb{N}$. Set $\mathbb{E}_t := \mathbb{E}[\cdot \,|\, X_t]$. We claim that

$$\mathbb{E}[\mathbb{E}_t[h(X_{t+k})]] = \mathbb{E}[h(X_{t+k})] \quad \text{for any } h \in \mathbb{R}^{\mathsf{X}}. \tag{3.15}$$

To see this, recall that $\mathbb{E}\left[h(X_{t+k}) \mid X_t = x\right] = (P^k h)(x)$. Hence $\mathbb{E}\left[h(X_{t+k}) \mid X_t\right] = (P^k h)(X_t)$. Therefore,

$$\mathbb{E}\left[\mathbb{E}_t\left[h(X_{t+k})\right]\right] = \mathbb{E}\left[(P^k h)(X_t)\right] = \sum_{x'} (P^k h)(x') \psi_t(x') = \sum_{x'} (P^k h)(x')(\psi_0 P^t)(x').$$

Since $\psi_0 P^t$ is a row vector, we can write the last expression as

$$\psi_0 P^t P^k h = \psi_0 P^{t+k} h = \psi_{t+k} h = \mathbb{E}h(X_{t+k}).$$

Hence (3.15) holds.

### 3.2.1.3 Monotone Markov Chains

Next we connect Markov chains to order theory via stochastic dominance. These connections will have many applications below.

Let $\mathsf{X}$ be a finite set partially ordered by $\preceq$. A Markov operator $P \in \mathcal{M}(\mathbb{R}^{\mathsf{X}})$ is called **monotone increasing** if

$$x, y \in \mathsf{X} \text{ and } x \preceq y \quad \implies \quad P(x, \cdot) \preceq_F P(y, \cdot).$$

Thus, $P$ is monotone increasing if shifting up the current state shifts up the next period state, in the sense that its distribution increases in the stochastic dominance ordering (see §2.2.4) on $\mathcal{D}(\mathsf{X})$. Below, we will see that monotonicity of Markov operators is closely related to monotonicity of value functions in dynamic programming.

Monotonicity of Markov operators is related to positive autocorrelation. To illustrate the idea, consider the AR(1) model $X_{t+1} = \rho X_t + \sigma \varepsilon_{t+1}$ from §3.1.3 and suppose we apply Tauchen discretization, mapping the parameters $\rho, \sigma$ and a discretization size $n$ into a Markov operator $P$ on state space $\mathsf{X} = \{x_1, \ldots, x_n\} \subset \mathbb{R}$, totally ordered by $\leqslant$. If $\rho \geqslant 0$, so that positive autocorrelation holds, then $P$ is monotone increasing.

EXERCISE 3.2.2. Verify this claim.

EXERCISE 3.2.3. In §3.1.2.1 we discussed a Markov chain

$$\mathsf{X} = \{1, 2\} \quad \text{and} \quad P_w = \begin{pmatrix} 1 - \alpha & \alpha \\ \beta & 1 - \beta \end{pmatrix}$$

for some $\alpha, \beta \in [0, 1]$. Show that $P_w$ is monotone increasing if and only if $\alpha + \beta \leqslant 1$.

EXERCISE 3.2.4. Prove that $P$ is monotone increasing if and only if $P$ is invariant on $i\mathbb{R}^{\mathsf{X}}$; i.e., if $h \in i\mathbb{R}^{\mathsf{X}}$ implies $Ph \in i\mathbb{R}^{\mathsf{X}}$.

EXERCISE 3.2.5. Prove: If $P$ is monotone increasing then so is $P^t$ for all $t \in \mathbb{N}$.

### 3.2.2  Geometric Sums

Dynamic programs often form a lifetime value $V_0$ as a geometric sum of a reward sequence $(R_t)_{t \geqslant 0}$ with constant discount factor, so that $V_0 = \mathbb{E} \sum_{t=0}^{\infty} \beta^t R_t$ for some $\beta > 0$. We saw this in (1.1) on page 2, where we aggregated a profit stream $(\pi_t)_{t \geqslant 0}$ into an expected present value of the firm, and again in (1.6) on page 10, where a worker evaluates lifetime earnings. In this section we study expectations of geometric sums.

#### 3.2.2.1  Theory

Consider a conditional mathematical expectation of a discounted sum of future measurements:

$$v(x) := \mathbb{E}_x \sum_{t=0}^{\infty} \beta^t h(X_t) := \mathbb{E}\left[\sum_{t=0}^{\infty} \beta^t h(X_t) \mid X_0 = x\right] \qquad (3.16)$$

for some constant $\beta \in \mathbb{R}_+$ and $h \in \mathbb{R}^{\mathsf{X}}$. Here

- $(X_t)$ is $P$-Markov on some finite set $\mathsf{X}$,
- $v(x)$ is a **lifetime reward** starting from state $x$, and
- $\mathbb{E}_x$ indicates that we are conditioning on $X_0 = x$.

With $I$ as the identity matrix, the next result describes $v$ as function of $\beta$, $P$ and $h$.

**Lemma 3.2.1.** *If $\beta < 1$, then $I - \beta P$ is invertible and*

$$v = \sum_{t=0}^{\infty} (\beta P)^t h = (I - \beta P)^{-1} h. \qquad (3.17)$$

*Proof.* Under the stated conditions

$$\mathbb{E}_x \sum_{t=0}^{\infty} \beta^t h(X_t) = \sum_{t=0}^{\infty} \beta^t \mathbb{E}_x h(X_t) = \sum_{t=0}^{\infty} \beta^t (P^t h)(x), \qquad (3.18)$$

where the first equality in (3.18) uses linearity of expectations and the second follows from (3.14) and the assumption that $(X_t)$ is $P$-Markov starting at $x$.[2] Applying the Neumann series lemma (p. 18) to the matrix $\beta P$, we see that $\sum_{t=0}^{\infty}(\beta P)^t = (I - \beta P)^{-1}$. The lemma applies because $\rho(\beta P) = \beta \rho(P) = \beta < 1$, as follows from Exercise 2.3.2. □

### 3.2.2.2 Application: Valuation of Firms

Consider a firm that receives random profit stream $(\pi_t)_{t \geqslant 0}$. Supposes that the value of the firm equals the expected present value of its profit stream. Suppose for now that the interest rate is constant at $r > 0$. With $\beta := 1/(1+r)$, total valuation is

$$V_0 = \mathbb{E} \sum_{t=0}^{\infty} \beta^t \pi_t. \tag{3.19}$$

To compute this value, we need to know how profits evolve. A common strategy is to set $\pi_t = \pi(X_t)$ for some fixed $\pi \in \mathbb{R}^{\mathsf{X}}$, where $(X_t)_{t \geqslant 0}$ is a state process. For known dynamics of $(X_t)$ and function $\pi$, the value $V_0$ in (3.19) can be computed.

Here we assume that $(X_t)$ is $P$-Markov for $P \in \mathcal{M}(\mathbb{R}^{\mathsf{X}})$ with finite X. Then conditioning on $X_0 = x$, we can write the value as

$$v(x) := \mathbb{E}_x \sum_{t=0}^{\infty} \beta^t \pi_t := \mathbb{E}\left[\sum_{t=0}^{\infty} \beta^t \pi_t \mid X_0 = x\right].$$

By Lemma 3.2.1, the value $v(x)$ is finite and the function $v \in \mathbb{R}^{\mathsf{X}}$ can be obtained by

$$v = \sum_{t=0}^{\infty} \beta^t P^t \pi = (I - \beta P)^{-1} \pi.$$

It is plausible that the value of the firm will be higher for a return process in which higher states generate higher profits and predict higher future states. The next exercise confirms this.

EXERCISE 3.2.6. Let X be partially ordered and suppose that $\pi \in i\mathbb{R}^{\mathsf{X}}$ and that $P$ is monotone increasing. (See §3.2.1.3 for terminology and notation.) Prove that, under these conditions, $v$ is increasing on X.

---

[2]To justify the first equality, care must be taken when pushing expectations through infinite sums. In the present setting, justification can be provided via the dominated convergence theorem (see, e.g., Dudley (2002), Theorem 4.3.5). A proof of a more general result can be found in §B.2.

### 3.2.2.3 Application: Valuing Consumption Streams

To model consumption-saving choices we want to evaluate different consumption paths, where a **consumption path** is a nonnegative random sequence $(C_t)_{t \geqslant 0}$. In what follows we consider consumption paths such that $C_t = c(X_t)$ for all $t \geqslant 0$, where $c \in \mathbb{R}_+^{\mathsf{X}}$ and $(X_t)_{t \geqslant 0}$ is $P$-Markov on finite set $\mathsf{X}$. Thus, consumption streams are time-invariant functions of a finite state Markov chain.

In a standard "time additive" model of consumer preferences with constant geometric discounting, the time zero value of a consumption stream $(C_t)_{t \geqslant 0}$, given current state $X_0 = x \in \mathsf{X}$, is

$$v(x) = \mathbb{E}_x \sum_{t=0}^{\infty} \beta^t u(C_t), \tag{3.20}$$

where $\beta \in (0, 1)$ is a discount factor and $u \colon \mathbb{R}_+ \to \mathbb{R}$ is called the **flow utility function**. Dependence of $v(x)$ on $x$ comes from the initial condition $X_0 = x$ influencing the Markov state process and, therefore, the consumption path.

Using $C_t = c(X_t)$ and defining $r := u \circ c$ we can write $v(x) = \mathbb{E}_x \sum_{t \geqslant 0} \beta^t r(X_t)$. By Lemma 3.2.1, this sum is finite and $v$ can be expressed as

$$v = (I - \beta P)^{-1} r. \tag{3.21}$$

Figure 3.4 shows an example when $u$ has the constant relative risk aversion (**CRRA**) specification

$$u(c) = \frac{c^{1-\gamma}}{1 - \gamma} \qquad (c \geqslant 0, \ \gamma > 0), \tag{3.22}$$

while $c(x) = \exp(x)$, so that consumption takes the form $C_t = \exp(X_t)$, and $(X_t)_{t \geqslant 0}$ is a Tauchen discretization (see §3.1.3) of $X_{t+1} = \rho X_t + \nu W_{t+1}$ where $(W_t)_{t \geqslant 1}$ is IID and standard normal. Parameters are $n = 25$, $\beta = 0.98$, $\rho = 0.96$, $\nu = 0.05$ and $\gamma = 2$. We set $r = u \circ c$ and solved for $v$ via (3.21).

EXERCISE 3.2.7. Replicate Figure 3.4.

EXERCISE 3.2.8. The value function in Figure 3.4 appears to be increasing in the state $x$. Prove this for the CRRA model when $\rho \geqslant 0$.

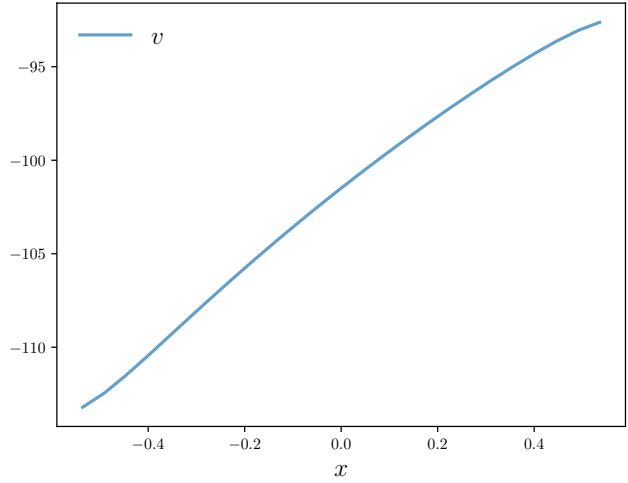

Figure 3.4: The value of $(C_t)_{t \geqslant 0}$ given $X_t = x$

## 3.3  Job Search Revisited

In this section we extend the job search problem studied in §1.3 to a setting with Markov wage offers. We discuss additional structure when the Markov operator for wage offers is monotone increasing. We will also allow job separations to occur.

### 3.3.1  Job Search with Markov State

We adopt the job search setting of §1.3 but assume now that the wage process $(W_t)$ is $P$-Markov on $W \subset \mathbb{R}_+$, where $P \in \mathcal{M}(\mathbb{R}^W)$ and $W$ is finite.

#### 3.3.1.1  Value Function Iteration

The **value function** $v^*$ for the Markov job search model is now defined as follows: $v^*(w)$ is the maximum lifetime value that can be obtained when the worker is unemployed with current wage offer is $w$ in hand. Value function $v^*$ satisfies Bellman equation

$$v^*(w) = \max \left\{ \frac{w}{1 - \beta}, \, c + \beta \sum_{w' \in W} v^*(w') P(w, w') \right\} \qquad (w \in W). \qquad (3.23)$$

We continue to assume that $c > 0$ and $\beta \in (0, 1)$.

Bellman equation (3.23) extends a corresponding Bellman equation for the IID case (cf. (1.25) on page 33). (A full proof is given in Chapter 4.) The Bellman operator corresponding to (3.23) is

$$(Tv)(w) = \max \left\{ \frac{w}{1-\beta}, \, c + \beta \sum_{w'} v(w')P(w, w') \right\} \qquad (w \in \mathsf{W}).$$

As before, $T$ is constructed so that $v^*$ is a fixed point (since (3.23) holds). We prove below that $v^*$ is the only fixed point of $T$ in $\mathbb{R}_+^{\mathsf{W}}$.

Extending the IID definition (cf. (1.29) on page 35), a policy $\sigma\colon \mathsf{W} \to \{0, 1\}$ is called $v$-**greedy** if

$$\sigma(w) = \mathbb{1}\left\{ \frac{w'}{1-\beta} \geqslant c + \beta \sum_{w'} v(w')P(w, w') \right\}$$

for all $w \in \mathsf{W}$.

Let $V := \mathbb{R}_+^{\mathsf{W}}$ and endow $V$ with the pointwise partial order $\leqslant$ and the supremum norm, so that $\|f - g\|_\infty = \max_{w \in \mathsf{W}} |f(w) - g(w)|$.

EXERCISE 3.3.1. Prove that

(i) $T$ is an order-preserving self-map on $V$.

(ii) $T$ is a contraction of modulus $\beta$ on $V$.

We recommend that you study the proof of the next lemma, since the same style of argument occurs often below.

**Lemma 3.3.1.** *$v^*$ is increasing on $(\mathsf{W}, \leqslant)$ whenever $P$ is monotone increasing.*

*Proof.* Let $iV$ be the increasing functions in $V$ and suppose that $P$ is monotone increasing. $T$ is a self-map on $iV$ in this setting, since $v \in iV$ implies $h(w) := c + \beta \sum_{w'} v(w')P(w, w')$ is in $iV$. Hence, for such a $v$, both $h$ and the stopping value function $e(w) := w/(1-\beta)$ are in $iV$. It follows that $Tv = h \vee e$ is in $iV$.

Since $iV$ is a closed subset of $V$ and $T$ is a self-map on $iV$, the fixed point $v^*$ is in $iV$ (see Exercise 1.2.18 on page 22). □

In view of the contraction property established in Exercise 3.3.1, we can use value function iteration (i) to compute an approximation $v$ to the value function and (ii) to calculate the $v$-greedy policy that approximates the optimal policy. Code for implementing this procedure is in Listing 10. The definition of a $v$-greedy policy resembles that for the IID case (see (1.29) on page 35).

```julia
using QuantEcon, LinearAlgebra
include("s_approx.jl")

"Creates an instance of the job search model with Markov wages."
function create_markov_js_model(;
        n=200,          # wage grid size
        ρ=0.9,          # wage persistence
        ν=0.2,          # wage volatility
        β=0.98,         # discount factor
        c=1.0           # unemployment compensation
    )
    mc = tauchen(n, ρ, ν)
    w_vals, P = exp.(mc.state_values), mc.p
    return (; n, w_vals, P, β, c)
end

" The Bellman operator Tv = max{e, c + β P v} with e(w) = w / (1-β)."
function T(v, model)
    (; n, w_vals, P, β, c) = model
    h = c .+ β * P * v
    e = w_vals ./ (1 - β)
    return max.(e, h)
end

" Get a v-greedy policy."
function get_greedy(v, model)
    (; n, w_vals, P, β, c) = model
    σ = w_vals / (1 - β) .>= c .+ β * P * v
    return σ
end

"Solve the infinite-horizon Markov job search model by VFI."
function vfi(model)
    v_init = zero(model.w_vals)
    v_star = successive_approx(v -> T(v, model), v_init)
    σ_star = get_greedy(v_star, model)
    return v_star, σ_star
end
```

Listing 10: Job search with Markov state (`markov_js.jl`)

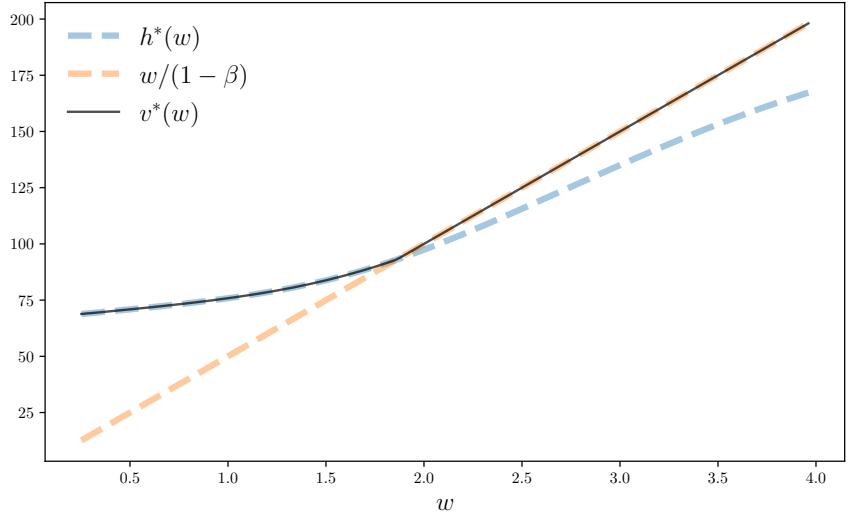

Figure 3.5: Value, stopping, and continuation for Markov job search

### 3.3.1.2 Continuation Values

The continuation value $h^*$ from the ɪɪᴅ case (see page 33) is now replaced by a **continuation value function**

$$h^*(w) := c + \beta \sum_{w'} v^*(w') P(w, w') \qquad (w \in \mathsf{W}).$$

The continuation value depends on $w$ because the current offer helps predict the offer next period, which in turn affects the value of continuing. The functions $w \mapsto w/(1-\beta)$, $h^*$ and $v^*$ corresponding to the default model in Listing 10 are shown in Figure 3.5.

EXERCISE 3.3.2. Explain why the continuation value function is increasing in Figure 3.5. If possible, provide a mathematical and economic explanation.

EXERCISE 3.3.3. Using the Bellman equation (3.23), show that $h^*$ obeys

$$h^*(w) := c + \beta \sum_{w'} \max\left\{ \frac{w'}{1-\beta}, \, h^*(w') \right\} P(w, w') \qquad (w \in \mathsf{W}).$$

EXERCISE 3.3.4. Let $Q$ be the operator on $V$ defined at $h \in V$ by

$$(Qh)(w) := c + \beta \sum_{w'} \max \left\{ \frac{w'}{1 - \beta}, h(w') \right\} P(w, w') \qquad (w \in \mathsf{W}). \tag{3.24}$$

Prove that $Q$ is (a) an order-preserving self-map on $V$ and (b) a contraction of modulus $\beta$ on $V$ under the supremum norm.

Exercise 3.3.4 suggests an alternative way to solve the job search problem: iterate with $Q$ to obtain the continuation value function $h^*$ and then use the policy

$$\sigma^*(w) = \mathbb{1} \left\{ \frac{w'}{1 - \beta} \geqslant h^*(w) \right\} \qquad (w \in \mathsf{W})$$

that tells the worker to accept when the current stopping value exceeds the current continuation value.

We saw that in the IID case a computational strategy based on continuation values is far more efficient than value function iteration (see §1.3.2.2). Since continuation values are functions rather than scalars, here the two approaches (iterating with $T$ vs iterating with $Q$) are more similar. In Chapter 4 we discuss alternative computational strategies in more detail, seeking conditions under which one approach will be more efficient than the other.

## 3.3.2  Job Search with Separation

We now modify the job search problem discussed in §3.3.1 by adding separations. In particular, an existing match between worker and firm terminates with probability $\alpha$ every period. (This is an extension because setting $\alpha = 0$ recovers the permanent job scenario from §3.3.1.)

The worker now views the loss of a job as a capital loss and a spell of unemployment as an investment. In what follows, the wage process and discount factor are unchanged from §3.3.1. As before, $V := \mathbb{R}_+^{\mathsf{W}}$ is endowed with the supremum norm.

The value function $v_u^*$ for an unemployed worker satisfies the recursion

$$v_u^*(w) = \max \left\{ v_e^*(w), \, c + \beta \sum_{w' \in \mathsf{W}} v_u^*(w') P(w, w') \right\} \qquad (w \in \mathsf{W}), \tag{3.25}$$

where $v_e^*$ is the value function for an employed worker, i.e., the lifetime value of a worker who starts the period employed at wage $w$. Value function $v_e^*$ satisfies

$$v_e^*(w) = w + \beta \left[ \alpha \sum_{w'} v_u^*(w') P(w, w') + (1 - \alpha) v_e^*(w) \right] \qquad (w \in \mathsf{W}). \qquad (3.26)$$

This equation states that value accruing to an employed worker is current wage plus the discounted expected value of being either employed or unemployed next period.

We claim that, when $0 < \alpha, \beta < 1$, the system (3.25)–(3.26) has a unique solution $(v_e^*, v_u^*)$ in $V \times V$. To show this we first solve (3.26) in terms of $v_e^*(w)$ to obtain

$$v_e^*(w) = \frac{1}{1 - \beta(1 - \alpha)} \left( w + \alpha \beta (P v_u^*)(w) \right). \qquad (3.27)$$

(Recall $(Ph)(w) := \sum_{w'} h(w') P(w, w')$ for $h \in \mathbb{R}^{\mathsf{W}}$.) Substituting into (3.25) yields

$$v_u^*(w) = \max \left\{ \frac{1}{1 - \beta(1 - \alpha)} \left( w + \alpha \beta (P v_u^*)(w) \right), c + \beta \, (P v_u^*)(w) \right\}. \qquad (3.28)$$

EXERCISE 3.3.5. Prove that there exists a unique $v_u^* \in V$ that solves (3.28). Propose a convergent method for computing both $v_u^*$ and $v_e^*$. [Hint: See Lemma 2.2.3 on page 59.]

Figure 3.6 shows the value function $v_u^*$ for an unemployed worker, which is the fixed point of (3.28), as well as the stopping and continuation values, which are given by

$$s^*(w) := \frac{1}{1 - \beta(1 - \alpha)} \left( w + \alpha \beta (P v_u^*)(w) \right) \quad \text{and} \quad h_e^*(w) := c + \beta \, (P v_u^*)(w)$$

respectively, for each $w \in \mathsf{W}$. Parameters are as in Listing 11. The value function $v_u^*$ is the pointwise maximum (i.e., $v_u^* = s^* \vee h^*$). The worker's optimal policy while unemployed is

$$\sigma^*(w) := \mathbb{1}\{s^*(w) \geqslant h^*(w)\}.$$

As before, the smallest $w$ such that $\sigma^*(w) = 1$ is called the **reservation wage**.

Figure 3.7 shows how the reservation wage changes with $\alpha$. To produce this figure we solved the model for the reservation wage at 10 values of $\alpha$ in an evenly spaced grid ranging 0 to 1. The reservation wage falls with $\alpha$, since time spent unemployed is a capital investment in better wages, and the value of this investment declines as the separation rate rises.

EXERCISE 3.3.6. Replicate Figure 3.7.

```julia
using QuantEcon, LinearAlgebra

"Creates an instance of the job search model with separation."
function create_js_with_sep_model(;
        n=200,                  # wage grid size
        ρ=0.9, ν=0.2,           # wage persistence and volatility
        β=0.98, α=0.1,          # discount factor and separation rate
        c=1.0)                  # unemployment compensation
    mc = tauchen(n, ρ, ν)
    w_vals, P = exp.(mc.state_values), mc.p
    return (; n, w_vals, P, β, c, α)
end
```

Listing 11: Job search with separation model (`markov_js_with_sep.jl`)

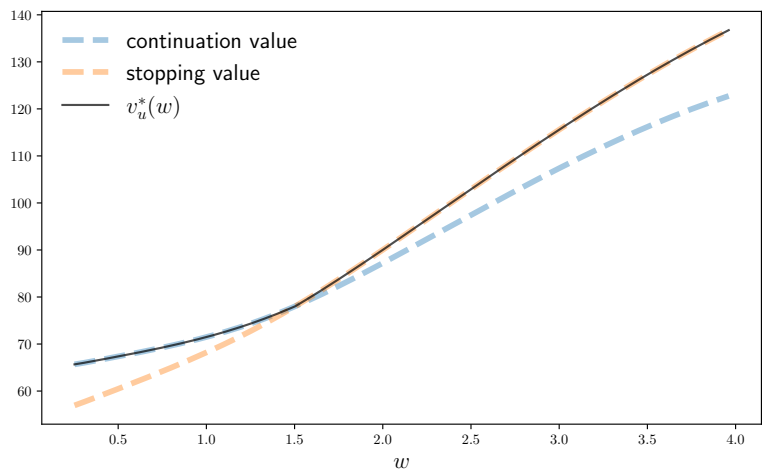

Figure 3.6: Value function with job separation

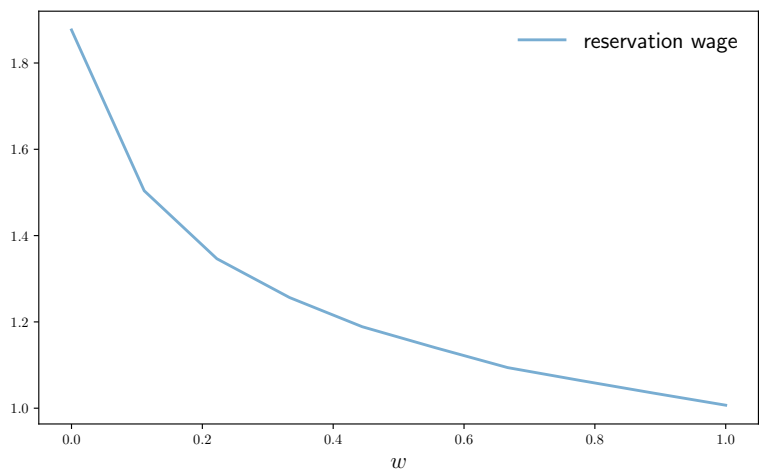

Figure 3.7: Reservation wage vs separation rate

## 3.4 Chapter Notes

Many good textbooks on Markov chains exist, including Norris (1998), Häggström et al. (2002) and Privault (2013). Sargent and Stachurski (2023b) provides a relatively comprehensive treatment from a network perspective that is a natural one for Markov chains. Other economic applications are discussed in Stokey and Lucas (1989) and Ljungqvist and Sargent (2018). Meyer (2000) gives a detailed account of the theory of nonnegative matrices. Another useful reference is Horn and Johnson (2012).

A systematic study of monotone Markov chains was initiated by Daley (1968). Monotone Markov methods have many important applications in economics. See, for example, Hopenhayn and Prescott (1992), Kamihigashi and Stachurski (2014), Jaśkiewicz and Nowak (2014), Balbus et al. (2014), Foss et al. (2018) and Hu and Shmaya (2019).

# Chapter 4

# Optimal Stopping

We study problems of maximizing lifetime rewards in settings in which decision makers face risks. The job search model studied in Chapters 1 and 3 is one example. Others include an entrepreneur who decides whether to exit or enter a market, a borrower who considers defaulting on a loan, a firm that contemplates introducing a new technology, or a portfolio manager deciding whether to exercise a real or financial option.

These can all be formulated as dynamic programming and have common features that facilitate sharp characterizations of optimality. They are all two-action (or binary choice) problems that provide good laboratories for studying some special dynamic programs in which recursive representations are particularly enlightening.

## 4.1   Introduction to Optimal Stopping

We begin with a standard theory of optimal stopping and then consider alternative approaches that feature continuation values and threshold policies. We aim to provide a rigorous discussion of optimality that refines our less formal analysis of job search in §1.3 and §3.3.1.

### 4.1.1   Theory

Our first step is to set out the fundamental theory of discrete time infinite-horizon optimal stopping problems.

### 4.1.1.1 The Stopping Problem

Let X be a finite set. Given X, an **optimal stopping problem** is a tuple $\mathcal{S} = (\beta, P, r, e)$ that consists of

   (i) a discount factor $\beta \in (0, 1)$,

  (ii) a Markov operator $P \in \mathcal{M}(\mathbb{R}^X)$,

 (iii) a **continuation reward function** $c \in \mathbb{R}^X$, and

  (iv) an **exit reward function** $e \in \mathbb{R}^X$.

Given a $P$-Markov chain $(X_t)_{t \geqslant 0}$, a decision maker observes the state $X_t$ in each period and decides whether to continue or stop. If she chooses to stop, she receives final reward $e(X_t)$ and the process terminates. If she decides to continue, then she receives $c(X_t)$ and the process repeats next period. Lifetime rewards are

$$\mathbb{E} \sum_{t \geqslant 0} \beta^t R_t,$$

where $R_t$ equals $c(X_t)$ while the agent continues, $e(X_t)$ when the agent stops, and zero thereafter.

**Example 4.1.1.** Consider the infinite-horizon job search problem from Chapter 1, where the wage offer process $(W_t)$ is IID with common distribution $\varphi$ on finite set W. This is an optimal stopping problem with state space X = W and $P \in \mathcal{M}(\mathbb{R}^X)$ having all rows equal to $\varphi$, so that all draws are IID from $\varphi$. The exit reward function is $e(x) = x/(1 - \beta)$ and the continuation reward function is constant and equal to unemployment compensation.

**Example 4.1.2.** Consider an infinite-horizon American call option that provides the right to buy a given asset at strike price $K$ at each future date. The market price of the asset is $S_t = s(X_t)$, where $(X_t)$ is $P$-Markov on finite set X and $s \in \mathbb{R}^X$. The interest rate is $r > 0$. Deciding when to exercise is an optimal stopping problem, with exit corresponding to exercising the option. The discount factor is $1/(1 + r)$, the exit reward function is $e(x) := s(x) - K$ and the continuation reward is zero.[1]

Optimal decisions are described by a **policy function**, which is a map $\sigma$ from X to $\{0, 1\}$. After observing state $x$ at any given time, the decision maker takes action $\sigma(x)$, where $0$ means "continue" and $1$ means "stop." Implicit in this formulation is the assumption that the current state contains enough information for the agent to decide whether or not to stop.

---

[1] We are studying American options in discrete time. Options with discrete exercise times are sometimes called **Bermudan options**. References for the continuous time case are provided in §4.3.

Let $\Sigma$ be the set of functions from $X$ to $\{0, 1\}$. Let $v_\sigma(x)$ denote the expected lifetime value of following policy $\sigma$ now and in every future period, given optimal stopping problem $S = (\beta, P, r, e)$ and current state $x \in X$. We call $v_\sigma$ the **$\sigma$-value function**. We also call $v_\sigma(x)$ the **lifetime value** of policy $\sigma$ conditional on initial state $x$. Section §4.1.1.2, shows that $v_\sigma$ is well defined and describes how to calculate it. A policy $\sigma^* \in \Sigma$ is called **optimal** for $S$ if

$$v_{\sigma^*}(x) = \max_{\sigma \in \Sigma} v_\sigma(x) \quad \text{for all } x \in X. \tag{4.1}$$

### 4.1.1.2 Lifetime Values

Fixing $\sigma \in \Sigma$, let us consider how to compute the lifetime value $v_\sigma(x)$ of following $\sigma$ conditional on $X_0 = x$. Evidently, $v_\sigma$ satisfies

$$v_\sigma(x) = \sigma(x)e(x) + (1 - \sigma(x)) \left[ c(x) + \beta \sum_{x' \in X} v_\sigma(x')P(x, x') \right] \quad \text{for all } x \in X. \tag{4.2}$$

Indeed, if $\sigma(x) = 1$, then (4.2) states that $v_\sigma(x) = e(x)$, which is what we expect: if we choose to stop at a given state, then lifetime value from that state equals the exit reward. If, instead, $\sigma(x) = 0$, then (4.2) becomes

$$v_\sigma(x) = c(x) + \beta \sum_{x'} v_\sigma(x')P(x, x'), \tag{4.3}$$

which is also what we expect: the value of continuing is the current reward plus the discounted expected reward obtained by continuing with policy $\sigma$ next period.

We want to solve (4.2) for $v_\sigma$. To this end, we define $r_\sigma \in \mathbb{R}^X$ and $L_\sigma \in \mathcal{L}(\mathbb{R}^X)$ via

$$r_\sigma(x) := \sigma(x)e(x) + (1 - \sigma(x))c(x) \quad \text{and} \quad L_\sigma(x, x') := \beta(1 - \sigma(x))P(x, x').$$

With this notation, we can write (4.2) pointwise as $v_\sigma = r_\sigma + L_\sigma v_\sigma$. If $\rho(L_\sigma) < 1$, then

$$v_\sigma = (I - L_\sigma)^{-1} r_\sigma. \tag{4.4}$$

**EXERCISE 4.1.1.** Confirm that $\rho(L_\sigma) < 1$ holds for any optimal stopping problem.

By Exercise 4.1.1 and the Neumann series lemma, $v_\sigma$ is uniquely defined by (4.4).

### 4.1.1.3 Policy Operators

For the proofs below, it is helpful to view $v_\sigma$ as the fixed point of an operator. We associate each $\sigma \in \Sigma$ with an **policy operator** $T_\sigma$ defined at $v \in \mathbb{R}^X$ by

$$(T_\sigma v)(x) = \sigma(x)e(x) + (1 - \sigma(x))\left[c(x) + \beta \sum_{x'} v(x')P(x, x')\right] \tag{4.5}$$

for each $x \in X$. With this notation, (4.2) can be written as $v_\sigma = T_\sigma v_\sigma$.

Exercise 4.1.2. Prove that, for any $\sigma \in \Sigma$, the operator $T_\sigma$ is order-preserving with respect to the pointwise partial order $\leqslant$ on $\mathbb{R}^X$.

Using the notation in §4.1.1.2, we can also define $T_\sigma$ via

$$T_\sigma v = r_\sigma + L_\sigma v.$$

**Proposition 4.1.1.** *For any $\sigma \in \Sigma$, the policy operator $T_\sigma$ is a contraction of modulus $\beta$ on $\mathbb{R}^X$ under the supremum norm.*

The significance of Proposition 4.1.1 is that by construction $v_\sigma$ is a fixed point of $T_\sigma$. By the contraction property in Proposition 4.1.1, $v_\sigma$ is the only fixed point of $T_\sigma$ in $\mathbb{R}^X$ and, moreover, iterates of $T_\sigma$ always converge to $v_\sigma$.

Exercise 4.1.3. Prove Proposition 4.1.1.

### 4.1.1.4 The Value Function

In the job search problem in §3.3.1, we argued that the value function equals the fixed point of the Bellman operator. Here we make the same argument more formally in the more general setting of optimal stopping.

First, given an optimal stopping problem $\mathcal{S} = (\beta, P, r, e)$ with $\sigma$-value functions $\{v_\sigma\}_{\sigma \in \Sigma}$, we define the **value function** $v^*$ of $\mathcal{S}$ via

$$v^*(x) := \max_{\sigma \in \Sigma} v_\sigma(x) \qquad (x \in X), \tag{4.6}$$

so that $v^*(x)$ is the maximal lifetime value available to an agent facing current state $x$. Following notation in §2.2.2.1, we can also write $v^* = \vee_\sigma v_\sigma$.

Given that solving the maximization in (4.6) is, in general, a difficult problem, how can we obtain the value function? The following steps can do the job:

(i) formulate a Bellman equation for the value function of the optimal stopping problem, namely,

$$v(x) = \max\left\{e(x), c(x) + \beta \sum_{x'} v(x')P(x,x')\right\} \qquad (x \in \mathsf{X}), \qquad (4.7)$$

(ii) prove that this Bellman equation has a unique solution in $\mathbb{R}^{\mathsf{X}}$, and then

(iii) show that this solution equals the value function, as defined in (4.6).

We shall complete these steps in §4.1.1.5.

### 4.1.1.5 The Bellman Operator

Define the **Bellman operator** for the optimal stopping problem $\mathcal{S} = (\beta, P, r, e)$ as

$$(Tv)(x) = \max\left\{e(x), \, c(x) + \beta \sum_{x'} v(x')P(x,x')\right\} \qquad (4.8)$$

where $x \in \mathsf{X}$ and $v \in \mathbb{R}^{\mathsf{X}}$. By construction, any fixed point of $T$ solves the Bellman equation and vice versa. Pointwise, we can express $T$ via $Tv = e \vee (c + \beta Pv)$.

EXERCISE 4.1.4. Prove that $T$ is an order-preserving self-map on $\mathbb{R}^{\mathsf{X}}$.

Our main result for this section is:

**Proposition 4.1.2.** *If $\mathcal{S}$ is an optimal stopping problem with Bellman operator $T$ and value function $v^*$, then*

(i) *$T$ is a contraction map of modulus $\beta$ on $\mathbb{R}^{\mathsf{X}}$ under the supremum norm $\|\cdot\|_\infty$ and*

(ii) *the unique fixed point of $T$ on $\mathbb{R}^{\mathsf{X}}$ is the value function $v^*$.*

EXERCISE 4.1.5. Prove the claim in (i) of Proposition 4.1.2.

*Proof of Proposition 4.1.2.* With the result of Exercise 4.1.5 in hand, we need only show that the unique fixed point $\bar{v}$ of $T$ in $\mathbb{R}^{\mathsf{X}}$ is equal to $v^* = \vee_\sigma v_\sigma$. We show $\bar{v} \leqslant v^*$ and then $\bar{v} \geqslant v^*$.

For the first inequality, let $\sigma \in \Sigma$ be defined by

$$\sigma(x) = \mathbb{1}\left\{e(x) \geqslant c(x) + \beta \sum_{x'} \bar{v}(x')P(x,x')\right\} \quad \text{for all } x \in \mathsf{X}.$$

Observe that for this choice of $\sigma$ we have, for any $x \in \mathsf{X}$,

$$(T_\sigma \bar{v})(x) = \sigma(x)e(x) + (1 - \sigma(x))\left[c(x) + \beta \sum_{x'} \bar{v}(x')P(x,x')\right]$$

$$= \max\left\{e(x),\ c(x) + \beta \sum_{x'} \bar{v}(x')P(x,x')\right\} = (T\bar{v})(x) = \bar{v}(x).$$

In particular, $T_\sigma \bar{v} = \bar{v}$. But the only fixed point of $T_\sigma$ in $\mathbb{R}^\mathsf{X}$ is $v_\sigma$, so $\bar{v} = v_\sigma$. But then $\bar{v} \leqslant v^*$, by the definition of $v^*$. This is our first inequality.

Regarding the second, fix $\sigma \in \Sigma$ and observe that $Tv \geqslant T_\sigma v$ for all $v \in \mathbb{R}^\mathsf{X}$. Since $T$ is order-preserving and globally stable, Proposition 2.2.7 on page 66 implies that $v_\sigma \leqslant \bar{v}$. Taking the maximum over $\sigma \in \Sigma$ yields $v^* \leqslant \bar{v}$. $\qquad\square$

### 4.1.1.6 Optimal Policies

Paralleling the definition provided in the discussion of job search (§1.3), for each $v \in \mathbb{R}^\mathsf{X}$, we call $\sigma \in \Sigma$ **$v$-greedy** if, for all $x \in \mathsf{X}$,

$$\sigma(x) \in \operatorname*{argmax}_{a \in \{0,1\}}\left\{ae(x) + (1 - a)\left[c(x) + \beta \sum_{x'} v(x')P(x,x')\right]\right\}. \tag{4.9}$$

A $v$-greedy policy uses $v$ to assign values to states and then chooses to stop or continue based on the action that generates a higher payoff.

With this language in place, the next proposition makes precise our informal §1.1.2 argument that optimal choices can be made using the value function.

**Proposition 4.1.3.** *Policy $\sigma \in \Sigma$ is optimal if and only if it is $v^*$-greedy.*

Proposition 4.1.3 is a version of **Bellman's principle of optimality**. We shall prove this principle in a more general setting in Chapter 5.

#### 4.1.1.7 Value Function Iteration

The theory stated above tells us that successive approximation using the Bellman operator converges to $v^*$ and $v^*$-greedy policies are optimal. These facts make value function iteration (VFI) a natural algorithm for solving optimal stopping problems. (VFI for optimal stopping problems is corresponds to VFI for job search, as shown on page 37.) Later, in Theorem 8.1.1, we will show that when the number of iterates is sufficiently large, VFI produces an optimal policy.

### 4.1.2 Firm Valuation with Exit

In §3.2.2.2 we discussed firm valuation using expected present value of the cash flow generated by profits. This is a standard approach. However, it ignores that firms have the option to cease operations and sell all remaining assets. In this section, we consider firm valuation in the presence of an exit option.

#### 4.1.2.1 Optional Exit

Consider a firm whose productivity is exogenous and evolves according to a $Q$-Markov chain $(Z_t)$ on finite set $\mathsf{Z} \subset \mathbb{R}$. Profits are given by $\pi_t = \pi(Z_t)$ for some fixed $\pi \in \mathbb{R}^\mathsf{Z}$. At the start of each period, the firm decides whether to remain in operation and receive current profit $\pi_t$, or to exit and receive scrap value $s > 0$ for sale of physical assets. Discounting is at fixed rate $r$ and $\beta := 1/(1+r)$. We assume that $r > 0$.

Let $\Sigma$ be all $\sigma \colon \mathsf{Z} \to \{0, 1\}$. For given $\sigma \in \Sigma$ and $v \in \mathbb{R}^\mathsf{Z}$, the corresponding policy operator is

$$(T_\sigma v)(z) = \sigma(z)s + (1 - \sigma(z)) \left[ \pi(z) + \beta \sum_{z'} v(z')Q(z, z') \right] \qquad (z \in \mathsf{Z}).$$

We saw in §4.1.1.2–§4.1.1.3 that $T_\sigma$ has a unique fixed point $v_\sigma$ and that $v_\sigma(z)$ represents the value of following policy $\sigma$ forever, conditional on $Z_0 = z$.

The Bellman operator for the firm's problem is the order-preserving self-map $T$ on $\mathbb{R}^\mathsf{Z}$ defined by

$$(Tv)(z) = \max \left\{ s, \pi(z) + \beta \sum_{z'} v(z')Q(z, z') \right\} \qquad (z \in \mathsf{Z}).$$

Pointwise, $T$ can be written as $Tv = s \vee (\pi + \beta Qv)$.

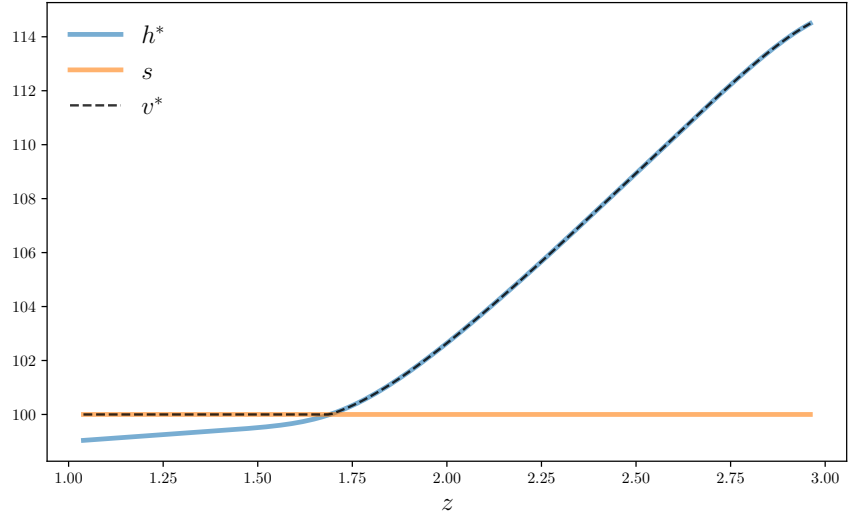

Figure 4.1:  Value function for firms with exit option

Let $v^*$ be the value function for this problem.  By Proposition 4.1.2, $v^*$ is the unique fixed point of $T$ in $\mathbb{R}^\mathsf{Z}$ and the unique solution to the Bellman equation.  Moreover, successive approximation from any $v \in \mathbb{R}^\mathsf{Z}$ converges to $v^*$.  Finally, by Proposition 4.1.3, a policy is optimal if and only if it is $v^*$-greedy.

Figure 4.1 plots $v^*$, computed via value function iteration (i.e., successive approximation using $T$, along with the stopping value $s$ and the continuation value function $h^* = \pi + \beta Q v^*$, under the parameterization given in Listing 12.  As implied by the Bellman equation, $v^*$ is the pointwise maximum of $s$ and $h^*$.  The $v^*$-greedy policy instruct the firm to exit when the continuation value of the firm falls below the scrap value.

**Exercise** 4.1.6.  Replicate Figure 4.1 by using the parameters in Listing 12 and applying value function iteration.  Reviewing the code for job search on page 100 should be helpful.

### 4.1.2.2    Exit vs No-Exit

If we define $w$ by $w(z) = \mathbb{E}_z \sum_{t \geqslant 0} \beta^t \pi_t$ for all $z \in \mathsf{Z}$, then $w(z)$ is the value of the firm given $Z_0 = z$ when the firm never exits so that $w$ evaluates the firm according to expected present value of the profit stream.  Figure 4.2 shows the no-exit value $w$ based on the parameterization in Listing 12.

In Figure 4.2, we see that $w \leqslant v^*$ on Z.  Let's now prove that this is always true.

```julia
"Creates an instance of the firm exit model."
function create_exit_model(;
        n=200,                      # productivity grid size
        ρ=0.95, μ=0.1, ν=0.1,       # persistence, mean and volatility
        β=0.98, s=100.0             # discount factor and scrap value
    )
    mc = tauchen(n, ρ, ν, μ)
    z_vals, Q = mc.state_values, mc.p
    return (; n, z_vals, Q, β, s)
end
```

Listing 12: Firm exit model (`firm_exit.jl`)

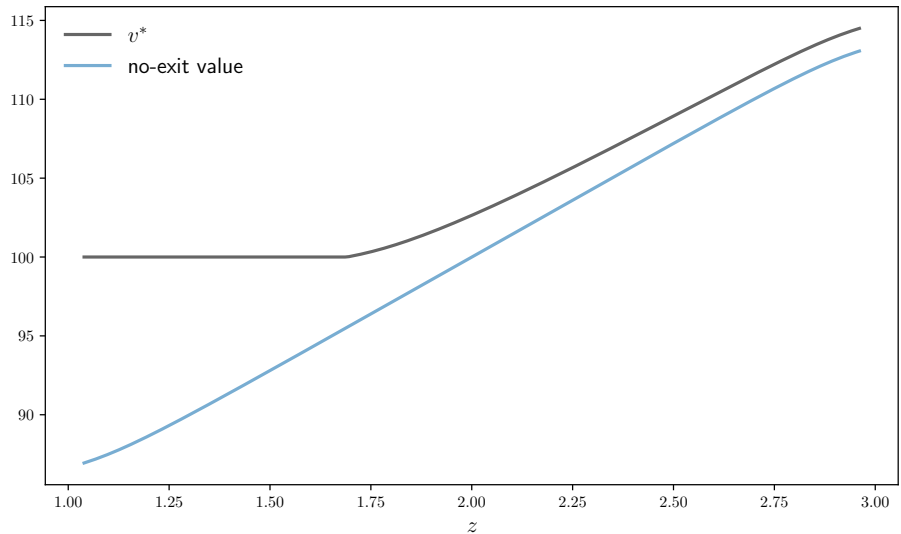

Figure 4.2: Firm value with and without exit

To show $w \leqslant v^*$, first observe that $w = (I - \beta Q)^{-1} \pi$, by $\beta < 1$ and Lemma 3.2.1 on page 95. Rearranging gives $w = \pi + \beta Q w$.

Now note that under the policy $\sigma \equiv 0$, where the firm chooses never to exit, we have $T_\sigma v = \pi + \beta Q v$. Hence the unique fixed point of $T_\sigma$ is $w$. As a result, $w = v_\sigma$ for $\sigma \equiv 0$. But $v^* \geqslant v_\sigma$ for all $\sigma \in \Sigma$. This proves that $w \leqslant v^*$.

Choosing never to exit is a feasible policy. Since $v^*$ involves maximization of firm value over the set of all feasible policies, it must be at least as large as the value of never exiting.

EXERCISE 4.1.7. Prove the following: If $Q \gg 0$ and $s > w(z)$ for at least one $z \in \mathsf{Z}$, then $w \ll v^*$. Provide some intuition for this result.

EXERCISE 4.1.8. Consider a version of the model of firm value with exit where productivity is constant but prices are stochastic. In particular, the price process $(P_t)$ for the final good is $Q$-Markov. Suppose further that one-period profits for a given price $p$ are $\max_{\ell \geqslant 0} \pi(\ell, p)$, where $\ell$ is labor input. Suppose that $\pi(\ell, p) = p\ell^{1/2} - w\ell$, where the wage rate $w$ is constant. Formulate the Bellman equation.

### 4.1.3 Monotonicity

We study monotonicity in values and actions in the general optimal stopping problem described in §4.1.1, with X as the state space, $e$ as the exit reward function and $c$ as the continuation reward function.

#### 4.1.3.1 Monotone Values

Let $v^*$ be the value function of an optimal stopping problem defined by X, $P$, $\beta$, $c$ and $e$ and define a **continuation value function** $h^*$

$$h^*(x) := c(x) + \beta \sum_{x'} v^*(x') P(x, x') \qquad (x \in \mathsf{X}). \tag{4.10}$$

(The continuation reward function $c$ and the continuation value function $h^*$ are distinct objects.)

Let X be partially ordered and let $i\mathbb{R}^{\mathsf{X}}$ be the increasing functions in $\mathbb{R}^{\mathsf{X}}$.

**Lemma 4.1.4.** *If $e, c \in i\mathbb{R}^{\mathsf{X}}$ and $P$ is monotone increasing, then $h^*$ and $v^*$ are both increasing.*

*Proof.* Let the stated conditions hold. The Bellman operator can be written pointwise as $Tv = e \vee (c + \beta Pv)$. Since $P$ is monotone increasing, $P$ is invariant on $i\mathbb{R}^{\mathsf{X}}$. It follows from this fact and the conditions on $e$ and $c$ that $T$ is invariant on $i\mathbb{R}^{\mathsf{X}}$. Hence, by Exercise 1.2.18 on page 22, $v^*$ is in $i\mathbb{R}^{\mathsf{X}}$. Since $h^* = c + \beta Pv^*$, the same is true for $h^*$. □

**Example 4.1.3.** Consider the §4.1.2 firm problem with exit with Bellman operator $Tv = s \vee (\pi + \beta Qv)$. Since $s$ is constant, it follows directly that $v^*$ and $h^*$ are both increasing functions when $\pi \in i\mathbb{R}^{\mathsf{Z}}$ and $Q$ is monotone increasing.

### 4.1.3.2 Monotone Actions

The optimal policy in the ııd job search problem takes the form $\sigma^*(w) = \mathbb{1}\{w \geqslant w^*\}$ for all $w \in \mathsf{W}$, where $w^* := (1 - \beta)h^*$ is the reservation wage and $h^*$ is the continuation value (see page 36). This optimal policy is of threshold type: once the wage offer exceeds the threshold, the decision is to stop.

Since threshold policies are convenient, let us now try to characterize them.

Throughout this section, we take $\mathsf{X}$ to be a subset of $\mathbb{R}$. Elements of $\mathsf{X}$ are ordered by $\leqslant$, the usual order on $\mathbb{R}$.

EXERCISE 4.1.9. Prove that the optimal policy $\sigma^*$ is decreasing on $\mathsf{X}$ whenever $e$ is decreasing on $\mathsf{X}$ and $h^*$ is increasing on $\mathsf{X}$.

For a binary function on $\mathsf{X} \subset \mathbb{R}$, the condition that $\sigma^*$ is decreasing means that the decision maker chooses to exit when $x$ is sufficiently small.

**Example 4.1.4.** In the firm problem with exit, as described in §4.1.2, $h^*$ is increasing whenever $\pi \in i\mathbb{R}^{\mathsf{Z}}$ and $Q$ is monotone increasing. Since the scrap value is constant, Exercise 4.1.9 applies under these conditions. Hence the optimal policy is decreasing. This reasoning agrees with Figure 4.1, where exit is optimal when the state is small and continuing is optimal when $z$ is large. This makes sense: since $Q$ is monotone increasing, low current values of $z$ predict low future values of $z$, so profits associated with continuing can be anticipated to be low.

EXERCISE 4.1.10. Show that the conditions of Exercise 4.1.9 hold when $e$ is constant on $\mathsf{X}$, $c$ is increasing on $\mathsf{X}$, and $P$ is monotone increasing.

EXERCISE 4.1.11. Prove that the optimal policy $\sigma^*$ is increasing on $\mathsf{X}$ whenever $e$ is increasing on $\mathsf{X}$ and $h^*$ is decreasing on $\mathsf{X}$.

**Example 4.1.5.** In the IID job search problem, $e(w) = w/(1 - \beta)$ is increasing and $h^*$ is constant. Hence the result in Exercise 4.1.11 applies. This is why the optimal policy $\sigma^*(w) = \mathbb{1}\{w \geqslant (1 - \beta)h^*\}$ is increasing. The agent accepts all sufficiently large wage offers.

In the settings of Exercises 4.1.9–4.1.11, the optimal policy is either increasing or decreasing. Since X is totally ordered, monotonicity implies that a threshold policy is optimal. For example, if $\sigma^*$ is increasing, then we take $x^*$ to be the smallest $x \in \mathsf{X}$ such that $\sigma^*(x) = 1$. For such an $x^*$ we have

$$x < x^* \implies \sigma^*(x) = 0 \quad \text{and} \quad x \geqslant x^* \implies \sigma^*(x) = 1.$$

**Remark 4.1.1.** Conditions in Exercises 4.1.9–4.1.11 are sufficient but not necessary for monotone policies. Figure 3.5 on 101 provides an example of a setting where the policy is increasing (the agent accepts for sufficiently large wage offers) even though both $e(x) = x/(1 - \beta)$ and $h^*$ are strictly increasing.

### 4.1.4 Continuation Values

In §1.3.2.2 we solved the job search problem with IID draws by computing the continuation value $h^*$ directly and then setting the optimal policy to $\sigma^*(w) = \mathbb{1}\{w/(1 - \beta) \geqslant h^*\}$. We saw that this approach is more efficient than first computing the value function, since the continuation value is one-dimensional rather than |W|-dimensional.

In §3.3.1.2, we tried the same approach for the job search problem with Markov state, where wage draws are correlated. We gathered fewer benefits from using the continuation value approach in that setting, since the continuation value function has the same dimensionality as the value function.

These observations motivate us to explore continuation value methods more carefully. In this section, we formulate a continuation value approach for the general optimal stopping problem and verify convergence. We will see that, while all relevant state components must be included in the value function, purely transitory components do not affect continuation values. Hence the continuation value approach is at least as efficient and sometimes substantially more so.

Another asymmetry between value functions and continuation value functions is that the latter are typically smoother. For example, in job search problems, the value function is usually kinked at the reservation wage, while the continuation value function is smooth. Greater smoothness comes from taking expectations over stochastic transitions: integration acts as a smoothing operation. Like lower dimensionality, increased smoothness facilitates analysis and computation.

### 4.1.4.1  The Continuation Value Operator

Let $h^*$ be the continuation value function for the optimal stopping problem defined in (4.10). To compute $h^*$ directly we begin with the optimal stopping version of the Bellman equation evaluated at $v^*$ and rewrite it as

$$v^*(x') = \max\{e(x'), h^*(x')\} \qquad (x' \in \mathsf{X}). \tag{4.11}$$

Taking expectations of both sides of the equation conditional on current state $x$ produces $\sum_{x'} v^*(x') P(x, x') = \sum_{x'} \max\{e(x'), h^*(x')\} P(x, x')$. Multiplying by $\beta$, adding $c(x)$, and using the definition of $h^*$, we get

$$h^*(x) = c(x) + \beta \sum_{x'} \max\{e(x'), h^*(x')\} P(x, x') \qquad (x \in \mathsf{X}). \tag{4.12}$$

This expression motivates us to introduce a **continuation value operator** $C \colon \mathbb{R}^\mathsf{X} \to \mathbb{R}^\mathsf{X}$ via

$$(Ch)(x) = c(x) + \beta \sum_{x'} \max\{e(x'), h(x')\} P(x, x') \qquad (x \in \mathsf{X}). \tag{4.13}$$

**Proposition 4.1.5.** *The operator $C$ is a contraction of modulus $\beta$ on $\mathbb{R}^\mathsf{X}$ with the unique fixed point $h^*$ in $\mathbb{R}^\mathsf{X}$.*

Proposition 4.1.5 provides the following alternative method to compute the optimal policy that does not involve value function iteration:

(i) Use successive approximations to $h^*$ with $C$ and

(ii) Calculate $\sigma^*$ via $\sigma^*(x) = \mathbb{1}\{e(x) \geqslant h^*(x)\}$ for each $x \in \mathsf{X}$.

In §4.1.4.2 we discuss settings where this approach is advantageous.

*Proof of Proposition 4.1.5.* Fix $f, g \in \mathbb{R}^\mathsf{X}$ and $x \in \mathsf{X}$. By the triangle inequality and the bound $|\alpha \vee x - \alpha \vee y| \leqslant |x - y|$ from page 34, we have

$$|(Cf)(x) - (Cg)(x)| \leqslant \beta \sum_{x'} |\max\{e(x'), f(x')\} - \max\{e(x'), g(x')\}| P(x, x')$$

$$\leqslant \beta \sum_{x'} |f(x') - g(x')| P(x, x').$$

The right-hand side is dominated by $\beta \|f - g\|_\infty$. Taking the maximum on the left-hand side gives

$$\|Cf - Cg\|_\infty \leqslant \beta \|f - g\|_\infty,$$

which confirms that $C$ is a contraction of modulus $\beta$ on $\mathbb{R}^{\mathsf{X}}$.

From the contraction property, we know that $C$ has exactly one fixed point in $\mathbb{R}^{\mathsf{X}}$; (4.12) implies that $h^*$ is that fixed point. $\qquad\square$

### 4.1.4.2 Dimensionality Reduction

The beginning of §4.1.4 mentioned that switching from value function iteration to continuation value iteration can substantially reduce the dimensionality of the problem in some cases. Here we describe situations where this works.

To begin, let W and Z be two finite sets and suppose that $\varphi \in \mathcal{D}(\mathsf{W})$ and $Q \in \mathcal{M}(\mathbb{R}^{\mathsf{Z}})$. Let $(W_t)$ be IID with distribution $\varphi$ and let $(Z_t)$ be an $Q$-Markov chain on Z. If $(W_t)$ and $(Z_t)$ are independent, then $(X_t)$ defined by $X_t = (W_t, Z_t)$ is $P$-Markov on X, where

$$P(x, x') = P((w, z), (w', z')) = \varphi(w')Q(z, z').$$

Suppose that the continuation reward depends only on $z$ so that we can write the Bellman operator as

$$(Tv)(w, z) = \max \left\{ e(w, z),\ c(z) + \beta \sum_{w' \in \mathsf{W}} \sum_{z' \in \mathsf{Z}} v(w', z')\varphi(w')Q(z, z') \right\}. \qquad (4.14)$$

Since the right-hand side depends on both $w$ and $z$, the Bellman operator acts on an $n$-dimensional space, where $n \coloneqq |\mathsf{X}| = |\mathsf{W}| \times |\mathsf{Z}|$.

However, if we inspect the right-hand side of (4.14), we see that the continuation value function depends only on $z$. Dependence on $w$ vanishes because $w$ does not help predict $w'$. Thus, the continuation value function is an object in $|\mathsf{Z}|$-dimensional space. The continuation value operator

$$(Ch)(z) = c(z) + \beta \sum_{w'} \sum_{z'} \max \{ e(w', z'), h(z') \}\, \varphi(w')Q(z, z') \qquad (z \in \mathsf{Z}) \qquad (4.15)$$

acts on this lower dimensional-space.

**Example 4.1.6.** We can embed the IID the job search problem into this setting by taking $(W_t)$ to be the wage offer process and $(Z_t)$ to be constant. This is why the IID case offers a large dimensionality reduction when we switch to continuation values.

More examples of dimensionality reduction are illustrated in the applications below.

### 4.1.4.3   Application to Firm Value

Consider the firm valuation problem from §4.1.2 but suppose now that scrap value fluctuates with prices of underlying assets. For simplicity let's assume that scrap value at each time $t$ is given by the IID sequence $(S_t)$, where each $S_t$ has density $\varphi$ on $\mathbb{R}_+$. The corresponding Bellman operator is

$$(Tv)(z,s) = \max \left\{ s, \pi(z) + \beta \sum_{z'} \int v(z',s') \varphi(s') \, \mathrm{d}s' Q(z,z') \right\}.$$

We can convert this problem to a finite state space optimal stopping problem by discretizing the density $\varphi$ onto a finite grid contained in $\mathbb{R}_+$. However, since continuation values depend only on $z$, a better approach is to switch to a continuation value operator.

EXERCISE 4.1.12. Write down the continuation value operator for this function as a mapping from $\mathbb{R}^Z$ to itself.

EXERCISE 4.1.13. In §2.2.4 we defined stochastic dominance for distributions on finite sets. For densities $\varphi$ and $\psi$ on $\mathbb{R}_+$, the definition is similar: we say that $\psi$ stochastically dominates $\varphi$ and write $\varphi \preceq_F \psi$ if $\int u(x) \varphi(x) \, \mathrm{d}x \leqslant \int u(x) \psi(x) \, \mathrm{d}x$ for every $u$ in $i\mathbb{R}^X$.[2] With this definition, show that if $\varphi_a$ and $\varphi_b$ are two alternative densities for scrap value and $\varphi_a \preceq_F \varphi_b$, then $\sigma_a^* \geqslant \sigma_b^*$ pointwise on Z, where $\sigma_i^*$ is the optimal policy corresponding to density $\varphi_i$ for $i \in \{a, b\}$. Interpret this result.

## 4.2   Further Applications

In this section we discuss some applications of optimal stopping and apply the results described above.

### 4.2.1   American Options

We discussed American options briefly in Example 4.1.2 on page 107. Here we investigate this class of derivatives more carefully. We focus on American call options that provide the right to buy a particular stock or bond at a fixed **strike price** $K$ at any time before a set expiration date. The market price of the asset at time $t$ is denoted by $S_t$.

---

[2]Actually, in most definitions, $u$ is also restricted to be bounded and measurable, in order to ensure that the integrals are finite. These technicalities can be ignored in the exercise.

We discussed a case in which the expiration date is infinity in Example 4.1.2. However, options without termination dates – also called perpetual options – are rare in practice. Hence we focus on the finite-horizon case. We are interested in computing the expected value of holding the option when discounting with a fixed interest rate, a typical assumption when pricing American options.

Finite horizon American options can be priced by backward induction in an approach like the one we used for the finite horizon job search problem discussed in Chapter 1. Alternatively, we can embed finite horizon optimal stopping into the theory of infinite-horizon optimal stopping. We use the second approach here, since we have just presented a theory for infinite-horizon optimal stopping.

To this end, we take $T \in \mathbb{N}$ to be a fixed integer indicating the date of expiration. The option is purchased at $t = 0$ and can be exercised at any $t \in \mathbb{N}$ with $t \leqslant T$. To include $t$ in the current state, we set

$$\mathsf{T} := \{1, \dots, T+1\} \quad \text{and} \quad m(t) := \min\{t+1, T+1\} \text{ for all } t \in \mathsf{T}.$$

The idea is that time is updated via $t' = m(t)$, so that time increments at each update until $t = T + 1$. After that we hold $t$ constant. Bounding time at $T + 1$ keeps the state space finite.

We assume that a stock price $S_t$ evolves according to

$$S_t = Z_t + W_t \quad \text{where} \quad (W_t)_{t \geqslant 0} \overset{\text{IID}}{\sim} \varphi \in \mathcal{D}(\mathsf{W}).$$

Here $(Z_t)_{t \geqslant 0}$ is $Q$-Markov on finite set $\mathsf{Z}$ for some $Q \in \mathcal{M}(\mathbb{R}^{\mathsf{Z}})$ and $\mathsf{W}$ is also finite. This means that the share price has both persistent and transient stochastic components. If we set parameters so that $(Z_t)_{t \geqslant 0}$ resembles a random walk, price changes will be difficult to predict.

To form a §4.1.1.1 optimal stopping problem, we must specify the state and clarify the $P \in \mathcal{M}(\mathbb{R}^{\mathsf{X}})$ that maps to the state process. We set the state space to $\mathsf{X} := \mathsf{T} \times \mathsf{W} \times \mathsf{Z}$ and

$$P((t, w, z), (t', w', z')) := \mathbb{1}\{t' = m(t)\}\varphi(w')Q(z, z').$$

Thus, time updates deterministically via $t' = m(t)$ and $z'$ and $w'$ are drawn independently from $Q(z, \cdot)$ and $\varphi$ respectively.

As for a perpetual option, the continuation reward is zero and the discount factor is $\beta := 1/(1+r)$, where $r > 0$ is a fixed risk-free rate. The exit reward can be expressed as $\mathbb{1}\{t \leqslant T\}(S_t - K)$ so that exercising at time $t$ earns the owner $S_t - K$ up to expiry and zero thereafter. In terms of the state $(t, z)$, the exit reward is

$$e(t, w, z) := \mathbb{1}\{t \leqslant T\}[z + w - K].$$

The Bellman equation can be written

$$v(t, w, z) = \max \left\{ e(t, w, z), \, \beta \sum_{w'} \sum_{z'} v(t', w', z') \varphi(w') Q(z, z') \right\},$$

where $t' = m(t)$. This value function $v(t, w, z)$ neatly captures the value of the option: It is the maximum of current exercise value and the discounted expected value of carrying the option over to the next period.

Since the problem described above is an optimal stopping problem in the sense of §4.1.1.1, all of the optimality results attained for that problem apply. In particular, iterates of the Bellman operator converge to the value function $v^*$ and, moreover, a policy is optimal if and only if it is $v^*$-greedy.

We can do better than value function iteration. Since $(W_t)_{t \geqslant 0}$ is IID and appears only in the exit reward, we can reduce dimensionality by switching to the continuation value operator, which, in this case, can be expressed as

$$(Ch)(t, z) = \beta \sum_{z'} \sum_{w'} \max \{e(t', w', z'), \, h(t', z')\} \, \varphi(w') Q(z, z'). \tag{4.16}$$

As proved in §4.1.4, the unique fixed point of $C$ is the continuation value function $h^*$, and $C^k h \to h^*$ as $k \to \infty$ for all $h \in \mathbb{R}^{\mathsf{X}}$. With the fixed point in hand, we can compute the optimal policy as

$$\sigma^*(t, w, z) = \mathbb{1}\left\{ e(t, w, z) \geqslant h^*(t, z) \right\}.$$

Here $\sigma^*(t, w, z) = 1$ prescribes exercising the option at time $t$.

Figure 4.3 provides a visual representation of optimal actions under the default parameterization described in Listing 13. Each of the three figures show contour lines of the net exit reward $f(t, w, z) := e(t, w, z) - h^*(w, z)$, viewed as a function of $(w, z)$, when $t$ is held fixed. The date $t$ for each subfigure is shown in the title. The optimal policy exercises the option when $f(t, w, z) \geqslant 0$.

In each subfigure, the **exercise region**, which is the set $(w, z)$ such that $f(t, w, z) \geqslant 0$, correspond to the northeast part of the figure, where $w$ and $z$ are both large. The boundary between exercise and continuing is the zero contour line, which is shown

```julia
using QuantEcon, LinearAlgebra, IterTools

"Creates an instance of the option model with log S_t = Z_t + W_t."
function create_american_option_model(;
        n=100, μ=10.0,      # Markov state grid size and mean value
        ρ=0.98, ν=0.2,      # persistence and volatility for Markov state
        s=0.3,              # volatility parameter for W_t
        r=0.01,             # interest rate
        K=10.0, T=200)      # strike price and expiration date
    t_vals = collect(1:T+1)
    mc = tauchen(n, ρ, ν)
    z_vals, Q = mc.state_values .+ μ, mc.p
    w_vals, φ, β = [-s, s], [0.5, 0.5], 1 / (1 + r)
    e(t, i_w, i_z) = (t ≤ T) * (z_vals[i_z] + w_vals[i_w] - K)
    return (; t_vals, z_vals, w_vals, Q, φ, T, β, K, e)
end
```

Listing 13:  Pricing and American option (`american_option.jl`)

in black. Notice that the size of the exercise region expands with $t$. This is because the value of waiting decreases when the set of possible exercise dates declines.

Figure 4.4 provides some simulations of the stock price process $(S_t)_{t \geqslant 0}$ over the lifetime of the option, again using the default parameterization described in Listing 13. The blue region in the top part of each subfigure contains values of the stock price $S_t = Z_t + W_t$ such that $S_t \geqslant K$. In this configuration in which the price of the underlying exceeds the strike price, the option is said to be "in the money." The figure also shows an optimal exercise date that is the first $t$ such that $e(t, W_t, Z_t) \geqslant h^*(W_t, Z_t)$ in a simulation.

### 4.2.2   Research and Development

Consider a firm that engages in costly research and development (R&D) in order to develop a new product. The firm decides whether to continue developing the product before starting to market it or to stop developing and start marketing it. For simplicity, we assume that the value of bringing the product to market is a one-time lump sum payment $\pi_t = \pi(X_t)$, where $(X_t)$ is a $P$-Markov chain on finite set $X$ with $P \in \mathcal{M}(\mathbb{R}^X)$. The flow cost of investing in R&D is $C_t$ per period, where $(C_t)$ is a stochastic process. Future payoffs are discounted at rate $r > 0$ and we set $\beta := 1/(1 + r)$.

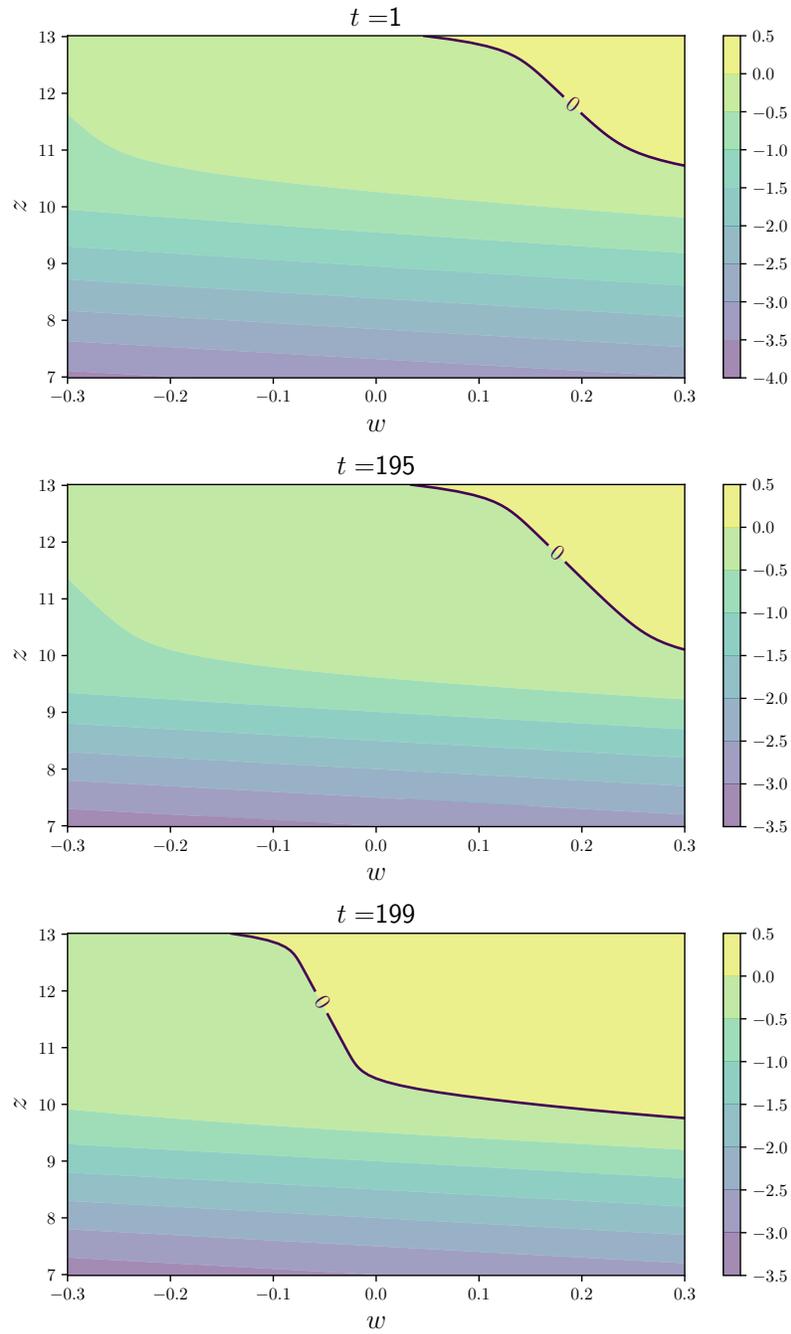

Figure 4.3: Exercise region for the American option

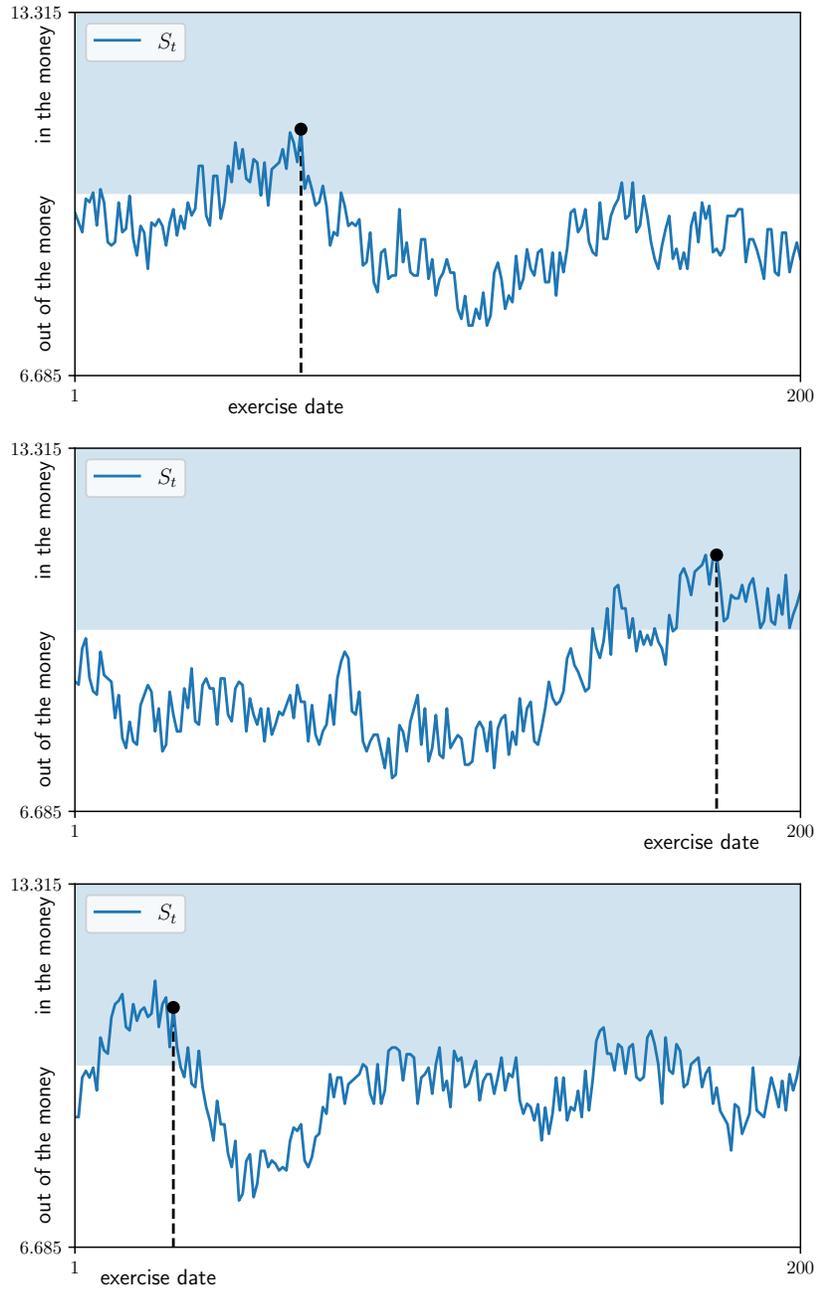

Figure 4.4: Simulations for the American option process

#### 4.2.2.1 Constant R&D Costs

As a first take on this problem, suppose that $C_t \equiv c \in \mathbb{R}_+$ for all $t$ and formulate an optimal stopping problem with exit reward $e = \pi$ and constant continuation reward $-c$. The Bellman equation is

$$v(x) = \max\left\{\pi(x), -c + \beta \sum_{x'} v(x')P(x,x')\right\} \qquad (x \in \mathsf{X}). \qquad (4.17)$$

Exercise 4.2.1. Write down the continuation value operator for this problem. Prove that the continuation value function $h^*$ is increasing in $x$ whenever $\pi \in i\mathbb{R}^{\mathsf{X}}$ and $P$ is monotone increasing.

Exercise 4.2.2. Prove that the optimal policy $\sigma^*$ is increasing whenever $\pi$ is increasing and $(X_t)$ is IID (so that all rows of $P$ are identical). Provide economic intuition for this result.

#### 4.2.2.2 IID R&D Costs

Let's suppose now that $(C_t)_{t \geqslant 0}$ is IID with common distribution $\varphi \in \mathcal{D}(\mathsf{W})$. The Bellman equation is

$$v(c,x) = \max\left\{\pi(x), -c + \beta \sum_{x'} \sum_{c'} v(c',x')\varphi(c')P(x,x')\right\}. \qquad (4.18)$$

Since $(C_t)$ is IID, we would ideally like to integrate it out in the manner of §4.1.4.2, thereby lowering the dimensionality of the problem. However, note that the continuation value associated with (4.18) is

$$h(c,x) := -c + \beta \sum_{x'} \sum_{c'} v(c',x')\varphi(c')P(x,x'),$$

which still depends on $c$.

Fortunately, there is a way to eliminate $c$. Define the expected discounted value $g(x)$ in state $x$

$$g(x) := \sum_{x'} \sum_{c'} v(c',x')\varphi(c')P(x,x'). \qquad (4.19)$$

Rewrite the Bellman equation using $g$ and replacing $(c, x)$ with $(c', x')$ to get

$$v(c', x') = \max\{\pi(x'), -c' + \beta g(x')\}.$$

Averaging over $(c', x')$ and using the definition of $g$ again gives

$$g(x) = \sum_{x'}\sum_{c'}\max\{\pi(x'), -c' + \beta g(x')\}\,\varphi(c')P(x, x'). \qquad (4.20)$$

This is a functional equation in $g$ that depends only on $x$. To solve it, we introduce the operator $R$ defined by

$$(Rg)(x) = \sum_{x'}\sum_{c'}\max\{\pi(x'), -c' + \beta g(x')\}\,\varphi(c')P(x, x') \quad (x \in \mathsf{X}).$$

EXERCISE 4.2.3. Prove that $R$ is a contraction of modulus $\beta$ on $\mathbb{R}^{\mathsf{X}}$.

From Exercise 4.2.3, we see that (4.20) has a unique solution $g^*$ in $\mathbb{R}^{\mathsf{X}}$ that can be computed by successive approximation. With $g^*$ in hand, we can compute the optimal policy via

$$\sigma^*(c, x) = \mathbb{1}\{\pi(x), -c + \beta g^*(x)\}.$$

**Remark 4.2.1.** This technique solves for the expected value function defined in (4.19). In §5.3 we shall discuss this method and its convergence properties in a more general setting.

## 4.3   Chapter Notes

Various textbooks treat optimal stopping in depth, although most use continuous time. Peskir and Shiryaev (2006) and Shiryaev (2007) are good examples.

There are many applications of optimal stopping in economics and finance, with influential early research papers including McCall (1970), Jovanovic (1982), Hopenhayn (1992), and Ericson and Pakes (1995). Arellano (2008) considers borrowing on international financial markets with the option of sovereign default (see §8.2.1.5). Riedel (2009) studies optimal stopping under Knightian uncertainty. Fajgelbaum et al. (2017) include an optimal stopping problem for firms in a model of uncertainty traps.

The firm problem with optimal exit has been used to analyze firm dynamics and firm size distributions in equilibrium models with heterogeneous firms. Hopenhayn (1992) is the classic reference. Perla and Tonetti (2014) construct a growth model in which firms at the bottom of the productivity distribution imitate more productive firms. Carvalho and Grassi (2019) analyze business cycles in a setting of firm growth with exit and a Pareto distribution of firms.

Infinite duration American options are analyzed in Mordecki (2002). Practical methods for pricing American options are provided by Longstaff and Schwartz (2001), Rogers (2002), and Kohler et al. (2010).

Replacement problems are an important optimal stopping problem not treated in this chapter. An important early paper by Rust (1987) uses dynamic programming to find optimal replacement policies for of engine parts and goes on to fit the model to data. §5.3.1 discusses structural estimation in the style of Rust Rust (1987) and others.

# Chapter 5

# Markov Decision Processes

In this chapter we study a class of discrete time, infinite horizon dynamic programs called Markov decision processes (MDPs). This standard class of problems is broad enough to encompass many applications, including the optimal stopping problems in Chapter 4. MDPs can also be combined with reinforcement learning to tackle settings where important inputs to an MDP are not known.

## 5.1  Definition and Properties

In this section we define MDPs and investigate optimality.

### 5.1.1  The MDP Model

We study a controller who interacts with a state process $(X_t)_{t \geqslant 0}$ by choosing an action path $(A_t)_{t \geqslant 0}$ to maximize expected discounted rewards

$$\mathbb{E} \sum_{t \geqslant 0} \beta^t r(X_t, A_t), \tag{5.1}$$

taking an initial state $X_0$ as given. As with all dynamic programs, we insist that the controller is not clairvoyant: he or she cannot choose actions that depend on future states.

To formalize the problem, we fix a finite set X, henceforth called the **state space**, and a finite set A, henceforth called the **action space**. In what follows, a **correspondence** $\Gamma$ from X to A is a function from X into $\wp(A)$, the set of all subsets of A. The correspondence is called **nonempty** if $\Gamma(x) \neq \emptyset$ for all $x \in \mathsf{X}$. For example, the map $\Gamma$ defined by $\Gamma(x) = [-x, x]$ is a nonempty correspondence from $\mathbb{R}$ to $\mathbb{R}$.

Given $\mathsf{X}$ and $\mathsf{A}$, we define a **Markov decision process** (**MDP**) to be a tuple $\mathcal{M} = (\Gamma, \beta, r, P)$ consisting of

(i) a nonempty correspondence $\Gamma$ from $\mathsf{X}$ to $\mathsf{A}$, referred to as the **feasible correspondence**, which in turn defines the **feasible state-action pairs**

$$\mathsf{G} := \{(x, a) \in \mathsf{X} \times \mathsf{A} : a \in \Gamma(x)\},$$

(ii) a constant $\beta$ in $(0, 1)$, referred to as the **discount factor**,

(iii) a function $r$ from $\mathsf{G}$ to $\mathbb{R}$, referred to as the **reward function**, and

(iv) a **stochastic kernel** $P$ from $\mathsf{G}$ to $\mathsf{X}$; that is, $P$ is a map from $\mathsf{G} \times \mathsf{X}$ to $\mathbb{R}_+$ satisfying

$$\sum_{x' \in \mathsf{X}} P(x, a, x') = 1 \quad \text{for all } (x, a) \text{ in } \mathsf{G}.$$

Here $\Gamma(x) \subset \mathsf{A}$ is the set of actions available to the controller in state $x$. Given a feasible state-action pair $(x, a)$, reward $r(x, a)$ is received and the next period state $x'$ is randomly drawn from $P(x, a, \cdot)$, which is an element of $\mathcal{D}(\mathsf{X})$. The dynamics and reward flow are summarized in Algorithm 5.1.

---
**Algorithm 5.1:** MDP dynamics: states, actions, and rewards
---
1 $t \leftarrow 0$
2 input $X_0$
3 **while** $t < \infty$ **do**
4     observe $X_t$
5     choose action $A_t$
6     receive reward $r(X_t, A_t)$
7     draw $X_{t+1}$ from $P(X_t, A_t, \cdot)$
8     $t \leftarrow t + 1$
9 **end**
---

The **Bellman equation** corresponding to $\mathcal{M}$ is

$$v(x) = \max_{a \in \Gamma(x)} \left\{ r(x, a) + \beta \sum_{x'} v(x') P(x, a, x') \right\} \quad \text{for all } x \in \mathsf{X}. \tag{5.2}$$

This can be understood as an equation in the unknown function $v \in \mathbb{R}^{\mathsf{X}}$. Below we define the value function $v^*$ as maximal lifetime rewards and show that $v^*$ is the unique solution to the Bellman equation in $\mathbb{R}^{\mathsf{X}}$.

We can understand the Bellman equation as reducing an infinite-horizon problem to a two period problem involving the present and the future. Current actions influence (i) current rewards and (ii) expected discounted value from future states. In every case we examine, there is a trade-off between maximizing current rewards and shifting probability mass towards states with high future rewards.

## 5.1.2 Examples

Here we list examples of MDPs. We will see that some models neatly fit the MDP structure, while others can be coaxed into the MDP framework by adding states or applying other tricks.

### 5.1.2.1 A Renewal Problem

Rust (1987) ignited the field of dynamic structural estimation by examining an engine replacement problem for a bus workshop. In each period the superintendent decides whether or not to replace the engine of a given bus. Replacement is costly but delaying risks unexpected failure. Rust (1987) solved this trade-off using dynamic programming.

We consider an abstract version of Rust's problem with binary action $A_t$. When $A_t = 1$, the state resets to some fixed **renewal state** $\bar{x}$ in a finite set $\mathsf{X}$ (e.g., mileage resets to zero when an engine is replaced). When $A_t = 0$, the state updates according to $Q \in \mathcal{M}(\mathbb{R}^{\mathsf{X}})$ (e.g., mileage increases stochastically when the engine is not replaced). Given current state $x$ and action $a$, current reward $r(x, a)$ is received. The discount factor is $\beta \in (0, 1)$.

For this problem, the Bellman equation has the form

$$v(x) = \max \left\{ r(x, 1) + \beta v(\bar{x}),\ r(x, 0) + \beta \sum_{x'} v(x') Q(x, x') \right\} \qquad (x \in \mathsf{X}), \qquad (5.3)$$

where the first term is the value from action 1 and the second is the value of action 0.

To set the problem up as an MDP we set $\mathsf{A} = \{0, 1\}$ and $\Gamma(x) = \mathsf{A}$ for all $x \in \mathsf{X}$. We define

$$P(x, a, x') := a \mathbb{1}\{x' = \bar{x}\} + (1 - a) Q(x, x') \qquad ((x, a) \in \mathsf{G},\ x' \in \mathsf{X}). \qquad (5.4)$$

EXERCISE 5.1.1. Prove that $P$ is a stochastic kernel from G to X.

The primitives $(\Gamma, \beta, r, P)$ form an MDP. Moreover, the renewal Bellman equation (5.3) is a special case of the MDP Bellman equation (5.2). To verify this we rewrite (5.3) as

$$v(x) = \max_{a \in \{0,1\}} \left\{ r(x, a) + \beta \left[ a v(\bar{x}) + (1-a) \sum_{x'} v(x') Q(x, x') \right] \right\},$$

Inserting $P$ from (5.4) into the right-hand side of the last equation recovers the MDP Bellman equation (5.2).

### 5.1.2.2  Optimal Inventory Management

We study a firm where a manager maximizes shareholder value. To simplify the problem, we ignore exit options (so that firm value is the expected present value of profits) and assume that the firm only sells one product. Letting $\pi_t$ be profits at time $t$ and $r > 0$ be the interest rate, the value of the firm is

$$V_0 = \mathbb{E} \sum_{t \geq 0} \beta^t \pi_t \qquad \text{where} \quad \beta := \frac{1}{1+r}. \tag{5.5}$$

The firm faces exogenous demand process $(D_t)_{t \geq 0} \overset{\text{IID}}{\sim} \varphi \in \mathcal{D}(\mathbb{Z}_+)$. Inventory $(X_t)_{t \geq 0}$ of the product obeys

$$X_{t+1} = f(X_t, D_{t+1}, A_t) \qquad \text{where} \quad f(x, a, d) := (x - d) \vee 0 + a. \tag{5.6}$$

The term $A_t$ is units of stock ordered this period, which take one period to arrive. The definition of $f$ imposes the assumption that firms cannot sell more stock than they have on hand. We assume that the firm can store at most $K$ items at one time.

With the price of the firm's product set to one, current profits are given by

$$\pi_t := X_t \wedge D_{t+1} - c A_t - \kappa \mathbb{1}\{A_t > 0\}.$$

Here $c$ is unit product cost and $\kappa$ is a fixed cost of ordering inventory. We take the minimum $X_t \wedge D_{t+1}$ because orders in excess of inventory are assumed to be lost rather than back-filled.

We can map our inventory problem into an MDP with state space $X := \{0, \dots, K\}$ and action space $A := X$. The feasible correspondence $\Gamma$ is

$$\Gamma(x) := \{0, \dots, K - x\}, \tag{5.7}$$

which represents the set of feasible orders when the current inventory state is $x$. The reward function is expected current profits, or

$$r(x, a) := \sum_{d \geqslant 0} (x \wedge d) \varphi(d) - ca - \kappa \mathbb{1}\{a > 0\}. \tag{5.8}$$

The stochastic kernel from the set of feasible state-action pairs G induced by $\Gamma$ is, in view of (5.6),

$$P(x, a, x') := \mathbb{P}\{f(x, a, D) = x'\} \qquad \text{when} \quad D \sim \varphi. \tag{5.9}$$

EXERCISE 5.1.2. Suppose that $\varphi$ is the geometric distribution on $\mathbb{Z}_+$ with parameter $p$. Write down an expression for the stochastic kernel (5.9) using only $x, a, x'$ and the parameters of the model.

The Bellman equation for this optimal inventory problem is

$$v(x) = \max_{a \in \Gamma(x)} \left\{ r(x, a) + \beta \sum_{d \geqslant 0} v(f(x, a, d)) \varphi(d) \right\} \tag{5.11}$$

at each $x \in \mathsf{X}$, where $r(x, a)$ is as given in (5.8) and the aim is to solve for $v$. We introduce the Bellman operator

$$(Tv)(x) = \max_{a \in \Gamma(x)} \left\{ r(x, a) + \beta \sum_{d \geqslant 0} v(f(x, a, d)) \varphi(d) \right\}. \tag{5.12}$$

This operator maps $\mathbb{R}^{\mathsf{X}}$ to itself and is designed so that its set of fixed points in $\mathbb{R}^{\mathsf{X}}$ coincide with solutions to (5.11) in $\mathbb{R}^{\mathsf{X}}$.

EXERCISE 5.1.3. Prove that $T$ is a contraction of modulus $\beta$ on $\mathbb{R}^{\mathsf{X}}$ when paired with the supremum norm $\|v\|_\infty := \max_{x \in \mathsf{X}} |v(x)|$.

### 5.1.2.3  Example: Cake Eating

Many dynamic programming problems in economics involve a trade-off between current and future consumption. The simplest example in this class is the "cake eating" problem, where initial household wealth is given but no labor income is received.

Wealth evolves according to

$$W_{t+1} = R(W_t - C_t) \qquad (t \geqslant 0)$$

where $C_t$ is current consumption and $R$ is the gross interest rate. The agent seeks to maximize

$$\mathbb{E} \sum_{t \geqslant 0} \beta^t u(C_t) \quad \text{given } W_0 = w$$

subject to $0 \leqslant C_t \leqslant W_t$ (implying that the agent cannot borrow). Consumption level $C_t$ generates utility $u(C_t)$. Assuming that wealth takes values in a finite set $W \subset \mathbb{R}_+$, the Bellman equation for this problem can be written as

$$v(w) = \max_{0 \leqslant w' \leqslant w} \left\{ u(w - w'/R) + \beta v(w') \right\}. \tag{5.13}$$

In (5.13) we are using $w' = R(w - c)$ to obtain $c = (w - w'/R)$. The household uses (5.13) to trade-off current utility of consumption against the value of future wealth.

EXERCISE 5.1.4. Frame this model as an MDP with $W$ as the state space.

### 5.1.2.4 Example: Optimal Stopping

The optimal stopping problem we studied in Chapter 4 can be framed as an MDP. On one hand, doing so allows us to apply results obtained for MDPs to optimal stopping. On the other hand, expressing an optimal stopping problem as an MDP requires an additional state variable, which complicates the exposition. The exercise below helps to illustrate the key ideas.

**Remark 5.1.1.** While readers interested in the connection between optimal stopping and MDPs will benefit from this section, others can freely skip to §5.1.3 without losing continuity. Later, in Chapter 8, we will show that optimal stopping problems can be embedded in a very general framework (which includes MDPs) without adding extra state variables.

Let's focus on the job search problem with Markov state discussed in §3.3.1 (although the arguments for the general optimal stopping problem in §4.1.1.1 are very similar). As before, $W$ is the set of wage outcomes. Since we need the symbol $P$ for other purposes, we let $Q$ be the Markov matrix for wages, so that $(W_t)_{t \geqslant 0}$ is $Q$-Markov on $W$.

To express the job search problem as an MDP, let $X = \{0, 1\} \times W$ be a state space whose typical element is $(e, w)$, with $e$ representing either unemployment ($e = 0$) or employment ($e = 1$) and $w$ being the current wage offer. An action $a \in A := \{0, 1\}$ indicates rejection or acceptance of the current wage offer.

EXERCISE 5.1.5. Express the job search problem as an MDP, with state space X and action space A as described in the previous paragraph.

### 5.1.3 Optimality

In this section we return to the general MDP setting of §5.1.1, define optimal policies and state our main optimality result. As was the case for job search, actions are governed by policies, which are maps from states to actions (see, in particular, §1.3.1.3, where policies were introduced).

#### 5.1.3.1 Policies and Lifetime Values

Let $\mathcal{M} = (\Gamma, \beta, r, P)$ be an MDP. The set of **feasible policies** corresponding to $\mathcal{M}$ is

$$\Sigma := \{\sigma \in A^X : \sigma(x) \in \Gamma(x) \text{ for all } x \in X\}. \tag{5.16}$$

If we select a policy $\sigma$ from $\Sigma$, it is understood that we respond to state $X_t$ with action $A_t := \sigma(X_t)$ at every date $t$. As a result, the state evolves by drawing $X_{t+1}$ from $P(X_t, \sigma(X_t), \cdot)$ at each $t \geqslant 0$. In other words, $(X_t)_{t \geqslant 0}$ is $P_\sigma$-Markov when

$$P_\sigma(x, x') := P(x, \sigma(x), x') \qquad (x, x' \in X).$$

Note that $P_\sigma \in \mathcal{M}(\mathbb{R}^X)$. Fixing a policy "closes the loop" in the state transition process and defines a Markov chain for the state.

Under the policy $\sigma$, rewards at state $x$ are $r(x, \sigma(x))$. If

$$r_\sigma(x) := r(x, \sigma(x)) \quad \text{and} \quad \mathbb{E}_x := \mathbb{E}[\,\cdot\,|X_0 = x]$$

then the lifetime value of following $\sigma$ starting from state $x$ can be written as

$$v_\sigma(x) = \mathbb{E}_x \sum_{t \geqslant 0} \beta^t r_\sigma(X_t) \quad \text{where } (X_t) \text{ is } P_\sigma\text{-Markov with } X_0 = x. \tag{5.17}$$

Since $\beta < 1$, applying Lemma 3.2.1 on page 95 to this expression yields

$$v_\sigma = \sum_{t \geqslant 0} \beta^t P_\sigma^t r_\sigma = (I - \beta P_\sigma)^{-1} r_\sigma. \tag{5.18}$$

Analogous to the optimal stopping case, we call $v_\sigma$ the **$\sigma$-value function**. We also call $v_\sigma(x)$ the **lifetime value** of policy $\sigma$ conditional on initial state $x$.

EXERCISE 5.1.6. Prove that $v_1 \leqslant v_\sigma \leqslant v_2$ when $v_2 := \|r\|_\infty/(1 - \beta)$ and $v_1 := -v_2$.

Another way to compute $v_\sigma$ is to use the **policy operator** $T_\sigma$ corresponding to $\sigma$, which is defined at $v \in \mathbb{R}^X$ by

$$(T_\sigma v)(x) = r(x, \sigma(x)) + \beta \sum_{x'} v(x') P(x, \sigma(x), x') \qquad (x \in X). \tag{5.19}$$

($T_\sigma$ is analogous to the policy operator defined for the optimal stopping problem in §4.1.1.3.) In vector notation,

$$T_\sigma v = r_\sigma + \beta P_\sigma v. \tag{5.20}$$

The next exercise shows how $T_\sigma$ can be put to work.

EXERCISE 5.1.7. Fixing $\sigma$ in $\Sigma$, prove that

(i) $T_\sigma$ is an order-preserving self-map on $\mathbb{R}^X$,

(ii) $T_\sigma$ is a contraction on $\mathbb{R}^X$ of modulus $\beta$ under the norm $\|\cdot\|_\infty$,

(iii) the $\sigma$-value function $v_\sigma$ is the unique fixed point of $T_\sigma$ in $\mathbb{R}^X$, and

(iv) $T_\sigma^k v \to v_\sigma$ as $k \to \infty$ for all $v \in \mathbb{R}^X$.

Computationally, this means that we can pick $v \in \mathbb{R}^X$ and iterate with $T_\sigma$ to obtain an approximation to $v_\sigma$.

EXERCISE 5.1.8. Prove that, when the initial condition for iteration is $v \equiv 0 \in \mathbb{R}^X$, the $k$-th iterate $T_\sigma^k v$ is equal to the truncated sum $\sum_{t=0}^{k-1} \beta^t P_\sigma^t r_\sigma$.

**Remark 5.1.2.** To compute $v_\sigma$, should we use the expression $(I - \beta P_\sigma)^{-1} r_\sigma$ in (5.18) or iterate with $T_\sigma$? For small state spaces, the first option is typically faster. However, it is easy to write down dynamic programming problems where $X$ is very large (see, e.g., Example 1.0.2 on page 2). If, say, $X$ has $10^9$ elements, then $I - \beta P_\sigma$ is $10^9 \times 10^9$. Matrices of this size are difficult to invert – or even store in memory. In such settings, iterating with $T_\sigma$ might be preferred.

The next exercise extends Exercise 5.1.8 and aids interpretation of policy operators. It tells us that $(T_\sigma^k v)(x)$ is the payoff from following policy $\sigma$ and starting in state $x$ when lifetime is truncated to the finite horizon $k$ and $v$ provides a terminal payoff in each state.

EXERCISE 5.1.9. Fix $\sigma \in \Sigma$ and let $(X_t)$ be $P_\sigma$-Markov with initial condition $x \in \mathsf{X}$. Prove that, for given $v \in \mathbb{R}^{\mathsf{X}}$, and $k \in \mathbb{N}$, we have

$$(T_\sigma^k v)(x) = \mathbb{E}_x \left[ \sum_{t=1}^{k-1} \beta^t r(X_t, \sigma(X_t)) + \beta^k v(X_k) \right].$$

### 5.1.3.2 Defining Optimality

Given MDP $\mathcal{M} = (\Gamma, \beta, r, P)$ with $\sigma$-value functions $\{v_\sigma\}_{\sigma \in \Sigma}$, the **value function** corresponding to $\mathcal{M}$ is defined as $v^* := \vee_{\sigma \in \Sigma} v_\sigma$, where, as usual, the maximum is pointwise. More explicitly,

$$v^*(x) = \max_{\sigma \in \Sigma} v_\sigma(x) \qquad (x \in \mathsf{X}). \tag{5.21}$$

This is consistent with our definition of the value function in the optimal stopping case (see page 109). It is the maximal lifetime value we can extract from each state using feasible behavior. The maximum in (5.21) exists at each $x$ because $\Sigma$ is finite.

A policy $\sigma \in \Sigma$ is called **optimal** for $\mathcal{M}$ if $v_\sigma = v^*$. In other words, a policy is optimal if its lifetime value is maximal at each state.

**Example 5.1.1.** Consider again Figure 2.5 on page 57, supposing that $\Sigma = \{\sigma', \sigma''\}$. As drawn, there is no optimal policy, since $v^*$ differs from both $v_{\sigma'}$ and $v_{\sigma''}$. Below, in Proposition 5.1.1, we will show that such an outcome is *not* possible for MDPs.

Our optimality results are easier to follow with some additional terminology. To start, given $v \in \mathbb{R}^{\mathsf{X}}$, we define a policy $\sigma \in \Sigma$ to be $v$-**greedy** if

$$\sigma(x) \in \operatorname*{argmax}_{a \in \Gamma(x)} \left\{ r(x, a) + \beta \sum_{x'} v(x') P(x, a, x') \right\} \quad \text{for all } x \in \mathsf{X}. \tag{5.22}$$

In essence, a $v$-greedy policy treats $v$ as the correct value function and sets all actions accordingly.

EXERCISE 5.1.10. Fix $\sigma \in \Sigma$ and $v \in \mathbb{R}^{\mathsf{X}}$. Prove that the set $\{T_\sigma v\}_{\sigma \in \Sigma}$ has a least and greatest element.

**Bellman's principle of optimality** is said to hold for the MDP $\mathcal{M}$ if

$$\sigma \in \Sigma \text{ is optimal for } \mathcal{M} \quad \Longleftrightarrow \quad \sigma \text{ is } v^*\text{-greedy.}$$

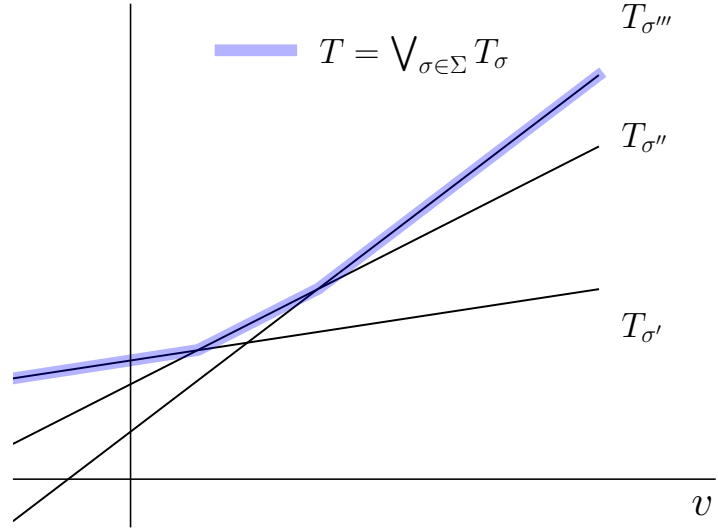

Figure 5.1: $T$ is the pointwise maximum of $\{T_\sigma\}_{\sigma \in \Sigma}$ (one-dimensional setting)

The **Bellman operator** corresponding to $\mathcal{M}$ is the self-map $T$ on $\mathbb{R}^{\mathsf{X}}$ defined by

$$(Tv)(x) = \max_{a \in \Gamma(x)} \left\{ r(x,a) + \beta \sum_{x'} v(x') P(x,a,x') \right\} \qquad (x \in \mathsf{X}). \tag{5.24}$$

Obviously $Tv = v$ if and only if $v$ satisfies the Bellman equation (5.2).

EXERCISE 5.1.11. Given $v \in \mathbb{R}^{\mathsf{X}}$, prove that

(i) at least one $v$-greedy policy exists,

(ii) $\sigma \in \Sigma$ is $v$-greedy if and only if $T_\sigma v = Tv$, and

(iii) $(Tv)(x) = \max_{\sigma \in \Sigma}(T_\sigma v)(x)$ for all $x \in \mathsf{X}$.

The last part of Exercise 5.1.11 tells us that $T$ is the pointwise maximum of $\{T_\sigma\}_{\sigma \in \Sigma}$, which can be expressed as $T = \vee_\sigma T_\sigma$. Figure 5.1 illustrates this relationship in one dimension.

EXERCISE 5.1.12. Prove: $T$ is a contraction of modulus $\beta$ on $\mathbb{R}^{\mathsf{X}}$ under norm $\|\cdot\|_\infty$.

### 5.1.3.3 Optimality Theory

We can now state our main optimality result for MDPs.

**Proposition 5.1.1.** *If $\mathcal{M} = (\Gamma, \beta, r, P)$ is an MDP with value function $v^*$ and Bellman operator $T$, then*

(i) *$v^*$ is the unique solution to the Bellman equation in $\mathbb{R}^{\mathsf{X}}$,*

(ii) *$\lim_{k \to \infty} T^k v = v^*$ for all $v \in \mathbb{R}^{\mathsf{X}}$,*

(iii) *Bellman's principle of optimality holds for $\mathcal{M}$, and*

(iv) *at least one optimal policy exists.*

While Proposition 5.1.1 is a special case of later results (see §8.1.3.3), a direct proof is not difficult and we provide one below for interested readers.

*Proof of Proposition 5.1.1.* In Exercise 5.1.12 we showed that $T$ is a contraction mapping on the closed set $\mathbb{R}^{\mathsf{X}}$. Hence $T$ is globally stable on $\mathbb{R}^{\mathsf{X}}$ and therefore has a unique fixed point $\bar{v} \in \mathbb{R}^{\mathsf{X}}$. Our first claim is that $\bar{v} = v^*$. We show $\bar{v} \leqslant v^*$ and then $\bar{v} \geqslant v^*$.

For the first inequality, let $\sigma \in \Sigma$ be $\bar{v}$-greedy. Recalling Exercise 5.1.11, we have $T_\sigma \bar{v} = T\bar{v} = \bar{v}$. Hence $\bar{v}$ is also a fixed point of $T_\sigma$. But the only fixed point of $T_\sigma$ in $\mathbb{R}^{\mathsf{X}}$ is $v_\sigma$, so $\bar{v} = v_\sigma$. But then $\bar{v} \leqslant v^*$, since, by definition, $v^* = \vee_\sigma v_\sigma$. This is our first inequality.

As for the second inequality, fix $\sigma \in \Sigma$ and observe that $T_\sigma v \leqslant Tv$ for all $v \in \mathbb{R}^{\mathsf{X}}$. Since $T$ is order-preserving and globally stable, Proposition 2.2.7 on page 66 implies that $v_\sigma \leqslant \bar{v}$. Taking the supremum over $\sigma \in \Sigma$ yields $v^* \leqslant \bar{v}$.

Hence $v^*$ is a fixed point of $T$ in $\mathbb{R}^{\mathsf{X}}$. Since $T$ is globally stable on $\mathbb{R}^{\mathsf{X}}$, the remaining claims in parts (i)–(ii) follow immediately.

As for part (iii), it follows from Exercise 5.1.11 and part (i) of this theorem that

$$\sigma \text{ is } v^*\text{-greedy} \quad \Longleftrightarrow \quad T_\sigma v^* = Tv^* \quad \Longleftrightarrow \quad T_\sigma v^* = v^*.$$

The right hand side of this expression tells us that $v^*$ is a fixed point of $T_\sigma$. But the only fixed point of $T_\sigma$ is $v_\sigma$, so the right hand side is equivalent to the statement $v_\sigma = v^*$. Hence, by this chain of logic and the definition of optimality,

$$\sigma \text{ is } v^*\text{-greedy} \iff v^* = v_\sigma \iff \sigma \text{ is optimal} \qquad (5.25)$$

Hence (iii) holds.

Part (iv) is left as an exercise immediately below. □

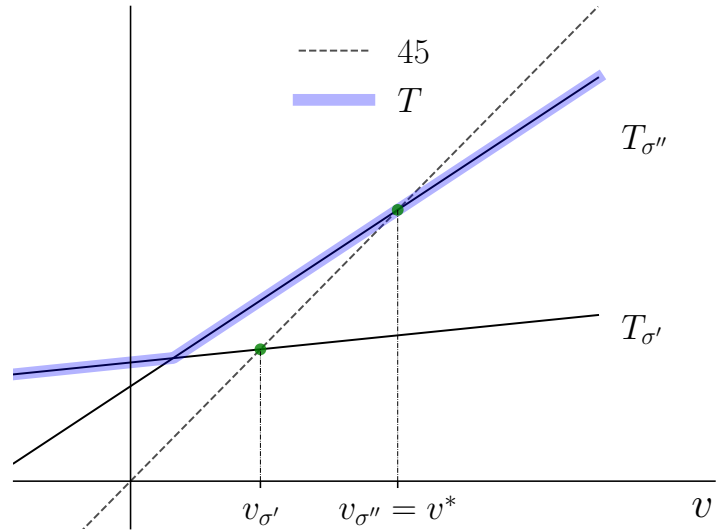

Figure 5.2: Illustration of optimality for MDPs

EXERCISE 5.1.13. Prove that, in Proposition 5.1.1, (iii) implies (iv).

Figure 5.2 illustrates Proposition 5.1.1 in an abstract case, where X is a singleton $\{x\}$. We write $v$ instead of $v(x)$ for the value of state $x$ and place $v$ on the horizontal axis. In the figure, the set of policies is $\Sigma = \{\sigma', \sigma''\}$. For given $\sigma \in \Sigma$, the map $T_\sigma$ is an affine function $T_\sigma v = r_\sigma + \beta P_\sigma v$ and the fixed point is $v_\sigma$. The Bellman operator $T$ is the upper envelope of the functions $\{T_\sigma\}$, as shown in (ii) of Exercise 5.1.11. By definition,

(i) $v^*$ is the largest of these fixed points, which equals $v_{\sigma''}$, and

(ii) $\sigma''$ is the optimal policy, since $v_{\sigma''} = v^*$.

In accordance with Proposition 5.1.1, $v^*$ is also the fixed point of the Bellman operator.

It is important to understand the significance of (iii) in Proposition 5.1.1. Greedy policies are relatively easy to compute, in the sense that solving (5.22) at each $x$ is easier than trying to directly solve the problem of maximizing lifetime value, since $\Sigma$ is in general far larger than $\Gamma(x)$. Part (iii) tells us that solving the overall problem reduces to computing a $v$-greedy policy with the right choice of $v$. For optimal stopping problems, that choice is the value function $v^*$. Intuitively, $v^*$ assigns a "correct" value to each state, in the sense of maximal lifetime value the controller can extract, so using $v^*$ to calculate greedy policies leads to the optimal outcome.

## 5.1.4  Algorithms

In previous chapters we solved job search and optimal stopping problems using value function iteration. In this section we present a generalization suitable for arbitrary MDPs and then discuss two important alternatives.

### 5.1.4.1  Value Function Iteration

**Value function iteration (VFI)** for MDPs is similar to VFI for the job search model (see page 37): we use successive approximation on $T$ to compute an approximation $v_k$ to the value function $v^*$ and then take a $v_k$-greedy policy. The general procedure is given by Algorithm 5.2.

---
**Algorithm 5.2:** Value function iteration for MDPs
---
1  input $v_0 \in \mathbb{R}^{\mathsf{X}}$, an initial guess of $v^*$
2  input $\tau$, a tolerance level for error
3  $\varepsilon \leftarrow +\infty$ and $k \leftarrow 0$
4  **while** $\varepsilon > \tau$ **do**
5  $\quad \Big|\quad v_{k+1} \leftarrow Tv_k$
6  $\quad \Big|\quad \varepsilon \leftarrow \|v_k - v_{k+1}\|_\infty$
7  $\quad \Big|\quad k \leftarrow k+1$
8  **end**
9  **return** a $v_k$-greedy policy $\sigma$
---

The fact that the sequence $(v_k)_{k \geqslant 0}$ produced by VFI converges to $v^*$ is immediate from Proposition 5.1.1 (as the tolerance $\tau$ is taken toward zero). It is also true that the greedy policy produced in the last step is approximately optimal when $\tau$ is small, and exactly optimal when $k$ is sufficiently large. Proofs are given in Chapter 8, where we examine VFI in a more general setting.

VFI is robust, easy to understand and easy to implement. These properties explain its enduring popularity. At the same time, in terms of efficiency, VFI is often dominated by alternative algorithms, two of which are discussed below.

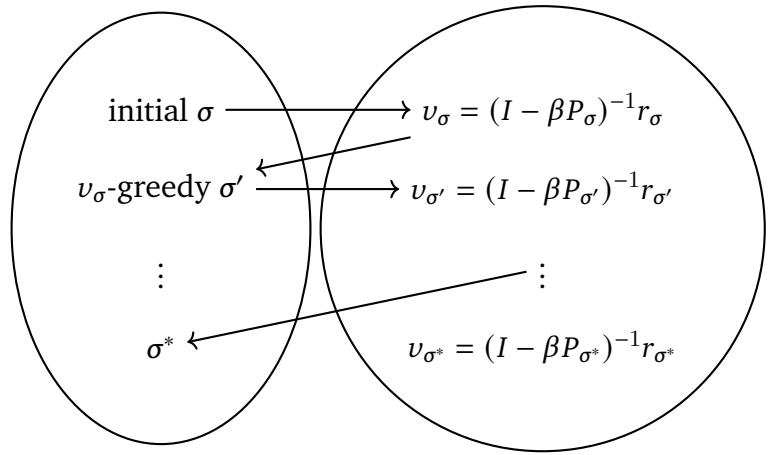

Figure 5.3: Howard policy iteration algorithm (HPI)

#### 5.1.4.2 Howard Policy Iteration

Unlike VFI, **Howard policy iteration (HPI)** computes optimal policies by iterating between computing the value of a given policy and computing the greedy policy associated with that value. The full technique is described in Algorithm 5.3.

---

**Algorithm 5.3:** Howard policy iteration for MDPs

1   input $\sigma \in \Sigma$
2   $v_0 \leftarrow v_\sigma$ and $k \leftarrow 0$
3   **repeat**
4      $\sigma_k \leftarrow$ a $v_k$-greedy policy
5      $v_{k+1} \leftarrow (I - \beta P_{\sigma_k})^{-1} r_{\sigma_k}$
6      **if** $v_{k+1} = v_k$ **then** break
7      $k \leftarrow k + 1$
8   **return** $\sigma_k$

---

A visualization of HPI is given in Figure 5.3, where $\sigma$ is the initial choice. Next we compute the lifetime value $v_\sigma$, and then the $v_\sigma$-greedy policy $\sigma'$, and so on. The computation of lifetime value is called the **policy evaluation** step, while the computation of greedy policies is called **policy improvement**.

HPI has two very attractive features. One is that, in a finite state setting, the algorithm always converges to an exact optimal policy in a finite number of steps, regardless of the initial condition. We prove this fact in a more general setting in

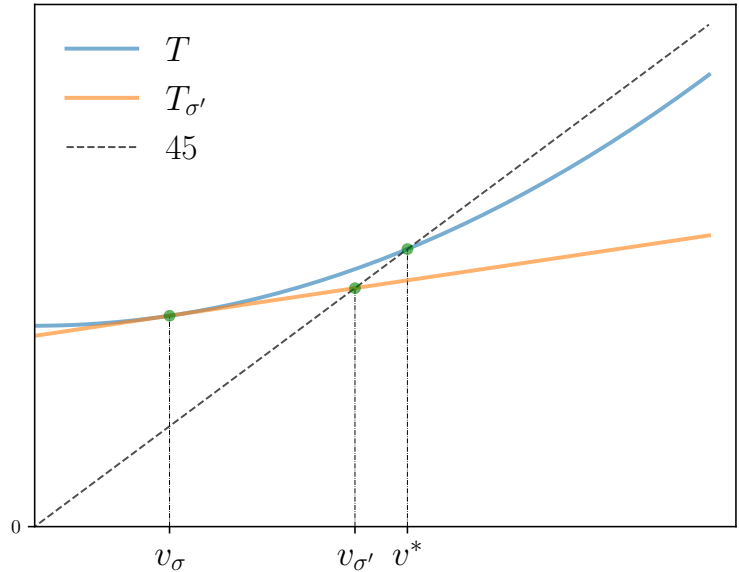

Figure 5.4: HPI as a version of Newton's method

Chapter 8. The second is that the rate of convergence is faster than VFI, as will be shown in §5.1.4.3.

Figure 5.4 gives another illustration, presented in the one-dimensional setting that we used for Figure 5.2. In this illustration, we imagine that there are many optimal policies, and hence many functions in $\{T_\sigma\}$, so that their upper envelope, which is the Bellman operator, becomes a smoother curve. The figure shows the update from $v_\sigma$ to the next lifetime value $v_{\sigma'}$, via the following two steps:

(i) Take $\sigma'$ to be $v_\sigma$-greedy, which means that $T_{\sigma'} v_\sigma = T v_\sigma$ (see Exercise 5.1.11).

(ii) Take $v_{\sigma'}$ to be the fixed point of $T_{\sigma'}$.

The next step, from $v_{\sigma'}$ to $v_{\sigma''}$ is analogous.

Comparison of this figure with Figure 2.1 on page 48 suggests that HPI is an implementation of Newton's method, applied to the Bellman operator. We confirm this in §5.1.4.3.

### 5.1.4.3 HPI as Newton Iteration

In discussing the connection between HPI and Newton iteration, one issue is that $T$ is not always differentiable, as seen in Figure 5.2. But $T$ is convex, and this lets us substitute *subgradients* for derivatives. Once we make this modification, HPI and Newton iteration are identical, as we now show.

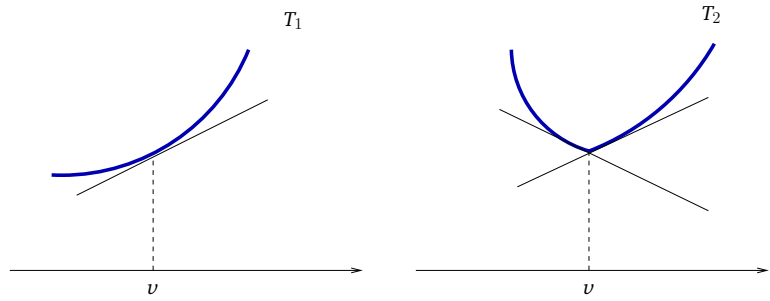

Figure 5.5: Subgradients of convex functions

First, recall that, given a self-map $T$ from $S \subset \mathbb{R}^n$ to itself, an $n \times n$ matrix $D$ is called a **subgradient** of $T$ at $v \in S$ if

$$Tu \geqslant Tv + D(u - v) \quad \text{for all } u \in S. \tag{5.26}$$

Figure 5.5 illustrates the definition in one dimension, where $D$ is just a scalar determining the slope of a tangent line at $v$. In the left subfigure, $T_1$ is convex and differentiable at $v$, which means that only one subgradient exists (since any other choice of slope implies that the inequality in (5.26) will fail for some $u$). In the right subfigure, $T_2$ is convex but nondifferentiable at $v$, so multiple subgradients exist.

In the next result, we take $(\Gamma, \beta, r, P)$ to be a given MDP and let $T$ be the associated Bellman operator.

**Lemma 5.1.2.** *If $v \in \mathbb{R}^{\mathsf{X}}$ and $\sigma \in \Sigma$ is $v$-greedy, then $\beta P_\sigma$ is a subgradient of $T$ at $v$.*

*Proof.* Fix $v \in \mathbb{R}^{\mathsf{X}}$ and let $\sigma$ be $v$-greedy. Using $T \geqslant T_\sigma$ and $T_\sigma v = Tv$, we have

$$Tu = Tv + Tu - Tv \geqslant Tv + T_\sigma u - T_\sigma v.$$

Applying the definition of $T_\sigma$ now gives

$$Tu \geqslant Tv + \beta P_\sigma u - \beta P_\sigma v = Tv + \beta P_\sigma(u - v).$$

Hence $\beta P_\sigma$ is a subgradient of $T$ at $v$, as claimed. $\qquad\square$

Now let's consider Newton's method applied to the problem of finding the fixed point of $T$. Since $T$ is nondifferentiable and convex, we replace the Jacobian in Newton's method (see (2.2) on page 48) with the subdifferential. This leads us to iterate

on

$$v_{k+1} = Q v_k \quad \text{where} \quad Q v := (I - \beta P_\sigma)^{-1} (T v - \beta P_\sigma v).$$

In the definition of $Q$, the policy $\sigma$ is $v$-greedy. Using $T v = T_\sigma v$, the map $Q$ reduces to $Q v := (I - \beta P_\sigma)^{-1} r_\sigma$, which is exactly the update step to produce the next $\sigma$-value function in HPI (i.e., the lifetime value of a $v$-greedy policy).

The fact that HPI is a version of Newton's method suggests that its iterates $(v_k)_{k \geqslant 0}$ enjoy quadratic convergence. This is indeed the case: under mild conditions one can show there exists a constant $N$ such that, for all $k \geqslant 0$,

$$\|v_{k+1} - v_k\| \leqslant N \|v_k - v_{k-1}\|^2 \tag{5.27}$$

(see, e.g., Puterman (2005), Theorem 6.4.8). Hence HPI enjoys both a fast convergence rate and the robustness of global convergence.

However, HPI is not always optimal in terms of efficiency, since the size of the constant term in (5.27) also matters. This term can be large because, at each step, the update from $v_\sigma$ to $v_{\sigma'}$ requires computing the exact lifetime value $v_{\sigma'}$ of the $v_\sigma$-greedy policy $\sigma'$. Computing this fixed point exactly can be computationally expensive in high dimensions.

One way around this issue is to forgo computing the fixed point $v_{\sigma'}$ exactly, replacing it with an approximation. The next section takes up this idea.

### 5.1.4.4 Optimistic Policy Iteration

Optimistic policy iteration (OPI) is an algorithm that borrows from both VFI and HPI. In essence, the algorithm is the same as HPI except that, instead of computing the full value $v_\sigma$ of a given policy, the approximation $T_\sigma^m v$ from Exercise 5.1.7 is used instead. Algorithm 5.4 clarifies.

In the algorithm, the policy operator $T_{\sigma_k}$ is applied $m$ times to generate an approximation of $v_{\sigma_k}$. The constant step size $m$ can also be replaced with a sequence $(m_k) \subset \mathbb{N}$. In either case, for MDPs, convergence to an optimal policy is guaranteed. We prove this in a more general setting in Chapter 8.

Notice that, as $m \to \infty$, the algorithm increasingly approximates Howard policy iteration, since $T_{\sigma_k}^m v_k$ converges to $v_{\sigma_k}$. At the same time, if $m = 1$, the reduces to VFI. This follows from Exercise 5.1.11, which tells us that, when $\sigma_k$ is $v_k$-greedy, $T_{\sigma_k} v_k = T v_k$. Hence, with intermediate $m$, OPI can be seen as a "convex combination" of HPI and VFI.

---

**Algorithm 5.4:** Optimistic policy iteration for MDPs

---

**1** input $m \in \mathbb{N}$ and tolerance $\tau \geqslant 0$
**2** input $\sigma \in \Sigma$ and set $v_0 \leftarrow v_\sigma$
**3** $k \leftarrow 0$
**4 repeat**
**5**     $\sigma_k \leftarrow$ a $v_k$-greedy policy
**6**     $v_{k+1} \leftarrow T_{\sigma_k}^m v_k$
**7**     **if** $\|v_{k+1} - v_k\| \leqslant \tau$ **then** break
**8**     $k \leftarrow k + 1$
**9 return** $\sigma_k$

---

In almost all dynamic programming applications, there exist choices of $m > 1$ such that OPI converges faster than VFI. We investigate these ideas in the applications below. In some cases, there exist values of $m$ such that OPI dominates HPI. However, this depends on the structure of the problem and the software and hardware platforms being employed – see 2.1.4.4 and the applications below for additional discussion.

## 5.2 Applications

This section gives several applications of the MDP model to economic problems. The applications illustrate the ease with which MDPs can be implemented on a computer (provided that the state and action spaces are not too large).

### 5.2.1 Optimal Inventories

In §3.1.1.2 we studied a firm whose inventory behavior was specified to follow S-s dynamics. In §5.1.2.2 we introduced a model where investment behavior is endogenous, determined by the desire to maximize firm value. In this section, we show that this endogenous inventory behavior can replicate the S-s dynamics from §3.1.1.2.

We saw in §5.1.2.2 that the optimal inventory model is an MDP, so the Proposition 5.1.1 optimality and convergence results apply. In particular, the unique fixed point of the Bellman operator is the value function $v^*$, and a policy $\sigma^*$ is optimal if and only if $\sigma^*$ is $v^*$-greedy.

We solve the model numerically using VFI. As in Exercise 5.1.2, we take $\varphi$ to be the geometric distribution on $\mathbb{Z}_+$ with parameter $p$. We use the default parameter values shown in Listing 14. The code listing also presents an implementation of the Bellman operator.

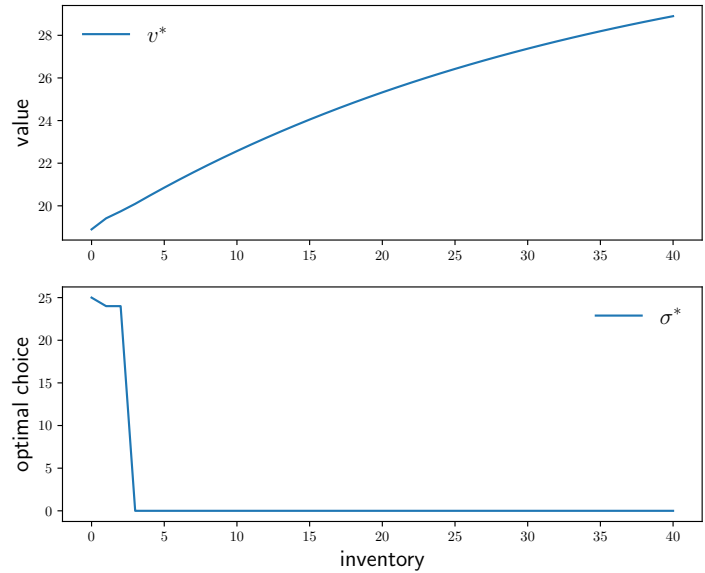

Figure 5.6: The value function and optimal policy for the inventory problem

Figure 5.6 exhibits an approximation of the value function $v^*$, computed by iterating with $T$ starting at $v \equiv 1$. Figure 5.6 also shows the approximate optimal policy, obtained as a $v^*$-greedy policy:

$$\sigma^*(x) \in \operatorname*{argmax}_{a \in \Gamma(x)} \left\{ r(x, a) + \beta \sum_{d \geqslant 0} v^*(f(x, a, d)) \varphi(d) \right\}$$

The plot of the optimal policy shows that there is a threshold region below which the firm orders large batches and above which the firm orders nothing. This makes sense, since the firm wishes to economize on the fixed cost of ordering. Figure 5.7 shows a simulation of inventory dynamics under the optimal policy, starting from $X_0 = 0$. The time path closely approximates the S-s dynamics discussed in §3.1.1.2.

EXERCISE 5.2.1. Compute the optimal policy by extending the code given in Listing 14. Replicate Figure 5.7, modulo randomness, by sampling from a geometric distribution and implementing the dynamics in (5.6). At each $X_t$, the action $A_t$ should be chosen according to the optimal policy $\sigma^*(X_t)$.

```julia
using Distributions

f(x, a, d) = max(x - d, 0) + a  # Inventory update

function create_inventory_model(; β=0.98,      # discount factor
                                  K=40,         # maximum inventory
                                  c=0.2, κ=2,   # cost paramters
                                  p=0.6)        # demand parameter
    φ(d) = (1 - p)^d * p        # demand distribution
    x_vals = collect(0:K)       # set of inventory levels
    return (; β, K, c, κ, p, φ, x_vals)
end

"The function B(x, a, v) = r(x, a) + β Σ_x′ v(x′) P(x, a, x′)."
function B(x, a, v, model; d_max=100)
    (; β, K, c, κ, p, φ, x_vals) = model
    revenue = sum(min(x, d) * φ(d) for d in 0:d_max)
    current_profit = revenue - c * a - κ * (a > 0)
    next_value = sum(v[f(x, a, d) + 1] * φ(d) for d in 0:d_max)
    return current_profit + β * next_value
end

"The Bellman operator."
function T(v, model)
    (; β, K, c, κ, p, φ, x_vals) = model
    new_v = similar(v)
    for (x_idx, x) in enumerate(x_vals)
        Γx = 0:(K - x)
        new_v[x_idx], _ = findmax(B(x, a, v, model) for a in Γx)
    end
    return new_v
end
```

Listing 14:  Solving the optimal inventory model (`inventory_dp.jl`)

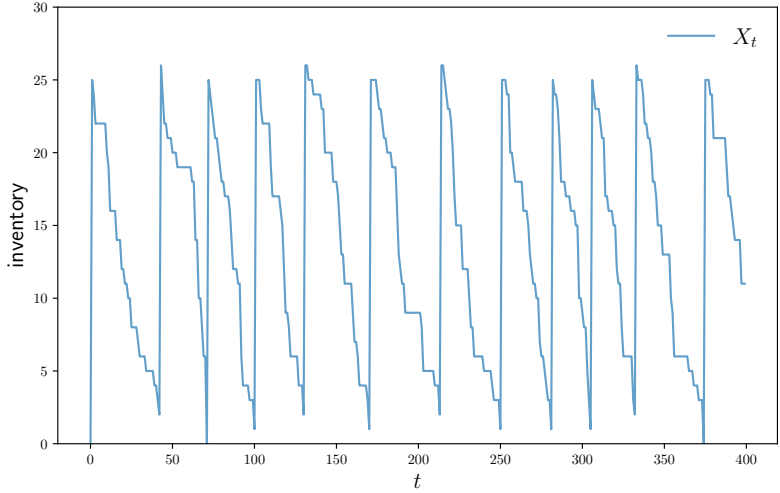

Figure 5.7: Optimal inventory dynamics

## 5.2.2 Optimal Savings with Labor Income

As our next example of an MDP, we modify the cake eating problem in §5.1.2.3 to add labor income. Wealth evolves according to

$$W_{t+1} = R(W_t + Y_t - C_t) \qquad (t = 0, 1, \ldots), \tag{5.28}$$

where $(W_t)$ takes values in finite set $\mathsf{W} \subset \mathbb{R}_+$ and labor income $(Y_t)$ is a Markov chain on finite set $\mathsf{Y} \subset \mathbb{R}_+$ with transition matrix $Q$.[1] $R$ is a gross rate of interest, so that investing $d$ dollars today returns $Rd$ next period. Other parts of the problem are unchanged. The Bellman operator can be written as

$$(Tv)(w, y) = \max_{w' \in \Gamma(w,y)} \left\{ u \left( w + y - \frac{w'}{R} \right) + \beta \sum_{y'} v(w', y') Q(y, y') \right\}. \tag{5.29}$$

### 5.2.2.1 MDP Representation

To frame this problem as an MDP, we set the state to $x := (w, y)$, representing current wealth and income, taking values in the state space $\mathsf{X} := \mathsf{W} \times \mathsf{Y}$. The action is savings $s$, which takes values in W and equals $w'$. The feasible correspondence is the set of feasible savings values

$$\Gamma(w, y) = \{s \in W : s \leqslant R(w + y)\}.$$

---

[1]See Marcet et al. (2007) and Zhu (2020) for more extensive analysis of how adding a labor supply choice can affect outcomes in a consumption-savings model.

The current reward is utility of consumption $r(w, s) = u(w + y - s/R)$. The stochastic kernel is

$$P((w, y), s, (w', y')) = \mathbb{1}\{w' = s\}Q(y, y').$$

Having framed an MDP, the Proposition 5.1.1 optimality results apply.

### 5.2.2.2 Implementation

To implement the algorithms discussed in §5.1.4, we use the Bellman operator (5.29), and the corresponding definition of a $v$-greedy policy, which is

$$\sigma(w, y) \in \operatorname*{argmax}_{w' \in \Gamma(w, y)} \left\{ u\left(w + y - \frac{w'}{R}\right) + \beta \sum_{y'} v(w', y')Q(y, y') \right\}$$

for all $(w, y)$. The policy operator for given $\sigma \in \Sigma$ is

$$(T_\sigma v)(w, y) = u\left(w + y - \frac{\sigma(w, y)}{R}\right) + \beta \sum_{y'} v(\sigma(w, y), y')Q(y, y'). \tag{5.30}$$

Code for implementing the model and these two operators is given in Listing 15. Income is constructed as a discretized AR(1) process using the method from §3.1.3. Exponentiation is applied to the grid so that income takes positive values.

The function `get_value` in Listing 16 uses the expression $v_\sigma = (I - \beta P_\sigma)^{-1}r_\sigma$ from (5.18) to obtain the value of a given policy $\sigma$. The matrix $P_\sigma$ and vector $r_\sigma$ take the form

$$P_\sigma((w, y), (w', y')) = \mathbb{1}\{\sigma(w, y) = w'\}Q(y, y')$$
$$r_\sigma(w, y) = u(w + y - \sigma(w, y)/R).$$

### 5.2.2.3 Timing

Since all results for MDPs apply, we know that the value function $v^*$ is the unique fixed point of the Bellman operator in $\mathbb{R}^X$, and that value function iteration, Howard policy

```julia
using QuantEcon, LinearAlgebra, IterTools

function create_savings_model(; R=1.01, β=0.98, γ=2.5,
                                w_min=0.01, w_max=20.0, w_size=200,
                                ρ=0.9, ν=0.1, y_size=5)
    w_grid = LinRange(w_min, w_max, w_size)
    mc = tauchen(y_size, ρ, ν)
    y_grid, Q = exp.(mc.state_values), mc.p
    return (; β, R, γ, w_grid, y_grid, Q)
end

"B(w, y, w′, v) = u(R*w + y - w′) + β Σ_y′ v(w′, y′) Q(y, y′)."
function B(i, j, k, v, model)
    (; β, R, γ, w_grid, y_grid, Q) = model
    w, y, w′ = w_grid[i], y_grid[j], w_grid[k]
    u(c) = c^(1-γ) / (1-γ)
    c = w + y - (w′ / R)
    @views value = c > 0 ? u(c) + β * dot(v[k, :], Q[j, :]) : -Inf
    return value
end

"The Bellman operator."
function T(v, model)
    w_idx, y_idx = (eachindex(g) for g in (model.w_grid, model.y_grid))
    v_new = similar(v)
    for (i, j) in product(w_idx, y_idx)
        v_new[i, j] = maximum(B(i, j, k, v, model) for k in w_idx)
    end
    return v_new
end

"The policy operator."
function T_σ(v, σ, model)
    w_idx, y_idx = (eachindex(g) for g in (model.w_grid, model.y_grid))
    v_new = similar(v)
    for (i, j) in product(w_idx, y_idx)
        v_new[i, j] = B(i, j, σ[i, j], v, model)
    end
    return v_new
end
```

Listing 15: Discrete optimal savings model (`finite_opt_saving_0.jl`)

```julia
include("finite_opt_saving_0.jl")

"Compute a v-greedy policy."
function get_greedy(v, model)
    w_idx, y_idx = (eachindex(g) for g in (model.w_grid, model.y_grid))
    σ = Matrix{Int32}(undef, length(w_idx), length(y_idx))
    for (i, j) in product(w_idx, y_idx)
        _, σ[i, j] = findmax(B(i, j, k, v, model) for k in w_idx)
    end
    return σ
end

"Get the value v_σ of policy σ."
function get_value(σ, model)
    # Unpack and set up
    (; β, R, γ, w_grid, y_grid, Q) = model
    w_idx, y_idx = (eachindex(g) for g in (w_grid, y_grid))
    wn, yn = length(w_idx), length(y_idx)
    n = wn * yn
    u(c) = c^(1-γ) / (1-γ)
    # Build P_σ and r_σ as multi-index arrays
    P_σ = zeros(wn, yn, wn, yn)
    r_σ = zeros(wn, yn)
    for (i, j) in product(w_idx, y_idx)
            w, y, w′ = w_grid[i], y_grid[j], w_grid[σ[i, j]]
            r_σ[i, j] = u(w + y - w′/R)
        for j′ in y_idx
            P_σ[i, j, σ[i, j], j′] = Q[j, j′]
        end
    end
    # Reshape for matrix algebra
    P_σ = reshape(P_σ, n, n)
    r_σ = reshape(r_σ, n)
    # Apply matrix operations --- solve for the value of σ
    v_σ = (I - β * P_σ) \ r_σ
    # Return as multi-index array
    return reshape(v_σ, wn, yn)
end
```

Listing 16:  Discrete optimal savings model (`finite_opt_saving_1.jl`)

```julia
include("s_approx.jl")
include("finite_opt_saving_1.jl")

"Value function iteration routine."
function value_iteration(model, tol=1e-5)
    vz = zeros(length(model.w_grid), length(model.y_grid))
    v_star = successive_approx(v -> T(v, model), vz, tolerance=tol)
    return get_greedy(v_star, model)
end

"Howard policy iteration routine."
function policy_iteration(model)
    wn, yn = length(model.w_grid), length(model.y_grid)
    σ = ones(Int32, wn, yn)
    i, error = 0, 1.0
    while error > 0
        v_σ = get_value(σ, model)
        σ_new = get_greedy(v_σ, model)
        error = maximum(abs.(σ_new - σ))
        σ = σ_new
        i = i + 1
        println("Concluded loop $i with error $error.")
    end
    return σ
end

"Optimistic policy iteration routine."
function optimistic_policy_iteration(model; tolerance=1e-5, m=100)
    v = zeros(length(model.w_grid), length(model.y_grid))
    error = tolerance + 1
    while error > tolerance
        last_v = v
        σ = get_greedy(v, model)
        for i in 1:m
            v = T_σ(v, σ, model)
        end
        error = maximum(abs.(v - last_v))
    end
    return get_greedy(v, model)
end
```

Listing 17:  Discrete optimal savings model (`finite_opt_saving_2.jl`)

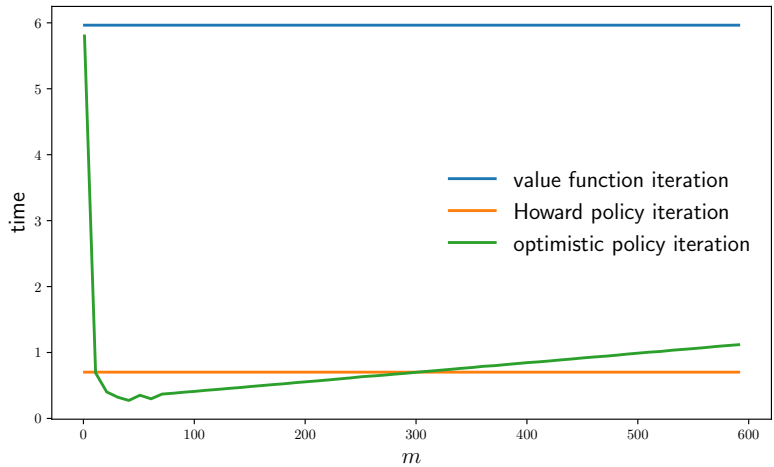

Figure 5.8:  Timings for alternative algorithms, savings model

iteration and optimistic policy iteration all converge.  Listing 17 implements these three algorithms.  Since the state and action space are finite, Howard policy iteration is guaranteed to return an exact optimal policy.

Figure 5.8 shows the number of seconds taken to solve the finite optimal savings model under the default parameters when executed on a laptop machine with 20 CPUs running at around 4GHz.  The horizontal axis corresponds to the step parameter $m$ in OPI (Algorithm 5.4).  The two other algorithms do not depend on $m$ and hence their timings are constant.  The figure shows that HPI is an order of magnitude faster than VFI and that optimistic policy iteration is even faster for moderate values of $m$.

One reason VFI is slow is that the discount factor is close to one.  This matters because the convergence rate for VFI is linear with error size decreasing geometrically in $\beta$.  In contrast, HPI, being an instance of Newton iteration, converges quadratically (see §2.1.4.2).  As a result, HPI tends to dominate VFI when the discount factor approaches unity.

Run-times are also dependent on implementation, and relative speed varies significantly with coding style, software and hardware platforms.  In our implementation, the main deficiency is that parallelization is under-utilized.  Better exploitation of parallelization tends to favor HPI, as discussed in §2.1.4.4.

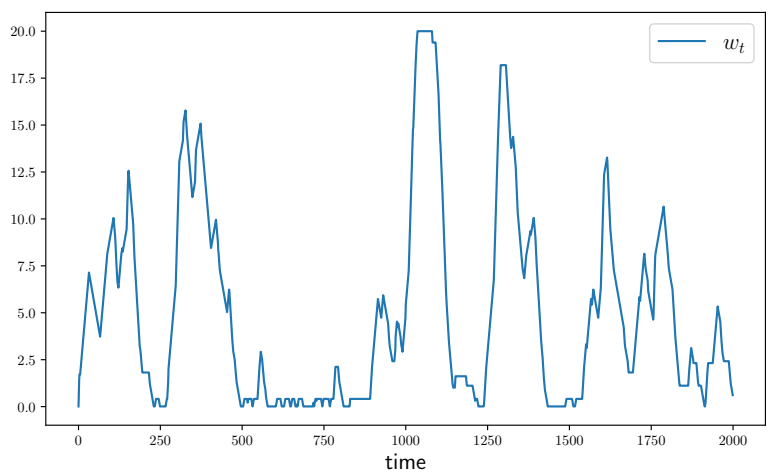

Figure 5.9: Time series for wealth

#### 5.2.2.4 Outputs

Figure 5.9 shows a typical time series for the wealth of a single household under the optimal policy. The series is created by computing an optimal policy $\sigma^*$, generating $(Y_t)_{t=0}^{m-1}$ as a $Q$-Markov chain on $\mathsf{Y}$ and then computing $(W_t)_{t=0}^{m}$ via $W_{t+1} = \sigma^*(W_t, Y_t)$ for $t$ running from $0$ to $m-1$. Initial wealth $W_0$ is set to $1.0$ and $m = 2000$.

Figure 5.10 shows the result of computing and histogramming a longer time series, with $m$ set to 1,000,000. This histogram approximates the stationary distribution of wealth for a large population, each updating via $\sigma^*$ and each with independently generated labor income series $(Y_t)_{t=0}^{m-1}$. (This is due to ergodicity of the wealth-income process. For a discussion of the connection between stationary distributions and time series under ergodicity see, for example, Sargent and Stachurski (2023b).)

The shape of the wealth distribution in Figure 5.10 is unrealistic. In almost all countries, the wealth distribution has a very long right tail. The Gini coefficient of the distribution in Figure 5.10 is 0.54, which is too low. For example, World Bank data for 2019 produces a wealth Gini for the US equal to 0.852. For Germany and Japan the figures are 0.816 and 0.627 respectively.

In §5.3.3 we discuss a variation on the optimal savings model that can produce a more realistic wealth distribution.

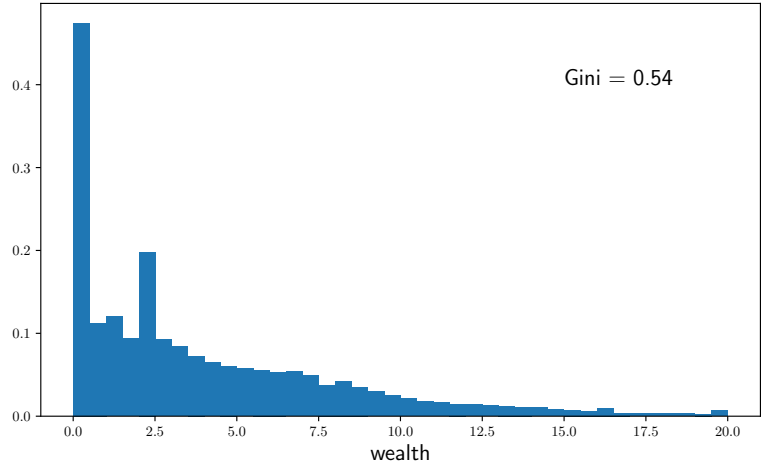

Figure 5.10: Histogram of wealth

## 5.2.3 Optimal Investment

As our next application, we consider a monopolist facing adjustment costs and stochastically evolving demand. The monopolist balances setting enough capacity to meet demand against costs of adjusting capacity.

### 5.2.3.1 Problem Description

We assume that the monopolist produces a single product and faces an inverse demand function of the form

$$P_t = a_0 - a_1 Y_t + Z_t,$$

where $a_0, a_1$ are positive parameters, $Y_t$ is output, $P_t$ is price and the demand shock $Z_t$ follows

$$Z_{t+1} = \rho Z_t + \sigma \eta_{t+1}, \qquad \{\eta_t\} \overset{\text{IID}}{\sim} N(0, 1).$$

Current profits are

$$\pi_t := P_t Y_t - c Y_t - \gamma (Y_{t+1} - Y_t)^2.$$

Here $\gamma(Y_{t+1} - Y_t)^2$ represents costs associated with adjusting production scale, parameterized by $\gamma$, and $c$ is unit cost of current production. Costs are convex, so rapid changes to capacity are expensive.

The monopolist chooses $(Y_t)$ to maximize the expected discounted value of its profit

flow, which we write as

$$\mathbb{E} \sum_{t=0}^{\infty} \beta^t \pi_t. \tag{5.31}$$

Here $\beta = 1/(1+r)$, where $r > 0$ is a fixed interest rate.

A way to start thinking about the optimal time path of output is to consider what would happen if $\gamma = 0$. Without adjustment costs there is no intertemporal trade-off, so the monopolist should choose output to maximize current profit in each period. The implied level of output at time $t$ is

$$\bar{Y}_t := \frac{a_0 - c + Z_t}{2a_1}. \tag{5.32}$$

EXERCISE 5.2.2. Show that $\bar{Y}_t$ maximizes current profit when $\gamma = 0$.

For $\gamma > 0$, we expect the following behavior.

- If $\gamma$ is close to zero, then the optimal output path $Y_t$ will track the time path of $\bar{Y}_t$ relatively closely, while

- if $\gamma$ is larger, then $Y_t$ will be significantly smoother than $\bar{Y}_t$, as the monopolist seeks to avoid adjustment costs.

### 5.2.3.2 MDP Representation

We can represent this problem as an MDP. To do so we let $\mathsf{Y}$ be a grid contained in $\mathbb{R}_+$ that lists possible output values. To conform to the finite state setting, we discretize the shock process $(Z_t)$ using Tauchen's method, as described in §3.1.3. For convenience we again use $(Z_t)$ to represent the discrete process, which is a finite Markov chain on $\mathsf{Z} \subset \mathbb{R}$ with transition matrix $Q$.

The state space for this MDP is $\mathsf{X} = \mathsf{Y} \times \mathsf{Z}$, while the action space is $\mathsf{Y}$. The feasible correspondence is defined by $\Gamma(x) = \mathsf{Y}$, meaning that choice of output is not restricted by the state. Thus, the feasible policy set $\Sigma$ is all $\sigma \colon \mathsf{Y} \times \mathsf{Z} \to \mathsf{Y}$.

We write $(y, z)$ for the current state, $q$ for the action (which chooses next period output) and $(y', z')$ for the next period state. The current reward function is current profits, which we can write as

$$r((y, z), q) = (a_0 - a_1 y + z - c)y - \gamma(q - y)^2.$$

The stochastic kernel is

$$P((y, z), q, (y', z')) = \mathbb{1}\{y' = q\}Q(z, z').$$

The term $\mathbb{1}\{y' = q\}$ states that next period output $y'$ is equal to our current choice $q$ for next period output. With these definitions, the problem defines an MDP and all of the optimality theory for MDPs applies.

### 5.2.3.3 Implementation

The Bellman operator can be expressed as

$$(Tv)(y, z) = \max_{y' \in \mathbb{R}} \left\{ r(y, z, y') + \beta \sum_{z'} v(y', z')Q(z, z') \right\}.$$

Given $\sigma \in \Sigma$, we can express the policy operator as

$$(T_\sigma v)(y, z) = r(y, z, \sigma(y, z)) + \beta \sum_{z'} v(\sigma(y, z), z')Q(z, z').$$

A $v$-greedy policy is a $\sigma \in \Sigma$ that obeys

$$\sigma(y, z) \in \operatorname*{argmax}_{y' \in \mathsf{Y}} \left\{ r(y, z, y') + \beta \sum_{z'} v(y', z')Q(z, z') \right\} \quad \text{for all } (y, z) \in \mathsf{X}.$$

By combining iteration with the policy operator and computation of greedy policies, we can implement optimistic policy iteration, compute the optimal policy $\sigma^*$, and study output choices generated by this policy. We are particularly interested in how output responds over time to randomly generated demand shocks.

Figure 5.11 shows the result of a simulation designed to shed light on how output responds to demand. After choosing initial values $(Y_1, Z_1)$ and generating a $Q$-Markov chain $(Z_t)_{t=1}^T$, we simulated optimal output via $Y_{t+1} = \sigma^*(Y_t, Z_t)$. The default parameters are shown in Listing 18. In the figure, the adjustment cost parameter $\gamma$ is varied as shown in the title. In addition to the optimal output path, the path of $(\bar{Y}_t)$ as defined in (5.32) is also presented.

The figure shows how increasing $\gamma$ promotes smoothing, as predicted in our discussion above. For small $\gamma$, adjustment costs have only minor effects on choices, so output closely follows $(\bar{Y}_t)$, the optimal path when output responds immediately to demand shocks. Conversely, larger values of $\gamma$ make adjustment expensive, so the operator responds relatively slowly to changes in demand.

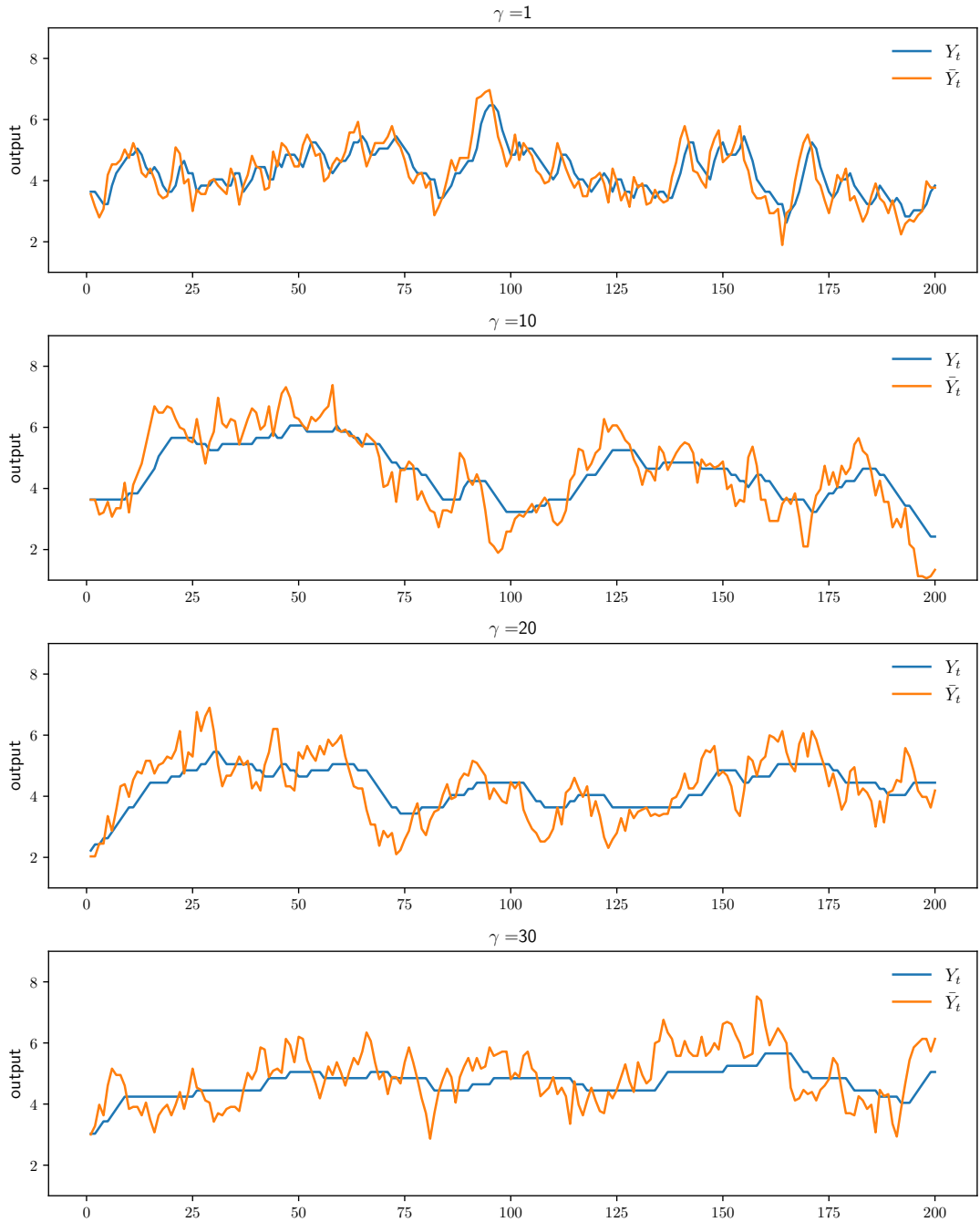

Figure 5.11: Simulation of optimal output with different adjustment costs

```julia
using QuantEcon, LinearAlgebra, IterTools
include("s_approx.jl")

function create_investment_model(;
        r=0.04,                              # Interest rate
        a_0=10.0, a_1=1.0,                   # Demand parameters
        γ=25.0, c=1.0,                       # Adjustment and unit cost
        y_min=0.0, y_max=20.0, y_size=100,   # Grid for output
        ρ=0.9, ν=1.0,                        # AR(1) parameters
        z_size=25)                           # Grid size for shock
    β = 1/(1+r)
    y_grid = LinRange(y_min, y_max, y_size)
    mc = tauchen(z_size, ρ, ν)
    z_grid, Q = mc.state_values, mc.p
    return (; β, a_0, a_1, γ, c, y_grid, z_grid, Q)
end
```

Listing 18: Optimal investment model (`finite_lq.jl`)

Figure 5.12 compares timings for VFI, HPI and OPI. Parameters are as in Listing 18. As in Figure 5.8, which gave timings for the optimal savings model, the horizontal axis shows $m$, which is the step parameter in OPI (see Algorithm 5.4). VFI and HPI do not depend on $m$ and hence their timings are constant. The vertical axis is time in seconds.

HPI is faster than VFI, although the difference is not as dramatic as was the case for optimal savings. One reason is that the discount factor is relatively small for the optimal investment model ($r = 0.04$ and $β = 1/(1 + r)$, so $β \approx 0.96$). Since $β$ is the modulus of contraction for the Bellman operator, this means that VFI converges relatively quickly. Another observation is that, for many values of $m$, OPI dominates both VFI and HPI in terms of speed, which is consistent with our findings for the optimal savings model. At $m = 70$, OPI is around 20 times faster than VFI.

EXERCISE 5.2.3. Consider a firm that maximizes expected discounted value in a setting where future profits are discounted at rate $β = 1/(1+r)$, the only production input is labor and hiring involves fixed costs. Let $\ell_t$ be employment at the firm at time $t$. Current profits are

$$\pi_t = p Z_t \ell_t^\alpha - w \ell_t - \kappa \mathbb{1}\{\ell_{t+1} \neq \ell_t\},$$

where $p$ is the output price, $w$ is the wage rate, $\alpha$ is a production parameter, the productivity shock is $Q$-Markov on Z and $\kappa$ is a fixed cost of hiring and firing. This fixed cost induces lumpy adjustment, as shown in Figure 5.13. Show that this model is an MDP. Write the Bellman equation and the procedure for optimistic policy iteration in the context of this model. Replicate Figure 5.13, modulo randomness, using the parameters shown in Listing 19.

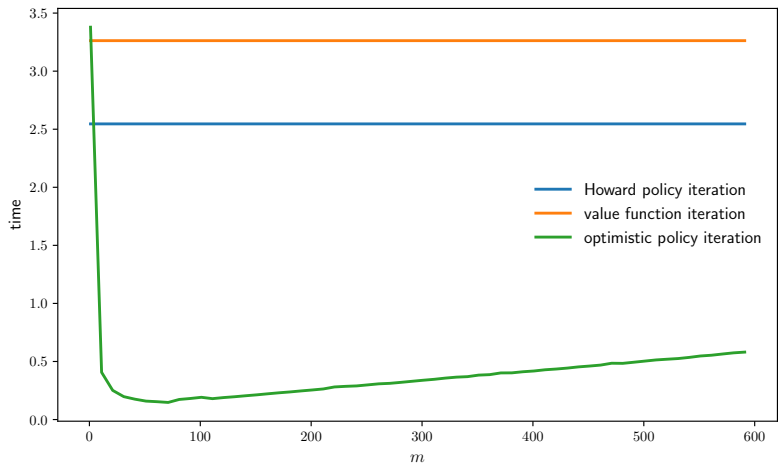

Figure 5.12: Timings for alternative algorithms, investment model

## 5.3 Modified Bellman Equations

Direct application of MDP theory is sometimes suboptimal. For example, we saw in §1.3.2.2 that solving the job search problem with IID wage draws is best accomplished by generating a recursion on the continuation value, which reduces dimensionality for iterative solution methods. Separately, in §4.2.2.2, we saw how a different manipulation of the Bellman equation also increased efficiency.

Now we aim to study such modifications systematically. We begin by providing other examples of how manipulating a Bellman equation can facilitate computation and analysis. Then we establish a theoretical foundation for this line of analysis, and show how similar ideas can also be applied to policy operators and greedy policies.

(We also treat similar topics at a more advanced and abstract level in §9.3.)

```julia
using QuantEcon, LinearAlgebra, IterTools

function create_hiring_model(;
        r=0.04,                                    # Interest rate
        κ=1.0,                                     # Adjustment cost
        α=0.4,                                     # Production parameter
        p=1.0, w=1.0,                             # Price and wage
        l_min=0.0, l_max=30.0, l_size=100,        # Grid for labor
        ρ=0.9, ν=0.4, b=1.0,                      # AR(1) parameters
        z_size=100)                                # Grid size for shock
    β = 1/(1+r)
    l_grid = LinRange(l_min, l_max, l_size)
    mc = tauchen(z_size, ρ, ν, b, 6)
    z_grid, Q = mc.state_values, mc.p
    return (; β, κ, α, p, w, l_grid, z_grid, Q)
end
```

Listing 19: Firm hiring model (`firm_hiring.jl`)

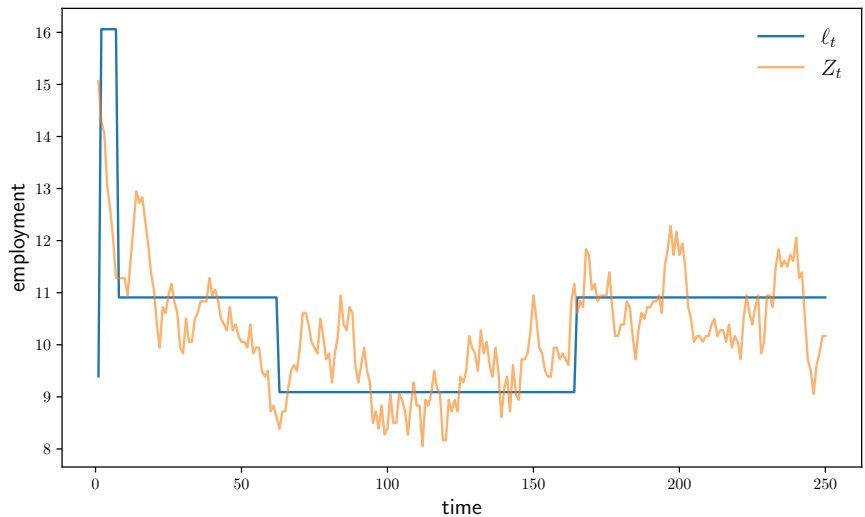

Figure 5.13: Optimal shifts in the stock of labor

## 5.3.1   Structural Estimation

As a first illustration of the ideas in this section, we discuss a connection between econometric estimation and dynamic programs. Our focus is on some modifications that econometricians often make to Bellman equations and how they affect computation and optimality.

### 5.3.1.1   What is Structural Estimation?

Structural estimation is a branch of quantitative social science in which, in a quest to understand observed quantities and prices, researchers attribute Markov decision problems to economic agents. A key step in this approach is to formulate dynamic programs in terms of functional forms and parameters. The econometric challenge is to infer parameters that bring the model outputs as close as possible to actual data.

Structural estimation aims to discover objects that are invariant to hypothetical interventions that the analysis wants to investigate. Examples of such invariant objects are parameters of utility functions, discount factors, and production technologies. Agents inside model solve their MDPs. A policy intervention that systematically alters the Markov processes that they face will alter agents' optimal policies, i.e., their decision rules. Various examples of such interventions involving aspects of fiscal and monetary policy are described in various chapters of Lucas and Sargent (1981) a compendium of early papers that were written in response to the Lucas (1976) Critique of then prevailing dynamic econometric models.[2]

**Example 5.3.1.** Gillingham et al. (2022) study the used car market in Denmark by modeling consumers who trade cars in the new and used car markets. By modeling consumers' decision problems, the authors are able to investigate how consumers would react to a hypothetical modification in automobile taxes. The study finds that automobile taxes were too high in the sense that the government could have raised more tax revenue by lowering tax rates.

Efficient solution methods are essential in structural estimation because the underlying dynamic program must be solved repeatedly in order to search the parameter space for a good fit to data. Moreover, these dynamic programs are often high-dimensional, due to shocks to preferences and other random variables that the agents inside the model are assumed to see but that the econometrician does not. When these shocks are persistent, the dimension of the state grows.[3]

---

[2]Rational expectations econometrics was a response to that Critique. While early work on rational expectations originated from the macroeconomics community (e.g. Hansen and Sargent (1980), Hansen and Sargent (1990)), many of their examples were actually about industrial organization and other microeconomic models. This work was part of a broad process that erased many boundaries between micro and macro theory.

In order to maintain focus on dynamic programming, we will not describe the details of the estimation step required for structural estimation (although §5.4 contains references for those who wish to learn about that). Instead, we focus on the kinds of dynamic programs treated in structural estimation and techniques for solving them efficiently.

### 5.3.1.2  An Illustration

Let us look at an example of a dynamic program with preference shocks used in structural estimation, which is taken from a study of labor supply by married women (Keane et al., 2011). The husband of the decision maker, a married woman, is already working. The couple has young children and the mother is deciding whether to work. Her utility function is

$$u(c, d, \xi) = c + (\alpha n + \xi)(1 - d),$$

where $c$ is consumption, $\alpha$ is a parameter, $n$ is the number of children, $\xi$ is a preference shock and $d$ is the action variable. The action is binary, with $d = 1$ representing the decision to work in the current period and $d = 0$ representing the decision not to work.[4]

The budget constraint for the household is

$$c_t = f_t + w_t d_t - \pi n d_t,$$

where $f_t$ is the father's income, $w_t$ is the mother's wage and $\pi$ is the cost of child care. Wages depend on human capital $h_t$, which increases with experience. In particular,

$$w_t = \gamma h_t + \eta_t, \quad \text{with} \quad h_t = h_{t-1} + d_{t-1}.$$

Here $\eta_t$ is random and $\gamma$ is a parameter. We assume that $(f_t)_{t \geqslant 0}$ is $F$-Markov on some finite set. In the model, $(\xi_t)_{t \geqslant 0}$ and $(\eta_t)_{t \geqslant 0}$ are IID. We denote their joint distribution by $\varphi$.

---

[3]Hansen and Sargent (1980) analyze the implications of such "Shiller errors" for efficient estimation procedures in a class of linear structural models.

[4]Here, the woman is the primary carer of the child; she derives no utility from children in periods in which she works. See Keane et al. (2011) for further discussion.

With constant discount factor $\beta$ and implied utility

$$r(f, h, \xi, \eta, d) := f + (\gamma h + \eta)d - \pi n d + (\alpha n + \xi)(1 - d),$$

the problem of maximizing expected discounted utility is an MDP with the Bellman equation

$$v(f, h, \xi, \eta) = \max_d \left\{ r(f, h, \xi, \eta, d) + \beta \sum_{f', \xi', \eta'} v(f', h + d, \xi', \eta')F(f, f')\varphi(\xi', \eta') \right\}.$$

While we can proceed directly with a technique such as VFI to obtain optimal choices, we can simplify.

One way is by reducing the number of states. A hint comes from looking at the expected value function

$$g(f, h, d) := \sum_{f', \xi', \eta'} v(f', h + d, \xi', \eta')F(f, f')\varphi(\xi', \eta')$$

This function depends only on three arguments and, moreover, the choice variable $d$ is binary. Hence we can break $g$ down into two functions $g(f, h, 0)$ and $g(f, h, 1)$, each of which depends only on the pair $(f, h)$. These functions are substantially simpler than $v$ when the domain of $(\xi, \eta)$ is large. Hence, it is natural to consider whether we can solve our problem using $g$ rather than $v$.

### 5.3.1.3 Expected Value Functions

Rather than address this question within the context of the preceding model, let's shift to a generic version of the dynamic program used in structural estimation and how it can be solved using expected value methods. Our generic version takes the form

$$v(y, \varepsilon) = \max_{a \in \Gamma(y)} \left\{ r(y, \varepsilon, a) + \beta \sum_{y'} \int v(y', \varepsilon')P(y, a, y')\varphi(\varepsilon')\, d\varepsilon' \right\} \tag{5.33}$$

for all $y \in \mathsf{Y}$ and $\varepsilon \in \mathsf{E}$. Here $\mathsf{Y}$ is a finite set, often determined by discretization of a continuous space, while $\mathsf{E}$, the outcome space for $\varepsilon$, is allowed to be continuous. The state $y$ will be called the endogenous state, while $\varepsilon$ is the preference shock. In practice, $\varepsilon$ will often be a vector of shocks that affect current rewards. The integral can therefore be multivariate and is over all of $\mathsf{E}$.

The problem represented by (5.33) is a version of a regular MDP, with state $x = (y, \varepsilon)$ taking values in $\mathsf{X} := \mathsf{Y} \times \mathsf{E}$. If we discretize the space $\mathsf{E}$, then all the optimality theory for MDPs applies. Instead of taking this approach, however, we draw on our discussion of labor choice in §5.3.1.2. In particular, to enhance efficiency, we will work with the **expected value function**

$$g(y, a) := \sum_{y'} \int v(y', \varepsilon') P(y, a, y') \varphi(\varepsilon') \, \mathrm{d}\varepsilon' \tag{5.34}$$

There are several potential advantages associated with working with $g$ rather than $v$. One is that the set of actions $\mathsf{A}$ can be much smaller than the set of states that would be created by discretization of the preference shock space $\mathsf{E}$ (especially if $\varepsilon_t$ takes values in a high-dimensional space). Another is that the integral provides smoothing, so that $g$ is typically a smooth function. This can accelerate structural estimation procedures.

### 5.3.1.4 Optimality via EV Methods

To exploit the relative simplicity of the expected value function, we rewrite the Bellman equation (5.33) as

$$v(y, \varepsilon) = \max_{a \in \Gamma(y)} \left\{ r(y, \varepsilon, a) + \beta g(y, a) \right\}.$$

Taking expectations of both sides and using (5.34) again gives

$$g(y, a) = \sum_{y'} \int \max_{a' \in \Gamma(y')} \left\{ r(y', \varepsilon', a') + \beta g(y', a') \right\} \varphi(\varepsilon') \, \mathrm{d}\varepsilon' P(y, a, y').$$

To solve this functional equation we introduce the **expected value Bellman operator** $R$ defined at $g \in \mathbb{R}^\mathsf{G}$ by

$$(Rg)(y, a) = \sum_{y'} \int \max_{a' \in \Gamma(y')} \left\{ r(y', \varepsilon', a') + \beta g(y', a') \right\} \varphi(\varepsilon') \, \mathrm{d}\varepsilon' P(y, a, y'). \tag{5.35}$$

Here $\mathsf{G}$ is the set of feasible state-action pairs $(y, a)$.

EXERCISE 5.3.1. Prove that $R$ is order-preserving and a contraction of modulus $\beta$ on $\mathbb{R}^\mathsf{G}$ (with respect to the supremum norm).

In what follows, we let $g^*$ be the fixed point of $R$ in $\mathbb{R}^\mathsf{G}$. Since $R$ is a contraction map, $g^*$ can be computed by successive approximation. The next result shows that knowing this fixed point is enough to solve the dynamic program.

**Proposition 5.3.1.** *A policy $\sigma \in \Sigma$ is optimal if and only if*

$$\sigma(y, \varepsilon) \in \operatorname*{argmax}_{a \in \Gamma(y)} \{r(y, \varepsilon, a) + \beta g^*(y, a)\} \qquad \text{for all } (y, \varepsilon) \in \mathsf{Y} \times \mathsf{E}.$$

We postpone proving Proposition 5.3.1 until §5.3.5, where we prove a more general result.

**Example 5.3.2.** In the labor supply problem in §5.3.1.2, the expected value Bellman operator becomes

$$(Rg)(f, h, d) = \sum_{f', \xi', \eta'} \max_{d'} \{r(f', h + d, \xi', \eta', d') \beta g(f', h + d, d')\} F(f, f') \varphi(\xi', \eta').$$

Iterating from an arbitrary guess of $g$ converges to the unique fixed point $g^*$ of $R$. By Proposition 5.3.1, we can then compute the optimal policy $\sigma^*$ at $(f, h, \xi, \eta)$ by taking

$$\sigma^*(f, h, \xi, \eta) \in \operatorname*{argmax}_{d} \{r(f, h, \xi, \eta, d) + \beta g^*(f, h, d)\}.$$

### 5.3.2   The Gumbel Max Trick

§5.3.1.3 described how using expected values can reduce dimensionality by smoothing. But there is another feature of an expected value formulation of a Bellman equation that we can take advantage of when we are prepared to impose extra structure on preference shocks. This section provides details.

A real-valued random variable $Z$ is said to have a **Gumbel distribution** (or a "type 1 generalized extreme value distribution") with mean $\mu \in \mathbb{R}$ if its cumulative distribution function takes the form $F(z) = \exp(-\exp(z - \mu))$. To denote a random variable with a Gumbel distribution, we write $Z \sim G(\mu)$. The expectation of $Z$ is $\mu + \gamma$, where $\gamma \approx 0.577$ is the **Euler–Mascheroni constant**.

EXERCISE 5.3.2. Prove: if $Z \sim G(\mu)$ and $\lambda \in \mathbb{R}$, then $Z + \lambda \sim G(\mu + \lambda)$.

The Gumbel distribution has the following useful stability property, a proof of which can be found in Huijben et al. (2022).

**Lemma 5.3.2.** *If $Z_1, \ldots, Z_k \overset{\text{IID}}{\sim} G(0)$ and $c_1, \ldots, c_k$ are real numbers, then*

$$\max_{1 \leqslant i \leqslant k} (Z_i + c_i) \sim G\left\{-\gamma + \ln\left[\sum_{i=1}^{k} \exp(c_i)\right]\right\}.$$

To exploit Lemma 5.3.2, we continue the discussion in the previous section but assume now that $A = \{a_1, \ldots, a_k\}$, that $\Gamma(y') = A$ for all $y'$ (so that actions are unrestricted), that $\varepsilon'$ in (5.35) is additive in rewards and indexed by actions, so that $r(y', \varepsilon', a') = r(y', a') + \varepsilon'(a')$ for all feasible $(y', a')$, and that, conditional on $y'$, the vector $(\varepsilon(a_1), \ldots, \varepsilon(a_k))$ consists of $k$ independent $G(0)$ shocks. Thus, each feasible choice returns a rewards perturbed by an independent Gumbel shock.

From these assumptions and Lemma 5.3.2, the term inside the integral in (5.35) satisfies

$$\max_{a'} \{r(y', \varepsilon', a') + \beta g(y', a')\} = \max_{a'} \{r(y', a') + \varepsilon'(a') + \beta g(y', a')\}$$
$$\sim G \left\{ -\gamma + \ln \left[ \sum_{a'} \exp\left(r(y', a') + \beta g(y', a')\right) \right] \right\}$$

Recalling our rule for computing mathematical expectations of Gumbel distributed random variables, the expected value Bellman operator $R$ in (5.35) becomes

$$(Rg)(y, a) = \sum_{y'} \ln \left[ \sum_{a'} \exp\left(r(y', a') + \beta g(y', a')\right) \right] P(y, a, y'). \quad (5.36)$$

This operator is convenient because the absence of a max operator permits fast evaluation. Notice also that $R$ is smooth in $g$, which suggests that we can use gradient information to compute its fixed points.

**Proposition 5.3.3.** *The operator $R$ in (5.36) is a contraction of modulus $\beta$ on $\mathbb{R}^G$.*

*Proof.* The operator $R$ is order-preserving on $\mathbb{R}^G$. Straightforward algebra shows that, for $c \in \mathbb{R}_+$ and $g \in \mathbb{R}^G$, we have $R(g + c) = Rg + \beta c$. The claim now follows from Blackwell's sufficient condition for a contraction (page 63). □

Notice how the Gumbel max trick that exploits Lemma 5.3.2 depends crucially on the expected value formulation of the Bellman equation, rather than the standard formulation (5.33). This is because the expected value formulation puts the max inside the expectation operator, unlike the standard formulation, where the max is on the outside.

Variations of the Gumbel max trick have many uses in structural econometrics (see §5.4).

### 5.3.3 Optimal Savings with Stochastic Returns on Wealth

We modify the §5.2.2 optimal savings problem by replacing a constant gross rate of interest $R$ by an IID sequence $(\eta_t)_{t \geqslant 0}$ with common distribution $\varphi$ on finite set E. So the consumer faces a fluctuating rate of returns on financial wealth. In each period $t$, the consumer knows $\eta_t$, the gross rate of interest between $t$ and $t+1$, before deciding how much to consume and how much to save. Other aspects of the problem are unchanged.

We have two motivations. One is computational, namely, to illustrate how framing a decision in terms of expected values can reduce dimensionality, analogous to the results in §5.3.1.4. The other is to generate a more realistic wealth distribution than that generated by the §5.2.2.4 optimal savings model.

With stochastic returns on wealth, the Bellman equation becomes

$$v(w, y, \eta) = \max_{w' \leqslant \eta(w+y)} \left\{ u\left(w + y - \frac{w'}{\eta}\right) + \beta \sum_{y', \eta'} v(w', y', \eta') Q(y, y') \varphi(\eta') \right\}.$$

Both $w$ and $w'$ are constrained to a finite set $W \subset \mathbb{R}_+$. The expected value function can be expressed as

$$g(y, w') := \sum_{y', \eta'} v(w', y', \eta') Q(y, y') \varphi(\eta'). \tag{5.37}$$

In the remainder of this section, we will say that a savings policy $\sigma$ is *g*-**greedy** if

$$\sigma(y, w, \eta) \in \operatorname*{argmax}_{w' \leqslant \eta(w+y)} \left\{ u\left(w + y - \frac{w'}{\eta}\right) + \beta g(y, w') \right\}.$$

Since it is an MDP, we can see immediately that if we replace $v$ in (5.37) with the value function $v^*$, then a $g$-greedy policy will be an optimal one.

Using manipulations analogous to those we used in §5.3.1.4, we can rewrite the Bellman equation in terms of expected value functions via

$$g(y, w') = \sum_{y', \eta'} \max_{w'' \leqslant \eta'(w'+y')} \left\{ u\left(w' + y' - \frac{w''}{\eta'}\right) + \beta g(y', w'') \right\} Q(y, y') \varphi(\eta').$$

From here we could proceed by introducing an expected value Bellman operator analogous to $\eta$ in (5.35), proving it to be a contraction map and then showing that greedy policies taken with respect to the fixed point are optimal. All of this can be accomplished without too much difficulty – we prove more general results in §5.3.5.

```
using QuantEcon, LinearAlgebra, IterTools

function create_savings_model(; β=0.98, γ=2.5,
                                w_min=0.01, w_max=20.0, w_size=100,
                                ρ=0.9, ν=0.1, y_size=20,
                                η_min=0.75, η_max=1.25, η_size=2)
    η_grid = LinRange(η_min, η_max, η_size)
    ϕ = ones(η_size) * (1 / η_size)  # Uniform distributoin
    w_grid = LinRange(w_min, w_max, w_size)
    mc = tauchen(y_size, ρ, ν)
    y_grid, Q = exp.(mc.state_values), mc.p
    return (; β, γ, η_grid, ϕ, w_grid, y_grid, Q)
end
```

Listing 20: Optimal savings parameters (`modified_opt_savings.jl`)

However, we also know that optimistic policy iteration (OPI) is, in general, more efficient than value function iteration. This motivates us to introduce the modified $\sigma$-value operator

$$(R_\sigma g)(y, w') = \sum_{y', \eta'} \left\{ u\left(w' + y' - \frac{\sigma(w', y', \eta')}{\eta'}\right) + \beta g(y', \sigma(w', y', \eta')) \right\} Q(y, y') \varphi(\eta').$$

This is a modification of the regular $\sigma$-value operator $T_\sigma$ that makes it act on expected value functions.

A suitably modified OPI routine that is adapted from the regular OPI algorithm in §5.1.4.4 can be found in Algorithm 5.5 on page 178. The routine is convergent. We discuss this in greater detail in §5.3.5.

Figure 5.14 shows a histogram of a long wealth time series that parallels Figure 5.10 on page 156. The only significant difference is the switch to stochastic returns (as described above). Parameters are as in Listing 20. Now the wealth distribution has a more realistic long right tail (a few observations are in the far right tail, although they are difficult to see). The Gini coefficient is 0.72, which is closer to typical country values recorded in World Bank data (but still lower than the US). In essence, this occurs because return shocks have multiplicative rather than additive effects on wealth, so a sequence of high draws compounds to make wealth grow fast.

EXERCISE 5.3.3. Consider a version of the optimal savings problem from §5.2.2

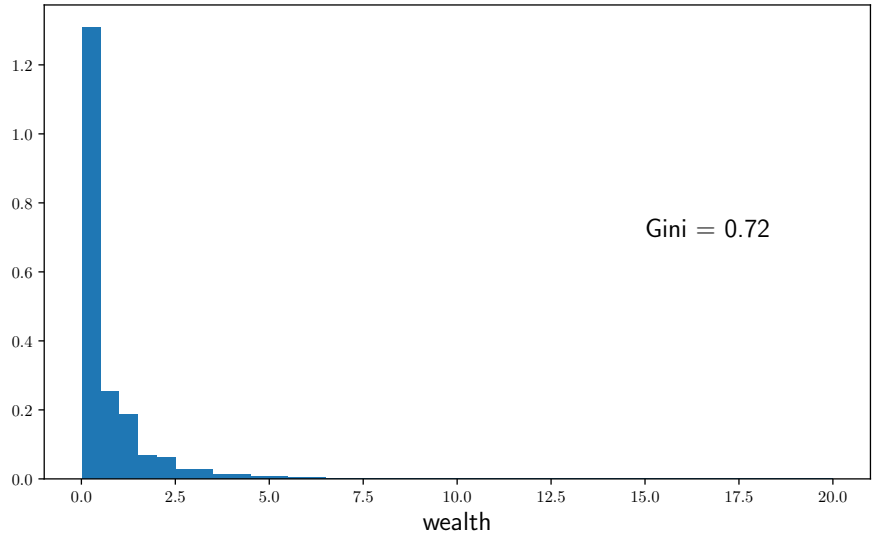

Figure 5.14: Histogram of wealth (stochastic returns)

where labor income has both persistent and transient components. In particular, assume that $Y_t = Z_t + \varepsilon_t$ for all $t$, where $(\varepsilon_t)_{t \geqslant 0}$ is IID with common distribution $\varphi$ on E, while $(Z_t)_{t \geqslant 0}$ is Q-Markov on Z. Such a specification of labor income can capture how households should react differently to transient and "permanent" shocks (see §5.4 for more discussion). Following the pattern developed for the savings model with stochastic returns on wealth, write down both the Bellman equation and the Bellman equation in terms of expected value functions.

### 5.3.4 Q-Factors

$Q$-factors assign values to state-action pairs. They set the stage for $Q$-learning, an application of reinforcement learning, a recursive algorithm for estimating parameters. $Q$-learning uses stochastic approximation techniques to learn $Q$-factors. Under special conditions $Q$-learning eventually learns optimal $Q$-factors for a finite MDP.

$Q$-learning is connected to the topic of this chapter because it relies on a Bellman operator for the $Q$-factor. We discuss that Bellman operator, but we don't discuss $Q$-learning here.

To begin, we fix an MDP $(\Gamma, \beta, r, P)$ with state space X and action space A. For each

$v \in \mathbb{R}^{\mathsf{X}}$, the **Q-factor** corresponding to $v$ is the function

$$q(x, a) = r(x, a) + \beta \sum_{x'} v(x') P(x, a, x') \qquad ((x, a) \in \mathsf{G}).$$

We can convert the Bellman equation into an equation in $Q$-factors by observing that, given such a $q$, the Bellman equation can be written as $v(x) = \max_{a \in \Gamma(x)} q(x, a)$. Taking the mean and discounting on both sides of this equation gives

$$\beta \sum_{x'} v(x') P(x, a, x') = \beta \sum_{x'} \max_{a' \in \Gamma(x')} q(x', a') P(x, a, x').$$

Adding $r(x, a)$ and using the definition of $q$ again gives

$$q(x, a) = r(x, a) + \beta \sum_{x'} \max_{a' \in \Gamma(x')} q(x', a') P(x, a, x').$$

This functional equation motivates us to introduce the **Q-factor Bellman operator**

$$(Sq)(x, a) = r(x, a) + \beta \sum_{x'} \max_{a' \in \Gamma(x')} q(x', a') P(x, a, x') \qquad ((x, a) \in \mathsf{G}). \tag{5.39}$$

EXERCISE 5.3.4. Prove that $S$ is order-preserving and a contraction of modulus $\beta$ on $\mathbb{R}^{\mathsf{G}}$ (with respect to the supremum norm).

Let $q^*$ be the unique fixed point of $S$ in $\mathbb{R}^{\mathsf{G}}$.

**Proposition 5.3.4.** *A policy $\sigma \in \Sigma$ is optimal if and only if*

$$\sigma(x) \in \operatorname*{argmax}_{a \in \Gamma(x)} q^*(x, a) \qquad \textit{for all } (x, a) \in \mathsf{G}.$$

Enthusiastic readers might like to try to prove Proposition 5.3.4 directly. We defer the proof until §5.3.5, where a more general result is obtained.

### 5.3.5 Operator Factorizations

Our study of structural estimation in §5.3.1, optimal savings in §5.3.3 and $Q$-factors in §5.3.4 all involved manipulations of the Bellman and policy operators that presented alternative perspectives on the respective optimization problems. Rather than offering additional applications that apply such ideas, we now develop a general theoretical framework from which to understand manipulations of the Bellman and policy operators for general MDPs. The framework clarifies when and how these techniques can be applied.

### 5.3.5.1 Refactoring the Bellman Operator

Fix an MDP $(\Gamma, \beta, r, P)$ with state space X and action space A. As usual, $\Sigma$ is the set of feasible policies, G is the set of feasible state, action pairs, $T$ is the Bellman operator and $v^*$ denotes the value function. Our first step is to decompose $T$. To do this we introduce three auxiliary operators:

- $E \colon \mathbb{R}^X \to \mathbb{R}^G$ defined by $(Ev)(x, a) = \sum_{x'} v(x') P(x, a, x')$,

- $D \colon \mathbb{R}^G \to \mathbb{R}^G$ defined by $(Dg)(x, a) = r(x, a) + \beta g(x, a)$ and

- $M \colon \mathbb{R}^G \to \mathbb{R}^X$ defined by $(Mq)(x) = \max_{a \in \Gamma(x)} q(x, a)$.

Evidently the action of the Bellman operator $T$ on a given $v \in \mathbb{R}^X$ is the composition of these three steps:

 (i) take conditional expectations given $(x, a) \in G$ (applying $E$),

 (ii) discount and adding current rewards (applying $D$), and

 (iii) maximize with respect to current action (applying $M$).

As a result, we can write $T = MDE := M \circ D \circ E$ (apply $E$ first, $D$ second, and $M$ third). This decomposition is visualized in Figure 5.15. The action of $T$ is a round trip from the top node, which is the set of value functions.

If we stare at Figure 5.15, we can imagine two other round trips. One is a round trip from the set of expected value functions, obtained by the sequence $EMD$. The other is a round trip from the set of $Q$-factors, obtained by the sequence $DEM$. Let's name these additional round trips $R$ and $S$ respectively, so that, collecting all three,

$$R = EMD, \quad S = DEM, \quad T = MDE. \tag{5.40}$$

Both $R$ and $S$ act on functions in $\mathbb{R}^G$. The next exercise provides an explicit representation of these operators.

EXERCISE 5.3.5. Show that for any $g, q \in \mathbb{R}^G$ and $(x, a) \in G$ we have

$$(Rg)(x, a) = \sum_{x'} \max_{a' \in \Gamma(x')} \{r(x', a') + \beta g(x', a')\} P(x, a, x') \quad \text{and}$$

$$(Sq)(x, a) = r(x, a) + \beta \sum_{x'} \max_{a' \in \Gamma(x')} q(x', a') P(x, a, x').$$

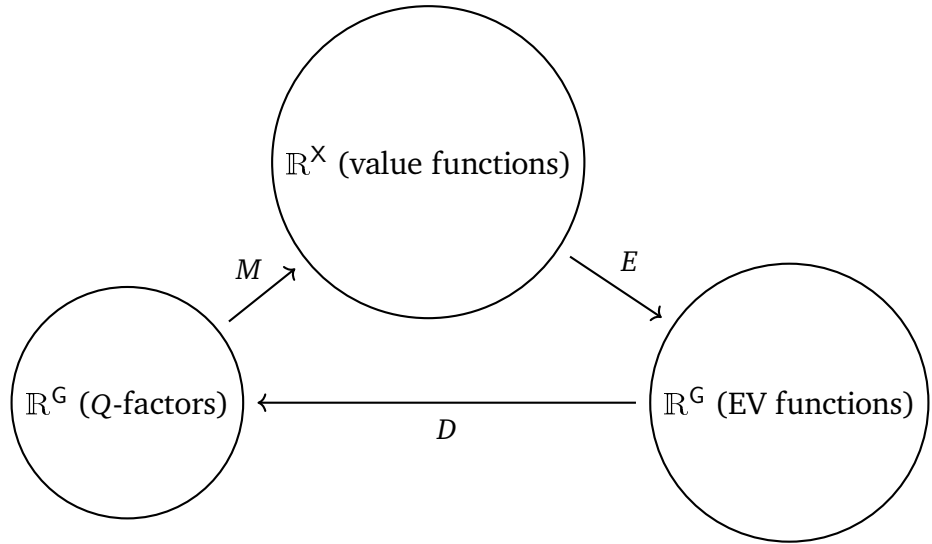

Figure 5.15: Multiple Bellman operators (EV = expected value)

Let's connect our "refactored" Bellman operators $R$ and $S$ to our preceding examples. Inspection of (5.39) confirms that $S$ is exactly the $Q$-factor Bellman operator. In addition, $R$ is a general version of the expected value Bellman operator defined in (5.35).

EXERCISE 5.3.6. Show that, for all $k \in \mathbb{N}$, the following relationships hold

- $R^k = ET^{k-1}MD = EMS^{k-1}D$

- $S^k = DR^{k-1}ME = MT^{k-1}ED$

- $T^k = MS^{k-1}DE = MDR^{k-1}E$

(Here, for any operator $A$, we take $A^0$ to be the identity map.)

While the equalities in Exercise 5.3.6 can be proved by induction via the logic revealed by (5.40), the intuition is straightforward from Figure 5.15. For example, the relationship $R^k = ET^{k-1}MD$ states that round-tripping $k$ times from the space of expected values (EV function space) is the same as shifting to value function space by applying $MD$, round-tripping $k-1$ times using $T$, and then shifting one more step to EV function space via $E$.

Although the relationships in Exercise 5.3.6 are easy to prove, they are already useful. For example, suppose that in a computational setting $R$ is easier to iterate than $T$. Then to iterate with $T$ $k$ times, we can instead use $T^k = MDR^{k-1}E$: We apply $E$ once, $R$ $k-1$ times, and $M$ and $D$ once each. If $k$ is large, this might be more efficient.

In the next exercise and the next section, we let $\|\cdot\| := \|\cdot\|_\infty$, the supremum norm on either $\mathbb{R}^X$ or $\mathbb{R}^G$.

EXERCISE 5.3.7. Prove the following facts:

(i) $\|Ev - Ev'\| \leqslant \|v - v'\|$ for all $v, v' \in \mathbb{R}^X$,

(ii) $\|Mg - Mg'\| \leqslant \|g - g'\|$ for all $g, g' \in \mathbb{R}^G$, and

(iii) $\|Dq - Dq'\| \leqslant \beta\|q - q'\|$ for all $q, q' \in \mathbb{R}^G$.

We can say that $E$ and $M$ are **nonexpansive** on $\mathbb{R}^X$ and $\mathbb{R}^G$ respectively, while $D$ is a contraction on $\mathbb{R}^G$.

**Lemma 5.3.5.** *The operators $R$, $S$ and $T$ are all contraction maps of modulus $\beta$ under the supremum norm.*

*Proof.* That $T$ is a contraction of modulus $\beta$ was proved in Proposition 5.1.1, on page 139. We can prove this more easily now by applying Exercise 5.3.7, which, for arbitrary $v, v' \in \mathbb{R}^X$, gives

$$\|Tv - Tv'\| = \|MDEv - MDEv'\| \leqslant \|MDv - MDv'\| \leqslant \beta\|Mv - Mv'\| \leqslant \beta\|v - v'\|.$$

Proofs for $R = EMD$ and $S = DEM$ are similar. □

In the next section we clarify relationships between these operators and prove Propositions 5.3.1 and 5.3.4.

### 5.3.5.2 Refactorizations and Optimality

From Lemma 5.3.5 we see that $R$, $S$ and $T$ all have unique fixed points. We denote them by $g^*$, $q^*$ and $v^*$ respectively, so that

$$Rg^* = g^*, \quad Sq^* = q^*, \quad \text{and} \quad Tv^* = v^*.$$

We already know that $v^*$ is the value function (Proposition 5.1.1). The results below show that the other two fixed points are, like the value function, sufficient to determine optimality.

**Proposition 5.3.6.** *The fixed points of $R$, $S$ and $T$ are connected by the following relationships:*

   (i)  $g^* = Ev^*$,

  (ii)  $q^* = Dg^*$, and

 (iii)  $v^* = Mq^*$.

*Proof.* To prove (i), first observe that, in the notation of (5.40), we have $Ev^* = ETv^* = EMDEv^* = REv^*$. Hence $Ev^*$ is a fixed point of $R$. But $R$ has only one fixed point, which is $g^*$. Therefore, $g^* = Ev^*$. The proofs of (ii) and (iii) are analogous. □

The results in Proposition 5.3.6 can be written more explicitly as

- $g^*(x, a) = \sum_{x'} v^*(x') P(x, a, x')$ for all $(x, a) \in \mathsf{G}$,

- $q^*(x, a) = r(x, a) + \beta g^*(x, a)$ for all $(x, a) \in \mathsf{G}$, and

- $v^*(x) = \max_{a \in \Gamma(x)} q^*(x, a)$ for all $x \in \mathsf{X}$.

In the next result and the discussion that follows, given $g, q \in \mathbb{R}^{\mathsf{G}}$, we will call $\sigma \in \Sigma$

- $g$-**greedy** if $\sigma(x) \in \operatorname{argmax}_{a \in \Gamma(x)} \{r(x, a) + \beta g(x, a)\}$ for all $x \in \mathsf{X}$, and

- $q$-**greedy** if $\sigma(x) \in \operatorname{argmax}_{a \in \Gamma(x)} q(x, a)$ for all $x \in \mathsf{X}$.

These definitions are exact analogs of the $v$-greedy concept, applied to expected value functions and $Q$-factors respectively.

**Corollary 5.3.7.** *For $\sigma \in \Sigma$, the following statements are equivalent:*

   (i)  *$\sigma$ is $v$-greedy when $v = v^*$.*

  (ii)  *$\sigma$ is $g$-greedy when $g = g^*$.*

 (iii)  *$\sigma$ is $q$-greedy when $q = q^*$.*

*In particular, $\sigma$ is optimal if and only if any one (and hence all) of (i)–(iii) holds.*

*Proof.* To see that (i) implies (ii), suppose that $\sigma$ is $v$-greedy when $v = v^*$. Then for arbitrary $x \in \mathsf{X}$

$$\sigma(x) \in \operatorname*{argmax}_{a \in \Gamma(x)} \left\{ r(x, a) + \beta \sum_{x'} v^*(x') P(x, a, x') \right\} = \operatorname*{argmax}_{a \in \Gamma(x)} \{r(x, a) + \beta g^*(x, a)\}.$$

Hence $\sigma$ is $g$-greedy when $g = g^*$, and (i) implies (ii). The proofs of the remaining equivalences (ii) $\implies$ (iii) $\implies$ (i) are similar. The claim that $\sigma$ is optimal if and only if any one of (i)–(iii) holds now follows from Proposition 5.1.1, which asserts that $\sigma$ is optimal if and only if $\sigma$ is $v^*$-greedy. □

Notice that Proposition 5.3.4 is a special case of Corollary 5.3.7.

The results in Corollary 5.3.7 can be understood as "refactored" versions of Bellman's principle of optimality. A consequence of these results is that we can solve a given MDP by modifying VFI to operate either on expected value functions or on $Q$-factors. For example, if we find it more convenient to iterate in expected value space, then (informally) we can proceed as follows:

  (i) Fix $g \in \mathbb{R}^{\mathsf{G}}$.

 (ii) Iterate with $R$ to obtain $g_k := R^k g \approx g^*$.

(iii) Compute a $g_k$-greedy policy.

Since $g_k \approx g^*$, the resulting policy will be approximately optimal.

### 5.3.5.3 Refactored OPI

In Chapter 5 we found that VFI is often outperformed by HPI or OPI. Our next step is to apply these methods to modified versions of the Bellman equation, as discussed in the previous section. This allows us to combine advantages of HPI/OPI with the potential efficiency gains obtained by refactoring the Bellman equation.

We illustrate these ideas below by producing a version of OPI that can compute $Q$-factors and expected value functions. (The same is true for HPI, although we leave details of that construction to interested readers.)

To begin, we introduce a new operator, denoted $M_\sigma$, that, for fixed $\sigma \in \Sigma$ and $q \in \mathbb{R}^{\mathsf{G}}$, produces

$$(M_\sigma q)(x) := q(x, \sigma(x)) \qquad (x \in \mathsf{X}).$$

This operator is the policy analog of the maximization operator $M$ defined by $(Mq)(x) = \max_{a \in \Gamma(x)} q(x, a)$ in §5.3.5.1. Analogous to (5.40), we set

$$R_\sigma := E M_\sigma D, \quad S_\sigma := D E M_\sigma, \quad T_\sigma := M_\sigma D E.$$

You can verify that $T_\sigma$ is the ordinary $\sigma$-policy operator (defined in (5.19)). The operators $R_\sigma$ and $S_\sigma$ are the expected value and $Q$-factor equivalents.

EXERCISE 5.3.8. The relationships in Exercise 5.3.6 continue to hold after we swap $R, S, T, M$ with $R_\sigma, S_\sigma, T_\sigma, M_\sigma$. Confirm the first of these relationships, showing in particular that

$$R_\sigma^k = E T_\sigma^{k-1} M_\sigma D \qquad \text{for all } k \in \mathbb{N}. \tag{5.41}$$

Let's now show that OPI can be successfully modified via these alternative operators. We will focus on the expected value viewpoint (value functions are replaced by expected value functions), which is often helpful in the applications we wish to consider.

Our modified OPI routine is given in Algorithm 5.5. It makes the obvious modifications to regular OPI, switching to working with expected value functions in $\mathbb{R}^\mathsf{G}$ and from iteration with $T_\sigma$ to iteration with $R_\sigma$.

---

**Algorithm 5.5:** Refactored OPI for expected value functions

---
1   input $g_0 \in \mathbb{R}^\mathsf{G}$, an initial guess of $g^*$
2   input $\tau$, a tolerance level for error
3   input $m \in \mathbb{N}$, a step size
4   $k \leftarrow 0$
5   $\varepsilon \leftarrow \tau + 1$
6   **while** $\varepsilon > \tau$ **do**
7      $\sigma_k \leftarrow$ a $g_k$-greedy policy
8      $g_{k+1} \leftarrow R_{\sigma_k}^m g_k$
9      $\varepsilon \leftarrow \|g_k - g_{k+1}\|_\infty$
10     $k \leftarrow k + 1$
11 **end**
12 **return** $\sigma_k$

---

Algorithm 5.5 is globally convergent in the same sense as regular OPI (Algorithm 5.4 on page 146). In fact, if we pick $v_0 \in \mathbb{R}^\mathsf{X}$ and apply regular OPI with this initial condition, as well as applying Algorithm 5.5 with initial condition $g_0 := Ev_0$, then the sequences $(v_k)_{k \geqslant 0}$ and $(g_k)_{k \geqslant 0}$ generated by the two algorithms are connected via $g_k = Ev_k$ for all $k \geqslant 0$. If greedy policies are unique, then it is also true that the policy sequences generated by the two algorithms are identical.

Let's prove these claims, assuming for convenience that greedy policies are unique. Consider first the claim that $g_k = Ev_k$ for all $k \geqslant 0$. This is true by assumption when $k = 0$. Suppose, as an induction hypothesis, that $g_k = Ev_k$ holds at arbitrary $k$. Let $\sigma$ be $g_k$-greedy. Then

$$\sigma(x) = \underset{a \in \Gamma(x)}{\operatorname{argmax}} \{r(x, a) + \beta g_k(x, a)\} = \underset{a \in \Gamma(x)}{\operatorname{argmax}} \left\{ r(x, a) + \beta \sum_{x'} v_k(x') P(x, a, x') \right\},$$

where the second equality is implied by $g_k = Ev_k$. Hence $\sigma$ is both $g_k$-greedy and $v_k$-greedy and so is the next policy selected by both modified and regular OPI. Moreover, updating via Algorithm 5.5 and applying (5.41), we have

$$g_{k+1} = R_\sigma^m g_k = E T_\sigma^{m-1} M_\sigma D g_k = E T_\sigma^{m-1} M_\sigma D E v_k = E T_\sigma^m v_k.$$

Since $\sigma$ is $v_k$-greedy, $T_\sigma^m v_k$ is the next function selected by regular OPI. Hence $v_{k+1} = T_\sigma^m v_k$. Connecting with the last chain of equalities yields $g_{k+1} = E v_{k+1}$. This completes the proof that $g_k = E v_k$ for all $k$. Policy functions generated by the algorithms are identical as well.

The preceding discussion provides a justification for the modified OPI algorithm we adopted in §5.3.3.

## 5.4 Chapter Notes

Detailed treatment of MDPs can be found in books by Bellman (1957), Howard (1960), Denardo (1981), Puterman (2005), Bertsekas (2012), Hernández-Lerma and Lasserre (2012a, 2012b), and Kochenderfer et al. (2022). The books by Hernández-Lerma and Lasserre (2012a, 2012b) provide excellent coverage of theory, while Puterman (2005) gives a clear and detailed exposition of algorithms and techniques. Further discussion of the connection between HPI and Newton iteration can be found in Section 6.4 of Puterman (2005).

HPI is routinely used in artificial intelligence applications, including during the training of AlphaZero by DeepMind. Further discussion of these variants of HPI and their connection to Newton iteration can be found in Bertsekas (2021) and Bertsekas (2022a).

There are several methods for available for accelerating value function iteration, including asynchronous VFI and Anderson acceleration. Due to space constraints, we omit discussion of these topics. Interested readers can find a treatment of asynchronous VFI in Bertsekas (2022b). For discussion of Anderson acceleration see, for example, Walker and Ni (2011) or Geist and Scherrer (2018). First order methods for accelerating VFI are presented in Goyal and Grand-Clement (2023).

Other methods for computing solutions to MDPs include the linear programming (LP) approach and the policy gradient technique, both of which solve a problem of the form

$$\max_{\sigma \in \Sigma} \sum_x w(x) v(x) \quad \text{s.t.} \quad v = r_\sigma + \beta P_\sigma v \tag{5.42}$$

for some chosen weight function $w$. The LP approach views (5.42) as a linear program and applies various algorithms to the primal and dual problems. See, for example, Puterman (2005) or Ying and Zhu (2020).

The policy gradient method involves approximating $\sigma$ and $\nu$ in (5.42) using smooth functions with finitely many parameters. These parameters are then adjusted via some version of gradient ascent. A recent trend for high-dimensional MDPs is to approximate the value and policy functions with neural nets. An early exposition can be found in Bertsekas and Tsitsiklis (1996). A more recent monograph is Bertsekas (2021). For research along these lines in the context of economic applications see, for example, Maliar et al. (2021), Hill et al. (2021), Han et al. (2021), Kahou et al. (2021), Kase et al. (2022), and Azinovic et al. (2022).

In some versions of these algorithms, as well as in VFI and HPI, the expectations associated with dynamic programs are computed using Monte Carlo sampling methods. See, for example, Rust (1997), Powell (2007), and Bertsekas (2021). Sidford et al. (2023) combine linear programming and sampling approaches.

The optimal savings problem is a workhorse in macroeconomics and has been treated extensively in the literature. Early references include Brock and Mirman (1972), Mirman and Zilcha (1975), Schechtman (1976), Deaton and Laroque (1992), and Carroll (1997). For more recent studies, see, for example, Li and Stachurski (2014), Açıkgöz (2018), Light (2018), Lehrer and Light (2018), or Ma et al. (2020). Recent applications involving optimal savings in a representative agent framework include Bianchi (2011), Paciello and Wiederholt (2014), Rendahl (2016), Heathcote and Perri (2018), Paroussos et al. (2019), Erosa and González (2019), Herrendorf et al. (2021), and Michelacci et al. (2022). For more on the long right tail of the wealth distribution (as discussed in §5.3.3), see, for example, Benhabib et al. (2015), Krueger et al. (2016), or Stachurski and Toda (2019).

Households solving optimal savings problems are often embedded in heterogeneous agent models in order to study income distributions, wealth distributions, business cycles and other macroeconomic phenomena. Representative examples include Aiyagari (1994), Huggett (1993), Krusell and Smith (1998), Miao (2006), Algan et al. (2014), Toda (2014), Benhabib et al. (2015), Stachurski and Toda (2019), Toda (2019), Light (2020), Hubmer et al. (2020), or Cao (2020).

Exercise §5.3.3 considered optimal savings and consumption in the presence of transient and persistent shocks to labor income. For research in this vein, see, for example, Quah (1990), Carroll (2009), De Nardi et al. (2010), or Lettau and Ludvigson (2014). For empirical work on labor income dynamics, see, for example, Newhouse (2005), Guvenen (2007), Guvenen (2009), or Blundell et al. (2015). For analysis of optimal savings in a very general setting, see Ma et al. (2020) or Ma and Toda (2021).

The optimal investment problem dates back to Lucas and Prescott (1971). Textbook treatments can be found in Stokey and Lucas (1989) and Dixit and Pindyck (2012). Sargent (1980) and Hayashi (1982) used optimal investment problems to connect optimal capital accumulation with Tobin's $q$ (which is the ratio between a physical asset's market value and its replacement value). Other influential papers in the field include Lee and Shin (2000), Hassett and Hubbard (2002), Bloom et al. (2007), Bond and Van Reenen (2007), Bloom (2009), and Wang and Wen (2012). Carruth et al. (2000) contains a survey.

Classic papers about S-s inventory models include Arrow et al. (1951) and Dvoretzky et al. (1952). Optimality of S-s policies under certain conditions was first established by Scarf (1960). Kelle and Milne (1999) study the impact of S-s inventory policies on the supply chain, including connection to the "bullwhip" effect. The connection between S-s inventory policies and macroeconomic fluctuations is studied in Nirei (2006).

The model in Exercise 5.2.3 is loosely adapted from Bagliano and Bertola (2004).

Rust (1994) is a classic and highly readable reference in the area of structural estimation of MDPs. Keane and Wolpin (1997) provides an influential study of the career choices of young men. Roberts and Tybout (1997) analyze the decision to export in the presence of sunk costs. Keane et al. (2011) provide an overview of structural estimation applied to labor market problems. Gentry et al. (2018) review analysis of auctions using structural estimation. Legrand (2019) surveys the use of structural models to study the dynamics of commodity prices. Calsamiglia et al. (2020) use structural estimation to study school choices. Iskhakov et al. (2020) provide a thoughtful discussion on the differences between structural estimation and machine learning. Luo and Sang (2022) propose structural estimation via sieves.

Theoretical analysis of expected value functions in discrete choice models and other settings can be found in Rust (1994), Norets (2010), Mogensen (2018) and Kristensen et al. (2021). The expected value Gumbel max trick is due to Rust (1987) and builds on work by McFadden (1974). The Gumbel max trick is also used in machine learning methods (see, e.g., Jang et al. (2016)).

In §5.3.4 we mentioned $Q$-learning, which was originally proposed by Watkins (1989). Tsitsiklis (1994) and Melo (2001) studied convergence of $Q$-learning. In related work, Esponda and Pouzo (2021) study Markov decision processes where dynamics are unknown, and where agents update their understanding of transition laws via Bayesian updating.

The theory in §5.3.5 on optimality under modifications of the Bellman equation is loosely based on Ma and Stachurski (2021). That paper considers arbitrary modifications in a very general setting.

# Chapter 6

# Stochastic Discounting

In this chapter we describe how to extend the MDP model to handle time-varying discount factors, a specification now widely used in macroeconomics and finance.

## 6.1 Time-Varying Discount Factors

We introduce formulas for infinite-horizon lifetime valuations under stochastic discounting and provide necessary and sufficient conditions for existence of finite solutions.

### 6.1.1 Valuation

Our first step is to motivate and understand lifetime valuation when discount factors vary over time.

#### 6.1.1.1 Motivation

In §3.2.2.2 we discussed firm valuation in a setting where the interest rate is constant. But data show that interest rates are time-varying, even for safe assets like US Treasury bills. Figure 6.1 shows nominal interest rate on 1 Year Treasury bills since the 1950s, while Figure 6.2 shows an estimate of the real interest rate for 10 year T-bills since 2012. Both nominal and real interest rates are evidently time varying.

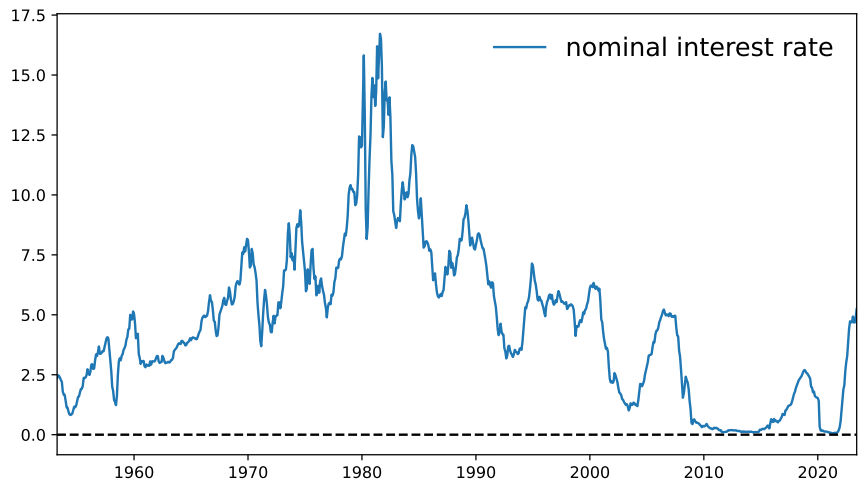

Figure 6.1: Nominal US interest rates (`plot_interest_rates_nominal.jl`)

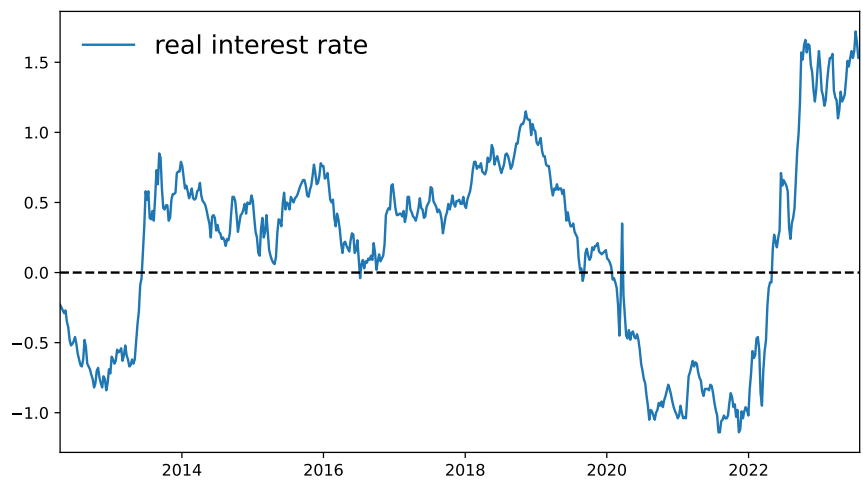

Figure 6.2: Real US interest rates (`plot_interest_rates_real.jl`)

**Example 6.1.1.** Consider a firm valuation problem where interest rates $(r_t)_{t \geqslant 0}$ are stochastic. The time zero expected present value of time $t$ profit $\pi_t$ is

$$\mathbb{E}\left[\beta_1 \cdots \beta_t \cdot \pi_t\right] \quad \text{where} \quad \beta_t := \frac{1}{1 + r_t}.$$

The lifetime value of the firm is then

$$V_0 = \mathbb{E} \sum_{t=0}^{\infty} \left[\prod_{i=0}^{t} \beta_i\right] \pi_t. \tag{6.1}$$

**Remark 6.1.1.** Time-varying discount factors are found in extensions of the Section §3.2.2.3 household consumption-saving problem that appear in modern models of business cycle dynamics, asset prices, and wealth distributions. For just one important example, see Krusell and Smith (1998). Marimon (1984) included random discount factors in his thorough analysis of growth and turnpike properties of general equilibrium models, unfortunately only parts of which were include in Marimon (1989). Exogenous impatience shocks have been used as demand shocks in some dynamic models. For more citations see §7.4.

### 6.1.1.2 Theory

The aim of this section is to understand and evaluate expressions such as (6.1). Throughout,

- X is a finite set, $P \in \mathcal{M}(\mathbb{R}^{\mathsf{X}})$, and $(X_t)_{t \geqslant 0}$ is $P$-Markov.

- $h$ is an element of $\mathbb{R}^{\mathsf{X}}$, with $h(X_t)$ typically interpreted as a payoff or reward at time $t$ in state $X_t$.

- $b$ is a map from $\mathsf{X} \times \mathsf{X}$ to $(0, \infty)$ and

$$\beta_t := b(X_{t-1}, X_t) \text{ for } t \in \mathbb{N} \quad \text{with} \quad \beta_0 := 1 \tag{6.2}$$

The sequence $(\beta_t)_{t \geqslant 0}$ is called a **discount factor process** and $\prod_{i=0}^{t} \beta_i$ is the discount factor for period $t$ payoffs evaluated at time zero. We are interested in expected discounted sums of the form

$$v(x) := \mathbb{E}_x \sum_{t=0}^{\infty} \left[\prod_{i=0}^{t} \beta_i\right] h(X_t) \qquad (x \in \mathsf{X}). \tag{6.3}$$

**Theorem 6.1.1.** *Let $L \in \mathcal{L}(\mathbb{R}^{\mathsf{X}})$ be the discount operator defined by*

$$L(x, x') := b(x, x')P(x, x') \tag{6.4}$$

*for $(x, x') \in \mathsf{X} \times \mathsf{X}$. If $\rho(L) < 1$, then $v$ in (6.3) is finite for all $x \in \mathsf{X}$ and, moreover,*

$$v = (I - L)^{-1} h = \sum_{t=0}^{\infty} L^t h. \tag{6.5}$$

Theorem 6.1.1 generalizes Lemma 3.2.1 on page 95. Indeed, if $b \equiv \beta \in (0, 1)$, then $L = \beta P$ and $\rho(L) = \beta \rho(P) = \beta < 1$, so the result in Theorem 6.1.1 reduces to Lemma 3.2.1.

*Proof of Theorem 6.1.1.* To verify Theorem 6.1.1, we first prove that

$$\mathbb{E}_x \left[ \prod_{i=0}^{t} \beta_i \right] h(X_t) = (L^t h)(x) \quad \text{for all } t \in \mathbb{N}, h \in \mathbb{R}^{\mathsf{X}} \text{ and } x \in \mathsf{X}. \tag{6.6}$$

We establish (6.6) using induction on $t$. It is easy to see that (6.6) holds at $t = 1$. Now suppose it holds at $t$. We claim it also holds at $t + 1$. To show this we fix $h \in \mathbb{R}^{\mathsf{X}}$ and set $\delta_t := \prod_{i=0}^{t} \beta_i$. Applying the law of iterated expectations (see §3.2.1.2) yields

$$\mathbb{E}_x \, \delta_{t+1} \, h(X_{t+1}) = \mathbb{E}_x \, \mathbb{E}_t \, b(X_t, X_{t+1}) \delta_t \, h(X_{t+1}) = \mathbb{E}_x \, \delta_t \, \mathbb{E}_t \, b(X_t, X_{t+1}) h(X_{t+1}).$$

Since

$$\mathbb{E}_t \, b(X_t, X_{t+1}) h(X_{t+1}) = \sum_{x'} b(X_t, x') h(x') P(X_t, x') = \sum_{x'} L(X_t, x') h(x') = (Lh)(X_t),$$

we can now write

$$\mathbb{E}_x \, \delta_{t+1} h(X_{t+1}) = \mathbb{E}_x \, \delta_t f(X_t) \quad \text{where} \quad f(x) := (Lh)(x). \tag{6.7}$$

Applying the induction hypothesis to (6.7) yields $\mathbb{E}_x \, \delta_{t+1} h(X_{t+1}) = (L^t f)(x)$. But $L^t f = L^t L h = L^{t+1} h$, so $\mathbb{E}_x \, \delta_{t+1} h(X_{t+1}) = (L^{t+1} h)(x)$. This completes the induction step and hence the proof of (6.6)

Now we can complete the proof of Theorem 6.1.1. To this end, we fix $x \in \mathsf{X}$ and use (6.6) to obtain

$$v(x) = \mathbb{E}_x \sum_{t=0}^{\infty} \left[ \prod_{i=0}^{t} \beta_i \right] h(X_t) = \sum_{t=0}^{\infty} \mathbb{E}_x \left[ \prod_{i=0}^{t} \beta_i \right] h(X_t) = \sum_{t=0}^{\infty} (L^t h)(x). \tag{6.8}$$

Pointwise, this is $v = \sum_{t \geq 0} L^t h$. By the Neumann series lemma and $\rho(L) < 1$, this sum converges and equals $(I - L)^{-1} h$. □

In (6.8) we passed expectations through an infinite sum. This operation is valid under the assumption $\rho(L) < 1$. A complete proof can be found in §B.2.

**EXERCISE 6.1.1.** Consider Example 6.1.1 again but now assume that $(X_t)$ is $P$-Markov, $\pi_t = \pi(X_t)$, and $r_t = r(X_t)$ for some $r, \pi \in \mathbb{R}^{\mathsf{X}}$.[1] The expected present value of the firm given current state $X_0 = x$ is

$$v(x) = \mathbb{E}_x \sum_{t=0}^{\infty} \left[ \prod_{i=0}^{t} \beta_i \right] \pi_t \tag{6.9}$$

Suggest a condition under which $v(x)$ is finite and discuss how to compute it.

**EXERCISE 6.1.2.** Let $\mathsf{X}$ be partially ordered and assume $\rho(L) < 1$. Prove that $v$ is increasing on $\mathsf{X}$ whenever $P$ is monotone increasing, $\pi$ is increasing on $\mathsf{X}$, and $r$ is decreasing on $\mathsf{X}$.

## 6.1.2 Testing the Spectral Radius Condition

In Theorem 6.1.1 the condition $\rho(L) < 1$ drives stability. In this section we develop necessary and sufficient conditions for $\rho(L) < 1$ to hold.

### 6.1.2.1 Spectral Radii via Expectations

First we develop an alternative representation of the spectral radius based on expectations. The result below is proved via a local spectral radius argument (see page 71). In the statement, $\beta_t$ is as defined in (6.2) and $L$ is the operator in (6.4).

**Lemma 6.1.2.** *Let* $(X_t)$ *be P-Markov starting at x. The spectral radius of L obeys*

$$\rho(L) = \lim_{t \to \infty} \ell_t^{1/t} \quad when \quad \ell_t := \max_{x \in \mathsf{X}} \mathbb{E}_x \prod_{i=0}^{t} \beta_i. \tag{6.10}$$

---

[1] We are assuming that randomness in interest rates is a function of the same Markov state that influences profits. There is very little loss of generality in making this assumption. In fact, the two processes can still be statistically independent. For example, if we take $X_t$ to have the form $X_t = (Y_t, Z_t)$, where $(Y_t)$ and $(Z_t)$ are independent Markov chains, then we can take $\beta_t$ to be a function of $Y_t$ and $\pi_t$ to be a function of $Z_t$. The resulting interest and profit processes are statistically independent.

*Moreover, $\rho(L) < 1$ if and only if there exists a $t \in \mathbb{N}$ such that $\ell_t < 1$.*

*Proof.* Let $\mathbb{1}$ be an $n$-vector of ones. In view of (6.6), for fixed $t \in \mathbb{N}$, we have

$$\ell_t^{1/t} = \left( \max_{x \in \mathsf{X}} (L^t \mathbb{1})(x) \right)^{1/t} = \| L^t \mathbb{1} \|_\infty^{1/t}.$$

Since $\mathbb{1} \gg 0$, Lemma 2.3.3 yields (6.10). For a proof of the second claim in Lemma 6.1.2, see Proposition 4.1 of Stachurski and Zhang (2021). □

The expression in (6.10) connects the spectral radius with the long run properties of the discount factor process. The connection becomes even simpler when $P$ is irreducible, as the next exercise asks you to show.

**EXERCISE 6.1.3.** Let $P$ be irreducible. Show that, when $(X_t)$ is $P$-Markov with $X_0$ drawn from the unique stationary distribution $\psi^*$ of $P$, we also have

$$\rho(L) = \lim_{t \to \infty} \left( \mathbb{E} \prod_{i=0}^{t} \beta_i \right)^{1/t}. \tag{6.11}$$

(Hint: Try replacing $\| \cdot \|_\infty$ in the proof of Lemma 2.3.3 with $\| h \|_* := \sum_x |h(x)| \psi^*(x)$. We showed that $\| \cdot \|_*$ is a norm on $\mathbb{R}^\mathsf{X}$ in Exercise 1.2.3 on page 14.)

Exercise 6.1.3 shows that the spectral radius is a long run (geometric) average of the discount factor process. For the conclusions of Theorem 6.1.1 to hold, we need this long run average to be less than unity.

Figure 6.3 illustrates the condition $\rho(L) < 1$ when $\beta_t = X_t$ and $P$ is a Markov matrix produced by discretization of the AR1 process

$$X_{t+1} = \mu(1-a) + aX_t + s(1-a^2)^{1/2}\varepsilon_{t+1} \qquad (\varepsilon_t) \overset{\text{IID}}{\sim} N(0,1). \tag{6.12}$$

The discussion in §3.1.3 tells us that the stationary distribution $\psi^*$ of (6.12) is normally distributed with mean $\mu$ and standard deviation $s$. The parameter $a$ controls autocorrelation. In the figure we set $\mu$ to 0.96, which, since $\beta_t = X_t$, is the stationary mean of the discount factor process. The parameters $a$ and $s$ are varied in the figure, and the contour plot shows the corresponding value of $\rho(L)$. The process (6.12) is discretized via the Tauchen method with the size of the state space set to 6 (which avoids negative values for $\beta(x)$).

The figure shows that $\rho(L)$ tends to increase with both the volatility and the autocorrelation of the state process. This seems natural given the expression on the right hand side of (6.11), since sequences of large values of $\beta_i$ compound in the product $\prod_{i=0}^{t} \beta_i$, pushing up the long run average value, and such sequences occur more often when autocorrelation and volatility are large.

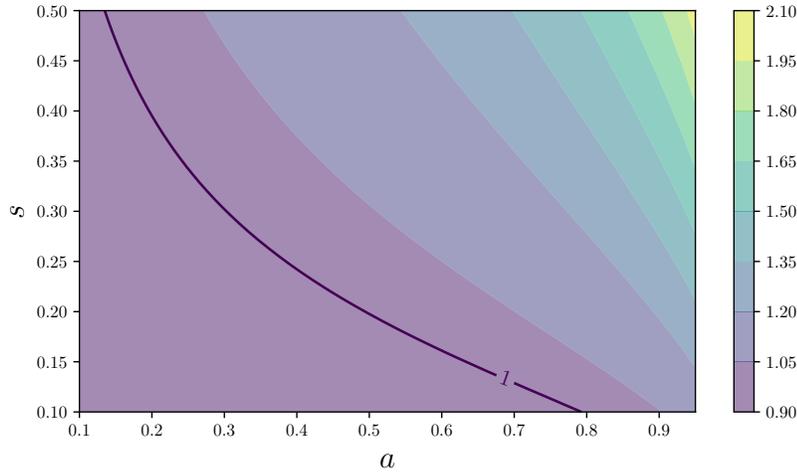

Figure 6.3: $\rho(L)$ for different values of $(a, s)$ (`discount_spec_rad.jl`)

We finish this section with a lemma that simplifies computation of the spectral radius in settings where the process $(\beta_t)$ depends only on a subset of the state variables – a setting that is common in applications. In the statement of the lemma, the state space X takes the form $\mathsf{X} = \mathsf{Y} \times \mathsf{Z}$. We fix $Q \in \mathcal{M}(\mathbb{R}^{\mathsf{Z}})$ and $R \in \mathcal{M}(\mathbb{R}^{\mathsf{Y}})$. The discount operator $L$ is

$$L(x, x') = b(z, z')Q(z, z')R(y, y') \quad \text{with} \quad b \colon \mathsf{Z} \times \mathsf{Z} \to \mathbb{R}_+.$$

Let $(Z_t)$ and $(Y_t)$ be $Q$-Markov and $R$-Markov respectively, so that, with $P$ as the pointwise product of $Q$ and $R$, the process $(X_t) \coloneqq ((Z_t, Y_t))$ is $P$-Markov. We set $L_{\mathsf{Z}}(z, z') \coloneqq b(z, z')Q(z, z')$.

**Lemma 6.1.3.** *The operators $L$ and $L_{\mathsf{Z}}$ obey $\rho(L_{\mathsf{Z}}) = \rho(L)$, where the first spectral radius is taken in $\mathcal{L}(\mathbb{R}^{\mathsf{X}})$ and the second is taken in $\mathcal{L}(\mathbb{R}^{\mathsf{Z}})$.*

*Proof.* Let $\beta_t = b(Z_t, Z_{t+1})$. Since $(\beta_t)$ depends only on $(Z_t)$ and, in addition, $(Y_t)$ and $(Z_t)$ are independent, for $x = (y, z) \in \mathsf{X}$ we have $\mathbb{E}_x \prod_{i=0}^{t} \beta_i = \mathbb{E}_{(y,z)} \prod_{i=0}^{t} \beta_i = \mathbb{E}_z \prod_{i=0}^{t} \beta_i$. Hence

$$\left( \max_{x \in \mathsf{X}} \mathbb{E}_x \prod_{i=0}^{t} \beta_i \right)^{1/t} = \left( \max_{z \in \mathsf{Z}} \mathbb{E}_z \prod_{i=0}^{t} \beta_i \right)^{1/t}.$$

Taking the limit and using Lemma 6.1.2 gives $\rho(L) = \rho(L_Z)$, where the first spectral radius is taken in $\mathcal{L}(\mathbb{R}^X)$ and the second is taken in $\mathcal{L}(\mathbb{R}^Z)$. □

### 6.1.2.2 Necessary Conditions

In §6.1.1 we studied settings where lifetime value is a function $v$ on a state space $X$ that satisfies an equation of the form $v = h + Lv$. The unknown is $v \in \mathbb{R}^X$ where $X$ is a finite set, $h \in \mathbb{R}^X$ is given and $L$ is a linear operator from $\mathbb{R}^X$ to itself. We discussed the fact that $\rho(L) < 1$ is sufficient for $v = h + Lv$ to have a unique solution.

In some settings the condition $\rho(L) < 1$ is also necessary. For example, let

- $V = (0, \infty)^X$
- $L$ be a positive linear operator on $\mathbb{R}^X$.

In this setting we have the following result:

**Lemma 6.1.4.** *If $h \in V$, then the next two statements are equivalent.*

(i) $\rho(L) < 1$.

(ii) *The equation $v = h + Lv$ has a unique solution in $V$.*

*Proof.* Regarding (i) $\implies$ (ii), existence of a unique $v \in \mathbb{R}^X$ satisfying $v = h + Lv$ follows from the Neumann series lemma. Since $v = \sum_{t \geq 0} L^t h \geq h \gg 0$, we have $v \in V$.

For (ii) $\implies$ (i), let $v$ be any solution to $v = h + Lv$ in $V$. By the Perron–Frobenius theorem (page 70), we can select a left eigenvector $e$ such that $e \geq 0$ and $e^\top L = \rho(L)e^\top$. For this $e$, we have $e^\top v = e^\top Lv + e^\top h = \rho(L)e^\top v + e^\top h$. Since $e \geq 0$, $e \neq 0$ and $v, h \gg 0$, it must be that $e^\top h > 0$ and $e^\top v > 0$. Therefore $\rho(L)$ satisfies $(1 - \rho(L))\alpha = \beta$ for $\alpha, \beta > 0$. Hence $\rho(L) < 1$. □

In §7.1.3 we will extend lemma 6.1.4 to handle certain nonlinear equations.

## 6.1.3 Fixed Point Results

State-dependent discounting breaks contractivity properties that we exploited in Chapter 5, when we studied optimality of MDPs (see, e.g., the proof of Proposition 5.1.1). Here we introduce a generalization of Banach's fixed point theorem that can deliver global stability under weaker conditions. For the remainder of this section, $X$ is any finite set.

### 6.1.3.1 Long Run Contractions

Fix $U \subset \mathbb{R}^\mathsf{X}$. We call a self-map $T$ on $U$ **eventually contracting** if there exists a $k \in \mathbb{N}$ and a norm $\| \cdot \|$ on $\mathbb{R}^\mathsf{X}$ such that $T^k$ is a contraction on $U$ under $\| \cdot \|$.

**Theorem 6.1.5.** *Let $U$ be a closed subset of $\mathbb{R}^\mathsf{X}$ and let $T$ be a self-map on $U$. If $T$ is eventually contracting on $U$, then $T$ is globally stable on $U$.*

EXERCISE 6.1.4. Prove Theorem 6.1.5. [Hint: Theorem 1.2.3 is self-improving, in the sense that it implies this seemingly stronger result.]

The next example illustrates Theorem 6.1.5 by proving a result similar to Exercise 1.2.17 on page 22.

**Example 6.1.2.** If $Tu = Au + b$ for some $b \in \mathbb{R}^\mathsf{X}$ and $A \in \mathcal{L}(\mathbb{R}^\mathsf{X})$ with $\rho(A) < 1$, then, under the Euclidean norm,

$$\|T^k u - T^k v\| = \|A^k u - A^k v\| = \|A^k(u - v)\| \leqslant \|A^k\|\|u - v\|,$$

where the last line is by the submuliplicative property of the operator norm (page 16). Since $\rho(A) < 1$, we can choose a $k \in \mathbb{N}$ such that $\|A^k\| < 1$ (see Exercise 1.2.11). Hence $T$ is eventually contracting and Theorem 6.1.5 yields global stability. The unique fixed point satisfies $u = Au + b$ and, since $\rho(A) < 1$, we can use the Neumann series lemma (page 18) to write it as $u = (I - A)^{-1}b$.

Example 6.1.2 illustrates the connection between Theorem 6.1.5 and the Neumann series lemma. Theorem 6.1.5 is more general because it can be applied in nonlinear settings. But the Neumann series lemma remains imporant because, when applicable, it provides inverse and power series representations of the fixed point.

On one hand, if $T$ is a contraction map on $U \subset \mathbb{R}^\mathsf{X}$ with respect to a given norm $\| \cdot \|_a$, we cannot necessarily say that $T$ is a contraction with respect to some other norm $\| \cdot \|_b$ on $\mathbb{R}^\mathsf{X}$. On the other hand, if $T$ is an eventual contraction on $U$ with respect to some given norm on $\mathbb{R}^\mathsf{X}$, then $T$ is eventually contracting with respect to every norm on $\mathbb{R}^\mathsf{X}$. The next exercise asks you to verify this.

EXERCISE 6.1.5. Let $\| \cdot \|_a$ and $\| \cdot \|_b$ be norms on $\mathbb{R}^\mathsf{X}$ and let $T$ be a self-map on $U \subset \mathbb{R}^\mathsf{X}$ such that $T^k$ is a contraction on $U$ with respect to $\| \cdot \|_a$ for some $k \in \mathbb{N}$. Prove that there exists an $\ell \in \mathbb{N}$ such that $T^\ell$ is a contraction on $U$ with respect to $\| \cdot \|_b$.

### 6.1.3.2 A Spectral Radius Condition

The following sufficient condition for eventual contractivity will be helpful when we study dynamic programs with state-dependent discounting.

**Proposition 6.1.6.** *Let $T$ be a self-map on $U \subset \mathbb{R}^{\mathsf{X}}$. If there exists a positive linear operator $L$ on $\mathbb{R}^{\mathsf{X}}$ such that $\rho(L) < 1$ and*

$$|Tv - Tw| \leqslant L|v - w| \tag{6.13}$$

*for all $v, w \in U$, then $T$ is an eventual contraction on $U$.*

*Proof.* Fix $v, w \in U$. Pick any $k \in \mathbb{N}$. We have $|T^k v - T^k w| \leqslant L|T^{k-1}v - T^{k-1}w|$, or

$$e_k \leqslant L e_{k-1} \quad \text{where} \quad e_k := |T^k v - T^k w|. \tag{6.14}$$

Since $L$ is positive, $L$ is order-preserving on $U$ by Exercise 2.3.11. As a result, we can iterate on (6.14) to obtain $e_k \leqslant L^k e_0$, or

$$|T^k v - T^k w| \leqslant L^k |v - w|.$$

Let $\|\cdot\|$ be the Euclidean norm. Since $0 \leqslant a \leqslant b$ implies $\|a\| \leqslant \|b\|$, we get

$$\|T^k v - T^k w\| \leqslant \|L^k|v - w|\| \leqslant \|L^k\| \|v - w\|,$$

where $\|L^k\|$ is the operator norm (see §1.2.1.4). Since $\rho(L) < 1$, we have $\|L^k\| \to 0$ as $k \to \infty$. (Exercise 1.2.11 on page 19) Hence $T$ is eventually contracting on $U$. □

### 6.1.3.3 A Generalized Blackwell Condition

In §2.2.3.4 we studied a sufficient condition for order-preserving self maps to be contractions. The next proposition provides an analogous result for eventual contractions. In the statement of the proposition, $U$ is a subset of $\mathbb{R}^{\mathsf{X}}$ such that $v, c \in U$ and $c \geqslant 0$ implies $v + c \in U$.

**Proposition 6.1.7.** *Let $T$ be an order-preserving self-map on $U$. If there exists a positive linear operator $L$ on $\mathbb{R}^{\mathsf{X}}$ such that $\rho(L) < 1$ and*

$$T(v + c) \leqslant Tv + Lc \quad \text{for all } c, v \in \mathbb{R}^{\mathsf{X}} \text{ with } c \geqslant 0,$$

*then $T$ is eventually contracting on $U$.*

*Proof.* Fix $v, w \in U$ and let $T$ and $L$ be as in the statement of the proposition. By the assumed properties on $T$, we have

$$Tv = T(v + w - w) \leqslant T(w + |v - w|) \leqslant Tw + L|v - w|.$$

Rearranging gives $Tv - Tw \leqslant L|v - w|$. Reversing the roles of $v$ and $w$ yields $|Tv - Tw| \leqslant L|v - w|$. The claim in Proposition 6.1.7 now follows from Proposition 6.1.6. $\qquad\square$

## 6.2 Optimality with State-Dependent Discounting

We can now turn to dynamic programs in which the objective is to maximize a lifetime value in the presence of state-dependent discounting. First we present an extension of the MDP model from Chapter 5 that admits state-dependent discounting. Then we provide weak conditions under which optimal policies exist and Bellman's principle of optimality holds.

### 6.2.1 MDPs with State-Dependent Discounting

We are ready to extend the MDP model to include state-dependent discounting. We construct a framework and then provide weak conditions for optimality based on spectral radius methods discussed above.

#### 6.2.1.1 Setup

To provide a framework for dynamic programs with state-dependent discounting, we begin with an MDP $(\Gamma, \beta, r, P)$ with state space $\mathsf{X}$, action space $\mathsf{A}$ and feasible state-action pairs $\mathsf{G}$. We then replace the constant discount factor $\beta$ with a function $\beta$ from $\mathsf{G} \times \mathsf{X}$ to $\mathbb{R}_+$. We call the resulting model an **MDP with state-dependent discounting**. The **Bellman equation** takes the form

$$v(x) = \max_{a \in \Gamma(x)} \left\{ r(x, a) + \sum_{x'} v(x')\beta(x, a, x')P(x, a, x') \right\} \tag{6.15}$$

where $x \in \mathsf{X}$ and $v \in \mathbb{R}^{\mathsf{X}}$. Notice that the discount factor depends on all relevant information: the current action, the current state and the stochastically determined next period state.

For MDPs with state-dependent discounting, we can obtain standard optimality results by assuming a that there exists a $b < 1$ such that $\beta(x, a, x') \leqslant b$ for all $(x, a, x') \in$ G × X. In this setting it is easy to show that lifetime values are finite, and to extend the optimality results for regular MDPs found in Proposition 5.1.1 on page 139.

Unfortunately, the assumption discussed in the previous paragraph is too strict for many applications. (We return to this point in §6.2.1.6.) We will state an optimality result under weaker conditions.

### 6.2.1.2   Finite Lifetime Values

Let $\Sigma$ be the set of all feasible policies, defined as for regular MDPs. The **policy operator** $T_\sigma$ corresponding to $\sigma \in \Sigma$ is represented by

$$(T_\sigma v)(x) = r(x, \sigma(x)) + \sum_{x'} v(x')\beta(x, \sigma(x), x')P(x, \sigma(x), x'). \qquad (6.16)$$

Following Chapter 5, we set $r_\sigma(x) := r(x, \sigma(x))$. We define $L_\sigma \in \mathcal{L}(\mathbb{R}^{\mathsf{X}})$ via

$$L_\sigma(x, x') := \beta(x, \sigma(x), x')P(x, \sigma(x), x'). \qquad (6.17)$$

Notice that we can now write (6.16) as $T_\sigma v = r_\sigma + L_\sigma v$. In line with our discussion of MDPs in Chapter 5, when $T_\sigma$ has a unique fixed point we denote it by $v_\sigma$ and interpret it as lifetime value.

**Assumption 6.2.1** (SD). For all $\sigma \in \Sigma$ we have $\rho(L_\sigma) < 1$.

**Lemma 6.2.1.** *If Assumption 6.2.1 holds, then, for each $\sigma \in \Sigma$, the linear operator $I - L_\sigma$ is invertible and, in $\mathbb{R}^{\mathsf{X}}$, the policy operator $T_\sigma$ has a unique fixed point*

$$v_\sigma = (I - L_\sigma)^{-1}r_\sigma. \qquad (6.18)$$

*Proof.* Fix $\sigma \in \Sigma$. By the Neumann series lemma, $I - L_\sigma$ is invertible. Any fixed point of $T_\sigma$ obeys $v = r_\sigma + L_\sigma v$, which, given invertibility of $I - L_\sigma$, is equivalent to (6.18). □

As discussed, the value $v_\sigma(x)$ has the interpretation of lifetime value of policy $\sigma$ conditional on initial state $x$. We can reinforce this interpretation by connecting Lemma 6.2.1 to Theorem 6.1.1 on page 184. The next exercise asks you to work through all the steps.

EXERCISE 6.2.1. Fix $\sigma \in \Sigma$, set $\beta_t := \beta(X_{t-1}, \sigma(X_{t-1}), X_t)$ for $t \geqslant 1$ and $\beta_0 := 1$. Let $(X_t)$ be $P_\sigma$-Markov with initial condition $x$. (As before, $P_\sigma(x, x') := P(x, \sigma(x), x')$.) Prove that, under Assumption 6.2.1, the function $v_\sigma$ obeys

$$v_\sigma(x) = \mathbb{E}_x \sum_{t=0}^{\infty} \left[ \prod_{i=0}^{t} \beta_i \right] r_\sigma(X_t) \qquad (x \in \mathsf{X}). \tag{6.19}$$

EXERCISE 6.2.2. Show that, under Assumption 6.2.1, the operator $T_\sigma$ is globally stable on $\mathbb{R}^\mathsf{X}$.

EXERCISE 6.2.3. Show that Assumption 6.2.1 holds whenever there exists an $L \in \mathcal{L}(\mathbb{R}^\mathsf{X})$ such that $\rho(L) < 1$ and

$$\beta(x, a, x')P(x, a, x') \leqslant L(x, x') \quad \text{for all } (x, a) \in \mathsf{G} \text{ and } x' \in \mathsf{X}. \tag{6.20}$$

### 6.2.1.3 Optimality

The **Bellman operator** takes the form

$$(Tv)(x) = \max_{a \in \Gamma(x)} \left\{ r(x, a) + \sum_{x'} v(x')\beta(x, a, x')P(x, a, x') \right\} \tag{6.21}$$

where $x \in \mathsf{X}$ and $v \in \mathbb{R}^\mathsf{X}$.

Given $v \in \mathbb{R}^\mathsf{X}$, a policy $\sigma$ is called $v$-**greedy** if $\sigma(x)$ is a maximizer of the right-hand side of (6.21) for all $x$ in $\mathsf{X}$. Equivalently, $\sigma$ is $v$-greedy whenever $T_\sigma v = Tv$.

When Assumption 6.2.1 holds and, as a result, $T_\sigma$ has a unique fixed point $v_\sigma$ for each $\sigma \in \Sigma$, we let $v^*$ denote the **value function**, which is defined as $v^* := \vee_{\sigma \in \Sigma} v_\sigma$. As for the regular MDP case, a policy $\sigma$ is called **optimal** if $v_\sigma = v^*$.

We can now state our main optimality result for MDPs with state-dependent discounting.

**Proposition 6.2.2.** *If Assumption 6.2.1 holds, then*

(i) *the value function $v^*$ is the unique solution to the Bellman equation in $\mathbb{R}^\mathsf{X}$,*

(ii) *a policy $\sigma \in \Sigma$ is optimal if and only if it is $v^*$-greedy, and*

(iii) *at least one optimal policy exists.*

In §8.2.2 we prove a result that includes Proposition 6.2.2 as a special case.

### 6.2.1.4 Algorithms

Algorithms for solving an MDP with state-dependent discounting include value function iteration (VFI), Howard policy iteration (HPI), and optimistic policy iteration (OPI). The algorithms for VFI and OPI are identical to those given for regular MDPs (see §5.1.4), provided that the correct operators $T$ and $T_\sigma$ are used, and that the definition of a $v$-greedy policy is as given in §6.2.1.1. The algorithm for HPI is almost identical, with the only change being that computation of lifetime values involves $L_\sigma$. Details are given in Algorithm 6.1.

---

**Algorithm 6.1:** HPI for MDPs with state-dependent discounting

---
**1** input $\sigma \in \Sigma$
**2** $v_0 \leftarrow v_\sigma$ and $k \leftarrow 0$
**3 repeat**
**4** $\quad$ $\sigma_k \leftarrow$ a $v_k$-greedy policy
**5** $\quad$ $v_{k+1} \leftarrow (I - L_{\sigma_k})^{-1} r_{\sigma_k}$
**6** $\quad$ **if** $v_{k+1} = v_k$ **then** break
**7** $\quad$ $k \leftarrow k + 1$
**8 return** $\sigma_k$

---

We prove in Chapter 8 that, under the conditions of Assumption 6.2.1, VFI, OPI and HPI are all convergent, and that HPI converges to an exact optimal policy in a finite number of steps.

### 6.2.1.5 Exogenous Discounting

Some applications use an exogenous state component to drive a discount factor process. In this section we set up such a model and obtain optimality conditions by applying Proposition 6.2.2.

The first step is to decompose the state $X_t$ into a pair $(Y_t, Z_t)$, where $(Y_t)_{t \geqslant 0}$ is endogenous (i.e., affected by the actions of the controller) and $(Z_t)_{t \geqslant 0}$ is purely exogenous. In particular, the primitives consist of

(i) a nonempty correspondence $\Gamma$ from $\mathsf{Y} \times \mathsf{Z}$ to $\mathsf{A}$,

(ii) a function $\beta$ from $\mathsf{Z}$ to $\mathbb{R}_+$,

(iii) a function $r$ from $\mathsf{G} := \{(y, a) \in \mathsf{Y} \times \mathsf{A} : a \in \Gamma(y)\}$ to $\mathbb{R}$,

(iv) a stochastic matrix $Q$ on $\mathsf{Z}$ and

(v) a stochastic kernel $R$ from $\mathsf{G}$ to $\mathsf{Y}$.

The corresponding Bellman equation is

$$v(y, z) = \max_{a \in \Gamma(y,z)} \left\{ r(y, a) + \beta(z) \sum_{z', y'} v(y', z') Q(z, z') R(y, a, y') \right\} \tag{6.22}$$

for all $(y, z) \in \mathsf{X}$. Given $v \in \mathbb{R}^\mathsf{X}$, a policy $\sigma \in \Sigma$ is called $v$-**greedy** if

$$\sigma(y, z) \in \underset{a \in \Gamma(y,z)}{\operatorname{argmax}} \left\{ r(y, a) + \beta(z) \sum_{z', y'} v(y', z') Q(z, z') R(y, a, y') \right\} \tag{6.23}$$

for all $(y, z) \in \mathsf{X}$.

This exogenous discount model is a special case of the general MDP with state-dependent discounting. Indeed, we can write (6.22) as (6.21) by setting $x := (y, z)$ and defining

$$P(x, a, x') := P((y, z), a, (y', z')) := Q(z, z') R(y, a, y').$$

The following proposition provides a relatively simple sufficient condition for the core optimality results in the setting of the exogenous discount model.

**Proposition 6.2.3.** *Let $L$ be the operator in $\mathcal{L}(\mathbb{R}^\mathsf{Z})$ defined by $L(z, z') := \beta(z) Q(z, z')$. If $\rho(L) < 1$, then all of the optimality results in Proposition 6.2.2 hold.*

EXERCISE 6.2.4. Prove Proposition 6.2.3. (Hint: Use Lemma 6.1.3.)

### 6.2.1.6 Comments on the Spectral Radius Condition

In §6.2.1.2 we mentioned that requiring $\sup \beta < 1$ is too strict for some applications. For example, the real interest rate $r_t$ shown in Figure 6.2 is sometimes negative. Using long historical records, Farmer et al. (2023) find that the discount rate is negative around 1/3 of the time. This means that the associated discount factor $\beta_t = 1/(1 + r_t)$ is sometimes greater than 1 and $\sup \beta < 1$ fails.

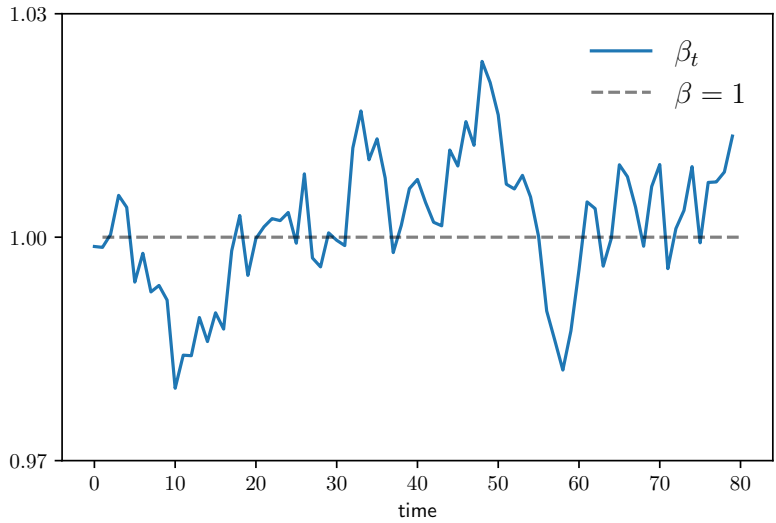

Figure 6.4: Discount factor process $(\beta)_{t \geqslant 0}$ in Hills et al. (2019).

In macroeconomics, empirically motivated time-varying discount factor specifications lead to models where $\beta_t > 1$ occurs with positive probability. For example, Hills et al. (2019) study a model that can be embedded in the MDP framework just described. Figure 6.4 shows a simulation of one of the discount factor processes used in their model, prior to discretization. The exogenous state and discount factor process takes the form $\beta_t = bZ_t$, where $(Z_t)$ is an exogenous state obeying $Z_{t+1} = 1 - \rho + \rho Z_t + \sigma \varepsilon_{t+1}$ with $(\varepsilon_t)$ IID and standard normal. Clearly $\sup \beta < 1$ fails for this model too.

Let's now consider the weaker condition $\rho(L) < 1$ described in Proposition 6.2.3 and check whether it holds. Following Hills et al. (2019), we discretize the dynamics of $(Z_t)$ via a Tauchen approximation, producing a stochastic matrix $Q$ on a finite set Z.[2] The set of values for $\beta_t$ ranges between $0.95$ and $1.04$, so that $\beta_t > 1$ remains possible. Nonetheless, with $L(z, z') = \beta(z)Q(z, z')$ we obtain $\rho(L) = 0.9996$. Hence Proposition 6.2.3 applies.

## 6.2.2 Inventory Management Revisited

In this section, we modify the inventory management model from §5.2.1 to include time-varying interest rates.

---

[2]The parameters are $\rho = 0.85$, $\sigma = 0.0062$, and $b = 0.99875$. In line with Hills et al. (2019), we discretize the model via `mc = tauchen(n, ρ, σ, 1 - ρ, m)` with $m = 4.5$ and $n = 15$.

Recall that, in the model of §5.2.1, the Bellman equation takes the form

$$v(x) = \max_{a \in \Gamma(x)} \left\{ r(x, a) + \beta \sum_{d \geqslant 0} v(f(x, a, d)) \varphi(d) \right\} \tag{6.24}$$

at each $x \in \mathsf{X}$, where $\mathsf{X} := \{0, \ldots, K\}$, $x$ is the current inventory level, $a$ is the current inventory order, $r(x, a)$ is current profits (defined in (5.8)), $f(x, a, d) := (x - d) \vee 0 + a$ and $d$ is an IID demand shock with distribution $\varphi$. Let's now add a time-varying discount rate and investigate its impact on optimal choices.

We add time-varying discounting by replacing the constant $\beta$ in (6.24) with a stochastic process $(\beta_t)$ where $\beta_t = 1/(1 + r_t)$. We suppose that the dynamics can be expressed as $\beta_t = \beta(Z_t)$, where the exogenous process $(Z_t)_{t \geqslant 0}$ is $Q$-Markov on $\mathsf{Z}$. After relabeling the endogenous state $X_t$ as $Y_t$ and $x$ as $y$, in line with the notation in §6.2.1.5, the Bellman equation becomes $v(y, z) = \max_{a \in \Gamma(y, z)} B((y, z), a, v)$ where

$$B((y, z), a, v) = r(y, a) + \beta(z) \sum_{d, z'} v(f(y, a, d), z') \varphi(d) Q(z, z'). \tag{6.25}$$

If we set

$$R(y, a, y') := \mathbb{P}\{f(y, a, d) = y'\} \quad \text{when} \quad D \sim \varphi,$$

then $R(y, a, y')$ is the probability of realizing next period inventory level $y'$ when the current level is $y$ and the action is $a$. Hence we can rewrite (6.25) as

$$B((y, z), a, v) = r(y, a) + \beta(z) \sum_{y', z'} v(y', z') Q(z, z') R(y, a, y'). \tag{6.26}$$

We have now created a version of the MDP with exogenous state-dependent discounting described in §6.2.1.5. Letting $L(z, z') := \beta(z) Q(z, z')$ and applying Proposition 6.2.3, we see that all of the standard optimality results hold whenever $\rho(L) < 1$.

Figure 6.5 shows how inventory evolves under an optimal program when the parameters of the problem are as given in Listing 21. (The code preallocates and computes arrays representing $r$, $R$ and $Q$ in (6.26) and includes a test for $\rho(L) < 1$.) We set $\beta(z) = z$ and take $(Z_t)$ to be a discretization of an AR(1) process. Figure 6.5 was created by simulating $(Z_t)$ according to $Q$ and inventory $(Y_t)$ according to $Y_{t+1} = (Y_t - D_{t+1}) \vee 0 + A_t$, where $A_t$ follows the optimal policy.

The outcome is similar to Figure 5.7, in the sense that inventory falls slowly and then jumps up. As before, fixed costs induce this lumpy behavior. However, a new phenomenon is now present: inventories trend up when interest rates fall and down when they rise. (The interest rate $r_t$ is calculated via $\beta_t = 1/(1 + r_t)$ at each $t$.) High interest rates foreshadow high interest rates due to positive autocorrelation ($\rho > 0$), which in turn devalue future profits and hence encourage managers to economize on stock.

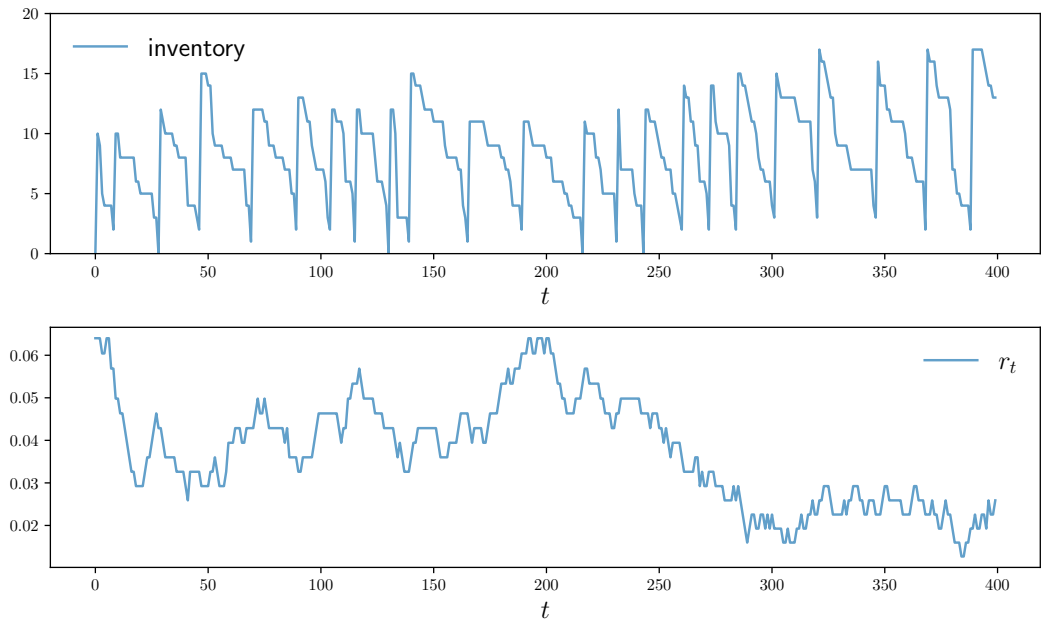

Figure 6.5: Inventory dynamics with time-varying interest rates

Figure 6.6 shows execution time for VFI and OPI at different choices of $m$ (see §6.2.1.3 for the interpretation of $m$). As for the optimal savings problem we studied in Chapter 5, OPI is around 1 order of magnitude faster when $m$ is close to 50 (cf. Figure 5.8 on page 154).

## 6.3  Asset Pricing

This section provides a brief introduction to asset pricing in a Markov environment. While the topic of asset pricing is fascinating in its own right, our main aim is to provide additional practice in handling linear valuation problems. (Readers who wish to push ahead with their study of dynamic programming can safely skip to Chapter 7.)

```julia
using LinearAlgebra, Random, Distributions, QuantEcon

f(y, a, d) = max(y - d, 0) + a   # Inventory update

function create_sdd_inventory_model(;
                ρ=0.98, ν=0.002, n_z=20, b=0.97,   # Z state parameters
                K=40, c=0.2, κ=0.8, p=0.6,          # firm and demand parameters
                d_max=100)                          # truncation of demand shock

    ϕ(d) = (1 - p)^d * p                            # demand distribution
    d_vals = collect(0:d_max)
    ϕ_vals = ϕ.(d_vals)
    y_vals = collect(0:K)                           # inventory levels
    n_y = length(y_vals)
    mc = tauchen(n_z, ρ, ν)
    z_vals, Q = mc.state_values .+ b, mc.p
    ρL = maximum(abs.(eigvals(z_vals .* Q)))
    @assert  ρL < 1 "Error: ρ(L) ≥ 1."

    R = zeros(n_y, n_y, n_y)
    for (i_y, y) in enumerate(y_vals)
        for (i_y′, y′) in enumerate(y_vals)
            for (i_a, a) in enumerate(0:(K - y))
                hits = f.(y, a, d_vals) .== y′
                R[i_y, i_a, i_y′] = dot(hits, ϕ_vals)
            end
        end
    end

    r = fill(-Inf, n_y, n_y)
    for (i_y, y) in enumerate(y_vals)
        for (i_a, a) in enumerate(0:(K - y))
                cost = c * a + κ * (a > 0)
                r[i_y, i_a] = dot(min.(y, d_vals), ϕ_vals) - cost
        end
    end

    return (; K, c, κ, p, r, R, y_vals, z_vals, Q)
end
```

Listing 21: Investment model with time-varying discounting (`inventory_sdd.jl`)

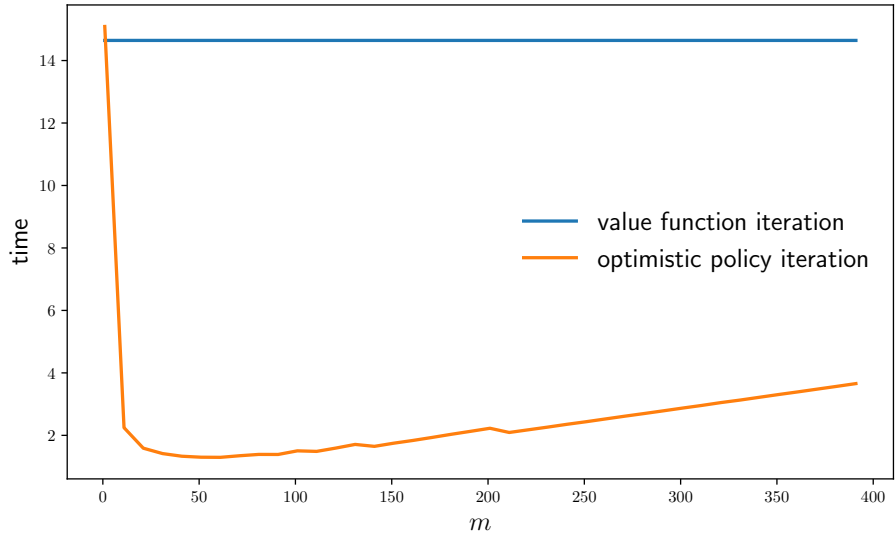

Figure 6.6: OPV vs VFI timings for the inventory problem

## 6.3.1 Introduction to Asset Pricing

We first discuss risk-neutral pricing and show why this assumption is typically implausible. Next, we introduce stochastic discount factors and stationary asset pricing.

### 6.3.1.1 Risk-Neutral Pricing?

Consider the problem of assigning a current price $\Pi_t$ to an asset that confers on its owner the right to payoff $G_{t+1}$. The payoff is stochastic and realized next period. One simple idea is to use **risk-neutral pricing,** which implies that

$$\Pi_t = \mathbb{E}_t \, \beta \, G_{t+1} \qquad (6.27)$$

for some constant discount factor $\beta \in (0,1)$. If the payoff is in $k$ periods, then we modify the price to $\mathbb{E}_t \, \beta^k \, G_{t+k}$. In essence, risk-neutral pricing says that cost equals expected reward, discounted to present value by compounding a constant rate of discount. (A *rate* of discount, say $\rho$, is linked to a discount *factor*, say $\beta$, by $\beta = 1/(1+\rho) \approx \exp(\rho)$.)

**Example 6.3.1.** Let $S_t$ be the price of a stock at each point in time $t$. A **European call option** gives its owner the right to purchase the stock at price $K$ at time $t+k$. There is no obligation to exercise the option, so the payoff at $t + k$ is $\max\{S_{t+k} - K, 0\}$. Under risk-neutral pricing, the time $t$ price of this option is

$$\Pi_t = \mathbb{E}_t\, \beta^k\, \max\{S_{t+k} - K, 0\}.$$

Although risk neutrality allows for simple pricing, assuming risk neutrality for all investors is *not* plausible.

To give one example, suppose that we take the asset that pays $G_{t+1}$ in (6.27) and replace it with another asset that pays $H_{t+1} = G_{t+1} + \varepsilon_{t+1}$, where $\varepsilon_{t+1}$ is independent of $G_{t+1}$, $\mathbb{E}_t\, \varepsilon_{t+1} = 0$ and $\operatorname{Var} \varepsilon_{t+1} > 0$. In effect, we are adding risk to the original payoff without changing its mean.

Under risk neutrality, the price of this new asset is

$$\Pi_t^H = \mathbb{E}_t\, \beta\, [G_{t+1} + \varepsilon_{t+1}] = \Pi_t + \beta\, \mathbb{E}_t\, \varepsilon_{t+1} = \Pi_t.$$

Thus, $H_{t+1}$ and $G_{t+1}$ are priced identically, even though their means are both $\mathbb{E}_t G_{t+1}$ and their variances satisfy

$$\operatorname{Var} H_{t+1} = \operatorname{Var} G_{t+1} + \operatorname{Var} \varepsilon_{t+1} > \operatorname{Var} G_{t+1}.$$

This outcome contradicts the idea that investors typically want compensation for bearing risk.

A helpful way to think about the same point is to consider the rate of return $r_{t+1} := (G_{t+1} - \Pi_t)/\Pi_t$ on holding an asset with payoff $G_{t+1}$. From (6.27) we have $\mathbb{E}_t\, \beta\,(1 + r_{t+1}) = 1$, or

$$\mathbb{E}_t\, r_{t+1} = \frac{1 - \beta}{\beta}.$$

Since the right-hand side does not depend on $G_{t+1}$, risk neutrality implies that all assets have the same expected rate of return. But this contradicts the finding that, on average, riskier assets tend to have higher rates of return that compensate investors for bearing risk.

**Example 6.3.2.** The risk premium on a given asset is defined as the expected rate of return minus the rate of return on a risk-free asset. If we assume risk neutrality then, by the preceding discussion, the risk premium is zero for all assets. However, calculations based on post-war US data show that the average return premium on equities over safe assets is around 8% per annum (see, e.g., Cochrane (2009)).

### 6.3.1.2 A Stochastic Discount Factor

To go beyond risk neutral-pricing, let's start with a model containing one asset and one agent. It is straightforward to price the asset and compare it to the risk neutral case.

A representative agent takes the price $\Pi_t$ of a risky asset as given and solves

$$\max_{0 \leqslant \alpha \leqslant 1} \{u(C_t) + \beta \mathbb{E}_t u(C_{t+1})\}$$

subject to $\quad C_t = E_t - \Pi_t \alpha \quad$ and $\quad C_{t+1} = E_{t+1} + \alpha G_{t+1}.$

Here

- $u$ is a flow utility function,

- $G_{t+1}$ is the payoff of the asset and $\Pi_t$ is the time-$t$ price,

- $\beta$ is a constant discount factor measuring impatience of the agent,

- $E_t$ and $E_{t+1}$ are endowments and

- $\alpha$ is the share of the asset purchased by the agent.

Rewriting as $\max_\alpha \{u(E_t - \Pi_t \alpha) + \beta \mathbb{E}_t u(E_{t+1} + \alpha G_{t+1})\}$ and differentiating with respect to $\alpha$ leads to the first order condition

$$u'(E_t - \Pi_t \alpha)\Pi_t = \beta \mathbb{E}_t u'(E_{t+1} + \alpha G_{t+1})G_{t+1}.$$

Rearranging gives us

$$\Pi_t = \mathbb{E}_t \left[ \beta \frac{u'(C_{t+1})}{u'(C_t)} G_{t+1} \right]. \tag{6.28}$$

Comparing (6.28) with (6.27), we see that the payoff is now multiplied by a positive random variable rather than a constant. The random variable

$$M_{t+1} := \beta \frac{u'(C_{t+1})}{u'(C_t)} \tag{6.29}$$

is called the **stochastic discount factor** or **pricing kernel**. We call this particular form of the pricing kernel shown in (6.29) **Lucas stochastic discount factor** (Lucas SDF) in honor of Lucas (1978a).

**Example 6.3.3.** If $u$ is linear, so that $u(c) = ac + b$ for some $a, b \in \mathbb{R}$, then $u'(c) = a$ for all $c$, so $M_{t+1} = \beta$. If the utility function has no curvature, then pricing is risk neutral.

**Example 6.3.4.** If utility has the CRRA form $u(c) = c^{1-\gamma}/(1-\gamma)$ for some $\gamma > 0$, then the Lucas SDF takes the form

$$M_{t+1} = \beta \left( \frac{C_{t+1}}{C_t} \right)^{-\gamma}, \tag{6.30}$$

which we can also write as $M_{t+1} = \beta \exp(-\gamma g_{t+1})$ when $g_{t+1} := \ln(C_{t+1}/C_t)$ is the growth rate of consumption.

In the CRRA case, the Lucas SDF applies heavier discounting to assets that concentrate payoffs in states of the world where the agent is already enjoying strong consumption growth. Conversely, the SDF attaches higher weights to future payoffs that occur when consumption growth is low because such payoffs hedge against the risk of drawing low consumption states.

### 6.3.1.3 A General Specification

The standard neoclassical theory of asset pricing generalizes the Lucas discounting specification by assuming only that there exists a positive random variable $M_{t+1}$ such that the price of an asset with payoff $G_{t+1}$ is

$$\Pi_t = \mathbb{E}_t M_{t+1} G_{t+1} \qquad (t \geqslant 0). \tag{6.31}$$

As above, $M_{t+1}$ is called a **stochastic discount factor** (SDF). Equation 6.31 generalizes (6.28) by refraining from restricting the SDF (apart from assuming positivity).

Actually, it can be shown that there exists an SDF $M_{t+1}$ such that (6.31) is always valid under relatively weak assumptions. In particular, a single SDF $M_{t+1}$ can be used to price *any* asset in the market, so if $H_{t+1}$ is a another stochastic payoff then the current price of an asset with this payoff is $\mathbb{E}_t M_{t+1} H_{t+1}$.

We do not prove these claims, since our interest is in understanding forward-looking equations in Markov environments. Some relevant references are listed in §6.4.

### 6.3.1.4 Markov Pricing

A common assumption in quantitative applications is that all underlying randomness is driven by a Markov model. In this spirit, we take $(X_t)$ to be $P$-Markov on finite state X, where $P \in \mathcal{M}(\mathbb{R}^X)$, and suppose further that the SDF and payoff have the forms

$$M_{t+1} = m(X_t, X_{t+1}) \quad \text{and} \quad G_{t+1} = g(X_t, X_{t+1})$$

for fixed functions $m, g$ mapping $\mathsf{X} \times \mathsf{X}$ to $\mathbb{R}_+$. Since $m$ is arbitrary at this point, we don't assume a particular specification for the SDF.

In this setting, conditioning on $X_t = x$, the standard asset pricing equation $\Pi_t = \mathbb{E}_t \, M_{t+1} \, G_{t+1}$ becomes

$$\pi(x) = \sum_{x'} m(x, x') g(x, x') P(x, x') \qquad (x \in \mathsf{X}), \tag{6.32}$$

where $\pi(x)$ is the price of the asset conditional on $X_t = x$. (That is, $\Pi_t = \pi(X_t)$.)

### 6.3.1.5 Pricing a Stationary Dividend Stream

Now we are ready to look at pricing a stationary cash flow over an infinite horizon, a basic problem in asset pricing. We will apply the Markov structure assumed in §6.3.1.4. In all that follows, $(X_t)$ is $P$-Markov on $\mathsf{X}$ and $M_{t+1}$ is defined as in §6.3.1.4.

We seek the time $t$ price, denoted by $\Pi_t$, for an **ex-dividend contract** on the dividend stream $(D_t)_{t \geqslant 0}$. The contract provides the owner with the right to the dividend stream. The "ex-dividend" component means that, should the dividend stream be traded at time $t$, the dividend paid at time $t$ goes to the seller rather than the buyer. As a result, purchasing at $t$ and selling at $t + 1$ pays $\Pi_{t+1} + D_{t+1}$. Hence, applying the asset pricing rule (6.31), at time $t$ price $\Pi_t$ of the contract must satisfy

$$\Pi_t = \mathbb{E}_t \, M_{t+1} (\Pi_{t+1} + D_{t+1}). \tag{6.33}$$

We assume the existence of a $d \in \mathbb{R}_+^{\mathsf{X}}$ such that $D_t = d(X_t)$ for all $t$. Using (6.32), we can write this as

$$\pi(x) = \sum_{x'} m(x, x') (\pi(x') + d(x')) P(x, x') \qquad (x \in \mathsf{X}), \tag{6.34}$$

or, equivalently,

$$\pi = A\pi + Ad \quad \text{when } A(x, x') := m(x, x') P(x, x'). \tag{6.35}$$

By the Neumann series lemma, $\rho(A) < 1$ implies (6.35) has unique solution

$$\pi^* := (I - A)^{-1} Ad = \sum_{k=1}^{\infty} A^k d.$$

The vector $\pi^*$ is called an **equilibrium price function**.

EXERCISE 6.3.1. Show that $\rho(A) < 1$ is both necessary and sufficient for existence of a unique solution to (6.34) in $(0, \infty)^X$ whenever $m, d \gg 0$.

EXERCISE 6.3.2. As discussed in §6.3.1.1, the case $m \equiv \beta$ for some $\beta \in \mathbb{R}_+$ is called the risk-neutral case. Provide a condition on $\beta$ under which $\rho(A) < 1$.

EXERCISE 6.3.3. Confirm that $(\Pi_t)_{t \geqslant 0}$ generated by $\Pi_t = \pi^*(X_t)$ solves (6.33).

**Remark 6.3.1.** We can call $A$ an **Arrow–Debreu discount operator**. Its powers apply discounting: the valuation of any random payoff $g$ in $k$ periods is $A^k g$.

EXERCISE 6.3.4. Derive the price for a **cum-dividend contract** on the dividend stream $(D_t)_{t \geqslant 0}$, with the model otherwise unchanged. Under this contract, should the right to the dividend stream be traded at time $t$, the dividend paid at time $t$ goes to the buyer rather than the seller.

### 6.3.1.6 Forward Sum Representation

Asset prices can be expressed as infinite sums under the assumptions stated above. Let's show this for cum-dividend contracts (although the case of ex-dividend contracts is similar). In Exercise 6.3.4 you found that the state-contingent price vector $\pi$ for a cum-dividend contract on the dividend stream $(D_t)_{t \geqslant 0}$ obeys

$$\pi = d + A\pi \quad \text{when } A(x, x') := m(x, x')P(x, x') \tag{6.37}$$

and $\rho(A) < 1$. As before, $D_t = d(X_t)$ and $(X_t)_{t \geqslant 0}$ is $P$-Markov on X. Applying the uniqueness component of the Neumann series lemma (page 18) and Theorem 6.1.1, we see that the function $\pi$ also obeys

$$\pi(x) = \mathbb{E}_x \sum_{t=0}^{\infty} \left[ \prod_{i=0}^{t} M_i \right] D_t \qquad (x \in X),$$

where $M_{t+1} := m(X_t, X_{t+1})$ for $t \geqslant 0$ and $M_0 := 1$. This expression agrees with our intuition: The price of the contract is the expected present value of the dividend stream, with the time $t$ dividend discounted by the composite factor $M_1 \cdots M_t$.

## 6.3.2 Nonstationary Dividends

Until now, our discussion of asset pricing has assumed that dividends are stationary. However, dividends typically grow over time, along with other economic measures such as GDP. In this section we solve for the price of a dividend stream when dividends exhibit random growth.

### 6.3.2.1 Price-Dividend Ratios

A standard model of dividend growth is

$$\ln \frac{D_{t+1}}{D_t} = \kappa(X_t, \eta_{t+1}) \qquad t = 0, 1, \dots,$$

where $\kappa$ is a fixed function, $(X_t)$ is the state process and $(\eta_t)$ is IID. We let $\varphi$ be the density of each $\eta_t$ and assume that $(X_t)$ is $P$-Markov on a finite set X. Let's suppose as before that the SDF obeys $M_{t+1} = m(X_t, X_{t+1})$ for some positive function $m$.

Since dividends grow over time, so will the price of the asset. As such, we should no longer seek a fixed function $\pi$ such that $\Pi_t = \pi(X_t)$ for all $t$, since the resulting price process $(\Pi_t)$ will fail to grow. Instead, we try to solve for the **price-dividend ratio** $V_t := \Pi_t/D_t$, which we hope will be stationary.

EXERCISE 6.3.5. Using $\Pi_t = \mathbb{E}_t [M_{t+1}(D_{t+1} + \Pi_{t+1})]$, show that

$$V_t = \mathbb{E}_t \left[ M_{t+1} \exp(\kappa(X_t, \eta_{t+1})) \left(1 + V_{t+1}\right) \right]. \tag{6.38}$$

After conditioning on $X_t = x$, (6.38) leads us to conjecture existence of a function $v$ such that

$$v(x) = \sum_{x'} m(x, x') \int \exp(\kappa(x, \eta)) \varphi(\mathrm{d}\eta) \left[1 + v(x')\right] P(x, x') \tag{6.39}$$

for all $x \in$ X. We understand (6.39) as an equation to be solved for the unknown object $v \in \mathbb{R}^{\mathsf{X}}$. If we can find a solution $v^*$ to (6.39), then setting $V_t = v^*(X_t)$ yields a process $(V_t)$ that obeys (6.38).

EXERCISE 6.3.6. Let

$$A(x, x') := m(x, x') \int \exp(\kappa(x, \eta)) \varphi(\mathrm{d}\eta) P(x, x') \qquad (x, x' \in \mathsf{X}). \tag{6.40}$$

Show that (6.38) has a unique solution $v^*$ in $\mathbb{R}^{\mathsf{X}}$ when $\rho(A) < 1$, and

$$v^* = (I - A)^{-1} A \mathbb{1} = \sum_{t \geqslant 1} A^t \mathbb{1}. \tag{6.41}$$

The price-dividend process $(V_t^*)$ defined by $V_t^* = v^*(X_t)$ solves (6.38).  The price can be recovered via $\Pi_t = V_t^* D_t$.

### 6.3.2.2  Application: Markov Growth with a Lucas SDF

As an example, suppose that dividend growth obeys

$$\kappa(X_t, \eta_{d,t+1}) = \mu_d + X_t + \sigma_d \, \eta_{d,t+1}$$

where $(\eta_{d,t})_{t \geqslant 0}$ is IID and standard normal.  Consumption growth is given by

$$\ln \frac{C_{t+1}}{C_t} = \mu_c + X_t + \sigma_c \, \eta_{c,t+1},$$

where $(\eta_{c,t})_{t \geqslant 0}$ is also IID and standard normal.  We use the Lucas SDF in (6.30), implying that

$$M_{t+1} = \beta \left( \frac{C_{t+1}}{C_t} \right)^{-\gamma} = \beta \exp(-\gamma(\mu_c + X_t + \sigma_c \eta_{c,t+1}))$$

EXERCISE 6.3.7.  Using (6.40), show that

$$A(x, x') = \beta \exp \left( -\gamma \mu_c + \mu_d + (1 - \gamma)x + \frac{\gamma^2 \sigma_c^2 + \sigma_d^2}{2} \right) P(x, x').$$

Figure 6.7 shows the price-dividend ratio function $v^*$ for the specification given in Listing 22, as well as for an alternative mean dividend growth rate $\mu_d$.  The state process is a Tauchen discretization of an AR(1) process with positive autocorrelation.  An increase in the state predicts higher dividends, which tends to increase the price.  At the same time, higher $x$ also predicts higher consumption growth, which acts negatively on the price.  For values of $\gamma$ greater than 1, the second effect dominates and the price-dividend ratio slopes down.

EXERCISE 6.3.8.  Complete the code in Listing 22 and replicate Figure 6.7.  Add a test to your code that checks $\rho(A) < 1$ before computing the price-dividend ratio.

```julia
using QuantEcon, LinearAlgebra

"Creates an instance of the asset pricing model with Markov state."
function create_asset_pricing_model(;
        n=200,                  # state grid size
        ρ=0.9, ν=0.2,           # state persistence and volatility
        β=0.99, γ=2.5,          # discount and preference parameter
        μ_c=0.01, σ_c=0.02,     # consumption growth mean and volatility
        μ_d=0.02, σ_d=0.1)      # dividend growth mean and volatility
    mc = tauchen(n, ρ, ν)
    x_vals, P = exp.(mc.state_values), mc.p
    return (; x_vals, P, β, γ, μ_c, σ_c, μ_d, σ_d)
end
```

Listing 22: Asset pricing model with Lucas SDF (`pd_ratio.jl`)

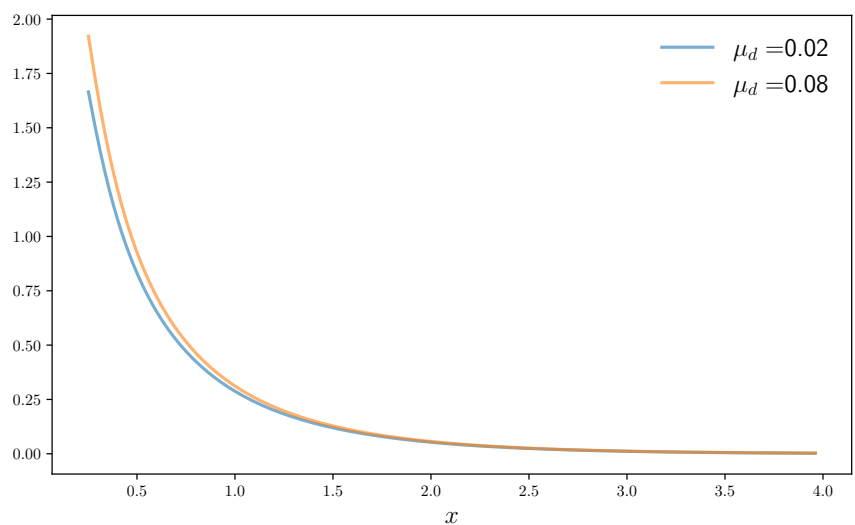

Figure 6.7: Price-dividend ratio as a function of the state

### 6.3.3 Incomplete Markets

In §6.3.1.5 we used the Neumann series lemma to solve for the equilibrium price vector $\pi$. However, some modifications to the basic model introduce nonlinearities that render the Neumann series lemma inapplicable. For example, Harrison and Kreps (1978) analyze a setting with heterogeneous beliefs and incomplete markets, leading to failure of the standard asset pricing equation. This results in a nonlinear equation for prices.

We treat the Harrison and Kreps model only briefly. There are two types of agents. Type $i$ believes that the state updates according to stochastic matrix $P_i$ for $i = 1, 2$. Agents are risk-neutral, so $m(x, y) \equiv \beta \in (0, 1)$. Harrison and Kreps (1978) show that, for their model, the equilibrium condition (6.34) becomes

$$\pi(x) = \max_i \beta \sum_{x'} [\pi(x') + d(x')] P_i(x, x') \tag{6.42}$$

for $x \in \mathsf{X}$ and $i \in \{1, 2\}$. Setting aside the details that lead to this equation, our objective is simply to obtain a vector of prices $\pi$ that solves (6.42).

As a first step, we introduce an operator $T \colon \mathbb{R}_+^\mathsf{X} \to \mathbb{R}_+^\mathsf{X}$ that maps $\pi$ to $T\pi$ via

$$(T\pi)(x) = \max_i \beta \sum_{x'} [\pi(x') + d(x')] P_i(x, x') \qquad (x \in \mathsf{X}). \tag{6.43}$$

We are assuming $d \geqslant 0$, so $T$ is indeed a self-map on $\mathbb{R}_+^\mathsf{X}$.

By construction, a vector $\pi \in \mathbb{R}_+^\mathsf{X}$ is a fixed point of $T$ if and only if it is a vector of prices that solves (6.42). Hence, we have successfully converted our equilibrium problem into a fixed point problem.

We aim to show that $T$ is a contraction. To this end, pick any $p, q \in \mathbb{R}_+^\mathsf{X}$. Applying the inequality from Lemma 2.2.2 on page 58, we obtain

$$|(Tp)(x) - (Tq)(x)| \leqslant \beta \max_i \left| \sum_{x'} [p(x') + d(x')] P_i(x, x') - \sum_{x'} [q(x') - d(x')] P_i(x, x') \right|.$$

Using the triangle inequality and canceling terms leads to

$$|(Tp)(x) - (Tq)(x)| \leqslant \beta \max_{i \in \{1, 2\}} \sum_{x'} |p(x') - q(x')| P_i(x, x') \leqslant \beta \|p - q\|_\infty.$$

Since this bound holds for all $x$, we can take the maximum with respect to $x$ and

obtain

$$\|Tp - Tq\|_\infty \leqslant \beta\|p - q\|_\infty.$$

Thus, on $\mathbb{R}_+^X$, the map $T$ is a contraction of modulus $\beta$ with respect to the sup norm.

Since $\mathbb{R}_+^X$ is a closed subset of $\mathbb{R}^X$, we conclude that $T$ has a unique fixed point in this set. Hence, the system (6.42) has a unique solution $\pi^*$ in $\mathbb{R}_+^X$, representing equilibrium prices. This fixed point can be computed by successive approximation.

EXERCISE 6.3.9. Provide an alternative proof of contractivity of $T$ on $\mathbb{R}_+^X$ using Blackwell's condition (§2.2.3.4).

## 6.4 Chapter Notes

Asset pricing is discussed in many sources, including Hansen and Renault (2010), Ross (2009), Cochrane (2009), Duffie (2010) and Campbell (2017). Asset pricing is part of many applications and extensions in macroeconomics, public finance, international economics, and other fields. Some of these are described in Ljungqvist and Sargent (2018).

Dynamic programming with state-dependent discounting is becoming more common in macroeconomics and finance. Representative examples include Krusell and Smith (1998), Woodford (2011), Christiano et al. (2014), Albuquerque et al. (2016), Saijo (2017), Basu and Bundick (2017), de Groot et al. (2018), Schorfheide et al. (2018), Hills et al. (2019), Toda (2019), Fagereng et al. (2019), Hubmer et al. (2020) and Cao (2020). For more on the theory of state-dependent discounting, see Jasso-Fuentes et al. (2020), Toda (2021) or Stachurski and Zhang (2021). An analysis of sovereign default with time-varying interest rates is provided by Bloise and Vailakis (2022).

Another challenge to the standard model with constant discount rates comes from empirical and experimental studies that find evidence of "hyperbolic discounting," where valuations across time fall rapidly at first and then more slowly. Provocative reviews of hyperbolic and quasi-hyperbolic discounting can be found in Frederick et al. (2002) and Rubinstein (2003). Cao and Werning (2018) provide conditions under which predictions from optimal savings models with quasi-hyperbolic discounting are robust. Balbus et al. (2018) analyze uniqueness of time-consistent stationary Markov policies for quasi-hyperbolic households under uncertainty. Balbus et al. (2022) study equilibria in dynamic models with recursive payoffs and generalized discounting. Noor and Takeoka (2022) addresses the topic of optimal discounting. Additional references include Diamond and Köszegi (2003), Dasgupta and Maskin (2005), Karp (2005), Amador et al. (2006), Balbus et al. (2018), Fedus et al. (2019), Hens and Schindler (2020), Jaśkiewicz and Nowak (2021), and Drugeon and Wigniolle (2021).

This chapter focused on time additive models with state-dependent discounting. More general preference specifications with this feature include Albuquerque et al. (2016), Schorfheide et al. (2018), Pohl et al. (2018), Gomez-Cram and Yaron (2020), and de Groot et al. (2022). In Chapter 8 we consider state-dependent discounting in general settings that accommodate such nonlinearities.

# Chapter 7

# Nonlinear Valuation

Dynamic programs are optimization problems where the objective to be maximized is lifetime value. As such, one key topic is how to combine a sequence of rewards into a corresponding lifetime value. So far we have considered linear valuation based on summation over expected discounted rewards, using either constant discount rates (Chapters 1–5) or state-dependent discounting (Chapter 6). In this chapter we consider extensions, where lifetime value is computed from a recursion over the reward sequence instead of a discounted sum. This "recursive preference" approach permits far more general specifications of lifetime value, and is becoming increasingly popular in economics, finance and computer science (see, e.g., §6.4).

This chapter focuses purely on valuation (i.e., combining reward sequences into lifetime values), rather than optimization. Later, in Chapter 8, we will show how to maximize lifetime value in settings where recursive preferences are adopted.

Throughout this chapter, the symbol X always represents a finite set.

## 7.1   Beyond Contraction Maps

The most natural way to express lifetime value in recursive preference environments is as a fixed point of a (typically nonlinear) operator. One challenge is that some recursive preference specifications induce operators that fail to be contractions. For this reason, we now invest in additional fixed point theory. All of this theory concerns order-preserving maps, since the operators we consider always inherit monotonicity from underlying preferences.

## 7.1.1   Knaster–Tarski for Function Space

If you try to draw an increasing function that maps $[0, 1]$ to itself without touching the 45 degree line you will find it impossible. Below we state a famous fixed point theorem due to Bronislaw Knaster (1893–1980) and Alfred Tarski (1901–1983) that generalizes this idea. In the statement, X is a finite set and $V := [v_1, v_2]$, where $v_1, v_2$ are functions in $\mathbb{R}^X$ with $v_1 \leqslant v_2$.

**Theorem 7.1.1** (Knaster–Tarski)**.** *If $T$ is an order-preserving self-map on $V$, then the set of fixed points of $T$ is nonempty and contains least and greatest elements $a \leqslant b$. Moreover,*

$$T^k v_1 \leqslant a \leqslant b \leqslant T^k v_2 \quad \text{for all } k \geqslant 0.$$

Unlike, say, the fixed point theorem of Banach (§1.2.2.3), Theorem 7.1.1 only yields existence. Uniqueness does not hold in general, as you can easily confirm by sketching the one-dimensional case or completing the following exercise.

EXERCISE 7.1.1. Consider the setting of Theorem 7.1.1 and suppose in addition that $v_1 \neq v_2$. Show that there exists an order-preserving self-map on $V$ with a continuum of fixed points.

## 7.1.2   Concavity, Convexity and Stability

In this section we study sufficient conditions for global stability that replace contractivity with shape properties such as concavity and monotonicity. To build intuition, we start with the one-dimensional case and show how these properties can be combined to achieve stability. Readers focused on results can safely skip to §7.1.2.2.

### 7.1.2.1   The One-Dimensional Case

In §1.2.3.2 we showed that concavity and monotonicity can yield global stability for the Solow–Swan model. Here is a more general result.

**Proposition 7.1.2.** *If $g$ is an increasing concave self-map on $U := (0, \infty)$ and, for all $x \in U$, there exist $a, b \in U$ with $a \leqslant x \leqslant b$, $a < g(a)$ and $g(b) \leqslant b$, then $g$ is globally stable on $U$.*

*Proof.* Regarding existence, fix $x \in U$ and suppose first that $x \leqslant g(x)$. Since $g$ is increasing, we have $g(x) \leqslant g^2(x)$. Continuing in this fashion shows that $(g^k(x))_{k \geqslant 0}$ is monotone increasing. Moreover, there exists a $b \in U$ such that $x \leqslant b$ and $g(b) \leqslant b$. Hence $g(x) \leqslant g(b) \leqslant b$. Iterating yields $g^k(x) \leqslant b$ for all $k$, so $(g^k(x))_{k \geqslant 0}$ is increasing and bounded above. Thus, there exists an $x^* \in U$ such that $x_k := g^k(x)$ converges to $x^*$ (by Theorem A.2.1 and Exercise A.2.4). Since $g$ is concave and hence continuous on any open set (see, e.g., Barbu and Precupanu (2012)), the result in Exercise 1.2.16 (page 21) implies that $x^* = g(x^*)$.

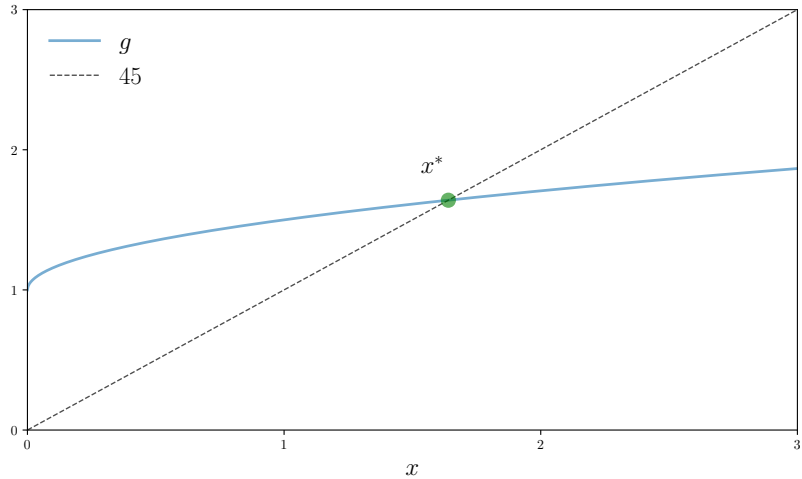

Figure 7.1: Global stability induced by increasing concave functions

If, instead, $g(x) \leqslant x$, then a similar argument shows that $(g^k(x))_{k \geqslant 0}$ is decreasing and bounded. Using analogous reasoning, we obtain a fixed point $x^*$ in $U$ with $g^k(x) \to x^*$.

To show the uniqueness of the fixed point, assume $g(x) = x$ and $g(y) = y$ for some $x, y \in U$. We claim that $x = y$. To see this, suppose without loss of generality that $x \leqslant y$. By assumption, there exists an $a \in U$ such that $a \leqslant x \leqslant y$ and $g(a) > a$. Because $a \leqslant x \leqslant y$, we can take $\lambda \in [0, 1]$ such that $x = \lambda a + (1 - \lambda)y$. If $\lambda > 0$, then concavity of $g$ and $g(a) > a$ implies the contradiction

$$g(x) = g\left(\lambda a + (1 - \lambda)y\right) \geqslant \lambda g(a) + (1 - \lambda)g(y) > \lambda a + (1 - \lambda)y = x = g(x).$$

Hence $\lambda = 0$. Since $x = \lambda a + (1 - \lambda)y$, this yields $x = y$. □

Figure 7.1 gives one example, where $g(x) = 1 + \sqrt{x}/2$. The conditions of Proposition 7.1.2 hold because, given any $x > 0$, we can find an $a$ in $(0, x)$ that gets mapped strictly up (i.e., $g(a)$ is above the 45 line) and a point $b > x$ that gets mapped down (i.e., $g(b)$ is below the 45 degree line).

EXERCISE 7.1.2. Prove that the map $g$ and set $U$ defined in the discussion of the Solow-Swan model above Proposition 7.1.2 satisfies the conditions of the proposition.

EXERCISE 7.1.3. Show that the condition $a < g(a)$ in Proposition 7.1.2 cannot be dropped without weakening the conclusion.

EXERCISE 7.1.4. Dropping the Cobb-Douglas specification on production, suppose $g(k) = sf(k) + (1 - \delta)k$ where $0 < s, \delta < 1$ and $f$ is a strictly positive increasing concave production function on $U = (0, \infty)$ satisfying the **Inada conditions**

$$f'(k) \to \infty \text{ as } k \to 0 \quad \text{and} \quad f'(k) \to 0 \text{ as } k \to \infty,$$

Use Proposition 7.1.2 to prove that $g$ is globally stable on $U$.

EXERCISE 7.1.5. Fajgelbaum et al. (2017) study a law of motion for aggregate uncertainty given by

$$s_{t+1} = g(s_t) \quad \text{where} \quad g(s) := \rho^2 \left[ \frac{1}{s} + a^2 \frac{1}{\eta} \right]^{-1} + \gamma.$$

Let $a$, $\eta$ and $\gamma$ be positive constants and assume $0 < \rho < 1$. Prove that $g$ is globally stable on $M := (0, \infty)$.

### 7.1.2.2 The Multidimensional Case

Proposition 7.1.2 extends to multiple dimensions. In this section we present a multi-dimensional version that covers both convex and concave functions.

To state our result we extend the definition of convexity and concavity to vector-valued self-maps. The definitions mirror those for scalar-valued functions: a self-map $T$ on a convex subset $D$ of $\mathbb{R}^X$ is called **convex** if

$$T(\lambda u + (1 - \lambda)v) \leqslant \lambda Tu + (1 - \lambda)Tv \text{ whenever } u, v \in D \text{ and } \lambda \in [0, 1];$$

and **concave** if

$$\lambda Tu + (1 - \lambda)Tv \leqslant T(\lambda u + (1 - \lambda)v) \text{ whenever } u, v \in D \text{ and } \lambda \in [0, 1].$$

Here $\leqslant$ is, as usual, the pointwise order.

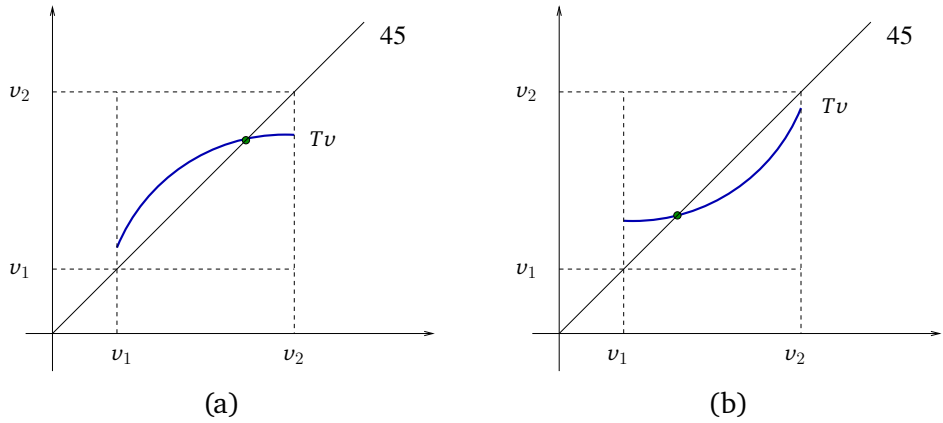

Figure 7.2: Du's theorem: convex and concave cases

We are now ready to state our next fixed point result, which was first proved in an infinite-dimensional setting by Du (1990). In the statement, $\mathsf{X}$ is a finite set, $V := [v_1, v_2]$ is a nonempty order interval in $(\mathbb{R}^{\mathsf{X}}, \leqslant)$, and $T$ is a self-map on $V$.

**Theorem 7.1.3** (Du)**.** *If $T$ is order-preserving on $V$, then $T$ is globally stable on $V$ under any one of (i)–(iv) below.*

(i) *$T$ is concave and $Tv_1 \gg v_1$, or*

(ii) *$T$ is concave and there exists a $\delta > 0$ such that $Tv_1 \geqslant v_1 + \delta(v_2 - v_1)$, or*

(iii) *$T$ is convex and $Tv_2 \ll v_2$, or*

(iv) *$T$ is convex and there exists a $\delta > 0$ such that $Tv_2 \leqslant v_2 - \delta(v_2 - v_1)$.*

Conditions (i) and (ii) are similar – in fact (ii) holds whenever (i) holds, so (ii) is the weaker (but slightly more complicated) condition. Conditions (iii) and (iv) are similar in the same sense. Figure 7.2 illustrates the convex and the concave versions of the result in one dimension. We encourage you to sketch your own variations to understand the roles that different conditions play.

EXERCISE 7.1.6. Let $F$ and $G$ be self-maps on convex $D \subset \mathbb{R}^n$. Show that $T := F \circ G$ is concave on $D$ whenever $F$ and $G$ are order-preserving and concave on $D$.

A full proof of Theorem 7.1.3 can be found in Du (1990) or Theorem 2.1.2 and Corollary 2.1.1 of Zhang (2012). In our setting, existence follows from the Knaster–Tarski theorem on page 214. We prove uniqueness on page 365.

### 7.1.3   A Power-Transformed Affine Equation

Du's theorem provides conditions under which concave or convex order-preserving self-maps on order intervals attain global stability. In this section we study maps of this type that have additional structure. While this additional structure is restrictive, it allows us to obtain global stability on unbounded subsets rather than order intervals.

To begin, let $\mathsf{X}$ be a finite set and consider the equation

$$v = [h + (Av)^{1/\theta}]^{\theta} \qquad (v \in V), \qquad (7.1)$$

where $\theta$ is a nonzero parameter, $A \in \mathcal{L}(\mathbb{R}^{\mathsf{X}})$ with $A \geqslant 0$, $V = (0, \infty)^{\mathsf{X}}$, and $h \in V$. This system reduces to the affine model studied in Lemma 6.1.4 (page 189) when $\theta = 1$.

To analyze (7.1), we introduce the self-map

$$Gv = [h + (Av)^{1/\theta}]^{\theta} \qquad (v \in V). \qquad (7.2)$$

Continuing to assume that $h \gg 0$ and $A$ is a positive linear operator, we can use Du's theorem to establish the next result (which generalizes Lemma 6.1.4 on page 189).

**Theorem 7.1.4.** *If $A$ is irreducible, then the following statements are equivalent.*

  (i)  $\rho(A)^{1/\theta} < 1$.

  (ii)  *$G$ is globally stable on $V$.*

*In the case $\rho(A)^{1/\theta} \geqslant 1$, the map $G$ has no fixed point in $V$.*

The key to proving (i) implies (ii) is that $G$ is order-preserving and either convex or concave, depending on the value of $\theta$. The remaining conditions in Du's theorem are established over order intervals using $\rho(A)^{1/\theta} < 1$. By applying an approximation argument, global stability is extended from order intervals to all of $V$. Some of these details are contained in the following exercises and a full proof can be found in Stachurski et al. (2022).

Let

$$F_x(t) = \left\{ h(x) + t^{1/\theta} \right\}^{\theta} \qquad (t > 0).$$

EXERCISE 7.1.7. Prove that, for all $x \in \mathsf{X}$, the function $F_x$ is increasing,

  (i)  convex whenever $\theta \in (0, 1]$, and

  (ii)  concave otherwise (i.e., for other nonzero $\theta$).

EXERCISE 7.1.8. Using Exercise 7.1.7, prove that $G$ is order-preserving on $V$, convex on $V$ whenever $\theta \in (0, 1]$, and concave otherwise.

EXERCISE 7.1.9. Kleinman et al. (2023) study a dynamic discrete choice model of migration with savings and capital accumulation. They show that optimal consumption for landlords in their model is $c_t = \sigma_t R_t k_t$, where $k_t$ is capital, $R_t$ is the gross rate of return on capital and $\sigma_t$ is a state-dependent process obeying

$$\sigma_t^{-1} = 1 + \beta^\psi \left[ \mathbb{E}_t R_{t+1}^{(\psi-1)/\psi} \sigma_{t+1}^{-1/\psi} \right]^\psi. \tag{7.3}$$

Here $\beta$ is a discount factor and $\psi$ is a utility parameter. Assume $R_t = f(X_t)$ where $\mathsf{X}$ is finite, $f \in \mathbb{R}^\mathsf{X}$, and $(X_t)$ is $P$-Markov for some $P \in \mathcal{M}(\mathbb{R}^\mathsf{X})$. Let $A \in \mathcal{L}(\mathbb{R}^\mathsf{X})$ be defined by

$$(Av)(x) = \beta \sum_{x'} f(x')^{(\psi-1)/\psi} v(x') P(x, x').$$

Prove that there exists a unique solution to (7.3) of the form $\sigma_t = \sigma(X_t)$ for some $\sigma \in \mathbb{R}^\mathsf{X}$ with $\sigma \gg 0$ if and only if $\rho(A)^\psi < 1$.

## 7.2 Recursive Preferences

In this section we compute lifetime values associated with given reward processes in settings that involve nonlinear recursions. These nonlinear recursions are called *recursive preferences*. We will show how some common specifications of recursive preferences can be translated into lifetime valuations via the fixed point methods introduced in Chapter 2 and §7.1.

### 7.2.1 Motivation: Optimal Savings

We motivate recursive preference models by analyzing consumption decisions.

#### 7.2.1.1 A Recursive View of a Standard Model

The time additive model of valuation in §3.2.2.3 can be studied from a purely recursive point of view. As a starting point, we state that the value $V_t$ of current and future consumption is defined at each point in time $t$ by the recursion

$$V_t = u(C_t) + \beta \, \mathbb{E}_t V_{t+1}. \tag{7.5}$$

The random variables $V_t$ and $V_{t+1}$ are the unknown objects in this expression. The expectation $\mathbb{E}_t$ conditions on $X_0, \ldots, X_t$ and $C_t = c(X_t)$. The process $(X_t)_{t \geqslant 0}$ is $P$-Markov.

Since consumption is a function of $(X_t)_{t \geqslant 0}$ and knowledge of the current state $X_t$ is sufficient to forecast future values (by the Markov property), it is natural to guess that $V_t$ will depend on the Markov chain only through $X_t$. Hence we guess there is a solution of (7.5) takes the form $V_t = v(X_t)$ for some $v \in \mathbb{R}^{\mathsf{X}}$.

(Here $v$ is an *ansatz*, meaning "educated guess." First we guess the form of a solution and then we try to verify that the guess is correct. So long as we carry out the second step, starting with a guess brings no loss of rigor.)

Under this conjecture, (7.5) can be rewritten as $v(X_t) = u(c(X_t)) + \beta \mathbb{E}_t v(X_{t+1})$. Conditioning on $X_t = x$ and setting $r := u \circ c$, this becomes

$$v(x) = r(x) + \beta \, \mathbb{E}_x \, v(X_{t+1}) = r(x) + \beta (Pv)(x) \qquad (x \in \mathsf{X}). \tag{7.6}$$

In vector form, we get $v = r + \beta P v$. From the Neumann series lemma, the solution is $v^* = (I - \beta P)^{-1} r$, which is identical to (3.21) on page 97.

EXERCISE 7.2.1. Verify our guess: Show $(V_t^*)$ obeys (7.5) when $V_t^* := v^*(X_t)$.

In summary, (7.5) and the sequential representation (3.20) specify the same lifetime value for consumption paths.

While the recursive formulation in (7.5) now seems redundant, since it produces the same specification that we obtained from the sequential approach, the recursive set up gives us a formula to build on, and hence a pathway to overcoming limitations of the time additive approach. Most of the rest of this chapter will be focused on this agenda.

Pursuing this agenda will produce preferences over consumption paths where the sequential approach has no natural counterpart. This occurs when current value $V_t$ is nonlinear in current rewards and continuation values (unlike the linear specification (7.5)). Such specifications are called **recursive preferences**. When dealing with recursive preference models, the lack of a sequential counterpart means that we are *forced* to proceed recursively.

**Remark 7.2.1.** The term "recursive preferences" is confusing, since, as we have just agreed, time additive preferences also admit the recursive specification (7.5). Nonetheless, when economists say "recursive preferences," they almost always refer to settings where lifetime utility can *only* be expressed recursively. We follow this convention.

### 7.2.1.2 Limitations of Time Additive Preferences

In the previous section we discussed how the time additive preference specification

$$v(x) = \mathbb{E}_x \sum_{t \geqslant 0} \beta^t u(C_t) \tag{7.7}$$

also called the **discounted expected utility model**, can be framed recursively, and how this provides a pathway to go beyond the time additive specification. We are motivated to do so because the time additive specification has been rejected by experimental and observational data in many settings.

In this section we highlight some of the limitations of time additive preferences. While our discussion is only brief, more background and a list of references can be found in §7.4.

One issue with (7.7) is the assumption of a constant positive discount rate, which has been refuted by a long list of empirical studies. This issue was discussed in §6.4.

Another limitation of time additive preferences is that agents are risk-neutral in future utility (see, e.g., (7.5), where current value depends linearly on future value). Although risk aversion over consumption can be built in through curvature of $u$, this same curvature also determines the elasticity of intertemporal substitution, meaning that the two aspects of preferences cannot be separated. We elaborate on this point in §7.3.1.4.

A third issue with time additivity is that agents with such preferences are indifferent to any variation in the joint distribution of rewards that leaves marginal distributions unchanged. To get a sense of what this means, suppose you accept a new job and will be employed by this firm for the rest of your life. Your daily consumption will be entirely determined by your daily wage. Your boss offers you two options:

(A) Your boss will flip a coin at the start of your first day on the job. If the coin is heads, you will receive $10,000 a day for the rest of your life. If the coin is tails, you will receive $1 per day for the rest of your life.

(B) Your boss will flip a coin at the start of every day. If the coin is heads, you will receive $10,000. If the coin is tails, you will receive $1.

If you have a strict preference between options A and B, then your choice cannot be rationalized with time additive preferences.

To see why, let $\varphi$ be a probability distribution that represents the lottery described above, putting mass 0.5 on 10,000 and mass 0.5 on 1. Under option A, consumption $(C_t)_{t \geqslant 1}$ is given by $C_t = C_1$ for all $t$, where $C_1 \sim \varphi$. Under option B, consumption $(C_t)_{t \geqslant 1}$ is an IID sequence drawn from $\varphi$. Either way, lifetime utility is

$$\mathbb{E} \sum_{t \geqslant 1} \beta^t u(C_t) = \sum_{t \geqslant 1} \beta^t \mathbb{E} u(C_t) = \frac{\beta \bar{u}}{1 - \beta},$$

where $\bar{u} := \mathbb{E} u(C_1) = u(1)/2 + u(10,000)/2$.

The critical part of this argument is the passing of expectations through the sum, which uses time additivity . The implication is that lifetime utility depends only on the marginal distribution of each $C_t$, rather than on the joint distribution of the stochastic process $(C_t)_{t \geqslant 0}$.

## 7.2.2 Risk-Sensitive Preferences

Having motivated recursive preferences, let's turn to our first example: **risk-sensitive preferences**. For the consumption problem described in §7.2.1.1, imposing risk-sensitive preferences means replacing the recursion $v = r + \beta P v$ for $v$ with

$$v(x) = r(x) + \beta \frac{1}{\theta} \ln \left\{ \sum_{x'} \exp(\theta v(x')) P(x, x') \right\} \qquad (x \in \mathsf{X}). \qquad (7.8)$$

As before, $r(x) = u(c(x))$ represents current utility when the current state is $x$. The parameter $\theta$ is a nonzero constant in $\mathbb{R}$.

In (7.8), the transform $f(v) = \exp(\theta v)$ is applied to $v$ before expectation is taken. After the expectation is computed, the transform is undone via $f^{-1}(v) = (1/\theta) \ln(v)$. We show below that the agent can be either risk-averse or risk-loving with respect to future outcomes, depending on the value of $\theta$.

### 7.2.2.1 Lifetime Utility

We understand the functional equation (7.8) as "defining" lifetime utility under risk-sensitive preferences. A function $v$ solving (7.8) gives a lifetime valuation $v(x)$ to each $x \in \mathsf{X}$, with the interpretation that $v(x)$ is lifetime utility conditional on initial state $x$. This definition of lifetime value is by analogy to the time additive case studied in §7.2.1.1, where the function $v$ solving $v = r + \beta P v$ measures lifetime utility from each initial state.

In the previous paragraph we wrote "defining" in scare quotes because we can't be sure we have a definition at this point. Just because we write down a recursive expression for lifetime utility doesn't mean that corresponding lifetime utility is actually well defined. (For example, we can happily write down the recursive vector equation $v = v + \mathbb{1}$ but no vector $v$ solving this equation exists.) One aim of this chapter is to provide conditions under which recursions like (7.8) have solutions.

Another issue is uniqueness. Suppose that (7.8) has many solutions. In this case the predictions of the utility model are ambiguous. Our perspective is that the recursive preference specification (7.8) is not correctly formulated unless existence and uniqueness hold. We return to this point in §7.2.2.3.

One final comment: even if we can find a $v$ that solves (7.8), the nonlinearities introduced by risk sensitivity imply that there will be no neat sequential representation analogous to $v(x) = \mathbb{E}_x \sum_t \beta^t u(C_t)$ from the time additive case. (This connects to Remark 7.2.1, where we discuss recursive preference terminology.)

### 7.2.2.2 Risk-Adjusted Expectation

We want to understand the "expectation-like" expression on the right hand side of (7.8) that replaces the ordinary conditional expectation $\sum_{x'} v(x')P(x, x')$ from the time additive case. To this end, we define, for arbitrary random variable $\xi$ and $\theta \in \mathbb{R}$,

$$\mathcal{E}_\theta[\xi] = \frac{1}{\theta} \ln \{\mathbb{E}[\exp(\theta \xi)]\}.$$

The value $\mathcal{E}_\theta[\xi]$ is called the **entropic risk-adjusted expectation** of $\xi$ given $\theta$.

EXERCISE 7.2.2. Prove that, for any random variable $\xi$ any nonzero $\theta$ and any constant $c$, we have $\mathcal{E}_\theta[\xi + c] = \mathcal{E}_\theta[\xi] + c$.

The key idea behind the entropic risk-adjusted expectation is that decreasing $\theta$ lowers appetite for risk and increasing $\theta$ does the opposite.

EXERCISE 7.2.3. Prove that, if $\xi$ is normally distributed, then

$$\mathcal{E}_\theta[\xi] = \mathbb{E}[\xi] + \theta \frac{\text{Var}[\xi]}{2}. \tag{7.9}$$

[Hint: Look up the moment generating function of a normal distribution.]

Expression (7.9) above shows that, for the Gaussian case, $\mathcal{E}_\theta[\xi]$ equals the mean plus a term that penalizes variance when $\theta < 0$ and rewards it when $\theta > 0$.

More generally, we have the following result.

**Lemma 7.2.1.** *For any random variable ξ taking values in* X*, we have*

(i) $\mathcal{E}_\theta[\xi] \leqslant \mathbb{E}[\xi]$ *for all* $\theta < 0$.

(ii) $\mathcal{E}_\theta[\xi] \geqslant \mathbb{E}[\xi]$ *for all* $\theta > 0$.

*Moreover, both of these inequalities are strict if and only if* $\mathrm{Var}[\xi] > 0$.

*Proof.* Fix $\theta \in \mathbb{R}$ and let $f\colon \mathbb{R} \to (0, \infty)$ be defined by $f(x) = \exp(\theta x)$. Note that $f'(x) = \theta \exp(\theta x)$ and $f''(x) = \theta^2 \exp(\theta x)$. Thus $f$ is convex and either increasing or decreasing depending on whether $\theta$ is positive or negative. Then $\mathcal{E}_\theta[\xi] = f^{-1}(\mathbb{E}f(\xi))$. By Jensen's inequality,

$$\mathbb{E}[f(\xi)] \geqslant f(\mathbb{E}[\xi]).$$

If $\theta > 0$, then $f^{-1}$ is increasing, so applying $f^{-1}$ to both sides gives $\mathcal{E}_\theta[\xi] \geqslant \mathbb{E}[\xi]$. If $\theta < 0$, then $f^{-1}$ is decreasing, so applying $f^{-1}$ to both sides gives $\mathcal{E}_\theta[\xi] \leqslant \mathbb{E}[\xi]$. This proves the two weak inequalities in Lemma 7.2.1. To obtain strict inequalities we can apply the same argument using a strict version of Jensen's inequality (see, e.g., Liao and Berg (2018)), which is valid when $\mathrm{Var}[\xi] > 0$. □

### 7.2.2.3 Existence and Uniqueness

Let's return to investigating lifetime utility under risk-sensitive preferences. To this end, we introduce the **risk-sensitive Koopmans operator** $K_\theta$ on $\mathbb{R}^\mathsf{X}$ via

$$(K_\theta v)(x) = r(x) + \beta \frac{1}{\theta} \ln \left\{ \sum_{x'} \exp(\theta v(x')) P(x, x') \right\} \qquad (x \in \mathsf{X}). \tag{7.10}$$

Evidently, for given nonzero $\theta$, a function $v \in \mathbb{R}^\mathsf{X}$ solves the risk-sensitive preference lifetime utility specification (7.8) if and only if $v$ is a fixed point of $K_\theta$. This explains the significance of the following result:

**Proposition 7.2.2.** *If $\beta < 1$, then $K_\theta$ is globally stable on $\mathbb{R}^\mathsf{X}$.*

We postpone a proof of Proposition 7.2.2 because we will prove a more general result in §7.3.2.2. For now we note the following implications.

(i) For each nonzero $\theta$, lifetime utility is both well-defined and uniquely defined for risk-sensitive preferences (i.e., (7.8) has a unique solution).

(ii) The unique solution, denoted henceforth by $v^*$, can be computed by successive approximation using $K_\theta$.

#### 7.2.2.4 The Gaussian Case

As a tractable case, let's suppose that $r(x) = x$ and that $X_{t+1} = \rho X_t + \sigma W_{t+1}$ where $(W_t)_{t \geqslant 1}$ is IID and standard normal. Here $|\rho| < 1$ and $\sigma \geqslant 0$ controls volatility of the state. Rather than discretizing the state process, we leave it as continuous and proceed by hand.

In this setting, the functional equation (7.8) for $v$ becomes

$$v(x) = x + \beta \mathcal{E}_\theta [v(\rho x + \sigma W)] \tag{7.11}$$

for each $x \in \mathsf{X}$, where $W$ is standard normal.

Since $\rho x + \sigma W$ is Gaussian, the expression (7.9) for the risk-adjusted expectation of a normal random variable leads us to conjecture that the solution $v$ will be affine, i.e., $v(x) = ax + b$ for some $a, b \in \mathbb{R}$. This conjecture turns out to be correct:

EXERCISE 7.2.4. Verify that $v(x) = ax + b$ solves (7.11) when

$$a := \frac{1}{1 - \rho\beta} \quad \text{and} \quad b := \theta \frac{\beta}{1 - \beta} \frac{(a\sigma)^2}{2}.$$

We can see that, under the stated assumptions, lifetime value $v$ is increasing in the state variable $x$. However, impacts of the parameters generally depend on $\theta$. For example, if $\theta > 0$, increasing $\sigma$ shifts up lifetime utility. If $\theta < 0$, then lifetime value decreases with $\sigma$. This is as we expect: lifetime utility is affected positively or negatively by volatility, depending on whether or not the agent is risk averse or risk loving.

Figure 7.3 shows the true solution $v(x) = ax + b$ to the risk-sensitive lifetime utility model, as well as an approximate fixed point from a discrete approximation. The discrete approximation is computed by applying successive approximation to $K_\theta$ after discretizing the state process via Tauchen's method. The parameters and discretization are shown in Listing 23.

EXERCISE 7.2.5. Replicate Figure 7.3.

EXERCISE 7.2.6. Dropping the Gaussian assumption, suppose now that consumption is IID with $C_t = c(X_t)$ where $(X_t)_{t \geqslant 0}$ is IID with distribution $\varphi$ on finite set $\mathsf{X}$. Now the operator $K_\theta$ becomes

$$(K_\theta v)(x) = r(x) + \beta \frac{1}{\theta} \ln \left\{ \sum_{x'} \exp(\theta v(x')) \varphi(x') \right\} \qquad (x \in \mathsf{X}).$$

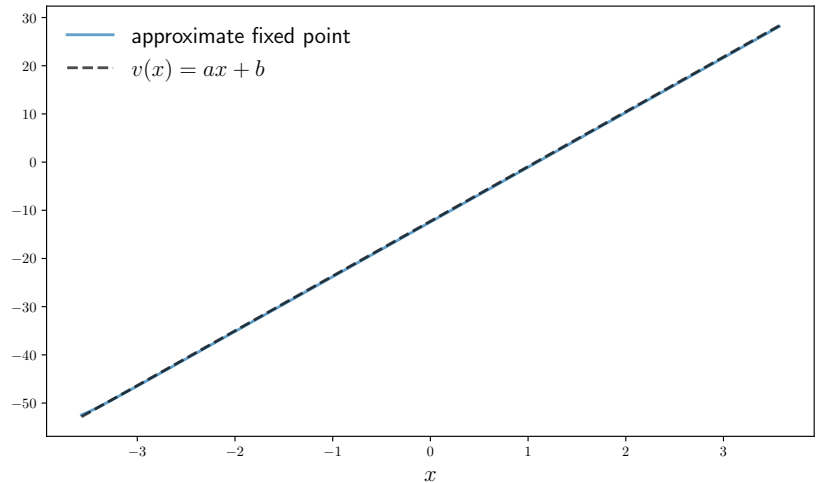

Figure 7.3:  Approximate and true solutions in the Gaussian case

```julia
using LinearAlgebra, QuantEcon

function create_rs_utility_model(;
        n=180,         # size of state space
        β=0.95,        # time discount factor
        ρ=0.96,        # correlation coef in AR(1)
        σ=0.1,         # volatility
        θ=-1.0)        # risk aversion
    mc = tauchen(n, ρ, σ, 0, 10)  # n_std = 10
    x_vals, P = mc.state_values, mc.p
    r = x_vals         # special case u(c(x)) = x
    return (; β, θ, ρ, σ, r, x_vals, P)
end
```

Listing 23:  Risk sensitive utility model parameters (`rs_utility.jl`)

Although iterating on $K_\theta$ is convergent, there is a more efficient method that reduces to solving a one-dimensional equation. Propose such a method and confirm that it is convergent. [Hint: Consider reviewing §4.2.2.2.]

## 7.2.3 Epstein–Zin Preferences

One of the most popular specifications of recursive preferences in quantitative research is Epstein–Zin utility.[1] This class of preferences has been used to study asset pricing, business cycles, monetary policy, fiscal policy, optimal taxation, climate policy, pension plans, and other topics. In this section we introduce the Epstein–Zin specification and discuss how to solve it. We will see that the specification, while highly nonlinear, is nonetheless well behaved.

### 7.2.3.1 Specification

With **Epstein–Zin** preferences, the relationship $V_t = u(C_t) + \beta \mathbb{E}_t V_{t+1}$ is replaced by

$$V_t = \left\{ (1 - \beta) C_t^\alpha + \beta \left[ \mathbb{E}_t V_{t+1}^\gamma \right]^{\alpha/\gamma} \right\}^{1/\alpha}, \tag{7.12}$$

where $\gamma$, $\alpha$ are nonzero parameters and $\beta \in (0, 1)$. As for risk-sensitive preferences, lack of time additivity implies that there is no neat sequential representation for lifetime value. As a result, we must work directly with the recursive expression (7.12).

Assume as before that $C_t = c(X_t)$, where $c \in \mathbb{R}_+^X$ and $(X_t)_{t \geq 0}$ is $P$-Markov on finite set X. We conjecture a solution of the form $V_t = v(X_t)$ for some $v \in V := \mathbb{R}_+^X$. Under this conjecture, the **Epstein–Zin Koopmans operator** corresponding to (7.12) is

$$(Kv)(x) = \left\{ (1 - \beta) c(x)^\alpha + \beta \left[ \sum_{x'} v(x')^\gamma P(x, x') \right]^{\alpha/\gamma} \right\}^{1/\alpha}. \tag{7.13}$$

As will be discussed further in §7.3.1.1, the parameter $\gamma$ governs risk aversion with respect to temporal gambles (where outcomes are resolved in the next period), while $\beta$ controls impatience and $\alpha$ parametrizes the intertemporal elasticity of substitution. The fact that all three parameters have distinct effects helps fit data. For example, see Tallarini Jr (2000) and Barillas et al. (2009).

[1]Epstein–Zin preferences were popularized in Epstein and Zin (1989). They are a special case of preferences defined by Kreps and Porteus (1978). Further discussion can be found in §7.4.

An important question is whether Epstein–Zin preferences are well defined. In particular, what conditions do we need on primitives such that the Koopmans operator $K$ in (7.13) has a unique fixed point?

### 7.2.3.2 Properties of the Koopmans Operator

To address this question we rewrite (7.13) in vector form as

$$Kv = \left\{ h + \beta \left[ Pv^{\gamma} \right]^{\alpha/\gamma} \right\}^{1/\alpha} \tag{7.14}$$

where $h \in \mathbb{R}^{\mathsf{X}}$. This is equivalent to (7.13) when $h = (1 - \beta)c^{\alpha}$. To avoid fractional powers of negative numbers, we assume throughout that $h \geqslant 0$.

EXERCISE 7.2.7. Prove that, under this assumption, $K$ is a self-map on $V := (0, \infty)^{\mathsf{X}}$.

The set $V$ is called the **interior of the positive cone** of $\mathbb{R}^{\mathsf{X}}$.

The operator $K$ is difficult to work with for two reasons. First, linear and nonlinear transformations are intertwined. Second, there are several cases for the parameters that we need to handle in order to understand stability. Nonetheless, by applying a smooth transformation, we will find it easy to show that the Epstein–Zin Koopmans operator $K$ is globally stable under mild conditions. In particular,

**Proposition 7.2.3.** *If $P$ is irreducible and $h \gg 0$, then $K$ is globally stable on $V$.*

A proof of Proposition 7.2.3 is provided in §7.2.3.3.

Proposition 7.2.3 implies that Epstein–Zin utility is well-defined under the stated conditions and, moreover, that the solution can be computed via successive approximation on $K$. Listing 24 provides code for performing this operation. Figure 7.4 shows convergence of the sequence of iterates to the fixed point $v^*$, under the parameters in Listing 24, given an initial condition $v_0$. The figure plots every 10th iterate, repeated 100 times.

### 7.2.3.3 Proof of the Stability Result

We prove Proposition 7.2.3 by

(i) introducing an operator $\hat{K}$ obtained from $K$ via a smooth transformation,

(ii) proving that $(\hat{V}, \hat{K})$ and $(V, K)$ are topologically conjugate, and

```julia
include("s_approx.jl")
using LinearAlgebra, QuantEcon

function create_ez_utility_model(;
        n=200,        # size of state space
        ρ=0.96,       # correlation coef in AR(1)
        σ=0.1,        # volatility
        β=0.99,       # time discount factor
        α=0.75,       # EIS parameter
        γ=-2.0)       # risk aversion parameter

    mc = tauchen(n, ρ, σ, 0, 5)
    x_vals, P = mc.state_values, mc.p
    c = exp.(x_vals)

    return (; β, ρ, σ, α, γ, c, x_vals, P)
end

function K(v, model)
    (; β, ρ, σ, α, γ, c, x_vals, P) = model

    R = (P * (v.^γ)).^(1/γ)
    return ((1 - β) * c.^α + β * R.^α).^(1/α)
end

function compute_ez_utility(model)
    v_init = ones(length(model.x_vals))
    v_star = successive_approx(v -> K(v, model),
                               v_init,
                               tolerance=1e-10)
    return v_star
end
```

Listing 24: Epstein–Zin utility model and Koopmans operator (`ez_utility.jl`)

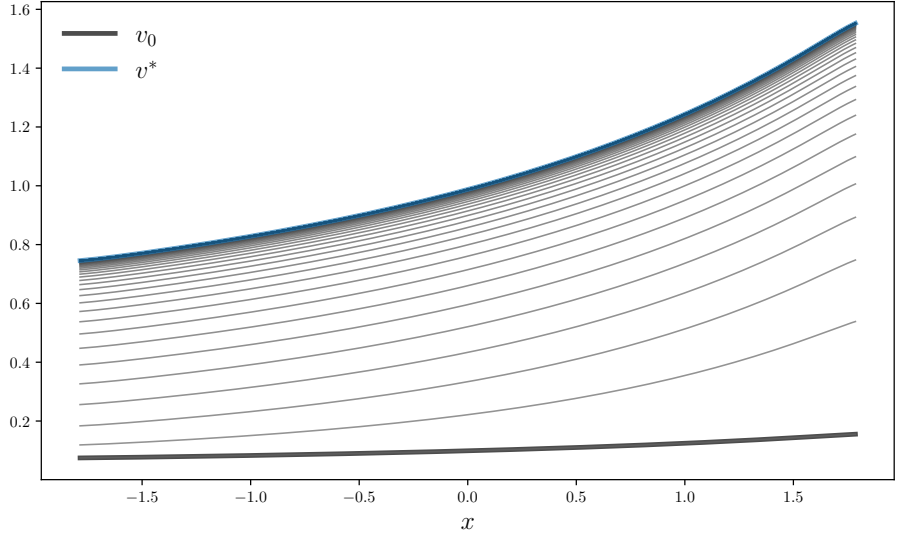

Figure 7.4: Convergence of Koopmans iterates for Epstein–Zin utility

(iii) obtaining conditions under which $\hat{K}$ is globally stable on $V$.

Throughout this section, the assumptions of Proposition 7.2.3 are in force.

To begin we define $\hat{K}$ via

$$\hat{K}v = \left\{ h + \beta(Pv)^{1/\theta} \right\}^{\theta} \qquad \text{where} \quad \theta := \frac{\gamma}{\alpha}. \tag{7.15}$$

The operator $\hat{K}$ is simpler to work with than $K$ because it unifies $\alpha, \gamma$ into a single parameter $\theta$ and decomposes the Epstein–Zin update rule into two parts: a linear map $P$ and a separate nonlinear component.

EXERCISE 7.2.8. Prove that

(i) $\hat{K}$ is a self-map on $V$ and

(ii) $v \in V$ is a fixed point of $K$ if and only if $v^{\gamma}$ is a fixed point of $\hat{K}$.

**Lemma 7.2.4.** *Let $\Phi$ be defined by $\Phi v = v^{\gamma}$. The map $\Phi$ is a homeomorphism from $V$ to itself and $(V, K)$ and $(V, \hat{K})$ are topologically conjugate under $\Phi$.*

*Proof.* Evidently $\Phi$ is a continuous bijection from $V$ to itself, with continuous inverse

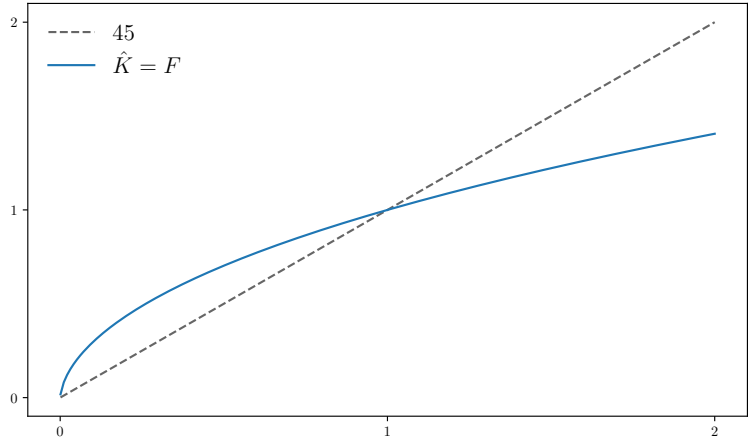

Figure 7.5: Shape properties of $\hat{K}$ in one dimension

$\Phi^{-1}v = v^{1/\gamma}$. Hence $\Phi$ is a homeomorphism. In addition, for $v \in V$,

$$\hat{K}\Phi v = \left\{ h + \beta(P\Phi v)^{1/\theta} \right\}^{\theta} = \left\{ h + \beta(Pv^{\gamma})^{\alpha/\gamma} \right\}^{\gamma/\alpha} = \Phi Kv.$$

This shows that $(V, K)$ and $(V, \hat{K})$ are topologically conjugate, as claimed. □

*Proof of Proposition 7.2.3.* Set $A = \beta^{\theta} P$. With this notation we have $\hat{K}v = [h+(Av)^{1/\theta}]^{\theta}$. In view of in Theorem 7.1.4 on page 218, this operator is globally stable on $V$ whenever $\rho(A)^{1/\theta} < 1$. In our case $\rho(A) = \rho(\beta^{\theta}P) = \beta^{\theta}$, so $\rho(A)^{1/\theta} = \beta$. It follows that $\hat{K}$ is globally stable on $V$ whenever $\beta < 1$. Since $(V, K)$ and $(V, \hat{K})$ are topologically conjugate, the proof of Proposition 7.2.3 is complete. □

### 7.2.3.4 Why Not Use Contractivity?

While we can consider studying stability of $\hat{K}$ using contraction arguments, this approach fails under useful parameterizations. To illustrate, suppose that $\mathsf{X} = \{x_1\}$. Then $h$ is a constant, $P$ is the identity, $v$ is a scalar and $\hat{K}v = F(v)$ with $F(v) = \left\{ h + \beta v^{1/\theta} \right\}^{\theta}$, as shown in Figure 7.5. Here $\theta = 5$, $h = 0.5$ and $\beta = 0.5$. We see that $\hat{K}$ has infinite slope at zero, so the contraction property fails.[2]

---

[2]We could try to truncate the interval to a neighborhood of the fixed point and hope that $\hat{K}$ is a contraction when restricted to this interval. But in higher dimensions we are not sure that a fixed point exists for a broad range of parameters, which makes this idea hard to implement.

EXERCISE 7.2.9. Prove that, given the parameter values used for Figure 7.5, the function $F$ satisfies $F'(t) \to \infty$ as $t \downarrow 0$.

# 7.3  General Representations

We have discussed two well-known examples of recursive preferences. In this section we build a general representation. While various constructions can be found in the decision theory literature, many are not well suited to quantitative work. Here we give a relatively parsimonious operator-theoretic definition.

## 7.3.1  Koopmans Operators

In §7.2.2.3 and §7.2.3.1 we met risk-sensitive and Epstein–Zin Koopmans operators respectively. In this section we provide a general definition of a Koopmans operator that will contain these two examples as special cases.

We begin by outlining structure that can be combined to generate Koopmans operators in a Markov environment. The two key components are an aggregation function and a certainty equivalent operator. We then build Koopmans operators from these primitives and connect them to applications. In every setting we consider, lifetime value is identified with the unique fixed point of the Koopmans operator (whenever it exists).

### 7.3.1.1  Certainty Equivalents

The first primitive we consider is a generalization of conditional expectations: Given $V \subset \mathbb{R}^\mathsf{X}$, we define a **certainty equivalent operator** on $V$ to be a self-map $R$ on $V$ such that

(i) $R$ is order-preserving on $V$ and

(ii) all constants are fixed under $R$ (i.e., $R(\lambda\mathbb{1}) = \lambda\mathbb{1}$ for all $\lambda \in \mathbb{R}$ with $\lambda\mathbb{1} \in V$).

**Example 7.3.1.** The usual conditional expectations operator is a certainty equivalent operator. To see this, set $V = \mathbb{R}^\mathsf{X}$ and fix $P \in \mathcal{M}(\mathbb{R}^\mathsf{X})$. Since $f, g \in \mathbb{R}^\mathsf{X}$ with $f \leqslant g$ implies $Pf \leqslant Pg$ and $P(\lambda\mathbb{1}) = \lambda P\mathbb{1} = \lambda\mathbb{1}$, we see that $P$ satisfies (i)–(ii) above.

EXERCISE 7.3.1. In the last example, the certainty equivalent $R = P$ is linear. Prove that this is the only linear case. In particular, prove the following: if $\mathbf{R}(\mathsf{X})$ is the set of all certainty equivalent operators on $\mathbb{R}^{\mathsf{X}}$, then $\mathbf{R}(\mathsf{X}) \cap \mathcal{L}(\mathbb{R}^{\mathsf{X}}) = \mathcal{M}(\mathbb{R}^{\mathsf{X}})$.

The next example is nonlinear. It treats the risk-adjusted expectation that appears in risk-sensitive preferences.

**Example 7.3.2.** Let $V = \mathbb{R}^{\mathsf{X}}$ and fix nonzero $\theta$ and $P \in \mathcal{M}(\mathbb{R}^{\mathsf{X}})$. The **entropic certainty equivalent operator** is the operator $R_\theta$ on $V$ defined by

$$(R_\theta v)(x) = \frac{1}{\theta} \ln \left\{ \sum_{x'} \exp(\theta v(x')) P(x, x') \right\} \qquad (v \in V, \ x \in \mathsf{X}).$$

EXERCISE 7.3.2. Show that $R_\theta$ is in fact a certainty equivalent operator.

**Example 7.3.3.** As a third example, let $V$ be the interior of the positive cone, as in §7.2.3.2, and fix $P \in \mathcal{M}(\mathbb{R}^{\mathsf{X}})$. The operator

$$(R_\gamma v)(x) = \left\{ \sum_{x'} v(x')^\gamma P(x, x') \right\}^{1/\gamma} \qquad (v \in V, \ x \in \mathsf{X}, \ \gamma \neq 0) \tag{7.16}$$

is a certainty equivalent operator on $V$. The map $R_\gamma$ is sometimes called the **Kreps-Porteus certainty equivalent operator** in honor of Kreps and Porteus (1978). We met $R_\gamma$ in §7.2.3, when we discussed Epstein–Zin preferences.

EXERCISE 7.3.3. Confirm that $R_\gamma$ is a certainty equivalent operator.

EXERCISE 7.3.4. Let $V = \mathbb{R}^{\mathsf{X}}$ and fix $P \in \mathcal{M}(\mathbb{R}^{\mathsf{X}})$ and $\tau \in [0, 1]$. Let $R_\tau$ be the **quantile certainty equivalent**. That is, $(R_\tau v)(x) = Q_\tau v(X)$ where $X \sim P(x, \cdot)$ and $Q_\tau$ is the quantile functional (see page 32). More specifically,

$$(R_\tau v)(x) = \min \left\{ y \in \mathbb{R} \ \Big| \ \sum_{x'} \mathbb{1}\{v(x') \leqslant y\} P(x, x') \geqslant \tau \right\} \qquad (v \in V, \ x \in \mathsf{X}).$$

Confirm that $R_\tau$ defines a certainty equivalent operator on $V$.

EXERCISE 7.3.5. Let $R$ be a certainty equivalent operator on $V \subset \mathbb{R}_+^{\mathsf{X}}$, where $\lambda \mathbb{1} \in V$ for all $\lambda \geqslant 0$. Prove that $R0 = 0$ and $Rv \geqslant 0$ whenever $v \geqslant 0$.

The set of certainty equivalent operators on $\mathbb{R}^X$ is invariant under convex combinations, as the next exercise asks you to confirm.

EXERCISE 7.3.6. Let $\mathbf{R}(X)$ be the set of certainty equivalent operators on $\mathbb{R}^X$ and prove the following:

$$R_a, R_b \in \mathbf{R}(X) \text{ and } 0 \leqslant \lambda \leqslant 1 \implies \lambda R_a + (1 - \lambda)R_b \in \mathbf{R}(X).$$

### 7.3.1.2 Properties

A certainty equivalent operator $R$ on $V$ is called

- **positive homogeneous** on $V$ if $R(\lambda v) = \lambda R v$ for all $v \in V$ and $\lambda \geqslant 0$ with $\lambda v \in V$,

- **superadditive** on $V$ if $R(v + w) \geqslant Rv + Rw$ for all $v, w \in V$ with $v + w \in V$,

- **subadditive** on $V$ if $R(v + w) \leqslant Rv + Rw$ for all $v, w \in V$ with $v + w \in V$,

- **constant-subadditive** on $V$ if $R(v + \lambda \mathbb{1}) \leqslant Rv + \lambda \mathbb{1}$ for all $v \in V$ and $\lambda \geqslant 0$ with $v + \lambda \mathbb{1} \in V$.

**Example 7.3.4.** Given $P \in \mathcal{M}(\mathbb{R}^X)$, the linear certainty equivalent $R = P$ is positive homogeneous and both superadditive and subadditive on $\mathbb{R}^X$.

**Example 7.3.5.** Let $V$ be the interior of the positive cone of $\mathbb{R}^X$ and fix $P \in \mathcal{M}(\mathbb{R}^X)$. In this setting, the Kreps–Porteus certainty equivalent operator $R_\gamma$ in (7.16) is subadditive on $V$ when $\gamma \geqslant 1$ and superadditive on $V$ when $\gamma \leqslant 1$ (and, as usual, $\gamma \neq 0$). The subadditive case follows directly from Minkowski's inequality, while the superadditive case follows from the mean inequalities in Bullen (2003) (p. 202).

EXERCISE 7.3.7. Prove that the quantile certainty equivalent operator $R_\tau$ from Exercise 7.3.4 is constant-subadditive.

EXERCISE 7.3.8. Show that the entropic certainty equivalent operator $R_\theta$ from Example 7.3.2 is constant-subadditive.

EXERCISE 7.3.9. Prove: If $R$ is constant-subadditive on $V$, then $R$ is nonexpansive with respect to the supremum norm. That is,

$$\|Rv - Rw\|_\infty \leqslant \|v - w\|_\infty \quad \text{for all } v, w \in \mathbb{R}^X.$$

In some instances, a certainty equivalent operator is either convex or concave in the sense of §7.1.2.2.

**Example 7.3.6.** The entropic certainty equivalent operator $R_\theta$ in Example 7.3.2 is concave on $\mathbb{R}^{\mathsf{X}}$ whenever $\theta < 0$. To prove this we use the result in Föllmer and Knispel (2011) which states that, for $\theta < 0$, $0 \leqslant \alpha \leqslant 1$ and finite-valued random variables $Z, Z'$, we have

$$\mathcal{E}_\theta(\alpha Z + (1-\alpha)Z') \geqslant \alpha \mathcal{E}_\theta(Z) + (1-\alpha)\mathcal{E}_\theta(Z') \tag{7.17}$$

where $\mathcal{E}_\theta$ is as defined in §7.2.2.2.

EXERCISE 7.3.10. Using (7.17), show that $R_\theta$ is concave on $V = \mathbb{R}^{\mathsf{X}}$ when $\theta < 0$.

EXERCISE 7.3.11. Let $V$ be convex and let $R$ be a certainty equivalent operator on $V$. Prove the following:

  (i)  $R$ is convex on $V$ whenever $R$ is subadditive and positive homogeneous on $V$.

  (ii)  $R$ is concave on $V$ whenever $R$ is superadditive and positive homogeneous on $V$.

Combining Exercise 7.3.11 and Example 7.3.5, we have proved

**Lemma 7.3.1.** *The Kreps–Porteus certainty equivalent operator $R_\gamma$ in (7.16) is convex on $V$ when $\gamma \geqslant 1$ and concave on $V$ when $\gamma \leqslant 1$.*

Later we will combine Lemma 7.3.1 with the fixed point results for convex and concave operators in §7.1.2.2 to establish existence and uniqueness of lifetime values for certain kinds of Koopmans operators.

### 7.3.1.3  Monotonicity

Let $\mathsf{X}$ be partially ordered and let $i\mathbb{R}^{\mathsf{X}}$ be the set of increasing functions in $\mathbb{R}^{\mathsf{X}}$. Let $V$ be such that $i\mathbb{R}^{\mathsf{X}} \subset V \subset \mathbb{R}^{\mathsf{X}}$ and let $R$ be a certainty equivalent on $V$. We call $R$ **monotone increasing** if $R$ is invariant on $i\mathbb{R}^{\mathsf{X}}$. This extends the terminology in §3.2.1.3, where we applied it to Markov operators (cf., Exercise 3.2.4 on page 95).

As shown below, the concept of monotone increasing certainty equivalent operators is connected to outcomes where lifetime preferences are increasing in the state.

EXERCISE 7.3.12. Show that the entropic certainty equivalent operator in Example 7.3.2 is monotone increasing on $V = \mathbb{R}^{\mathsf{X}}$ whenever $P$ is monotone increasing, for all nonzero values of $\theta$.

EXERCISE 7.3.13. Show that the Kreps–Porteus certainty equivalent operator in Example 7.3.3 is monotone increasing on $V = (0, \infty)^{\mathsf{X}}$ whenever $P$ is monotone increasing, for all nonzero values of $\gamma$.

### 7.3.1.4  Aggregation

We mentioned above that Koopmans operators are typically constructed by combining a certainty equivalent operator and an aggregation function. Let's now discuss the second of these components.

Given $V \subset \mathbb{R}^{\mathsf{X}}$, an **aggregator** $A$ on $V$ is a map $A$ from $\mathsf{X} \times \mathbb{R}$ to $\mathbb{R}$ such that

 (i) $w(x) = A(x, v(x))$ is in $V$ whenever $v \in V$ and

 (ii) $y \mapsto A(x, y)$ is increasing for all $x \in \mathsf{X}$.

Intuitively, an aggregator combines current state and continuation values to measure lifetime value.

Common types of aggregators include the

- **Leontief aggregator** $A_{\text{MIN}}(x, y) = \min\{r(x), \beta y\}$ with $r \in \mathbb{R}^{\mathsf{X}}$ and $\beta \geqslant 0$,

- **Uzawa aggregator** $A_{\text{UZAWA}}(x, y) = r(x) + b(x)y$ with $r \in \mathbb{R}^{\mathsf{X}}$ and $b \in \mathbb{R}_+^{\mathsf{X}}$, and

- **CES aggregator** $A_{\text{CES}}(x, y) = \{r(x)^\alpha + \beta y^\alpha\}^{1/\alpha}$ with $r \in (0, \infty)^{\mathsf{X}}$, $\beta \geqslant 0$ and $\alpha \neq 0$.

Here CES stands for "constant elasticity of substitution." An important special case of both the CES and Uzawa aggregators is the

- **additive aggregator** $A_{\text{ADD}}(x, y) = r(x) + \beta y$ with $r \in \mathbb{R}^{\mathsf{X}}$ and $\beta \geqslant 0$.

From these basic types we can also build composite aggregators. For example, we might consider a CES-Uzawa aggregator of the form $A(x, y) = \{r(x)^\alpha + b(x)y^\alpha\}^{1/\alpha}$ with $r, b \in \mathbb{R}^{\mathsf{X}}$, $b \geqslant 0$ and $\alpha \neq 0$. As we will see in §7.3.3.3, the CES-Uzawa aggregator can be used to construct models with both Epstein–Zin utility and state-dependent discounting (as in, say, Albuquerque et al. (2016) or Schorfheide et al. (2018).)

### 7.3.1.5  Building Koopmans Operators

We are now ready to build Koopmans operators by combining certainty equivalents and aggregators. Given $V \subset \mathbb{R}^{\mathsf{X}}$, we call a self-map $K$ on $V$ a **Koopmans operator** if

$$K = A \circ R \tag{7.18}$$

for some aggregator $A$ and certainty equivalent operator $R$ on $V$. The expression in (7.18) means that $(Kv)(x) = A(x, (Rv)(x))$ at $v \in V$ and $x \in \mathsf{X}$.

It is generally appropriate to suppose that a uniform increase in continuation values will increase current value. This property holds for $K$ in (7.18). In particular, it follows from the definitions of $A$ and $R$ that $K$ is an order-preserving self-map on $V$.

**Example 7.3.7.** For risk-sensitive preferences, the Koopmans operator (page 224) can be expressed as $K_\theta = A_{\text{ADD}} \circ R_\theta$, where $R_\theta$ is the entropic certainty equivalent operator.

**Example 7.3.8.** The Epstein–Zin Koopmans operator can be expressed as $K = A_{\text{CES}} \circ R_\gamma$, where $R_\gamma$ is the Kreps–Porteus expectations operator, as defined in (7.16). This is a version of (7.18) under the CES aggregator.

**Remark 7.3.1.** We defined time additive preferences somewhat loosely in §3.2.2.3. Here is a better definition: The Koopmans operator $K = A \circ R$ is **time additive** if $A = A_{\text{ADD}}$ and $R$ is ordinary condition expectations (as in Example 7.3.1).

### 7.3.1.6 Comments on CES Aggregation

The CES aggregator is so-named because, in a static utility maximization problem where $c$ and $y$ are two goods and utility is $U(c, y) = ((1-\beta)c^\alpha + \beta y^\alpha)^{1/\alpha}$, the elasticity of substitution is constant and given by $1/(1-\alpha)$. In the present setting, where aggregation is across time, $1/(1-\alpha)$ is usually called the **elasticity of intertemporal substitution** (EIS). The next exercise explains.

EXERCISE 7.3.14. Consider $U(c, y) = ((1-\beta)c^\alpha + \beta y^\alpha)^{1/\alpha}$ as a utility function over current and future goods $c$ and $y$. Then

$$\text{EIS} = \frac{d\ln(y/c)}{d\ln(U_c/U_y)} \quad \text{where} \quad U_c := \frac{\partial U(c, y)}{\partial c} \quad \text{and} \quad U_y := \frac{\partial U(c, y)}{\partial y}.$$

Confirm that $\text{EIS} = 1/(1-\alpha)$.

The fact that $\text{EIS} = 1/(1-\alpha)$ under the CES aggregator is significant because the EIS can be measured from data using regression and other techniques. While estimates vary significantly, the detailed meta-analysis by Havranek et al. (2015) suggests 0.5 as a plausible average value for international studies, with rich countries tending slightly higher. Basu and Bundick (2017) use 0.8 when calibrating to US data. Under these estimates, the relationship $\text{EIS} = 1/(1-\alpha)$ implies a value for $\alpha$ between -1.0 and -0.25.

### 7.3.1.7 Lifetime Value

In §7.3.1.5 we constructed a generic Koopmans operator using an aggregator and a certainty equivalent operator. In this section we connect this Koopmans operator to lifetime values and discuss the significance of global stability.

To begin, fix set $\mathsf{X}$ and function class $V \subset \mathbb{R}^\mathsf{X}$. Let $K = A \circ R$ be a Koopmans operator for some aggregator $A$ and certainty equivalent operator $R$ on $V$. The **lifetime value** generated by $K$ is the unique fixed point of $K$ in $V$, whenever it exists. Given such a $v$, the value $v(x)$ is interpreted as lifetime value conditional on initial state $x$.

**Example 7.3.9.** In the case of time additive preferences, lifetime value was defined in (3.21) by $v = (I - \beta P)^{-1} r$. Equivalently, $v$ is the fixed point of the operator $K$ defined by $Kv = r + \beta Pv$. Since $K$ is globally stable, the fixed point is unique. In view of Lemma 3.2.1 on page 95, it satisfies

$$v(x) = \mathbb{E} \sum_{t \geqslant 0} \beta^t r(X_t) \quad \text{when } (X_t) \text{ is } P\text{-Markov and } X_0 = x.$$

**Example 7.3.10.** By Proposition 7.2.2, the risk-sensitive Koopmans operator $K_\theta = A_{\mathrm{ADD}} \circ R_\theta$ is globally stable on $V = \mathbb{R}^\mathsf{X}$ when $\beta \in (0, 1)$. In this setting, the unique fixed point of $K_\theta$ in $V$ is interpreted as lifetime value under the risk-sensitive preferences described in §7.2.2.

In many applications, our existence and uniqueness proofs for fixed points of $K$ will also establish global stability. For Koopmans operators, global stability has the following interpretation: for $w \in V$, $m \in \mathbb{N}$ and $x \in \mathsf{X}$, the value $(K^m w)(x)$ gives total finite-horizon utility over periods $0, \ldots, m$ under the preferences embedded in $K$, with initial state $x$ and terminal condition $w$. Hence global stability implies that, for any choice of terminal condition, finite-horizon utility converges to infinite-horizon utility as the time horizon converges to infinity. The next exercise helps to illustrate this point.

EXERCISE 7.3.15. Consider again the time additive preferences $V_t = u(C_t) + \beta \mathbb{E}_t V_{t+1}$ in Example 7.3.9. Suppose that the time horizon is finite, with some exogenous terminal value $V_m = w(X_m)$ at time $m$. Letting $v_m(x)$ represent lifetime value up until time $m$, conditional on initial state $x$, show that

(i) $v_m = \sum_{t=0}^{m-1} (\beta P)^t r + (\beta P)^m w$,

(ii) $v_m = K^m w$, where $K$ is the associated Koopmans operator $Kv = r + \beta Pv$ and,

(iii) $K^m w \to v^* := (I - \beta P)^{-1} r$ as $m \to \infty$.

Exercise 7.3.15 confirms that, at least for the time additive dcase, global stability of $K$ is equivalent to the statement that a finite-horizon valuation with arbitrary terminal condition $w$ converges to the infinite-horizon valuation.

### 7.3.1.8 Monotone Lifetime Values

Let $\mathsf{X} = (X, \preceq)$ be partially ordered, let $i\mathbb{R}^{\mathsf{X}}$ be the set of increasing functions in $\mathbb{R}^{\mathsf{X}}$, and let $V$ be such that $i\mathbb{R}^{\mathsf{X}} \subset V \subset \mathbb{R}^{\mathsf{X}}$. Let $K$ be a Koopmans operator on $V$, so that $Kv = A \circ R$ for some aggregator $A$ and certainty equivalent operator $R$ on $V$. Suppose that $K$ has a unique fixed point $v^* \in V$. A natural question is: when is $v^*$ increasing in the state?

**Lemma 7.3.2.** *If $K$ is globally stable, then $v^*$ is increasing on $\mathsf{X}$ whenever the following two conditions hold:*

(i) *$A(x, v) \leqslant A(x', v)$ whenever $v \in V$ and $x \preceq x'$, and*

(ii) *$R$ is monotone increasing on $V$.*

*Proof.* It is not difficult to check that, under the stated conditions, $K$ is invariant on $i\mathbb{R}^{\mathsf{X}}$. It follows from Exercise 1.2.18 on page 22 that $v^*$ is increasing on $\mathsf{X}$. $\qquad\square$

EXERCISE 7.3.16. Consider the Epstein–Zin Koopmans operator $K = A_{\text{CES}} \circ R_\gamma$ on $V$, where $V := (0, \infty)^{\mathsf{X}}$ and the primitives is as in (7.13). Assume the conditions of Proposition 7.2.3, so that $K$ has a unique fixed point $v^*$ in $V$. Given $P \in \mathcal{M}(\mathbb{R}^{\mathsf{X}})$, we can write $R_\gamma$ as $R_\gamma v = (P v^\gamma)^{1/\gamma}$ at each $v \in V$. Prove that $v^*$ is increasing in $\mathsf{X}$ whenever $P$ is monotone increasing and $c \in i\mathbb{R}^{\mathsf{X}}$.

## 7.3.2 A Blackwell-Type Condition

Let $R$ be a certainty equivalent operator on $V = \mathbb{R}^{\mathsf{X}}$ and let $A$ be an aggregator on $V$. Let $K$ be the Koopmans operator on $V$ defined by $(Kv)(x) = A(x, (Rv)(x))$. When $R$ is constant-subadditive, we can often establish global stability of $K$ on $V$ via a contraction mapping argument. This section gives details.

### 7.3.2.1 Blackwell Aggregators

We call an aggregator $A$ on $V$ a **Blackwell aggregator** if there exists a $\beta \in (0, 1)$ such that

$$A(x, y + \lambda) \leqslant A(x, y) + \beta\lambda \tag{7.19}$$

for all $x \in \mathsf{X}$, $y \in \mathbb{R}$ and $\lambda \in \mathbb{R}_+$.

EXERCISE 7.3.17. Fix $\beta \in \mathbb{R}_+$ and $r \in \mathbb{R}^\mathsf{X}$. Show that the additive aggregator $A(x, y) = r(x) + \beta y$ and the Leontief aggregator $A(x, y) = \min\{r(x), \beta y\}$ are Blackwell aggregators when $\beta < 1$.

The next proposition states conditions for global stability in settings where aggregators have the Blackwell property.

**Proposition 7.3.3.** *If $A$ is a Blackwell aggregator and $R$ is constant-subadditive, then the Koopmans operator $K := A \circ R$ is a contraction on $V$ with respect to $\|\cdot\|_\infty$.*

*Proof.* Let the primitives be as stated. In view of Lemma 2.2.4 on page 63, and taking into account the fact that $K$ is order-preserving, we need only show that there exists a $\beta \in (0, 1)$ with $K(v + \lambda) \leqslant Kv + \beta\lambda$ for all $v \in V$ and $\lambda \in \mathbb{R}_+$. To see this, fix $v \in V$ and $\lambda \in \mathbb{R}_+$. Applying constant-subadditivity of $R$ and monotonicity of $A$, we have

$$K(v + \lambda) = A(\cdot, R(v + \lambda)) \leqslant A(\cdot, Rv + \lambda)$$

Since $A$ is a Blackwell aggregator, the last term is bounded by $A(\cdot, Rv) + \beta\lambda$ with $\beta < 1$. Hence $K(v + \lambda) \leqslant Kv + \beta\lambda$, and $K$ is a contraction of modulus $\beta$ on $V$. □

The stability of time additive preferences is a special case of Proposition 7.3.3.

### 7.3.2.2 The Risk-Sensitive Case

We can now complete the proof of Proposition 7.2.2, which concerned global stability of the Koopmans operator generated by risk-sensitive preferences.

*Proof of Proposition 7.2.2.* Fix $\theta \neq 0$ and recall that $K_\theta$ in (7.10) can be expressed as $K_\theta = A_{\text{ADD}} \circ R_\theta$ when $R_\theta$ is the entropic certainty equivalent. Since $A_{\text{ADD}}$ is a Blackwell aggregator and $R_\theta$ is constant-subadditive (Exercise 7.3.8), Proposition 7.3.3 applies. In particular, $K_\theta$ is globally stable on $\mathbb{R}^\mathsf{X}$. □

EXERCISE 7.3.18. Let $K = A_{\text{MIN}} \circ R$ on $V = \mathbb{R}^\mathsf{X}$. Prove that $K$ is globally stable on $V$ whenever $R$ is constant-subadditive and $A_{\text{MIN}}(x, y) = \min\{r(x), \beta y\}$ with $\beta \in (0, 1)$.

### 7.3.2.3 Quantile Preferences

Consider a setting where $V = \mathbb{R}^X$ and $K_\tau := A_{\text{ADD}} \circ R_\tau$. That is,

$$(K_\tau v)(x) = r(x) + \beta(R_\tau v)(x) \qquad (x \in X) \tag{7.20}$$

for $\beta \in (0,1)$, $\tau \in [0,1]$, $r \in \mathbb{R}^X$ and $R_\tau$ as described in Exercise 7.3.4. Since $R_\tau$ is constant-subadditive (Exercise 7.3.7) and the additive aggregator is Blackwell, $K_\tau$ is globally stable (Proposition 7.3.3). The operator $K_\tau$ represents quantile preferences, as described in de Castro and Galvao (2019) and other studies (see 7.4). The value $\tau$ parameterizes attitude to risk, a point we return to in §8.2.1.4.

**EXERCISE 7.3.19.** Consider replacing the operator $K_\tau$ in (7.20) with $K = A_{\text{MIN}} \circ R_\tau$. Under the same assumptions as above (apart from the switch to Leontief aggregator), prove that $K$ is globally stable.

## 7.3.3 Uzawa Aggregation

Let's consider the Koopmans operator $K = A_{\text{UZAWA}} \circ R$, where $V$ is some subset of $\mathbb{R}^X$ and $R$ is a certainty equivalent operator on $V$. In particular,

$$(Kv)(x) = r(x) + b(x)(Rv)(x) \qquad (x \in X, \ v \in V) \tag{7.21}$$

with $r, b \in \mathbb{R}^X$ and $b \geqslant 0$. We are interested in conditions that imply $K$ is globally stable on $V$.

### 7.3.3.1 The Case of Conditional Expectation

Let $V = \mathbb{R}^X$ and suppose $R = P$ for some $P \in \mathcal{M}(\mathbb{R}^X)$, so that $R$ is ordinary conditional expectations. Then $K$ becomes $Kv = r + Lv$ where $L \in \mathcal{L}(\mathbb{R}^X)$ with $L(x, x') = b(x)P(x, x')$. By Exercise 1.2.17 on page 22, $K$ is globally stable on $V$ whenever $\rho(L) < 1$.

This kind of structure arises when households derive utility from a consumption path while their discount factor fluctuates according to some state variable (see, e.g., Krusell and Smith (1998), Toda (2019), Cao (2020), and Hubmer et al. (2020)). For a given consumption path $(C_t)$, lifetime values takes the form

$$v(x) = \mathbb{E}_x \sum_{t=0}^{\infty} \left( \prod_{i=1}^{t} \beta_i \right) u(C_t) \tag{7.22}$$

where $u$ is a flow utility function and $\{\beta_t\}$ is a discount factor process. Suppose $C_t = c(X_t)$ and $\beta_t = b(X_t)$ where $b \geqslant 0$ and $(X_t)$ is $P$-Markov for some $P \in \mathcal{M}(\mathbb{R}^\mathsf{X})$. Set $r := u \circ c$ and $L(x, x') := b(x)P(x, x')$. By Theorem 6.1.1 on page 184, the condition $\rho(L) < 1$ implies that $v$ in (7.22) is the unique fixed point of $Kv = r + Lv = r + bPv$. In other words, lifetime value under (7.22) is the unique fixed point of the Koopmans operator when the aggregator is of Uzawa type and the certainty equivalent is conditional expectation.

How does this relate to optimization? Recall our discussion of state-dependent MDPs in Chapter 6. There, the policy operator $T_\sigma$ in (6.16) on page 193 is a special case of (7.21) when the discount factor depends only on the current state and action.

With some additional requirements, the condition $\rho(L) < 1$ is necessary as well as sufficient for existence of a unique fixed point for $Kv = r + Lv$. Indeed, if $b \gg 0$ and $P$ is irreducible, then $L$ is also irreducible and a positive linear operator. Applying Lemma 6.1.4, we see that $r \gg 0$ and $\rho(L) \geqslant 1$ implies $Kv = r + Lv$ has no fixed point in $V := \{v \in \mathbb{R}^\mathsf{X} : v \gg 0\}$.

EXERCISE 7.3.20. Confirm that $L$ is irreducible when $b \gg 0$ and $P$ is irreducible.

### 7.3.3.2 Stability via Concavity

Now consider $Kv = r + bRv$ from (7.21) when $R$ is not in $\mathcal{M}(\mathbb{R}^\mathsf{X})$. Here $bRv$ is the pointwise product, so that $(bRv)(x) = b(x)(Rv)(x)$ for all $x$.

We cannot use Proposition 7.3.3 to prove stability of $K$ unless $b(x) < 1$ for all $x \in \mathsf{X}$. Since this condition is rather strict, we now study weaker conditions that can be valid even when $b$ exceeds 1 in some states. Specifically, we consider

(a) $bRv \leqslant c + Lv$ for some $c \in \mathbb{R}^\mathsf{X}$ and $L \in \mathcal{L}(\mathbb{R}^\mathsf{X})$ with $\rho(L) < 1$.

(b) $r \gg 0$ and $R$ is concave on $\mathbb{R}_+^\mathsf{X}$.

Let $V = [0, \bar{v}]$ where $\bar{v} := (I - L)^{-1}(r + c)$.

**Proposition 7.3.4.** *If conditions* (a)–(b) *hold, then $K$ is globally stable on $V$.*

*Proof.* Under (a)–(b), $K$ is concave on $\mathbb{R}_+^\mathsf{X}$, with

$$0 \ll r = r + bR0 = K0 \quad \text{and} \quad K\bar{v} = r + bR\bar{v} \leqslant r + c + L\bar{v} = r + c - (I - L)\bar{v} + \bar{v} = \bar{v}.$$

The claim now follows from Du's theorem (page 217). $\qquad \square$

### 7.3.3.3   Epstein–Zin Preferences with State-Dependent Discounting

Combining the CES-Uzawa aggregator $A(x, y) = \{r(x)^\alpha + b(x)y^\alpha\}^{1/\alpha}$ with the Kreps–Porteus certainty equivalent operator leads to the Koopmans operator

$$Kv = \left\{ h + b \left[ Pv^\gamma \right]^{\alpha/\gamma} \right\}^{1/\alpha}, \quad \text{with} \quad h, b \in \mathbb{R}_+^{\mathsf{X}}. \tag{7.23}$$

A fixed point of $K$ corresponds to lifetime value for an agent with Epstein–Zin preferences and state-dependent discounting. (Such set ups are used in research on macroeconomic dynamics and asset pricing – see §7.4 for more details).

In what follows we take $V = (0, \infty)^{\mathsf{X}}$ and assume that $h, b \in V$ and $P$ is irreducible.

EXERCISE 7.3.21.  Show that $K$ is self-map on $V$.

To discuss stability of $K$ we introduce the operator $A \in \mathcal{L}(\mathbb{R}^{\mathsf{X}})$ defined by

$$(Av)(x) := b(x)^\theta \sum_{x'} v(x') P(x, x') \quad \text{where} \quad \theta := \frac{\gamma}{\alpha}.$$

**Proposition 7.3.5.**  *$K$ is globally stable on $V$ if and only if $\rho(A)^{\alpha/\gamma} < 1$.*

To prove Proposition 7.3.5, we proceed as in §7.2.3.3, constructing a conjugate operator $\hat{K}$ and proving stability of the latter. For this purpose, we introduce

$$\hat{K}v = \left\{ h + (Av)^{1/\theta} \right\}^\theta \qquad (v \in V), \tag{7.24}$$

Also, let $\Phi$ be defined by $\Phi v = v^\gamma$.

EXERCISE 7.3.22.  Prove: $(V, K)$ and $(V, \hat{K})$ are topologically conjugate under $\Phi$.

*Proof of Proposition 7.3.5.*  In view of Exercise 7.3.22, it suffices to show that $\hat{K}$ is globally stable on $V$ if and only if $\rho(A)^{\alpha/\gamma} < 1$. This is implied by Theorem 7.1.4, since $A$ is irreducible (see Exercise 7.3.20 on page 242) and $\rho(A)^{1/\theta} = \rho(A)^{\alpha/\gamma}$.                    □

## 7.4   Chapter Notes

The time additive preference structure in §7.2.1 was popularized by Samuelson (1939), who built on earlier work by Fisher (1930) and Ramsey (1928). An axiomatic foun-dation was supplied by Koopmans (1960). Bastianello and Faro (2022) study the foundations of discounted expected utility from a purely subjective framework.

Problems with the time additive discounted utility model include non-constant discounting, as discussed in §6.4, as well as sign effects (gains being discounted more than losses) and magnitude effects (small outcomes being discounted more than large ones). See, for example, Thaler (1981) and Benzion et al. (1989). A critical review of the time additive model and a list of many references can be found in Frederick et al. (2002).

In the stochastic setting, the time additive framework is a subset of the expected utility model (Von Neumann and Morgenstern (1944), Friedman (1956), Savage (1951)). There are many well documented departures from expected utility in experimental data. See the start of Andreoni and Sprenger (2012) and the article Ericson and Laibson (2019) for an introduction to the literature. An interesting historical discussion of time additive expected utility can be found in Becker et al. (1989).

(It is ironic that those most responsible for popularizing the time additive discounted expected utility (DEU) framework have also been among the most critical. For example, Samuelson (1939) stated that it is "completely arbitrary" to assume that the DEU specification holds. He goes on to claim that, in the analysis of savings and consumption, it is "extremely doubtful whether we can learn much from considering such an economic man." In addition, Stokey and Lucas (1989), whose work helped to standardize DEU as a methodology for quantitative analysis, argued in a separate study that DEU is attractive only because of its relative simplicity (Lucas and Stokey, 1984).)

Do the departures from time additive expected utility found in experimental data actually matter for quantitative work? Evidence suggests that the answer is affirmative. In macroeconomics and asset pricing in particular, researchers increasingly use non-additive preferences in order to bring model outputs closer to the data. For example, many quantitative models of asset pricing rely heavily on Epstein–Zin preferences. Representative examples include Epstein and Zin (1991), Tallarini Jr (2000), Bansal and Yaron (2004), Hansen et al. (2008), Bansal et al. (2012), Schorfheide et al. (2018), and de Groot et al. (2022). Alternative numerical solution methods are discussed in Pohl et al. (2018).

An excellent introduction to recursive preference models can be found in Backus et al. (2004). Our use of the term "Koopmans operator," which is not entirely standard, honors early contributions by Nobel laureate Tjalling Koopmans on recursive preferences (see Koopmans (1960) and Koopmans et al. (1964)).

Theoretical properties of recursive preference models have been studied in many papers, including Epstein and Zin (1989), Weil (1990), Boyd (1990), Hansen and Scheinkman (2009), Marinacci and Montrucchio (2010), Bommier et al. (2017), Bloise and Vailakis (2018), Marinacci and Montrucchio (2019), Pohl et al. (2019), Balbus (2020), Borovička and Stachurski (2020), DeJarnette et al. (2020), Christensen (2022), and Becker and Rincon-Zapatero (2023). The paper by Marinacci and Montrucchio (2019) provides a useful alternative approach to existence of unique fixed points in the setting of order-preserving maps. Experimental results on Epstein–Zin preferences can be found in Meissner and Pfeiffer (2022).

There is a strong connection between risk-sensitive preferences and the literature on robust control. See, for example, Cagetti et al. (2002), Hansen and Sargent (2007), and Barillas et al. (2009). We return to this point in Chapter 8.

The quantile preferences we considered in §7.3.2.3 have been analyzed in static and dynamic settings by Giovannetti (2013), de Castro and Galvao (2019), de Castro and Galvao (2022) and de Castro et al. (2022). Recursive components of the analysis of quantile and Uzawa preference models build on the study of monotone preferences in Bommier et al. (2017).

Some recursive preference specifications involve ambiguity aversion. An introduction to this literature and its applications can be found in Klibanoff et al. (2009), Hayashi and Miao (2011), Hansen and Miao (2018), Bommier et al. (2019) and Hansen and Sargent (2020). Marinacci et al. (2023) discuss the connection between recursivity and attitudes to uncertainty. We discuss ambiguity again in Chapter 8.

Recursive preferences are increasingly applied outside the field of asset pricing, where they first came to prominence. See, for example, Bommier and Villeneuve (2012), Colacito et al. (2018), Jensen (2019), or Augeraud-Véron et al. (2019).

The coin flip application in §7.2.1.2 is related to correlation aversion, as discussed in Stanca (2023), and preference for "consumption spreads" as reviewed in Frederick et al. (2002).

Some applications of Theorem 7.1.3 to network analysis can be found in Sargent and Stachurski (2023b).

# Chapter 8

# Recursive Decision Processes

While the MDP model from Chapters 5–6 is elegant and widely used, researchers in economics, finance, and other fields are working to extend it. Reasons include:

(i) MDP theory cannot be applied to settings where lifetime values are described by the kinds of nonlinear recursions discussed in Chapter 7.

(ii) Equilibria in some models of production and economic geography can be computed using dynamic programming but not all such programming problems fit within the MDP framework.

(iii) Dynamic programming problems that include adversarial agents to promote robust decision rules can fail to be MDPs.

To handle such departures from the MDP assumptions, we now construct a more abstract dynamic programming framework, building on an approach to optimization initially developed by Denardo (1967) and greatly extended by Bertsekas (2022b). Further references are provided in §8.4.

Apart from generality, another motivation for greater abstraction is a desire for clarity: while we have already provided some optimality proofs (such as the proof of Proposition 5.1.1 on optimality for MDPs), the abstract DP framework allows us to strip away extraneous structure and isolate roles of key assumptions.

Our discussion of abstract dynamic programming begins with this chapter and continues in Chapter 9. We start with a framework that centers on an abstract representation of the Bellman equation (§8.1). We then state optimality results and show how they can be verified in a range of applications. We defer proofs of optimality results to Chapter 9, where we strip dynamic programs down to their essence by adopting a purely operator-theoretic approach. This Puritanical approach simplifies proofs and allows us to extend dynamic programming to a broader class of applications.

# 8.1 Definition and Properties

In this section we introduce and analyze optimality conditions for recursive decision processes that include and extend all dynamic programming frameworks discussed so far. Throughout this chapter, $\mathsf{X}$ denotes a finite set.

## 8.1.1 Defining RDPs

Consider a generic Bellman equation of the form

$$v(x) = \max_{a \in \Gamma(x)} B(x, a, v). \tag{8.1}$$

Here $x$ is the state, $a$ is an action, $\Gamma$ is a feasible correspondence, and $B$ is an "aggregator" function. We understand $\Gamma(x)$ as all actions available to the controller in state $x$. The function $v$ assigns values to states and is a member of some class $V \subset \mathbb{R}^{\mathsf{X}}$. This "abstract" Bellman equation generalizes all of the Bellman equations presented in previous chapters.

Our plan is to analyze the Bellman equation (8.1) and state conditions on $B$ and the other primitives that make strong optimality properties hold. As a first step, we introduce two finite sets,

- an **action space** $\mathsf{A}$ and
- a **state space** $\mathsf{X}$.

Given $\mathsf{X}$ and $\mathsf{A}$, we define a **recursive decision process** (RDP) to be a triple $\mathcal{R} = (\Gamma, V, B)$ consisting of

(i) a **feasible correspondence** $\Gamma$ that is a nonempty correspondence from $\mathsf{X}$ to $\mathsf{A}$, which in turn defines

- the feasible state-action pairs

$$\mathsf{G} := \{(x, a) \in \mathsf{X} \times \mathsf{A} : a \in \Gamma(x)\}$$

- and the set of feasible policies

$$\Sigma := \{\sigma \in \mathsf{A}^{\mathsf{X}} : \sigma(x) \in \Gamma(x) \text{ for all } x \in \mathsf{X}\},$$

(ii) a subset $V$ of $\mathbb{R}^{\mathsf{X}}$ called the **value space**, and

(iii) a **value aggregator** $B$ that maps $\mathsf{G} \times V$ to $\mathbb{R}$ and satisfies both the monotonicity condition

$$v, w \in V \text{ and } v \leqslant w \implies B(x, a, v) \leqslant B(x, a, w) \text{ for all } (x, a) \in \mathsf{G}, \qquad (8.2)$$

and the consistency condition

$$w \in V \text{ whenever } w(x) = B(x, \sigma(x), v) \text{ for some } \sigma \in \Sigma \text{ and } v \in V. \qquad (8.3)$$

Throughout, $\leqslant$ represents the pointwise order on $\mathbb{R}^{\mathsf{X}}$.

The definition of the feasible correspondence in (i) is identical to that for the MDP in Chapter 5. As for (ii), we understand $V$ to be a class of functions that assign values to states. In (iii), the interpretation of the aggregator $B$ is:

$B(x, a, v) =$ total lifetime rewards, contingent on current action $a$, current state $x$, and using $v$ to evaluate future states.

As we will see, determining optimal choices is greatly simplified if we can find a "good" $v$ for evaluating states. Locating and calculating this $v$ is one of our major concerns.

The monotonicity condition (8.2) is natural: if, relative to $v$, rewards are at least as high for $w$ in every future state, then the total rewards one can extract under $w$ should be at least as high. The consistency condition in (8.3) ensures that as we consider values of different policies we remain within the value space $V$.

Every MDP generates an RDP with the same Bellman equation, as the next example clarifies.

**Example 8.1.1.** Consider an arbitrary MDP $\mathcal{M} = (\Gamma, \beta, r, P)$ with state space $\mathsf{X}$ and action space $\mathsf{A}$ (see, e.g., §5.1.1). We can frame $\mathcal{M}$ as an RDP by taking $\Gamma$ as unchanged, $V = \mathbb{R}^{\mathsf{X}}$, and

$$B(x, a, v) = r(x, a) + \beta \sum_{x'} v(x') P(x, a, x') \qquad ((x, a) \in \mathsf{G}, \ v \in V). \qquad (8.4)$$

Now $(\Gamma, V, B)$ forms an RDP. The monotonicity condition (8.2) clearly holds and the consistency condition (8.3) is trivial, since $V$ is all of $\mathbb{R}^{\mathsf{X}}$. Inserting (8.4) into the abstract Bellman equation (8.1) recovers the MDP Bellman equation ((5.2) on page 130).

**Example 8.1.2.** Consider a basic cake eating problem (see §5.1.2.3), where $\mathsf{X}$ is a finite subset of $\mathbb{R}_+$ and $x \in \mathsf{X}$ is understood to be the number of remaining slices of cake today. Let $x'$ be the number of remaining slices next period and $u(x - x')$ be the utility from slices enjoyed today. The utility function $u$ maps $\mathbb{R}_+$ to $\mathbb{R}$. Let $V = \mathbb{R}^{\mathsf{X}}$, let $\Gamma$ be defined by $\Gamma(x) = \{x' \in \mathsf{X} : x' \leqslant x\}$ and let

$$B(x, x', v) = u(x - x') + \beta v(x').$$

Then $(\Gamma, V, B)$ is an RDP with Bellman equation identical to that of the original cake eating problem in §5.1.2.3. The monotonicity condition (8.2) and the consistency condition (8.3) are easy to verify.

The last example is a special case of Example 8.1.1, since the cake eating problem is an MDP (see §5.1.2.3). Nonetheless, Example 8.1.2 is instructive because, for cake eating, the MDP construction is tedious (e.g., we need to define a stochastic kernel $P$ even though transitions are deterministic), while the RDP construction is straightforward.

The next example makes a related point.

**Example 8.1.3.** In §5.1.2.4 we showed that the job search model is an MDP but the construction was tedious. But we can also represent job search as an RDP and the embedding is straightforward. To see this, recall that, for an arbitrary optimal stopping problem with primitives as described in Chapter 4, the Bellman equation is

$$v(x) = \max \left\{ e(x), c(x) + \beta \sum_{x'} v(x')P(x, x') \right\} \qquad (x \in \mathsf{X}). \tag{8.5}$$

Let $V = \mathbb{R}^{\mathsf{X}}$ and $\Gamma(x) = \{0, 1\}$ for all $x$. Let

$$B(x, a, v) = ae(x) + (1 - a)\left[ c(x) + \beta \sum_{x'} v(x')P(x, x') \right] \tag{8.6}$$

for $x \in \mathsf{X}$ and $a \in \mathsf{A} := \{0, 1\}$. Then $(\Gamma, V, B)$ is an RDP (Exercise 8.1.1) and setting $v(x) = \max_{a \in \Gamma(x)} B(x, a, v)$ reproduces the Bellman equation (8.5).

EXERCISE 8.1.1. Verify that conditions (8.2)–(8.3) hold for this RDP.

**Example 8.1.4.** The dynamic programming framework popularized by Stokey and Lucas (1989) is characterized by two features: First, the state is divided into an exogenous process ($Z_t$) and an endogenous process ($Y_t$). In addition, the next period endogenous state is directly chosen by the current action. The Bellman equation

takes the form

$$v(y, z) = \max_{y' \in \Gamma(y,z)} \left\{ F(y, z, y') + \beta \sum_{z'} v(y', z') Q(z, z') \right\}. \tag{8.7}$$

We assume that $(Z_t)$ is $Q$-Markov on finite set $\mathsf{Z}$ and $(Y_t)$ takes values in finite set $\mathsf{Y}$. With state space $\mathsf{X} := \mathsf{Y} \times \mathsf{Z}$, action space $\mathsf{Y}$, feasible correspondence $x \mapsto \Gamma(x)$, value space $V = \mathbb{R}^{\mathsf{X}}$ and aggregator

$$B(x, a, v) = B((y, z), y', v) = F(y, z, y') + \beta \sum_{z'} v(y', z') Q(z, z'),$$

we obtain an RDP with Bellman equation identical to (8.7).

EXERCISE 8.1.2. Show that this RDP can also be expressed as an MDP.

Examples 8.1.1–8.1.4 treated RDPs that can be embedded into the MDP framework. In the remaining examples, we consider models that cannot be represented as MDPs.

**Example 8.1.5** (State-Dependent Discounting). We can add state-dependent discounting to Example 8.1.1 by changing the aggregator to

$$B(x, a, v) = r(x, a) + \sum_{x'} v(x') \beta(x, a, x') P(x, a, x'). \tag{8.8}$$

Here $\beta$ is a map from $\mathsf{G} \times \mathsf{X}$ to $\mathbb{R}_+$. With $\Gamma$ and $V$ unchanged, $(\Gamma, V, B)$ is an RDP with Bellman equation identical to that of the MDP with state-dependent discounting we analyzed on Chapter 6. In §8.2.2 we will use RDP theory developed in this chapter to verify the optimality results claimed in Chapter 6.

EXERCISE 8.1.3. Verify that $(\Gamma, V, B)$ as defined in Example 8.1.5 is an RDP.

**Example 8.1.6.** We can modify the MDP in Example 8.1.1 to use risk-sensitive preferences. We do this by taking $\Gamma, V$ to be the same as the MDP example and setting

$$B(x, a, v) = r(x, a) + \beta \frac{1}{\theta} \ln \left\{ \sum_{x'} \exp(\theta v(x')) P(x, a, x') \right\} \tag{8.9}$$

for all $(x, a) \in \mathsf{G}$ and $v \in V$. Here $\mathsf{G}$ is generated by the feasible correspondence $\Gamma$.

EXERCISE 8.1.4. Confirm that the risk-sensitive model $(\Gamma, V, B)$ in Example 8.1.6 is an RDP for all nonzero $\theta$.

**Example 8.1.7** (Epstein–Zin Preferences)**.** We can also modify the MDP in Example 8.1.1 to use the Epstein–Zin specification (see (7.12) on page 227) by setting

$$B(x, a, v) = \left\{ r(x, a) + \beta \left[ \sum_{x'} v(x')^\gamma P(x, a, x') \right]^{\alpha/\gamma} \right\}^{1/\alpha}, \tag{8.10}$$

where $\beta \in (0, 1)$, $\gamma$ and $\alpha$ are nonzero parameters, and $r \gg 0$. Let $V$ be all strictly positive functions in $\mathbb{R}^{\mathsf{X}}$. Then $(\Gamma, V, B)$ is an RDP.

EXERCISE 8.1.5. Confirm the last claim in Example 8.1.7.

**Example 8.1.8.** The **shortest path problem** considers optimal traversal of a directed graph $\mathcal{G} = (\mathsf{X}, E)$, where $\mathsf{X}$ is the vertices of the graph and $E$ is the edges. A weight function $c \colon E \to \mathbb{R}_+$ associates cost to each edge $(x, x') \in E$. The aim is to find the minimum cost path from $x$ to a specified vertex $d$ for every $x \in \mathsf{X}$. Under some conditions, the problem can be solved by applying a Bellman operator of the form

$$(Tv)(x) = \min_{x' \in \mathcal{O}(x)} \{c(x, x') + v(x')\} \qquad (x \in \mathsf{X}), \tag{8.11}$$

where $\mathcal{O}(x) := \{x' \in \mathsf{X} : (x, x') \in E\}$ is the direct successors of $x$ and $v(x')$ is the minimum cost-to-go from state $x'$. The problem is not an MDP because future values are not discounted. It can be framed as an RDP, however, by setting $\Gamma(x) = \mathcal{O}(x)$, $B(x, x', v) = c(x, x') + v(x')$ and $V = \mathbb{R}^{\mathsf{X}}$.

Example 8.1.8 is a minimization problem. We treat minimization explicitly in §8.3.5, although the shortest path setting can be converted maximization by replacing $c(x, x')$ with $-c(x, x')$. This produces an application similar to the cake eating problem in Example 8.1.2 (although discounting is eliminated and network structure shows up in the constraint).

## 8.1.2 Lifetime Value

We aim to discuss optimality of RDPs. To prepare for this topic, we now clarify lifetime values associated with different policy choices in the RDP setting.

### 8.1.2.1 Policies and Value

Let $\mathcal{R} = (\Gamma, V, B)$ be an RDP with state and action spaces X and A, and let $\Sigma$ be the set of all feasible policies. For each $\sigma \in \Sigma$ we introduce the **policy operator** $T_\sigma$ as a self-map on $V$ defined by

$$(T_\sigma v)(x) = B(x, \sigma(x), v) \qquad (x \in \mathsf{X}). \tag{8.12}$$

The RDP policy operator is a direct generalization of the MDP policy operator defined on page 136, as well as the optimal stopping policy operator defined on page 109.

EXERCISE 8.1.6. Show that $T_\sigma$ is an order-preserving self-map on $V$ for all $\sigma \in \Sigma$.

Consider a given RDP $(\Gamma, V, B)$ and fix $\sigma \in \Sigma$. If $T_\sigma$ has a unique fixed point in $V$, we denote this fixed point by $v_\sigma$ and call it the $\sigma$-**value function**. It is natural to interpret $v_\sigma$ as representing the lifetime value of following policy $\sigma$.

**Example 8.1.9.** For the optimal stopping problem discussed in Chapter 4, the function $v_\sigma$ that records the lifetime value of a policy $\sigma$ from any given state is the unique fixed point of the optimal stopping policy operator $T_\sigma$. See §4.1.1.3.

**Example 8.1.10.** For the MDP model discussed in Chapter 5, lifetime value of policy $\sigma$ is given by $v_\sigma = (I - \beta P_\sigma)^{-1} r_\sigma$. As discussed in Exercise 5.1.7, $v_\sigma$ is the unique fixed point of the MDP policy operator $T_\sigma$ defined by $T_\sigma v = r_\sigma + \beta P_\sigma v$.

**Example 8.1.11.** For the MDP model with state-dependent discounting introduced in Chapter 6, Exercise 6.2.1 shows that the lifetime value of following policy $\sigma$ is the unique fixed point of the policy operator $T_\sigma$ defined in (6.16) on page 193.

The previous examples are linear but the same idea extends to nonlinear recursive preference models as well. To see this, recall the generic Koopmans operator $(Kv)(x) = A(x, (Rv)(x))$ introduced in §7.3.1. Lifetime value is the unique fixed point of this operator whenever it exists. In all of the RDP examples we have considered, the policy operator can be expressed as $(T_\sigma v)(x) = A_\sigma(x, (R_\sigma v)(x))$ for some aggregator $A_\sigma$ and certainty equivalent operator $R_\sigma$. Hence $T_\sigma$ is a Koopmans operator and lifetime value associated with policy $\sigma$ is the fixed point of this operator.

### 8.1.2.2 Uniqueness and Stability

Let $\mathcal{R} = (\Gamma, V, B)$ be a given RDP with policy operators $\{T_\sigma\}$. Given that our objective is to maximize lifetime value over the set of policies in $\Sigma$, we need to assume at the very least that lifetime value is well defined at each policy. To this end, we say that $\mathcal{R}$ is **well-posed** whenever $T_\sigma$ has a unique fixed point $v_\sigma$ in $V$ for all $\sigma \in \Sigma$.

**Example 8.1.12.** The optimal stopping RDP we introduced in Example 8.1.3 is well-posed. Indeed, for each $\sigma \in \Sigma$, the policy operator $T_\sigma$ has a unique fixed point in $\mathbb{R}^{\mathsf{X}}$ by Proposition 4.1.1 on page 109.

**Example 8.1.13.** The RDP generated by the MDP model in Example 8.1.1 is well-posed, since, for each $\sigma \in \Sigma$, the operator $T_\sigma = r_\sigma + \beta P_\sigma$ has the unique fixed point $v_\sigma = (I - \beta P_\sigma)^{-1} r_\sigma$ in $\mathbb{R}^{\mathsf{X}}$.

**Example 8.1.14.** The shortest path problem discussed in Example 8.1.8 is not well-posed without further assumptions. For example, consider a graph that contains two vertices $x$ and $y$, with $x \in \mathcal{O}(y)$, $y \in \mathcal{O}(x)$, and $c(x, y) + c(y, x) > 0$. Then, for any policy $\sigma$ that maps $x$ to $y$ and $y$ to $x$, we have

$$(T_\sigma v)(x) = c(x, y) + v(y) \quad \text{and} \quad (T_\sigma v)(y) = c(y, x) + v(x).$$

Hence, if $v \in \mathbb{R}^{\mathsf{X}}$ is a fixed point of $T_\sigma$, we obtain $v(x) = c(x, y) + v(y)$ and $v(y) = c(y, x) + v(x)$. Substition yields $v(x) = c(x, y) + c(y, x) + v(x)$, which is a contradiction.

Let $\mathcal{R}$ be an RDP with policy operators $\{T_\sigma\}_{\sigma \in \Sigma}$. In what follows, we call $\mathcal{R}$ **globally stable** if $T_\sigma$ is globally stable on $V$ for all $\sigma \in \Sigma$.

**Example 8.1.15.** The optimal stopping RDP we introduced in Example 8.1.3 is globally stable, since, for each $\sigma \in \Sigma$, the policy operator $T_\sigma$ is globally stable on $\mathbb{R}^{\mathsf{X}}$ by Proposition 4.1.1 on page 109.

**Example 8.1.16.** The RDP generated by the MDP model in Example 8.1.1 is globally stable. See Exercise 5.1.7 on page 136.

Obviously every globally stable RDP is well-posed.

**Remark 8.1.1.** In line with our discussion of stability of Koopmans operators in §7.3.1.7, for $v \in V$, $\sigma \in \Sigma$, $m \in \mathbb{N}$ and $x \in \mathsf{X}$, it is natural to interpret $(T_\sigma^m v)(x)$ as total (finite horizon) utility over periods $0, \ldots, m$ under policy $\sigma$, with initial state $x$ and terminal condition $v \in V$. Hence global stability implies that, for any choice of terminal condition, finite horizon valuations always converge to their infinite horizon counterparts.

In §8.1.3 we will see that global stability yields strong optimality properties.

### 8.1.2.3 Continuity

Let $\mathcal{R} = (\Gamma, V, B)$ be an RDP. We call $\mathcal{R}$ **continuous** if $B(x, a, v)$ is continuous in $v$ for all $(x, a) \in \mathsf{G}$. In other words, $\mathcal{R}$ is continuous if, for any $v \in V$, any $(x, a) \in \mathsf{G}$ and any sequence $(v_k)_{k \geq 1}$ in $V$, we have

$$\lim_{k \to \infty} B(x, a, v_k) = B(x, a, v) \quad \text{whenever} \quad \lim_{k \to \infty} v_k = v.$$

Continuity is satisfied by all applications considered in this text. For example, for the RDP generated by an MDP (Example 8.1.1), the deviation $|B(x, a, v_k) - B(x, a, v)|$ is dominated by $\beta \|v_k - v\|_\infty$ for all $(x, a) \in \mathsf{G}$. Hence continuity holds.

Below we will see that continuity is useful when considering covergence of certain algorithms.

## 8.1.3 Optimality

In this section we present optimality theory for RDPs.

### 8.1.3.1 Greedy Policies

Given an RDP $\mathcal{R} = (\Gamma, V, B)$ and $v \in V$, a policy $\sigma \in \Sigma$ is called $v$-**greedy** if

$$\sigma(x) \in \operatorname*{argmax}_{a \in \Gamma(x)} B(x, a, v) \quad \text{for all } x \in \mathsf{X}. \tag{8.13}$$

Since $\Gamma(x)$ is finite and nonempty at each $x \in \mathsf{X}$, at least one such policy exists. As with policy operators, the notion of greedy policies extends existing definitions from earlier chapters.

EXERCISE 8.1.7. Show that, for each $v \in V$, the set $\{T_\sigma v\}_{\sigma \in \Sigma} \subset V$ contains a least and greatest element (see §2.2.1.2 for the definitions). Explain the connection between the greatest element and $v$-greedy policies.

Given RDP $\mathcal{R} = (\Gamma, V, B)$, we say that $v \in V$ satisfies the **Bellman equation** if $v(x) = \max_{a \in \Gamma(x)} B(x, a, v)$ for all $x \in \mathsf{X}$. The **Bellman operator** corresponding to $\mathcal{R}$ is the map $T$ on $V$ defined by

$$(Tv)(x) = \max_{a \in \Gamma(x)} B(x, a, v) \qquad (x \in \mathsf{X}).$$

We say that $v \in V$ satisfies the **Bellman equation** if $Tv = v$.

**Example 8.1.17.** For the Epstein–Zin RDP in (8.10), the Bellman operator is given by

$$(Tv)(x) = \max_{a \in \Gamma(x)} \left\{ r(x,a)^\alpha + \beta \left[ \sum_{x'} v(x')^\gamma P(x,a,x') \right]^{\alpha/\gamma} \right\}^{1/\alpha} \quad (x \in \mathsf{X}).$$

EXERCISE 8.1.8. Given RDP $\mathcal{R} = (\Gamma, V, B)$ with policy operators $\{T_\sigma\}$ and Bellman operator $T$, show that, for each $v \in V$,

(i) $Tv = \bigvee_\sigma T_\sigma v := \bigvee_{\sigma \in \Sigma} (T_\sigma v)$ and

(ii) $\sigma$ is $v$-greedy if and only if $Tv = T_\sigma v$.

(iii) $T$ is an order-preserving self-map on $V$.

EXERCISE 8.1.9. Show that, for a given RDP $(\Gamma, V, B)$ and fixed $v \in V$, the Bellman operator $T$ obeys

$$(T^k v)(x) = \max_{a \in \Gamma(x)} B(x, a, T^{k-1}v) \tag{8.14}$$

for all $k \in \mathbb{Z}_+$ and all $x \in \mathsf{X}$. Show, in addition, that for any policy $\sigma \in \Sigma$, the policy operator $T_\sigma$ obeys

$$(T_\sigma^k v)(x) = B(x, \sigma(x), T_\sigma^{k-1}v) \tag{8.15}$$

for all $k \in \mathbb{Z}_+$ and all $x \in \mathsf{X}$.

### 8.1.3.2 Algorithms

To solve RDPs for optimal policies, we use two core algorithms: Howard policy iteration (HPI) and optimistic policy iteration (OPI). As in previous chapters, OPI includes VFI as a special case.

To describe HPI we take $\mathcal{R} = (\Gamma, V, B)$ to be a well-posed RDP with feasible policy set $\Sigma$, policy operators $\{T_\sigma\}$, and Bellman operator $T$. In this setting, the HPI algorithm is essentially identical to the one given for MDPs in §5.1.4.2, except that $v_\sigma$ is calculated as the fixed point of $T_\sigma$, rather than taking the specific form $(I - \beta P_\sigma)^{-1} r_\sigma$. The details are in Algorithm 8.1.

Algorithm 8.1 is somewhat ambiguous, since it is not always clear how to implement the instruction "$v_k \leftarrow$ the fixed point of $T_{\sigma_k}$". However, if $\mathcal{R}$ is globally stable, then each $T_{\sigma_k}$ is globally stable, so an approximation of the fixed point can be calculated by iterating with $T_{\sigma_k}$. This line of thought leads us to consider optimistic policy

---

**Algorithm 8.1:** Howard policy iteration for RDPs

---
1  input $\sigma \in \Sigma$
2  $v_0 \leftarrow v_\sigma$ and $k \leftarrow 0$
3  **repeat**
4  $\quad$ $\sigma_k \leftarrow$ a $v_k$-greedy policy
5  $\quad$ $v_{k+1} \leftarrow$ the fixed point of $T_{\sigma_k}$
6  $\quad$ **if** $v_{k+1} = v_k$ **then** break
7  $\quad$ $k \leftarrow k + 1$
8  **return** $\sigma_k$

---

iterating (OPI) as a more practical alternative. Algorithm 8.2 states an OPI routine for solving $\mathcal{R}$ that generalizes the MDP OPI routine in §5.1.4.

---

**Algorithm 8.2:** Optimistic policy iteration for RDPs

---
1  input $m \in \mathbb{N}$ and tolerance $\tau \geqslant 0$
2  input $\sigma \in \Sigma$ and set $v_0 \leftarrow v_\sigma$
3  $k \leftarrow 0$
4  **repeat**
5  $\quad$ $\sigma_k \leftarrow$ a $v_k$-greedy policy
6  $\quad$ $v_{k+1} \leftarrow T_{\sigma_k}^m v_k$
7  $\quad$ **if** $\|v_{k+1} - v_k\| \leqslant \tau$ **then** break
8  $\quad$ $k \leftarrow k + 1$
9  **return** $\sigma_k$

---

In Algorithm 8.2 we require that $v_0 = v_\sigma$ for some $\sigma \in \Sigma$. This assumption can be dropped in some settings. For practical purposes, however, it is almost always straightforward to initialize OPI with $v_0 = v_\sigma$ for some simple choice of $\sigma$.

EXERCISE 8.1.10. Prove that, for the sequence $(v_k)$ in the OPI algorithm 8.2, we have $v_k = T^k v_0$ whenever $m = 1$. (In other words, OPI reduces to VFI when $m = 1$.)

When we turn to proofs, it will help to have an operator-theoretic description of HPI and OPI. To this end, we define two operators. The first is $H \colon V \to \{v_\sigma\}$, which is defined via

$$Hv = v_\sigma \text{ where } \sigma \text{ is } v\text{-greedy.} \tag{8.17}$$

We call $H$ the **Howard operator** generated by $\mathcal{R}$. Iterating with $H$ implements HPI. In particular, if we fix $\sigma \in \Sigma$ and set $v_k = H^k v_\sigma$, then $(v_k)_{k \geqslant 0}$ is the sequence of $\sigma$-value functions generated by HPI.[1]

Next, fixing $m \in \mathbb{N}$ and $\sigma \in \Sigma$, we define the operator $W_m$ from $V$ to itself via

$$W_m v := T_\sigma^m v \quad \text{where} \quad \sigma \text{ is } v\text{-greedy.}$$

(See footnote 1 on the choice of $v$-greedy policies.) The operator $W_m$ is an approximation of $H$, since $T_\sigma^m v \to v_\sigma = Hv$ as $m \to \infty$. Iterating with $W_m$ generates the value sequence in OPI. More specifically, we take $v_0 \in \{v_\sigma\}$ and generate

$$(v_k, \sigma_k)_{k \geqslant 0} \quad \text{where } v_k = W_m^k v_0 \text{ and } \sigma_k \text{ is } v_k\text{-greedy.} \tag{8.18}$$

This produces an infinite sequence of OPI value and policy iterates.

### 8.1.3.3 Optimality

Let $\mathcal{R}$ be a well-posed RDP with policy operators $\{T_\sigma\}$ and $\sigma$-value functions $\{v_\sigma\}$. In this context, we set $v^* := \bigvee_\sigma v_\sigma \in \mathbb{R}^\mathsf{X}$ and call $v^*$ the **value function** of $\mathcal{R}$. By definition, $v^*$ satisfies

$$v^*(x) = \max_{\sigma \in \Sigma} v_\sigma(x) \qquad \text{for all} \quad x \in \mathsf{X}. \tag{8.19}$$

A policy $\sigma$ is called **optimal** for $\mathcal{R}$ if $v_\sigma = v^*$; that is, if

$$v_\sigma(x) \geqslant v_\tau(x) \quad \text{for all } \tau \in \Sigma \text{ and all } x \in \mathsf{X}.$$

Both of these definitions generalize the definitions we used for MDPs and optimal stopping. In particular, optimality of a policy means that it generates maximum possible lifetime value from every state.

We say that $\mathcal{R}$ satisfies **Bellman's principle of optimality** if

$$\sigma \in \Sigma \text{ is optimal for } \mathcal{R} \quad \Longleftrightarrow \quad \sigma \text{ is } v^*\text{-greedy.}$$

We can now state our main optimality result for RDPs. In the statement, $\mathcal{R}$ is a well-posed RDP with value function $v^*$.

**Theorem 8.1.1.** *If $\mathcal{R}$ is globally stable, then*

---

[1]For $H$ to be well-defined, we must always select the same $v$-greedy policy when the operator is applied to $v$. To this end, we enumerate the policy set $\Sigma$ and choose the first $v$-greedy policy. This choice of convention has no effect on convergence results.

(i) $v^*$ *is the unique solution to the Bellman equation in* $V$,

(ii) $\mathscr{R}$ *satisfies Bellman's principle of optimality,*

(iii) $\mathscr{R}$ *has at least one optimal policy,*

(iv) *HPI returns an optimal policy in finitely many steps, and*

(v) *the OPI sequence in* (8.18) *is such that* $v_k \to v^*$ *as* $k \to \infty$ *and, moreover, there exists a* $K \in \mathbb{N}$ *such that* $\sigma_k$ *is optimal for all* $k \geqslant K$.

As OPI includes VFI as a special case ($m = 1$), Theorem 8.1.1 also implies convergence of VFI under the stated conditions.

In terms of applications, Theorem 8.1.1 is the most important optimality result in this book. It provides the core optimality results from dynamic programming and a broadly convergent algorithm for computing optimal policies.

The proof of Theorem 8.1.1 is deferred to §9.1.

**Example 8.1.18.** The optimality results for optimal stopping problems we presented in Chapter 4 are a special case of Theorem 8.1.1, since such optimal stopping problems generate globally stable RDPs (as discussed in Example 8.1.15).

**Example 8.1.19.** The optimality results for MDPs we presented in Chapter 5 are a special case of Theorem 8.1.1, since MDPs generate globally stable RDPs (as discussed in Example 8.1.16).

Examples 8.1.18–8.1.16 are relatively elementary. More complex models will be handled in §8.2.

### 8.1.3.4 Comments on the Optimality Theorem

Many traditional treatments of dynamic programming build optimality theory around contractivity (see, e.g., Puterman (2005) or Stokey and Lucas (1989), Section 4.2). Assumptions are constructed so that the policy operators and Bellman operator are all contraction mappings.

While such assumptions are sufficient for Theorem 8.1.1 (since contractivity of the policy operators implies stability), they are not necessary. There are a variety of ways to prove uniqueness and stability of fixed points, including the monotonicity-based methods discussed in §7.1.2 and the spectral methods in §6.1.3.2. These alternatives will prove useful in settings where contractivity fails, as we shall see in §8.2.

Another point worth noting about the conditions in Theorem 8.1.1 is that no assumptions are placed on the Bellman operator. Rather, one only needs to check properties of the policy operators. This is advantageous because, unlike the Bellman operator, the policy operators do not involve maximization.

### 8.1.3.5 Nonstationary Policies

Up until now we have focused entirely on stationary policies, in the sense that the same policy is used at every point in time. What if we drop this assumption and admit the option to change policies? Might this lead to higher lifetime values?

In this section, we show that for globally stable RDPs the answer is negative. This finding justifies our focus on stationary policies.

To begin, let $\mathcal{R} = (\Gamma, V, B)$ be a globally stable RDP. Recall from Remark 8.1.1 that, given $v \in V$, $\sigma \in \Sigma$, $k \in \mathbb{N}$ and $x \in \mathsf{X}$, the value $(T_\sigma^k v)(x)$ gives finite horizon utility over periods $0, \ldots, k$ under policy $\sigma$, with initial state $x$ and terminal condition $v$. Extending this idea, it is natural to understand $T_{\sigma_k} T_{\sigma_{k-1}} \cdots T_{\sigma_1} v$ as providing finite horizon utility values for the nonstationary policy sequence $(\sigma_k)_{k \in \mathbb{N}} \subset \Sigma$, given terminal condition $v \in V$. For the same policy sequence, we define its lifetime value via

$$\bar{v} := \limsup_{k \to \infty} v_k \quad \text{with } v_k := T_{\sigma_k} T_{\sigma_{k-1}} \cdots T_{\sigma_1} v$$

whenever the limsup is finite and independent of the terminal condition $v$.

Suppose that this is the case, and hence $\bar{v}$ is well defined. We claim that $\bar{v} \leqslant v^*$.

EXERCISE 8.1.11. Show that, under the stated conditions, $v_k \leqslant T^k v$ for all $k \in \mathbb{N}$.

Since $\bar{v}$ is independent of the terminal condition $v$, we can assume without loss of generality that $v \in V_\Sigma$. By Theorem 8.1.1, we have $T^k v \to v^*$ as $k \to \infty$. Hence, by Exercise 8.1.11,

$$\bar{v} = \limsup_{k \to \infty} v_k \leqslant \limsup_{k \to \infty} T^k v = \lim_{k \to \infty} T^k v = v^*,$$

as was to be shown.

### 8.1.3.6 Bounded RDPs

We call an RDP $\mathcal{R} = (\Gamma, V, B)$ **bounded** if $V$ is convex and, moreover, there exist functions $v_1, v_2 \in V$ such that $v_1 \leqslant v_2$,

$$v_1(x) \leqslant B(x, a, v_1) \quad \text{and} \quad B(x, a, v_2) \leqslant v_2(x) \quad \text{for all } (x, a) \in \mathsf{G}. \tag{8.20}$$

We show below that boundedness can be used to obtain optimality results for well-posed RDPs, even without global stability.

Another attractive feature of boundedness is that it permits a reduction of value space, as illustrated by the next two exercises.

EXERCISE 8.1.12. Let $(\Gamma, V, B)$ be bounded and let $v_1, v_2 \in V$ be such that (8.20) holds. Prove that, in this setting, $(\Gamma, \hat{V}, B)$ is also an RDP when $\hat{V} := [v_1, v_2]$.

EXERCISE 8.1.13. Adopt the setting of Exercise 8.1.12 and suppose, in addition, that the RDP is well-posed. Show that $v_\sigma \in \hat{V}$ for all $\sigma \in \Sigma$.

Exercise 8.1.13 implies the reduced RDP $(\Gamma, \hat{V}, B)$ is also well-posed under the stated conditions, and that it contains all the $\sigma$-value functions and the value function from the original RDP $(\Gamma, V, B)$. Hence any optimality results for $(\Gamma, \hat{V}, B)$ carry over to $(\Gamma, V, B)$.

EXERCISE 8.1.14. Show that the RDP generated by an MDP in Example 8.1.1 is bounded.

EXERCISE 8.1.15. Show that the optimal stopping RDP from Example 8.1.3 is bounded.

EXERCISE 8.1.16. Consider the RDP $(\Gamma, V, B)$ generated by an MDP with stochastic discounting in Example 8.1.5. Prove that this RDP is bounded whenever the conditions of Proposition 6.2.2 on page 194 hold.

EXERCISE 8.1.17. Consider the shortest path RDP $(\Gamma, V, B)$ in Example 8.1.8 and assume in addition that the graph $\mathcal{G}$ contains only one cycle, which is a self-loop at $d$, that $d$ is accessible from every vertex $x \in \mathsf{X}$, and that $c(d, d) = 0$. (These assumptions imply that every path leads to $d$ in finite time and that travelers reaching $d$ remain there forever at zero cost.) Let $C(x)$ be the maximum cost of traveling to $d$ from $x$, which is finite by the stated assumptions. Show that (8.20) holds when $v_1 := 0$ and $v_2 := C$.

The next result shows that, when considering optimality, stability can be replaced by boundedness.

**Theorem 8.1.2.** *If $\mathcal{R}$ is well-posed and bounded, then* (i)–(iv) *of Theorem 8.1.1 hold.*

### 8.1.4   Topologically Conjugate RDPs

Sometimes RDP models can be simplified by transformations over value space. In this section we investigate such transformations. The underlying ideas are related to topological conjugacy of dynamical systems, which we introduced in §2.1.1.2.

To begin, let $\mathcal{R} = (\Gamma, V, B)$ and $\hat{\mathcal{R}} = (\Gamma, \hat{V}, \hat{B})$ be two RDPs with identical state space X, action space A and feasible correspondence $\Gamma$. We consider settings where

$$V = \mathbb{M}^{\mathsf{X}} \quad \text{and} \quad \hat{V} = \hat{\mathbb{M}}^{\mathsf{X}} \quad \text{where } \mathbb{M}, \hat{\mathbb{M}} \subset \mathbb{R},$$

and, in addition, that there exists a homeomorphism $\varphi$ from $\mathbb{M}$ onto $\hat{\mathbb{M}}$ such that

$$B(x, a, v) = \varphi^{-1}[\hat{B}(x, a, \varphi \circ v)] \quad \text{for all } v \in V \text{ and } (x, a) \in \mathsf{G}. \tag{8.21}$$

We call $\mathcal{R}$ and $\hat{\mathcal{R}}$ **topologically conjugate** under $\varphi$ if $\varphi$ is a homeomorphism $\varphi$ from $\mathbb{M}$ to $\hat{\mathbb{M}}$ and (8.21) holds.

Here is our main result for this section.

**Proposition 8.1.3.** *If $\mathcal{R}$ and $\hat{\mathcal{R}}$ are topologically conjugate, then $\mathcal{R}$ is globally stable if and only if $\hat{\mathcal{R}}$ is globally stable.*

The benefit of Proposition 8.1.3 is that one of these models might be easier to analyze than the other. We apply the proposition to the Epstein–Zin specification in §8.1.4.1 and to a smooth ambiguity model in §8.3.4. The next exercise will be useful for the proof.

EXERCISE 8.1.18. Prove the following: If $\varphi$ is a homeomorphism from $\mathbb{M}$ to $\hat{\mathbb{M}}$ and $\Phi v := \varphi \circ v$, then $\Phi$ is a homeomorphism from $V$ to $\hat{V}$.

*Proof of Proposition 8.1.3.* By Exercise 8.1.18, $\Phi v := \varphi \circ v$ is a homeomorphism from $V$ to $\hat{V}$. Moreover, for any $\sigma \in \Sigma$, the respective policy operators $T_\sigma$ and $\hat{T}_\sigma$ are linked by

$$(T_\sigma v)(x) = B(x, \sigma(x), v) = \varphi^{-1}[\hat{B}(x, \sigma(x), \varphi \circ v)] = \varphi^{-1}[(\hat{T}_\sigma \varphi \circ v)(x)].$$

This shows that $T_\sigma = \Phi^{-1} \circ \hat{T}_\sigma \circ \Phi$ on $V$. Hence $(V, T_\sigma)$ and $(\hat{V}, \hat{T}_\sigma)$ are topologically conjugate dynamical systems, from which it follows that $T_\sigma$ is globally stable if and only if $\hat{T}_\sigma$ is globally stable (see page 45). This completes the proof of Proposition 8.1.3. □

In the next section we will see how these ideas can simplify optimality analysis.

### 8.1.4.1 Application: Epstein–Zin RDPs

In this section we show how the preceding optimality results and the notion of topologically conjugacy can be deployed to analyze the Epstein–Zin RDP from Example 8.1.7. (Later, when we have more tools in hand, we will return to the Epstein–Zin problem and establish similar results under weaker conditions (§9.2.3).)

Recall that the aggregator in Example 8.1.7 is

$$B(x, a, v) = \left\{ r(x, a) + \beta \left( \sum_{x'} v(x')^\gamma P(x, a, x') \right)^{\alpha/\gamma} \right\}^{1/\alpha}. \tag{8.22}$$

Let $V = (0, \infty)^X$. We assume that $r \gg 0$ and take a nonempty feasible correspondence $\Gamma$ as given. Exercise 8.1.5 on page 251 confirmed that $\mathcal{R} := (\Gamma, V, B)$ is an RDP.

We will call the stochastic kernel $P$ **irreducible** if $P(x, \sigma(x), x')$ is irreducible for all $\sigma \in \Sigma$. Below we establish stability of $\mathcal{R}$ under irreducibility.

**Proposition 8.1.4.** *If $P$ is irreducible, then $\mathcal{R}$ is globally stable.*

To prove Proposition 8.1.4, we set up a simpler and more tractable model. Our first step is to introduce another RDP by setting

$$\hat{B}(x, a, v) = B\left(x, a, v^{1/\gamma}\right)^\gamma \tag{8.23}$$

We set $\mathcal{R} := (\Gamma, B, V)$ and $\hat{\mathcal{R}} := (\Gamma, \hat{B}, V)$. Notice that $\hat{B}$ can also be expressed as

$$\hat{B}(x, a, v) = \left\{ r(x, a) + \beta \left( \sum_{x'} v(x') P(x, a, x') \right)^{1/\theta} \right\}^\theta, \tag{8.24}$$

where $\theta := \gamma/\alpha$.

The value of of introducing $\hat{\mathcal{R}}$ comes from the fact that $\hat{\mathcal{R}}$ is easier to work with than $\mathcal{R}$ (just as the modified Epstein–Zin Koopmans operator $\hat{K}$ defined in §7.2.3.3 turned out to be easier to work with than the original Epstein–Zin Koopmans operator $K$ introduced in §7.2.3.2).

EXERCISE 8.1.19. Prove that $\mathcal{R}$ and $\hat{\mathcal{R}}$ are topologically conjugate RDPs (see §8.1.4).

Now we investigate the properties of the simpler RDP $\hat{\mathcal{R}}$.

**Lemma 8.1.5.** *If $P$ is irreducible, then $\hat{\mathcal{R}}$ is a globally stable RDP.*

*Proof.* In view of (8.24), each policy operator $\hat{T}_\sigma$ associated with $\hat{\mathcal{R}}$ takes the form

$$(\hat{T}_\sigma v)(x) = \left\{ r(x, \sigma(x)) + \beta \left( \sum_{x'} w(x') P(x, \sigma(x), x') \right)^{1/\theta} \right\}^{\theta} \tag{8.25}$$

Each such $\hat{T}_\sigma$ is a special case of $\hat{K}$ defined on page 230 by $\hat{K}v = \left\{ h + \beta(Pv)^{1/\theta} \right\}^{\theta}$ (see (7.15)). We saw in §7.2.3.3 that this operator is globally stable under the stated assumptions. Hence $\hat{\mathcal{R}}$ is a globally stable RDP. □

Now we can complete the

*Proof of Proposition 8.1.4.* Exercise 8.1.19 and Proposition 8.1.3 on page 261 together imply that $\mathcal{R}$ is globally stable if and only if $\hat{\mathcal{R}}$ is globally stable. The claim that $\mathcal{R}$ is globally stable now follows from and Lemma 8.1.5. □

# 8.2 Types of RDPs

In §8.1 we showed that well-posed RDPs have strong optimality properties whenever they are globally stable or bounded, and that VFI and OPI converge whenever they are globally stable. But what conditions are sufficient for these properties? We start with a relatively strict condition based on contractivity and then progress to models that fail to be contractive.

## 8.2.1 Contracting RDPs

In this section we study RDPs with strong contraction properties. Many traditional dynamic programs fit into this framework.

### 8.2.1.1 Definition and Examples

Let $\mathcal{R} = (\Gamma, V, B)$ be an RDP with state space X, action space A, and feasible state-action pair set G. We call $\mathcal{R}$ **contracting** if there exists a $\beta < 1$ such that

$$|B(x, a, v) - B(x, a, w)| \leqslant \beta \|v - w\|_\infty \quad \text{for all } (x, a) \in \mathsf{G} \text{ and } v, w \in V. \tag{8.26}$$

In line with the terminology for contraction maps, we call $\beta$ the **modulus of contraction** for $\mathcal{R}$ when (8.26) holds.

**Example 8.2.1.** The optimal stopping RDP from Example 8.1.3 is contracting with modulus $\beta$, since, for $B$ in (8.6), an application of the triangle inequality gives

$$|B(x, a, v) - B(x, a, w)| = (1-a)\beta \left| \sum_{x'} [v(x') - w(x')] P(x, x') \right| \leqslant \beta \|v - w\|_\infty.$$

EXERCISE 8.2.1. Show that any MDP is a contracting RDP.

EXERCISE 8.2.2. Show that every contracting RDP is value-continuous.

**Proposition 8.2.1.** *If $\mathcal{R}$ is contracting with modulus $\beta$, then $T$ and $\{T_\sigma\}_{\sigma \in \Sigma}$ are all contractions of modulus $\beta$ on $V$ under the norm $\|\cdot\|_\infty$.*

*Proof.* Let $\mathcal{R} = (\Gamma, V, B)$ be contracting with modulus $\beta$. Fix $\sigma \in \Sigma$ and let $v$ and $w$ be elements of $V$. By (8.26) we have

$$|(T_\sigma v)(x) - (T_\sigma w)(x)| = |B(x, \sigma(x), v) - B(x, \sigma(x), w)| \leqslant \beta \|v - w\|_\infty \qquad (8.27)$$

for every $x \in \mathsf{X}$. Maximizing over $x$ proves that $T_\sigma$ is a contraction of modulus $\beta$ with respect to the supremum norm.

Contractivity of $T$ now follows from Lemma 2.2.3 on page 59. □

The following corollary to Proposition 8.2.1 is immediate from Banach's contraction mapping theorem.

**Corollary 8.2.2.** *If $\mathcal{R}$ is contracting and $V$ is closed in $\mathbb{R}^{\mathsf{X}}$, then $\mathcal{R}$ is globally stable and hence all of the optimality results in Theorem 8.1.1 apply.*

### 8.2.1.2 Error Bounds

Corollary 8.2.2 tells us that contracting RDPs are globally stable and, as a result, the sequence of functions in $V$ generated by VFI (Algorithm 8.2 with $m = 1$) converges to $v^*$. However this result is asymptotic and conditions on $v_0 = v_\sigma$ for some $\sigma \in \Sigma$. We can improve this result in the current setting by leveraging the contraction property:

**Proposition 8.2.3.** *Let $(\Gamma, V, B)$ be a contracting RDP with modulus of contraction $\beta$ and Bellman operator $T$. Fix $v \in V$ and let $v_k = T^k v$. If $\sigma$ is $v_k$-greedy, then*

$$\|v^* - v_\sigma\|_\infty \leqslant \frac{2\beta}{1 - \beta} \|v_k - v_{k-1}\|_\infty \quad \text{for all } k \in \mathbb{N}. \tag{8.28}$$

Since the VFI algorithm terminates when $\|v_k - v_{k-1}\|_\infty$ falls below a given tolerance, the result in (8.28) directly provides a quantitative bound on the performance of the policy returned by VFI.

*Proof of Proposition 8.2.3.* Let $(\Gamma, V, B)$ and $v$ be as stated and let $v^*$ be the value function. Note that

$$\|v^* - v_\sigma\|_\infty \leqslant \|v^* - v_k\|_\infty + \|v_k - v_\sigma\|_\infty. \tag{8.29}$$

To bound the first term on the right-hand side of (8.29), we use the fact that $v^*$ is a fixed point of $T$, obtaining

$$\|v^* - v_k\|_\infty \leqslant \|v^* - T v_k\|_\infty + \|T v_k - v_k\|_\infty \leqslant \beta \|v^* - v_k\|_\infty + \beta \|v_k - v_{k-1}\|_\infty.$$

Hence

$$\|v^* - v_k\|_\infty \leqslant \frac{\beta}{1 - \beta} \|v_k - v_{k-1}\|_\infty. \tag{8.30}$$

Now consider the second term on the right-hand side of (8.29). Since $\sigma$ is $v_k$-greedy, we have $T v_k = T_\sigma v_k$, and

$$\|v_k - v_\sigma\|_\infty \leqslant \|v_k - T v_k\|_\infty + \|T v_k - v_\sigma\|_\infty = \|T v_{k-1} - T v_k\|_\infty + \|T_\sigma v_k - T_\sigma v_\sigma\|_\infty.$$

$$\therefore \quad \|v_k - v_\sigma\|_\infty \leqslant \beta \|v_{k-1} - v_k\|_\infty + \beta \|v_k - v_\sigma\|_\infty.$$

$$\therefore \quad \|v_k - v_\sigma\|_\infty \leqslant \frac{\beta}{1 - \beta} \|v_k - v_{k-1}\|_\infty. \tag{8.31}$$

Together, (8.29), (8.30), and (8.31) give us (8.28). □

### 8.2.1.3 A Blackwell-Type Condition

Next we state a useful condition for contractivity that is related to Blackwell's sufficient condition discussed in §2.2.3.4. We say that RDP $(\Gamma, V, B)$ satisfies **Blackwell's condition** if $v \in V$ implies $v + \lambda := v + \lambda \mathbb{1}$ is in $V$ for every $\lambda \geqslant 0$ and, in addition, there exists a $\beta \in [0, 1)$ such that

$$B(x, a, v + \lambda) \leqslant B(x, a, v) + \beta \lambda \qquad \text{for all } (x, a) \in \mathsf{G}, \, v \in V \text{ and } \lambda \in \mathbb{R}_+.$$

EXERCISE 8.2.3. Prove the following: If $\mathcal{R}$ satisfies Blackwell's condition, then $\mathcal{R}$ is contracting with modulus $\beta$.

EXERCISE 8.2.4. Prove that the RDP for the state-dependent discounting model in Example 8.1.5 is contracting on $V = \mathbb{R}^X$ whenever there exists a $b < 1$ with $\beta(x, a, x') \leqslant b$ for all $(x, a) \in G$ and $x' \in X$.

EXERCISE 8.2.5. Prove that the discrete optimal savings model from §5.2.2 satisfies Blackwell's condition.

### 8.2.1.4 Application: Job Search with Quantile Preferences

Consider the job search problem with correlated wage draws first investigated in §3.3.1. With finite wage offer set W, wage offer process generated by $P \in \mathcal{M}(\mathbb{R}^W)$ and $\beta \in (0, 1)$, we can frame this as an RDP $(\Gamma, V, B)$ with $V = \mathbb{R}^W$, $\Gamma(w) = \{0, 1\}$ for $w \in W$ and

$$B(w, a, v) := a \frac{w}{1 - \beta} + (1 - a)[c + \beta(Pv)(w)].$$

Since the model just described is an optimal stopping problem, Example 8.2.1 tells us that $(V, \Gamma, B)$ is contracting.

Now consider the following modification, where $\Gamma$ and $V$ are as before but $B$ is replaced by

$$B_\tau(w, a, v) := a \frac{w}{1 - \beta} + (1 - a)[c + \beta(R_\tau v)(w)],$$

where $\tau \in [0, 1]$ and $R_\tau$ is the quantile certainty equivalent operator described in Exercise 7.3.4 (page 233).

EXERCISE 8.2.6. Prove that $(V, \Gamma, B_\tau)$ is a contracting RDP.

Figure 8.1 shows the reservation wage for a range of $\tau$ values, computed using optimistic policy iteration (and taking the smallest $w \in W$ such that $\sigma^*(w) = 1$). The stationary distribution of $P$ is also shown in the figure, tilted 90 degrees.

The parameters and the code for applying $T_\sigma$ and evaluating greedy functions is shown in Listing 26. That listing includes the quantile operator $R_\tau$, which is implemented in Listing 25. (Quantiles of discrete random variables can also be computed using functionality contained in `Distributions.jl`.)

The main message of Figure 8.1 is that the reservation wage rises in $\tau$. In essence, higher $\tau$ focuses the attention of the worker on the right tail of the distribution of continuation values. This encourages the worker to take on more risk, which leads to a higher reservation wage (i.e., reluctance to accept a given current offer).

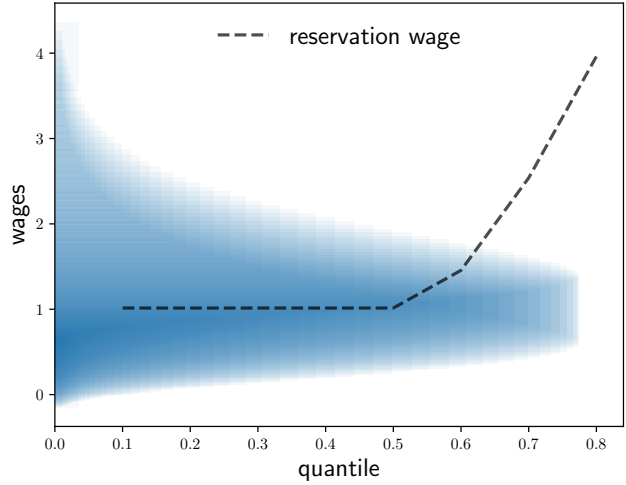

Figure 8.1: The reservation wage as a function of $\tau$

### 8.2.1.5 Application: Optimal Default

In this section we consider a small open economy that borrows in international financial markets in order to smooth consumption and has the option to default. We show that the model is a contractive RDP.

Income $(Y_t)_{t \geqslant 0}$ is exogenous and $Q$-Markov on finite set $\mathsf{Y}$. A representative household faces budget constraint

$$C_t = Y_t + b_t - q b_{t+1} \qquad (t \geqslant 0),$$

where $C_t$ is consumption at time $t$, $q$ is the price at time $t$ of a risk-free claim on one unit of time $t+1$ consumption; $q$ is determined outside the model, say international markets; $b_t$ measures foreign lending. Purchasing a claim on $b_{t+1}$ units of time $t+1$ consumption costs $q b_{t+1}$. Purchasing bond with *negative* face value $b_{t+1}$ pays $q b_{t+1}$ in current consumption goods and promises to deliver $b_{t+1}$ next period.

Trading bonds is managed by a benevolent government that wants to maximize household utility. Households discount future utility at rate $\beta \in (0, 1)$ and current consumption $C_t$ generates current utility $u(C_t)$. The government faces borrowing constraint $b_t \geqslant -m$ where $m \geqslant 0$. The government maximizes expected discounted utility

```julia
"Compute the τ-th quantile of v(X) when X ~ φ and v = sort(v)."
function quantile(τ, v, φ)
    for (i, v_value) in enumerate(v)
        p = sum(φ[1:i])   # sum all φ[j] s.t. v[j] ≤ v_value
        if p ≥ τ          # exit and return v_value if prob ≥ τ
            return v_value
        end
    end
end

"For each i, compute the τ-th quantile of v(Y) when Y ~ P(i, ·)"
function R(τ, v, P)
    return [quantile(τ, v, P[i, :]) for i in eachindex(v)]
end
```

Listing 25: Conditional quantile operator (`quantile_function.jl`)

for the households.

The government can default on foreign loans. In this case, output available for consumption drops from $Y_t$ to $h(Y_t)$, where $h$ is a function satisfying $h(y) < y$ for all $y$. After a country defaults, it temporarily loses access to the international credit market.

At the end of each period during which the country is in default, it regains access to international credit markets with probability $\theta \in (0, 1)$. With probability $1 - \theta$ it remains in financial autarky. When a country regains access to foreign borrowing, its debt is reset to zero.

We can cast this as an RDP by considering the value of each state and action. We set the state space X to be the set of all $(y, b, d)$ in $\mathsf{Y} \times \mathsf{B} \times \{0, 1\}$, where B is a finite subset of $[-m, \infty)$ indicating possible choices for bond holdings $b_t$ and $d$ is a binary variable indicating whether the country is in default ($d = 0$ means not in default and $d = 1$ means in default).

The value space $V$ is all of $\mathbb{R}^{\mathsf{X}}$. The action space is $(b_a, d_a) \in \mathsf{B} \times \{0, 1\}$ indicating choices for bond holdings and default. The feasible correspondence specifies feasible $(b_a, d_a)$ at given state $(y, b, d)$ and is given by

$$\Gamma(y, b, d) = \begin{cases} \mathsf{B} \times \{0, 1\} & \text{if } d = 0 \text{ and} \\ \{0\} \times \{1\} & \text{if } d = 1. \end{cases}$$

```julia
using QuantEcon
include("quantile_function.jl")

"Creates an instance of the job search model."
function create_markov_js_model(;
        n=100,         # wage grid size
        ρ=0.9,         # wage persistence
        ν=0.2,         # wage volatility
        β=0.98,        # discount factor
        c=1.0,         # unemployment compensation
        τ=0.5          # quantile parameter
    )
    mc = tauchen(n, ρ, ν)
    w_vals, P = exp.(mc.state_values), mc.p
    return (; n, w_vals, P, β, c, τ)
end

"""
The policy operator

    (T_σ v)(w) = σ(w) (w / (1-β)) + (1 - σ(w))(c + β (R_τ v)(w))

"""
function T_σ(v, σ, model)
    (; n, w_vals, P, β, c, τ) = model
    h = c .+ β * R(τ, v, P)
    e = w_vals ./ (1 - β)
    return σ .* e + (1 .- σ) .* h
end

" Get a v-greedy policy."
function get_greedy(v, model)
    (; n, w_vals, P, β, c, τ) = model
    σ = w_vals / (1 - β) .≥ c .+ β * R(τ, v, P)
    return σ
end
```

Listing 26: Job search with quantile operator (`quantile_js.jl`)

In other words, if $d = 0$, so the country is not in default, the government can choose any $b_a \in B$ and also any $d_a \in \{0, 1\}$ (i.e., default or not default). If $d = 1$, however, the government has no choices. We represent this situation by $b_a = 0$ and $d_a = 1$.

The value aggregator takes the form

$$B((y, b, d), (b_a, d_a), v) = \text{value in state } (y, b, d) \text{ under action } (b_a, d_a).$$

To specify it we decompose the problem across cases for $d$ and $d_a$. First consider the case where $d = 0$ (not currently in default) and $d_a = 0$ (the government chooses not to default). For this case $y + b - qb_a$ is current consumption, so we set

$$B((y, b, 0), (b_a, 0), v) = u(y + b - qb_a) + \beta \sum_{y'} v(y', b_a, 0) Q(y, y') \qquad (8.32)$$

Now consider the case where $d = 0$ and $d_a = 1$, so the government chooses to default. Then current consumption is $h(y)$ and we set

$$B((y, b, 0), (b_a, 1), v) = u(h(y)) + \beta$$
$$\left[ \theta \sum_{y'} v(y', 0, 0) Q(y, y') + (1 - \theta) \sum_{y'} v(y', 0, 1) Q(y, y') \right]. \qquad (8.33)$$

The term $\sum_{y'} v(y', 0, 0) Q(y, y')$ is the expected value next period when the country is readmitted to international financial markets (with $b' = 0$ and $d' = 0$), while the term $\sum_{y'} v(y', 0, 1) Q(y, y')$ is the expected value next period when default continues (with $b' = 0$ and $d' = 1$).

Since $B((y, b, 1), (b_a, 0), v)$ is not feasible (a defaulted country cannot itself directly choose to reenter financial markets), the only other possibility is $B((y, b, 1), (b_a, 1), v)$, which is the expected value when the country remains in default. But this is the same as $B((y, b, 0), (b_a, 1), v)$ specified above: the value for a country that stays in default is the same as that for a country that newly enters default.

EXERCISE 8.2.7. By working through cases (8.32)–(8.33) for the value aggregator $B$, show that the model described above is a contractive RDP.

## 8.2.2 Eventually Contracting RDPs

Many RDPs are not contracting. There is no single method for handling all types of non-contractive RDPs, so we introduce alternative techniques over the next few sections. The first such technique, treated in this section, handles RDPs that contract

"eventually," even though they may fail to contract in one step. We show that these eventually contracting RDPs are globally stable, so all of the fundamental optimality results apply.

One application for these results is the MDP model with state-dependent discounting treated in Chapter 6. This section contains a proof of the main optimality result in that chapter (Proposition 6.2.2 on page 194).

### 8.2.2.1 Definition and Properties

Let $\mathcal{R} = (\Gamma, V, B)$ be an RDP with policy set $\Sigma$. We call $\mathcal{R}$ **eventually contracting** if there is a map $L$ from $\mathsf{G} \times \mathsf{X}$ to $\mathbb{R}_+$ such that

$$|B(x, a, v) - B(x, a, w)| \leqslant \sum_{x'} |v(x') - w(x')| L(x, a, x') \tag{8.34}$$

for all $(x, a) \in \mathsf{G}$ and all $v, w \in V$, and moreover,

$$\sigma \in \Sigma \implies \rho(L_\sigma) < 1 \quad \text{where} \quad L_\sigma(x, x') := L(x, \sigma(x), x').$$

**Proposition 8.2.4.** *Let $\mathcal{R} = (\Gamma, V, B)$ be an RDP. If $\mathcal{R}$ is eventually contracting and $V$ is closed in $\mathbb{R}^\mathsf{X}$, then $\mathcal{R}$ is also globally stable and hence all of the optimality and convergence results in Theorem 8.1.1 apply.*

*Proof.* Let $\mathcal{R}$ be as stated and fix $\sigma \in \Sigma$. Let $T_\sigma$ be the associated policy operator and let $L_\sigma$ be the linear operator in (8.34). For fixed $v, w \in V$ we have

$$|(T_\sigma v)(x) - (T_\sigma w)(x)| = |B(x, \sigma(x), v) - B(x, \sigma(x), w)|$$
$$\leqslant \sum_{x'} |v(x') - w(x')| \, L_\sigma(x, x').$$

Since $L_\sigma \geqslant 0$ and $\rho(L_\sigma) < 1$, Proposition 6.1.6 on page 191 implies that $T_\sigma$ is eventually contracting on $V$. Since $V$ is closed in $\mathbb{R}^\mathsf{X}$, it follows that $T_\sigma$ is globally stable (Theorem 6.1.5, page 190). Hence $\mathcal{R}$ is globally stable, as claimed. $\qquad\square$

EXERCISE 8.2.8. Prove: If $\mathcal{R} = (\Gamma, V, B)$ is eventually contracting, $V$ is closed in $\mathbb{R}^\mathsf{X}$ and $T$ is the Bellman operator generated by $\mathcal{R}$, then $T$ is globally stable on $V$.

EXERCISE 8.2.9. In §4.1.2 we studied an optimal exit problem for a firm. We can modify this problem to handle stochastic interest rates by introducing the RDP $\mathcal{R} = (\Gamma, V, B)$ on state space $\mathsf{X}$ with $\Gamma(x) = \{0, 1\}$, $V = \mathbb{R}^{\mathsf{X}}$ and

$$B(x, a, v) = as + (1 - a) \left[ \pi(x) + \beta(x) \sum_{x'} v(x') Q(x, x') \right]$$

for some $\beta \in \mathbb{R}_+^{\mathsf{X}}$. (We suppose that state-dependence of $\beta$ is generated by state-dependent interest rates.) State the Bellman equation for this problem. Prove that $\mathcal{R}$ is globally stable whenever there exists an $L \in \mathcal{L}(\mathbb{R}^{\mathsf{X}})$ such that $\rho(L) < 1$ and $\beta(x) Q(x, x') \leqslant L(x, x')$ for all $x, x' \in \mathsf{X}$.

### 8.2.2.2 Optimality for MDPs with State-Dependent Discounting

With Proposition 8.2.4 in hand, we can complete the proof of Proposition 6.2.2 on page 194, which pertained to optimality properties for MDPs with state-dependent discounting.

Let $(\Gamma, \beta, r, P)$ be an MDP with state-dependent discounting, as defined in §6.2.1.1. The state space is $\mathsf{X}$ and the action space is $\mathsf{A}$. The function $\beta$ maps $\mathsf{G} \times \mathsf{X}$ to $\mathbb{R}_+$. Set

$$L(x, a, x') := \beta(x, a, x') P(x, a, x') \quad \text{and} \quad L_\sigma(x, x') := L(x, \sigma(x), x')$$

for all $(x, a, x') \in \mathsf{G} \times \mathsf{X}$ and $\sigma \in \Sigma$.

Assume the conditions of Proposition 6.2.2, so that $\rho(L_\sigma) < 1$ for all $\sigma \in \Sigma$.

If we set

$$B(x, a, v) := r(x, a) + \sum_{x'} v(x') \beta(x, a, x') P(x, a, x') \tag{8.35}$$

and take $V$ to be all of $\mathbb{R}^{\mathsf{X}}$, then $\mathcal{R} := (\Gamma, V, B)$ forms an RDP, as discussed in Exercise 8.2.4. We claim that $\mathcal{R}$ is an eventually contracting RDP.

To see this, fix $v, w \in V$ and $(x, a) \in \mathsf{G}$. Applying the definition (8.35) and the triangle inequality, we have

$$|B(x, a, v) - B(x, a, w)| \leqslant \left| \sum_{x'} [v(x') - w(x')] \beta(x, a, x') P(x, a, x') \right|$$
$$\leqslant \sum_{x'} |v(x') - w(x')| L(x, a, x'),$$

Under the stated assumptions, for each $\sigma \in \Sigma$, the operator $L_\sigma(x, x') = L(x, \sigma(x), x')$ satisfies $\rho(L_\sigma) < 1$. Hence $\mathcal{R}$ is eventually contracting, as claimed. Since $V = \mathbb{R}^{\mathsf{X}}$ is closed, Proposition 8.2.4 implies that $\mathcal{R}$ is a globally stable RDP. The claims in Proposition 6.2.2 now follow from Theorem 8.1.1.

## 8.2.3 Convex and Concave RDPs

Theorem 8.1.1 shows that RDPs have excellent optimality properties when all policy operators are globally stable on value space. So far we have looked at conditions for stability based on contractions (§8.2.1) and eventual contractions (§8.2.2). But sometimes both of these approaches fail and we need alternative conditions.

In this section we explore alternative conditions based on Du's theorem (page 217). Du's theorem is well suited to the task of studying stability of policy operators, since it leverages the fact that all policy operators are order-preserving.

### 8.2.3.1 Definitions and Optimality

Let $\mathcal{R} = (\Gamma, V, B)$ be an RDP with $V = [v_1, v_2]$ for some $v_1 \leqslant v_2$ in $\mathbb{R}^X$. We call $\mathcal{R}$ **convex** if

(i) for all $(x, a) \in \mathsf{G}$, $\lambda \in [0, 1]$ and $v, w$ in $V$, we have

$$B(x, a, \lambda v + (1 - \lambda)w) \leqslant \lambda B(x, a, v) + (1 - \lambda)B(x, a, w) \quad \text{and,} \qquad (8.36)$$

(ii) there exists a $\delta > 0$ such that

$$B(x, a, v_2) \leqslant v_2(x) - \delta[v_2(x) - v_1(x)] \text{ for all } (x, a) \in \mathsf{G}. \qquad (8.37)$$

Analogous to the convex case, we call $\mathcal{R}$ **concave** if

(i) for all $(x, a) \in \mathsf{G}$, $\lambda \in [0, 1]$ and $v, w$ in $V$, we have

$$B(x, a, \lambda v + (1 - \lambda)w) \geqslant \lambda B(x, a, v) + (1 - \lambda)B(x, a, w) \quad \text{and,} \qquad (8.38)$$

(ii) there exists a $\delta > 0$ such that

$$B(x, a, v_1) \geqslant v_1(x) + \delta[v_2(x) - v_1(x)] \text{ for all } (x, a) \in \mathsf{G}. \qquad (8.39)$$

In both of the definitions above, condition (ii) is rather complex. The next exercise provides simpler sufficient conditions.

EXERCISE 8.2.10. Prove that (8.37) holds whenever

$$B(x, a, v_2) < v_2(x) \text{ for all } (x, a) \in \mathsf{G}. \tag{8.40}$$

Similarly, prove that (8.39) holds whenever

$$B(x, a, v_1) > v_1(x) \text{ for all } (x, a) \in \mathsf{G}. \tag{8.41}$$

Both convexity and concavity yield stability, as the next proposition shows.

**Proposition 8.2.5.** *If $\mathcal{R}$ is either convex or concave, then $\mathcal{R}$ is globally stable.*

*Proof.* We begin with the convex case. Fix $\sigma \in \Sigma$. By the monotonicity property of RDPs, $T_\sigma$ is an order-preserving self-map on $V$. Since (8.36) holds, $T_\sigma$ is also a convex operator on $V$. Moreover, $T_\sigma v_1 \geqslant v_1$ because $T_\sigma \colon V \to V$ and, by (8.40), $T_\sigma v_2 \leqslant v_2 - \delta(v_2 - v_1)$. Hence Du's theorem on page 217 applies and $T_\sigma$ is globally stable on $V$. This shows that $\mathcal{R}$ is a globally stable RDP.

The proof of the concave case is analogous (using Du's theorem applied to order-preserving concave operators). □

It follows from Proposition 8.2.5 that, for convex and concave RDPs, all of the optimality and convergence results in Theorem 8.1.1 apply.

### 8.2.3.2 Application to MDPs

Proposition 8.2.5 can be applied to establish optimality properties of regular MDPs. This exercise is redundant in the sense that optimality properties of regular MDPs have already been established using other means. At the same time, some of the arguments developed here will be helpful when we face more sophisticated problems below.

To sketch the argument, let $\mathcal{R} = (\Gamma, B, V)$ be an RDP generated by an ordinary MDP $(\Gamma, \beta, r, P)$, as discussed in Example 8.1.1 on page 248. In particular, $V = \mathbb{R}^{\mathsf{X}}$, and $B(x, a, v) = r(x, a) + \beta \sum_{x'} v(x') P(x, a, x')$. We set $r_1 := \min r$ and $r_2 := \max r$. Then we fix $\varepsilon > 0$ and define $V$ via

$$\hat{V} := [v_1, v_2] \quad \text{where} \quad v_1 := \frac{r_1 - \varepsilon}{1 - \beta} \text{ and } v_2 := \frac{r_2 + \varepsilon}{1 - \beta}. \tag{8.42}$$

(The functions $v_1$ and $v_2$ are constant.) We claim that the RDP $\hat{\mathcal{R}} := (\Gamma, \hat{V}, B)$ is both convex *and* concave.

EXERCISE 8.2.11. Prove that (8.40) and (8.41) both hold for $\hat{\mathcal{R}}$.

EXERCISE 8.2.12. Complete the proof that $\hat{\mathcal{R}}$ is both concave and convex.

# 8.3 Further Applications

In this section we consider some applications of the optimality results in §8.2.

## 8.3.1 Risk-Sensitive RDPs

In §7.2.2 we introduced risk-sensitive preferences and discussed a recursive utility problem. Now we embed risk-sensitive preferences into a dynamic program and apply the preceding optimality results to compute optimal policies.

### 8.3.1.1 Optimality Results

Consider the risk-sensitive preference RDP in Example 8.1.6, with state space $\mathsf{X}$ and action space $\mathsf{A}$. Let $V = \mathbb{R}^{\mathsf{X}}$. For $(x, a) \in \mathsf{G}$ and $v \in V$, we can express the aggregator as

$$B(x, a, v) := r(x, a) + \beta (R_\theta^a v)(x)$$

where $\theta$ is a nonzero constant and

$$(R_\theta^a v)(x) := \frac{1}{\theta} \ln \left\{ \sum_{x'} \exp(\theta v(x')) P(x, a, x') \right\}.$$

Notice that, for each fixed $a \in \Gamma(x)$, the operator $R_\theta^a$ is an entropic certainty equivalent operator on $V$ (see Example 7.3.2 on page 233).

**Proposition 8.3.1.** *If $\beta < 1$, then $(\Gamma, V, B)$ is contracting.*

*Proof.* Fix $\beta < 1$. We show that $(\Gamma, V, B)$ obeys Blackwell's condition (§8.2.1.3). To this end, fix $v \in V$, $(x, a) \in \mathsf{G}$, and $\lambda \geqslant 0$. Since $R_\theta^a$ is constant-subadditive (Exercise 7.3.8 on page 234), we have

$$B(x, a, v + \lambda) = r(x, a) + \beta [R_\theta^a (v + \lambda)](x) \leqslant r(x, a) + \beta (R_\theta^a v)(x) + \beta \lambda.$$

The right-hand side equals $B(x, a, v) + \beta \lambda$, so Blackwell's condition holds. The claim in Proposition 8.3.1 now follows from Exercise 8.2.3. □

The next exercise pertains to quantile preferences rather than risk-sensitive preferences, but the result can be obtained via a relatively straightforward modification of the proof of Proposition 8.3.1.

EXERCISE 8.3.1. Let $\mathcal{R} := (\Gamma, V, B)$ be an RDP with $V = \mathbb{R}^{\mathsf{X}}$ and fix $\tau \in [0, 1]$. Let $B(x, a, v) = r(x, a) + \beta (R_\tau^a v)(x)$ where, for each $a \in \Gamma(x)$, the map $R_\tau^a$ is given by

$$(R_\tau^a v)(x) = \min \left\{ y \in \mathbb{R} \,\Big|\, \sum_{x'} \mathbb{1}\{v(x') \leqslant y\} P(x, a, x') \geqslant \tau \right\} \qquad (v \in V, \ x \in \mathsf{X}).$$

Prove that $\mathcal{R}$ is globally stable whenever $\beta < 1$.

### 8.3.1.2 Risk-Sensitive Job Search

Let's consider a job search problem where future wage outcomes are evaluated via risk-sensitive expectations. The associated Bellman operator is

$$(Tv)(w) = \max \left\{ \frac{w}{1 - \beta}, \ c + \frac{\beta}{\theta} \ln \left[ \sum_{w'} \exp(\theta v(w')) P(w, w') \right] \right\} \qquad (w \in \mathsf{W}).$$

Here $\theta$ is a nonzero parameter and other details are as in §3.3.1. We can represent the problem as an RDP with state space W, action space $\mathsf{A} = \{0, 1\}$, feasible correspondence $\Gamma(w) = \mathsf{A}$, value space $V := \mathbb{R}^{\mathsf{W}}$, and value aggregator

$$B(w, a, v) = a \frac{w}{1 - \beta} + (1 - a) \left\{ c + \frac{\beta}{\theta} \ln \left[ \sum_{w'} \exp(\theta v(w')) P(w, w') \right] \right\}.$$

If $\theta < 0$, then the agent is risk-averse with respect to the gamble associated with continuing and waiting for new wage draws. If $\theta > 0$ then the agent is risk-loving with respect to such gambles. For $\theta \approx 0$, the agent is close to risk-neutral.

Figure 8.2 shows how the continuation value, value function and optimal decision vary with $\theta$. Apart from $\theta$, parameters are identical to those in Listing 10 on page 100. Indeed, for $\theta$ close to zero, as in the middle sub-figure of Figure 8.2, we see that the value function and reservation wage are almost identical to those from the risk-neutral model in Figure 3.5 on page 101.

As expected, a negative value of $\theta$ tends to reduce the continuation value and hence the reservation wage, since the agent's dislike of risk encourages early acceptance of an offer. For positive values of $\theta$ the reverse is true, as seen in the bottom sub-figure.

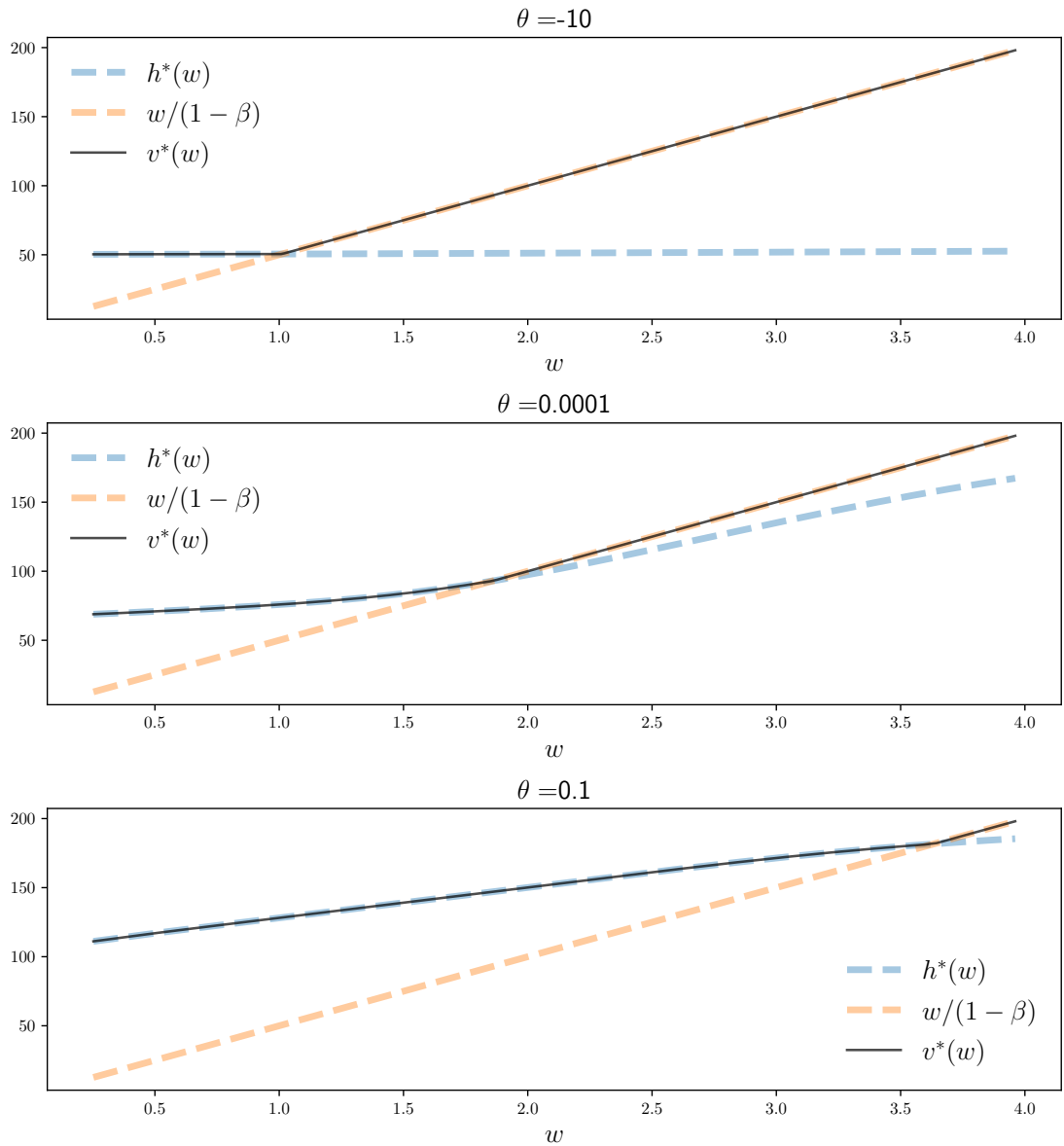

Figure 8.2:  Job search with risk-sensitive preferences

EXERCISE 8.3.2. Replicate Figure 8.2. The simplest method is to modify the code in Listing 10 and use value function iteration.

## 8.3.2 Adversarial Agents

Some problems in economics, finance and artificial intelligence assume that decisions emerge from a dynamic two-person zero sum game in which the two agents' preferences are perfectly *mis*aligned. This can lead to a dynamic program where the Bellman equation takes the form

$$v(x) = \max_{a \in \Gamma(x)} \inf_{d \in D(x,a)} B(x, a, d, v) \qquad (x \in \mathsf{X}, \ v \in \mathbb{R}^{\mathsf{X}}), \tag{8.43}$$

where $B(x, a, d, v)$ represents lifetime value for the decision maker conditional on her current action $a$ and her adversary's action $d$. The decision maker chooses action $a \in \Gamma(x)$ with the knowledge that the opponent will then choose $d \in D(x, a)$ to minimize her lifetime value.

**Remark 8.3.1.** In some settings we can replace the inf in (8.43) with min. In other settings this is not so obvious. For this reason we use inf throughout, paired with the assumption that $B$ is bounded below. This means that the infimum is always well-defined and finite.

### 8.3.2.1 Optimality

To establish optimality properties in the setting of (8.43), we introduce the following assumptions:

(a) If $v, w \in \mathbb{R}^{\mathsf{X}}$ with $v \leqslant w$, then

$$B(x, a, d, v) \leqslant B(x, a, d, w) \quad \text{for all } x \in \mathsf{X}, \ a \in \Gamma(x), \ d \in D(x, a).$$

(b) There exists a $v_1 \in \mathbb{R}^{\mathsf{X}}$ and $\varepsilon > 0$ such that

$$v_1(x) + \varepsilon \leqslant B(x, a, d, v_1) \quad \text{for all } x \in \mathsf{X}, \ a \in \Gamma(x), \ d \in D(x, a).$$

(c) There exists a $v_2 \in \mathbb{R}^{\mathsf{X}}$ such that $v_1 \leqslant v_2$ and

$$B(x, a, d, v_2) \leqslant v_2(x) \quad \text{for all } x \in \mathsf{X}, \ a \in \Gamma(x), \ d \in D(x, a).$$

(d) If $\lambda \in [0, 1]$ and $v, w \in \mathbb{R}^{\mathsf{X}}$, then

$$B(x, a, d, \lambda v + (1 - \lambda)w) \geqslant \lambda B(x, a, d, v) + (1 - \lambda)B(x, a, d, w)$$

for all $x \in \mathsf{X}$, $a \in \Gamma(x)$ and $d \in D(x, a)$.

Condition (a) is a natural monotonicity condition: a uniform increase in continuation values increases current value at all states and actions. Conditions (b) and (c) provide upper and lower bounds. Condition (d) is a concavity condition.

To analyze the decision maker's problem, we set $V := [v_1, v_2]$ and

$$\hat{B}(x, a, v) := \inf_{d \in D(x, a)} B(x, a, d, v) \qquad ((x, a) \in \mathsf{G}, \ v \in V),$$

We consider $\mathcal{R} = (\Gamma, V, \hat{B})$.

**Proposition 8.3.2.** *If conditions* (a)–(d) *hold, then $\mathcal{R}$ is a concave RDP.*

An immediate corollary of Proposition 8.3.2 is that, under the stated conditions, the decision maker's problem is a globally stable RDP, and hence the fundamental optimality properties in Theorem 8.1.1 all hold.

In the proof of Proposition 8.3.2, we use the following exercise.

EXERCISE 8.3.3. Let $f$ and $g$ map nonempty set $D$ into $\mathbb{R}$. Assume that both $f$ and $g$ are bounded below. Prove that, in this setting,

$$\inf_{d \in D}(f(d) + g(d)) \geqslant \inf_{d \in D} f(d) + \inf_{d \in D} g(d).$$

*Proof of Proposition 8.3.2.* First we need to check that $\mathcal{R}$ is an RDP. In view (a) we have $\hat{B}(x, a, v) \leqslant \hat{B}(x, a, w)$ whenever $(x, a) \in \mathsf{G}$ and $v, w \in V$ and $v \leqslant w$. Also, by (b) and (c),

$$v_1(x) < \hat{B}(x, a, v_1) \ \text{ and } \hat{B}(x, a, v_2) \leqslant v_2(x) \quad \text{for all } (x, a) \in \mathsf{G}. \tag{8.44}$$

As a result, $v_1(x) \leqslant \hat{B}(x, \sigma(x), v) \leqslant v_2(x)$ for all $x \in \mathsf{X}$ and $v \in V$. Together, these facts imply the monotonicity and consistency conditions required of an RDP.

In view of (8.44) and Exercise 8.2.10 on page 274, to establish that $\mathcal{R}$ is concave, we need only show that, for fixed $\lambda \in [0, 1]$ and $v, w \in V$,

$$\hat{B}(x, a, \lambda v + (1 - \lambda)w) \geqslant \lambda \hat{B}(x, a, v) + (1 - \lambda)\hat{B}(x, a, w) \tag{8.45}$$

for all $(x, a) \in \mathsf{G}$. This holds because, given $(x, a) \in \mathsf{G}$, $\lambda \in [0, 1]$ and $v, w \in V$,

$$
\begin{aligned}
\hat{B}(x, a, \lambda v + (1 - \lambda)w) &= \inf_{d \in D(x,a)} B(x, a, d, \lambda v + (1 - \lambda)w) \\
&\geqslant \inf_{d \in D(x,a)} [\lambda B(x, a, d, v) + (1 - \lambda)B(x, a, d, w)] \\
&\geqslant \lambda \inf_{d \in D(x,a)} B(x, a, d, v) + (1 - \lambda) \inf_{d \in D(x,a)} B(x, a, d, w),
\end{aligned}
$$

where the first inequality is by condition (d) above and the second is by Exercise 8.3.3. This proves (8.45), so $\mathcal{R}$ is a concave RDP. $\qquad\square$

### 8.3.2.2 A Perturbed MDP Problem

In this section we provide a relatively abstract application of Proposition 8.3.2. Later, in §8.3.3, we will see more concrete applications.

The setting we consider is a modified MDP where the adversarial agent's actions affect the reward function and transition kernel. This leads to a Bellman equation of the form

$$
v(x) = \max_{a \in \Gamma(x)} \inf_{d \in D(x,a)} \left\{ r(x, a, d) + \beta \sum_{x'} v(x')P(x, a, d, x') \right\} \qquad (x \in \mathsf{X}) \qquad (8.46)
$$

The choice perturbation $d \in D(x, a)$ is made by the adversary. The object $P$ is a stochastic kernel, in the sense that $P(x, a, d, \cdot)$ is a distribution over $\mathsf{X}$ for each feasible $(x, a, d)$. We assume that $\Gamma$ is a nonempty correspondence from $\mathsf{X}$ to $\mathsf{A}$ and $D(x, a)$ is nonempty for all $(x, a) \in \mathsf{G}$. Let

$$
\hat{B}(x, a, v) = \inf_{d \in D(x,a)} \left\{ r(x, a, d) + \beta \sum_{x'} v(x')P(x, a, d, x') \right\} \qquad ((x, a) \in \mathsf{G}).
$$

To construct the value space $V$, we let $r_1 = \min r$ and $r_2 = \max r$, and set

$$
V = [v_1, v_2] \quad \text{where} \quad v_1 := \frac{r_1 - \varepsilon}{1 - \beta} \quad \text{and} \quad v_2 := \frac{r_2}{1 - \beta}. \qquad (8.47)
$$

(These constant functions are similar to $v_1, v_2$ in (8.42) on page 274.)

EXERCISE 8.3.4. Prove: For $v_1, v_2$ in (8.47), conditions (b)–(c) on page 278 hold.

**Lemma 8.3.3.** *The perturbed MDP model* $\mathcal{R} := (\Gamma, V, \hat{B})$ *is a concave RDP.*

An immediate corollary of Lemma 8.3.3 is that $\mathcal{R}$ is globally stable (via Proposition 8.2.5) and all optimality results in Theorem 8.1.1 apply.

*Proof of Lemma 8.3.3.* It suffices to show that $\mathcal{R}$ obeys (a)–(d) on page 278. Condition (a) and (d) are elementary in this setting. Conditions (b) and (c) were established in Exercise 8.3.4. □

### 8.3.3 Ambiguity and Robustness

Until now we have considered agents facing decision problems where outcomes are uncertain but probabilities are known. For example, while the job seeker introduced in Chapter 1 does not know the next period wage offer when choosing her current action, she does know the distribution of that offer. She uses this distribution to determine an optimal course of action. Similarly, the controllers in our discussion of optimal stopping and MDPs used their knowledge of the Markov transition law to determine an optimal policy.

In many cases, the assumption that the decision maker knows all probability distributions that govern outcomes under different actions is debatable. In this section we study lifetime valuations in settings of **Knightian uncertainty** (Knight, 1921), which means that outcome distributions are themselves unknown. Some authors refer to Knightian uncertainty as **ambiguity**.

Below we consider some dynamic problems where decision makers face Knightian uncertainty.

#### 8.3.3.1 Robust Control

First we study the choices of a decision maker who knows her reward function but distrusts her specification of the stochastic kernel $P$ that describes the evolution of the state. This distrust is expressed by assuming that she knows that $P$ belongs to some class of stochastic kernels from $\mathsf{G} \times \mathsf{X}$ to $\mathsf{X}$. This can lead to aggregators of the form

$$B(x, a, v) = r(x, a) + \beta \inf_{P \in \mathcal{P}(x,a)} \left\{ \sum_{x'} v(x') P(x, a, x') \right\} \tag{8.48}$$

for $(x, a) \in \mathsf{G}$. As usual, $r$ maps $\mathsf{G}$ to $\mathbb{R}$ and $\beta \in (0, 1)$. The decision maker can construct a policy that is robust to her distrust of the stochastic kernel by using this aggregator $B$. Such aggregators arise in the field of robust control.

Positing that the decision maker knows a nontrivial set of stochastic kernels is a way of modeling Knightian uncertainty, as distinguished from risks that are described by known probability distributions.

**Example 8.3.1.** Consider the simple job search problem from Chapter 1. Suppose that the worker believes that the wage offer distribution lies in some subset $\mathcal{P}$ of $\mathcal{D}(W)$. She can seek a decision rule that is robust to worst-case beliefs by optimizing with aggregator

$$B(w, a, v) = a\frac{w}{1-\beta} + (1-a)\inf_{\varphi \in \mathcal{P}}\sum_{w'}v(x')\varphi(w').$$

Returning to the robust control model with aggregator $B$ in (8.48), we take $V$ is as defined in (8.47) and set $\mathcal{R} = (\Gamma, V, B)$. The set $\mathcal{P}$ of stochastic kernels is entirely arbitrary.

**Proposition 8.3.4.** $\mathcal{R}$ *is a concave RDP.*

*Proof.* Writing $B$ as

$$B(x, a, v) = \inf_{P \in \mathcal{P}(x,a)}\left\{r(x, a) + \beta\sum_{x'}v(x')P(x, a, x')\right\}, \tag{8.49}$$

we see that $\mathcal{R}$ is a special case of the perturbed MDP model in §8.3.2.2. Concavity now follows from Lemma 8.3.3. □

We conclude from the discussion above that the robust control RDP is globally stable. Hence all of the fundamental optimality properties hold.

### 8.3.3.2 Robustness and Adversarial Agents

A more general way to implement robustness is via the aggregator

$$B(x, a, v) = r(x, a) + \beta\inf_{P \in \mathcal{P}(x,a)}\left\{\sum_{x'}v(x')P(x, a, x') + d(P(x, a, \cdot), \bar{P}(x, a, \cdot))\right\}. \tag{8.50}$$

In this set up, $\mathcal{P}(x, a)$ is often large, weakening the constraint on $P$. At the same time, we introduce the penalty term $d(P(x, a, \cdot), \bar{P}(x, a, \cdot))$, which can be understood as recording the deviation between a given kernel $P$ and some baseline specification $\bar{P}$.

One interpretation of this setting is that the decision maker begins with a baseline specification of dynamics but lacks confidence in its accuracy. In her desire to choose a robust policy, she imagines herself playing against an adversarial agent. Her adversary can choose transition kernels that deviate from the baseline, but the presence of the penalty term means that extreme deviations are curbed.

If we define

$$\hat{r}(x, a) = r(x, a) + d(P(x, a, \cdot), \bar{P}(x, a, \cdot)),$$

then (8.50) can be expressed as

$$B(x, a, v) = \inf_{P} \left\{ \hat{r}(x, a) + \beta \sum_{x'} v(x') P(x, a, x') \right\}.$$

This is a special case of (8.49), so the same optimality theory applies.

### 8.3.3.3 Connection to Risk-Sensitive Preferences

A useful measure of discrepancy between two probability distributions is the **Kullback–Liebler divergence** (KL divergence)

$$d_{KL}(q \mid p) := \sum_{x} q(x) \ln \left( \frac{q(x)}{p(x)} \right) \quad \text{for } q, p \in \mathcal{D}(\mathsf{X}).$$

It is assumed here that $q \prec_{\mathrm{ac}} p$, which means that $q(x) = 0$ whenever $p(x) = 0$. We note for future reference that $d_{KL}$ obeys the **duality formula for variational inference**, which states that, given $h \in \mathbb{R}^{\mathsf{X}}$,

$$\ln \sum_{x} \exp(h(x)) p(x) = \sup_{q \prec_{\mathrm{ac}} p} \left\{ \sum_{x} h(x) q(x) - d_{KL}(q \mid p) \right\}. \tag{8.51}$$

(See, e.g., Dupuis and Ellis (2011), Proposition 1.4.2.)

In robust control, KL divergence can be used measure deviation between the baseline specification and alternative specifications. It turns out that, under this measure of divergence, there is a tight relationship between robust control and risk-sensitive preferences.

To illustrate this relationship, we fix $\theta < 0$ and set $d_{\theta} := -(1/\theta)d_{KL}$, so that $d_{\theta}$ is a simple positive rescaling of the Kullback–Leibler divergence. Using $d_{\theta}$ in (8.50) leads

to

$$B(x, a, \nu) = r(x, a) + \beta \inf_{P \in \mathcal{P}(x,a)} \left\{ \sum_{x'} \nu(x') P(x, a, x') + d_\theta(P(x, a, \cdot) \,|\, \bar{P}(x, a, \cdot)) \right\}.$$

The constraint set $\mathcal{P}(x, a)$ is all $P \in \mathcal{M}(\mathbb{R}^\mathsf{X})$ such that $P(x, a, \cdot) \prec_{\mathrm{ac}} \bar{P}(x, a, \cdot)$.

If we multiply both sides of the variational formula (8.51) by $(1/\theta)$ and set $h = \theta \nu$ we get

$$\frac{1}{\theta} \ln \sum_x \exp(\theta \nu(x)) p(x) = \inf_{q \prec_{\mathrm{ac}} p} \left\{ \sum_x \nu(x) q(x) - \frac{1}{\theta} d_{KL}(q \,|\, p) \right\}.$$

This allows us to rewrite $B$ as

$$B(x, a, \nu) = r(x, a) + \beta \frac{1}{\theta} \ln \left\{ \sum_{x'} \exp(\theta \nu(x')) \bar{P}(x, a, x') \right\}.$$

Hence, for this choice of deviation, the robust control aggregator (8.50) reduces to the risk-sensitive aggregator (see Example 8.1.6 on page 250) under the baseline transition kernel.

### 8.3.4 Smooth Ambiguity

Ju and Miao (2012) propose and study a recursive smooth ambiguity model in the context of asset pricing. A generic discrete formulation of their optimization problem can be expressed in terms of the aggregator

$$B(x, a, \nu) = \left\{ r(x, a) + \beta \left\{ \int \left[ \sum_{x'} \nu(x')^\gamma P_\theta(x, a, x') \right]^{\kappa/\gamma} \mu(x, \mathrm{d}\theta) \right\}^{\alpha/\kappa} \right\}^{1/\alpha}, \tag{8.52}$$

where $\alpha, \kappa, \gamma$ are nonzero parameters, $P_\theta$ is a stochastic kernel from $\mathsf{G}$ to $\mathsf{X}$ for each $\theta$ in a finite dimensional parameter space $\Theta$, and $\mu(x, \cdot)$ is a probability distribution over $\Theta$ for each $x \in \mathsf{X}$. The distribution $\mu(x, \cdot)$ represents subjective beliefs over the transition rule for the state.

The aggregator $B$ in (8.52) is defined for $x \in \mathsf{X}$, $a \in \Gamma(x)$ and $\nu \in I$, where $I$ is be the interior of the positive cone of $\mathbb{R}^\mathsf{X}$. To ensure finite real values, we assume $r \gg 0$.

As with the Epstein–Zin case, $\alpha$ parameterizes the elasticity of intertemporal substitution and $\gamma$ governs risk aversion. The parameter $\kappa$ captures ambiguity aversion. If $\kappa = \gamma$, the agent is said to be ambiguity neutral.

EXERCISE 8.3.5. Show that the smooth ambiguity aggregator $B$ reduces to the Epstein–Zin aggregator when the agent is ambiguity neutral.

Returning to (8.52), we focus on the case $\kappa < \gamma < 0 < \alpha < 1$, which includes the calibration used in Ju and Miao (2012). (Other cases can be handled using similar methods and details are left to the reader.) After constructing a suitable value space, we will show that the resulting RDP is globally stable.

As a first step, set $r_1 := \min r$, $r_2 := \max r$ and fix $\varepsilon > 0$. Consider the constant functions

$$v_1 := \left( \frac{r_1}{1 - \beta} \right)^{1/\alpha} \quad \text{and} \quad v_2 := \left( \frac{r_2 + \varepsilon}{1 - \beta} \right)^{1/\alpha}.$$

EXERCISE 8.3.6. Prove that

$$v_1 \leqslant B(x, a, v_1) \leqslant B(x, a, v_2) < v_2 \quad \text{for all } (x, a) \in \mathsf{G}. \tag{8.53}$$

In the remainder of this section on smooth ambiguity we set $V = [v_1, v_2]$.

EXERCISE 8.3.7. Prove that $\mathcal{R} := (\Gamma, V, B)$ is an RDP.

Here is our main result for this section. It implies that all optimality and convergence results for $\mathcal{R}$ are valid (see, in particular, Theorem 8.1.1).

**Proposition 8.3.5.** *Under the stated assumptions, the RDP $\mathcal{R}$ is a globally stable.*

To prove Proposition 8.3.5, we use a transformation, just as we did with the Epstein–Zin case in §8.1.4.1. To this end we introduce the composite parameters

$$\xi := \frac{\gamma}{\kappa} \in (0, 1) \quad \text{and} \quad \zeta := \frac{\kappa}{\alpha} < 0.$$

Then we define

$$\hat{B}(x, a, v) = \left\{ r(x, a) + \beta \left\{ \int \left[ \sum_{x'} v(x')^{\xi} P_{\theta}(x, a, x') \right]^{1/\xi} \mu(x, \mathrm{d}\theta) \right\}^{\zeta} \right\}^{1/\zeta} \tag{8.54}$$

and

$$\hat{V} = [\hat{v}_1, \hat{v}_2] \quad \text{where} \quad \hat{v}_1 := v_2^{1/\kappa} \text{ and } \hat{v}_2 := v_1^{1/\kappa}.$$

Note that $\hat{V}$ is a nonempty order interval of strictly positive real-valued functions, since $0 < v_1 < v_2$ and $\kappa < 0$. We set $\hat{\mathcal{R}} = (\Gamma, \hat{V}, \hat{B})$.

EXERCISE 8.3.8. Prove that $\hat{\mathcal{R}}$ is an RDP satisfying

$$\hat{v}_1 < \hat{B}(x, a, \hat{v}_1) \quad \text{and} \quad \hat{B}(x, a, \hat{v}_2) \leqslant \hat{v}_2 \quad \text{for all } (x, a) \in \mathsf{G}.$$

The next exercise shows that $\mathcal{R}$ and $\hat{\mathcal{R}}$ are topologically conjugate (see §8.1.4).

EXERCISE 8.3.9. Let $\varphi$ be defined on $(0, \infty)$ by $\varphi(t) = t^\kappa$. Show that

(i) $B(x, a, v) = \varphi^{-1}[\hat{B}(x, a, \varphi \circ v)]$ for all $v \in V$ and $(x, a) \in \mathsf{G}$, and

(ii) $\varphi$ is a homeomorphism from $[v_1, v_2]$ to $[\hat{v}_1, \hat{v}_2]$ (as subsets of $\mathbb{R}$).

**Lemma 8.3.6.** *For each $(x, a) \in \mathsf{G}$, the function $\hat{v} \mapsto \hat{B}(x, a, \hat{v})$ is concave on $\hat{V}$.*

*Proof.* Fix $(x, a) \in \mathsf{G}$. We write $\hat{B}(x, a, \hat{v})$ has

$$\hat{B}(x, a, \hat{v}) = \psi \left( \int f(\theta, v) \mu(x, \mathrm{d}\theta) \right)$$

where

$$f(\theta, v) := \left[ \sum_{x'} v(x')^\xi P_\theta(x, a, x') \right]^{1/\xi} \quad \text{and} \quad \psi(t) := \left\{ r(x, a) + \beta t^\zeta \right\}^{1/\zeta}$$

For fixed $\theta$, the function $v \mapsto f(\theta, v)$ is concave over all $v$ in the interior of the positive cone of $\mathbb{R}^\mathsf{X}$ by Lemma 7.3.1 on page 235. The real-valued function $\psi$ satisfies $\psi' > 0$ and $\psi'' < 0$ over $t \in (0, \infty)$. Since we are composing order-preserving concave functions, it follows that $\hat{B}(x, a, \hat{v})$ is concave on $\hat{V}$. □

*Proof of Proposition 8.3.5.* To prove that $\mathcal{R}$ is globally stable it suffices to prove that $\hat{\mathcal{R}}$ is globally stable (see Exercise 8.3.9 and Proposition 8.1.3 on page 261). Given the results of Exercise 8.3.8 and Lemma 8.3.6, the RDP $\hat{\mathcal{R}}$ is concave. But then $\hat{\mathcal{R}}$ is globally stable, by Proposition 8.2.5. □

## 8.3.5 Minimization

Until now, all theory and applications have concerned maximization of lifetime values. Now is a good time to treat minimization. Throughout this section, $\mathcal{R}$ is a well-posed RDP. The pointwise minimum $v^*_\downarrow := \bigwedge_\sigma v_\sigma$ is called the **min-value function** generated by $\mathcal{R}$. We call a policy $\sigma \in \Sigma$ **min-optimal** for $\mathcal{R}$ if $v_\sigma = v_\downarrow^*$. A policy $\sigma \in \Sigma$ is called $v$-**min-greedy** for $\mathcal{R}$ if

$$\sigma(x) \in \operatorname*{argmin}_{a \in \Gamma(x)} B(x, a, v) \quad \text{for all } x \in \mathsf{X}.$$

We say that $\mathcal{R}$ obeys **Bellman's principle of min-optimality** if

$$\sigma \in \Sigma \text{ is min-optimal for } \mathcal{R} \quad \Longleftrightarrow \quad \sigma \text{ is } v_\downarrow^*\text{-min-greedy.}$$

The **Bellman min-operator** $T_\downarrow$ is defined by

$$(T_\downarrow v)(x) = \min_{a \in \Gamma(x)} B(x, a, v) \qquad (x \in \mathsf{X}).$$

We say that $v \in V$ obeys the **min-Bellman equation** if $T_\downarrow v = v$. The algorithm defined by replacing "$v$-greedy" with "$v$-min-greedy" in Algorithm 8.1 (HPI) will be called **min-HPI**.

We can now state the following result, which is analogous to Theorem 8.1.1. In the statement, $\mathcal{R}$ is a well-posed RDP with min-value function $v_\downarrow^*$.

**Theorem 8.3.7** (Min-optimality)**.** *If $\mathcal{R}$ is globally stable, then*

(i) *$v_\downarrow^*$ is the unique solution to the min-Bellman equation in $V$,*

(ii) *$\mathcal{R}$ satisfies Bellman's principle of min-optimality,*

(iii) *$\mathcal{R}$ has at least one min-optimal policy, and*

(iv) *min-HPI returns an exact optimal policy in finitely many steps.*

Although we omit the details, a min-OPI convergence result directly analogous to the OPI convergence result in (v) of Theorem 8.1.1 also holds (after replacing maximization-based $v$-greedy policies with $v$-min-greedy policies).

Theorem 8.3.7 is proved in §9.1.5. For now we consider two applications that involve minimization.

### 8.3.5.1 Application: Shortest Paths

Recall the shortest path problem introduced in Example 8.1.8, where $\mathsf{X}$ is the vertices of a graph, $E$ is the edges, $c \colon E \to \mathbb{R}_+$ maps a travel cost to each edge $(x, x') \in E$, and $\mathcal{O}(x)$ is the set of direct successors of $x$. The aim is to minimize total travel cost to a destination node $d$. We adopt all assumptions from Exercise 8.1.17 and assume in addition that $c(x, x') = 0$ implies $x = d$. As in Exercise 8.1.17, we let $C(x)$ be the maximum cost of traveling to $d$ from $x$ along any directed path.

We regard the problem as an RDP $\mathcal{R} = (\mathcal{O}, V, B)$ with $V = [0, C]$ and

$$B(x, x', v) = c(x, x') + v(x') \qquad (x \in \mathsf{X}). \tag{8.55}$$

In the present setting, the function $v$ in (8.55) is often called the **cost-to-go** function, with $v(x')$ in (8.57) understood as remaining costs after moving to state $x'$.

While the value aggregator $B$ in (8.55) is simple, the absence of discounting (which is standard in the shortest path literature) means that $\mathcal{R}$ is not contracting. Fortunately, $\mathcal{R}$ turns out to be concave (in the sense of §8.2.3), which allows us to prove

**Proposition 8.3.8.** *Under the stated conditions, the shortest path RDP is globally stable and the min-value function $v_\downarrow^*$ is the unique solution to*

$$v_\downarrow^*(x) = \min_{x' \in \Gamma(x)} \left\{ c(x, x') + v_\downarrow^*(x') \right\} \qquad (x \in \mathsf{X})$$

*in $V$. A policy $\sigma \in \Sigma$ is min-optimal if and only if*

$$\sigma(x) \in \underset{x' \in \Gamma(x)}{\operatorname{argmin}} \left\{ c(x, x') + v_\downarrow^*(x') \right\} \quad \text{for all } x \in \mathsf{X}.$$

(In the present context, $v_\downarrow^*$ is also known as the **minimum cost-to-go** function.)

*Proof.* We first show that $\mathcal{R}$ is concave. By the definition of concave RDPs in §8.2.3, and given that $B(x, x', v)$ is affine in $v$ (and hence concave), it suffices to prove that there exists a $\delta > 0$ such that

$$c(x, x') \geqslant \delta C(x) \text{ for all } x \in \mathsf{X} \text{ and } x' \in \mathcal{O}(x). \tag{8.56}$$

(This corresponds to (8.39) on page 273 when $v_1 = 0$ and $v_2 = C$.)

To this end, we set

$$\delta = \min_{x \neq d} \min_{x' \in \mathcal{O}(x)} \frac{c(x, x')}{C(x)}.$$

By the stated cost assumptions, we have $c(x, x') > 0$ when $x \neq d$ and $x' \in \mathcal{O}(x)$, while $C(x) > 0$ when $x \neq d$. Since $\mathsf{X}$ is finite, it follows that $\delta$ is finite and positive. Evidently, with this definition, the bound (8.56) holds for all $x \neq d$. In addition, (8.56) holds trivially when $x = d$, since $C(d) = 0$. Hence (8.56) is valid for all $x \in \mathsf{X}$.

Concavity of $\mathcal{R}$ implies global stability by Proposition 8.2.5. The remaining claims now follow from Theorem 8.3.7. □

### 8.3.5.2 Application: Negative Discount Rate Optimality

When discussing MDPs we used $\beta$ to represent the discount factor. Given $\beta$, the **discount rate** or **rate of time preference** is the value $\rho$ that solves $\beta = 1/(1 + \rho)$. The standard MDP assumption $\beta < 1$ implies this rate is positive. You will recall from Chapter 5 that the condition $\beta < 1$ is central to the general theory of MDPs, since it yields global stability of the Bellman and policy operators on $\mathbb{R}^{\mathsf{X}}$ (via the Neumann series lemma or Banach's fixed point theorem).

In the previous section, on shortest paths, we studied an RDP with a zero discount rate. Now we go one step further and consider problems with negative rates of time preference. Such preference are commonly inferred when people face unpleasant tasks. Subjects of studies often prefer getting such tasks "over and done with" rather than postponing them. (Negative discount rates are inferred in other settings as well. §9.4 provides background and references.)

In this section, we model optimal choice under a negative discount rate. Taking our cue from the discussion above, we consider a scenario where a task generates disutility but has to be completed. In particular, we assume that

$$B(x, x', v) = c(x, x') + \beta v(x') \qquad (x, x' \in \mathsf{X}) \tag{8.57}$$

where $\mathsf{X}$ is a finite set and $\beta > 1$ is some positive constant. The function $c$ gives the cost of transitioning from $x$ to the new state $x'$

The value aggregator $B$ in (8.57) is the same as the shortest path aggregator (8.55), except for the constant $\beta$. To keep the discussion simple, we adopt all other assumptions from the shortest path discussion in §8.3.5.1.

EXERCISE 8.3.10. Let $C(x)$ be the maximum cost of traveling from $x \in \mathsf{X}$ to the destination node $d$ under any feasible policy. Prove that $C(x) < \infty$ for all $x$.

We now have an $\mathcal{R} = (\Gamma, B, V)$ with $\Gamma = \mathcal{O}$, $B$ as in (8.57) and $V = [0, C]$. The policy operators map $V$ into itself because, for $v \in V$, we clearly have $0 \leqslant T_\sigma v$ and, in addition,

$$(T_\sigma v)(x) = c(x, \sigma(x)) + \beta v(\sigma(x)) \leqslant c(x, \sigma(x)) + \beta C(\sigma(x)) \leqslant C(x).$$

The last bound holds because $C(x)$ is, by definition, greater than the cost of traveling from $x$ to $\sigma(x)$ and then following the most expensive path.

**Proposition 8.3.9.** *Under the stated conditions, the negative discount rate RDP is glob-*

*ally stable, the min-value function $v_\downarrow^*$ is the unique solution to*

$$v_\downarrow^*(x) = \min_{x' \in \Gamma(x)} \left\{ c(x, x') + \beta v_\downarrow^*(x') \right\} \qquad (x \in \mathsf{X})$$

*in $V$ and a policy $\sigma \in \Sigma$ is min-optimal if and only if*

$$\sigma(x) \in \operatorname*{argmin}_{x' \in \Gamma(x)} \left\{ c(x, x') + \beta v_\downarrow^*(x') \right\} \quad \text{for all } x \in \mathsf{X}.$$

*Proof.* The proof of Proposition 8.3.9 is essentially identical to the proof of Proposition 8.3.8. Readers are invited to confirm this. □

## 8.4 Chapter Notes

The RDP framework adopted in this chapter is inspired by Bertsekas (2022b), who in turn credits Mitten (1964) as the first research paper to frame Richard Bellman's dynamic programming problems in an abstract setting. Denardo (1967) describes key ideas including what we call contracting RDPs (see §8.2.1). Denardo credits Shapley (1953) for inspiring his contraction-based arguments.

The key optimality results from this chapter (Theoremm 9.1.6 and 8.1.1) are somewhat new, although closely related results appear in Bertsekas (2022b). See, in addition, Bloise et al. (2023), which builds on Bertsekas (2022b) and Ren and Stachurski (2021).

The job search application with quantile preferences in §8.2.1.4 is based on de Castro et al. (2022). The same reference includes a general theory of dynamic programming when certainty equivalents are computed using quantile operators and aggregation is time additive.

The optimal default application in §8.2.1.5 is loosely based on Arellano (2008). Influential contributions to this line of work include, Yue (2010), Chatterjee and Eyigungor (2012), Arellano and Ramanarayanan (2012), Cruces and Trebesch (2013), Ghosh et al. (2013), Gennaioli et al. (2014), and Bocola et al. (2019).

At the start of the chapter we motivated RDPs by mentioning that equilibria in some models of production and economic geography can be computed using dynamic programming. Examples include Hsu (2012), Hsu et al. (2014), Antràs and De Gortari (2020), Kikuchi et al. (2021) and Tyazhelnikov (2022).

Early references for dynamic programming with risk-sensitive preferences include Jacobson (1973), Whittle (1981), and Hansen and Sargent (1995). Elegant modern treatments can be found in Asienkiewicz and Jaśkiewicz (2017) and Bäuerle and Jaśkiewicz (2023), and an extension to general static risk measures is available in Bäuerle and Glauner (2022). Risk-sensitivity is applied to the study of optimal growth in Bäuerle and Jaśkiewicz (2018), and to optimal divided payouts in Bäuerle and Jaśkiewicz (2017). Risk-sensitivity is also used in applications of reinforcement learning, where the underlying state process is not known. See, for example, Shen et al. (2014), Majumdar et al. (2017) or Gao et al. (2021).

Dynamic programming problems that acknowledge model uncertainty by including adversarial agents to promote robust decision rules can be found in Cagetti et al. (2002), Hansen and Sargent (2011), and other related papers. Al-Najjar and Shmaya (2019) study the connection between Epstein–Zin utility and parameter uncertainty. Ruszczyński (2010) considers risk averse dynamic programming and time consistency.

The smooth ambiguity model in §8.3.4 is loosely adapted from Klibanoff et al. (2009) and Ju and Miao (2012). For applications of optimization under smooth ambiguity, see, for example, Guan and Wang (2020) or Yu et al. (2023). Zhao (2020) studies yield curves in a setting where ambiguity-averse agents face varying amounts of Knightian uncertainty over the short and long run.

An algorithm that we neglected to discuss is stochastic gradient descent (or ascent) in policy space. Typically policies are parameterized via an approximation architecture that consists of basis functions, activation functions, and compositions of them (e.g., a neural network). In large models, such approximation is used even when the state and action spaces are finite, simply because the curse of dimensionality makes exact representations infeasible. For recent discussions of gradient descent in policy spaces see Nota and Thomas (2019), Mei et al. (2020), and Bhandari and Russo (2022).

# Chapter 9

# Abstract Dynamic Programming

In Chapter 8 we introduced RDPs, stated their optimality properties and investigated applications that satisfy optimality conditions. But we have yet to prove the core optimality and convergence results in Theorems 8.1.1.

Rather than proving these result directly, we now present a completely abstract version of a dynamic programming problem that consist of a family of self-maps on a partially ordered set. Doing so allows us to simplify proofs and extend the reach of dynamic programming theory.

## 9.1 Abstract Dynamic Programs

In this section we define abstract dynamic programs and prove optimality results under a set of high level assumptions. Then we connect these results to our Chapter 8 optimality claims for RDPs.

### 9.1.1 Preliminaries

Before defining abstract dynamic programs, let's cover some fundamental concepts that we'll use. We begin with suprema and infima and then discuss notions of stability in partially ordered sets.

#### 9.1.1.1 Order Stability

We first discuss stability of maps over partially ordered spaces. Our aim is to provide a weak notion of operator stability that can be applied in any partially ordered set

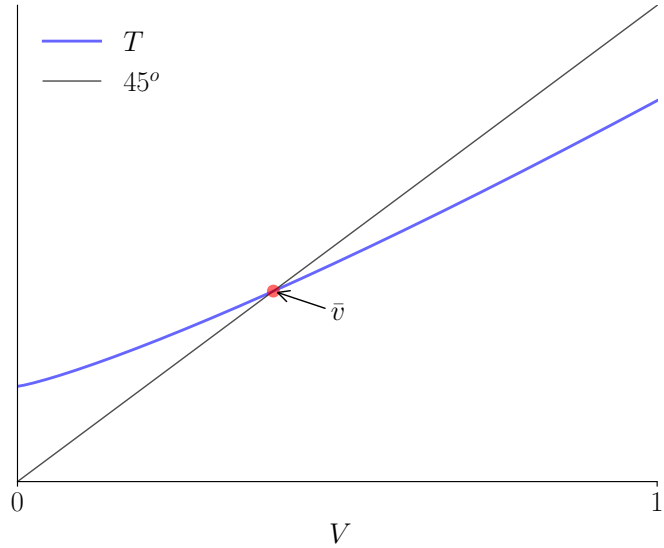

Figure 9.1: Order-stability

(without any concept of convergence).

Let $V$ be a partially ordered set and let $T$ be a self-map on $V$ with exactly one fixed point $\bar{v}$ in $V$. In this setting, we call $T$

- **upward stable** on $V$ if $v \in V$ and $v \preceq T v$ implies $v \preceq \bar{v}$,

- **downward stable** on $V$ if $v \in V$ and $T v \preceq v$ implies $\bar{v} \preceq v$, and

- **order stable** on $V$ if $T$ is both upward and downward stable.

Figure 9.1 gives an illustration of a map $T$ that on $V = [0, 1]$ that is order stable, since all points mapped up by $T$ lie below its fixed point, while all points mapped down by $T$ lie above its fixed point. The figure suggests that order stability is related to global stability, as defined in §1.2.2.2. We return to this relationship in Lemma 9.1.1 below.

**EXERCISE 9.1.1.** Let $\mathsf{X}$ be finite and consider the self-map on $V := (\mathbb{R}^{\mathsf{X}}, \leqslant)$ defined by $Tv = r + Av$ for some $r \in \mathbb{R}^{\mathsf{X}}$ and $A \in \mathcal{L}(\mathbb{R}^{\mathsf{X}})$ with $0 \leqslant A$ and $\rho(A) < 1$. Prove that $T$ is order stable on $V$.

**Lemma 9.1.1.** *Let $\mathsf{X}$ be finite, let $V$ be a subset of $\mathbb{R}^{\mathsf{X}}$, and let $T$ be an order-preserving self-map on $V$. If $T$ is globally stable on $V$, then $T$ is order stable on $V$.*

*Proof.* Assume the stated conditions. By global stability, $T$ has a unique fixed point $\bar{v}$ in $V$. If $v \in V$ and $v \leqslant T v$, then iterating on this inequality and using the fact that $T$ is order-preserving yields $v \leqslant T^k v$ for all $k \in \mathbb{N}$. Applying global stability and taking the limit gives $v \leqslant \bar{v}$. Hence upward stability holds. The proof of downward stability is similar. □

#### 9.1.1.2 Order Duals

Given partially ordered set $V$, let $V^\partial = (V, \preceq^\partial)$ be the **order dual**, so that, for $u, v \in V$, we have $u \preceq^\partial v$ if and only if $v \preceq u$. The following result will be useful.

**Lemma 9.1.2.** *$S$ is order stable on $V$ if and only if $S$ is order stable on $V^\partial$.*

*Proof.* Let $S$ be as stated. By definition, $S$ has a unique fixed point $\bar{v} \in V$. Hence it remains only to check that $S$ is upward and downward stable on $V^\partial$. Regarding upward stability, suppose $v \in V$ and $v \preceq^\partial Sv$. Then $Sv \preceq v$ and hence $\bar{v} \preceq v$, by downward stability of $S$ on $V$. But then $v \preceq^\partial \bar{v}$, so $S$ is upward stable on $V^\partial$. The proof of downward stability is similar.

We have shown that $S$ is order stable on $V^\partial$ whenever $S$ is order stable on $V$. The reverse implication holds because the dual of $V^\partial$ is $V$. □

EXERCISE 9.1.2. Let $V$ and $\hat{V}$ be partially ordered and let $F$ be an order isomorphism from $V$ to $\hat{V}$. Prove that $F$ is an order anti-isomorphism from $V$ to $\hat{V}^\partial$.

### 9.1.2 Abstract Dynamic Programs

In this section we formalize abstract dynamic programs and present fundamental optimality results. §9.1.2.1 starts the ball rolling with an informal overview.

#### 9.1.2.1 Prelude

We saw in §8.1 that a globally stable RDP yields a set of feasible policies $\Sigma$ and, for each $\sigma \in \Sigma$, a policy operator $T_\sigma$ defined on the value space $V \subset \hat{\mathbb{R}}^X$. Notice that the dynamic programming problem is fully specified by the family of operators $\{T_\sigma\}_{\sigma \in \Sigma}$ and the space $V$ that they act on. From this set of operators we obtain the set of lifetime values $\{v_\sigma\}_{\sigma \in \Sigma}$, with each $v_\sigma$ uniquely identified as a fixed point of $T_\sigma$. These lifetime values define the value function $v^*$ as the pointwise maximum $v^* = \vee_\sigma v_\sigma$. An optimal policy is then defined as a $\sigma \in \Sigma$ obeying $v_\sigma = v^*$.

To shed unnecessary structure before the main optimality proofs, a natural idea is to start directly with an abstract set of "policy operators" $\{T_\sigma\}$ acting on some set $V$. One can then define lifetime values and optimality as in the previous paragraph and start to investigate conditions on the family of operators $\{T_\sigma\}$ that lead to characterizations of optimality that we hope to obtain.

We use these ideas as our starting point, beginning with an arbitrary family $\{T_\sigma\}$ of operators on a partially ordered set.

### 9.1.2.2  Defining ADPs

An **abstract dynamic program** (**ADP**) is a pair $\mathcal{A} = (V, \{T_\sigma\}_{\sigma \in \Sigma})$ such that

(i) $V = (V, \preceq)$ is a partially ordered set,

(ii) $\{T_\sigma\} := \{T_\sigma\}_{\sigma \in \Sigma}$ is a family of self-maps on $V$, and

(iii) for all $v \in V$, the set $\{T_\sigma v\}_{\sigma \in \Sigma}$ has both a least and greatest element.

Elements of the index set $\Sigma$ are called **policies** and elements of $\{T_\sigma\}$ are called **policy operators**. Given $v \in V$, a policy $\sigma$ in $\Sigma$ is called $v$-**greedy** if $T_\sigma v \succeq T_\tau v$ for all $\tau \in \Sigma$. Existence of a greatest element in (iii) of the definition above is equivalent to the statement that each $v \in V$ has at least one $v$-greedy policy.

Existence of a least element in (iii) is needed only because we wish to consider minimization as well as maximization. For settings where only maximization is considered, this can be dropped from the list of assumptions. (An analogous statement holds for minimization and greatest elements.) We mention least elements in Example 9.1.1 below and then disregard them until we treat minimization in §9.1.5.

**Example 9.1.1** (RDPs generate ADPs). Let $\mathcal{R} = (\Gamma, V, B)$ be an RDP with finite state space $\mathsf{X}$, as defined in §8.1.1, and, for each $\sigma$ in the feasible policy set $\Sigma$, let $T_\sigma$ be the corresponding policy operator, defined at $v \in V$ by $(T_\sigma v)(x) = B(x, \sigma(x), v)$. The pair $\mathcal{A}_{\mathcal{R}} := (V, \{T_\sigma\})$ is an ADP, since $V$ is partially ordered by $\leqslant$, $T_\sigma$ is a self-map on $V$ for all $\sigma \in \Sigma$, and, given $v \in V$, choosing $\bar{\sigma} \in \Sigma$ such that $\bar{\sigma}(x) \in \operatorname{argmax}_{a \in \Gamma(x)} B(x, a, v)$ for all $x \in \mathsf{X}$ produces a $v$-greedy policy and a greatest element for $\{T_\sigma v\}$ (cf., Exercise 8.1.7 on page 254). A least element of $\{T_\sigma v\}$ can be generated by replacing "argmax" with "argmin."

In the setting of Example 9.1.1, we call $\mathcal{A}_{\mathcal{R}}$ the ADP **generated by** $\mathcal{R}$.

**Example 9.1.2.** Let $\mathcal{M} = (\Gamma, \beta, r, P)$ be an MDP, as defined in §5.1.1, with policy operators $\{T_\sigma\}$ defined by $T_\sigma v = r_\sigma + \beta P_\sigma v$ (as in (5.19)). $\mathcal{A}_{\mathcal{M}} := (\mathbb{R}^{\mathsf{X}}, \{T_\sigma\})$ is an ADP (as a special case of Example 9.1.1). We call $\mathcal{A}_{\mathcal{M}}$ the ADP **generated by** $\mathcal{M}$.

We have just shown that RDPs are ADPs. But there are also ADPs that do not fit naturally into the RDP framework (although they might be derived from RDPs, as will be discussed in §9.3). The next two examples illustrate. In these examples, the Bellman equation does not match the RDP Bellman equation $v(x) = \max_{a \in \Gamma(x)} B(x, a, v)$ due to the inverted order of expectation and maximization.

**Example 9.1.3.** Recall the $Q$-factor MDP Bellman operator, which takes the form

$$(Sq)(x, a) = r(x, a) + \beta \sum_{x'} \max_{a' \in \Gamma(x')} q(x', a') P(x, a, x') \tag{9.1}$$

with $q \in \mathbb{R}^\mathsf{G}$ and $(x, a) \in \mathsf{G}$ (We are repeating (5.39) on page 172.) The $Q$-factor policy operators $\{S_\sigma\}$ corresponding to (9.1) are given by

$$(S_\sigma q)(x, a) = r(x, a) + \beta \sum_{x'} q(x', \sigma(x')) P(x, a, x') \qquad ((x, a) \in \mathsf{G}). \tag{9.2}$$

Each $S_\sigma$ is a self-map on $\mathbb{R}^\mathsf{G} = (\mathbb{R}^\mathsf{G}, \leqslant)$. If $q \in \mathbb{R}^\mathsf{G}$ and $\sigma \in \Sigma$ is such that $\sigma(x) \in \operatorname{argmax}_{a \in \Gamma(x)} q(x, a)$ for all $x \in \mathsf{X}$, then $S_\sigma q \geqslant S_\tau q$ on $\mathsf{G}$ for all $\tau \in \Sigma$. Hence $\sigma$ is $q$-greedy and $\mathcal{A} := (\mathbb{R}^\mathsf{G}, \{S_\sigma\})$ is an ADP.

**Example 9.1.4.** In reinforcement learning and related fields the $Q$-factor approach from Example 9.1.3 has been extended to risk-sensitive decision processes (see, e.g., Fei et al. (2021)). The corresponding $Q$-factor Bellman equation is given by

$$f(x, a) = r(x, a) + \frac{\beta}{\theta} \ln \left\{ \sum_{x'} \exp \left[ \theta \max_{a' \in \Gamma(x')} f(x', a') \right] P(x, a, x') \right\} \qquad ((x, a) \in \mathsf{G}). \tag{9.3}$$

The policy operators over risk-sensitive $Q$-factors take the form

$$(Q_\sigma f)(x, a) = r(x, a) + \frac{\beta}{\theta} \ln \left[ \sum_{x'} \exp \left[ \theta f(x', \sigma(x')) \right] P(x, a, x') \right] \tag{9.4}$$

where $f \in \mathbb{R}^\mathsf{G}$ and $\sigma \in \Sigma$. An argument similar to the one given in Example 9.1.3 confirms that each $f \in \mathbb{R}^\mathsf{G}$ has an $f$-greedy policy. Hence $(\mathbb{R}^\mathsf{G}, \{Q_\sigma\})$ is an ADP.

In Chapter 10 we will see that continuous time dynamic programs can also be viewed as ADPs.

### 9.1.3 Optimality

In this section we study optimality properties of ADPs, aiming for generalizations of the foundational results of dynamic programming. To achieve this aim we need to define optimality and provide sufficient conditions.

#### 9.1.3.1 Lifetime Values

The objective of dynamic programming is to optimize lifetime value. But what is lifetime value in this abstract context? In general, for an ADP $(V, \{T_\sigma\})$ and fixed $\sigma \in \Sigma$, we write $v_\sigma$ for the fixed point of $T_\sigma$ and call it the $\sigma$-value function. We interpret it as the lifetime value of policy $\sigma$ whenever it is uniquely defined. This interpretation was discussed at length for RDPs in §8.1.2.1 and the situation here is analogous.

**Example 9.1.5.** Let $\mathcal{M}$ be an MDP. If $\mathcal{A}_{\mathcal{M}}$ is the ADP generated by $\mathcal{M}$, as in Example 9.1.2, then the unique fixed point of $T_\sigma$ is $v_\sigma = (I - \beta P_\sigma)^{-1} r_\sigma$. This accords with our interpretation of fixed points of $T_\sigma$ as lifetime values, since $(I - \beta P_\sigma)^{-1} r_\sigma$ is precisely the lifetime value of $\sigma$ under the MDP assumptions (see §5.1.3.1).

**Example 9.1.6.** Let $\mathcal{A} = (V, \{T_\sigma\})$ when each $T_\sigma$ is a Koopmans operator on $V$, as defined in §7.3.1. A fixed point of a Koopmans operator is interpreted as lifetime utility under the preferences it represents (see §7.3.1). Thus $v_\sigma$, when well-defined, is the lifetime value associated with policy $\sigma$ and the preferences embedded in $T_\sigma$.

We call a given ADP $\mathcal{A} := (V, \{T_\sigma\})$ well-posed if every policy operator $T_\sigma$ has a unique fixed point in $V$. Obviously well-posedness is a minimum requirement for constructing an optimality theory around ADPs.

#### 9.1.3.2 Operators

Let $\mathcal{A} = (V, \{T_\sigma\})$ be an ADP. We set

$$T v := \bigvee_\sigma T_\sigma v \qquad (v \in V) \tag{9.5}$$

and call $T$ the Bellman operator generated by $\mathcal{A}$. Note that $T$ is a well-defined self-map on $V$ by part (iii) of the definition of ADPs (existence of greedy policies). A function $v \in V$ is said to satisfy the Bellman equation if it is a fixed point of $T$.

The definition of $T$ in (9.5) includes all of the Bellman operators we have met as special cases. For example, consider an RDP $\mathcal{R} = (\Gamma, V, B)$ with Bellman operator $(Tv)(x) = \max_{a \in \Gamma(x)} B(x, a, v)$. We can write $T$ as $\bigvee_\sigma T_\sigma v$, as shown in Exercise 8.1.8 on page 255. Thus, the Bellman operator of the RDP agrees with the Bellman operator $T$ of the corresponding ADP $\mathcal{A}_\mathcal{R}$.

EXERCISE 9.1.3. Show that

(i) $\sigma \in \Sigma$ is $v$-greedy if and only if $T_\sigma v = T v$, and

(ii) $T$ in (9.5) is order-preserving whenever $T_\sigma$ is order-preserving for all $\sigma \in \Sigma$.

Below we consider Howard policy iteration (HPI) as an algorithm for solving for optimal policies of ADPs. We use precisely the same instruction set as for the RDP case, as shown in Algorithm 8.1 on page 256. To further clarify the algorithm, we define a map $H$ from $V$ to $\{v_\sigma\}$ via $H v = v_\sigma$ where $\sigma$ is $v$-max-greedy. Iterating with $H$ generates the value sequence associated with Howard policy iteration.[1] In what follows, we call $H$ the **Howard operator** generated by the ADP.

### 9.1.3.3 Properties

Let $\mathcal{A} := (V, \{T_\sigma\}_{\sigma \in \Sigma})$ be an ADP. We call $\mathcal{A}$

- **finite** if $\Sigma$ is a finite set,

- **order stable** if every policy operator $T_\sigma$ is order stable on $V$,

- **max-stable** if $\mathcal{A}$ is order stable and $T$ has at least one fixed point in $V$.

Obviously max-stable $\implies$ order stable $\implies$ well-posed.

Regarding the definition of max-stability, existence of a fixed point of $T$ in $V$ is a high level assumption that can be challenging to verify in applications. At the same time, our main concern in the present volume is the case where $\mathcal{A}$ is finite, and, in this setting, order stability is enough:

**Proposition 9.1.3.** *If $\mathcal{A}$ is order stable and finite, then $\mathcal{A}$ is max-stable.*

Proposition 9.1.3 is proved in §B.4.

**Corollary 9.1.4.** *Let $\mathcal{R}$ be an RDP and let $\mathcal{A}_\mathcal{R}$ be the ADP generated by $\mathcal{R}$. If $\mathcal{R}$ is globally stable, then $\mathcal{A}_\mathcal{R}$ is max-stable.*

---

[1] For $H$ to be well-defined, we must always select the same $v$-greedy policy when the operator is applied to $v$. We can use the axiom of choice to assign to each $v$ a designated $v$-greedy policy, although, in applications, a simple rule usually suffices. For example, if $\Sigma$ is finite, we can enumerate the policy set $\Sigma$ and choose the first $v$-greedy policy.

*Proof.* Let $\mathcal{R}$ and $\mathcal{A}_{\mathcal{R}}$ be as stated, and suppose that $\mathcal{R}$ is globally stable. In view of Lemma 9.1.1 on page 293, each policy operator is order stable. Hence $\mathcal{A}_{\mathcal{R}}$ is order stable. Since $\Sigma$ is finite, Proposition 9.1.3 implies that $\mathcal{A}_{\mathcal{R}}$ is also max-stable. $\square$

EXERCISE 9.1.4. Show that the ADP described in Example 9.1.3 is max-stable.

Order stability is central to the optimality results stated below. While order stability is a somewhat nonstandard condition, the next result shows that, at least in simple settings, order stability is necessary for any discussion of optimality.

**Proposition 9.1.5.** *Let $\mathcal{A} = (V, \{T_\sigma\})$ be an ADP generated by an RDP $\mathcal{R} = (\Gamma, V, B)$. If $V$ is an order interval in $\mathbb{R}^{\mathsf{X}}$, then the following statements are equivalent:*

(i) *$\mathcal{A}$ is well-posed.*

(ii) *$\mathcal{A}$ is order stable.*

*Proof.* Let $\mathcal{A}$ be as stated, with $V = [v_1, v_2]$ for some $v_1, v_2$ in $\mathbb{R}^{\mathsf{X}}$ with $v_1 \leqslant v_2$. Obviously (ii) $\implies$ (i). Regarding (i) $\implies$ (ii), let $\mathcal{A}$ be well-posed and pick any policy operator $T_\sigma$. Since $\mathcal{A}$ is well-posed, $T_\sigma$ has a unique fixed point $v_\sigma$ in $V$. Suppose $v \in V$ with $T_\sigma v \leqslant v$. Since, $T_\sigma$ is order-preserving, $T_\sigma$ is a self-map on $[v_1, v]$. By the Knaster–Tarski theorem (p. 214), $T_\sigma$ has at least one fixed point in $[v_1, v]$. By uniqueness, that fixed point is $v_\sigma$. Hence $v_\sigma \leqslant v$ and downward stability holds. Upward stability can be confirmed via a similar argument. Hence $\mathcal{A}$ is order stable. $\square$

### 9.1.3.4   Max-Optimality Results

Let $\mathcal{A} = (V, \{T_\sigma\})$ be a well-posed ADP with $\sigma$-value functions $\{v_\sigma\}_{\sigma \in \Sigma}$. We define

$$V_\Sigma := \{v_\sigma\}_{\sigma \in \Sigma} \quad \text{and} \quad V_u := \{v \in V : v \leq Tv\}.$$

EXERCISE 9.1.5. Prove that $V_\Sigma \subset V_u$.

If $V_\Sigma$ has a greatest element, then we denote it by $v^*$ and call it the **value function** generated by $\mathcal{A}$. In this setting, a policy $\sigma \in \Sigma$ is called **optimal** for $\mathcal{A}$ if $v_\sigma = v^*$. We say that $\mathcal{A}$ obeys **Bellman's principle of optimality** if

$$\sigma \in \Sigma \text{ is optimal for } \mathcal{A} \quad \Longleftrightarrow \quad \sigma \text{ is } v^*\text{-greedy.}$$

These definitions are direct generalizations of of the corresponding definitions for RDPs discussed in Chapter 8.

We can now state our main optimality result for ADPs.

**Theorem 9.1.6** (Max-optimality). *If $\mathscr{A}$ is finite and order stable, then*

(i) *the set of $\sigma$-value functions $V_\Sigma$ has a greatest element $v^*$,*

(ii) *$v^*$ is the unique solution to the Bellman equation in $V$,*

(iii) *$\mathscr{A}$ obeys Bellman's principle of optimality,*

(iv) *$\mathscr{A}$ has at least one optimal policy, and*

(v) *HPI returns an exact optimal policy in finitely many steps.*

Theorem 9.1.6 informs us that finite well-posed ADPs have first-rate optimality properties under a relatively mild stability condition. In §9.1.4 we use Theorem 9.1.6 to prove all optimality results for RDPs stated in Chapter 8. The proof of Theorem 9.1.6 is given in §B.4 (see page 365). Note that (iv) follows directly from (i) and is included only for completeness.

### 9.1.3.5 General States

This volume focuses on dynamic programming problems with finite states. Here we restrict ourselves to one high-level result for general state spaces.

**Proposition 9.1.7.** *If $\mathscr{A}$ is max-stable, then* (i)–(iv) *of Theorem 9.1.6 hold.*

Proposition 9.1.7 tells us that we can drop finiteness of policy set $\Sigma$ (which is implied by finite states and actions) whenever the Bellman operator has at least one fixed point. Various fixed point methods are available for establishing this existence. We defer further details until Volume II. Proposition 9.1.7 is proved in in §B.4.

### 9.1.3.6 Application: Mixed Strategies

This section discusses adding mixed strategies to an RDP. We will need to apply Proposition 9.1.7 to discuss optimality because the set of mixed strategies is not finite.

Let $\mathscr{R} = (\Gamma, V, B)$ be an RDP with finite state space $\mathsf{X}$, finite action space $\mathsf{A}$, policy set $\Sigma$ and Bellman operator $T$ (see 8.1.3). A **mixed strategy** for $\mathscr{R}$ is a map $\varphi$ sending $x \in \mathsf{X}$ into a distribution $\varphi_x \in \mathcal{D}(\mathsf{A})$ supported on $\Gamma(x)$. In other words, for each $x \in \mathsf{X}$,

$$\varphi_x \in \mathcal{D}(\mathsf{A}) \quad \text{and} \quad \sum_{a \in \Gamma(x)} \varphi_x(a) = 1.$$

Let $\Phi$ be the set of all mixed strategies for $\mathcal{R}$. For each mixed strategy $\varphi \in \Phi$, we introduce the policy operator on $V$ defined by

$$(\hat{T}_\varphi v)(x) = \sum_{a \in \mathsf{A}} B(x, a, v)\varphi_x(a) \qquad (v \in V, \ x \in \mathsf{X}).$$

Intuitively, the right hand side is the expected lifetime value from current state $x$, when the current action is drawn from $\varphi_x$ and future states are evaluated via $v$.

EXERCISE 9.1.6. Fix $v \in V$. Prove: If $\varphi \in \Phi$ and, for each $x \in \mathsf{X}$, the distribution $\varphi_x$ is supported on $\mathrm{argmax}_{a \in \Gamma(x)} B(x, a, v)$, then $\hat{T}_\varphi v \geqslant \hat{T}_\psi v$ for all $\psi \in \Phi$.

EXERCISE 9.1.7. Show that, given $v \in V$ and $x \in \mathsf{X}$ we have

$$\max_{\varphi \in \Phi} (\hat{T}_\varphi v)(x) = \max_{a \in \Gamma(x)} B(x, a, v).$$

It follows from the discussion above that $\mathcal{A}_M := (V, \{\hat{T}_\varphi\}_{\varphi \in \Phi})$ is an ADP (where "M" stands for "mixed"), and that the Bellman operator $\hat{T}$ associated with the ADP $\mathcal{A}_M$ is given by

$$(\hat{T}v)(x) = \max_{a \in \Gamma(x)} B(x, a, v) = (Tv)(x) \qquad (v \in V, \ x \in \mathsf{X}). \tag{9.6}$$

Let us assume for simplicity that $\mathcal{R}$ is contracting (see §8.2.1), with modulus of contraction $\beta \in (0, 1)$. Assume also that $V$ is closed in $\mathbb{R}^{\mathsf{X}}$. As a result, the value function $v^*$ for $\mathcal{R}$ exists in $V$ and is the unique fixed point of $T$ in $V$ (Corollary 8.2.2).

EXERCISE 9.1.8. Show that, under the assumptions stated above, $\{\hat{T}_\varphi\}_{\varphi \in \Phi}$ and $\hat{T}$ are all contraction mappings.

By Exercise 9.1.8, the ADP $\mathcal{A}_M$ is max-stable (since globally stable operators are order stable – see Lemma 9.1.1 on page 293 – and the Bellman operator $\hat{T}$ has a fixed point). Hence, by Proposition 9.1.7, the value function $\hat{v}^*$ for $\mathcal{A}_M$ exists in $V$ and is the unique fixed point of $\hat{T}$ in $V$. But, by (9.6), $\hat{T}$ and $T$ agree on $V$. Hence $\hat{v}^* = v^*$. Thus, for RDPs, we conclude as follows: although the set of mixed strategies is larger than the set of pure strategies (i.e., deterministic policies), the maximal lifetime value from each state is the same.

## 9.1.4 Optimality Results for RDPs

In this section we return to the optimality properties of RDPs, as first discussed in §8.1.3.3. Our aim is to connect the ADP optimality results from §9.1.3.4 to the special case of RDPs and, through this process, complete the proofs of our key RDP optimality results from Chapter 8.

### 9.1.4.1 OPI Convergence

In this section we provide some preliminary results related to OPI convergence, where OPI obeys the algorithm given on page 256. Throughout, $\mathcal{R} = (\Gamma, V, B)$ is a globally stable RDP with policy set $\Sigma$, policy operators $\{T_\sigma\}$, Bellman operator $T$, and value function $v^*$. As usual, $v_\sigma$ denotes the unique fixed point of $T_\sigma$ for all $\sigma \in \Sigma$. In the results stated below, $m$ is a fixed natural number indicating the OPI step size and $H$ and $W_m$ are as defined in §8.1.3.2.

**Lemma 9.1.8.** *If $v \in V_\Sigma$, then $T^k v \to v^*$ as $k \to \infty$.*

*Proof.* Fix $v \in V_\Sigma$. On one hand, $v \leqslant v^*$ and hence $T^k v \leqslant T^k v^* = v^*$ for all $k$. On the other hand, if $\sigma$ is any policy, then $T_\sigma^k v \leqslant T^k v$ for all $k$. Hence $T_\sigma^k v \leqslant T^k v \leqslant v^*$ for all $k$. If we now take $\sigma$ to be an optimal policy, which exists under the stated assumptions, we have $T_\sigma^k v \to v_\sigma = v^*$ as $k \to \infty$. Hence $T^k v \to v^*$, as required. □

**Lemma 9.1.9.** *$W_m$ is an order-preserving self-map on $V_u$. Moreover,*

$$v \in V_u \implies Tv \leqslant W_m v \leqslant T^m v.$$

*Proof.* As for the self-map property, pick any $v \in V_u$. Since $T$ and $T_\sigma$ are order-preserving, $v \leqslant Tv$ and $\sigma$ is $v$-greedy, we have

$$W_m v = T_\sigma T_\sigma^{m-1} v \leqslant T T_\sigma^{m-1} v \leqslant T T_\sigma^{m-1} T v = T T_\sigma^m v = T W_m v.$$

Hence $W_m v \in V_u$ and $W_m$ is invariant on $V_u$.

As for the order-preserving property, this is immediate from the definition, since powers of order-preserving self-maps are order-preserving (Exercise 2.2.26).

As for the inequality $Tv \leqslant W_m v$, fix $v \in V_u$. Since $T_\sigma$ is order-preserving, $v \leqslant Tv$ and $\sigma$ is $v$-greedy, we have

$$T_\sigma^{m-1} v \leqslant T_\sigma^{m-1} T v = T_\sigma^{m-1} T_\sigma v = W_m v.$$

Continuing in the same manner gives $T_\sigma^{m-j} v \leqslant W_m v$ for $j < m$ and, in particular, $T_\sigma v \leqslant W_m v$. Because $\sigma$ is $v$-greedy, this yields $Tv \leqslant W_m v$.

Regarding the second inequality, we use the fact that $T_\sigma \leqslant T$ on $V$ and $T$ and $T_\sigma$ are both order-preserving to obtain $W_m v = T_\sigma^m v \leqslant T^m v$ (see Exercise 2.2.39 on page 66). □

**Lemma 9.1.10.** *For each $v_0 \in V_\Sigma$ we have*

$$T^k v_0 \leqslant W_m^k v_0 \leqslant T^{km} v_0 \quad \text{for all } k \in \mathbb{N}. \tag{9.7}$$

*Proof.* The bounds in (9.7) hold for $k = 1$ by $v_0 \in V_u$ and Lemma 9.1.9. Since all operators are order-preserving and invariant on $V_u$, the extension to arbitrary $k$ follows from Exercise 2.2.39 on page 66. □

**Lemma 9.1.11.** *Let $v_0$ be any element of $V_\Sigma$ and let $v_k = W_m^k v_0$ for all $k \in \mathbb{N}$. If $v_k = v_{k+1}$ for some $k \in \mathbb{N}$, then $v_k = v^*$ and every $v_k$-greedy policy is optimal.*

*Proof.* Let the sequence $(v_k)$ be as stated and suppose that $v_k = v_{k+1}$. Let $\sigma$ be $v_k$-greedy. It follows that $T_\sigma^m v_k = v_k$ and, moreover, $v_k \leqslant T v_k = T_\sigma v_k$, where the last inequality is by $v_k \in V_u$. As a result,

$$v_k \leqslant T_\sigma v_k \leqslant T_\sigma^m v_k = v_k.$$

In particular, $T v_k = T_\sigma v_k = v_k$, which in turn gives $v_k = v^*$. Bellman's principle of optimality now implies that every $v_k$-greedy policy is optimal. □

**Lemma 9.1.12.** *If $(v_k) \subset V_u$ and $v_k \to v^*$ as $k \to \infty$, then there exists a $K \in \mathbb{N}$ such that*

$$k \geqslant K \implies \text{every } v_k\text{-greedy policy is optimal.}$$

*Proof.* Let $\mathcal{R}$ be as stated and fix $(v_k) \subset V_u$ with $v_k \to v^*$ as $k \to \infty$. Let $\Sigma^*$ be the set of optimal policies and let $\Sigma' := \Sigma \setminus \Sigma^*$. Since $\Sigma'$ is finite, we have

$$e := \min_{\sigma \in \Sigma'} \|v_\sigma - v^*\|_\infty > 0.$$

Choose $K \in \mathbb{N}$ such that $\|v_k - v^*\|_\infty < e$ for all $k \geqslant K$. Fix $k \geqslant K$ and let $\sigma$ be $v_k$-greedy. We claim that $\sigma$ is optimal. Indeed, since $v_k \subset V_u$, we have $v_k \leqslant T v_k = T_\sigma v_k$, so, by upward stability, $v_k \leqslant v_\sigma$. As a result,

$$|v^* - v_\sigma| = v^* - v_\sigma \leqslant v^* - v_k.$$

Hence $\|v^* - v_\sigma\|_\infty \leqslant \|v^* - v_k\|_\infty < e$. But then $\sigma \notin \Sigma'$, so $\sigma$ is optimal. □

### 9.1.4.2  Proofs of RDP Results

In §8.1.3 we stated two key optimality results for RDPs, the first concerning globally stable RDPs (Theorem 8.1.1 on page 257) and the second concerning bounded RDPs (Theorem 8.1.2 on page 260). Let's now prove them. In what follows, $\mathcal{R} = (\Gamma, V, B)$ is a well-posed RDP and $\mathcal{A}_\mathcal{R} := (V, \{T_\sigma\})$ is the ADP generated by $\mathcal{R}$.

*Proof of Theorem 8.1.1.* Let $\mathcal{R}$ be globally stable. Then $\mathcal{A}_\mathcal{R}$ is finite and max-stable, by Corollary 9.1.4. Hence the optimality and HPI convergence claims in Theorem 8.1.1 follow from Theorem 9.1.6.

Regarding OPI convergence, let $(v_k, \sigma_k)$ be as given in (8.18) on page 257. Together, Lemma 9.1.8 and Lemma 9.1.10 imply that $v_k \to v^*$ as $k \to \infty$. Given such convergence, Lemma 9.1.12 implies that there exists a $K \in \mathbb{N}$ such that $\sigma_k$ is optimal whenever $k \geqslant K$. □

*Proof of Theorem 8.1.2.* Let $\mathcal{R} = (\Gamma, V, B)$ be a bounded and well-posed. In view of Exercises 8.1.12–8.1.13 (see page 260), it suffices to prove the optimality claims in Theorem 8.1.2 for the reduced RDP $\hat{\mathcal{R}} = (\Gamma, \hat{V}, B)$, where $\hat{V}$ is the order interval in $\mathbb{R}^\mathsf{X}$ generated by the bounding functions (i.e., $\hat{V} = [v_1, v_2]$).

Let $\mathcal{A}$ be the ADP generated by $\hat{\mathcal{R}}$ By Proposition 9.1.5, $\mathcal{A}$ is order stable. Corollary 9.1.4 now implies that $\mathcal{A}$ is max-stable. Hence the claims in Theorem 8.1.2 follow from Theorem 9.1.6. □

## 9.1.5  Min-Optimality

Until now, our ADP theory has focused on maximization of lifetime values. Now we turn to minimization. One of our aims is to prove the RDP minimization results in §8.3.5. We will see that ADP minimization results are easily recovered from ADP maximization results via order duality.

Let $\mathcal{A} = (V, \{T_\sigma\})$ be a well-posed ADP and let $V_\Sigma := \{v_\sigma\}$ be the set of $\sigma$-value functions. We call $\sigma \in \Sigma$ **min-optimal** for $\mathcal{A}$ if $v_\sigma$ is a least element of $V_\Sigma$. When $V_\Sigma$ has a least element we denote it by $v_\downarrow^*$ and call it the **min-value function** generated by $\mathcal{A}$. A policy is called $v$-**min-greedy** if $T_\sigma v \leq T_\tau v$ for all $\tau \in \Sigma$. Existence of a $v$-min-greedy policy for each $v \in V$ is guaranteed by the definition of ADPs.

We say that $\mathcal{A}$ obeys **Bellman's principle of min-optimality** if

$$\sigma \in \Sigma \text{ is min-optimal for } \mathcal{A} \quad \Longleftrightarrow \quad \sigma \text{ is } v_\downarrow^*\text{-min-greedy.}$$

We define the **Bellman min-operator** corresponding to $\mathcal{A}$ as the self-map $T_\downarrow$ on $V$ defined by $T_\downarrow v = \bigwedge_\sigma T_\sigma v$. This map is well-defined because $\{T_\sigma v\}_{\sigma \in \Sigma}$ has a least element (by the definition of ADPs) and, moreover, $\sigma \in \Sigma$ is $v$-min-greedy if and only if $T_\sigma v = T_\downarrow v$.

We say that $v$ satisfies the **Bellman min-equation** if $T_\downarrow v = v$. We call $\mathcal{A}$ **min-stable** if $\mathcal{A}$ is order stable and $T_\downarrow$ has at least one fixed point in $V$. We define $H_\downarrow$ from $V$ to $\{v_\sigma\}$ via $H_\downarrow v = v_\sigma$ where $\sigma$ is $v$-min-greedy and call $H_\downarrow$ the **Howard min-operator** generated by $\mathcal{A}$. Iterating with $H_\downarrow$ is called **min-HPI**.

To make our terminology more symmetric, in this section and below we refer to maximization-based optimal policies as **max-optimal,** the Bellman operator $T = \bigvee_\sigma T_\sigma v$ as the **Bellman max-operator**, and so on.

Results analogous to Theorem 9.1.6 hold for the minimization case.

**Theorem 9.1.13** (Min-optimality)**.** *If $\mathcal{A}$ is min-stable, then*

(i) *the min-value function $v_\downarrow^*$ generated by $\mathcal{A}$ exists in $V$,*

(ii) *$v_\downarrow^*$ is the unique solution to the Bellman min-equation in $V$,*

(iii) *$\mathcal{A}$ obeys Bellman's principle of min-optimality, and*

(iv) *$\mathcal{A}$ has at least one min-optimal policy.*

*If, in addition, $\Sigma$ is finite, then min-HPI converges to $v_\downarrow^*$ in finitely many steps.*

To prove Theorem 9.1.13 we use order duality. Below, if $\mathcal{A} := (V, \{T_\sigma\})$ is an ADP then its **dual** is

$$\mathcal{A}^\partial := (V^\partial, \{T_\sigma\}) \quad \text{where } V^\partial \text{ is the order dual of } V.$$

In this setting, we let $T^\partial$ be the Bellman operator for $\mathcal{A}^\partial$, $(v^*)^\partial$ be the value function for $\mathcal{A}^\partial$, and so on. We note that $\mathcal{A}$ is self-dual, in the sense that $(\mathcal{A}^\partial)^\partial = \mathcal{A}$, since the same is true for $V$.

EXERCISE 9.1.9. Let $\mathcal{A}$ be a well-posed ADP with dual $\mathcal{A}^\partial$. Verify the following.

(i) Given $v \in V$, $\sigma \in \Sigma$ is $v$-min-greedy for $\mathcal{A}$ if and only if $\sigma$ is $v$-max-greedy for $\mathcal{A}^\partial$,

(ii) $T_\downarrow = T^\partial$ and $T_\downarrow^\partial = T$,

(iii) $H_\downarrow = H^\partial$ and $H_\downarrow^\partial = H$,

(iv) $\mathcal{A}$ is order stable if and only if $\mathcal{A}^\partial$ is order stable,

(v) $\mathcal{A}$ is min-stable if and only if $\mathcal{A}^\partial$ is max-stable, and, in this case, $v_\downarrow^* = (v^*)^\partial$.

(vi)  $\sigma \in \Sigma$ is max-optimal for $\mathcal{A}$ if and only if $\sigma$ is min-optimal for $\mathcal{A}^{\partial}$.

Self-duality implies corollaries to Exercise 9.1.9 that we treat as self-evident. For example, if $\mathcal{A}$ is max-stable if and only if $\mathcal{A}^{\partial}$ is min-stable, which follows from part (v) and the fact that $(\mathcal{A}^{\partial})^{\partial} = \mathcal{A}$.

*Proof of Theorem 9.1.13.* Let $\mathcal{A}$ be min-stable. By Exercise 9.1.9, the dual $\mathcal{A}^{\partial}$ is max-stable. Hence all of the conclusions of the max-optimality result in Theorem 9.1.6 apply to $\mathcal{A}^{\partial}$. All that remains is to translate these max-optimality results for $\mathcal{A}^{\partial}$ back to min-optimality results for $\mathcal{A}$.

Regarding claim (i) of the min-optimality results, max-optimality of $\mathcal{A}^{\partial}$ implies that $(v^*)^{\partial}$ exists in $V$. But then $v_{\downarrow}^*$ exists in $V$, since, by Exercise 9.1.9, $v_{\downarrow}^* = (v^*)^{\partial}$.

Regarding (ii), we know that $(v^*)^{\partial}$ is the unique solution to $T^{\partial}(v^*)^{\partial} = (v^*)^{\partial}$, so, applying Exercise 9.1.9 again, we have $T_{\downarrow} v_{\downarrow}^* = v_{\downarrow}^*$.

The remaining steps of the proof are similar and left to the reader.  □

## 9.2   Isomorphic Dynamic Programs

It can be useful to understand when two superficially different problems are actually the same. For example, students of algebra study group isomorphisms that tell when two groups are identical in terms of their characteristics under group operations. Topological isomorphisms play a similar role for topological spaces – these kinds of isomorphisms are also called homeomorphisms, which we introduced in §2.1.1. In turn, homeomorphisms can identify topologically conjugate dynamical systems, allowing us to connect essentially identical dynamic models (see §2.1.1.2).

It is also useful to have a notion of "isomorphic" dynamic programs. In this section describe an isomorphic relationship that leads to essentially equivalent optimality properties. We exploit these ideas to study additional optimization problems, including decision problems with ambiguity.

### 9.2.1   Isomorphic ADPs

To define isomorphic ADPs and relationships between them, we will use a similarity notion for dynamical systems called "order conjugacy" that is based on the idea of order isomorphisms.

### 9.2.1.1   Order Conjugacy

Let $(V, T)$ and $(\hat{V}, \hat{T})$ be two dynamical systems, where $V$ and $\hat{V}$ are partially ordered. Recall that $(V, T)$ and $(\hat{V}, \hat{T})$ are called conjugate if there exists a bijection $F$ from $V$ to $\hat{V}$ such that $F \circ T = \hat{T} \circ F$ (see §2.1.1.1). We call $(V, T)$ and $(\hat{V}, \hat{T})$ **order conjugate** if, in addition, $F$ is an order isomorphism. When $F$ needs to be specified we will say that $(V, T)$ and $(\hat{V}, \hat{T})$ are order conjugate **under** $F$.

EXERCISE 9.2.1. Verify that order conjugacy is an equivalence relation (see §A.1) on the set of dynamical systems over partially ordered sets.

The next lemma modifies Proposition 2.1.2 on page 45 to an order setting.

**Lemma 9.2.1.** *If $(V, T)$ and $(\hat{V}, \hat{T})$ are order conjugate under $F$ and $v$ is the unique fixed point of $T$ in $V$, then $Fv$ is the unique fixed point of $\hat{T}$ in $\hat{V}$. If, in addition, $T$ is is order stable on $V$, then $\hat{T}$ is order stable on $\hat{V}$.*

*Proof.* Let $(V, T)$ and $(\hat{V}, \hat{T})$ be order conjugate under $F$, let $\{v\}$ be the unique fixed point of $T$ and let $\hat{v} = Fv$. The claim that $\hat{v}$ is the unique fixed point of $\hat{T}$ in $\hat{V}$ follows from Proposition 2.1.1. The remainder of the proof is left to Exercise 9.2.2.                      □

EXERCISE 9.2.2. Prove: The operator $\hat{T}$ is order stable on $\hat{V}$ under the assumptions of Lemma 9.2.1.

### 9.2.1.2   Definitions and Relationships

Let $\mathcal{A} = (V, \{T_\sigma\})$ and $\hat{\mathcal{A}} = (\hat{V}, \{\hat{T}_\sigma\})$ be two ADPs. We call $\mathcal{A}$ and $\hat{\mathcal{A}}$ **isomorphic** under $F$ if these two ADPs have the same policy set $\Sigma$ and $F$ is an order isomorphism from $V$ to $\hat{V}$ such that

$$F \circ T_\sigma = \hat{T}_\sigma \circ F \quad \text{on } V \text{ for all } \sigma \in \Sigma. \tag{9.8}$$

In other words, $(V, T_\sigma)$ and $(\hat{V}, \hat{T}_\sigma)$ are order conjugate under $F$ for all $\sigma \in \Sigma$.[2]

**Example 9.2.1.** Let $\mathcal{R}$ and $\hat{\mathcal{R}}$ be RDPs, generating ADPs $\mathcal{A}$ and $\hat{\mathcal{A}}$ respectively. If $\mathcal{R}$ and $\hat{\mathcal{R}}$ are topologically conjugate RDPs (see §8.1.4) under $\varphi$ and, in addition, $\varphi$ is an increasing function, then $\mathcal{A}$ and $\hat{\mathcal{A}}$ are isomorphic. To see this, let $V$ and $\hat{V}$ be the respective value spaces for $\mathcal{R}$ and $\hat{\mathcal{R}}$, and let $v \mapsto Fv$ be defined by $Fv = \varphi \circ v$.

---

[2]While the definition require that the two ADPs have the same policy set $\Sigma$, it suffices that the policy sets can be put in one-to-one correspondence with each other.

Since $\varphi$ is an increasing bijection, $F$ is an order isomorphism from $V$ to $\hat{V}$. Moreover, the respective policy operators are linked by $T_\sigma = F^{-1} \circ \hat{T}_\sigma \circ F$ on $V$ (see the proof of Proposition 8.1.3). This confirms that $\mathcal{A}$ and $\hat{\mathcal{A}}$ are isomorphic.

**Example 9.2.2.** Fei et al. (2021) consider an "exponential" risk-sensitive $Q$-factor Bellman equation, which has policy operator

$$(M_\sigma h)(x, a) = \exp\left\{\theta r(x, a) + \beta \ln\left[\sum_{x'} h(x', \sigma(x')) P(x, a, x')\right]\right\}$$

mapping $(0, \infty)^\mathsf{G}$ into itself. Here $\theta$ is a nonzero constant and all other primitives are as in the risk-sensitive $Q$-factor ADP $\mathcal{A} := (\mathbb{R}^\mathsf{G}, \{Q_\sigma\})$ in Example 9.1.4. Let $F$ be the order isomorphism from $\mathbb{R}^\mathsf{G}$ to $(0, \infty)^\mathsf{G}$ defined by $(Fh)(x, a) = \exp(\theta h(x, a))$. Then, for $Q_\sigma$ defined in (9.4), $h \in \mathbb{R}^\mathsf{G}$ and $(x, a) \in \mathsf{G}$,

$$(FQ_\sigma h)(x, a) = \exp\left\{\theta r(x, a) + \beta \ln\left[\sum_{x'} \exp\left[\theta h(x', \sigma(x'))\right] P(x, a, x')\right]\right\},$$

which is equal to $(M_\sigma Fh)(x, a)$. In other words, $F \circ Q_\sigma = M_\sigma \circ F$ on $\mathbb{R}^\mathsf{G}$. Hence $\hat{\mathcal{A}} := ((0, \infty)^\mathsf{G}, \{M_\sigma\})$ and $\mathcal{A}$ are isomorphic ADPs.

**Lemma 9.2.2.** *Isomorphism between ADPs is an equivalence relation on the set of ADPs.*

In other words, if $\mathbf{A}$ is the set of all ADPs and, for $\mathcal{A}, \hat{\mathcal{A}} \in \mathbf{A}$, the symbol $\mathcal{A} \sim \hat{\mathcal{A}}$ means $\mathcal{A}$ and $\hat{\mathcal{A}}$ are isomorphic, then $\sim$ is reflexive, symmetric and transitive.

Exercise 9.2.3. Prove Lemma 9.2.2. [Hint: review Exercise 2.1.5 on page 44.]

### 9.2.1.3 Isomorphisms and Optimality

We seek relationships between optimality properties of isomorphic ADPs. For all of this section, we take $\mathcal{A} = (V, \{T_\sigma\})$ and $\hat{\mathcal{A}} = (\hat{V}, \{\hat{T}_\sigma\})$ to be two ADPs with the same policy set. When they exist, we let

- $v_\sigma$ (resp., $\hat{v}_\sigma$) be the unique fixed point of $T_\sigma$ (resp., $\hat{T}_\sigma$)
- $T$ (resp., $\hat{T}$) be the Bellman operator of $\mathcal{A}$ (resp., $\hat{\mathcal{A}}$)
- $T_\downarrow$ (resp., $\hat{T}_\downarrow$) be the min-Bellman operator of $\mathcal{A}$ (resp., $\hat{\mathcal{A}}$)
- $v^*$ (resp., $\hat{v}^*$) be the value function of $\mathcal{A}$ (resp., $\hat{\mathcal{A}}$)
- $v_\downarrow^*$ (resp., $\hat{v}_\downarrow^*$) be the min-value function of $\mathcal{A}$ (resp., $\hat{\mathcal{A}}$)

The next theorem shows that isomorphic ADPs share the same regularity and optimality properties:

**Theorem 9.2.3.** *If $\mathcal{A}$ and $\hat{\mathcal{A}}$ are isomorphic under $F$, then*

(i) *$\mathcal{A}$ is well-posed if and only if $\hat{\mathcal{A}}$ is well-posed,*

(ii) *$\mathcal{A}$ is order stable if and only if $\hat{\mathcal{A}}$ is order stable,*

(iii) *the Bellman max- and min-operators of $\mathcal{A}$ and $\hat{\mathcal{A}}$ are related by*

$$F \circ T = \hat{T} \circ F \quad and \quad F \circ T_\downarrow = \hat{T}_\downarrow \circ F, \tag{9.9}$$

(iv) *$\mathcal{A}$ is max-stable if and only if $\hat{\mathcal{A}}$ is max-stable and, in this case, $\hat{v}^* = F v^*$.*

(v) *$\mathcal{A}$ is min-stable if and only if $\hat{\mathcal{A}}$ is min-stable and, in this case, $\hat{v}_\downarrow^* = F v_\downarrow^*$.*

(vi) *$\mathcal{A}$ and $\hat{\mathcal{A}}$ have the same max- and min-optimal policies.*

*Proof.* Let $\mathcal{A}$ and $\hat{\mathcal{A}}$ be as stated. Claims (i)–(ii) follow directly from order conjugacy of the policy operators (as in (9.8)) and Lemma 9.2.1. Regarding (iii), we fix $v \in V$ and apply (9.8) to obtain

$$Tv = \bigvee_\sigma T_\sigma v = \bigvee_\sigma F^{-1} \hat{T}_\sigma F v = F^{-1} \bigvee_\sigma \hat{T}_\sigma F v = F^{-1} \hat{T} F v.$$

(The third inequality is justified by existence of greedy policies – see, in particular, Exercises 2.2.25 and 2.2.34.) This is equivalent to $F \circ T = \hat{T} \circ F$, which confirms the first equality in (9.9). The second equality can be obtained in the same way, after replacing $\bigvee_\sigma$ with $\bigwedge_\sigma$.

Regarding (iv), note that $(V, T)$ and $(\hat{V}, \hat{T})$ are order conjugate under $F$ by (9.9). It now follows from Lemma 9.2.1 that $\mathcal{A}$ is max-stable if and only if $\hat{\mathcal{A}}$ is max-stable, and, when this max-stability holds, that the unique fixed points of $T$ and $\hat{T}$ are related by $\hat{v}^* = F v^*$. The proof of (v) is analogous.

Regarding (vi), we use order conjugacy of the policy operators to obtain $F v_\sigma = \hat{v}_\sigma$ for all $\sigma \in \Sigma$, from which it follows that

$$v_\sigma = \bigvee_\tau v_\tau \quad \Longleftrightarrow \quad F v_\sigma = F \bigvee_\tau v_\tau = \bigvee_\tau F v_\tau \quad \Longleftrightarrow \quad \hat{v}_\sigma = \bigvee_\tau \hat{v}_\tau$$

In other words, $\sigma$ is max-optimal for $\mathcal{A}$ if and only if $\sigma$ is max-optimal for $\hat{\mathcal{A}}$. Replacing $\bigvee$ with $\bigwedge$ proves that the same statement is true for min-optimal policies. □

## 9.2.2 Anti-Isomorphic ADPs

Let $\mathcal{A} = (V, \{T_\sigma\})$ and $\hat{\mathcal{A}} = (\hat{V}, \{\hat{T}_\sigma\})$ be ADPs. We call $\mathcal{A}$ and $\hat{\mathcal{A}}$ **anti-isomorphic** under $F$ if they have the same policy set $\Sigma$ and, in addition, $F$ is an anti-isomorphism from $V$ to $\hat{V}$ such that (9.8) holds. Equivalently, $\mathcal{A}$ and $\hat{\mathcal{A}}$ are anti-isomorphic if $\mathcal{A}$ is isomorphic to $\hat{\mathcal{A}}^\partial$, the dual of $\hat{\mathcal{A}}$ (see Exercise 9.1.2). If $\mathcal{A} \overset{a}{\sim} \hat{\mathcal{A}}$ means that $\mathcal{A}$ and $\hat{\mathcal{A}}$ are anti-isomorphic, then $\overset{a}{\sim}$ is symmetric and transitive but not necessarily reflexive.

**Example 9.2.3.** Let $\mathcal{R}$ and $\hat{\mathcal{R}}$ to be RDPs, generating ADPs $\mathcal{A}$ and $\hat{\mathcal{A}}$ respectively. If $\mathcal{R}$ and $\hat{\mathcal{R}}$ are topologically conjugate RDPs (see §8.1.4) under $\varphi$ and, in addition, $\varphi$ is a decreasing function, then $\mathcal{A}$ and $\hat{\mathcal{A}}$ is anti-isomorphic. This can be verified using arguments is very similar to those we used for the isomorphic case, in Example 9.2.1.

Here is an optimality result for anti-isomorphic ADPs that parallels Theorem 9.2.3.

**Theorem 9.2.4.** *If $\mathcal{A}$ and $\hat{\mathcal{A}}$ are anti-isomorphic under $F$, then*

(i) *$\mathcal{A}$ is well-posed if and only if $\hat{\mathcal{A}}$ is well-posed,*

(ii) *$\mathcal{A}$ is order stable if and only if $\hat{\mathcal{A}}$ is order stable,*

(iii) *the Bellman max- and min-operators obey $F \circ T = \hat{T}_\downarrow \circ F$,*

(iv) *$\mathcal{A}$ is max-stable if and only if $\hat{\mathcal{A}}$ is min-stable and, in this case, $\hat{v}_\downarrow^* = F v^*$, and*

(v) *$\sigma \in \Sigma$ is optimal for $\mathcal{A}$ if and only if $\sigma$ is min-optimal for $\hat{\mathcal{A}}$.*

*Proof.* Let $\mathcal{A}$ and $\hat{\mathcal{A}}$ be anti-isomorphic, so that $\mathcal{A}$ is isomorphic to $\hat{\mathcal{A}}^\partial$. By Theorem 9.2.3, $\mathcal{A}$ is well-posed if and only if $\hat{\mathcal{A}}^\partial$ is well-posed, which means that $\hat{T}_\sigma$ has a unique fixed point in $\hat{V}$ for all $\sigma \in \Sigma$. This is equivalent to the statement that that $\hat{\mathcal{A}}$ is well-posed, completing the proof of (i). Regarding (ii), Theorem 9.2.3 implies that $\mathcal{A}$ is order stable if and only if $\hat{\mathcal{A}}^\partial$ is order stable, which is equivalent to the statement that $\hat{\mathcal{A}}$ is order stable (by Lemma 9.1.2).

Regarding (iii), Theorem 9.2.3 yields $F \circ T = \hat{T}^\partial \circ F$. As in our discussion of duality in § 9.1.5, this is equivalent to $F \circ T = \hat{T}_\downarrow \circ F$.

Regarding (iv), Theorem 9.2.3 implies that $\mathcal{A}$ is max-stable if and only if $\hat{\mathcal{A}}^\partial$ is max-stable, and, in this setting, that $(\hat{v}^*)^\partial = F v^*$. Exercise 9.1.9 implies that max-stability of $\hat{\mathcal{A}}^\partial$ is equivalent to min-stability of $\hat{\mathcal{A}}$, and that $(\hat{v}^*)^\partial = \hat{v}_\downarrow^*$. Combining the last two equations gives and $\hat{v}_\downarrow^* = F v^*$.

Finally, regarding (v), Theorem 9.2.3 tells us that $\mathcal{A}$ and $\hat{\mathcal{A}}^\partial$ have the same max-optimal policies. Applying Exercise 9.1.9, we see that the max-optimal policies of $\mathcal{A}$ are the same as the min-optimal policies of $\hat{\mathcal{A}}$. □

### 9.2.3   Application: Epstein–Zin Without Irreducibility

In this section we illustrate how isomorphic relationships can be useful. To this end, we return to the Epstein–Zin specification studied in §8.1.4.1, where each policy operator $T_\sigma$ has the form

$$(T_\sigma v)(x) = \left\{ r(x, \sigma(x)) + \beta \left( \sum_{x'} v(x')^\gamma P(x, \sigma(x), x') \right)^{\alpha/\gamma} \right\}^{1/\alpha}. \qquad (9.10)$$

As in §8.1.4.1, we assume that $r \gg 0$, so $T_\sigma$ maps $(0, \infty)^{\mathsf{X}}$ into itself. In §8.1.4.1, we established optimality properties for this specification when $P_\sigma$ is irreducible for all $\sigma \in \Sigma$. Here we drop this assumption. With a little more effort, we can establish the same optimality properties without irreducibility.

To this end, let $\theta = \gamma/\alpha$, let $r_1 = \min r$ and let $r_2 = \max r$. Fix $\varepsilon > 0$ with $r_1 - \varepsilon > 0$. Consider the constant functions

$$v_1 = m_1 \wedge m_2 \text{ and } v_2 = m_1 \vee m_2 \quad \text{where } m_1 := \left( \frac{r_1 - \varepsilon}{1 - \beta} \right)^\theta \quad \text{and} \quad m_2 := \left( \frac{r_2 + \varepsilon}{1 - \beta} \right)^\theta.$$

Let $\hat{V} = [v_1, v_2]$. Let $F$ be defined at $v \in (0, \infty)^{\mathsf{X}}$ by $Fv = v^\gamma$ and let $V = F^{-1}\hat{V}$. Let $\mathcal{A} = (V, \{T_\sigma\})$ with $T_\sigma$ given by (9.10).

**Proposition 9.2.5.** *If $\beta \in (0, 1)$, then $\mathcal{A}$ is both min- and max-stable.*

To prove Proposition 9.2.5, we introduce the auxillary ADP $\hat{\mathcal{A}} = (\hat{V}, \{\hat{T}_\sigma\})$ as follows. First we define $\hat{B}$ by

$$\hat{B}(x, a, v) = \left\{ r(x, a) + \beta \left( \sum_{x'} v(x') P(x, a, x') \right)^{1/\theta} \right\}^\theta,$$

which is the same as as in (8.24) on page 262. Then we let $\hat{T}_\sigma$ be the corresponding policy operator, so that $(\hat{T}_\sigma \hat{v})(x) = \hat{B}(x, a, \hat{v})$ for all $\hat{v} \in \hat{V}$.

EXERCISE 9.2.4. Show that, for all $(x, a) \in \mathsf{G}$, we have

$$v_1(x) < \hat{B}(x, a, v_1) \quad \text{and} \quad \hat{B}(x, a, v_2) < v_2(x). \qquad (9.11)$$

Fix $\sigma \in \Sigma$ and let $\hat{T}_\sigma$ be the policy operator associated with $\hat{B}$ (see (8.25)).

EXERCISE 9.2.5. Confirm the following statements.

(i) If either $\theta < 0$ or $1 \leqslant \theta$, then $\hat{T}_\sigma$ is concave on $I$.

(ii) If $0 < \theta \leqslant 1$, then $\hat{T}_\sigma$ is convex on $I$.

EXERCISE 9.2.6. Show that, for all $\sigma \in \Sigma$, the operator $\hat{T}_\sigma$ is globally stable on $\hat{V}$.

EXERCISE 9.2.7. Show that, for all $\sigma \in \Sigma$, we have $F \circ T_\sigma = \hat{T}_\sigma \circ F$ on $I$.

**Lemma 9.2.6.** *The following statements are true:*

(i) *If $\gamma > 0$, then $\mathcal{A}$ and $\hat{\mathcal{A}}$ are isomorphic.*

(ii) *If $\gamma < 0$, then $\mathcal{A}$ and $\hat{\mathcal{A}}$ are anti-isomorphic.*

*Proof.* If $\gamma < 0$, then $F$ is an order anti-isomorphism from $V$ to $\hat{V}$. (Obviously $F$ is order-reversing. Also, $F$ is clearly one-to-one and, by construction, $F$ maps $V$ onto $\hat{V}$.) From this fact and Exercise 9.2.7, the ADPs $\mathcal{A}$ and $\hat{\mathcal{A}}$ are anti-isomorphic. If $\gamma > 0$, then $F$ is an order isomorphism from $V$ to $\hat{V}$, so $\mathcal{A}$ and $\hat{\mathcal{A}}$ are isomorphic. □

*Proof of Proposition 9.2.5.* By Exercise 9.2.6 and Corollary 9.1.4 on page 298, $\hat{\mathcal{A}}$ is both max- and min-stable. If $\gamma < 0$, then, by Lemma 9.2.6, $\mathcal{A}$ and $\hat{\mathcal{A}}$ are isomorphic, so Theorem 9.2.3 implies that $\hat{\mathcal{A}}$ is both min- and max-stable. If $\gamma > 0$, then, by Lemma 9.2.6, $\mathcal{A}$ and $\hat{\mathcal{A}}$ are anti-isomorphic. Now Theorem 9.2.4 implies that $\hat{\mathcal{A}}$ is both min- and max-stable. □

Notice that the relationship between $\mathcal{A}$ and $\hat{\mathcal{A}}$ allows us to use either one to solve for an optimal policy. For example, if $\gamma < 0$, then any min-optimal policy for $\hat{\mathcal{A}}$ will be max-optimal for $\mathcal{A}$. Hence we can solve for the Epstein–Zin max-optimal policy either by directly solving $\mathcal{A}$ or by solving $\hat{\mathcal{A}}$ for a min-optimal policy. The best choice typically depends on computational simplicity and numerical stability.

## 9.3 Subordinate ADPs

Next we introduce an asymmetric relationship between ADPs referred to as as subordination. In essence, a subordinate ADP is an ADP that is derived from another ADP, often via some rearrangement of the Bellman equation. In the applications we consider, the associated transformations are not bijective, which differentiates subordination from the isomorphic relationships considered in §9.2.1. (This is typically because one dynamic program evolves in a higher dimensional space than another.) Nonetheless, we show that subordination provides valuable connections between ADPs in terms of optimality.

Key examples include the dynamic programs associated with the modified Bellman equations we introduced in §5.3. In particular, in §5.3.5.3 we introduced the families of operators $\{R_\sigma\}$, $\{S_\sigma\}$ and $\{T_\sigma\}$, which represented policy operators for the expected value Bellman equation, the $Q$-factor Bellman equation and the regular MDP Bellman equation respectively. When paired with their respective domains, these operator families become ADPs. As we show below, the expected value and $Q$-factor ADPs are subordinate to the ADP generated by $\{T_\sigma\}$.

Relative to the theory provided in §5.3, the benefit of the exposition below is that the proofs are both more concise and more general in the current abstract setting. As a result, we can easily cover other variations on the MDP Bellman equation (of which there are many), as well as studying relationships between ADPs beyond the traditional MDP setting.

## 9.3.1   Definition and Properties

First we define subordination and investigate basic properties.

### 9.3.1.1   Semiconjugacy

The theory below uses a similarity notion for dynamical systems. To define this notion, we take $(V, S)$ and $(\hat{V}, \hat{S})$ to be dynamical systems, where $V$ and $\hat{V}$ are both partially ordered. In this setting, we call $(V, S)$ and $(\hat{V}, \hat{S})$ **mutually semiconjugate** if there exist order-preserving maps $F \colon V \to \hat{V}$ and $G \colon \hat{V} \to V$ such that

$$S = G \circ F \quad \text{and} \quad \hat{S} = F \circ G. \tag{9.12}$$

The "semiconjugate" terminology comes from the fact that, when (9.12) holds,

$$F \circ S = \hat{S} \circ F \quad \text{and} \quad G \circ \hat{S} = S \circ G. \tag{9.13}$$

It follows from (9.13) that, if either $F$ or $G$ is an order isomorphism, then $S$ and $\hat{S}$ are order conjugate. Our current focus is on settings where this is *not* the case.

**Lemma 9.3.1.** *Let* $(V, S)$ *and* $(\hat{V}, \hat{S})$ *be mutually semiconjugate under the maps* $F, G$ *in* (9.12). *In this this setting,*

(i) *if* $v$ *is a fixed point of* $S$ *in* $V$, *then* $Fv$ *is a fixed point of* $\hat{S}$ *in* $\hat{V}$.

(ii) *if* $\hat{v}$ *is a fixed point of* $\hat{S}$ *in* $\hat{V}$, *then* $G\hat{v}$ *is a fixed point of* $S$ *in* $V$.

(iii) $S$ *has a unique fixed point in* $V$ *if and only if* $\hat{S}$ *has a unique fixed point in* $\hat{V}$.

(iv) $S$ *is order stable on* $V$ *if and only if* $\hat{S}$ *is order stable on* $\hat{V}$.

*Proof.* Let $(V, S)$ and $(\hat{V}, \hat{S})$ be as stated. If $v$ is a fixed point of $S$ in $V$, then $\hat{S}Fv = FSv = Fv$, so $Fv$ is a fixed point of $\hat{S}$ in $\hat{V}$. Similarly, if $\hat{v}$ is a fixed point of $\hat{S}$ in $\hat{V}$, then $SG\hat{v} = G\hat{S}\hat{v} = G\hat{v}$, so $G\hat{v}$ is a fixed point of $S$ in $V$. This proves (i)–(ii).

Regarding (iii), suppose that $v$ is the only fixed point of $S$ in $v$. We know that $Fv$ is a fixed point of $\hat{S}$ in $\hat{V}$. Suppose in addition that $\hat{v}$ is fixed for $\hat{S}$. Then $FG\hat{v} = \hat{v}$ and hence $GFG\hat{v} = G\hat{v}$, or $SG\hat{v} = G\hat{v}$. Since $v$ is the only fixed point of $S$ in $V$, we have $G\hat{v} = v$. Applying $F$ gives $\hat{S}\hat{v} = Fv$. But $\hat{v}$ is fixed for $\hat{S}$, so $\hat{v} = Fv$. This shows that $\hat{S}$ has exactly one fixed point in $\hat{V}$. The reverse implication holds by symmetry.

Regarding (iv), suppose that $S$ is order stable on $V$, with unique fixed point $v \in V$. Then, by the preceding argument, $Fv$ is the unique fixed point of $\hat{S}$ in $\hat{V}$. The map $\hat{S}$ is upward stable because if $\hat{v} \in \hat{V}$ and $\hat{v} \leq \hat{S}\hat{v}$, then $G\hat{v} \leq G\hat{S}\hat{v} = SG\hat{v}$ and so, by upward stability of $S$, $G\hat{v} \leq v$. Applying $F$ gives $\hat{S}\hat{v} \leq Fv$, so $\hat{S}$ is upward stable. The proof of downward stability is similar. Hence $\hat{S}$ is order stable on $\hat{V}$. The reverse implication holds by symmetry. □

### 9.3.1.2 Subordination

Let $\mathcal{A} := (V, \{T_\sigma\})$ and $\hat{\mathcal{A}} := (\hat{V}, \{\hat{T}_\sigma\})$ be ADPs with common policy set $\Sigma$. We say that $\hat{\mathcal{A}}$ is **subordinate** to $\mathcal{A}$ if there exists an order-preserving map $F$ from $V$ onto $\hat{V}$ and a family of order-preserving maps $\{G_\sigma\}_{\sigma \in \Sigma}$ from $\hat{V}$ to $V$, such that

$$T_\sigma = G_\sigma \circ F \quad \text{and} \quad \hat{T}_\sigma = F \circ G_\sigma \quad \text{for all } \sigma \in \Sigma. \quad (9.14)$$

Note that, in this setting, the dynamical systems $(V, T_\sigma)$ and $(\hat{V}, \hat{T}_\sigma)$ are mutually semi-conjugate (see §9.3.1.1) for all $\sigma \in \Sigma$.

The following examples show that some dynamic programs investigated in the recent literature are subordinate to a more traditional dynamic program.

**Example 9.3.1.** Let $\mathcal{A}$ be the ADP generated by MDP $\mathcal{M} = (\Gamma, \beta, r, P)$. Set

$$(Fv)(x, a) \coloneqq r(x, a) + \beta \sum_{x'} v(x')P(x, a, x') \tag{9.15}$$

for $v \in \mathbb{R}^{\mathsf{X}}$ and $\hat{V} \coloneqq F(\mathbb{R}^{\mathsf{X}}) \subset \mathbb{R}^{\mathsf{G}}$. Let $\{G_\sigma\}$ defined by

$$(G_\sigma f)(x) = f(x, \sigma(x)) \qquad \left(f \in \mathbb{R}^{\mathsf{G}}\right)$$

and let $\{S_\sigma\}$ be the policy operators of the $Q$-factor MDP (see (9.2)). Then, given $f \in \hat{V}$, each $S_\sigma$ can be expressed as

$$(S_\sigma f)(x, a) = r(x, a) + \beta \sum_{x'} f(x', \sigma(x'))P(x, a, x') = (F G_\sigma f)(x, a).$$

The pair $\hat{\mathcal{A}} \coloneqq (\hat{V}, \{S_\sigma\})$ is an ADP. The map $F$ is the order-preserving and, by construction, maps $V \coloneqq \mathbb{R}^{\mathsf{X}}$ onto $\hat{V}$. Also, each $G_\sigma$ is order-preserving and $T_\sigma$ satisfies

$$(T_\sigma v)(x) = r(x, \sigma(x) + \beta \sum_{x'} v(x')P(x, \sigma(x), x') = (G_\sigma F v)(x)$$

at $v \in \mathbb{R}^{\mathsf{X}}$. Hence $\hat{\mathcal{A}}$ is subordinate to $\mathcal{A}$.

**Example 9.3.2.** Let $\mathcal{A}$ be the ADP generated by MDP $\mathcal{M} = (\Gamma, \beta, r, P)$, with fintie state space $\mathsf{X}$, and let $F$ be the order-preserving map from $\mathbb{R}^{\mathsf{X}}$ to $\mathbb{R}^{\mathsf{G}}$ defined by

$$(Fv)(x, a) = \sum_{x'} v(x')P(x, a, x') \qquad (v \in \mathbb{R}^{\mathsf{X}}). \tag{9.16}$$

If $\hat{V} \coloneqq F(\mathbb{R}^{\mathsf{X}}) \subset \mathbb{R}^{\mathsf{G}}$ and $\{R_\sigma\}$ is the policy operators

$$(R_\sigma g)(x, a) = \sum_{x'} \left\{ r(x', \sigma(x')) + \beta g(x', \sigma(x')) \right\} P(x, a, x')$$

associated with the expected value MDP (cf., §5.3.1.4), then $\hat{\mathcal{A}} \coloneqq (\hat{V}, \{R_\sigma\})$ is subordinate to $\mathcal{A}$. Indeed, $F$ is order-preserving and, by construction, maps $V \coloneqq \mathbb{R}^{\mathsf{X}}$ onto $\hat{V}$. Moreover, the maps $\{G_\sigma\}$ defined by

$$(G_\sigma g)(x) = r(x, \sigma(x)) + \beta g(x, \sigma(x)) \qquad (g \in \mathbb{R}^{\mathsf{G}}). \tag{9.17}$$

are order-preserving and the MDP policy operator $T_\sigma$ satisfies

$$(T_\sigma v)(x) = r(x, \sigma(x)) + \beta \sum_{x'} v(x')P(x, \sigma(x), x') = (G_\sigma F v)(x),$$

while the expected value policy operator $R_\sigma$ satisfies $(R_\sigma g)(x, a) = (F G_\sigma g)(x, a)$.

**Example 9.3.3.** Let $\mathcal{R}$ be the risk-sensitive RDP in Example 8.1.6 and let $\mathcal{A}$ be the ADP generated by $\mathcal{R}$. The risk-sensitive $Q$-factor ADP in Example 9.1.4 is subordinate to $\mathcal{A}$. The proof is almost identical to the argument in Example 9.3.1, after replacing $F$ in (9.15) with

$$(Fv)(x, a) = r(x, a) + \frac{\beta}{\theta} \ln \left[ \sum_{x'} \exp(\theta v(x')) P(x, a, x') \right].$$

Let $\mathcal{A}$ and $\hat{\mathcal{A}}$ represent ADPs with with respective $\sigma$-value functions $\{v_\sigma\}$ and $\{\hat{v}_\sigma\}$. Below, if $\hat{\mathcal{A}}$ is subordinate to $\mathcal{A}$, then $F$ and $\{G_\sigma\}$ always represent the order-preserving maps in (9.14).

**Proposition 9.3.2.** *If $\hat{\mathcal{A}}$ is subordinate to $\mathcal{A}$, then*

(i) *$\hat{A}$ is well-posed if and only if $\mathcal{A}$ is well-posed, and*

(ii) *$\hat{A}$ is order stable if and only if $\mathcal{A}$ is order stable.*

*In either case, the $\sigma$-value functions are linked by*

$$\hat{v}_\sigma = Fv_\sigma \quad and \quad v_\sigma = G_\sigma \hat{v}_\sigma \quad for\ all \quad \sigma \in \Sigma.$$

*Proof.* All claims follow from Lemma 9.3.1 and the observation that $(V, T_\sigma)$ and $(\hat{V}, \hat{T}_\sigma)$ are mutually semiconjugate at every $\sigma \in \Sigma$. □

## 9.3.2 Optimality

In this section we study the extent to which optimality properties, such as those in Theorem 9.1.6, are transferred under a subordinate relationship.

### 9.3.2.1 Maximization

Let $\mathcal{A} = (V, \{T_\sigma\})$ and $\hat{\mathcal{A}} = (\hat{V}, \{\hat{T}_\sigma\})$ be two ADPs. When $\mathcal{A}$ and $\hat{\mathcal{A}}$ are max-stable, $v^*$ and $\hat{v}^*$ will represent their max-value functions, while $T$ and $\hat{T}$ denote their Bellman max-operators.

If $T_\sigma = G_\sigma \circ F$, as in (9.14), then $\{G_\sigma \hat{v}\}_{\sigma \in \Sigma}$ has a greatest element for every $\hat{v} \in \hat{V}$. Indeed, if $\hat{v} \in \hat{V}$, then, since $F$ is onto, there exists a $v \in V$ with $Fv = \hat{v}$. Moreover, by existence of greedy policies, there exists a $\sigma \in \Sigma$ such that $T_\sigma v = G_\sigma \hat{v}$ dominates $T_\tau v = G_\tau \hat{v}$ for all $\tau \in \Sigma$. This confirms that $\{G_\sigma \hat{v}\}_{\sigma \in \Sigma}$ has a greatest element. As a consequence,

$$G\hat{v} := \bigvee_\sigma G_\sigma \hat{v} \tag{9.18}$$

is a well-defined map from $\hat{V}$ to $V$, and, for each $\hat{v} \in \hat{V}$, there is a $\sigma \in \Sigma$ with $G\hat{v} = G_\sigma \hat{v}$.

**Theorem 9.3.3.** *If $\mathcal{A}$ is max-stable and $\hat{\mathcal{A}}$ is subordinate to $\mathcal{A}$, then $\hat{\mathcal{A}}$ is also max-stable and the Bellman max-operators are related by*

$$T = G \circ F \quad and \quad \hat{T} = F \circ G, \tag{9.19}$$

*while the max-value functions are related by*

$$v^* = G\hat{v}^* \quad and \quad \hat{v}^* = Fv^*. \tag{9.20}$$

*Moreover,*

(i) *if $\sigma$ is max-optimal for $\mathcal{A}$, then $\sigma$ is max-optimal for $\hat{\mathcal{A}}$, and*

(ii) *if $G_\sigma \hat{v}^* = G\hat{v}^*$, then $\sigma$ is max-optimal for $\mathcal{A}$.*

Part (i) of Theorem 9.3.3 shows us how to use $\mathcal{A}$ to solve $\hat{\mathcal{A}}$, while part (ii) tells us how to use $\hat{\mathcal{A}}$ to solve $\mathcal{A}$.

*Proof of Theorem 9.3.3.* Let $\mathcal{A}$ be max-stable and let $\hat{\mathcal{A}}$ be subordinate to $\mathcal{A}$. We begin by proving (9.19). To this end, fix $v \in V$ and observe that $Tv = \bigvee_\sigma G_\sigma Fv = GFv$. This proves the first claim in (9.19). The second claim in (9.19) follows from

$$\hat{T}\hat{v} = \bigvee_\sigma FG_\sigma \hat{v} = F \bigvee_\sigma G_\sigma \hat{v} = FG\hat{v}.$$

(The second equality holds because $\{G_\sigma \hat{v}\}_{\sigma \in \Sigma}$ has a greatest element.) This proves (9.19), so $(V, T)$ and $(\hat{V}, \hat{T})$ are mutually semiconjugate under $F, G$.

Regarding max-stability, since $\mathcal{A}$ is max-stable it is also order stable, and hence $\hat{\mathcal{A}}$ is likewise order stable (Proposition 9.3.2). Also, $T$ has a fixed point in $V$, so $\hat{T}$ must have a fixed point in $\hat{V}$ (Lemma 9.3.1). Hence $\hat{\mathcal{A}}$ is max-stable.

In this setting, applying Theorem 9.1.6 to the max-stable ADP $\hat{\mathcal{A}}$, we see that the value function $\hat{v}^*$ is the unique fixed point of $\hat{T}$ in $\hat{V}$. The equalities in (9.20) now follow from mutual semiconjugacy of $(V, T)$ and $(\hat{V}, \hat{T})$ under $F, G$ and Lemma 9.3.1.

Regarding (i) of the last part of Theorem 9.3.3, let $\sigma$ be max-optimal for $\mathcal{A}$. Since $\mathcal{A}$ is max-stable, Theorem 9.1.6 implies that $\sigma$ is $v^*$-max-greedy (i.e., $T_\sigma v^* = Tv^*$) and, in addition, $v^* = Tv^*$. Also, by (9.20), we have $\hat{v}^* = Fv^*$. Therefore,

$$\hat{T}_\sigma \hat{v}^* = \hat{T}_\sigma Fv^* = F T_\sigma v^* = Fv^* = \hat{v}^* = \hat{T} \hat{v}^*.$$

(The last equality uses the fact that stability of $\mathcal{A}$ implies stability of $\hat{\mathcal{A}}$ combined with Theorem 9.1.6.) Thus, $\sigma$ is $\hat{v}^*$-max-greedy for $\hat{\mathcal{A}}$. But $\hat{\mathcal{A}}$ is max-stable, so another application of Theorem 9.1.6 confirms that $\sigma$ is max-optimal for $\hat{\mathcal{A}}$.

Regarding (ii) of the last part of Theorem 9.3.3, let $\sigma \in \Sigma$ be such that $G_\sigma \hat{v}^* = G \hat{v}^*$. Applying (9.20) yields $G_\sigma Fv^* = GFv^*$, or $T_\sigma v^* = T v^*$. So $\sigma$ is $v^*$-max-greedy for $\mathcal{A}$. Since $\mathcal{A}$ is max-stable, Theorem 9.1.6 implies that $\sigma$ is max-optimal for $\mathcal{A}$. $\qquad\square$

**Example 9.3.4.** Let $\mathcal{A}$ be the ADP generated by MDP $\mathcal{M} = (\Gamma, \beta, r, P)$ with finite state space $\mathsf{X}$. To find optimal policies for $\mathcal{A}$, we can study instead the subordinate ADP $\hat{\mathcal{A}} := (\hat{V}, \{R_\sigma\})$ from Example 9.3.2. In view of Theorem 9.3.3, we can do this by computing the fixed point $\bar{f}$ of the corresponding Bellman operator $R := \bigvee_\sigma R_\sigma$ and then finding a policy $\sigma$ obeying $G_\sigma \bar{f} = G \bar{f}$. By the definition of $G_\sigma$ in (9.17), this means that we solve for $\sigma$ satisfying

$$\sigma(x) \in \operatorname*{argmax}_{a \in \Gamma(x)} \left\{ r(x, a) + \beta \bar{f}(x, a) \right\} \qquad (x \in \mathsf{X}).$$

### 9.3.2.2 Minimization

**Example 9.3.5.** Let $\mathcal{A}$ be the ADP generated by RDP $\mathcal{R} = (\Gamma, V, B)$, as in Example 9.1.1. If $v \in V$, then a policy $\sigma \in \Sigma$ is $v$-min-greedy if and only if $\sigma(x)$ is in $\operatorname{argmin}_{a \in \Gamma(x)} B(x, a, v)$ at each $x \in \mathsf{X}$. Since $\Gamma(x)$ is always finite and nonempty, we can find at least one $v$-min-greedy policy. Hence $\mathcal{A}$ is min-stable whenever $\mathcal{R}$ is globally stable.

The minimization case is identical to the maximization setting after the obvious modifications. In the setting where $\mathcal{A}$ is min-stable, we take $G_\downarrow$ to the order-preserving map from $\hat{V}$ to $V$ defined by

$$G_\downarrow \hat{v} := \bigwedge_\sigma G_\sigma \hat{v}. \tag{9.21}$$

We can then state the following minimization version of Theorem 9.3.3.

**Theorem 9.3.4.** *If $\mathcal{A}$ is min-stable and $\hat{\mathcal{A}}$ is subordinate to $\mathcal{A}$, then $\hat{\mathcal{A}}$ is also min-stable and the Bellman min-operators are related by*

$$T_\downarrow = G_\downarrow \circ F \quad and \quad \hat{T}_\downarrow = F \circ G_\downarrow, \tag{9.22}$$

*while the min-value functions are related by*

$$v_\downarrow^* = G_\downarrow \, \hat{v}_\downarrow^* \quad and \quad \hat{v}_\downarrow^* = F \, v_\downarrow^*. \tag{9.23}$$

*Moreover,*

(i) *if $\sigma$ is min-optimal for $\mathcal{A}$, then $\sigma$ is min-optimal for $\hat{\mathcal{A}}$, and*

(ii) *if $G_\sigma \, \hat{v}_\downarrow^* = G_\downarrow \, \hat{v}_\downarrow^*$, then $\sigma$ is min-optimal for $\mathcal{A}$.*

The proof of Theorem 9.3.4 is the same as as that of Theorem 9.3.3, after replacing $\max$ with $\min$ and $\vee$ with $\wedge$. It can also be obtained by applying Theorem 9.3.3 to the duals of $\mathcal{A}$ and $\hat{\mathcal{A}}$.

### 9.3.3 Application

In this section we study a special case of an Epstein–Zin ADP $\mathcal{A} := (V, \{T_\sigma\})$ analyzed in § 9.2.3. The problem concerns optimal savings in the presence of an IID endowment shock. We will produce a subordinate ADP via a transformation reminiscent of the expected value transformation of an ordinary MDP in Example 9.3.2. This subordinate ADP is substantially easier to analyze.

To begin, we define

$$(T_\sigma v)(w, e) = \left\{ r(w, \sigma(w), e)^\alpha + \beta \left( \sum_{e'} v(\sigma(w), e')^\gamma \varphi(e') \right)^{\alpha/\gamma} \right\}^{1/\alpha}. \tag{9.24}$$

Here $w$ is current wealth, $s$ is savings and $e$ is an endowment shock which we take to be IID with common distribution $\varphi$ and range E. If $\mathsf{X} := \mathsf{W} \times \mathsf{E}$ and $V := (0, \infty)^{\mathsf{X}}$, then $\mathcal{A} = (V, \{T_\sigma\})$ is a special case of the ADP discussed in §9.2.3. Since $\beta$ is constant, we have $\mathcal{E}(\beta, Q, \theta)^{1/\theta} = \beta < 1$. Hence $\mathcal{A}$ is max-stable (Proposition 9.2.5).

Now consider the operator

$$(U_\sigma h)(w) = \left\{ \sum_e \{ r(w, \sigma(w), e)^\alpha + \beta h(\sigma(w))^\alpha \}^{\gamma/\alpha} \varphi(e) \right\}^{1/\gamma}, \tag{9.25}$$

where $h$ is an element of $(0, \infty)^{\mathsf{W}}$. If $F$ is defined at $v \in V$ by

$$(Fv)(w) = \left\{ \sum_e v(w, e)^\gamma \varphi(e) \right\}^{1/\gamma} \qquad (w \in \mathsf{W})$$

and $H := F(V) \subset (0, \infty)^{\mathsf{W}}$, then $\mathcal{U} = (H, \{U_\sigma\})$ is an ADP. Moreover, $\mathcal{U}$ is subordinate to $\mathcal{A}$, since, with $G_\sigma$ defined at $h \in H$ by

$$(G_\sigma h)(w, e) = \{r(w, \sigma(w), e)^\alpha + \beta h(\sigma(w))^\alpha\}^{1/\alpha} \qquad ((w, e) \in \mathsf{X}), \qquad (9.26)$$

we can see that both $F$ and $G_\sigma$ are order-preserving, that $T_\sigma$ in (9.24) is equal to $G_\sigma \circ F$, and that $U_\sigma$ in (9.25) is equal to $F \circ G_\sigma$.

The benefit of working with $\mathcal{U}$ is that $U_\sigma$ acts on functions that depend only on $w$ rather than on both $w$ and $e$ (as is the case for $T_\sigma$). These lower dimensional operations are significantly more efficient, even when the range $\mathsf{E}$ of $e$ is relatively small.

Since $\mathcal{U}$ is subordinate to $\mathcal{A}$, Theorem 9.3.3 implies that $\mathcal{U}$ is max-stable and we can obtain a max-optimal policy for $\mathcal{A}$ by finding the max-value function $h^*$ for $\mathcal{U}$ and then calculating a policy $\sigma$ obeying $G_\sigma h^* = G h^*$ (see (ii) in Theorem 9.3.3). By the definition of $G_\sigma$ in (9.26), this means that we solve for $\sigma$ satisfying

$$\sigma(w, e) \in \operatorname*{argmax}_{0 \leqslant s \leqslant w} \{r(w, s, e)^\alpha + \beta h(s)^\alpha\}^{1/\alpha} \qquad (9.27)$$

with $h = h^*$ at each $(w, e) \in \mathsf{X}$. To compute $h^*$, we can use Theorem 9.1.6, which tells us that Howard max-policy iteration converges to $h^*$ in finitely many steps. Summarizing this analysis, an optimal policy for $\mathcal{A}$ can be computed via Algorithm 9.1.

---

**Algorithm 9.1:** Solving $\mathcal{A}$ via $\mathcal{U}$

1   input $\sigma_0 \in \Sigma$, set $k \leftarrow 0$ and $\varepsilon \leftarrow 1$
2   **while** $\varepsilon > 0$ **do**
3     $h_k \leftarrow$ the fixed point of $U_{\sigma_k}$
4     $\sigma_{k+1} \leftarrow$ an $h_k$-max-greedy policy, satisfying

$$\sigma_{k+1}(w) \in \operatorname*{argmax}_{0 \leqslant s \leqslant w} \left\{ \sum_e \{r(w, s, e)^\alpha + \beta h(s)^\alpha\}^{\gamma/\alpha} \, \varphi(e) \right\}^{1/\gamma}$$

5     $\varepsilon \leftarrow \mathbb{1}\{\sigma_k \neq \sigma_{k+1}\}$
6     $k \leftarrow k + 1$
7   **end**
8   **return** the $\sigma$ in (9.27) with $h = h_k$

---

Figure 9.2 shows $w \mapsto \sigma^*(w, e)$ for two values of $e$ (smallest and largest) when $\sigma^*$ is the optimal policy, calculated using Algorithm 9.1. In the figure we set $r(w, s, e) = (w - s + e)$ and choose $\alpha$ and $\gamma$ to match the values used in Schorfheide et al. (2018). The full parameterization is shown in Listing 27. The two versions of B in the listing are used to evaluate the policy operators $T_\sigma$ and $U_\sigma$ respectively (with the correct method being selected by multiple dispatch in the calling code).

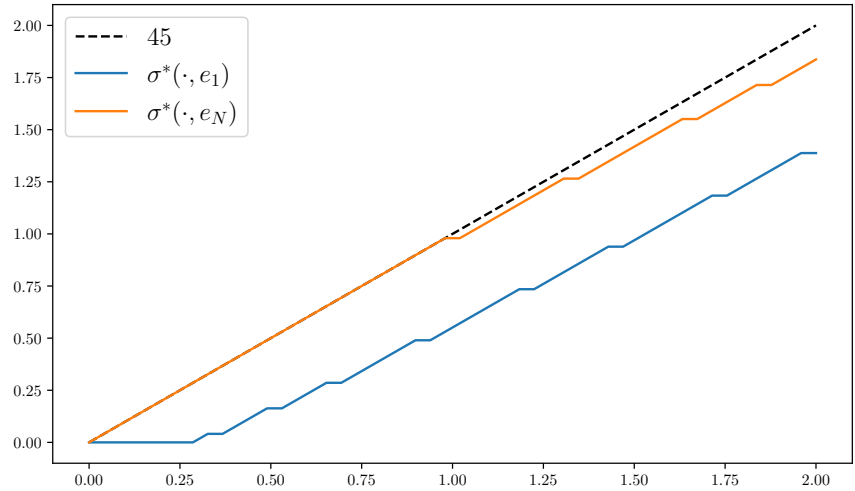

Figure 9.2: Optimal savings policy with Epstein–Zin preference

In Figure 9.3 we display the relative speed gain from using the lower-dimensional model $\mathcal{U}$ instead of $\mathcal{A}$ across multiple choices of |W| and |E|. The speed gain is the time required to solve an optimal policy for $\mathcal{A}$ using HPI applied to $\mathcal{A}$ (as in Theorem 9.1.6), divided by the time required to solve for the same optimal policy via Algorithm 9.1. The speed gain increases linearly in the size of E.

## 9.4 Chapter Notes

As indicated in notes for Chapter 8, our interest in abstract dynamic programming was inspired by Bertsekas (2022b). This chapter generalizes his framework by switching to a "completely abstract" setting based on analysis of self-maps on partially ordered space. The material here is based on Sargent and Stachurski (2023a). Earlier work on dynamic programming in a setting with no topology can be found in Kamihigashi (2014).

Readers who wish to see some motivation for the discussion of negative discounting in §8.3.5.2 can consult Loewenstein and Sicherman (1991), who found that the majority of workers they surveyed reported a preference for increasing wage profiles

```julia
using QuantEcon, Distributions, LinearAlgebra, IterTools

function create_ez_model(; ψ=1.97,    # elasticity of intertemp. substitution
                           β=0.96,    # discount factor
                           γ=-7.89,   # risk aversion parameter
                           n=80,      # size of range(e)
                           p=0.5,
                           e_max=0.5,
                           w_size=50, w_max=2)
    α = 1 - 1/ψ
    θ = γ / α
    b = Binomial(n - 1, p)
    φ = [pdf(b, k) for k in 0:(n-1)]
    e_grid = LinRange(1e-5, e_max, n)
    w_grid = LinRange(0, w_max, w_size)
    return (; α, β, γ, θ, φ, e_grid, w_grid)
end

"Action-value aggregator for the original model."
function B(i, j, k, v, model)
    (; α, β, γ, θ, φ, e_grid, w_grid) = model
    w, e, s = w_grid[i], e_grid[j], w_grid[k]
    value = -Inf
    if s <= w
        Rv = @views dot(v[k, :].^γ, φ)^(1/γ)
        value = ((w - s + e)^α + β * Rv^α)^(1/α)
    end
    return value
end

"Action-value aggregator for the subordinate model."
function B(i, k, h, model)
    (; α, β, γ, θ, φ, e_grid, w_grid) = model
    w, s = w_grid[i], w_grid[k]
    G(e) = ((w - s + e)^α + β * h[k]^α)^(1/α)
    value = s <= w ? dot(G.(e_grid).^γ, φ)^(1/γ) : -Inf
    return value
end
```

Listing 27: Epstein–Zin optimal savings model (`ez_model.jl`)

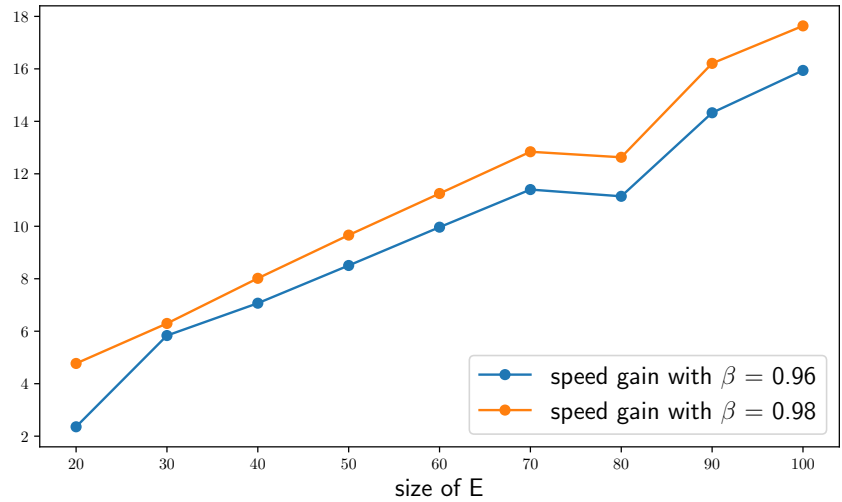

Figure 9.3: Speed gain from replacing $\mathcal{A}$ with subordinate model $\mathcal{U}$

over decreasing ones that yield the same undiscounted sum, even when it was pointed out that the latter could be used to construct a dominating consumption sequence. Loewenstein and Prelec (1991) obtained similar results. In summarizing their study, they argue that, in the context of the choice problems that they examined, "sequences of outcomes that decline in value are greatly disliked, indicating a negative rate of time preference" (Loewenstein and Prelec, 1991, p. 351).

In §8.3.5.2 we considered dynamic programs with negative discount rates. A more general treatment of such problems can be found in Kikuchi et al. (2021), which also shows how negative discount rate dynamic programs connect to static problems concerning equilibria in production networks and draws connections with Coase's theory of the firm.

# Chapter 10

# Continuous Time

Earlier chapters treated dynamics in discrete time. Now we switch to continuous time. We restrict ourselves to finite state spaces, where continuous time processes are pure jump processes. This allows us to provide a rigorous and self-contained treatment, while laying foundations for a treatment of general state problems.

## 10.1 Continuous Time Markov Chains

In this section we introduce continuous time Markov models. In §10.2, we will use them as components of continuous time Markov decision processes.

### 10.1.1 Background

In §3.1.1 we learned that if $(X_t) = (X_0, X_1, \ldots)$ is $P$-Markov, then the distributions $(\psi_t)$ of the state process obey $\psi_{t+1} = \psi_t P$ for all $t$. This update rule is a linear difference equation in distribution space, which in turn suggests that, once we switch to continuous time, distributions will evolve according to linear *differential* equations in distribution space.

This idea turns out to be correct. As such, we begin this chapter with some facts about linear differential equations.

### 10.1.1.1 Scalar Exponentials

Solutions to linear differential equations involve exponential functions. The real-valued **exponential function** can be defined by the power series

$$e^x =: \exp(x) := \sum_{k \geqslant 0} \frac{x^k}{k!} \qquad (x \in \mathbb{R}). \tag{10.1}$$

**Example 10.1.1.** If $u_t$ is the balance of a savings account that pays a continuously compounded interest rate $r$, then the balance evolves according to

$$\dot{u}_t := \frac{\mathrm{d}}{\mathrm{d}t} u_t = r u_t \quad \text{for all } t \geqslant 0 \quad \text{with initial balance } u_0 \text{ given.} \tag{10.2}$$

We understand (10.2) as a functional equation whose solution is an element $t \mapsto u_t$ of $C_1(\mathbb{R}_+, \mathbb{R})$, the set of continuously differentiable functions from $\mathbb{R}_+$ to $\mathbb{R}$, that satisfies (10.2). We claim that $u_t := e^{rt} u_0$ is the only solution to (10.2) in $C_1(\mathbb{R}_+, \mathbb{R})$. It is easy to check that this choice of $u_t$ obeys (10.2). As for uniqueness, suppose that $t \mapsto y_t$ is another solution in $C_1(\mathbb{R}_+, \mathbb{R})$, so that $\dot{y}_t = r y_t$ for all $t \geqslant 0$ and $y_0 = u_0$. Then

$$\frac{\mathrm{d}}{\mathrm{d}t} \left( y_t \, e^{-rt} \right) = \dot{y}_t \, e^{-rt} - r y_t \, e^{-rt} = r y_t \, e^{-rt} - r y_t \, e^{-rt} = 0,$$

so $t \mapsto y_t \, e^{-rt}$ is constant on $\mathbb{R}_+$, implying existence of a $c \in \mathbb{R}$ such that $y_t = c \, e^{rt}$ for all $t \geqslant 0$. Setting $t = 0$ and using the initial condition gives $c = u_0$. Hence, at any $t$, we have $y_t = e^{rt} u_0 = u_t$.

The continuous time system in Example 10.1.1 is closely related to the discrete time difference equation $u_{t+1} = e^r u_t$. Indeed, if we start at $u_0$, then the $t$-th iterate is $e^{rt} u_0$, so solutions agree at integer times. We can think of the continuous time system as one that interpolates between points in time of a corresponding discrete time system.

The exponential $e^\lambda$ of $\lambda = a + ib \in \mathbb{C}$ can also be defined via (10.1). From the identity $e^{ib} = \cos(b) + i \sin(b)$, we obtain

$$e^\lambda = e^{a+ib} = e^a (\cos(b) + i \sin(b)). \tag{10.3}$$

This equation will soon prove useful.

### 10.1.1.2 The Exponential Distribution

A random variable $W$ is said to be **exponentially distributed** with rate $\theta$, and we write $W \overset{d}{=} \mathrm{Exp}(\theta)$, when the counter CDF $G$ satisfies

$$G(t) := \mathbb{P}\{W > t\} = \mathrm{e}^{-\theta t} \qquad (t \geqslant 0).$$

Continuous time Markov chains have a close relationship with the exponential distribution, a fact that stems from its being the only distribution having the **memoryless** property

$$\mathbb{P}\{W > s + t \mid W > s\} = \mathbb{P}\{W > t\} \quad \text{for all } s, t > 0. \tag{10.4}$$

EXERCISE 10.1.1. Verify that (10.4) holds when $W \overset{d}{=} \mathrm{Exp}(\theta)$.

The memoryless property is special. For example, the probability that an individual human being lives 70 years from birth is not equal to the probability that he or she lives another 70 years conditional on having reached age 70. In fact the exponential distribution is the *only* memoryless distribution supported on the nonnegative reals:

**Lemma 10.1.1.** *If $W$ has counter CDF $G$ satisfying $0 < G(t) < 1$ for all $t > 0$, then the following statements are equivalent:*

(i) $W \overset{d}{=} \mathrm{Exp}(\theta)$ *for some $\theta > 0$.*

(ii) *$W$ satisfies the memoryless property in* (10.4).

*Proof.* Exercise (10.1.1) treats (i) $\Rightarrow$ (ii). As for (ii) $\Rightarrow$ (i), suppose (ii) holds. Then $G$ has three properties:

(a) $G$ is decreasing on $\mathbb{R}_+$ (as is any counter CDF),

(b) $0 < G(t) < 1$ for all $t > 0$, and

(c) $G(s + t) = G(s)G(t)$ for all $s, t > 0$.

From (a)–(c) we will show that

$$G(t) = G(1)^t \qquad \text{for all } t \geqslant 0. \tag{10.5}$$

This is sufficient to prove (i) because then $\theta := -\ln G(1)$ is a positive real number (by (b)) and, furthermore,

$$G(t) = \exp\{\ln[G(1)^t]\} = \exp\{\ln[G(1)]t\} = \exp(-\theta t).$$

To see that (10.5) holds, fix $m, n \in \mathbb{N}$. We can use (c) to obtain both $G(m/n) = G(1/n)^m$ and $G(1) = G(1/n)^n$. It follows that $G(m/n)^n = G(1/n)^{mn} = G(1)^m$ and, raising to the power of $1/n$, we get (10.5) when $t = m/n$.

The discussion so far confirms that (10.5) holds when $t$ is rational. So now take any $t \geqslant 0$ and rational sequences $(a_n)$ and $(b_n)$ converging to $t$ with $a_n \leqslant t \leqslant b_n$ for all $n$. By (a) we have $G(b_n) \leqslant G(t) \leqslant G(a_n)$ for all $n$, so $G(1)^{b_n} \leqslant G(t) \leqslant G(1)^{a_n}$. for all $n \in \mathbb{N}$. Taking the limit in $n$ completes the proof. $\qquad \square$

### 10.1.1.3 Extension to Matrices

The real exponential formula (10.1) extends to the **matrix exponential** via

$$e^A := I + A + \frac{A^2}{2!} + \cdots = \sum_{k \geqslant 0} \frac{A^k}{k!}, \tag{10.6}$$

where $A$ is any square matrix. As we will see, the matrix exponential plays a key role in the solution of vector-valued linear differential equations.

EXERCISE 10.1.2. Let $A$ be $n \times n$ and let $\| \cdot \|$ be the operator norm (see page 16). Show that (10.6) converges, in the sense that $\| \sum_{k=0}^{m} \frac{A^k}{k!} \|$ is bounded in $m$.

**Lemma 10.1.2** (Properties of the matrix exponential). *Let $A$ and $B$ be square matrices.*

(i) *If $A$ is diagonalizable with $A = PDP^{-1}$, then $e^A = Pe^D P^{-1}$.*

(ii) *If $A$ and $B$ commute (i.e. $AB = BA$), then $e^{A+B} = e^A e^B$.*

(iii) *If $m$ is any positive integer, then $e^{mA} = (e^A)^m$.*

(iv) *$\lambda$ is an eigenvalue of $A$ if and only if $e^\lambda$ is an eigenvalue of $e^A$.*

(v) *The function $\mathbb{R} \ni t \mapsto e^{tA}$ is differentiable in $t$, with*

$$\frac{\mathrm{d}}{\mathrm{d}t} e^{tA} = A e^{tA} = e^{tA} A. \tag{10.7}$$

(vi) *$e^{A^\top} = (e^A)^\top$.*

(vii) *The fundamental theorem of calculus holds, in the sense that*

$$e^{tA} - e^{sA} = \int_s^t e^{\tau A} A \, \mathrm{d}\tau \quad \text{for all } s \leqslant t. \tag{10.8}$$

In Lemma 10.1.2 and below, integration or differentiation of a vector- or matrix-valued function is carried out element by element. For example, to differentiate a matrix $B(t) = (b_{ij}(t))$ that depends on $t$, we form a new matrix by differentiating each element $b_{ij}(t)$ with respect to $t$. The integral $\int_a^b B(t)\, \mathrm{d}t$ is the matrix of integrals $\int_a^b b_{ij}(t)\, \mathrm{d}t$.

EXERCISE 10.1.3. Prove part (i) of Lemma 10.1.2.

The proof of part (ii) of Lemma 10.1.2 uses the definition of the exponential and the binomial formula. See, for example, Hirsh and Smale (1974). Part (iii) follows directly from part (ii). Part (iv) follows easily from part (i) when $A$ is diagonalizable (and can be proved more generally via the Jordan canonical form).

EXERCISE 10.1.4. Prove (v) of Lemma 10.1.2. A good starting point is to observe that, for any $t \in \mathbb{R}$,

$$\frac{\mathrm{d}}{\mathrm{d}t} \mathrm{e}^{tA} = \lim_{h \to 0} \frac{\mathrm{e}^{tA+hA} - \mathrm{e}^{tA}}{h} = \mathrm{e}^{tA} \lim_{h \to 0} \frac{\mathrm{e}^{hA} - I}{h}. \tag{10.9}$$

EXERCISE 10.1.5. Using Lemma 10.1.2, show that, for any $n \times n$ matrix $A$, the matrix $\mathrm{e}^A$ is invertible, with inverse $\mathrm{e}^{-A}$.

As for (vii), we are drawing an analogy with the fundamental theorem of calculus for scalar-valued functions, which states that $f(t) - f(s) = \int_s^t f'(\tau)\, \mathrm{d}\tau$ for all $s \leqslant t$, where $f'$ is the derivative of $f$.

EXERCISE 10.1.6. Prove part (vii) of Lemma 10.1.2. [Hint: use part (v).]

## 10.1.2 Continuous Time Flows

Next we study solutions of multivariate differential equations, with a focus on linear systems. These results lay foundations for our study of continuous time Markov dynamics in §10.1.3.

### 10.1.2.1 Continuous Time Dynamical Systems

Recall from §2.1.1 that a discrete dynamical system is a pair $(U, S)$, where $U$ is a set and $S$ is a self-map on $U$. Trajectories are sequences $(S^t u)_{t \geqslant 0} = (u, Su, S^2 u, \dots)$, where $u \in U$ is the initial condition. These ideas can be extended to continuous time by considering a pair $(U, (S_t)_{t \geqslant 0})$ where $U$ is any set and $S_t$ is a self-map on $U$ for each $t \in \mathbb{R}_+$. The interpretation is that if $u \in U$ is the current state of the system, then $S_t u$ will be the state after $t$ units of time.

**Example 10.1.2.** For the savings account in Example 10.1.1 with solution $u_t := e^{rt} u_0$, we can take $U = \mathbb{R}$ and $S_t u = e^{rt} u$. Then the state $S_t u$ at time $t$ is the balance at time $t$ associated with initial deposit $u$.

In general, to understand $(U, (S_t)_{t \geqslant 0})$ as a continuous time dynamical system, we require that (a) $S_0$ is the identity map, so that the state after zero units of time is just the initial condition, and (b) if we start at $u$, move forward to $u_s := S_s u$, and then move again to $S_t u_s$ after another $t$ units of time, the outcome should be the same as moving from $u$ to $S_{s+t} u$ directly. That is,

$$S_{s+t} = S_t \circ S_s \quad \text{for all } t, t' \geqslant 0.$$

This is the **semigroup property**.

One way that continuous time dynamical systems arise is via initial value problems. An **initial value problem** (IVP) in $\mathbb{R}^n$ consists of a differential equation $\dot{u}_t = f(u_t)$ paired with an initial condition $u_0 \in \mathbb{R}^n$, where $u_t \in \mathbb{R}^n$ and $f : \mathbb{R}^n \to \mathbb{R}^n$. Under suitable conditions on $f$, the solution $u_t := F(t, u_0)$ is uniquely defined for all $t \geqslant 0$, and, moreover,

$$F(0, u_0) = u_0 \quad \text{and} \quad F(s + t, u_0) = F(t, F(s, u_0)) \quad \text{for all } s, t \geqslant 0$$

(see, e.g., Hirsh and Smale (1974), Section 8.7). Hence $(S_t)_{t \geqslant 0}$ defined by $S_t u = F(t, u)$ satisfies the semigroup property and $(\mathbb{R}^n, (S_t)_{t \geqslant 0})$ is a continuous time dynamical system. The function $f$ is called the **vector field** of $(\mathbb{R}^n, (S_t)_{t \geqslant 0})$.

### 10.1.2.2 Linear Initial Value Problems

Given our interest in continuous time Markov chains and their connection to linear systems (see the comments at the start of §10.1.1), we focus primarily on linear differential equations. The next result discusses linear IVPs, illustrating the key role of the matrix exponential. In the statement, $A$ is $n \times n$ and both $\dot{u}_t$ and $u_t$ are column vectors in $\mathbb{R}^n$.

**Proposition 10.1.3.** *The unique solution of the $n$-dimensional IVP*

$$\dot{u}_t = A u_t, \qquad u_0 \in \mathbb{R}^n \text{ given} \tag{10.10}$$

*in the set of continuously differentialbe functions $t \mapsto u_t$ from $\mathbb{R}_+$ to $\mathbb{R}^n$ is*

$$u_t := \mathrm{e}^{tA} u_0 \qquad (t \geqslant 0). \tag{10.11}$$

(Here $\dot{u}_t := \mathrm{d}u_t/\mathrm{d}t$ is defined by differentiating the vector $u_t$ element-by-element, as discussed after Lemma 10.1.2.)

*Proof of Proposition 10.1.3.* That $u_t := \mathrm{e}^{tA} u_0$ solves (10.10) is immediate from Exercise 10.1.4. The proof of uniqueness is omitted, although the logic is very similar to the scalar case, which was discussed in Example 10.1.1. □

EXERCISE 10.1.7. Let $P$ be $n \times n$ and consider the IVP $\dot{\varphi}_t = \varphi_t P$ and $\varphi_0$ given, where each $\varphi_t$ is a *row* vector in $\mathbb{R}^n$. Prove that this IVP has the unique solution $\varphi_t := \varphi_0 \, \mathrm{e}^{tP}$.

Proposition 10.1.3 motivates us to study flows of the form

$$t \mapsto u_t, \quad u_t = \mathrm{e}^{tA} u_0 \qquad (t \geqslant 0) \tag{10.12}$$

where $A$ is $n \times n$, $u_0$ is a vector in $\mathbb{R}^n$ indicating the initial condition, and $u_t$ is the "state" of the system at time $t$. Figure 10.1 shows an example when

$$A = \begin{pmatrix} -2.0 & -0.4 & 0 \\ -1.4 & -1.0 & 2.2 \\ 0.0 & -2.0 & -0.6 \end{pmatrix}. \tag{10.13}$$

### 10.1.2.3  Stability in the Diagonalizable Case

For an exponential flow such as (10.12), a key question is whether or not $u_t \to 0$ as $t \to \infty$. (This will matter when we try to evaluate lifetime rewards over an infinite horizon in continuous time.) Rather than analyze these issues at every possible $u_0$, we directly consider the matrix-valued flow $t \mapsto \mathrm{e}^{tA}$ and study whether $\mathrm{e}^{tA} \to 0$.

The case where $A$ is diagonalizable provides a good starting point. Suppose $A = P^{-1} D P$ with $D = \mathrm{diag}_j(\lambda_j)$ containing the eigenvalues of $A$. Then, by Lemma 10.1.2, for any $t \geqslant 0$, we have

$$\mathrm{e}^{tA} = \mathrm{e}^{tP^{-1}DP} = P^{-1} \mathrm{e}^{tD} P. \tag{10.14}$$

EXERCISE 10.1.8. Prove that $\mathrm{e}^{tD} = \mathrm{diag}(\mathrm{e}^{t\lambda_1}, \ldots, \mathrm{e}^{t\lambda_n})$.

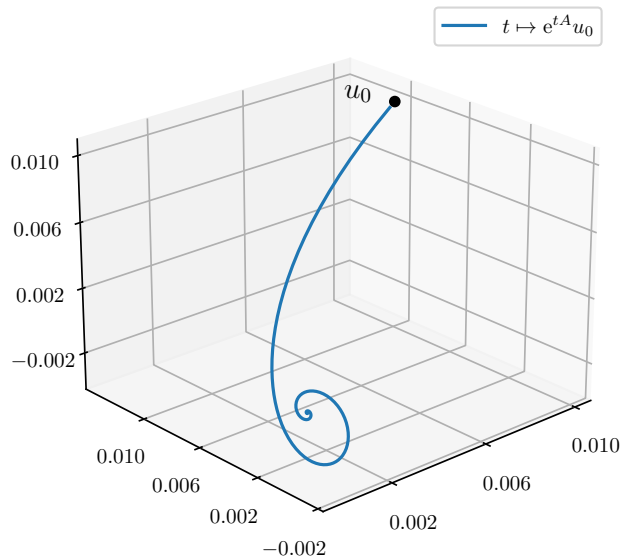

Figure 10.1: Exponential flow $t \mapsto \mathrm{e}^{tA} u_0$ starting from $u_0 \in \mathbb{R}^3$

Exercise 10.1.8 and (10.14) tell us that the long run dynamics of $\mathrm{e}^{tA}$ are determined by the scalar flows $t \mapsto \mathrm{e}^{t\lambda_j}$. How does $\mathrm{e}^{t\lambda}$ evolve over time when $\lambda \in \mathbb{C}$?

To answer this question we write $\lambda = a + ib$ and apply (10.3) to obtain

$$\mathrm{e}^{t\lambda} = \mathrm{e}^{ta}(\cos(tb) + i\sin(tb)).$$

This equation tells us that

$$\mathrm{e}^{t\lambda} \to 0 \text{ as } t \to \infty \quad \text{if and only if} \quad \mathrm{Re}\,\lambda < 0, \tag{10.15}$$

where $\mathrm{Re}\,\lambda$ is the **real part** of $\lambda$ (i.e., if $\lambda = a + ib$, then $\mathrm{Re}\,\lambda = a$).

From this analysis, we conclude that, when $A$ is diagonalizable, we have $\mathrm{e}^{tA} \to 0$ if and only if $\mathrm{Re}\,\lambda_j < 0$ for all $\lambda_j \in \sigma(A)$, where $\sigma(A)$ denotes the set of all eigenvalues (the **spectrum**) of $A$. Another way to put this is that $\mathrm{e}^{tA} \to 0$ if and only if $s(A) < 0$, where

$$s(A) := \max_{\lambda \in \sigma(A)} \mathrm{Re}\,\lambda, \tag{10.16}$$

is the **spectral bound** of $A$.

As the preceding analysis suggests, the spectral bound plays a key role in the asymptotics of exponential flows, just as a spectral radius governs asymptotics of trajectories of linear maps (see, e.g., Exercise 1.2.11 on page 19). The next section expands on this analysis, while dropping the assumption that $A$ is diagonalizable.

### 10.1.2.4 The General Case

Let $A$ be any square matrix. In the following statement about a spectral bound, $\| \cdot \|$ is the operator norm defined in §1.2.1.4.

**Lemma 10.1.4.** *If $\tau > 0$, then $\tau s(A) = s(\tau A)$. Moreover,*

$$e^{s(A)} = \rho(e^A) \quad and \quad s(A) = \lim_{t \to \infty} \frac{1}{t} \ln \|e^{tA}\|. \tag{10.17}$$

EXERCISE 10.1.9. Confirm that $\tau s(A) = s(\tau A)$ when $\tau > 0$.

EXERCISE 10.1.10. Prove the first equality in (10.17).

EXERCISE 10.1.11. The second equality in (10.17) is reminiscent of Gelfand's lemma (see page 19). Confirm that it holds when the limit is taken over $t \in \mathbb{N}$.

(The second equality in (10.17) also holds when the limit is taken over $t \in \mathbb{R}_+$. See, for example, Engel and Nagel (2006).)

The next theorem is a key stability result for exponential flows. Among other things, it extends to arbitrary $A$ the finding that $s(A) < 0$ is necessary and sufficient for stability.

**Theorem 10.1.5.** *For any square matrix $A$, the following statements are equivalent:*

(i) $s(A) < 0$.

(ii) $\|e^{tA}\| \to 0$ as $t \to \infty$.

(iii) *There exist $M, \omega > 0$ such that $\|e^{tA}\| \leqslant M e^{-t\omega}$ for all $t \geqslant 0$.*

(iv) $\int_0^\infty \|e^{tA} u_0\|^p \, dt < \infty$ for all $p \geqslant 1$ and $u_0 \in \mathbb{R}^n$.

A full proof of Theorem 10.1.5 in a general setting can be found in §V.II of Engel and Nagel (2006).

Theorem 10.1.5 tells us that the flow $t \mapsto e^{tA} u_0$ converges to the origin at an exponential rate if and only if $s(A) < 0$. The equivalence of (i) and (ii) was proved for the diagonalizable case in §10.1.2.3. It can be viewed as the continuous time analog of $\|B^k\| \to 0$ if and only if $\rho(B) < 1$ (see Exercise 1.2.11 on page 19).

EXERCISE 10.1.12. Prove that (i) implies (ii) without assuming that $A$ is diagonalizable. In addition, prove that (iii) implies (iv).

### 10.1.2.5 Semigroup Terminology

Advanced treatments of continuous time systems often begin with operator semigroups. Let's briefly describe these and connect them to things we have studied earlier. (If you prefer to skip this section on first reading, you can move to the next one after noting that, given an $n \times n$ matrix $A$, the family $(S_t)_{t \geqslant 0} = (e^{tA})_{t \geqslant 0}$ is called an **exponential semigroup** and that $A$ is called the **infinitesimal generator** of the semigroup.)

Let $\mathsf{X}$ be a finite set and let $(S_t)_{t \geqslant 0}$ be a subset of $\mathcal{L}(\mathbb{R}^{\mathsf{X}})$ indexed by $t \in \mathbb{R}_+$. The family $(S_t)_{t \geqslant 0}$ is called an **operator semigroup** on $\mathbb{R}^{\mathsf{X}}$ if

(i) $S_0 = I$, where $I$ is the identity,

(ii) $S_{t+t'} = S_t \circ S_{t'}$, and

(iii) $t \mapsto S_t$ is continuous as a map from $\mathbb{R}_+$ to $\mathcal{L}(\mathbb{R}^{\mathsf{X}})$.

In essence, an operator semigroup on $\mathbb{R}^{\mathsf{X}}$ is a continuous time dynamical system $(\mathbb{R}^{\mathsf{X}}, (S_t)_{t \geqslant 0})$ where each $S_t$ is a linear operator mapping $\mathbb{R}^{\mathsf{X}}$ to itself.

**Example 10.1.3** (Exponential semigroup). Fix $A$ in $\mathcal{L}(\mathbb{R}^{\mathsf{X}})$ and let $(S_t)_{t \geqslant 0}$ be defined by $S_t = e^{tA}$. Then $(S_t)_{t \geqslant 0}$ is an operator semigroup on $\mathbb{R}^{\mathsf{X}}$. To verify this we take $\mathsf{X} = \{x_1, \ldots, x_n\}$ and view the operator $S_t$ and $A$ as $n \times n$ matrices. The operator semigroup properties now follow directly from Lemma 10.1.2. For example, $S_t$ is continuous in $t$ because it is differentiable in $t$, by (v) of Lemma 10.1.2.

The operator semigroup perspective is important because it extends in a natural way to settings where $\mathsf{X}$ is not finite, in which case we replace the finite-dimensional set $\mathbb{R}^{\mathsf{X}}$ with some (typically infinite-dimensional) class of functions $\mathcal{B} \subset \mathbb{R}^{\mathsf{X}}$, and each $S_t$ becomes a linear operator mapping $\mathcal{B}$ into itself. At this level of generality, $S_t u$ can be the solution to a partial differential equation, or a stochastic differential equation (see, e.g., Engel and Nagel (2006) or Applebaum (2019)). Operator semigroup theory offers an elegant and powerful framework for handling such systems.

For operator semigroups in general settings we often have no analytical expressions for $S_t$. This situation is like the one we encountered in the continuous time system in §10.1.2.1, where $\dot{u}_t = f(u_t)$ and $f$ is potentially nonlinear. When no analytical solution $u_t$ exists, analyzing the dynamics requires us to try to infer its properties from the vector field $f$, so that $f$ becomes the primary focus of analysis.

A natural question, then, is, given an operator semigroup $(S_t)_{t \geqslant 0}$ on $\mathcal{L}(\mathbb{R}^\mathsf{X})$, does there always exist a "vector field" type object that "generates" $(S_t)_{t \geqslant 0}$? When $\mathsf{X}$ is finite, the answer is affirmative. This object, denoted below by $A$, is called the **infinitesimal generator** of the semigroup and is defined by

$$A = \lim_{t \downarrow 0} \frac{S_t - S_0}{t} = \lim_{t \downarrow 0} \frac{S_t - I}{t} \tag{10.19}$$

At $u \in U$, the vector $Au$ indicates the instantaneous change in the state.

More precisely, when $\mathsf{X}$ is finite, we have:

**Proposition 10.1.6.** *If $(S_t)_{t \geqslant 0}$ is an operator semigroup on $\mathbb{R}^\mathsf{X}$, then*

(i) *there exists an $A \in \mathcal{L}(\mathbb{R}^\mathsf{X})$ such that $S_t = \mathrm{e}^{tA}$ for all $t \geqslant 0$, and*

(ii) *$A$ is the infinitesimal generator of $(S_t)_{t \geqslant 0}$.*

Semigroups of the form described in Proposition 10.1.6 are called **exponential semigroups** (or "uniformly continuous" semigroups).

A full proof of Proposition 10.1.6 can be found in the discussion of Theorem 2.12 of Engel and Nagel (2006). The results are not surprising, since the main claim is that, in finite dimensions, solutions to linear differential equations have exponential form. The fact that $A$ is the infinitesimal generator of the semigroup $(S_t)_{t \geqslant 0} = (\mathrm{e}^{tA})_{t \geqslant 0}$ follows from Lemma 10.1.2, which gives

$$\lim_{t \downarrow 0} \frac{S_t - S_0}{t} = \lim_{t \downarrow 0} \frac{\mathrm{e}^{tA} - \mathrm{e}^0}{t} = \frac{\mathrm{d}}{\mathrm{d}t} \mathrm{e}^{tA} \Big|_{t=0} = A\mathrm{e}^{0A} = A.$$

The preceding discussion places our analysis in a wider context. To practice our new terminology, we can restate (i) $\iff$ (ii) from Theorem 10.1.5 by saying that the exponential semigroup $(S_t)_{t \geqslant 0} = (\mathrm{e}^{tA})_{t \geqslant 0}$ converges to zero if and only if the spectral bound of its infinitesimal generator is negative.

### 10.1.3 Markov Semigroups

Having studied multivariate linear dynamics, we are now ready to specialize to the Markov case, where dynamics evolve in distribution space. For the most part we now switch to operator-theoretic notation, where $\mathsf{X}$ is a finite set with $n$ elements, and an $n \times n$ matrix is identified with a linear operator on $\mathcal{L}(\mathbb{R}^\mathsf{X})$. As emphasized in §2.3.3.1, this is merely a change in terminology, and all preceding results for matrices extend directly to linear operators.

### 10.1.3.1 Intensity Matrices

If $(X_t)_{t \geqslant 0}$ is $P$-Markov on $\mathsf{X}$ for some $P \in \mathcal{M}(\mathbb{R}^\mathsf{X})$, then the marginal distributions of $(X_t)_{t \geqslant 0}$ evolve according to the linear difference system $\psi_{t+1} = \psi_t P$ (see §3.1.1). We now seek a continuous time analog in the form of a linear differential equation.

To this end we call $Q \in \mathcal{L}(\mathbb{R}^\mathsf{X})$ an **intensity operator** or **intensity matrix**[1] when

$$Q(x, x') \geqslant 0 \text{ whenever } x \neq x' \quad \text{and} \quad \sum_{x'} Q(x, x') = 0 \text{ for all } x \in \mathsf{X}. \qquad (10.20)$$

Let

$$\mathcal{I}(\mathbb{R}^\mathsf{X}) = \text{ the set of all intensity operators in } \mathcal{L}(\mathbb{R}^\mathsf{X}).$$

**Example 10.1.4.** The matrix

$$Q := \begin{pmatrix} -2 & 1 & 1 \\ 0 & -1 & 1 \\ 2 & 1 & -3 \end{pmatrix}$$

is an intensity matrix, since off-diagonal terms are nonnegative and rows sum to zero.

Consider the IVP

$$\dot{\psi}_t(x') = \sum_{x'} Q(x, x') \psi_t(x) \qquad (t \geqslant 0, \ x' \in \mathsf{X}),$$

which we can also write as

$$\dot{\psi}_t = \psi_t Q, \qquad \psi_0 \in \mathcal{D}(\mathsf{X}) \text{ given.} \qquad (10.21)$$

when $\psi_t$ and $\dot{\psi}_t$ are understood to be row vectors. We say that $\mathcal{D}(\mathsf{X})$ is **invariant** for the IVP (10.21) if the solution $(\psi_t)_{t \geqslant 0}$ remains in $\mathcal{D}(\mathsf{X})$ for all $t \geqslant 0$.

In view of Proposition 10.1.3, we can rephase this by stating that $\mathcal{D}(\mathsf{X})$ is invariant for (10.21) whenever

$$\psi_0 \in \mathcal{D}(\mathsf{X}) \quad \Longrightarrow \quad \psi_0 \, e^{tQ} \in \mathcal{D}(\mathsf{X}) \text{ for all } t \geqslant 0. \qquad (10.22)$$

Our key result for this section shows the central role of intensity matrices:

**Proposition 10.1.7.** *Fix $Q \in \mathcal{L}(\mathbb{R}^\mathsf{X})$ and set $P_t := e^{tQ}$ for each $t \geqslant 0$. The following statements are equivalent:*

---

[1]Other names for intensity matrices include "$Q$-matrices" (which is fine until you need to use another symbol), "Kolmogorov matrices", and "infinitesimal stochastic matrices."

(i) $Q \in \mathcal{I}(\mathbb{R}^{\mathsf{X}})$.

(ii) $P_t \in \mathcal{M}(\mathbb{R}^{\mathsf{X}})$ *for all $t \geqslant 0$.*

(iii) *the set of distributions $\mathcal{D}(\mathsf{X})$ is invariant for the IVP (10.21).*

Proposition 10.1.7 tells us that the set $\mathcal{I}(\mathbb{R}^{\mathsf{X}})$ coincides with the set of continuous time (and time-homogeneous) Markov models on $\mathsf{X}$. Any specification outside this class fails to generate flows in distribution space. The proof is completed in several steps below.

For Exercises 10.1.13–10.1.15, $Q \in \mathcal{I}(\mathbb{R}^{\mathsf{X}})$ and $P_t := \mathrm{e}^{tQ}$.

EXERCISE 10.1.13. Show that $P_t \mathbb{1} = \mathbb{1}$ for all $t \geqslant 0$.

EXERCISE 10.1.14. Set $\theta := \max_{x \in \mathsf{X}} |Q(x, x)|$ and $K := I + \frac{1}{\theta} Q$, where $I$ is the $n \times n$ identity. (If $\theta = 0$, then set $K := I$.) Prove that $K$ is a stochastic matrix and $Q = \theta(K - I)$.

EXERCISE 10.1.15. Using the representation for $Q$ obtained in Exercise 10.1.14 and the definition of the matrix exponential, show that $P_t$ is nonnegative for all $t \geqslant 0$.

For the proof of Proposition 10.1.7, we have now shown that (i) implies (ii). Evidently (ii) implies (iii), because if $\psi_0 \in \mathcal{D}$ and $\psi_t = \psi_0 P_t$ where $P_t$ is stochastic, then $\psi_t \in \mathcal{D}(\mathsf{X})$. Hence it remains only to show that (iii) implies (i).

EXERCISE 10.1.16. Let $Q$ be $n \times n$ and assume (iii). Fix $x \in \mathsf{X}$. By (iii) we have $\delta_x \mathrm{e}^{tQ} \mathbb{1} = 1$ for all $t \geqslant 0$, where $\mathbb{1}$ is a vector of ones. Show that $\sum_{x'} Q(x, x') = 0$ using this identity.

EXERCISE 10.1.17. Prove that $Q(x, x') \geqslant 0$ when (iii) holds and $x, x' \in \mathsf{X}$ with $x \neq x'$.

Returning to Proposition 10.1.7, the last two exercises confirm that (iii) implies (i). The proof is now complete.

### 10.1.3.2 Interpretation

The previous section covered the formal relationship between intensity matrices and Markov operators. Let's now discuss the connection more informally, in order to build intuition.

To this end, let $(X_t)_{t \geqslant 0}$ be $P_h$-Markov in discrete time. Here $h > 0$ is the length of the time step. We write the corresponding distribution sequence $\psi_{t+h} = \psi_t P_h$ in terms of change per unit of time, as in

$$\frac{\psi_{t+h} - \psi_t}{h} = \psi_t \frac{P_h - I}{h} \quad \text{where} \quad I \text{ is the } n \times n \text{ identity.} \tag{10.25}$$

Continuous time dynamics are obtained by taking the limit as $h \downarrow 0$. If we define

$$Q := \lim_{h \downarrow 0} \frac{P_h - I}{h}, \tag{10.26}$$

and assume that limits exist, then (10.25) becomes (10.21).

What properties does $Q$ have? Inspecting (10.26) implies

$$Q(x, x') \approx \frac{P_h(x, x') - \mathbb{1}\{x = x'\}}{h} \tag{10.27}$$

when $h$ is small and positive.

EXERCISE 10.1.18. Prove that, when $h > 0$ and $P_h$ is stochastic, the matrix on the right-hand side of (10.27) is an intensity matrix.

EXERCISE 10.1.19. To formalize (10.27), use the expression for the matrix exponential in (10.6) to prove that if $P_t = e^{tQ}$, then

$$P_h(x, x') = h\, Q(x, x') + o(h) \quad \text{whenever} \quad x \neq x'. \tag{10.28}$$

Equation (10.28) tells us that $Q(x, x')$ represents the instantaneous rate of flow out of state $x$ and into state $x'$. The on-diagonal value $P_h(x, x)$ just balances the off-diagonal probabilities.

### 10.1.3.3  Markov Semigroups

Fix $Q \in \mathcal{I}(\mathbb{R}^{\mathsf{X}})$. In the terminology of §10.1.2.5, the family of operators $(P_t)_{t \geqslant 0} = (e^{tQ})_{t \geqslant 0} \subset \mathcal{M}(\mathbb{R}^{\mathsf{X}})$ that solves $\dot\psi_t = \psi_t Q$ (see (10.22)) is an exponential semigroup. Since each $P_t$ is in $\mathcal{M}(\mathbb{R}^{\mathsf{X}})$, it is also called the **Markov semigroup** generated by $Q$. It satisfies the operator semigroup property $P_{s+t} = P_s P_t$ for all $s, t \geqslant 0$, which can be written more explicitly as

$$P_{s+t}(x, x') = \sum_z P_s(x, z) P_t(z, x') \qquad (s, t \geqslant 0, \ x, x' \in \mathsf{X}). \tag{10.29}$$

In the present setting, (10.29) is called the (continuous time) **Chapman–Kolmogorov equation**. It states that the probability of moving from $x$ to $x'$ over $s + t$ units of time equals the probability of moving from $x$ to $z$ over $s$ units of time, and then $z$ to $x'$ over $t$ units of time, summed over all $z$.

Again following the terminology in §10.1.2.5, the intensity matrix $Q$ that defines $(P_t)_{t \geqslant 0} = (\mathrm{e}^{tQ})_{t \geqslant 0}$ is also called the infinitesimal generator of $(P_t)_{t \geqslant 0}$.

From Lemma 10.1.2, the derivative of $\mathrm{e}^{tQ}$ is $Q\mathrm{e}^{tQ} = \mathrm{e}^{tQ}Q$. We can write this as

- $\dot{P}_t = QP_t$, which is called the **Kolmogorov backward equation**, and

- $\dot{P}_t = P_tQ$, which is called the **Kolmogorov forward equation**.

We can work in the other direction as well: if we can establish that a function $t \mapsto P_t$ from $\mathbb{R}_+$ to $\mathcal{L}(\mathbb{R}^{\mathsf{X}})$ satisfies either one of these equations, then $(P_t)_{t \geqslant 0}$ is a Markov semigroup with infinitesimal generator $Q$. The next proposition gives details.

**Proposition 10.1.8.** *Let $Q$ be an intensity matrix. If $t \mapsto P_t$ is a differentiable function from $\mathbb{R}_+$ to $\mathcal{L}(\mathbb{R}^{\mathsf{X}})$ such that $P_0 = I$ and either*

(i) $\dot{P}_t = QP_t$ *or*

(ii) $\dot{P}_t = P_tQ$,

*then $P_t = \mathrm{e}^{tQ}$ for all $t \geqslant 0$.*

Proposition 10.1.8 is a version of our result for linear IVPs in Proposition 10.1.3, except that the IVP is now defined in operator space, rather than vector space.

## 10.1.4   Continuous Time Markov Chains

We have discussed the one-to-one connection between intensity matrices and Markov semigroups, and how the dynamics generated by Markov semigroups trace out distribution flows. Let's now connect these objects to continuous time Markov chains.

### 10.1.4.1 Definition

Let $C(\mathbb{R}_+, \mathsf{X})$ be the set of right-continuous functions from $\mathbb{R}_+$ to $\mathsf{X}$ and let $(P_t)_{t \geqslant 0}$ be a Markov semigroup generated by some $Q \in \mathcal{I}(\mathbb{R}^\mathsf{X})$. A **continuous time Markov chain** generated by $(P_t)_{t \geqslant 0}$ is a random function $(X_t)_{t \geqslant 0}$ that takes values in $C(\mathbb{R}_+, \mathsf{X})$ and satisfies

$$\mathbb{P}\{X_{s+t} = x' \mid \mathcal{F}_s\} = P_t(X_s, x') \qquad \text{for all } s, t \geqslant 0 \text{ and } x' \in \mathsf{X}, \tag{10.30}$$

where $\mathcal{F}_s := (X_\tau)_{0 \leqslant \tau \leqslant s}$ is the history of the process up to time $s$. To update from time $s$ to time $t$ given this history, we simply take the last value $X_s$ and update using $P_t$. Conditioning on $X_s = x$, we get

$$P_t(x, x') = \mathbb{P}\{X_{s+t} = x' \mid X_s = x\} \qquad (s, t \geqslant 0, \ x, x' \in \mathsf{X}).$$

Mirroring terminology for discrete chains from §3.1.1.1, we will call a continuous time Markov chain $(X_t)_{t \geqslant 0}$ $Q$-**Markov** when (10.30) holds and $Q$ is the infinitesimal generator of $(P_t)_{t \geqslant 0}$.

In what follows, $\mathbb{P}_x$ and $\mathbb{E}_x$ denote probabilities and expectations conditional on $X_0 = x$. Given $h \in \mathbb{R}^\mathsf{X}$, we have

$$\mathbb{E}_x h(X_t) = \sum_{x'} P_t(x, x') h(x') =: (P_t h)(x).$$

This expression mirrors the discrete time case discussed in §3.2.1.1.

### 10.1.4.2 A Jump Chain Construction

In §10.1.4.1 we defined a continuous time Markov chain. In this section we describe a standard method for constructing one by using three components:

(i) an initial condition $\psi \in \mathcal{D}(\mathsf{X})$,

(ii) a **jump matrix** $\Pi \in \mathcal{M}(\mathbb{R}^\mathsf{X})$, and

(iii) a **rate function** $\lambda$ mapping $\mathsf{X}$ to $(0, \infty)$.

The process $(X_t)$ starts at state $x$, which is drawn from $\psi$, waits there for an exponential time $W$ with rate $\lambda(x)$, and then updates to a new state $x'$ drawn from $\Pi(x, \cdot)$. We take $x'$ as the new state for the process and repeat.

These ideas are restated in Algorithm 10.1. In the algorithm, $(W_k)$ and $(Y_k)$ are drawn independently. The process $(W_k)$ is called the sequence of **holding times** or

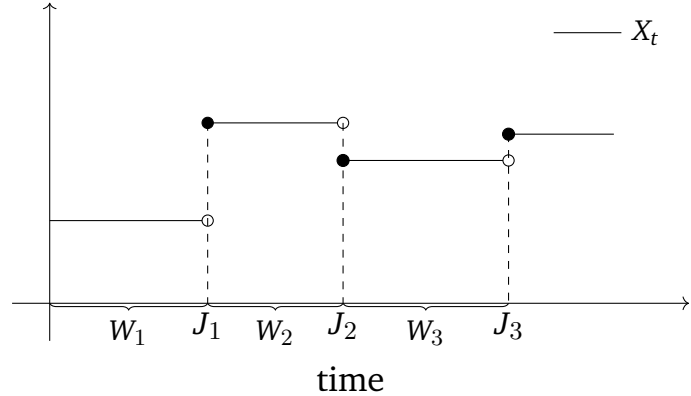

Figure 10.2: A jump chain sample path

**wait times**, the sums $J_k = \sum_{i=1}^{k} W_i$ are called the **jump times** and $(Y_k)$ is called the **embedded jump chain**. The jumps and the process $(X_t)_{t \geqslant 0}$ are illustrated in Figure 10.2.

---

**Algorithm 10.1:** Jump chain algorithm

**1** draw $Y_0$ from $\psi$, set $J_0 = 0$ and $k = 1$
**2 while** $t < \infty$ **do**
**3**     draw $W_k$ independently from $\text{Exp}(\lambda(Y_{k-1}))$
**4**     $J_k \leftarrow J_{k-1} + W_k$
**5**     $X_t \leftarrow Y_{k-1}$ for all $t$ in $[J_{k-1}, J_k)$
**6**     draw $Y_k$ from $\Pi(Y_{k-1}, \cdot)$
**7**     $k \leftarrow k + 1$
**8 end**

---

Let $I \in \mathcal{L}(\mathbb{R}^\mathsf{X})$ be the identity matrix, so $I(x, x') = \mathbb{1}\{x = x'\}$, and define $Q \in \mathcal{L}(\mathbb{R}^\mathsf{X})$ via

$$Q(x, x') = \lambda(x)(\Pi(x, x') - I(x, x')) \qquad (x, x' \in \mathsf{X}) \tag{10.31}$$

It is easy to verify that $Q$ is an intensity matrix. In fact $Q$ is the intensity matrix for the Markov semigroup associated with the process generated by Algorithm 10.1. For $x \neq x'$, it tells us that probability flows from $x$ to $x'$ at rate $\lambda(x)\Pi(x, x')$, which is the rate of leaving $x$ times the rate of moving from $x$ to $x'$. The next result formalizes these ideas.

**Proposition 10.1.9.** *The process* $(X_t)_{t \geqslant 0}$ *generated by Algorithm 10.1 is Q-Markov.*

To prove Proposition 10.1.9 we take $(X_t)_{t \geqslant 0}$ to be as in the statement of the proposition and define $(P_t)_{t \geqslant 0}$ by $P_t(x, x') = \mathbb{P}_x\{X_t = x'\}$ for all $x, x' \in \mathsf{X}$. The proof uses the following steps:

(i) Obtain an integral equation that $(P_t)_{t \geqslant 0}$ must satisfy.

(ii) Differentiate to obtain the Kolmogorov backward equation $\dot{P}_t = QP_t$.

(iii) Solve this differential equation to obtain $P_t = e^{tQ}$ for all $t$.

Here is the first step. In the statement, $\Pi P_{t-\tau}$ is the matrix product of $\Pi$ and $P_{t-\tau}$, while the equation in (10.32) is sometimes called the **integrated Kolmogorov backward equation**.

**Lemma 10.1.10.** *For all $t \geqslant 0$ and $x, x'$ in $\mathsf{X}$, the operator family $(P_t)_{t \geqslant 0}$ satisfies*

$$P_t(x, x') = e^{-t\lambda(x)} I(x, x') + \lambda(x) \int_0^t (\Pi P_{t-\tau})(x, x') e^{-\tau \lambda(x)} d\tau \tag{10.32}$$

*Proof.* Conditioning on $X_0 = x$, we have

$$P_t(x, x') := \mathbb{P}_x\{X_t = x'\} = \mathbb{P}_x\{X_t = x', \ J_1 > t\} + \mathbb{P}_x\{X_t = x', \ J_1 \leqslant t\}. \tag{10.33}$$

Regarding the first term on the right hand side of (10.33), we have

$$\mathbb{P}_x\{X_t = x', \ J_1 > t\} = I(x, x')P\{J_1 > t\} = I(x, x')e^{-t\lambda(x)}. \tag{10.34}$$

For the second term on the right hand side of (10.33), we obtain

$$\mathbb{P}_x\{X_t = x', \ J_1 \leqslant t\} = \mathbb{E}_x\left[\mathbb{1}\{J_1 \leqslant t\}\mathbb{P}_x\{X_t = x' \mid W_1, Y_1\}\right] = \mathbb{E}_x\left[\mathbb{1}\{J_1 \leqslant t\}P_{t-J_1}(Y_1, x')\right].$$

Evaluating the expectation and using the independence of $J_1$ and $Y_1$, this becomes

$$\mathbb{P}_x\{X_t = x', \ J_1 \leqslant t\} = \int_0^\infty \mathbb{1}\{\tau \leqslant t\} \sum_z \Pi(x, z) P_{t-\tau}(z, x') \lambda(x) e^{-\tau \lambda(x)} d\tau$$

$$= \lambda(x) \int_0^t \sum_z \Pi(x, z) P_{t-\tau}(z, x') e^{-\tau \lambda(x)} d\tau.$$

Combining this result with (10.33) and (10.34) gives (10.32). $\qquad\square$

Differentiating the integrated Kolmogorov backward equation produces the Kolmogorov backward equation:

**Lemma 10.1.11.** *If $(P_t)_{t \geqslant 0}$ satisfies (10.32), then $P_0 = I$ and $\dot{P}_t = QP_t$ for all $t \geqslant 0$*

*Proof.* The claim that $P_0 = I$ is obvious. For the second claim, one can easily verify that, when $f$ is a differentiable function and $\alpha > 0$, we have

$$g(t) = e^{-t\alpha} f(t) \quad \Longrightarrow \quad g'(t) = e^{-t\alpha} f'(t) - \alpha g(t) \tag{10.35}$$

Note also that, with the change of variable $s = t - \tau$, we can rewrite (10.32) as

$$P_t(x, x') = e^{-t\lambda(x)} \left\{ I(x, x') + \lambda(x) \int_0^t (\Pi P_s)(x, x') e^{s\lambda(x)} ds \right\}. \tag{10.36}$$

Applying (10.35) produces

$$P_t'(x, x') = e^{-t\lambda(x)} \left\{ \lambda(x)(\Pi P_t)(x, x') e^{t\lambda(x)} \right\} - \lambda(x) P_t(x, x').$$

Rearranging yields $\dot{P}_t(x, x') = \lambda(x)[(\Pi - I)P_t](x, x')$, which is identical to $\dot{P}_t = QP_t$. □

*Proof of Proposition 10.1.9.* Proposition 10.1.9 follows directly from Lemma 10.1.10 and Lemma 10.1.11, combined with Proposition 10.1.8. □

### 10.1.4.3 Application: Inventory Dynamics

Let $X_t$ be a firm's inventory at time $t$. When current stock is $x > 0$, customers arrive at rate $\lambda(x)$, so the wait time for the next customer is an independent draw from the $\text{Exp}(\lambda(x))$ distribution; $\lambda$ maps $\mathsf{X}$ to $(0, \infty)$.

The $k$-th customer demands $U_k$ units, where each $U_k$ is an independent draw from a fixed distribution $\varphi$ on $\mathbb{N}$. Purchases are constrained by inventory, however, so inventory falls by $U_k \wedge X_t$. When inventory hits zero the firm orders $b$ units of new stock. The wait time for new stock is also exponential, being an independent draw from $\text{Exp}(\lambda(0))$.

Let $Y$ represent the inventory size after the next jump (induced by either a customer purchase or ordering new stock), given current stock $x$. If $x > 0$, then $Y$ is a draw from the distribution of $x - U \wedge x$ where $U \sim \varphi$. If $x = 0$, then $Y \equiv b$. Hence $Y$ is a draw from $\Pi(x, \cdot)$, where $\Pi(0, y) = \mathbb{1}\{y = b\}$ and, for $0 < x \leqslant b$,

$$\Pi(x, y) = \begin{cases} 0 & \text{if } x \leqslant y \\ \mathbb{P}\{x - U = y\} & \text{if } 0 < y < x \\ \mathbb{P}\{U \geqslant x\} & \text{if } y = 0 \end{cases} \tag{10.37}$$

EXERCISE 10.1.20. Prove that $\Pi$ is a stochastic matrix on $\mathsf{X} := \{0, 1, \ldots, b\}$.

We can simulate the inventory process $(X_t)_{t \geqslant 0}$ via the jump chain algorithm on page 340. In this case, the wait time sequence $(W_k)$ is the wait time for customers (and for inventory when $X_t = 0$) and the jump sequence $(Y_k)$ is the level of inventory immediately after each jump. By Proposition 10.1.9, the inventory process is $Q$-Markov with $Q$ given by $Q(x, x') = \lambda(x)(\Pi(x, x') - I(x, x'))$.

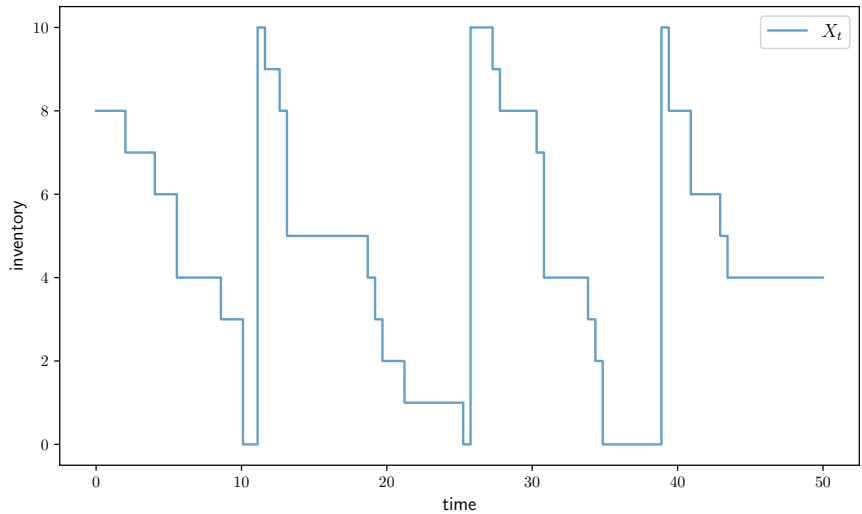

Figure 10.3: Continuous time inventory dynamics

Figure 10.3 shows a simulation when orders are geometric, so that

$$\varphi(k) = \mathbb{P}\{U = k\} = (1 - \alpha)^{k-1}\alpha \qquad (k \in \mathbb{N}, \ \alpha \in (0, 1)).$$

In the simulation we set $\alpha = 0.7$, $b = 10$ and $\lambda(x) \equiv 0.5$. The figure plots $X_t$ for $t \in [0, 50]$. Since each wait time $W_i$ is a draw from $\mathrm{Exp}(0.5)$ the mean wait time is 2.0. The function that produces the map $t \mapsto X_t$ is shown in Listing 28.

### 10.1.4.4 From Intensity Matrices to Jump Chains

If $Q \in \mathcal{L}(\mathbb{R}^X)$ is a given intensity matrix, how should we produce a continuous time $Q$-Markov chain? If we can construct a jump chain that is $Q$-Markov, then not only do we obtain existence of a $Q$-Markov chain but we also provide a way to simulate one (via Algorithm 10.1).

To construct such a jump chain we first fix an intensity matrix $Q \in \mathcal{L}(\mathbb{R}^X)$ and, to simply matters, assume that all rows of $Q$ are nonzero. This means that the process has no absorbing states (since nonzero rows is equivalent to $Q(x, x) < 0$ for all $x$, which in turn states that there is a nonzero outflow from each state).

```julia
using Random, Distributions

"""
Generate a path for inventory starting at b, up to time T.

Return the path as a function X(t) constructed from (J_k) and (Y_k).
"""
function sim_path(; T=10, seed=123, λ=0.5, α=0.7, b=10)

    J, Y = 0.0, b
    J_vals, Y_vals = [J], [Y]
    Random.seed!(seed)
    φ = Exponential(1/λ)      # Wait times are exponential
    G = Geometric(α)          # Orders are geometric

    while true
        W = rand(φ)
        J += W
        push!(J_vals, J)
        if Y == 0
            Y = b
        else
            U = rand(G) + 1    # Geometric on 1, 2,...
            Y = Y - min(Y, U)
        end
        push!(Y_vals, Y)
        if J > T
            break
        end
    end

    function X(t)
        k = searchsortedlast(J_vals, t)
        return Y_vals[k+1]
    end

    return X
end
```

Listing 28:  Continuous time inventory dynamics (`inventory_cont_time.jl`)

Then we set

$$\lambda(x) := -Q(x,x) \quad \text{and} \quad \Pi(x,x') := I(x,x') + \frac{Q(x,x')}{\lambda(x)}.$$

It is straightforward to confirm that $\Pi \in \mathcal{M}(\mathbb{R}^{\mathsf{X}})$ and that $Q$ satisfies (10.31). Hence, by Proposition 10.1.9, the process $(X_t)_{t \geqslant 0}$ generated by Algorithm 10.1 is $Q$-Markov.

## 10.2   Continuous Time Markov Decision Processes

We are ready to turn to dynamic programming in continuous time. As for the discrete time case, continuous time dynamic programs aim to maximize a measure of lifetime value. In §10.2.1 we study lifetime valuations. In §10.2.2 we learn how to maximize them.

### 10.2.1   Valuation

In this section we consider lifetime valuations associated with continuous reward flows, starting from a general semigroup perspective and then progressing to specific cases (such as expected lifetime value under constant discounting). Throughout, $\mathsf{X}$ is a finite set.

#### 10.2.1.1   A Semigroup Perspective

For the discrete time problems with state-dependent discounting that we studied in Chapter 6, lifetime valuations take the form $v = \sum_{t \geqslant 0} K^t h$ for some $h \in \mathbb{R}^{\mathsf{X}}$ and a positive linear operator $K$ on $\mathbb{R}^{\mathsf{X}}$. (See Theorem 6.1.1 and (6.18) on page 193.) For a continuous time version we fix $h \in \mathbb{R}^{\mathsf{X}}$, take $(K_t)_{t \geqslant 0}$ to be a **positive operator semigroup** in $\mathcal{L}(\mathbb{R}^{\mathsf{X}})$, where positive means $K_t \geqslant 0$ for all $t$, and set

$$v = \int_0^\infty K_t h \, dt. \tag{10.38}$$

Let $A \in \mathcal{L}(\mathbb{R}^{\mathsf{X}})$ be the infinitesimal generator of $(K_t)_{t \geqslant 0}$ (existence of which follows from Proposition 10.1.6). The next result provides a condition for finiteness of $v$ and several characterizations.

**Proposition 10.2.1.** *If $s(A) < 0$, then*

(i) *the integral in* (10.38) *is finite and*

$$\nu = \int_0^t K_\tau h \, d\tau + K_t \nu \quad \text{for all } t \geqslant 0, \tag{10.39}$$

(ii) *$A$ is bijective and $\nu = -A^{-1}h$,*

(iii) *$A^{-1} \leqslant 0$, and*

(iv) *the operator $U \colon \mathbb{R}^\mathsf{X} \to \mathbb{R}^\mathsf{X}$ defined by*

$$Uw = h + (I + A)w \qquad \left( w \in \mathbb{R}^\mathsf{X} \right) \tag{10.40}$$

*is order stable on $\mathbb{R}^\mathsf{X}$ and $\nu$ in* (10.38) *is the unique fixed point.*

A way to understand (10.39) is to view the valuation $\nu$ as a price that reflects prospective benefits from holding an asset. The asset yields a flow of benefits, where $h(x)$ is the instantaneous reward in state $x$. Rewards $t$ periods in the future are discounted by the pricing operator $K_t$. Thus, $(K_t h)(x)$ is the anticipated payoff $t$ periods ahead, discounted for the wait time and possibly also for risk as in (6.31) on page 204. The value $\nu(x)$ is then lifetime value, which equals the current price.

In this asset valuation setting, (10.39) is a natural consistency condition. It says that the price of purchasing the asset today is equal to the payouts obtained from holding the asset from now until time $t$ and then selling it for current discounted value $K_t \nu$. (This is the continuous time analog of (6.33) on page 205.)

The discussion above matches the semigroup perspective on asset pricing introduced in Garman (1985) and Duffie and Garman (1986). In addition to shedding light on (10.39), it also leads to the assertion that $\nu = -A^{-1}h$ in (ii), which is obtained by differentiating (10.39). Details are in the proof below.

*Proof of Proposition 10.2.1.* From Proposition 10.1.6, we have $K_t = e^{tA}$ for all $t \geqslant 0$. Since $s(A) < 0$, Theorem 10.1.5 implies that the integral in (10.38) is finite. For any $t \geqslant 0$,

$$\nu = \int_0^\infty K_\tau h \, d\tau = \int_0^t K_\tau h \, d\tau + \int_t^\infty K_\tau h \, d\tau.$$

Using the semigroup property and linearity of $K_t$, we can write the last term on the right hand side as

$$\int_t^\infty K_\tau h \, d\tau = \int_0^\infty K_{t+\tau} h \, d\tau = \int_0^\infty K_t K_\tau h \, d\tau = K_t \int_0^\infty K_\tau h \, d\tau = K_t \nu.$$

Combining this result with the expression for $v$ in the previous display proves (10.39). This proves part (i) of the proposition.

Turning to (ii), if we rearrange (10.39) and divide by $t > 0$, we get

$$-\frac{K_t - I}{t} v = \frac{1}{t} \int_0^t K_\tau h \, d\tau. \tag{10.41}$$

By the fundamental theorem of calculus,

$$\lim_{t \to 0} \frac{1}{t} \int_0^t K_\tau h \, d\tau = \frac{d}{dt} \int_0^t K_\tau h \, d\tau \Big|_{t=0} = K_0 \, h = I \, h = h.$$

As a result, taking $t \to 0$ in (10.41) and using the definition of the infinitesimal generator yields $-Av = h$. Moreover, since $s(A) < 0$, all eigenvalues of $A$ are nonzero. Hence $A$ has nonzero determinant and is therefore nonsingular (bijective). Combining these facts yields $v = -A^{-1}h$.

Regarding (iii), fix $g \in \mathbb{R}^{\mathsf{X}}$ with $g \geqslant 0$. From the preceding results, the function $w = \int_0^\infty K_t g \, dt$ is finite and equals $-A^{-1}g$. Since $K_t \geqslant 0$ for all $t$, we have $w \geqslant 0$. Thus, $-A^{-1}g \geqslant 0$ whenever $g \geqslant 0$. Hence $-A^{-1} \geqslant 0$, or $A^{-1} \leqslant 0$.

For (iv) we use the fact that $v$ obeys $-Av = h$ to obtain $v = h + (I + A)v$. Hence $v$ is a fixed point of $U$. Conversely, if $w$ is a fixed point of $U$, then $-Aw = h$. But $A$ is invertible, so then $w = -A^{-1}h = v$. Hence $v$ is the only fixed point of $U$ in $\mathbb{R}^{\mathsf{X}}$.

Order stability of $U$ requires upward and downward stability on $\mathbb{R}^{\mathsf{X}}$. For upward stability, suppose that $w \in \mathbb{R}^{\mathsf{X}}$ and $Uw \geqslant w$. Then $h + Aw \geqslant 0$, or $-Aw \leqslant h$. But $-A^{-1} \geqslant 0$, so $w \leqslant -A^{-1}h = v$ and upward stability holds. The proof of downward stability is similar. □

### 10.2.1.2 Valuations as Expectations

In applications, the expression $v = \int_0^\infty K_t h \, dt$ from (10.38) typically arises as a discounted expectation over a flow of rewards. When analyzing $v$ we wish to deploy Proposition 10.2.1, so we need to check that any expectation we propose results in $(K_t)$ being a semigroup. The next proposition provides one result along these lines.

**Proposition 10.2.2.** *If* $(X_t)_{t \geqslant 0}$ *is a continuous time Markov chain on* $\mathsf{X}$ *and* $\delta \in \mathbb{R}^{\mathsf{X}}$, *then the family of operators* $(K_t)_{t \geqslant 0} \subset \mathcal{L}(\mathbb{R}^{\mathsf{X}})$ *defined by*

$$(K_t \, h)(x) = \mathbb{E}_x \exp\left(-\int_0^t \delta(X_\tau) \, d\tau\right) h(X_t) \qquad (t \geqslant 0) \tag{10.42}$$

*is a positive operator semigroup.*

In the proof of Proposition 10.2.2, we will use the fact that $(X_t)_{t \geqslant 0}$ satisfies the **Markov property**. In particular, if $H$ is a real-valued function on the path space $C(\mathbb{R}_+, \mathsf{X})$, then

$$\mathbb{E}_x \left[ H((X_\tau)_{\tau \geqslant s}) \mid (X_\tau)_{\tau=0}^s \right] = \mathbb{E}_{X_s} H((X_\tau)_{\tau \geqslant 0}) \quad \text{for all } x \in \mathsf{X}. \tag{10.43}$$

For a proof of (10.43), see, for example, Chapter 2 of Liggett (2010).

**EXERCISE** 10.2.1. Let $(X_t)_{t \geqslant 0}$ be as stated and, for each $s, t \in \mathbb{R}_+$ with $s \leqslant t$, let $\delta(s, t)$ be the random variable defined by

$$\delta(s, t) = \exp \left( - \int_s^t \delta(X_\tau) \, d\tau \right).$$

Show that

(i)  $\delta(s, t) > 0$ for all $0 \leqslant s \leqslant t$,

(ii) $\delta(s, s) = 1$ for all $s \in \mathbb{R}_+$, and

(iii) $\delta(0, s+t) = \delta(0, s) \, \delta(s, s+t)$ for all $s, t \in \mathbb{R}_+$.

*Proof of Proposition 10.2.2.* Fix $h \in \mathbb{R}^{\mathsf{X}}$. Evidently $(K_0 h)(x) = h(x)$, so $K_0 = I$. Regarding the semigroup property, we fix $s \leqslant t$ and use Exercise 10.2.1 and the law of iterated expectations to obtain

$$(K_{s+t} h)(x) = \mathbb{E}_x \, \delta(0, s+t) \, h(X_{s+t}) = \mathbb{E}_x \, \delta(0, s) \, \mathbb{E} \left[ \delta(s, s+t) \, h(X_{s+t}) \mid (X_\tau)_{\tau=0}^s \right].$$

The inner expectation in the last display can be expressed as

$$\mathbb{E} \left[ \exp \left( - \int_s^{s+t} \delta(X_\tau) \, d\tau \right) h(X_{s+t}) \mid (X_\tau)_{\tau=0}^s \right] = \mathbb{E}_{X_s} \left[ \exp \left( - \int_0^t \delta(X_\tau) \, d\tau \right) h(X_t) \right],$$

where the second equality is by the Markov property (10.43). The term on the right hand side equals $(K_t h)(X_s)$. Combining the last two displays gives

$$(K_{s+t} h)(x) = \mathbb{E}_x \delta(0, s)(K_t h)(X_s) = \mathbb{E}_x \exp \left( - \int_0^s \delta(X_\tau) \, d\tau \right) (K_t h)(X_s) = (K_s K_t h)(x).$$

This argument confirms that $K_{s+t} = K_s \circ K_t$.

To see that $K_t$ is a positive operator for all $t$, observe that if $h \geqslant 0$, then the expectation in (10.42) is nonnegative. Hence $K_t h \geqslant 0$ whenever $h \geqslant 0$.

To prove continuity of $t \mapsto K_t$, it suffices to show that $(K_t h)(x) \to h(x)$ as $t \downarrow 0$ (see, e.g., Engel and Nagel (2006), Proposition 1.3). This holds by right-continuity of $X_t$, which gives $h(X_t) \to x$ as $t \downarrow 0$, and hence

$$\lim_{t \downarrow 0}(K_t h)(x) = \mathbb{E}_x \lim_{t \downarrow 0} \exp\left(-\int_0^t \delta(X_\tau)\,d\tau\right) h(X_t) = h(x).$$

(Readers familiar with measure theory can justify the change of limit and expectation via the dominated convergence theorem.) □

### 10.2.1.3 Constant Discounting

Many studies of continuous time dynamic programming with discounting use a constant discount rate. In this setting, lifetime value is given by

$$v(x) := \mathbb{E}_x \int_0^\infty e^{-t\delta} h(X_t)\,dt \tag{10.44}$$

for some $\delta \in \mathbb{R}$ and $h \in \mathbb{R}^{\mathsf{X}}$. Here $(X_t)_{t \geqslant 0}$ is a continuous time Markov chain on finite state $\mathsf{X}$ generated by Markov semigroup $(P_t)_{t \geqslant 0}$ with intensity operator $Q$. The idea is that $h(X_t)$ is an instantaneous reward at each time $t$, while $\delta$ is a fixed discount rate.

Equation 10.44 is the continuous time version of (3.16) on page 95.

**Proposition 10.2.3.** *If $\delta > 0$, then $v$ in (10.44) is finite, $\delta I - Q$ is bijective,*

$$(\delta I - Q)^{-1} \geqslant 0 \quad and \quad v = (\delta I - Q)^{-1} h. \tag{10.45}$$

*In addition, $v$ is the unique fixed point of*

$$Uw = h + (Q + (1 - \delta)I)w \qquad \left(w \in \mathbb{R}^{\mathsf{X}}\right) \tag{10.46}$$

*and $U$ is order stable on $\mathbb{R}^{\mathsf{X}}$.*

*Proof.* As a first step, we reverse the order of expectation and integration in (10.44) to get

$$v(x) = \int_0^\infty (K_t h)(x)\,dt \quad \text{where} \quad (K_t h)(x) := e^{-t\delta}\,\mathbb{E}_x\, h(X_t) = e^{-t\delta}(P_t\, h)(x).$$

(This change of order can be justified by Fubini's theorem, which can be applied when $\mathbb{E}_x \int_0^\infty e^{-t\delta}|h(X_t)|\,dt < \infty$. Since $\mathsf{X}$ is finite, we have $|h| \leqslant M < \infty$ for some constant $M$, and the double integral is dominated by $M \int_0^\infty e^{-t\delta}\,dt = M/\delta$.)

Note that $K_t$ is a special case of (10.42). Hence $(K_t)_{t \geqslant 0}$ is a positive operator semigroup. Its infinitesimal generator is $A := Q - \delta I$, since $K_t = \mathrm{e}^{-t\delta} P_t = \mathrm{e}^{t(Q - \delta I)}$. We claim that $s(A) < 0$. To see this, observe that (using (10.17)),

$$\mathrm{e}^{s(Q - \delta I)} = \rho(\mathrm{e}^{Q - \delta I}) = \rho(\mathrm{e}^Q \mathrm{e}^{-\delta I}) = \rho(\mathrm{e}^Q \mathrm{e}^{-\delta} I) = \mathrm{e}^{-\delta} \rho(\mathrm{e}^Q) = \mathrm{e}^{-\delta} \rho(P_1) = \mathrm{e}^{-\delta}.$$

Taking logs gives $s(Q - \delta I) = -\delta$. Since $\delta > 0$, we have $s(Q - \delta I) < 0$, as claimed.

We can now apply Proposition 10.2.1 with $A = Q - \delta I$ and $K_t = \mathrm{e}^{tA}$. The proposition tells us that that $A$ is bijective, and

$$\nu = -A^{-1} h = (-A)^{-1} h = (\delta I - Q)^{-1} h.$$

It also tells us that $-A^{-1} \geqslant 0$, so

$$(\delta I - Q)^{-1} = (-A)^{-1} = -A^{-1} \geqslant 0.$$

The operator (10.46) is a special case of (10.40), with $A = Q - \delta I$. All of the claims in Proposition 10.2.3 are now verified.  □

## 10.2.2  Constructing a Decision Process

In this section we define continuous time Markov decision processes, discuss optimality theory, and provide algorithms and applications.

### 10.2.2.1  Definition

Given two finite sets A and X, called the state and action spaces respectively, we define a **continuous time Markov decision process** (or **continuous time MDP**) to be a tuple $\mathcal{C} = (\Gamma, \delta, r, Q)$ consisting of

(i) a nonempty correspondence $\Gamma$ from X to A, referred to as the **feasible correspondence**, which in turn defines the **feasible state-action pairs**

$$\mathsf{G} := \{(x, a) \in \mathsf{X} \times \mathsf{A} : a \in \Gamma(x)\},$$

(ii) a constant $\delta > 0$, referred to as the **discount rate**,

(iii) a function $r$ from G to $\mathbb{R}$, referred to as the **reward function**, and

(iv) an **intensity kernel** $Q$ from G to X; that is, a map $Q$ from $G \times X$ to $\mathbb{R}$ satisfying

$$\sum_{x'} Q(x, a, x') = 0 \quad \text{for all } (x, a) \text{ in } G$$

and $Q(x, a, x') \geqslant 0$ whenever $x \neq x'$.

Informally, at state $x$ with action $a$ over the short interval from $t$ to $t + h$, the controller receives instantaneous reward $r(x, a)h$ and the state transitions to state $x'$ with probability $Q(x, a, x')h + o(h)$.

Paralleling our discussion of the discrete time case in Chapter 5, the set of **feasible policies** is

$$\Sigma := \{\sigma \in A^X : \sigma(x) \in \Gamma(x) \text{ for all } x \in X\}. \tag{10.47}$$

### 10.2.2.2  Lifetime Values

Choosing policy $\sigma$ from $\Sigma$ means that we respond to state $X_t$ with action $A_t := \sigma(X_t)$ at every $t \in \mathbb{R}_+$. The state then evolves according to the intensity operator

$$Q_\sigma(x, x') := Q(x, \sigma(x), x') \qquad (x, x' \in X).$$

Letting

$$P_t^\sigma := e^{tQ_\sigma} \quad \text{and} \quad r_\sigma(x) := r(x, \sigma(x)) \qquad (x \in X)$$

the **lifetime value** of following $\sigma$ starting from state $x$ is defined as

$$v_\sigma(x) = \mathbb{E}_x \int_0^\infty e^{-\delta t} r(X_t, \sigma(X_t)) \, dt = \mathbb{E}_x \int_0^\infty e^{-\delta t} r_\sigma(X_t) \, dt, \tag{10.48}$$

where $(X_t)_{t \geqslant 0}$ is $Q_\sigma$-Markov with initial condition $x$. We call $v_\sigma$ the **$\sigma$-value function**.

Since $\delta > 0$, we can apply Proposition 10.2.3 to obtain

$$v_\sigma = (\delta I - Q_\sigma)^{-1} r_\sigma. \tag{10.49}$$

Representation (10.49) provides a straightforward method for computing $v_\sigma$.

### 10.2.2.3  Greedy Policies

A policy $\sigma \in \Sigma$ is called **$v$-greedy** for $\mathcal{C}$ if

$$\sigma(x) \in \operatorname*{argmax}_{a \in \Gamma(x)} \left\{ r(x, a) + \sum_{x'} v(x') Q(x, a, x') \right\} \quad \text{for all } x \in X. \tag{10.50}$$

Like the discrete time case, a $v$-greedy policy chooses actions optimally to trade off high current rewards versus high rate of flow into future states with high values. Unlike the discrete time case, the discount factor does not appear in (10.50) because the trade-off is instantaneous.

#### 10.2.2.4 Policy Iteration

We introduce a continuous time policy iteration algorithm that parallels discrete time HPI for Markov decision processes, as described in §5.1.4.2.

The continuous time HPI routine is given in Algorithm 10.2, with the intuition being similar to that for the discrete time MDP version given on page 142. We provide convergence results in §10.2.3.

---

**Algorithm 10.2:** Continuous time Howard policy iteration

---

**1** input $\sigma_0 \in \Sigma$, an initial guess of $\sigma^*$
**2** $k \leftarrow 0$
**3** $\varepsilon \leftarrow 1$
**4** **while** $\varepsilon > 0$ **do**
**5**   $\quad v_k \leftarrow (\delta I - Q_{\sigma_k})^{-1} r_{\sigma_k}$
**6**   $\quad \sigma_{k+1} \leftarrow$ a $v_k$-greedy policy
**7**   $\quad \varepsilon \leftarrow \mathbb{1}\{\sigma_k \neq \sigma_{k+1}\}$
**8**   $\quad k \leftarrow k + 1$
**9** **end**
**10** **return** $\sigma_k$

---

#### 10.2.2.5 Policy Operators

For each $\sigma \in \Sigma$, let $T_\sigma$ be the operator defined at $v \in \mathbb{R}^X$ by

$$T_\sigma v = r_\sigma + (Q_\sigma + (1 - \delta)I)v. \tag{10.51}$$

As shown in Proposition 10.2.3, each $T_\sigma$ is order stable on $\mathbb{R}^X$, with unique fixed point $v_\sigma$. Hence $\mathcal{A} := (\mathbb{R}^X, \{T_\sigma\})$ is an order stable ADP.

EXERCISE 10.2.2. Show that $\sigma$ is $v$-greedy (i.e., (10.50) holds) if and only if $\sigma$ is $v$-greedy for $\mathcal{A}$ in the sense of §9.1.2.2.

## 10.2.3 Optimality

For a continuous time MDP $\mathcal{C} = (\Gamma, \delta, r, Q)$ with $\sigma$-value functions $\{v_\sigma\}$,

- the **value function** generated by $\mathcal{C}$ is $v^* := \bigvee_\sigma v_\sigma$, and

- a policy is called **optimal** for $\mathcal{C}$ if $v_\sigma = v^*$.

A function $v \in \mathbb{R}^{\mathsf{X}}$ is said to satisfy a **Hamilton–Jacobi–Bellman** (**HJB**) equation if

$$\delta v(x) = \max_{a \in \Gamma(x)} \left\{ r(x, a) + \sum_{x'} v(x') Q(x, a, x') \right\} \quad \text{for all } x \in \mathsf{X}. \tag{10.52}$$

We say that $\mathcal{C}$ obeys **Bellman's principle of optimality** if

$$\sigma \in \Sigma \text{ is optimal for } \mathcal{C} \quad \Longleftrightarrow \quad \sigma \text{ is } v^*\text{-greedy}.$$

Here is our main optimality result for continuous time MDPs.

**Theorem 10.2.4.** *If $\mathcal{C} = (\Gamma, \delta, r, Q)$ is a continuous time MDP, then*

(i) *the value function $v^*$ is the unique solution to the HJB equation in $\mathbb{R}^{\mathsf{X}}$,*

(ii) *$\mathcal{C}$ obeys Bellman's principle of optimality, and*

(iii) *$\mathcal{C}$ has at least one optimal policy.*

*In addition, continuous time HPI converges to an optimal policy in finitely many steps.*

*Proof.* Let $\mathcal{C} = (\Gamma, \delta, r, Q)$ be a fixed continuous time MDP with lifetime values $\{v_\sigma\}$ and value function $v^*$. Consider the order stable ADP $\mathcal{A} := (\mathbb{R}^{\mathsf{X}}, \{T_\sigma\})$ discussed in §10.2.2.5. The ADP Bellman max-operator is $T := \bigvee_\sigma T_\sigma$, which can be written more explicitly as

$$(Tv)(x) = \max_{a \in \Gamma(x)} \left\{ r(x, a) + \sum_{x'} v(x') Q(x, a, x') \right\} + (1 - \delta)v(x). \tag{10.53}$$

It is clear from (10.50) and Exercise 10.2.2 that, for each $v \in \mathbb{R}^{\mathsf{X}}$, the set of $v$-max-greedy policies is nonempty. Since $\Sigma$ is finite, it follows from Proposition 9.1.3 that $\mathcal{A}$ is max-stable. Hence, by Theorem 9.1.6, an optimal policy always exists and the value function $v^*$ is the unique fixed point of $T$ in $\mathbb{R}^{\mathsf{X}}$. The last statement is equivalent to the assertion that $v^*$ is the unique element of $\mathbb{R}^{\mathsf{X}}$ satisfying

$$v^*(x) = \max_{a \in \Gamma(x)} \left\{ r(x, a) + \sum_{x'} v^*(x') Q(x, a, x') \right\} + (1 - \delta)v^*(x).$$

Rearranging this expression confirms that $v^*$ is the unique solution to the HJB equation in $\mathbb{R}^X$.

Applying Theorem 9.1.6 again, a policy is optimal for $\mathcal{A}$ if and only if $T_\sigma v^* = Tv^*$. Since the definition of optimality for $\mathcal{A}$ coincides with the definition of optimality for $\mathcal{C}$, we see that $\mathcal{C}$ obeys Bellman's principle of optimality.

The continuous time HPI routine described in Algorithm 10.2 is just ADP max-HPI (see §9.1.3.4) specialized to the current setting. Hence, applying Theorem 9.1.6 once more, continuous time HPI converges to an optimal policy in finitely many steps. □

### 10.2.4 Application: Job Search

Here we study a continuous time version of the job search problem with separation considered in §3.3.2. As before, a worker can be either unemployed (state 0) or employed (state 1). When the worker is employed, she can be fired at any time. Firing occurs at rate $\alpha > 0$, meaning that the probability of being fired over the short interval from $t$ to $t + h$ is approximately $\alpha h$. When unemployed, the worker receives receives flow unemployment compensation $c$ and job offers at rate $\kappa$. She can choose either to accept or to reject an offer; she discounts the future at rate $\delta > 0$.

We assume that job offers are associated with wage offers that take values in finite set W. Let $P \in \mathcal{M}(\mathbb{R}^W)$ give probabilities for new wage draws, so that, conditional on previous draw $w$, the next offer is drawn from $P(w, \cdot)$.

For the state space we set $X = \{0, 1\} \times W$, with typical state $x = (s, w)$. Here $s$ is binary and indicates current employment status, while $w$ is the current wage. Let

$$\lambda(x) = \lambda(s, w) = \mathbb{1}\{s = 0\}\kappa + \mathbb{1}\{s = 1\}\alpha$$

denote the state-dependent jump rate, which switches between $\kappa$ and $\alpha$ depending on employment status.

Let $a \in A := \{0, 1\}$ indicate the action, where 0 means reject and 1 means accept. Let $\Pi(x, a, x')$ represent the jump probabilities, with

$$\Pi((0, w), a, (0, w')) = P(w, w')(1 - a) \qquad \text{(unemployed to unemployed)}$$
$$\Pi((0, w), a, (1, w')) = P(w, w')a \qquad \text{(unemployed to employed)}$$
$$\Pi((1, w), a, (0, w')) = P(w, w') \qquad \text{(employed to unemployed)}$$
$$\Pi((1, w), a, (1, w')) = 0. \qquad \text{(employed to employed)}$$

The first two lines consider jump probabilities for the state $(s, w)$ when unemployed and the action is $a$. The second two consider jump probabilities when employed. The reason that the probability assigned to the last line is zero is that a jump from $s = 1$ occurs because the worker is fired, so the value of $s$ after the jump is zero.

EXERCISE 10.2.3. Prove that $\Pi$ is a stochastic kernel, in the sense that $\Pi \geqslant 0$ and $\sum_{x'} \Pi(x, a, x') = 1$ for all possible $(x, a) = ((s, w), a)$ in $\mathsf{X} \times \{0, 1\}$.

Motivated by the jump chain construction of intensity matrices in (10.31) on page 340, we set

$$Q(x, a, x') = \lambda(x)(\Pi(x, a, x') - I(x, x')).$$

It follows that, for any $\sigma \in \Sigma := \{0, 1\}^{\mathsf{X}}$, the operator

$$Q_\sigma(x, x') := \lambda(x)(\Pi(x, \sigma(x), x') - I(x, x')),$$

is an intensity matrix for the jump chain under policy $\sigma$.

If we define

$$r(x, a) = r((s, w), a) = c\mathbb{1}\{s = 0\} + w\mathbb{1}\{s = 1\},$$

then lifetime value is given by (10.48), where $(X_t)_{t \geqslant 0}$ is $Q_\sigma$-Markov and $X_0 = x$.

With $\Gamma$ defined by $\Gamma(x) = \mathsf{A}$ for all $x \in \mathsf{X}$, the tuple $\mathcal{C} = (\Gamma, \delta, r, Q)$ is a continuous time MDP and Theorem 10.2.4 applies. In particular, an optimal policy exists and can be computed with HPI in a finite number of iterations.

Figure 10.4 shows an optimal policy computed in this way. (Code and parameter values can be found in `cont_time_js.jl`.) The policy is of threshold type, with a reservation wage of around 12. Figure 10.5 shows how this reservation wage changes with parameters. The reservation wage increases as the separation rate falls, as the offer rate increases, as the discount rate falls, and as unemployment compensation increases.

EXERCISE 10.2.4. Provide economic intuition for the monotone relationships between parameters and the reservation wage discussed in the preceding paragraph.

## 10.3   Chapter Notes

Applebaum (2019) and Engel and Nagel (2006) provide elegant introductions to semigroup theory and its applications in studying partial and stochastic differential equations. The beautiful book by Lasota and Mackey (1994) and covers connections among semigroups, Markov processes, and stochastic differential equations. Norris (1998) provides a good introduction to continuous time Markov chains, while Liggett (2010) is more advanced.

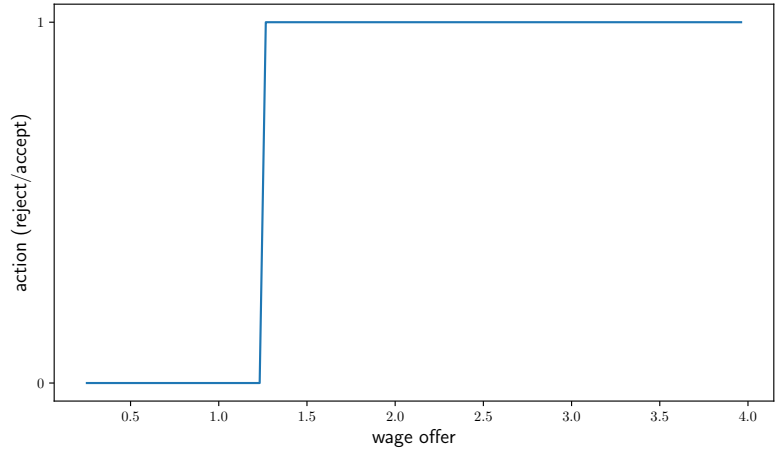

Figure 10.4:  Continuous time job search policy

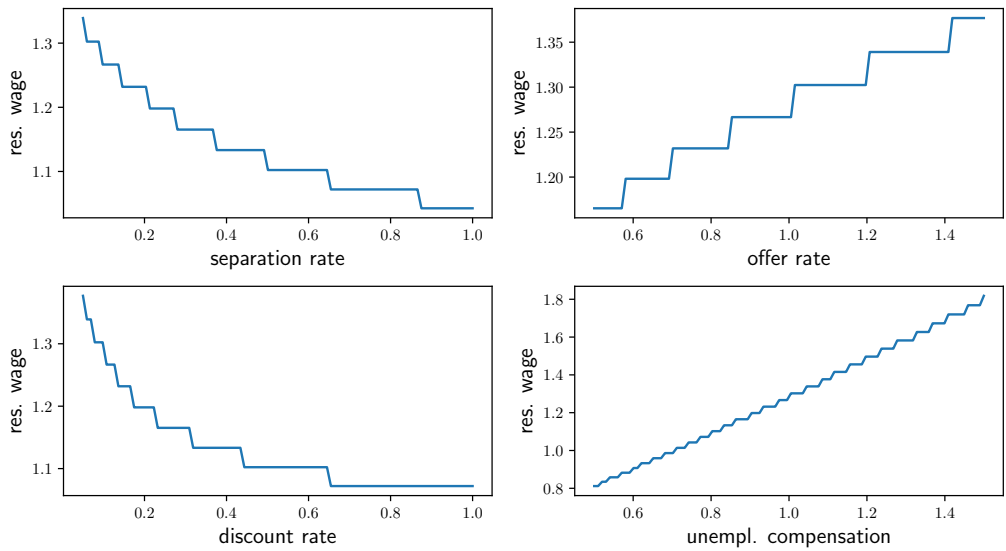

Figure 10.5:  Continuous time job search reservation wage

A rigorous treatment of continuous time MDPs can be found in Hernández-Lerma and Lasserre (2012b), which also handles the case where X is countably infinite. Our approach is somewhat different, since our main optimality results rest on the ADP theory in Chapter 9.

In recent years, continuous time dynamic programming has become more common in macroeconomic analysis. Influential references include Nuño and Moll (2018), Kaplan et al. (2018), Achdou et al. (2022), and Fernández-Villaverde et al. (2023). For computational aspects, see Duarte (2018), Ráfales and Vázquez (2021), Rendahl (2022), and Eslami and Phelan (2023).

# Part I

# Appendices

# Appendix A

# Suprema and Infima

This section of the appendix contains an extremely brief review of basic facts concerning sets, functions, suprema and infima. We recommend Bartle and Sherbert (2011) for those who wish to learn more.

## A.1   Sets and Functions

A **set** is a collection of objections viewed as a whole. Examples include the set of **natural numbers** $\mathbb{N} \coloneqq \{1, 2, \ldots\}$ and $[n] \coloneqq \{1, 2, \ldots, n\}$ when $n \in \mathbb{N}$. The set that contains no elements is called the **empty set** and denoted by $\emptyset$.

Let $A$ and $B$ be two sets and let $A \times B$ be their **Cartesian product**, defined as the set of all ordered pairs $(a, b)$ such that $a \in A$ and $b \in B$. A **binary relation** $\sim$ between two sets $A$ and $B$ is a subset of $A \times B$. If $(a, b)$ is in this subset we write $a \sim b$. An **equivalence relation** on $A$ is a binary relation $\sim$ between $A$ and itself that is reflexive, symmetric and transitive. That is,

(a)  $a \sim a$ for all $a \in A$,

(c)  $a \sim a'$ implies $a' \sim a$, and

(d)  $a \sim a'$ and $a' \sim a''$ implies $a \sim a''$.

A **function** $f$ from set $A$ to set $B$, written $A \ni x \mapsto f(x) \in B$ or $f \colon A \to B$, is a rule (in fact, a binary relation) associating to each and every element $a$ in $A$ one and only one element $b \in B$. The point $b$ is also written as $f(a)$, and called the **image** of $a$ under $f$. For $C \subset A$, the set $f(C)$ is the set of all images of points in $C$, and is called the image of $C$ under $f$. Also, for $D \subset B$, the set $f^{-1}(D)$ is all points in $A$ that map into $D$ under $f$, and is called the **preimage** of $D$ under $f$.

A function $f \colon A \to B$ is called **one-to-one** if distinct elements of $A$ are always mapped into distinct elements of $B$, and **onto** if every element of $B$ is the image under $f$ of at least one point in $A$. A **bijection** or **one-to-one correspondence** from $A$ to $B$ is a function $f$ from $A$ to $B$ that is both one-to-one and onto.

A set $\mathsf{X}$ is called **finite** if there exists a bijection from $\mathsf{X}$ to $[n] := \{1, \ldots, n\}$ for some $n \in \mathbb{N}$. In this case we can write $\mathsf{X} = \{x_1, \ldots, x_n\}$. The number $n$ is called the **cardinality** of $\mathsf{X}$. Note that, according to our definition, every finite set is automatically nonempty.

If $f \colon A \to B$ and $g \colon B \to C$, then the **composition** of $f$ and $g$ is the function $g \circ f$ from $A$ to $C$ defined at $a \in A$ by $(g \circ f)(a) := g(f(a))$.

## A.2   Some Properties of the Real Line

Given a subset $A$ of $\mathbb{R}$, we call $u \in \mathbb{R}$ an **upper bound** of $A$ if $a \leqslant u$ for all $a$ in $A$. A **lower bound** of $A$ is any number $\ell$ such that $\ell \leqslant a$ for all $a \in A$. If $A$ has both an upper and lower bound then $A$ is called **bounded**. Equivalently, $A$ is bounded whenever there exists an $n \in \mathbb{N}$ with $A \subset [-n, n]$.

Let $U(A)$ be the set of all upper bounds of $A$. An element $\bar{u}$ of $\mathbb{R}$ is called a **supremum** or **least upper bound** of $A$ if

(i)  $\bar{u} \in U(A)$ and

(ii)  $\bar{u} \leqslant u$ for every $u \in U(A)$.

When a supremum of $A$ exists in $\mathbb{R}$, we write it as $\sup A$.

**Example A.2.1.** For the set $I := [0, 1] \subset \mathbb{R}$, the number $1$ is an upper bound of $I$. Moreover, if $u$ is an upper bound of $I$, then $u \geqslant 1$. Hence $1$ is the supremum of $I$.

**Example A.2.2.** $\mathbb{N}$ has no supremum in $\mathbb{R}$, since the set of upper bounds is empty.

EXERCISE A.2.1. Show that, for all of the sets $(0, 1)$, $[0, 1)$ and $(0, 1]$, the number $1$ is the supremum of the set.

EXERCISE A.2.2. Fix $A \subset \mathbb{R}$. Prove that, for $s \in U(A)$, we have $s = \sup A$ if and only if, for all $\varepsilon > 0$, there exists a point $a \in A$ with $a > s - \varepsilon$.

EXERCISE A.2.3. Fix $A \subset \mathbb{R}$. Prove that $A$ has at most one supremum.

One of the most important properties of $\mathbb{R}$ is stated below.

**Theorem A.2.1** (Least upper bound property). *Every nonempty subset of $\mathbb{R}$ with an upper bound in $\mathbb{R}$ has a supremum in $\mathbb{R}$.*

Theorem A.2.1 is often taken as axiomatic in formal constructions of the real numbers. (Alternatively, one may assume completeness of the reals and then prove Theorem A.2.1 using this property. See, e.g., Bartle and Sherbert (2011).)

If $i \in \mathbb{R}$ is a lower bound for $A$ and also satisfies $i \geqslant \ell$ for every lower bound $\ell$ of $A$, then $i$ is called the **infimum** of $A$ and we write $i = \inf A$. At most one such $i$ exists, and every nonempty subset of $\mathbb{R}$ bounded from below has an infimum.

A **real sequence** is a map $x$ from $\mathbb{N}$ to $\mathbb{R}$, with the value of the function at $k \in \mathbb{N}$ typically denoted by $x_k$ rather than $x(k)$. A real sequence $x = (x_k)_{k \geqslant 1} := (x_k)_{k \in \mathbb{N}}$ is said to **converge** to $\bar{x} \in \mathbb{R}$ if, for each $\varepsilon > 0$, there exists an $N \in \mathbb{N}$ such that $|x_k - \bar{x}| < \varepsilon$ for all $k \geqslant N$. In this case we write $\lim_k x_k = \bar{x}$ or $x_k \to \bar{x}$. Bartle and Sherbert (2011) give an excellent introduction to real sequences and their basic properties.

A real sequence $(x_k)_{k \geqslant 1}$ is called **increasing** if $x_k \leqslant x_{k+1}$ for all $k$ and **decreasing** if $x_{k+1} \leqslant x_k$ for all $k$. If $(x_k)_{k \geqslant 1}$ is increasing (resp., decreasing) and $x_k \to x \in \mathbb{R}$ then we also write $x_k \uparrow x$ (resp., $x_k \downarrow x$).

EXERCISE A.2.4. Let $(x_k)$ be a bounded monotone increasing sequence in $\mathbb{R}$. Prove that $\sup_k x_k = \lim_k x_k$.

Let $(x_k)$ be a real sequence in $\mathbb{R}$ and set $s_n := \sum_{k=1}^{n} x_k$. If the sequence $(s_n)$ converges to some $s \in \mathbb{R}$, then we set

$$\sum_{k=1}^{\infty} x_k := \sum_{k \geqslant 1} x_k := s = \lim_{n \to \infty} s_n.$$

We say that the **series** $\sum_{k=1}^{n} x_k$ converges to $\sum_{k=1}^{\infty} x_k$.

## A.3 Max and Min

A number $m$ contained in a subset $A$ of $\mathbb{R}$ is called the **maximum** of $A$ and we write $m = \max A$ if $a \leqslant m$ for every $a \in A$. It is called the **minimum** of $A$ and we write $m = \min A$ if $a \geqslant m$ for every $a \in A$.

EXERCISE A.3.1. Prove: If $A$ is a finite subset of $\mathbb{R}$, then $\sup A = \max A$.

A subset $A$ of $\mathbb{R}$ is called **closed** if, for any sequence $(x_n)$ contained in $A$ and converging to some limit $x \in \mathbb{R}$, the limit $x$ is in $A$.

EXERCISE A.3.2. Show that, if $A$ is a closed and bounded subset of $\mathbb{R}$, then $A$ has both a maximum and a minimum.

EXERCISE A.3.3. Prove the following statements:

(i)  If $A \subset B$, then $\sup A \leqslant \sup B$.

(ii)  If $s = \sup A$ and $s \in A$, then $s = \max A$.

(iii)  If $i = \inf A$ and $i \in A$, then $i = \min A$.

Given an arbitrary set $D$ and a function $f \colon D \to \mathbb{R}$, define

$$\sup_{x \in D} f(x) := \sup\{f(x) : x \in D\} \quad \text{and} \quad \max_{x \in D} f(x) := \max\{f(x) : x \in D\}$$

whenever the latter exists. The terms $\inf_{x \in D} f(x)$ and $\min_{x \in D} f(x)$ are defined analogously. A point $x^* \in D$ is called a

- **maximizer** of $f$ on $D$ if $x^* \in D$ and $f(x^*) \geqslant f(x)$ for all $x \in D$, and a

- **minimizer** of $f$ on $D$ if $x^* \in D$ and $f(x^*) \leqslant f(x)$ for all $x \in D$.

Equivalently, $x^* \in D$ is a maximizer of $f$ on $D$ if $f(x^*) = \max_{x \in D} f(x)$, and a minimizer if $f(x^*) = \min_{x \in D} f(x)$. We define

$$\operatorname*{argmax}_{x \in D} f(x) := \{x^* \in X : f(x^*) \geqslant f(x) \text{ for all } x \in D\}.$$

The set $\operatorname{argmin}_{x \in D} f(x)$ is defined analogously.

# Appendix B

# Remaining Proofs

## B.1 Chapter 2 Results

*Proof of Lemma 2.2.5.* Regarding (i), fix $\varphi, \psi \in \mathcal{D}(\mathsf{X})$ with $\varphi \preceq_{\mathrm{F}} \psi$. Pick any $y \in \mathsf{X}$. By transitivity of partial orders, the function $u(x) := \mathbb{1}\{y \preceq x\}$ is in $i\mathbb{R}^{\mathsf{X}}$. Hence $\sum_x u(x)\varphi(x) \leqslant \sum_x u(x)\psi(x)$. Given the definition of $u$, this is equivalent to $G^\varphi(y) \leqslant G^\psi(y)$. As $y$ was chosen arbitrarily, we have $G^\varphi \leqslant G^\psi$ pointwise on $\mathsf{X}$.

Regarding (ii), let $\varphi, \psi \in \mathcal{D}(\mathsf{X})$ be such that $G^\varphi \leqslant G^\psi$ and let $\mathsf{X}$ be totally ordered by $\preceq$. We can write $\mathsf{X}$ as $\{x_1, \ldots, x_n\}$ with $x_i \preceq x_{i+1}$ for all $i$. Pick any $u \in i\mathbb{R}^{\mathsf{X}}$ and let $\alpha_i = u(x_i)$. By Exercise 2.2.32, we can write $u$ as $u(x) = \sum_{i=1}^n s_i \mathbb{1}\{x \succeq x_i\}$ at each $x \in \mathsf{X}$, where $s_i \geqslant 0$ for all $i$. Hence

$$\sum_{x \in \mathsf{X}} u(x)\varphi(x) = \sum_{x \in \mathsf{X}} \sum_{i=1}^n s_i \mathbb{1}\{x \succeq x_i\}\varphi(x) = \sum_{i=1}^n s_i \sum_{x \in \mathsf{X}} \mathbb{1}\{x \succeq x_i\}\varphi(x) = \sum_{i=1}^n s_i\, G^\varphi(x_i).$$

A similar argument gives $\sum_{x \in \mathsf{X}} u(x)\psi(x) = \sum_{i=1}^n s_i\, G^\psi(x_i)$. Since $G^\varphi \leqslant G^\psi$, we have

$$\sum_{x \in \mathsf{X}} u(x)\varphi(x) = \sum_{i=1}^n s_i\, G^\varphi(x_i) \leqslant \sum_{i=1}^n s_i\, G^\psi(x_i) = \sum_{x \in \mathsf{X}} u(x)\psi(x).$$

We conclude that $\varphi \preceq_{\mathrm{F}} \psi$, as was to be shown. □

# B.2 Chapter 6 Results

We adopt the setting of §6.1.1.2 and consider the claim

$$\mathbb{E}_x \sum_{t=0}^{\infty} \left[ \prod_{i=0}^{t} \beta_i \right] h(X_t) = \sum_{t=0}^{\infty} \mathbb{E}_x \left[ \prod_{i=0}^{t} \beta_i \right] h(X_t) \tag{B.1}$$

when $(X_t)$ is $P$-Markov with initial condition $x$ and $h \in \mathbb{R}^{\mathsf{X}}$. Throughout this discussion the assumption $\rho(L) < 1$ is in force (see Theorem 6.1.1). Unlike the rest of the book, we assume some familiarity with measure theory, at the level of, say, Dudley (2002), Chapters 3 and 4.

To begin the discussion we set

$$F_T := \sum_{t=0}^{T} \delta_t \, h(X_t) \quad \text{and} \quad F := \sum_{t=0}^{\infty} \delta_t \, h(X_t) \quad \text{where} \quad \delta_t := \prod_{i=0}^{t} \beta_i.$$

Our first aim is to show that $F$ is a well-defined random variable, in the sense that the sum converges almost surely. Since absolute convergence of real series implies convergence, and since finite expectation implies finiteness almost everywhere, it suffices to show that

$$\mathbb{E}_x \sum_{t=0}^{\infty} \delta_t \, |h(X_t)| < \infty. \tag{B.2}$$

By the monotone convergence theorem (see, e.g., Dudley (2002), Theorem 4.3.2), we have

$$\mathbb{E}_x \sum_{t=0}^{\infty} \delta_t \, |h(X_t)| = \sum_{t=0}^{\infty} \mathbb{E}_x \, \delta_t \, |h(X_t)| = \sum_{t=0}^{\infty} (L^t |h|)(x),$$

where the last equality is by (6.6). Since $\rho(L) < 1$, we have shown that (B.2) holds, which in turn confirms that $F$ is well-defined and finite almost surely.

Now observe that, on the probability one set where $F$ is finite, we have $F_T \to F$ as $T \to \infty$. Moreover,

$$|F_T| \leqslant \sum_{t=0}^{T} \delta_t \, |h(X_t)| \leqslant Y := \sum_{t=0}^{\infty} \delta_t \, |h(X_t)|,$$

and, as shown above, $\mathbb{E}_x \, Y < \infty$. By the dominated convergence theorem, we now have $\mathbb{E}_x \, F = \lim_{T \to \infty} \mathbb{E}_x \, F_T$, or, equivalently,

$$\mathbb{E}_x \sum_{t=0}^{\infty} \delta_t \, h(X_t) = \lim_{T \to \infty} \mathbb{E}_x \sum_{t=0}^{T} \delta_t \, h(X_t) = \lim_{T \to \infty} \sum_{t=0}^{T} \mathbb{E}_x \, \delta_t \, h(X_t) = \sum_{t=0}^{\infty} \mathbb{E}_x \, \delta_t \, h(X_t).$$

Hence (B.1) holds.

## B.3 Chapter 7 Results

*Proof of uniqueness for Theorem 7.1.3.* We focus on the concave case. Let $I$ be as in Theorem 7.1.3 and suppose that $T$ is an order-preserving concave self map on $I$ with $T\varphi \gg \varphi$. By Theorem 7.1.1, $T$ has least and greatest fixed points in $I$. We denote them by $a$ and $b$, respectively. Let

$$\lambda = \min_{x \in \mathsf{X}} \frac{a(x) - \varphi(x)}{b(x) - \varphi(x)}$$

and let $\bar{x}$ be a minimizer. It follows immediately from its definition that $\lambda$ obeys $0 \leqslant \lambda \leqslant 1$ and

$$a(x) \geqslant \lambda b(x) + (1-\lambda)\varphi(x) \quad \text{for all } x \in \mathsf{X} \text{ with equality at } \bar{x}.$$

As a result, applying the assumed properties of $T$, we have

$$a = Ta \geqslant T(\lambda b + (1-\lambda)\varphi) \geqslant \lambda b + (1-\lambda)T\varphi.$$

Suppose now that $\lambda < 1$. Since $T\varphi \gg \varphi$, we get $a \gg \lambda b + (1-\lambda)\varphi$ and evaluating this at $\bar{x}$ yields

$$a(\bar{x}) > \lambda b(\bar{x}) + (1-\lambda)\varphi(\bar{x}) = a(\bar{x}),$$

which is a contradiction. Hence $\lambda = 1$ and, therefore, $a \geqslant b$. Since all fixed points $\bar{u}$ of $T$ in $I$ obey $a \leqslant \bar{u} \leqslant b$, we see that $a = b$ is the unique fixed point of $T$ in $I$. □

## B.4 Chapter 9 Results

Let's now turn to the proof of the core optimality results for ADPs. In what follows, $\mathcal{A} = (V, \{T_\sigma\})$ is a well-posed ADP with Bellman operator $T$ and $\sigma$-value functions $\{v_\sigma\}_{\sigma \in \Sigma}$. We start with

**Lemma B.4.1.** *If $\mathcal{A}$ is order stable, then the following statements hold:*

(i) $v \in V_u \implies v \preceq Hv$.

(ii) *If $\sigma \in \Sigma$ and $Tv_\sigma = v_\sigma$, then $v_\sigma = v^*$.*

(iii) *If $v \in V$ and $Hv = v$, then $v = v^*$ and $Tv^* = v^*$.*

(iv) *If $\mathcal{A}$ is finite, then $v^*$ exists in $V$ and $H v^* = v^*$. Moreover, for all $v \in V$, the HPI sequence $(v_k)$ defined by $v_k = H^k v$ converges to $v^*$ in finitely many steps.*

(v) *Fix $v \in V$ and let $(v_k)$ be the HPI sequence defined by $v_k = H^k v$ for $k \in \mathbb{N}$. If $v_{k+1} = v_k$ for some $k \in \mathbb{N}$, then $v_k = v^*$ and every $v_k$-greedy policy is optimal.*

*Proof.* Regarding (i), fix $v \in V_u$ and let $\tau$ be $v$-greedy, with $H v = v_\tau$. Since $v \in V_u$, we have $v \leq T v = T_\tau v$. This inequality and upward stability of $T_\tau$ yield $v \leq v_\tau$. But then $v \leq H v$, as claimed.

Regarding (ii), suppose $\sigma \in \Sigma$ and $T v_\sigma = v_\sigma$. Fix $\tau \in \Sigma$ and note that $v_\sigma = T v_\sigma \geq T_\tau v_\sigma$. Downward stability of $T_\tau$ implies $v_\sigma \geq v_\tau$. Since $\tau \in \Sigma$ was arbitrary, $v_\sigma = v^*$.

Regarding (iii), fix $v \in V$ with $H v = v$ and let $\sigma$ be such that $H v = v_\sigma$. Then $v_\sigma = v$, and, since $\sigma$ is $v$-greedy, $T_\sigma v = T v$. But then $T_\sigma v_\sigma = T v_\sigma$, and, since $v_\sigma = T_\sigma v_\sigma$, we have $v_\sigma = T v_\sigma$. Part (ii) now implies $v = v_\sigma = v^*$. This proves the first claim. Regarding the second, substituting $v_\sigma = v^*$ into $v_\sigma = T v_\sigma$ yields $v^* = T v^*$.

For (iv), it suffices to show that $H v^* = v^*$ and there exists a $K \in \mathbb{N}$ such that $H^K v = v^*$. To this end, let $v_k = H^k v$ and note that $v_k \in V_\Sigma$ for all $k \geq 1$. Part (i) implies that $v_{k+1} \geq v_k$ for all $k \in \mathbb{N}$. Since the sequence $(v_k)$ is contained in the finite set $V_\Sigma$, it must be that $v_{K+1} = v_K$ for some $K \in \mathbb{N}$ (since otherwise $V_\Sigma$ contains an infinite sequence of distinct points). But then $H v_K = v_{K+1} = v_K$, so $v_K$ is a fixed point of $H$. Part (iii) now implies that $v_K = v^*$.

For (v), let $(v_k)$ be as stated and suppose that $v_{k+1} = v_k$ for some $k \in \mathbb{N}$. Then $v_k$ is a fixed point of $H$, so, by (iii) above, we have $v_k = v^*$. By Bellman's principle of optimality, every $v_k$-greedy policy is optimal. $\square$

*Proof of Proposition 9.1.3.* If $\mathcal{A}$ is finite, then, by (iii)–(iv) of Lemma B.4.1, the point $v^*$ exists in $V$ and is a fixed point of $T$. $\square$

We first prove Proposition 9.1.7 and then return to Theorem 9.1.6.

*Proof of Proposition 9.1.7.* Let $\mathcal{A}$ be max-stable. We need to establish the following claims.

(a) $V_\Sigma$ has a greatest element $v^*$ and

(b) $v^*$ is the unique fixed point of $T$ in $V$.

(c) a policy is optimal if and only if it is $v^*$-greedy.

(d) at least one optimal policy exists.

For claims (a)–(b), we observe that, by max-stability, $T$ has a fixed point $\bar{v}$ in $V$. By existence of greedy policies, we can find a $\sigma \in \Sigma$ such that $\bar{v} = T\bar{v} = T_\sigma \bar{v}$. But $T_\sigma$ has a unique fixed point in $V$, equal to $v_\sigma$, so $\bar{v} = v_\sigma$. Moreover, if $\tau$ is any policy, then $T_\tau \bar{v} \leq T\bar{v} = \bar{v}$ and hence, by downward stability, $v_\tau \leq \bar{v}$. These facts imply that $v^* := \bar{v}$ is the greatest element of $V_\Sigma$ and a fixed point of $T$. Since greatest elements are unique, $v^*$ is the only fixed point of $T$ in $V$.

Regarding (c), parts (a)–(b) give $v^* \in V$ and $Tv^* = v^*$. Now recall that $\sigma$ is optimal if and only if $v_\sigma = v^*$. Since $v_\sigma$ is the unique fixed point of $T_\sigma$, this is equivalent to $T_\sigma v^* = v^*$. Since $Tv^* = v^*$, the last statement is equivalent to $T_\sigma v^* = Tv^*$, which is, in turn equivalent to the statement that $\sigma$ is $v^*$-greedy.

Part (d) follows directly from (a). □

*Proof of Theorem 9.1.6.* Parts (i)–(iv) of Theorem 9.1.6 follow from Proposition 9.1.7, which provides optimality results for max-stable ADPs, and Proposition 9.1.3, which tells us that every finite order stable ADP is max-stable. Regarding the final claim in Theorem 9.1.6, on convergence of HPI, suppose that $\mathcal{A}$ is finite and order stable. If HPI terminates, then (v) of Lemma B.4.1 implies that it returns an optimal policy. Part (iv) of the same lemma implies that HPI terminates in finitely many steps. □

# Appendix C

# Solutions to Selected Exercises

**Solution to Exercise 1.1.1.** Here is one possible answer: On one hand, providing additional unemployment compensation is costly for taxpayers and tends to increase the unemployment rate. On the other hand, unemployment compensation encourages the worker to reject low initial offers, leading to a better lifetime wage. This can enhance worker welfare and expand the tax base. A larger model is needed to disentangle these effects.

**Solution to Exercise 1.2.1.** Fix $\alpha$, $s$ and $t$ with $s \geqslant 0$. Suppose first that $\alpha \geqslant s + t$. Then $\alpha \vee (s + t) = \alpha \leqslant \alpha \vee t \leqslant s + \alpha \vee t$, as claimed. Suppose next that $\alpha \leqslant s + t$. Then $\alpha \vee (s + t) = s + t \leqslant s + \alpha \vee t$, as required.

**Solution to Exercise 1.2.5.** For $\alpha > 0$ we always have $\|\alpha u\|_0 = \|u\|_0$, which violates absolute homogeneity.

**Solution to Exercise 1.2.15.** Let $T$ and $U$ be as stated in the exercise. Regarding uniqueness, suppose that $T$ has two distinct fixed points $u$ and $y$ in $U$. Since $T^m u = \bar{u}$ and $T^m y = \bar{u}$, we have $T^m u = T^m y$. But $u$ and $y$ are distinct fixed points, so $u = T^m u$ must be distinct from $y = T^m y$. Contradiction.

Regarding the claim that $\bar{u}$ is a fixed point, we recall that $T^k u = \bar{u}$ for $k \geqslant m$. Hence $T^m \bar{u} = \bar{u}$ and $T^{m+1} \bar{u} = \bar{u}$. But then

$$T\bar{u} = T T^m \bar{u} = T^{m+1} \bar{u} = \bar{u},$$

so $\bar{u}$ is a fixed point of $T$.

**Solution to Exercise 1.2.16.** Assume the hypotheses of the exercise and let $u_m :=$ $T^m u$ for all $m \in \mathbb{N}$. By continuity and $u_m \to u^*$ we have $Tu_m \to Tu^*$. But the sequence $(Tu_m)$ is just $(u_m)$ with the first element omitted, so, given that $u_m \to u^*$, we must have $Tu_m \to u^*$. Since limits are unique, it follows that $u^* = Tu^*$.

**Solution to Exercise 1.2.18.** Let the stated hypotheses hold and fix $u \in C$. By global stability we have $T^k u \to u^*$. Since $T$ is invariant on $C$ we have $(T^k u)_{k \in \mathbb{N}} \subset C$. Since $C$ is closed, this implies that the limit is in $C$. In other words, $u^* \in C$, as claimed.

**Solution to Exercise 1.2.20.** By the definition of the operator norm we have $\|Au\| \leqslant \|A\|\|u\|$ for all $u \in \mathbb{R}^n$. If $\|A\| < 1$, then $T$ is a contraction of modulus $\|A\|$, since, for any $x, y \in U$,

$$\|Ax + b - Ay - b\| = \|A(x - y)\| \leqslant \|A\|\|x - y\|.$$

**Solution to Exercise 1.2.22.** From the bound in Exercise 1.2.21, we obtain

$$\|u_m - u_k\| \leqslant \frac{\lambda^m - \lambda^k}{1 - \lambda} \|u_0 - u_1\| \qquad (m, k \in \mathbb{N} \text{ with } m < k).$$

Hence $(u_m)$ is Cauchy, as claimed.

**Solution to Exercise 1.2.24.** Fix $\alpha \in (0, 1]$ and let $F$ be defined by $Fu = (1 - \alpha)u + \alpha Tu$. Readers will be able to verify that $F$ is also a contraction with identical fixed point $\bar{u}$, and that damped iteration is just iteration with $F$. The claim follows.

**Solution to Exercise 1.2.25.** By the definition of the derivative, for any $x \in U :=$ $(0, \infty)$, we have

$$\lim_{y \to x} \left| \frac{g(y) - g(x)}{y - x} - g'(x) \right| = 0.$$

Hence, by the reverse triangle inequality, for fixed $\varepsilon > 0$, we can take a $\delta > 0$ such that

$$\left| \frac{g(y) - g(x)}{y - x} \right| > |g'(x)| - \varepsilon = g'(x) - \varepsilon$$

for all $y \in (x - \delta, x + \delta)$. Rearranging gives

$$|g(x) - g(y)| > [g'(x) - \varepsilon]|x - y|$$

for all $y \in (x - \delta, x + \delta)$. But $g'(x) = s\alpha x^{\alpha - 1} + 1 - \delta$, which converges to $+\infty$ as $x \to 0$. It

follows that, for any $\lambda \in [0, 1)$, we can find a pair $x, y$ such that $|g(x) - g(y)| > \lambda |x - y|$. Hence $g$ is not a contraction map under $|\cdot|$.

**Solution to Exercise 1.2.27.** Let $m = \max_{x \in \mathsf{X}} h(x)$ and $M = \operatorname{argmax}_{x \in \mathsf{X}} h(x)$. Suppose first that $\varphi^*$ is supported on $M$ and let $\varphi$ be any distribution on $\mathsf{X}$. Then $\langle h, \varphi^* \rangle = m \geq \langle h, \varphi \rangle$. Conversely, if $\varphi^*$ is not supported on $M$, then $\langle h, \varphi^* \rangle < m$. In other words, $\varphi^* \in \operatorname{argmax}_{\varphi \in \mathcal{D}(\mathsf{X})} \langle h, \varphi \rangle$ if and only if $\varphi^*$ is supported on $M$.

**Solution to Exercise 1.2.28.** Fix $\tau \in [0, 1]$, $X \sim \varphi \in \mathcal{D}(\mathsf{X})$ and $\alpha \in \mathbb{R}$. Let $\Phi_X$ be the CDF of $X$. Let $Y := X + \alpha$, let $\mathsf{Y} := \{x + \alpha : x \in \mathsf{X}\}$ and let $\Phi_Y$ the CDF of $Y$. Note that $\Phi_Y(y) = \mathbb{P}\{Y \leq y\} = \mathbb{P}\{X \leq y - \alpha\} = \Phi_X(y - \alpha)$ for all $y \in \mathsf{Y}$.

Let $x^* := Q_\tau X$ and let $y^* = Q_\tau(X + \alpha) = \min\{y \in \mathsf{Y} : \Phi_Y(y) \geq \tau\}$. We need to show that $y^* = x^* + \alpha$. We do this by proving $y^* \geq x^* + \alpha$ and $y^* \leq x^* + \alpha$.

For the first inequality, fix $y \in \mathsf{Y}$ such that $\Phi_Y(y) \geq \tau$. Let $x = y - \alpha$. We then have $\Phi_Y(x + \alpha) \geq \tau$ and hence $\Phi_X(x) \geq \tau$. Hence $x \geq x^*$, or $y \geq x^* + \alpha$. Since this last inequality holds for any $y \in \mathsf{Y}$ with $\Phi_Y(y) \geq \tau$, we have $y^* \geq x^* + \alpha$.

For the reverse inequality, fix $x \in \mathsf{X}$ with $\Phi_X(x) \geq \tau$ and set $y = x + \alpha$. We have $\Phi_Y(y) = \Phi_X(y - \alpha) = \Phi_X(x) \geq \tau$, so $y \geq y^*$, or $x \geq y^* - \alpha$. Since the last inequality holds for all $x \in \mathsf{X}$ with $\Phi_X(x) \geq \tau$, we have $x^* \geq y^* - \alpha$. Rearranging gives $y^* \leq x^* + \alpha$, as was to be shown.

**Solution to Exercise 1.3.1.** Fix $\alpha, x, y \in \mathbb{R}$. We have $x = x - y + y \leq |x - y| + y$. Applying the result in Exercise 1.2.1 yields

$$\alpha \vee x \leq |x - y| + \alpha \vee y \quad \Longleftrightarrow \quad \alpha \vee x - \alpha \vee y \leq |x - y|.$$

Reversing the roles of $x$ and $y$ completes the proof.

**Solution to Exercise 2.1.1.** Let $(U, T)$ and $(\hat{U}, \hat{T})$ be conjugate under $\Phi$, with $\hat{T} \circ \Phi = \Phi \circ T$. The stated equivalence holds because

$$Tu = u \quad \Longleftrightarrow \quad \Phi Tu = \Phi u \quad \Longleftrightarrow \quad \hat{T}\Phi u = \Phi u.$$

**Solution to Exercise 2.1.3.** To show that $T = \Phi^{-1} \circ \hat{T} \circ \Phi$ holds, we can equivalently prove that $\Phi \circ T = \hat{T} \circ \Phi$. For $u \in \mathbb{R}$, we have $\Phi Tu = \ln A + \alpha \ln u$ and $\hat{T}\Phi u = \ln A + \alpha \ln u$. Hence $\Phi \circ T = \hat{T} \circ \Phi$, as was to be shown.

**Solution to Exercise 2.1.4.** From $\hat{T} = \Phi \circ T \circ \Phi^{-1}$ we have $\hat{T}^2 = \Phi \circ T \circ \Phi^{-1} \circ \Phi \circ T \circ \Phi^{-1} = \Phi \circ T^2 \circ \Phi^{-1}$ and, continuing in the same way (or using induction), $\hat{T}^k = \Phi \circ T^k \circ \Phi^{-1}$ for all $k \in \mathbb{N}$. Equivalently, $\hat{T}^k \circ \Phi = \Phi \circ T^k$ for all $k \in \mathbb{N}$. Hence, using continuity of $\Phi$ and $\Phi^{-1}$,

$$T^k u \to u^* \iff \Phi T^k u \to \Phi u^* \iff \hat{T}^k \Phi u \to \Phi u^*.$$

**Solution to Exercise 2.1.5.** Let $\mathsf{U}$ be the set of all dynamical systems $(U, T)$ with $U \subset \mathbb{R}^n$ and write $(U, T) \sim (\hat{U}, \hat{T})$ if these systems are topologically conjugate. It is easy to see that $\sim$ is reflexive and symmetric. Regarding transitivity, suppose that $(U, T) \sim (U', T')$ and $(U', T') \sim (U'', T'')$. Let $F$ be the homeomorphism from $U$ to $U'$ and $G$ be the homeomorphism from $U'$ to $U''$. Then $H := G \circ F$ is a homeomorphism from $U$ to $U''$ with inverse $(F \circ G)^{-1}$. Moreover, on $U$, we have

$$T = F^{-1} \circ T' \circ F = F^{-1} \circ G^{-1} \circ T'' \circ G \circ F = (GF)^{-1} \circ T'' \circ G \circ F.$$

Hence $(U, T) \sim (U'', T'')$ and $\sim$ is transitive, as required.

**Solution to Exercise 2.1.7.** Since $u_{k+1} = Tu_k$, we have

$$\frac{u_{k+1} - u^*}{u_k - u^*} = T'u^* + \frac{T''v_k}{2}(u_k - u^*).$$

Since $T$ is twice continuously differentiable, $T''v_k$ is bounded on bounded sets. As a result, taking absolute values and using $u_k \to u^*$ confirms the linear rate claimed in the exercise.

**Solution to Exercise 2.2.4.** It is easy to confirm that $\ll$ violates reflexivity.

**Solution to Exercise 2.2.6.** Just set $\mathsf{X} = [n] \times [k]$.

**Solution to Exercise 2.2.7.** Regarding the first claim, fix $B \in \mathbb{M}^{m \times k}$ with $b_{ij} \geqslant 0$ for all $i, j$. Pick any $i \in [m]$ and $u \in \mathbb{R}^k$. By the triangle inequality, we have $|\sum_j b_{ij} u_j| \leqslant \sum_j b_{ij} |u_j|$. Stacking these inequalities yields $|Bu| \leqslant B|u|$, as was to be shown.

Regarding the second, let $A$ and $(u_k)$ be as stated, with $u_{k+1} \leqslant Au_k$ for all $k$. We aim to prove $u_k \leqslant A^k u_0$ for all $k$ using induction. In doing so, we observe that $u_1 \leqslant Au_0$, so the claim is true at $k = 1$. Suppose now that it holds at $k - 1$. Then $u_k \leqslant Au_{k-1} \leqslant AA^{k-1} u_0 = A^k u_0$, where the last step used nonnegativity of $A$ and the induction hypothesis. The claim is now proved.

**Solution to Exercise 2.2.8.** Assume the stated conditions. Let $h := v - u$ and let $a_{ij}$ be the $i, j$-th element of $A$. We have $h \geqslant 0$ and $h_j > 0$ at some $j$. Hence $\sum_j a_{ij} h_j > 0$. This says that every row of $Ah$ is strictly positive. In other words $Ah = A(v - u) \gg 0$. The claim follows.

**Solution to Exercise 2.2.9.** Let $M = \{1, 2\}$, let $A = \{1\}$ and let $B = \{2\}$. Then $A \subset B$ and $B \subset A$ both fail. Hence $\subset$ is not a total order on $\wp(M)$.

**Solution to Exercise 2.2.10.** We prove the claim concerning greatest elements: Suppose that $g$ and $g'$ are greatest elements of $A$. Then, since both are in $A$, we have $g \preceq g'$ and $g' \preceq g$. Hence, by antisymmetry, $g = g'$.

**Solution to Exercise 2.2.11.** To see this, suppose first that $S \in \{A_i\}$. Since $A_j \subset \cup_i A_i =: S$ for all $j \in I$, the set $S$ is a greatest element of $\{A_i\}$. Conversely, if $S$ is not in $\{A_i\}$, then $S$ is not a greatest element (since the definition directly requires that $S \in \{A_i\}$.

**Solution to Exercise 2.2.12.** Since the union of all bounded subsets of $\mathbb{R}^n$ is $\mathbb{R}^n$ (which is not bounded), $\{A_i\}$ has no greatest element. Indeed, if $G$ is the greatest element of $\{A_i\}$, then $G$ contains every bounded subset of $\mathbb{R}^n$. But then $G$ is not bounded. Contradiction.

**Solution to Exercise 2.2.13.** Suppose that $s$ and $s'$ are both suprema of $A$ in $P$. Then both $s$ and $s'$ are upper bounds, so $s \preceq s'$ and $s' \preceq s$. Hence $s = s'$.

**Solution to Exercise 2.2.16.** To see the former, observe that $A_j \subset \cup_i A_i$ for all $j \in I$. Hence $\cup_i A_i$ is an upper bound of $\{A_i\}$. Moreover, if $B \subset M$ and $A_j \subset B$ for all $i \in I$, then $\cup_i A_i \subset B$. This proves that $\cup_i A_i$ is the supremum. The proof of the infimum case is similar.

**Solution to Exercise 2.2.17.** Here is one possible answer. Let $P = (0, 1)$, partially ordered by $\leqslant$. The set $A = [1/2, 1)$ is bounded above in $\mathbb{R}$ (and hence has a supremum in $\mathbb{R}$) but has no supremum in $P$. Indeed, if $s = \bigvee A$, the $s \in P$ and $a \leqslant s$ for all $s \in A$. It is clear that no such element exists.

**Solution to Exercise 2.2.21.** Suppose first that $v^* \in \{v_\sigma\}$. Since $v_\sigma \leqslant v^*$ for all $\sigma$, the function $v^*$ is the greatest element. Regarding the second claim, suppose (seeking a contradiction), that $v^* \notin \{v_\sigma\}$ and $\bar{v}$ is a greatest element of $\{v_\sigma\}$. By definition, $v_\sigma \leqslant \bar{v}$ for all $\sigma$, so taking the pointwise maximum gives $v^* \leqslant \bar{v}$. At the same time, since $\bar{v}$ is a greatest element, we have $\bar{v} \in \{v_\sigma\}$, and therefore $\bar{v} \leqslant \max_\sigma v_\sigma = v^*$. Putting the two inequalities together gives $\bar{v} = v^*$, which in turn implies that $v^* \in \{v_\sigma\}$. Contradiction.

**Solution to Exercise 2.2.22.** Let $I_a := [a_1, a_2]$ and $I_b := [b_1, b_2]$ be two order intervals in $V$. Consider the order interval $I := [a_1 \wedge b_1, a_2 \vee b_2]$. If $h \in I$, then $h \geqslant a_1 \wedge b_1$, so $h \geqslant a_1$ and $h \geqslant b_1$. A similar argument gives $h \leqslant a_2$ and $h \leqslant b_2$. Hence $h \in I_a \cap I_b$. Working in the other direction, it is not difficult to show that $h \in I_a \cap I_b$ implies $h \in I$. Hence $I = I_a \cap I_b$. In particular, $I_a \cap I_b$ is an order interval in $V$.

**Solution to Exercise 2.2.23.** Fix $a, b \in \mathbb{R}_+$ and $c \in \mathbb{R}_+$. By (2.3), we have

$$a \wedge c = (a - b + b) \wedge c \leqslant (|a - b| + b) \wedge c \leqslant |a - b| \wedge c + b \wedge c.$$

Thus, $a \wedge c - b \wedge c \leqslant |a - b| \wedge c$. Reversing roles of $a$ and $b$ gives $b \wedge c - a \wedge c \leqslant |a - b| \wedge c$. This proves the claim in Exercise 2.2.23.

**Solution to Exercise 2.2.24.** Since $\min f = -\max(-f)$ and similarly for $g$, we can apply Lemma 2.2.2 to obtain

$$|\min f - \min g| = |\max(-g) - \max(-f)| \leqslant \max |(-g) - (-f)| = \max |f - g|.$$

**Solution to Exercise 2.2.25.** We prove the claim regarding greatest elements. To this end, let $\bar{u}$ be the greatest element of $\{u_i\}$. Then $Fu_i \leqslant F\bar{u}$ for all $i$, so $F\bar{u}$ is the greatest, and hence the supremum, of $\{Fu_i\}$. That is, $\bigvee_i Fu_i = F\bar{u} = F \bigvee_i u_i$.

**Solution to Exercise 2.2.26.** Let $A$ and $P$ be as stated. The claim that $A^k$ is order-preserving on $P$ holds at $k = 1$. Suppose now that it holds at $k$ and fix $p, q \in P$ with $p \leqslant q$. By the induction hypothesis and the fact that $A$ is order-preserving, we have $AA^k p \leqslant AA^k q$. Hence $A^{k+1} p \leqslant A^{k+1} q$. We conclude that $A^{k+1}$ is also order-preserving, as was to be shown.

**Solution to Exercise 2.2.27.** Fix an $n \times k$ matrix $A$ with $A \geqslant 0$, along with $u, v \in \mathbb{R}^k$. We need to show that $u \leqslant v$ implies $Au \leqslant Av$ for any conformable vectors $u, v$. This holds because if $u \leqslant v$ we have $v - u \geqslant 0$, so $A(v - u) \geqslant 0$. But then $Av - Au \geqslant 0$, or $Au \leqslant Av$.

**Solution to Exercise 2.2.28.** Fix square $A, B$ with $0 \leqslant A \leqslant B$. It follows from the rules of matrix multiplication that, for arbitrary nonnegative square matrices $E, F, G$ with $F \leqslant G$, we have $EF \leqslant EG$ and $FE \leqslant GE$. Hence, if $A^k \leqslant B^k$ for some $k \in \mathbb{N}$, then $A^{k+1} = AA^k \leqslant BA^k \leqslant BB^k = B^{k+1}$. Thus, by induction, $A^k \leqslant B^k$ for all $k \in \mathbb{N}$, which verifies the first claim. Regarding the second, it is clear that for nonnegative matrices $E, F$ with $E \leqslant F$ we have $\|E\|_\infty \leqslant \|F\|_\infty$. Hence $\|A^k\|_\infty \leqslant \|B^k\|_\infty$ for all $k \in \mathbb{N}$. Raising both sides to the power $1/k$ and applying Gelfand's lemma verifies $\rho(A) \leqslant \rho(B)$.

**Solution to Exercise 2.2.30.** Take $(f_k)_{k \geqslant 1}$ in $i\mathbb{R}^P$ and $f \in \mathbb{R}^P$ with $f_k \to f$ as $k \to \infty$. Since $f_k \to f$ we have $f_k(z) \to f(z)$ for all $z \in P$. (Norm convergence implies pointwise convergence.) Fix $x, y \in P$ with $x \leq y$. From $(f_k) \subset i\mathbb{R}^P$ we have $f_k(x) \leqslant f_k(y)$ for all $k$. Since weak inequalities are preserved under limits, $f(x) \leqslant f(y)$. Hence $f \in i\mathbb{R}^P$.

**Solution to Exercise 2.2.32.** Set $\alpha_k := u(x_k)$ for all $k$ and $s_k := \alpha_k - \alpha_{k-1}$ with $\alpha_0 := 0$. Fix $x_j \in \mathsf{X}$. Then

$$\sum_{k=1}^n s_k \mathbb{1}\{x_j \geqslant x_k\} = \sum_{k=1}^j s_k = (\alpha_1 - \alpha_0) + (\alpha_2 - \alpha_1) + \ldots + (\alpha_j - \alpha_{j-1}) = \alpha_j.$$

In other words, $\sum_{k=1}^n s_k \mathbb{1}\{x_j \geqslant x_k\} = u(x_j)$. This completes the proofs.

**Solution to Exercise 2.2.36.** Fix $\varphi, \psi \in \mathsf{X}$ and suppose that $\varphi \preceq_F \psi$. Let $u \in \mathbb{R}^\mathsf{X}$ be defined by $u(1) = 0$ and $u(2) = 1$. Then, by the definition of stochastic dominance, we have $\varphi(2) \leqslant \psi(2)$. Since $\varphi(1) = 1 - \varphi(2)$ and $\psi(1) = 1 - \psi(2)$, this inequality is equivalent to $\psi(1) \leqslant \varphi(1)$. Finally, suppose that $\psi(1) \leqslant \varphi(1)$ and fix $u \in i\mathbb{R}^\mathsf{X}$. Let $h = u(2) - u(1) \geqslant 0$. Then

$$\sum_x u(x)\varphi(x) = u(1)\varphi(1) + (u(1) + h)(1 - \varphi(1)) = u(1) + h(1 - \varphi(1)).$$

Similarly, $\sum_x u(x)\psi(x) = u(1) + h(1 - \psi(1))$. Since $h \geqslant 0$ and $\psi(1) \leqslant \varphi(1)$, we have $\sum_x u(x)\varphi(x) \leqslant \sum_x u(x)\psi(x)$. Thus, $\varphi \preceq_F \psi$. This chain of implications proves the equivalences in the exercise.

**Solution to Exercise 2.2.37.** Suppose $f, g, h \in \mathcal{D}(\mathsf{X})$ with $f \preceq_F g$ and $g \preceq_F h$. Fixing $u \in i\mathbb{R}^\mathsf{X}$, we have

$$\sum_x u(x)f(x) \leqslant \sum_x u(x)g(x) \quad \text{and} \quad \sum_x u(x)g(x) \leqslant \sum_x u(x)h(x)$$

Hence $\sum_x u(x)f(x) \leqslant \sum_x u(x)h(x)$. Since $u$ was arbitrary in $i\mathbb{R}^\mathsf{X}$, we are done.

**Solution to Exercise 2.2.38.** Let $F^\varphi := 1 - G^\varphi$ be the CDF of $\varphi$ and let $F^\psi$ be the CDF of $\psi$. In view of Lemma 2.2.5, we have $F^\psi \leqslant F^\varphi$. As a consequence,

$$\{x \in \mathsf{X} : F^\psi(x) \geqslant \tau\} \subset \{x \in \mathsf{X} : F^\varphi(x) \geqslant \tau\}.$$

It follows directly that

$$\min\{x \in \mathsf{X} : F^\varphi(x) \geqslant \tau\} \leqslant \min\{x \in \mathsf{X} : F^\psi(x) \geqslant \tau\}.$$

That is, $Q_\tau(X) \leqslant Q_\tau(Y)$.

**Solution to Exercise 2.2.39.** Let $P, S, T$ be as described in the exercise. We aim to show that $S^k \leqslant T^k$ holds for all $k \in \mathbb{N}$. Clearly it holds for $k = 1$. If it also holds at $k-1$, then, for any $u \in P$, we have $S^k u = SS^{k-1} u \leqslant ST^{k-1} u \leqslant TT^{k-1} u = T^k u$, where we used the induction hypothesis, the order-preserving property of $S$ and the assumption that $S \leqslant T$.

**Solution to Exercise 2.2.42.** Fix $\beta_1 \leqslant \beta_2$. Let $g_1$ and $g_2$ be the corresponding fixed point maps, as defined in (1.33). Since $\beta_1 \leqslant \beta_2$, we have $g_1(h) \leqslant g_2(h)$ for all $h \in \mathbb{R}_+$ and, in addition, $g_2$ is a contraction map (and hence globally stable), Proposition 2.2.7 applies. In particular, the fixed point $h_1^*$ corresponding to $\beta_1$ is less than or equal to $h_2^*$, the fixed point corresponding to $\beta_2$.

**Solution to Exercise 2.3.1.** Let $A$ be as stated and let $e$ be the right eigenvector in (2.10). Since $e$ is nonnegative and nonzero, and since eigenvectors are defined only up to constant multiples, we can and do assume that $\sum_j e_j = 1$. From $Ae = \rho(A)e$ we have $\sum_j a_{ij} e_j = \rho(A)e_i$ for all $i$. Summing with respect to $i$ gives $\sum_j \mathrm{colsum}_j(A)e_j = \rho(A)$. Since the elements of $e$ are nonnegative and sum to one, $\rho(A)$ is a weighted average of the column sums. Hence the second pair of bounds in Lemma 2.3.2 holds. The remaining proof is similar (use the left eigenvector).

**Solution to Exercise 2.3.2.** Let $P$ and $Q$ be as stated. Evidently $PQ \geqslant 0$. Moreover, $PQ\mathbb{1} = P\mathbb{1} = \mathbb{1}$, so $PQ$ is Markov. That $\rho(P) = 1$ follows directly from Lemma 2.3.2. By the Perron–Frobenius theorem, there exists a nonzero, nonnegative row vector $\varphi$ satisfying $\varphi P = \varphi$. Rescaling $\varphi$ to $\varphi/(\varphi\mathbb{1})$ gives the desired vector $\psi$.

The final positivity and uniqueness claim is also by the Perron–Frobenius theorem, and its consequences for irreducible matrices. Indeed, if $\varphi$ is another nonnegative vector satisfying $\varphi\mathbb{1} = 1$ and $\varphi P = \varphi$, then, by the Perron–Frobenius theorem, $\varphi = \alpha\psi$ for some $\alpha > 0$. But then $\alpha\psi\mathbb{1} = 1$ and $\psi\mathbb{1} = 1$, which gives $\alpha = 1$. Hence $\varphi = \psi$.

**Solution to Exercise 2.3.3.** Let $P$ and $\varepsilon$ have the stated properties. Fix $h \in \mathbb{R}^{\mathsf{X}}$. It suffices to show that for this arbitrary $h$ we can find an $x \in \mathsf{X}$ such that $(Ph)(x) < h(x) + \varepsilon$. This is easy to verify, since, for $\bar{x} \in \operatorname{argmax}_{x \in \mathsf{X}} h(x)$ we have $(Ph)(\bar{x}) = \sum_{x'} h(x') P(\bar{x}, x') \leqslant h(\bar{x})$.

**Solution to Exercise 2.3.4.** It is straightforward to confirm that both columns of $A$ sum to $1 + g$. As a result, with $\mathbb{1}^{\top}$ as a row vector of ones, we have

$$n_{t+1} = \mathbb{1}^{\top} x_{t+1} = \mathbb{1}^{\top} A x_t = (1+g)\mathbb{1}^{\top} x_t = (1+g)n_t,$$

as was to be shown.

**Solution to Exercise 2.3.10.** Fix $L \in \mathcal{L}(\mathbb{R}^{\mathsf{X}})$ with $(Lu)(x) = \sum_{x' \in \mathsf{X}} L(x, x')u(x')$ for all $x \in \mathsf{X}$ and $u \in \mathbb{R}^{\mathsf{X}}$. Positivity of $L$ requires that

$$u \geqslant 0 \implies \sum_{x' \in \mathsf{X}} L(x, x')u(x') \geqslant 0 \text{ for all } x \in \mathsf{X}.$$

Clearly, this holds whenever $L(x, x') \geqslant 0$ for all $x, x' \in \mathsf{X}$.

Regarding the converse, suppose that $L$ is positive. Seeking a contradiction, suppose in addition that we can find a pair $(x_a, x_b) \in \mathsf{X} \times \mathsf{X}$ such that $L(x_a, x_b) < 0$. With $u(x) := \mathbb{1}\{x = x_b\}$, we have $(Lu)(x_a) = \sum_{x' \in \mathsf{X}} L(x_a, x')u(x') = L(x_a, x_b) < 0$. This contradicts positivity of $L$.

**Solution to Exercise 2.3.11.** Suppose first that $L$ is positive. Fix $u \leqslant v$ in $\mathbb{R}^{\mathsf{X}}$ and observe that, by positivity, $L(v - u) \geqslant 0$. But then $Lv - Lu \geqslant 0$ and hence $Lu \leqslant Lv$. This shows that $L$ is order-preserving.

Regarding the converse, if $L$ is order-preserving, then $u \geqslant 0$ implies $Lu \geqslant L0$. But for every linear operator we have $L0 = 0$, and so $Lu \geqslant 0$. Hence $L$ is a positive operator.

**Solution to Exercise 2.3.12.** Fix $P \in \mathcal{L}(\mathbb{R}^{\mathsf{X}})$ and let $P(x, x')$ be the matrix representation, so that

$$(Pu)(x) = \sum_{x'} P(x, x')u(x') \qquad (x \in \mathsf{X})$$

for any $u \in \mathbb{R}^{\mathsf{X}}$. Suppose first that $P \in \mathcal{M}(\mathbb{R}^{\mathsf{X}})$. The statement that $P$ is a positive linear operator is equivalent to $P(x, x') \geqslant 0$ for all $x, x'$ by Lemma 2.3.6. Moreover, $P\mathbb{1} = \mathbb{1}$ is equivalent to $\sum_{x' \in \mathsf{X}} P(x, x') = 1$ for all $x \in \mathsf{X}$.

**Solution to Exercise 2.3.14.** In the solution below, we use the characterization in Exercise 2.3.12: $P \in \mathcal{M}(\mathbb{R}^{\mathsf{X}})$ if and only if $P(x, x') \geqslant 0$ for all $x, x' \in \mathsf{X}$ and $\sum_{x' \in \mathsf{X}} P(x, x') = 1$ for all $x \in \mathsf{X}$.

Fix $P \in \mathcal{L}(\mathbb{R}^{\mathsf{X}})$ and suppose first that $P \in \mathcal{M}(\mathbb{R}^{\mathsf{X}})$. Then

$$(\varphi P)(x') = \sum_x P(x, x') \varphi(x) \qquad (x' \in \mathsf{X}) \tag{2.20}$$

is in $\mathcal{D}(\mathsf{X})$ whenever $\varphi \in \mathcal{D}(\mathsf{X})$, since, for any such $\varphi$, the vector $\varphi P$ is clearly nonnegative and

$$\sum_{x'} (\varphi P)(x') = \sum_x \sum_{x'} P(x, x') \varphi(x) = \sum_x \varphi(x) = 1.$$

Now suppose instead that $P \in \mathcal{L}(\mathbb{R}^{\mathsf{X}})$ and $\varphi P \in \mathcal{D}(\mathsf{X})$ whenever $\varphi \in \mathcal{D}(\mathsf{X})$. It follows that $P(x, x')$ is nonnegative at arbitrary $(x, x')$, since $(\varphi P)(x') = P(x, x')$ when $\varphi$ is the distribution that puts all mass on $x$. Moreover, $P(x, \cdot)$ must sum to one at arbitrary $x$ because if $\varphi$ is the distribution that puts all mass on $x$, then

$$1 = \sum_{x'} (\varphi P)(x') = \sum_{x'} P(x, x').$$

**Solution to Exercise 3.1.1.** Let $X_t = x \in \mathsf{X}$, so that $X_{t+1} = \max\{x - D_{t+1}, 0\} + S\mathbb{1}\{x \leqslant s\}$. Evidently $X_{t+1}$ is integer-valued and nonnegative. If $x \leqslant s$, then $X_{t+1} \leqslant \max\{s - D_{t+1}, 0\} + S \leqslant s + S$. Similarly, if $s < x \leqslant S + s$, then $X_{t+1} \leqslant \max\{x - D_{t+1}, 0\} \leqslant S + s$. The claim is verified.

**Solution to Exercise 3.1.2.** Fixing $t \geqslant 0$ and $P \in \mathcal{M}(\mathbb{R}^{\mathsf{X}})$, this claim can be verified by induction over $k$. The claim is obviously true when $k = 0, 1$. Suppose the claim is also true at $k$ and now consider the case $k + 1$. By the law of total probability, for given $x, x' \in \mathsf{X}$, we have

$$\mathbb{P}\{X_{t+k+1} = x' \mid X_t = x\} = \sum_z \mathbb{P}\{X_{t+k+1} = x' \mid X_{t+k} = z\} \mathbb{P}\{X_{t+k} = z \mid X_t = x\}.$$

The induction hypothesis allows us to use (3.2) at $k$, so the last equation becomes

$$\mathbb{P}\{X_{t+k+1} = x' \mid X_t = x\} = \sum_z P^k(x, z) P(z, x') = P^{k+1}(x, x').$$

The law (3.2) is now verified at $k + 1$, completing our proof by induction.

**Solution to Exercise 3.1.3.** Let $x \in \mathsf{X}$ be the current state at time $t$ and suppose first that $s < x$. The next period state $X_{t+1}$ hits $s$ with positive probability, since $\varphi(d) > 0$ for all $d \in \mathbb{Z}_+$. The state $X_{t+2}$ hits $S + s$ with positive probability, since $\varphi(0) > 0$. From $S + s$, the inventory level reaches any point in $\mathsf{X} = \{0, \ldots, S + s\}$ in one step with positive probability. Hence, from current state $x$, inventory reaches any other state $y$ with positive probability in three steps.

The logic for the case $x \leqslant s$ is similar and left to the reader.

**Solution to Exercise 3.1.4.** Fix $t \in \mathbb{N}$. Under the stated hypotheses, we have $X_t \overset{d}{=} \psi_0 P^t$ (see (3.5)). Hence

$$\mathbb{E}h(X_t) = \sum_{x'} h(x') \mathbb{P}\{X_t = x'\} = \sum_{x'} h(x')(\psi_0 P^t)(x') = \langle \psi_0 P^t, h \rangle .$$

**Solution to Exercise 3.1.7.** Assume $P$ is everywhere positive with unique stationary distribution $\psi^*$. Since $\rho(P) = 1$, the last part of the Perron–Frobenius theorem tells us that $P^t \to e \varepsilon$ as $t \to \infty$, where $e$ and $\varepsilon$ are the dominant right and left eigenvectors, normalized such that $\langle e, \varepsilon \rangle = 1$. In this case we know $\psi^*$ is the dominant left eigenvector and $\mathbb{1}$ is the dominant right eigenvector. Moreover, $\psi^* \in \mathscr{D}(\mathsf{X})$ yields $\langle \psi^*, \mathbb{1} \rangle = 1$. Hence, for any $\psi \in \mathscr{D}(\mathsf{X})$, we have

$$\psi P^t \to \psi \mathbb{1} \psi^* = \psi^* \quad \text{as} \quad t \to \infty.$$

Hence global stability holds, as claimed.

**Solution to Exercise 3.1.11.** Since we are conditioning on $X_t = x$, we can replace $X_{t+1}$ with $\rho x + \nu \varepsilon_{t+1}$. The result then follows from $\mathbb{P}\{\alpha < \nu \varepsilon_{t+1} \leqslant \beta\} = F(\beta) - F(\alpha)$.

**Solution to Exercise 3.2.2.** Using Exercise 3.1.11 and the definition of $P$, it can be shown that

$$G(x, x_k) := \sum_{j=k}^{n} P(x, x_j) = \mathbb{P}\{x_k - s/2 < X_{t+1} \mid X_t = x\}.$$

Rewriting the probability in terms of $\varepsilon_{t+1}$, we get

$$G(x, x_k) = \mathbb{P}\{\varepsilon_{t+1} > (x_k - s/2 - \rho x)/\sigma\}.$$

Since $\rho \geqslant 0$, we can now see that $x \leqslant y$ implies $G(x, x_k) \leqslant G(y, x_k)$ for all $k$, or,

equivalently, $G(x, \cdot) \leqslant G(y, \cdot)$ pointwise on X. By Lemma 2.2.5, this is equivalent to the statement that $P(x, \cdot) \preceq_F P(y, \cdot)$, which confirms that $P$ is monotone increasing.

**Solution to Exercise 3.2.3.** This matrix $P_w$ is monotone increasing if and only if $(1 - \alpha, \alpha) \preceq_F (\beta, 1 - \beta)$. From Exercise 2.2.36, we know that this is equivalent to $\beta \leqslant 1 - \alpha$, or $\beta + \alpha \leqslant 1$.

**Solution to Exercise 3.2.4.** Suppose that $P$ is monotone increasing and fix $h \in i\mathbb{R}^X$. We claim that $Ph \in i\mathbb{R}^X$. To see this, pick any $x, y \in X$ with $x \preceq y$. Since $x \preceq y$ we have $P(x, \cdot) \preceq_F P(y, \cdot)$. Hence $\sum_{x'} h(x')P(x, x') \leqslant \sum_{x'} h(x')P(y, x')$. This shows that $Ph \in i\mathbb{R}^X$.

To see the converse, suppose that $P$ is invariant on $i\mathbb{R}^X$. Fix $x, y \in X$ with $x \preceq y$. We claim that $P(x, \cdot) \preceq_F P(y, \cdot)$. To see this, fix $u \in i\mathbb{R}^X$. $Pu \in i\mathbb{R}^X$ by invariance, so $(Pu)(x) \leqslant (Pu)(y)$ and hence $\sum_{x'} u(x')P(x, x') \leqslant \sum_{x'} u(x')P(y, x')$. Since $u$ was chosen arbitrarily from $i\mathbb{R}^X$, we have $P(x, \cdot) \preceq_F P(y, \cdot)$. Hence $P$ is monotone increasing, as was to be shown.

**Solution to Exercise 3.2.5.** Clearly this is true for $t = 1$. Suppose it is also true for arbitrary $t$. Then, for any $h \in i\mathbb{R}^X$, the function $P^t h$ is again in $i\mathbb{R}^X$. From this it follow that $P^{t+1}h = PP^t h$ is also in $i\mathbb{R}^X$, since $P$ is monotone increasing. This proves that $P^{t+1}$ is invariant on $i\mathbb{R}^X$, and therefore monotone increasing.

**Solution to Exercise 3.2.6.** Let $\pi$ and $P$ satisfy the stated conditions. By Exercise 3.2.5, $P^t$ is monotone increasing for all $t$. By this fact and the assumption $\pi \in i\mathbb{R}^X$, we see that $P^t \pi \in i\mathbb{R}^X$ for all $t$. Hence $v = \sum_{t \geqslant 0} \beta^t P^t \pi$ is also increasing.

**Solution to Exercise 3.2.8.** Both $u$ and $\exp$ are increasing on X, so $r$ is in $i\mathbb{R}^X$. Since $\rho \geqslant 0$, $P$ is monotone increasing (see §3.2.1.3). Clearly $\beta P$ shares this property. It follows that $\beta Pr \in i\mathbb{R}^X$. Applying $\beta P$ again, we have $(\beta P)^2 r \in i\mathbb{R}^X$. Continuing in this way, we see that $(\beta P)^k r$ is increasing for all $k$. Hence $\sum_{k \geqslant 0}(\beta P)^k r$ is increasing. By the Neumann series lemma, this sum is equal to $v$, so $v \in i\mathbb{R}^X$.

**Solution to Exercise 3.3.1.** We start with part (i). To show that $T$ is a self-map on $V := \mathbb{R}_+^W$, we just need to verify that $v \in V$ implies $Tv \in V$, which only requires us to verify that $T$ maps nonnegative functions into nonnegative functions. This is clear from the definition. Regarding the order-preserving property, fix $f, g \in V$ with $f \leqslant g$. We claim that $Tf \leqslant Tg$. Indeed, if $w \in W$, then $\sum_{w' \in W} f(w')P(w, w') \leqslant \sum_{w' \in W} g(w')P(w, w')$, which in turn implies that $(Tf)(w) \leqslant (Tg)(w)$. Since $w$ was an arbitrary wage value, we have $Tf \leqslant Tg$, so $T$ is order-preserving.

Regarding part (ii), let $e(w) := w/(1 - \beta)$ and fix $f, g$ in $V$. Writing the operators pointwise and applying the last result in Lemma 2.2.1 (page 58) gives

$$\begin{aligned}
|Tf - Tg| &= |e \vee (c + \beta Pf) - e \vee (c + \beta Pg)| \\
&\leqslant |\beta Pf - \beta Pg| \\
&= \beta |P(f - g)| \\
&\leqslant \beta P |f - g|.
\end{aligned}$$

(Here the last inequality uses the result in Exercise 2.2.7 on page 53.) Since $P \geqslant 0$ we have $P|f - g| \leqslant P\|f - g\|_\infty \mathbb{1} = \|f - g\|_\infty \mathbb{1}$, so

$$|Tf - Tg| \leqslant \beta \|f - g\|_\infty \mathbb{1}.$$

Taking the maximum on both sides gives $\|Tf - Tg\|_\infty \leqslant \beta \|f - g\|_\infty$. Since $f, g$ were arbitrary elements of $V$, the contraction claim is verified.

**Solution to Exercise 3.3.2.** The code in Listing 10 creates a Markov chain via Tauchen approximation of an AR(1) process with positive autocorrelation parameter. By Exercise 3.2.2, $P$ is monotone increasing. Hence, by Lemma 3.3.1, the value function is increasing. Since $h^* = c + \beta Pv^*$, it follows that $h^*$ is increasing. Regarding intuition, positive autocorrelation in wages means that high current wages predict high future wages. It follows that the value of waiting for future wages rises with current wages.

**Solution to Exercise 3.3.5.** Let $T$ be the operator on $V$ such that $(Tv_u)(w)$ is the right-hand side of (3.28). To solve the exercise, it suffices to prove that $T$ is a contraction map on $V$. (Then $v_u$ can be obtained, in the limit, by applying successive approximation to $T$ and, once the approximate fixed point is computed, $v_e$ can be obtained via (3.27).) To show that $T$ is a contraction, we let $T_1$ and $T_2$ be the operators on $V$ defined by

$$(T_1 v)(w) = \frac{1}{1 - \beta(1 - \alpha)} (w + \alpha\beta(Pv)(w)) \quad \text{and} \quad (T_2 v)(w) = c + \beta (Pv)(w).$$

Since $Tv = (T_1 v) \vee (T_2 v)$, Lemma 2.2.3 on page 59 tells us that $T$ will be a contraction provided that $T_1$ and $T_2$ are both contraction maps. For the case of $T_2$, we have

$$\|T_2 f - T_2 g\|_\infty = \max_w |c + \beta (Pf)(w) - c - \beta (Pg)(w)| \leqslant \max_w \beta \sum_{w'} |f(w') - g(w')| P(w, w').$$

The last term is dominated by $\beta \|f - g\|_\infty$, so $T_2$ is a contraction. The proof for $T_1$ is similar in spirit and hence left to the reader.

**Solution to Exercise 4.1.1.** Pointwise on X we have $1 - \sigma \leqslant 1$, so $L_\sigma \leqslant \beta P$. By Exercise 2.2.28 on page 60, we then have $\rho(L_\sigma) \leqslant \rho(\beta P) = \beta < 1$.

**Solution to Exercise 4.1.2.** Fix $\sigma \in \Sigma$. If $f, g \in \mathbb{R}^{\mathsf{X}}$, $f \leqslant g$ and $x \in \mathsf{X}$, then

$$(T_\sigma g)(x) - (T_\sigma f)(x) = (1 - \sigma(x)) \left[ \beta \sum_{x'} g(x') P(x, x') - \beta \sum_{x'} f(x') P(x, x') \right]$$

$$= (1 - \sigma(x)) \beta \sum_{x'} (g(x') - f(x')) P(x, x').$$

Since $g(x') \geqslant f(x')$ for all $x'$ we have $(T_\sigma g)(x) \geqslant (T_\sigma f)(x)$ for all $x$.

**Solution to Exercise 4.1.3.** Fix $\sigma \in \Sigma$. Given $f, g \in \mathbb{R}^{\mathsf{X}}$ and $x \in \mathsf{X}$, we have

$$|(T_\sigma f)(x) - (T_\sigma g)(x)| = \left| (1 - \sigma(x)) \beta \sum_{x'} (g(x') - f(x')) P(x, x') \right|$$

$$\leqslant \beta \left| \sum_{x'} [f(x') - g(x')] P(x, x') \right|.$$

Applying the triangle inequality and $\sum_{x'} P(x, x') = 1$, we obtain

$$|(T_\sigma f)(x) - (T_\sigma g)(x)| \leqslant \beta \sum_{x'} |f(x') - g(x')| P(x, x') \leqslant \beta \|f - g\|_\infty.$$

Taking the supremum over all $x$ on the left hand side of this expression leads to

$$\|T_\sigma f - T_\sigma g\|_\infty \leqslant \beta \|f - g\|_\infty.$$

Since $f, g$ were arbitrary elements of $\mathbb{R}^{\mathsf{X}}$, the contraction claim is proved.

**Solution to Exercise 4.1.4.** Fix $f, g \in \mathbb{R}^{\mathsf{X}}$ with $f \leqslant g$. Since $P \geqslant 0$, we have $Pf \leqslant Pg$. Hence $c + \beta Pf \leqslant c + \beta Pg$. As a result,

$$Tf = e \vee (c + \beta Pf) \leqslant e \vee (c + \beta Pg) = Tg.$$

**Solution to Exercise 4.1.5.** This result follows from Lemma 2.2.3 on page 59. For the sake of the exercise, we also provide a direct proof:

Take any $f, g$ in $\mathbb{R}^X$. Writing the operators pointwise and applying the last result in Lemma 2.2.1 (page 58) gives

$$
\begin{aligned}
|Tf - Tg| &= |e \vee (c + \beta Pf) - e \vee (c + \beta Pg)| \\
&\leqslant |\beta Pf - \beta Pg| \\
&= \beta |P(f - g)| \\
&\leqslant \beta P |f - g|.
\end{aligned}
$$

(Here the last inequality uses the result in Exercise 2.2.7 on page 53.) Since $P \geqslant 0$ we have $P|f - g| \leqslant P\|f - g\|_\infty \mathbb{1} = \|f - g\|_\infty \mathbb{1}$, so

$$
|Tf - Tg| \leqslant \beta \|f - g\|_\infty \mathbb{1}.
$$

Taking the maximum on both sides gives $\|Tf - Tg\|_\infty \leqslant \beta \|f - g\|_\infty$. Since $f, g$ were arbitrary elements of $\mathbb{R}^X$, the contraction claim is verified.

**Solution to Exercise 4.1.7.** First observe that, since $v^* \geqslant w$ and $T$ is order-preserving, we have $v^* = Tv^* \geqslant Tw = s \vee (\pi + \beta Qw) = s \vee w$. From this we get $v^* \geqslant s \vee w$ and applying $T$ to both sides gives $v^* \geqslant T(s \vee w)$.

Next, observe that

$$
T(s \vee w) = s \vee (\pi + \beta Q(s \vee w)) \geqslant \pi + \beta Q(s \vee w) \gg \pi + \beta Qw = w
$$

where the strict inequality is by Exercise 2.2.8 on page 53. We conclude that $v^* \geqslant T(s \vee w) \gg w$, as was to be shown.

Intuitively, the option to exit adds value to firms everywhere in the state space, since $Q \gg 0$ implies that the state can shift to a region of the state space where exit is optimal in a later period.

**Solution to Exercise 4.1.8.** For the model described, the Bellman equation takes the form

$$
v(p) = \max \left\{ s, \max_{\ell \geqslant 0} \pi(\ell, p) + \beta \sum_{p'} v(p') Q(p, p') \right\}.
$$

Straightforward calculus shows that maximized one-period profits are $\pi(p) = p^2/(4w)$. Hence the final expression is

$$
v(p) = \max \left\{ s, \frac{p^2}{4w} + \beta \sum_{p'} v(p') Q(p, p') \right\}
$$

**Solution to Exercise 4.1.9.** Fix $x, x' \in \mathsf{X}$ with $x \leqslant x'$. Since $\sigma^*$ is binary, to show $\sigma^*$ is decreasing it suffices to show that $\sigma^*(x) = 0$ implies $\sigma^*(x') = 0$. Hence we suppose that $\sigma^*(x) = 0$. This in turn implies that $e(x) < h^*(x)$. As $x \leqslant x'$, $e$ is decreasing and $h^*$ is increasing on $\mathsf{X}$, we have $e(x') < h^*(x')$. Hence $\sigma^*(x') = 0$. We conclude that $\sigma^*$ is decreasing on $\mathsf{X}$, as claimed.

**Solution to Exercise 4.1.11.** The solution to Exercise 4.1.11 is similar to that of Exercise 4.1.9 and hence omitted.

**Solution to Exercise 4.1.12.** Either by manipulating the Bellman equation or appealing to (4.15) on page 119, we see that the continuation value operator is defined at $h \in \mathbb{R}^{\mathsf{Z}}$ by

$$(Ch)(z) = \pi(z) + \beta \sum_{z'} \int \max\{s', h(z')\} \varphi(s') \, \mathrm{d}s' Q(z, z') \qquad (z \in \mathsf{Z}).$$

The next period scrap value $S_{t+1}$ is integrated out and the remaining function depends only on $z \in \mathsf{Z}$.

**Solution to Exercise 4.1.13.** Let $\varphi_a$ and $\varphi_b$ be as stated. For $i \in \{a, b\}$ and $h \in \mathbb{R}^{\mathsf{Z}}$, let

$$(C_i h)(z) = \pi(z) + \beta \sum_{z'} \int \max\{s', h(z')\} \varphi_i(s') \, \mathrm{d}s' Q(z, z').$$

Since, for each $z' \in \mathsf{Z}$, the function $s' \mapsto \max\{s', h(z')\}$ is increasing, we have

$$\sum_{z'} \int \max\{s', h(z')\} \varphi_a(s') \, \mathrm{d}s' Q(z, z') \leqslant \sum_{z'} \int \max\{s', h(z')\} \varphi_b(s') \, \mathrm{d}s' Q(z, z').$$

Hence $C_a h \leqslant C_b h$ for all $h \in \mathbb{R}^{\mathsf{Z}}$. As $C_b$ is order-preserving and globally stable, Proposition 2.2.7 on page 66 implies that the fixed point of $C_b$ dominates the fixed point of $C_a$. That is, $h_a^* \leqslant h_b^*$. But then, for any $z \in \mathsf{Z}$, we have $h_a^*(z) \leqslant h_b^*(z)$ and hence

$$\sigma_b^*(z) = \mathbb{1}\{s \geqslant h_b^*(z)\} \leqslant \mathbb{1}\{s \geqslant h_a^*(z)\} = \sigma_a^*(z).$$

The interpretation of $\sigma_b^* \leqslant \sigma_a^*$ is that firm exits at fewer states when the scrap value distribution is $\varphi_b^*$. This makes sense, since the current scrap value offer $s$ is already known, while future offers are more promising under $\varphi_b^*$ than $\varphi_a^*$. Hence continuing is more attractive.

**Solution to Exercise 4.2.1.** In view of (4.13), the continuation value operator for this problem is

$$(Ch)(x) = -c + \beta \sum_{x'} \max \{\pi(x'), h(x')\} P(x, x') \qquad (x \in \mathsf{X}).$$

The monotonicity result for $h^*$ follows from Lemma 4.1.4 on page 115.

**Solution to Exercise 4.2.2.** If $(X_t)$ is IID with common distribution $\varphi$, then the continuation value $h^*$ is constant; in particular, it is the unique solution to

$$h = -c + \beta \sum_{x'} \max \{\pi(x'), h(x')\} \varphi(x').$$

Since constant functions are (weakly) decreasing, Exercise 4.1.11 applies and $\sigma^*$ is increasing. Intuitively, the value of waiting is independent of the current state, while the value of bringing the product to market is increasing in the current state. Hence, if the firm brings to the product to market in state $x$, then it should also do so at any $x' \geqslant x$.

**Solution to Exercise 5.1.2.** The stochastic kernel is

$$P(x, a, x') = \begin{cases} 0 & \text{if } x' < a \\ (1-p)^x & \text{if } x' = a \\ (1-p)^{x+a-x'} p & \text{if } x' > a \end{cases} \qquad (5.10)$$

The middle case is obtained by observing that the next period state hits $x'$ when $x' = a$ if and only if $D_{t+1} \geqslant x$ and then using the expression for the geometric distribution.

**Solution to Exercise 5.1.3.** $T$ is a sup norm contraction mapping on $\mathbb{R}^{\mathsf{X}}$ because, in view of the max-inequality lemma (page 58), for any $v, w$ in $\mathbb{R}^{\mathsf{X}}$,

$$|(Tv)(x)| - (Tw)(x)| \leqslant \beta \max_{a \in \Gamma(x)} \left| \sum_{d \geqslant 0} [v(f(x, a, d)) - w(f(x, a, d))] \varphi(d) \right|$$

$$\leqslant \beta \max_{a \in \Gamma(x)} \sum_{d \geqslant 0} |v(f(x, a, d)) - w(f(x, a, d))| \varphi(d)$$

Since $\sum_{d \geqslant 0} \varphi(d) = 1$, it follows that, for arbitrary $x \in \mathsf{X}$,

$$|(Tv)(x) - (Tw)(x)| \leqslant \beta \|v - w\|_\infty$$

Taking the supremum over all $x \in \mathsf{X}$ yields the desired result.

**Solution to Exercise 5.1.4.** We take the action $A_t$ to be the choice of next period wealth $W_{t+1}$, so that the action space is also $\mathsf{W}$. The feasible correspondence is

$$\Gamma(w) = \{a \in \mathsf{W} : a \leqslant Rw\} \qquad (w \in \mathsf{W}),$$

implying that $\mathsf{G} = \{(w, a) \in \mathsf{W} \times \mathsf{W} : a \leqslant Rw\}$. The current reward is utility of consumption, or

$$r(w, a) = u\left(w - \frac{a}{R}\right) \qquad ((w, a) \in \mathsf{G}).$$

The stochastic kernel is $P(w, a, w') = \mathbb{1}\{w' = a\}$. This just states that next period wealth $w'$ is equal to the action $a$ with probability one.

**Solution to Exercise 5.1.5.** To impose that workers never leave the firm, we require $a \geqslant e$. Thus, the feasible correspondence is

$$\Gamma(x) = \Gamma(e, w) = \{a \in \{0, 1\} : a \geqslant e\}.$$

The set of feasible state-action pairs is $\mathsf{G} = \{((e, w), a) \in \mathsf{X} \times \mathsf{A} : a \geqslant e\}$. The reward function is

$$r(x, a) = r((e, w), a) = aw + (1 - a)c.$$

Regarding the stochastic kernel, we need to define state transition probabilities for all feasible state-action pairs. Letting $P[(e, w), a, (e', w')]$ be the probability of transitioning to state $(e', w')$ given current state $(e, w)$ and current action $a \leqslant e$, we set

$$P[(0, w), a, (e', w')] = \mathbb{1}\{e' = a\} \cdot [\, a\mathbb{1}\{w' = w\} + (1 - a)Q(w, w')\,] \qquad (5.14)$$

and $P[(1, w), 1, (e', w')] = \mathbb{1}\{e' = 1\}\mathbb{1}\{w' = w\}$. Equation (5.14) says that if $a = 0$ then $e' = 0$ and the next wage is drawn from $Q(w, w')$, while if $a = 1$ then $e' = 1$ and the next wage is $w$. You can verify that $P$ is a stochastic kernel from $\mathsf{G}$ to $\mathsf{X}$.

To double check that these definitions work, we can verify that they lead to the same Bellman equations that we saw in §3.3.1. Under the definitions of $\Gamma$, $r$ and $P$ just provided, we have $v(1, w) = w + \beta \mathbb{E}v(1, w)$. This implies that $v(1, w) = w/(1 - \beta)$, which is what we expect for lifetime value of an agent employed with wage $w$.

Moreover, the Bellman equation for $v(0, w)$ agrees with the one we obtained for an unemployed agent on page 98. To see this when $e = 0$, observe that the Bellman

equation is

$$v(0, w) = \max_{a \in \{0,1\}} \left\{ aw + (1-a)c + \beta \sum_{(e', w')} v(e', w') P[(0, w), a, (e', w')] \right\}$$

$$= \max_{a \in \{0,1\}} \left\{ aw + (1-a)c + \beta \left[ av(a, w) + (1-a) \sum_{w'} v(a, w') Q(w, w') \right] \right\},$$

where the second equation follows from (5.14). (You can see this by checking the cases $a = 0$ and $a = 1$.) Rearranging and using $v(1, w) = w/(1 - \beta)$ now gives

$$v(0, w) = \max \left\{ \frac{w}{1 - \beta}, \; c + \beta \sum_{w'} v(0, w') Q(w', w') \right\}. \tag{5.15}$$

This is the Bellman equation for an unemployed agent from the job search problem we saw previously on page 98.

**Solution to Exercise 5.1.6.** We need to show that $v_\sigma = (I - \beta P_\sigma)^{-1} r_\sigma$ obeys $v_1 \leqslant v_\sigma \leqslant v_2$ where $v_1, v_2$ are as defined in the exercise. Regarding the upper bound, let $\bar{r} := \|r\|_\infty$. We have

$$(I - \beta P_\sigma)^{-1} r_\sigma \leqslant (I - \beta P_\sigma)^{-1} \bar{r} \, \mathbb{1} = \bar{r} \sum_{t \geqslant 0} (\beta P)^t \mathbb{1} = \frac{\bar{r}}{1 - \beta} = v_2.$$

A similar argument shows that $v_1 \leqslant v_\sigma$.

**Solution to Exercise 5.1.7.** Fix $\sigma \in \Sigma$. It is obvious that $T_\sigma$ is a self-map on $\mathbb{R}^X$ and $T_\sigma$ is clearly order-preserving, since $v \leqslant w$ implies $P_\sigma v \leqslant P_\sigma w$ and hence $T_\sigma v \leqslant T_\sigma w$.

Also, $T_\sigma$ is a contraction of modulus $\beta$ on $\mathbb{R}^X$ under the supremum norm, since, for any $v, w$ in $\mathbb{R}^X$ we have

$$|(T_\sigma v)(x) - (T_\sigma w)(x)| = \beta \left| \sum_{x'} P(x, \sigma(x), x') v(x') - \sum_{x'} P(x, \sigma(x), x') w(x') \right|$$

$$\leqslant \sum_{x'} P(x, \sigma(x), x') \beta \; |v(x') - w(x')| \leqslant \beta \|v - w\|_\infty$$

Taking the supremum over all $x \in X$ yields the desired result. This contraction property combined with Banach's fixed point theorem implies that $T_\sigma$ has a unique fixed point.

Now suppose that $v$ is the unique fixed point of $T_\sigma$. Then $v = r_\sigma + \beta P_\sigma v$. But then $v = (I - \beta P_\sigma)^{-1} r_\sigma$. Hence $v = v_\sigma$. This establishes all claims in the lemma.

**Solution to Exercise 5.1.10.** Fix $v \in V$ and take $\hat\sigma$ to be $v$-greedy, so that

$$\hat\sigma(x) \in \operatorname*{argmax}_{a \in \Gamma(x)} \left\{ r(x, a) + \beta \sum_{x'} v(x') P(x, a, x') \right\} \quad \text{for all } x \in \mathsf{X} \tag{5.23}$$

If $\sigma$ is any other feasible policy, then

$$r(x, \hat\sigma(x)) + \beta \sum_{x'} v(x') P(x, \hat\sigma(x), x') \geqslant r(x, \sigma(x)) + \beta \sum_{x'} v(x') P(x, \sigma(x), x')$$

at all $x$. In operator form, this is $T_{\hat\sigma} v \geqslant T_\sigma v$. Since $\sigma$ is an arbitrary greedy policy, we have shown that $T_{\hat\sigma} v$ is the greatest element of $\{T_\sigma v\}_{\sigma \in \Sigma}$.

A similar argument replacing argmax with argmin in (5.23) shows that a least element also exists.

**Solution to Exercise 5.1.11.** Fix $v \in \mathbb{R}^{\mathsf{X}}$. Part (i) follows from the fact that $\Gamma(x)$ is finite and nonempty at each $x \in \mathsf{X}$. Hence we can select an element $a_x^*$ from the argmax in the definition of a $v$-greedy policy at each $x$ in $\mathsf{X}$. The resulting policy is $v$-greedy. For part (ii) we need to show that $\sigma \in \Sigma$ is $v$-greedy if and only if

$$r(x, \sigma(x)) + \beta \sum_{x'} v(x') P(x, \sigma(x), x') = \max_{a \in \Gamma(x)} \left\{ r(x, a) + \beta \sum_{x'} v(x') P(x, a, x') \right\}$$

for all $x \in \mathsf{X}$. But this immediate from the definition.

Regarding part (iii), it follows from the definitions that $(T_\sigma v)(x) \leqslant (Tv)(x)$ for all $x \in \mathsf{X}$. At the same time, for any $v$-greedy $\sigma \in \Sigma$, we have $(T_\sigma v)(x) = (Tv)(x)$ for all $x$. Hence $Tv = \vee_\sigma T_\sigma v$, as was to be shown.

**Solution to Exercise 5.1.12.** This result follows from Lemma 2.2.3 on page 59. For the sake of the exercise, we also provide a direct proof:

Fix $v, w \in \mathbb{R}^{\mathsf{X}}$ and $x \in \mathsf{X}$. By Exercise 5.1.11 and the max-inequality lemma (page 58), we have

$$|(Tv)(x) - (Tw)(x)| = \left| \max_{\sigma \in \Sigma} (T_\sigma v)(x) - \max_{\sigma \in \Sigma} (T_\sigma w)(x) \right|$$
$$\leqslant \max_{\sigma \in \Sigma} |(T_\sigma v)(x) - (T_\sigma w)(x)| = \|T_\sigma v - T_\sigma w\|_\infty.$$

Applying contractivity of $T_\sigma$ (Exercise 5.1.7), we get $\|Tv - Tw\|_\infty \leqslant \beta \|v - w\|_\infty$.

**Solution to Exercise 5.1.13.** Part (iii) of Proposition 5.1.1 implies (iv) because every $v \in \mathbb{R}^\mathsf{X}$ has at least one greedy policy (Exercise 5.1.11). In particular, at least one $v^*$-greedy policy exists.

**Solution to Exercise 5.3.3.** The Bellman equation becomes

$$v(w, z, \varepsilon) = \max_{w' \leqslant R(w + z + \varepsilon)} \left\{ u \left( w + z + \varepsilon - \frac{w'}{R} \right) + \beta \sum_{z', \varepsilon'} v(w', z', \varepsilon') Q(z, z') \varphi(\varepsilon') \right\}.$$

Both $w$ and $w'$ are constrained to a finite set $\mathsf{W} \subset \mathbb{R}_+$. The expected value function can be expressed as

$$g(z, w') := \sum_{z', \varepsilon'} v(w', z', \varepsilon') Q(z, z') \varphi(\varepsilon'). \tag{5.38}$$

In the remainder of this section, we will say that a savings policy $\sigma$ is $g$-**greedy** if

$$\sigma(z, w, \varepsilon) \in \operatorname*{argmax}_{w' \leqslant R(w + z + \varepsilon)} \left\{ u \left( w + z + \varepsilon - \frac{w'}{R} \right) + \beta g(z, w') \right\}.$$

Since it is an MDP, we can see immediately that if we replace $v$ in (5.38) with the value function $v^*$, then a $g$-greedy policy will be an optimal one. We can rewrite the Bellman equation in terms of expected value functions via

$$g(z, w') = \sum_{z', \varepsilon'} \max_{w'' \leqslant R(w' + z' + \varepsilon')} \left\{ u \left( w' + z' + \varepsilon' - \frac{w''}{R} \right) + \beta g(z', w'') \right\} Q(z, z') \varphi(\varepsilon').$$

**Solution to Exercise 6.1.1.** Set $L(x, x') := \beta(x) P(x, x')$ with $\beta(x) := 1/(1 + r(x))$. We claim that (6.9) is finite for all $x \in \mathsf{X}$ and satisfies $v = (I - L)^{-1} \pi$ whenever $\rho(L) < 1$. To see this, we apply Theorem 6.1.1 with $b(x, x') = \beta(x)$ and $h = \pi$.

Incidentally, to understand $v = \pi + Lv$, suppose we buy the firm now, hold it for one period and then sell it. The expected present value of the payoff is $\pi + Lv$. If expected benefit equals cost, then the value of (i.e., cost of buying) the firm now should equal $\pi + Lv$. That is, $v = \pi + Lv$. We expand on these ideas in §6.3.

**Solution to Exercise 6.1.3.** Let $(X_t)$ be $P$-Markov with $X_0$ drawn from $\psi^*$, let $\|\cdot\|_*$ be the norm defined in the exercise and let $\mathbb{1}$ be an $n$-vector of ones. In view of (6.6), for fixed $t \in \mathbb{N}$, we have

$$\mathbb{E} \prod_{i=0}^{t} \beta_i = \mathbb{E} \left[ \mathbb{E} \prod_{i=0}^{t} \beta_i \mid X_0 \right] = \mathbb{E}(L^t \mathbb{1})(X_0) = \|L^t \mathbb{1}\|_*.$$

Since $\mathbb{1} \gg 0$, the local spectral radius result on page 71 yields (6.11).

**Solution to Exercise 6.1.4.** Let $U$ be closed in $\mathbb{R}^X$ and let $T$ be a self-map on $U$ such that $T^k$ is a contraction. Let $u^*$ be the unique fixed point of $T^k$. Fix $\varepsilon > 0$. We can choose $m$ such that $\|T^{mk}Tu^* - u^*\| < \varepsilon$. Then

$$\|TT^{mk}u^* - u^*\| = \|Tu^* - u^*\| < \varepsilon.$$

Since $\varepsilon$ was arbitrary we have $\|Tu^* - u^*\| = 0$, implying that $u^*$ is a fixed point of $T$. The proof that $T^m u \to u^*$ for any $u$ is left to the reader.

**Solution to Exercise 6.2.2.** Fix $\sigma \in \Sigma$ and let Assumption 6.2.1 on page 193 hold. We saw in the proof of Lemma 6.2.1 that $T_\sigma v = r_\sigma + L_\sigma v$ and $v_\sigma = (I - L_\sigma)^{-1} r_\sigma$ is the unique fixed point in of this operator $\mathbb{R}^X$. Moreover, for fixed $v, w \in \mathbb{R}^X$, we have

$$|T_\sigma v - T_\sigma w| = |L_\sigma v - L_\sigma w| = |L_\sigma (v - w)| = L_\sigma |v - w|.$$

Hence, by Proposition 6.1.6, $T_\sigma$ is globally stable on $\mathbb{R}^X$.

**Solution to Exercise 6.2.3.** Fix $\sigma \in \Sigma$. When (6.20) holds, we have $0 \leqslant L_\sigma \leqslant L$. Exercise 2.2.28 on page 60 now implies that $\rho(L_\sigma) \leqslant \rho(L)$. Hence $\rho(L_\sigma) < 1$.

**Solution to Exercise 6.2.4.** Fix $\sigma \in \Sigma$. In the present setting, the discount operator $L_\sigma$ from (6.17) becomes

$$L_\sigma(x, x') = L_\sigma((y, z), (y', z')) = \beta(z)Q(z, z')R(y, \sigma(y), y').$$

In view of Lemma 6.1.3, the spectral radius of $L_\sigma$ on $\mathcal{L}(\mathbb{R}^X)$ is equal to the spectral radius of $L(z, z') = \beta(z)Q(z, z')$ on $\mathcal{L}(\mathbb{R}^Z)$. It follows that $\rho(L) < 1$ in $\mathcal{L}(\mathbb{R}^Z)$ implies $\rho(L_\sigma) < 1$ in $\mathcal{L}(\mathbb{R}^X)$, so Assumption 6.2.1 holds. Hence, under this condition, Proposition 6.2.2 is valid.

**Solution to Exercise 6.3.1.** Assume $m, d \gg 0$ and write (6.34) as $\pi = A\pi + h$, where $h := Ad$. A simple argument shows that $h \gg 0$ (see Exercise 2.3.13 on page 80 for a closely related claim.) The claim in Exercise 6.3.1 now follows directly from Lemma 6.1.4.

**Solution to Exercise 6.3.4.** Under a cum-dividend contract, purchasing at $t$ and selling at $t+1$ pays $D_t + \Pi_{t+1}$. Hence, applying the fundamental asset pricing equation, the time $t$ price $\Pi_t$ of the contract must satisfy

$$\Pi_t = D_t + \mathbb{E}_t M_{t+1} \Pi_{t+1}. \qquad (6.36)$$

Proceeding as for the ex-dividend contract, the price conditional on current state $x$ is $\pi(x) = d(x) + \sum_{x'} m(x, x') \pi(x') P(x, x')$. In vector form, this is $\pi = d + A\pi$. Solving out for prices gives $\pi^* = (I - A)^{-1} d$.

**Solution to Exercise 6.3.6.** We seek a $v$ that solves

$$v(x) = \sum_{x'} [1 + v(x')] A(x, x') \qquad (x, x' \in \mathsf{X}).$$

Treating $A$ as a matrix and $v$ as a column vector, this equation becomes $v = A\mathbb{1} + Av$, where $\mathbb{1}$ is a column vector of ones. By the Neumann series lemma, $\rho(A) < 1$ implies that this equation has the unique solution $v^* = (I - A)^{-1} A\mathbb{1}$. By the same lemma, $v^*$ has the alternative representation $v^* = \sum_{t \geqslant 0} A^t (A\mathbb{1}) = \sum_{t \geqslant 1} A^t \mathbb{1}$.

**Solution to Exercise 7.1.1.** Here is one possible answer: Let $v_1, v_2$ be distinct and let $T$ be the identity map on $V = [v_1, v_2]$. Then $T$ is order-preserving and every point in $V$ is fixed under $T$. The set $V$ is a continuum because it contains all points $v = \alpha v_1 + (1 - \alpha) v_2$ with $0 \leqslant \alpha \leqslant 1$.

**Solution to Exercise 7.1.3.** If the condition $a < g(a)$ in Proposition 7.1.2 is dropped then $g$ could be the identity map, which has multiple fixed points and is not globally stable.

**Solution to Exercise 7.1.7.** It is straightforward to show that $F'_x > 0$ on $(0, \infty)$, which proves that $F_x$ is increasing. Some additional algebra confirms that $\theta \in (0, 1]$ implies $F''_x > 0$, while $\theta < 0$ and $\theta > 1$ both imply $F''_x < 0$. Details are left to the reader.

**Solution to Exercise 7.1.8.** Observe that $(Gu)(x) = F_x[(Au)(x)]$. Since $A$ is order-preserving and $F$ is increasing, $u \leqslant v$ implies $Gu \leqslant Gv$. In particular, $G$ is order-preserving. If $\theta \in (0, 1]$, then $F$ is convex. Hence, fixing $u, v \in V$ and $\lambda \in [0, 1]$ (and dropping $x$ from our notation), we have

$$FA(\lambda u + (1 - \lambda) v) = F(\lambda Au + (1 - \lambda) Av) \leqslant \lambda FAu + (1 - \lambda) FAv.$$

Hence $G$ is convex. The proof that $G$ is concave for other values of $\theta$ is similar and omitted.

**Solution to Exercise 7.1.9.** Setting $v_t = \sigma_t^{-1/\psi}$, we can write (7.3) as

$$v_t = \left\{ 1 + \beta^\psi \left[ \mathbb{E}_t R_{t+1}^{(\psi-1)/\psi} v_{t+1} \right]^\psi \right\}^{1/\psi}. \tag{7.4}$$

We conjecture a stationary Markov solution $v_t = v(X_t)$ for some some $v \in \mathbb{R}^{\mathsf{X}}$ with $v \gg 0$. This $v$ must satisfy

$$v(x) = \left\{ 1 + \beta^\psi \left[ \sum_{x'} f(x')^{(\psi-1)/\psi} v(x') P(x, x') \right]^\psi \right\}^{1/\psi} \qquad (x \in \mathsf{X}).$$

Using the definition of $A$ in the exercise, we can write the equation in vector form as $v = [1 + (Av)^\psi]^{1/\psi}$. It follows from Theorem 7.1.4 that a unique strictly positive solution to this equation exists if and only if $\rho(A)^\psi < 1$. This proves the claim in the exercise.

**Solution to Exercise 7.2.1.** Fixing $t$, rearranging $v^* = (I - \beta P)^{-1} r$ to $v^* = r + \beta P v^*$ and evaluating at $X_t$ gives

$$V_t^* = v^*(X_t) = r(X_t) + \beta \sum_{x'} v^*(x') P(X_t, x')$$

$$= u(C_t) + \beta \mathbb{E} \left[ v^*(X_{t+1}) \mid X_t \right] = u(C_t) + \beta \mathbb{E}_t V_{t+1}^*.$$

Hence $(V_t^*)_{t \geq 0}$ obeys (7.5), as claimed.

**Solution to Exercise 7.2.7.** Let $K$ be as stated (see (7.14)) and fix $v \gg 0$. Clearly $v^\gamma \gg 0$ and hence $P v^\gamma \gg 0$ (see Exercise 3.2.5 on page 95). Since $h \geq 0$, it follows easily that $K v \gg 0$.

**Solution to Exercise 7.2.8.** If $v \gg 0$, then, since $h \gg 0$ also holds, we have

$$\hat{K} v = \left\{ h + \beta (P v)^{1/\theta} \right\}^\theta \geq h^\theta \gg 0.$$

Hence $\hat{K}$ is a self-map on $V$. In addition, the statement $K v = v$ is equivalent to $v^\alpha = h + \beta (P v^\alpha)^{\alpha/\gamma}$. Using $\theta = \gamma/\alpha$, we can rewrite the last equation as $v^\gamma = [h + \beta (P v^\gamma)^{1/\theta}]^\theta$. In other words, $v = K v$ if and only if $v^\gamma = \hat{K} v^\gamma$.

**Solution to Exercise 7.3.1.** Let $R$ be a linear operator on $\mathbb{R}^{\mathsf{X}}$ that is also a certainty equivalent. In particular, $R$ is order-preserving and $R\mathbb{1} = \mathbb{1}$. Since $R$ is order-preserving and linear, $R$ is a positive linear operator (see Exercise 2.3.11). Hence $R$ is a Markov operator.

**Solution to Exercise 7.3.4.** Since $V$ is all of $\mathbb{R}^{\mathsf{X}}$, the condition $R_\tau \colon V \to V$ is trivially satisfied. Regarding monotonicity, fix $v, w \in V$ with $v \leqslant w$ and $x \in \mathsf{X}$. Let $X$ be a draw from $P(x, \cdot)$. Then $(R_\tau v)(x) = Q_\tau[v(X)] \leqslant Q_\tau[w(X)] = R_\tau(v, \varphi)$, where the inequality is by Exercise 2.2.38 on page 66. Moreover, given $\lambda \in \mathbb{R}$ and a random variable $Y$ with $\mathbb{P}\{Y = \lambda\} = 1$, we clearly have $Q_\tau(Y) = \lambda$. It follows that $R_\tau \lambda \mathbb{1} = \lambda \mathbb{1}$. Hence $R_\tau$ is a certainty equivalent operator, as was to be shown.

**Solution to Exercise 7.3.7.** Fix $v \in V = \mathbb{R}^{\mathsf{X}}$, $\lambda \in \mathbb{R}_+$ and $x \in \mathsf{X}$. If $X \sim P(x, \cdot)$, then

$$(R_\tau (v + \lambda))(x) = Q_\tau (v(X) + \lambda) = Q_\tau (v(X)) + \lambda,$$

where the second equality is by Exercise 1.2.28 on page 32. Since $x$ was arbitrary, we have $R_\tau(v + \lambda) = R_\tau v + \lambda$. Hence $R_\tau$ is constant-subadditive, as claimed.

**Solution to Exercise 7.3.8.** Fix $v \in V$, $P \in \mathcal{M}(\mathbb{R}^{\mathsf{X}})$ and $\lambda \in \mathbb{R}_+$. Let $X$ be a draw from $P(x, \cdot)$. We have

$$\begin{aligned}(R_\theta(v + \lambda))(x) &= \frac{1}{\theta} \ln \{\mathbb{E} \exp[\theta(v(X) + \lambda)]\} \\ &= \frac{1}{\theta} \ln \{\mathbb{E} \exp[\theta v(X)] \cdot \exp(\theta \lambda)\} \\ &= \frac{1}{\theta} \ln \{\mathbb{E} \exp[\theta v(X)]\} + \lambda.\end{aligned}$$

Hence constant-subadditivity holds.

**Solution to Exercise 7.3.9.** Let the primitives be as stated. Fix $v, w \in V$. By monotonicity and constant-subadditivity of $R$, we have

$$Rv = R(v - w + w) \leqslant R(\|v - w\|_\infty \mathbb{1} + w) \leqslant Rw + \|v - w\|_\infty \mathbb{1}.$$

Hence $(Rv)(x) - (Rw)(x) \leqslant \|v - w\|_\infty$ for all $x \in \mathsf{X}$. Reversing the roles of $v$ and $w$ proves the claim.

**Solution to Exercise 7.3.10.** Fix $v, v' \in V$, $\theta < 0$, $x \in \mathsf{X}$ and $\alpha \in [0, 1]$. Letting

$X \sim P(x, \cdot)$, $Z = v(X)$ and $Z' = v'(X)$, we have

$$(R(v + v'))(x) = \mathcal{E}_\theta(\alpha Z + (1 - \alpha)Z') \geqslant \alpha \mathcal{E}_\theta(Z) + (1 - \alpha)\mathcal{E}_\theta(Z').$$

The last expression is just $\alpha(Rv)(x) + (1 - \alpha)(Rv')(x)$, so $R$ is concave on $V$, as claimed.

**Solution to Exercise 7.3.11.** Regarding (i), fix $\lambda \in [0, 1]$ and $v, w \in V$. Using subadditivity and positive homogeneity, we have

$$R(\lambda v + (1 - \lambda)w) \leqslant R(\lambda v) + R((1 - \lambda)w) = \lambda R v + (1 - \lambda)R w.$$

This proves that $R$ is convex on $V$. The proof of (ii) is similar.

**Solution to Exercise 7.3.14.** It is not difficult to show that $(U_c/U_y) = ((1 - \beta)/\beta)(y/c)^{1-\alpha}$. Taking logs and rearranging gives $\ln(y/c) = (1/(1 - \alpha))\ln(U_c/U_y) + k$, where $k$ is a constant. Using the definition in the exercise yields EIS $= 1/(1 - \alpha)$.

**Solution to Exercise 7.3.15.** Iterating forward from $V_0$ gives

$$V_0 = u(C_0) + \beta \mathbb{E}_0 V_1 = u(C_0) + \beta \mathbb{E}_0 \left[ u(C_1) + \beta \mathbb{E}_1 V_2 \right] = u(C_0) + \beta \mathbb{E}_0 u(C_1) + \beta^2 \mathbb{E}_0 V_2.$$

Continuing forward until time $m$ yields $V_0 = \sum_{t=0}^{m-1} \beta^t \mathbb{E}_0 u(C_t) + \beta^m \mathbb{E}_0 V_m$. Shifting to functional form and using $r = u \circ c$, the last expression becomes

$$v = \sum_{t=0}^{m-1} (\beta P)^t r + (\beta P)^m w.$$

By Exercise 1.2.17 on page 22, this is just $K^m w$ when $K$ is the associated Koopmans operator $K v = r + \beta P v$ and, moreover, $K^m w \to v^* := (I - \beta P)^{-1} r$.

**Solution to Exercise 7.3.17.** The additive case is obvious. Regarding the Leontief case, fix $x \in \mathsf{X}$, $y \in \mathbb{R}$ and $\lambda \in \mathbb{R}_+$. We have

$$\min\{r(x), \beta(y + \lambda)\} \leqslant \min\{r(x) + \beta\lambda, \beta y + \beta\lambda\} = \min\{r(x), \beta y\} + \beta\lambda.$$

That is, $A(x, y + \lambda) \leqslant A(x, y) + \beta\lambda$. Hence Blackwell's condition holds.

**Solution to Exercise 7.3.19.** We already know from Exercise 7.3.17 that the Leontief aggregator satisfies Blackwell's condition when $\beta \in (0, 1)$. Since $R_\tau$ is constant-subadditive, global stability follows from Proposition 7.3.3.

**Solution to Exercise 7.3.22.** We saw in Lemma 7.2.4 that $\Phi$ is a homeomorphism from $V$ to itself. For $v \in V$, we have

$$\hat{K}\Phi v = \left\{ h + (b^\theta P \Phi v)^{1/\theta} \right\}^\theta = \left\{ h + b(P v^\gamma)^{1/\theta} \right\}^\theta = \left\{ h + b(P v^\gamma)^{\alpha/\gamma} \right\}^{\gamma/\alpha} = \Phi K v.$$

Thus, $\hat{K}\Phi = \Phi K$ on $V$. Rearranging gives $K = \Phi^{-1}\hat{K}\Phi$, so $(V, K)$ and $(V, \hat{K})$ are topologically conjugate, as claimed.

**Solution to Exercise 8.1.2.** This problem can be formulated as an MDP by setting the state to $x := (y, z)$, taking values in $\mathsf{X} := \mathsf{Y} \times \mathsf{Z}$. The action space is $\mathsf{Y}$. The feasible correspondence is $x \mapsto \Gamma(x)$ and current reward is $r(x, a) = r((y, z), y') = F(y, z, y')$. The stochastic kernel is

$$P(x, a, x') = P((y, z), a, (y', z')) = \mathbb{1}\{y' = a\}Q(z, z').$$

This MDP $(\Gamma, \beta, r, P)$ has a Bellman equation identical to (8.7).

**Solution to Exercise 8.1.3.** We must check that $(\Gamma, V, B)$ satisfies conditions (8.2)–(8.3). The monotonicity condition holds because $\beta$ is nonnegative, so $w \leqslant v$ implies

$$\sum_{x'} w(x')\beta(x, a, x')P(x, a, x') \leqslant \sum_{x'} v(x')\beta(x, a, x')P(x, a, x') \quad \text{for all } (x, a) \in \mathsf{G}.$$

The consistency condition is trivial, since $V$ is all of $\mathbb{R}^\mathsf{X}$.

**Solution to Exercise 8.1.6.** Fix $\sigma \in \Sigma$. The claim that $T_\sigma$ is a self-map on $V$ follows immediately from the consistency condition in (8.3). The order-preserving property follows from the monotonicity condition in (8.2).

**Solution to Exercise 8.1.7.** Fix $v \in V$ and consider the set $\{T_\sigma v\}_{\sigma \in \Sigma} \subset V$. We first show that $\{T_\sigma v\}_{\sigma \in \Sigma}$ contains a greatest element. Suppose that $\bar{\sigma}$ is $v$-greedy. If $\sigma$ is any other policy, then

$$(T_\sigma v)(x) = B(x, \sigma(x), v) \leqslant (T_{\bar{\sigma}} v)(x) \quad \text{for all } x \in \mathsf{X}.$$

Hence $T_{\bar{\sigma}} v$ is a greatest element of $\{T_\sigma v\}_{\sigma \in \Sigma}$.

The proof that $\{T_\sigma v\}_{\sigma \in \Sigma}$ contains a least element is analogous, after replacing argmax with argmin.

**Solution to Exercise 8.1.8.** Regarding part (i), fix $v \in V$. For any $\sigma \in \Sigma$ and $x \in \mathsf{X}$, we have

$$(Tv)(x) = \max_{a \in \Gamma(x)} B(x, a, v) = \max_{\sigma \in \Sigma} B(x, \sigma(x), v) = \max_{\sigma \in \Sigma}(T_\sigma v)(x).$$

Since $x$ was chosen arbitrarily, we have confirmed that $Tv = \bigvee_{\sigma \in \Sigma} T_\sigma v$.

Regarding part (ii), $\sigma$ is $v$-greedy if and only if

$$\sigma(x) \in \operatorname*{argmax}_{a \in \Gamma(x)} B(x, a, v) \quad \text{for all } x \in \mathsf{X}.$$

This is equivalent to $B(x, \sigma(x), v) = \max_{a \in \Gamma(x)} B(x, a, v)$ for all $x \in \mathsf{X}$. Hence $\sigma$ is $v$-greedy if and only if $T_\sigma v = Tv$, as claimed.

Regarding (iii), to see that $T$ is a self-map on $V$, fix $v \in V$ and let $\sigma$ be $v$-greedy. Then, by (ii), $Tv = T_\sigma v \in V$. Hence $T$ is a self-map, as claimed. The fact that $T$ is order-preserving on $V$ follows immediately from the monotonicity property of $B$ in (8.2).

**Solution to Exercise 8.1.9.** Here's a proof for $T$ and fixed $k \in \mathbb{N}$: At arbitrary $x \in \mathsf{X}$,

$$(T^k v)(x) = (T(T^{k-1}v))(x) = \max_{a \in \Gamma(x)} B(x, a, T^{k-1}v) \tag{8.16}$$

This confirms the claim in (8.14).

**Solution to Exercise 8.1.13.** Let $\{T_\sigma\}$ be the policy operators associated with a bounded and well-posed RDP $(\Gamma, V, B)$. Let $\hat{V} := [v_1, v_2]$, where $v_1, v_2$ are as in (8.20). Fix $\sigma \in \Sigma$ and let $v_\sigma$ be the $\sigma$-value function of policy operator $T_\sigma$. It follows from (8.20) that $T_\sigma$ is a self-map on $\hat{V}$. By the Knaster–Tarski fixed point theorem (page 214), $T_\sigma$ has at least one fixed point in $\hat{V}$. By uniqueness, that fixed point is $v_\sigma$.

**Solution to Exercise 8.1.14.** Let $(\Gamma, V, B)$ be as stated. Let $\bar{r} = \|r\|_\infty$. We claim that (8.20) holds when $v_2 = \bar{r}/(1 - \beta)$ and $v_1 = -v_2$. (The functions $v_1$ and $v_2$ are constant.) To see this, observe that

$$B(x, a, v_2) = r(x, a) + \beta v_2 \leqslant \bar{r} + \beta v_2 = v_2.$$

for all $(x, a) \in \mathsf{G}$. This is the upper bound condition in (8.20). The proof of the lower bound condition is similar.

**Solution to Exercise 8.1.15.** Letting $\bar{r} = \|e\|_\infty + \|c\|_\infty$, it can be shown that (8.20) holds when $v_2 = \bar{r}/(1 - \beta)$ and $v_1 = -v_2$. The argument is similar to that provided for the MDP case in the solution to Exercise 8.1.14.

**Solution to Exercise 8.1.16.** Let $\bar{r} := \|r\|_\infty \mathbb{1}$. We claim that the functions $v_2 = (I - L)^{-1}\bar{r}$ and $v_1 = -v_2$ satisfy (8.20). To see this, observe that

$$\bar{r} + Lv_2 = \bar{r} + Lv_2 - v_2 + v_2 = \bar{r} - (I - L)v_2 + v_2 = \bar{r} - \bar{r} + v_2 = v_2.$$

Since $B(x, a, v_2) \leqslant \bar{r} + (Lv_2)(x)$, this proves that $v_2$ satisfies the upper bound condition in (8.20). The proof of the lower bound condition is similar.

**Solution to Exercise 8.1.17.** The only nontrivial condition to check is that the bound $B(x, x', v_2) \leqslant v_2(x)$ holds for all feasible $(x, x')$. In particular, we need to show that

$$c(x, x') + C(x') \leqslant C(x) \text{ whenever } x' \in \mathcal{O}(x) \text{ and } x' \neq x.$$

This is true by the definition of $C$, since $C(x)$ is the maximum travel cost to $d$ and $c(x, x') + C(x')$ is the cost of traveling to $d$ via $x'$ and then taking the most expensive path.

**Solution to Exercise 8.1.19.** The map $\varphi(m) = m^\gamma$ is a homeomorphism from $V$ to itself and (8.21) holds under $\varphi$. This implies the claim in the exercise.

**Solution to Exercise 8.2.3.** Let $\mathcal{R} = (\Gamma, V, B)$ satisfy Blackwell's condition. Fix $v, w \in V$ and $(x, a) \in G$. Observe that $v = w + v - w \leqslant w + \|v - w\|_\infty$. By monotonicity of $B$ and Blackwell's condition, we have

$$B(x, a, v) \leqslant B(x, a, w + \|v - w\|_\infty) \leqslant B(x, a, w) + \beta\|v - w\|_\infty.$$

As a result, $B(x, a, v) - B(x, a, w) \leqslant \beta\|v - w\|_\infty$. Reversing the roles of $v$ and $w$ yields

$$|B(x, a, v) - B(x, a, w)| \leqslant \beta\|v - w\|_\infty.$$

Since $\beta < 1$, the RDP $\mathcal{R}$ is contracting.

**Solution to Exercise 8.2.8.** Let $\mathcal{R} = (\Gamma, V, B)$ be as stated. Fix $v, w \in V$. Using the max-inequality (page 58), we obtain

$$|(Tv)(x) - (Tw)(x)| \leqslant \max_{a \in \Gamma(x)} |B(x, a, v) - B(x, a, w)| \qquad (x \in \mathsf{X}).$$

Let $\sigma$ be a map from X to A such that $\sigma(x)$ is a maximizer of the right hand side of this expression for all $x$. Clearly $\sigma \in \Sigma$ and

$$|(Tv)(x) - (Tw)(x)| \leqslant |B(x, \sigma(x), v) - B(x, \sigma(x), w)| \quad \text{for all } x \in \mathsf{X}.$$

Since $\mathcal{R}$ is eventually contracting, there is a positive linear operator $L_\sigma$ with $\rho(L_\sigma) < 1$ and $|Tv - Tw| \leqslant L_\sigma|v - w|$. Proposition 6.1.6 on page 191 implies that $T$ is eventually contracting on $V$. Since $V$ is closed, it follows that $T$ is globally stable (Theorem 6.1.5, page 190).

**Solution to Exercise 8.2.10.** We discuss the first case, regarding (8.37). When (8.40) holds, by finiteness of G, we can take an $\varepsilon > 0$ such that

$$B(x, a, v_2) \leqslant v_2(x) - \varepsilon \text{ for all } (x, a) \in \mathsf{G}.$$

We then have

$$\varepsilon \leqslant v_2(x) - B(x, a, v_2) \leqslant v_2(x) - B(x, a, v_1) \leqslant v_2(x) - v_1(x)$$

for all $x$, so $0 < \varepsilon \leqslant \|v_2 - v_1\|_\infty$. Set $\delta := \varepsilon / \|v_2 - v_1\|_\infty$. From (8.40) we get

$$B(x, a, v_2) \leqslant v_2(x) - \delta\|v_2 - v_1\|_\infty \leqslant v_2(x) - \delta[v_2(x) - v_1(x)]$$

for arbitrary $(x, a) \in \mathsf{G}$. Hence (8.37) holds.

**Solution to Exercise 8.2.11.** We prove (8.41) and leave (8.40) to the reader. For given $(x, a) \in \mathsf{G}$,

$$B(x, a, v_1) = r(x, a) + \beta \sum_{x'} v_1(x') P(x, a, x') \geqslant r_1 + \beta \frac{r_1 - \varepsilon}{1 - \beta} = \frac{r_1 - \beta r_1 + \beta r_1 - \beta \varepsilon}{1 - \beta} = v_1 + \varepsilon.$$

Hence (8.41) is confirmed.

**Solution to Exercise 8.3.1.** For each fixed $a \in \Gamma(x)$, the map $R_\tau^a$ is a version of the quantile certainty equivalent operator defined in Exercise 7.3.4 on page 233. With this observation, we can replicate the proof of Proposition 8.3.1, after replacing $R_\theta^a$ with $R_\tau^a$. The latter is also constant-subadditive, by Exercise 7.3.7 on page 234.

**Solution to Exercise 8.3.4.** Regarding (b), note that $v_1$ is constant. Hence, at fixed $(x, a) \in \mathsf{G}$ and $d \in D(x, a)$, we have

$$B(x, a, d, v_1) = r(x, a) + \beta \frac{r_1 - \varepsilon}{1 - \beta} \geqslant r_1 + \beta \frac{r_1 - \varepsilon}{1 - \beta} = \frac{r_1 - \beta r_1 + \beta r_1 - \beta \varepsilon}{1 - \beta} = v_1 + \varepsilon.$$

Hence (b) is confirmed. Regarding (c), we have

$$B(x, a, d, v_2) = r(x, a) + \beta \frac{r_2}{1 - \beta} \leqslant r_2 + \beta \frac{r_2}{1 - \beta} = v_2.$$

We have now verified conditions (a)–(d) on page 278.

**Solution to Exercise 8.3.5.** In this case, where $\kappa = \gamma$, $B$ reduces to

$$B(x, a, v) = \left\{ r(x, a) + \beta \left\{ \sum_{x'} v(x')^\gamma P(x, a, x') \right\}^{\alpha/\gamma} \right\}^{1/\alpha},$$

where $P(x, a, x') := \int P_\theta(x, a, x') \mu(x, \mathrm{d}\theta)$ is a weighted average over beliefs. This is identical to the Epstein–Zin aggregator (see Example 8.1.7).

**Solution to Exercise 8.3.10.** A feasible policy $\sigma$ is a map from $\mathsf{X}$ to itself satisfying $\sigma(x) \in \mathcal{O}(x)$ for all $x$. Recalling that $\mathsf{X}$ is finite and setting $n = |\mathsf{X}|$, the stated assumptions imply that $\sigma^k(x) = d$ for all $k \geqslant n$ (since all paths lead to $d$ in at most $n$ steps). Given that $c(d, d) = 0$, it follows that the lifetime cost of following $\sigma$ from initial condition $x$ is no more than

$$c(x, \sigma(x)) + \beta c(\sigma(x), \sigma^2(x)) + \beta^2 c(\sigma^2(x), \sigma^3(x)) + \cdots + \beta^{n-1} c(\sigma^{n-1}(x), \sigma^n(x))$$

With $c_{\mathsf{T}} := \max c$, we then have

$$C(x) \leqslant c_{\mathsf{T}} \frac{1 - \beta^n}{1 - \beta}.$$

**Solution to Exercise 9.1.1.** By the Neumann series lemma, $T$ has a unique fixed point in $V$ given by $\bar{v} := (I - A)^{-1} r$. $T$ is upward stable because, given $v \in \mathbb{R}^{\mathsf{X}}$ with $v \leqslant T v$, we have $v \leqslant r + A v$, or $(I - A)v \leqslant r$. By the Neumann series lemma, $(I - A)^{-1}$ is a positive linear operator (as the sum of nonnegative matrices), so we can multiply by this inverse to get $v \leqslant (I - A)^{-1} r = \bar{v}$. This proves upward stability. Reversing the inequalities shows that downward stability also holds.

**Solution to Exercise 9.1.3.** The first part of the exercise is immediate from the definitions. For the second, take $v, w \in V$ with $v \preceq w$. Since $T_\sigma$ is order-preserving, we have $T_\sigma v \preceq T_\sigma w$ for all $\sigma \in \Sigma$. Hence $T_\sigma v \preceq Tw$ for all $\sigma \in \Sigma$. Therefore $Tv \preceq Tw$.

**Solution to Exercise 9.1.5.** For all $v \in V_\Sigma$, we have $v = v_\sigma$ for some $\sigma$, and hence $Tv \succcurlyeq T_\sigma v = T_\sigma v_\sigma = v_\sigma = v$.

**Solution to Exercise 9.1.6.** This follows directly from Exercise 1.2.27 on page 31.

**Solution to Exercise 9.1.7.** This result follows from Exercise 9.1.6, since, at each $x$, the maximizing distribution $\varphi_x$ is supported on $\operatorname{argmax}_{a \in \Gamma(x)} B(x, a, v)$.

**Solution to Exercise 9.1.9.** Regarding (i), fix $v \in V$. Policy $\sigma$ is $v$-min-greedy for $\mathcal{A}$ if and only if $T_\sigma v \preceq T_\tau v$ for all $\tau \in \Sigma$, which is equivalent to $T_\sigma v \succeq^\partial T_\tau v$ for all $\tau \in \Sigma$. Hence $\sigma$ is $v$-min-greedy for $\mathcal{A}$ if and only if $\sigma$ is $v$-max-greedy for $\mathcal{A}^\partial$.

Regarding (ii)–(iii), fix $v \in V$ and let $\sigma$ be $v$-min-greedy for $\mathcal{A}$ (and hence $v$-max-greedy for $\hat{\mathcal{A}}$). We then have $T^\partial v = T_\sigma v = T_\downarrow v$. Hence $T^\partial = T_\downarrow$. Similarly, at the same $v$ and with the same policy $\sigma$, $H^\partial v$ is equal to $v_\sigma$ and so is $H_\downarrow$. A similar argument gives $T_\downarrow^\partial = T$ and $H_\downarrow^\partial = H$.

Regarding (iv), Lemma 9.1.2 implies that $\mathcal{A}$ is order stable if and only if $\mathcal{A}^\partial$ is order stable.

Regarding (v), $T_\downarrow = T^\partial$, so $T_\downarrow$ has a fixed point in $V$ if and only if $T^\partial$ has a fixed point in $V$. By this fact and (iv), $\mathcal{A}$ is min-stable if and only if $\mathcal{A}^\partial$ is max-stable. Moreover, in this setting, we have $v_\downarrow^* = \bigwedge_\sigma v_\sigma = \bigvee_\sigma^\partial v_\sigma = (v^*)^\partial$.

Part (vi) follows from similar analysis and details are left to the reader.

**Solution to Exercise 9.2.2.** Let $(V, T)$ and $(\hat{V}, \hat{T})$ be order conjugate under $F$, with respective fixed points $v$ and $\hat{v} = Fv$. Let $T$ be order stable on $V$ and let $\hat{w}$ be an element of $\hat{V}$ satisfying $\hat{T}\hat{w} \preceq \hat{w}$. Then $F^{-1}\hat{T}\hat{w} \preceq F^{-1}\hat{w}$ and hence $TF^{-1}\hat{w} \preceq F^{-1}\hat{w}$. But then $v \preceq F^{-1}\hat{w}$, by downward stability of $T$. Applying $F$ gives $\hat{v} \preceq \hat{w}$. Hence $\hat{T}$ is downward stable on $\hat{V}$. Similarly, if $\hat{w}$ is an element of $\hat{V}$ satisfying $\hat{w} \preceq \hat{T}\hat{w}$, then $F^{-1}\hat{w} \preceq F^{-1}\hat{T}\hat{w} = TF^{-1}\hat{w}$. By upward stability of $T$ on $V$, we have $F^{-1}\hat{w} \preceq v$. Applying $F$ gives $\hat{w} \preceq \hat{v}$, so $\hat{T}$ is upward stable on $\hat{V}$. Together, these results show that $\hat{T}$ is order stable on $\hat{V}$.

**Solution to Exercise 9.2.4.** Fix $(x, a) \in \mathsf{G}$. Suppose first that $\theta > 0$, so that $v_1 = m_1$ and $v_2 = m_2$. Fix $(x, a) \in \mathsf{G}$. We have

$$\hat{B}(x, a, v_1) = \left\{ r(x, a) + \beta \frac{r_1 - \varepsilon}{1 - \beta} \right\}^\theta \geqslant \left\{ r_1 + \beta \frac{r_1 - \varepsilon}{1 - \beta} \right\}^\theta = \left\{ \frac{r_1 - \beta\varepsilon}{1 - \beta} \right\}^\theta > m_1 = v_1.$$

In addition,

$$\hat{B}(x, a, v_2) = \left\{ r(x, a) + \beta \frac{r_2 + \varepsilon}{1 - \beta} \right\}^\theta \leqslant \left\{ r_2 + \beta \frac{r_2 + \varepsilon}{1 - \beta} \right\}^\theta = \left\{ \frac{r_2 + \beta\varepsilon}{1 - \beta} \right\}^\theta < m_2 = v_2.$$

Suppose next that $\theta < 0$, so that $v_1 = m_2$ and $v_2 = m_1$. Fix $(x, a) \in \mathsf{G}$. We have

$$\hat{B}(x, a, v_1) = \left\{ r(x, a) + \beta \frac{r_2 + \varepsilon}{1 - \beta} \right\}^\theta \geqslant \left\{ r_2 + \beta \frac{r_2 + \varepsilon}{1 - \beta} \right\}^\theta = \left\{ \frac{r_2 + \beta\varepsilon}{1 - \beta} \right\}^\theta > m_2 = v_1.$$

In addition,

$$\hat{B}(x, a, v_2) = \left\{ r(x, a) + \beta \frac{r_1 - \varepsilon}{1 - \beta} \right\}^\theta \leqslant \left\{ r_1 + \beta \frac{r_1 - \varepsilon}{1 - \beta} \right\}^\theta = \left\{ \frac{r_1 - \beta\varepsilon}{1 - \beta} \right\}^\theta < m_1 = v_2.$$

**Solution to Exercise 9.2.5.** For $c, t > 0$, let $f(t) := (c + \beta t^{1/\theta})^\theta$. Simple calculations show that $f' > 0$, that $f'' < 0$ when $\theta < 0$ or $1 \leqslant \theta$, and that $f'' > 0$ when $0 < \theta \leqslant 1$. Hence $f$ is concave when $\theta < 0$ or $1 \leqslant \theta$, and convex when $0 < \theta \leqslant 1$. The claim in the exercise follows easily from these facts.

**Solution to Exercise 9.2.6.** Fix $\sigma \in \Sigma$. It follows from Exercise 9.2.4 that $v_1 \ll \hat{T}_\sigma v \ll v_2$ for all $v \in \hat{V}$. Moreover, $T$ is order-preserving and, by Exercise 9.2.5, either concave or convex. Du's theorem (page 217) now implies that $\hat{T}_\sigma$ is globally stable.

**Solution to Exercise 9.2.7.** Fix $\sigma \in \Sigma$ and $v \in I$. On one hand,

$$F T_\sigma v = (T_\sigma v)^\gamma = \left\{ r_\sigma + \beta (P_\sigma v^\gamma)^{\alpha/\gamma} \right\}^{\gamma/\alpha} = \left\{ r_\sigma + \beta (P_\sigma v^\gamma)^{1/\theta} \right\}^\theta.$$

On the other,

$$\hat{T}_\sigma F v = \hat{T}_\sigma v^\gamma = \left\{ r_\sigma + \beta (P_\sigma v^\gamma)^{1/\theta} \right\}^\theta.$$

Hence $F \circ T_\sigma = \hat{T}_\sigma \circ F$ on $I$, as claimed.

**Solution to Exercise 10.1.1.** if $W \stackrel{d}{=} \mathrm{Exp}(\theta)$ and $s, t > 0$, then

$$\frac{\mathbb{P}\{W > s + t \text{ and } W > s\}}{\mathbb{P}\{W > s\}} = \frac{\mathbb{P}\{W > s + t\}}{\mathbb{P}\{W > s\}} = \frac{e^{-\theta s - \theta t}}{e^{-\theta s}} = e^{-\theta t}.$$

This is equivalent to (10.4).

**Solution to Exercise 10.1.2.** Using the triangle inequality and submultiplicative property of the matrix norm, we have

$$\left\| \sum_{k=0}^{m} \frac{A^k}{k!} \right\| \leqslant \sum_{k=0}^{m} \frac{\|A^k\|}{k!} \leqslant \sum_{k=0}^{m} \frac{\|A\|^k}{k!} \leqslant e^{\|A\|},$$

where the last term uses the ordinary (scalar) exponential function defined in (10.1). (If you also want to prove that the scalar series in (10.1) converges, you can do so via the ratio test.)

**Solution to Exercise 10.1.3.** Given $A = P^{-1}DP$ we have $A^k = P^{-1}D^k P$ for all $k$, so

$$e^A = \sum_{k \geqslant 0} \frac{A^k}{k!} = \sum_{k \geqslant 0} \frac{P^{-1}D^k P}{k!} = P^{-1} \sum_{k \geqslant 0} \frac{D^k}{k!} P = P^{-1} e^D P.$$

**Solution to Exercise 10.1.4.** We use the definition $e^A = \sum_{k \geqslant 0} \frac{A^k}{k!}$ for the proof and fix $t \in \mathbb{R}$. A common argument for differentiating $e^{tA}$ with respect to $t$ is to take the derivative through the infinite sum to get

$$\frac{\mathrm{d}}{\mathrm{d}t} e^{tA} = \left( A + t \frac{A^2}{1!} + t^2 \frac{A^3}{2!} + \cdots \right) = A e^{tA}.$$

But this is not fully rigorous, since we have not justified interchange of limits. A better answer is to start with (10.9), which gives

$$\frac{\mathrm{d}}{\mathrm{d}t} e^{tA} = e^{tA} \lim_{h \to 0} \frac{e^{hA} - I}{h}.$$

and note that

$$\frac{e^{hA} - I}{h} = A + \frac{1}{2!} h A^2 + \frac{1}{3!} h^2 A^3 + \cdots,$$

which converges to $A$ as $h \to 0$.

**Solution to Exercise 10.1.5.** Fix $A$ in $\mathbb{M}^{n\times n}$ and let $B = -A$. Evidently $AB = BA$, so $e^A e^B = e^{A-A} = e^0$. It is easy to check that $e^0 = I$, so $e^A e^{-A} = I$.

**Solution to Exercise 10.1.6.** Fix $i, j$ with $1 \leqslant i, j \leqslant n$, let $e_k$ be the $k$-th canonical basis vector and let $f$ be the function on $\mathbb{R}$ defined by $f(t) = \langle e_i, e^{tA} e_j \rangle$. Part (v) tells of Lemma 10.1.2 tells us that $f'(t) = \langle e_i, e^{tA} A e_j \rangle$. By the fundamental theorem of calculus, we have $f(t) - f(s) = \int_s^t f'(\tau)\,\mathrm{d}\tau$, or

$$\langle e_i, e^{tA} e_j \rangle - \langle e_i, e^{sA} e_j \rangle = \int_s^t \langle e_i, e^{\tau A} A e_j \rangle \,\mathrm{d}\tau.$$

As this is true for any $i$,, we have $e^{tA} - e^{sA} = \int e^{\tau A} A \,\mathrm{d}\tau$, which is what we need to prove.

**Solution to Exercise 10.1.7.** If we take $u_t = \varphi_t^\top$ and transpose $\dot{\varphi}_t = \varphi_t P$ we get (10.10) when $A = P^\top$. By Proposition 10.1.3, the unique solution is $u_t = e^{tP^\top} u_0 = (e^{tP})^\top u_0$. Transposing again gives $\varphi_t = \varphi_0 e^{tP}$, as was to be shown.

**Solution to Exercise 10.1.9.** For any $\tau > 0$, we have

$$\tau s(A) = \tau \max_{\lambda \in \sigma(A)} \operatorname{Re}\lambda = \max_{\tau\lambda \in \sigma(\tau A)} \tau \operatorname{Re}\lambda = s(\tau A). \tag{10.18}$$

**Solution to Exercise 10.1.10.** With $\xi := \max_{\lambda \in \sigma(A)} \operatorname{Re}\lambda$, we have

$$\rho(e^A) = \max_{\lambda \in \sigma(e^A)} |\lambda| = \max_{\lambda \in \sigma(A)} |e^\lambda| = \max_{\lambda \in \sigma(A)} e^{\operatorname{Re}\lambda} = e^\xi.$$

(The second equality is by Lemma 10.1.2.) Hence $\rho(e^A) = e^{s(A)}$, as was to be shown.

**Solution to Exercise 10.1.11.** For $t \in \mathbb{N}$, we have

$$\frac{1}{t} \ln \|e^{tA}\| = \ln\left(\|e^{tA}\|^{1/t}\right) = \ln\left(\|(e^A)^t\|^{1/t}\right).$$

Taking the limit $t \to \infty$ and applying Gelfand's lemma, this sequence converges to $\ln \rho(e^A)$. But $\ln \rho(e^A) = s(A)$, by the first equality in (10.17). This proves the second equality in (10.17).

**Solution to Exercise 10.1.12.** Let's start with (i) $\implies$ (ii), or $s(A) < 0$ implies $\|e^{tA}\| \to 0$ as $t \to \infty$.

Here is one proof that works for $t \in \mathbb{N}$ and $t \to 0$. Observe that, since $(e^A)^t = e^{tA}$, the powers $B^t$ of $B := e^A$ match the flow $t \mapsto e^{tA}$ at integer times. We have $B^t \to 0$ if and only if $\rho(B) < 1$. But, by Lemma 10.1.4, $\rho(B) = \rho(e^A) = e^{s(A)}$. Hence $\rho(B) < 1$ is equivalent to $s(A) < 0$. Thus, $s(A) < 0$ is the exact condition we need to obtain $B^t = e^{tA} \to 0$.

We can improve on this proof of (i) $\implies$ (ii) by allowing $t \in \mathbb{R}$ and $t \to \infty$ as follows. Suppose $s(A) < 0$. Fix $\varepsilon > 0$ such that $s(A) + \varepsilon < 0$ and use (10.17) to obtain a $T < \infty$ such that $(1/t) \ln \|e^{tA}\| \leqslant s(A) + \varepsilon$ for all $t \geqslant T$. Equivalently, for $t$ large, we have $\|e^{tA}\| \leqslant e^{t(s(A)+\varepsilon)}$. The claim follows.

That (iii) implies (iv) is immediate: Just substitute the bound in (iii) into the integral.

**Solution to Exercise 10.1.15.** Recalling that, for matrix exponentials, $e^{A+B} = e^A e^B$ whenever $AB = BA$, we have

$$e^{tQ} = e^{t\theta(K-I)} = e^{-t\theta I} e^{t\theta K} = e^{-t\theta}\left(I + t\theta K + \frac{(t\theta)^2}{2!}K^2 + \cdots\right).$$

It is clear from this representation that all entries of $P_t = e^{tQ}$ are nonnegative.

**Solution to Exercise 10.1.16.** Since $f(t) := \delta_x e^{tQ} \mathbb{1} = 1$ for all $t \geqslant 0$, we have $f'(t) = 0$. Recalling that $(e^{tQ})' = e^{tQ}Q$, this means that

$$\frac{d}{dt}\delta_x e^{tQ} \mathbb{1} = \delta_x \frac{d}{dt} e^{tQ} \mathbb{1} = \delta_x e^{tQ} Q \mathbb{1} = 0$$

for all $t \geqslant 0$. Evaluating at $t = 0$, we get $\delta_x Q \mathbb{1} = 0$. That is, $\sum_{x'} Q(x, x') = 0$.

**Solution to Exercise 10.1.17.** By Lemma 10.1.2, we have

$$\frac{d}{dt}e^{tQ} = Qe^{tQ} = e^{tQ}Q \quad \text{for all} \quad t \geqslant 0. \tag{10.23}$$

Evaluating (10.23) at $t = 0$ and recalling that $e^0 = I$ gives

$$Q = \lim_{h \downarrow 0} \frac{1}{h}\left(e^{hQ} - I\right). \tag{10.24}$$

Interpreting $\delta_x$ as a row vector and $\delta_{x'}$ as a column vector, while using the fact that

$x \neq x'$ combined with (10.24), we obtain

$$Q(x, x') = \delta_x Q \delta_{x'} = \delta_x \left[ \lim_{h \downarrow 0} \frac{e^{hQ}}{h} \right] \delta_{x'} = \lim_{h \downarrow 0} \delta_x \frac{e^{hQ}}{h} \delta_{x'}.$$

Hence we need only show that the $\delta_x e^{hQ} \delta_{x'} \geqslant 0$. By (ii), $\delta_x e^{hQ}$ is a distribution, so the inequality holds.

**Solution to Exercise 10.1.19.** Using the matrix exponential (10.6) and $P_t = e^{tQ}$ yields

$$P_t(x, x') = \mathbb{1}\{x = x'\} + tQ(x, x') + t^2 \frac{Q^2(x, x')}{2!} + t^3 \frac{Q^3(x, x')}{3!} + \cdots$$

Setting $t = h$ and using $o(h)$ to capture terms converging to zero faster than $h$ as $h \downarrow 0$ recovers (10.28).

**Solution to Exercise 10.2.2.** Fix $v \in \mathbb{R}^X$. Policy $\sigma$ is $v$-max-greedy for $\mathcal{A}$ if and only if $T_\sigma v \geqslant T_\tau v$ for all $\tau \in \Sigma$, which in turn holds if and only if

$$r(x, \sigma(x)) + \sum_{x'} v^*(x') Q(x, \sigma(x), x') + (1 - \delta) v^*(x)$$

$$= \max_{a \in \Gamma(x)} \left\{ r(x, a) + \sum_{x'} v^*(x') Q(x, a, x') \right\} + (1 - \delta) v^*(x)$$

for all $x \in X$. Canceling terms, this reduces to

$$r(x, \sigma(x)) + \sum_{x'} v^*(x') Q(x, \sigma(x), x') = \max_{a \in \Gamma(x)} \left\{ r(x, a) + \sum_{x'} v^*(x') Q(x, a, x') \right\}$$

for all $x \in X$, which is equivalent to the definition of $v^*$-greedy for $\mathcal{C}$ in (10.50).

# Author Index

# Subject Index